\documentclass[10pt,a4paper,twoside]{report}
\pdfoutput=1
\makeatletter
\title{Leptonic CP Violation and its Origin} \let\Title\@title
\author{Iv\'{a}n Esteban Mu\~{n}oz} \let\Author\@author
\makeatother

\usepackage{geometry} 

\usepackage[T1]{fontenc} 
\usepackage[utf8]{inputenc}
\usepackage{lmodern}
\usepackage[main=UKenglish, spanish]{babel} 

\usepackage{amsmath}
\usepackage{amsfonts}
\usepackage{amssymb}
\usepackage{braket}
\usepackage{mathtools} 
\usepackage{slashed}
\usepackage{bbm} 
\newcommand{\parenbar}[1]{\protect\overset{
            \raisebox{-1.059pt}[0pt][0pt]{\scalebox{.3}{\textbf{(}}}
            \raisebox{-0.706pt}[0pt][0pt]{{\hspace{.03em}\rule{3.594pt}{0.297pt}\hspace{.03em}}}
            \raisebox{-1.059pt}[0pt][0pt]{\scalebox{.3}{\textbf{)}}}} {#1}}

\newcommand{\Det}{\operatorname{Det}}
\newcommand{\Tr}{\operatorname{Tr}}
\renewcommand{\Re}{\operatorname{Re}}
\renewcommand{\Im}{\operatorname{Im}}
\newcommand{\diag}{\operatorname{diag}}
\newcommand{\dimension}{\operatorname{dim}}
\newcommand{\Ph}{\operatorname{Ph}}

\makeatletter
\renewcommand*\env@matrix[1][*\c@MaxMatrixCols c]{%
  \hskip -\arraycolsep
  \let\@ifnextchar\new@ifnextchar
  \array{#1}}
\makeatother

\def\NOvA/{\texorpdfstring{NO$\nu$A}{NOvA}}

\usepackage{sectsty} 
\allsectionsfont{\sffamily}

\usepackage{csquotes} 
\usepackage{cite} 
\usepackage[nottoc]{tocbibind} 
\addto{\captionsUKenglish}{%
  
} 
\usepackage{url} 

\makeatletter
\DeclareRobustCommand\recite[1]{\begingroup\@fileswfalse\cite{#1}\endgroup}
\makeatother

\usepackage{hyperref} 
\usepackage[spanish, english, capitalise]{cleveref} 
\usepackage{graphicx} 
\usepackage{subcaption} 
\usepackage{afterpage}
\newenvironment{pagefigure}{\begin{figure}[!p]}{\afterpage{\clearpage}\end{figure}} 

\usepackage{booktabs} 
\usepackage{multirow} 
\usepackage{rotating} 

\usepackage{siunitx}

\usepackage{fancyhdr} 

\usepackage{epigraph} 
\hypersetup
{	
	pdfauthor={\Author},
	pdftitle={\Title},
	pdfsubject={Leptonic CP Violation},
	pdfkeywords={Particle Physics} {CP Violation} {Neutrinos} {Beyond the Standard Model},
}

\usepackage{xcolor}
\definecolor{myRed}{RGB}{255,16,16}
\definecolor{myBlue}{RGB}{51,102,255}
\definecolor{myGreen}{RGB}{0,128,0}

\let\origdoublepage\cleardoublepage
\newcommand{\clearemptydoublepage}{%
  \clearpage
  {\pagestyle{empty}\origdoublepage}%
}

\begin{document}

\newgeometry{bottom=1cm,hmarginratio=1:1}
\begin{titlepage}

	\centering 
	
	{\scshape 
	
	\vspace*{\baselineskip} 
	
	
	\rule{\textwidth}{1.6pt}\vspace*{-\baselineskip}\vspace*{2pt} 
	\rule{\textwidth}{0.4pt} 
	
	\vspace{0.75\baselineskip} 
	
	{\Huge Leptonic CP Violation \\ and its Origin\\} 
	
	\vspace{0.75\baselineskip} 
	
	\rule{\textwidth}{0.4pt}\vspace*{-\baselineskip}\vspace{3.2pt} 
	\rule{\textwidth}{1.6pt} 
	
	\vspace{2\baselineskip} 
	
	
	\vspace*{2\baselineskip} 
	
			
	{\huge Iván Esteban Muñoz \\}
	
	\vspace*{4\baselineskip} 
	
	PhD advisor:\\
	\vspace*{0.2cm}
	{\Large
	 Prof. Dr. Mar\'ia Concepci\'on Gonz\'alez Garc\'ia\\
	}
	
	\vfill 
	
	
\begin{figure}[b]
\centering
\includegraphics[width=0.5\textwidth]{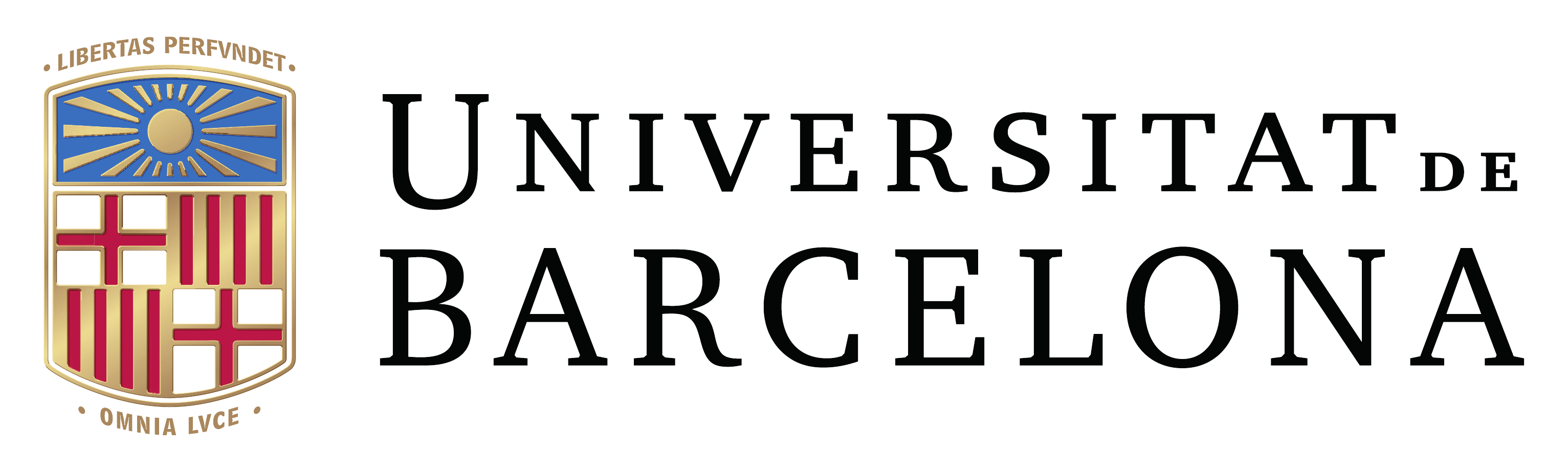}
\end{figure}
}
\end{titlepage}
\restoregeometry
\thispagestyle{empty}
\begin{center}
 
\vspace*{0.9cm}
{\Huge
 \textbf{Leptonic CP Violation \\ and its Origin}\\
}
\vspace*{1.2cm}
{\large Memoria presentada para optar al grado de doctor por la Universidad de Barcelona \\}
\vspace*{0.2cm}
{\LARGE (Programa de Doctorado en Física)\\}
\vspace*{1.2cm}
{\Large Autor:
Iván Esteban Muñoz \\
}
{\Large Directora:
Prof. Dr. Mar\'ia Concepci\'on Gonz\'alez Garc\'ia\\
}
{\Large Tutor:
Prof. Dr. Domènec Espriu Climent\\
}
\vspace*{1.2cm}

{\large Departament de Física Quàntica i Astrofísica\\
Institut de Ci\`encies del Cosmos\\
Universitat de Barcelona\\
}

\begin{figure}[hbtp]
\centering
\includegraphics[width=0.3\textwidth]{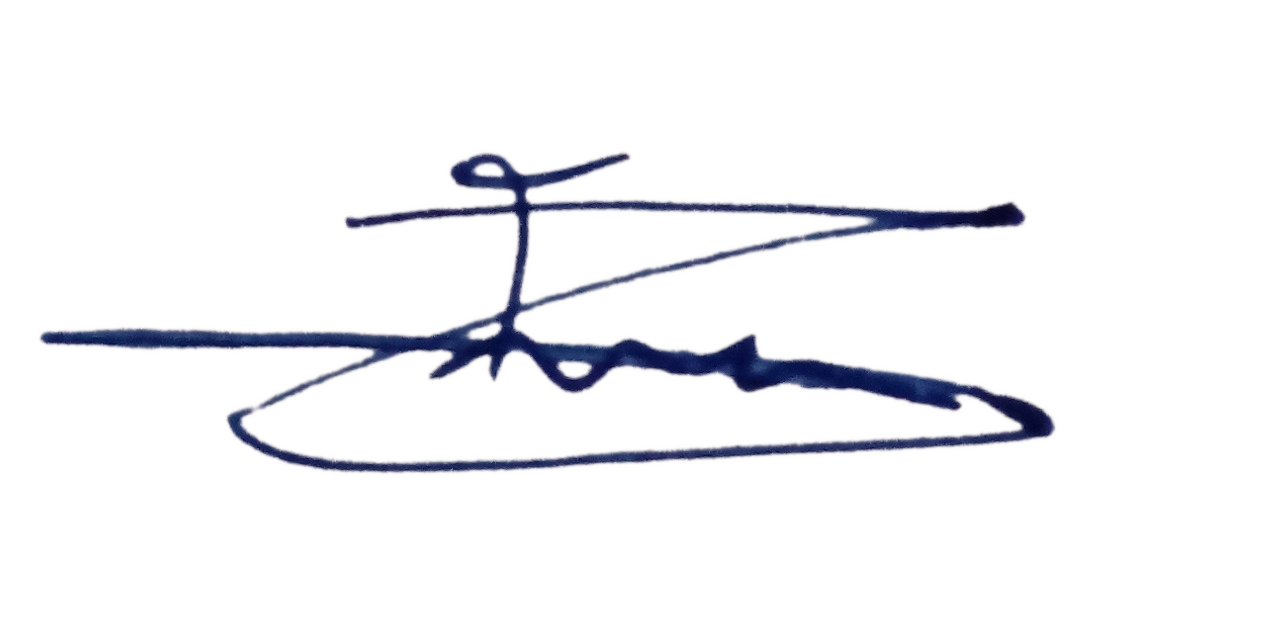}
\end{figure}

\begin{figure}[b]
\centering
\includegraphics[width=0.5\textwidth]{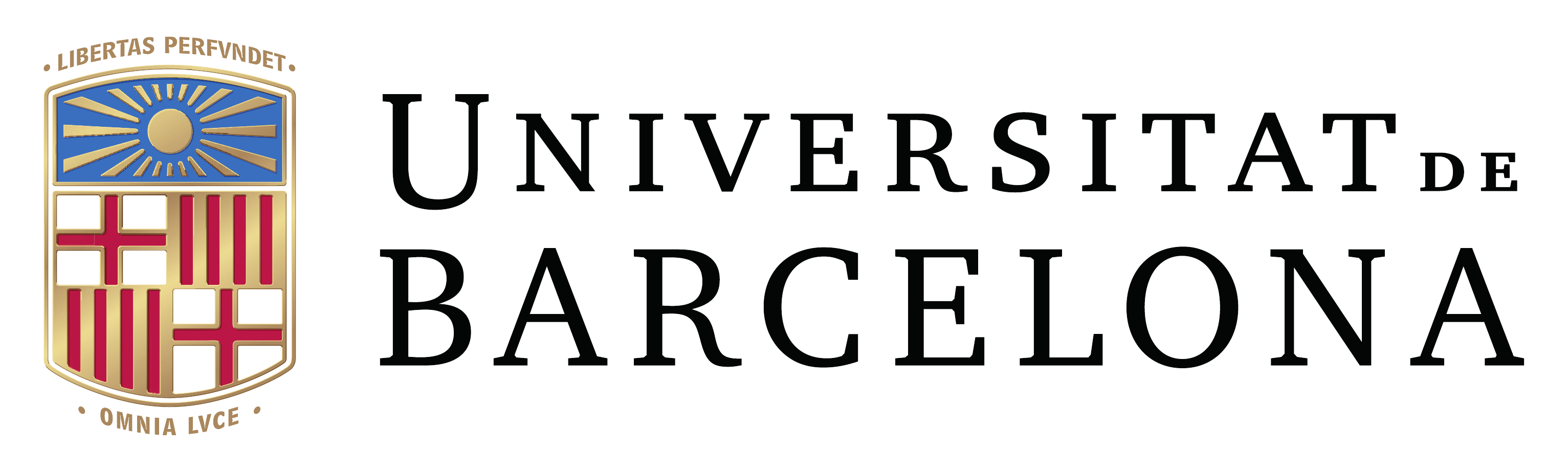}
\end{figure}\end{center}

\newpage
\thispagestyle{empty}

\cleardoublepage

\pagenumbering{roman}
\chapter*{Resumen}
\addcontentsline{toc}{chapter}{Resumen}  

\begin{otherlanguage}{spanish}
Esta tesis se centra en la cuantificación y caracterización de los
resultados experimentales que apuntan hacia la existencia de violación de la
simetría CP en el sector leptónico. Se hace especial énfasis en el
origen físico de esta señal, para cuya interpretación se emplean datos
recogidos por diversos experimentos de neutrinos. Los resultados
presentados en esta tesis se basan en trabajos publicados en revistas
internacionales de física de altas energías
(Refs.~\cite{Esteban:2016qun, Esteban:2018ppq, Esteban:2018azc,
  Esteban:2019lfo, Coloma:2019mbs, Baxter:2019mcx}).

El Modelo Estándar es el paradigma actual para describir los
componentes más fundamentales de la naturaleza y sus interacciones.
Para su construcción tanto teórica como experimental, los neutrinos 
desempeñaron un papel importante~\cite{Danby:1962nd,
Hasert:1973cr, Eichten:1973cs, Benvenuti:1975ru}. Pese a ello, los
neutrinos son las partículas elementales más elusivas que se han
logrado detectar. Cuando fueron inicialmente propuestos por Wolfgang Pauli,
para explicar la conservación de la energía en decaimientos nucleares
beta~\cite{Pauli:83282}, se pensó que su detección era imposible. El
motivo es su minúscula sección eficaz de interacción: el poder
penetrante en materia sólida de neutrinos emitidos en decaimientos
beta es del orden de $10^4$ años luz~\cite{Bethe:1934qn}. Por ello, no
fue hasta 26 años después de que fueran propuestos teóricamente
que se pudieron detectar en un experimento subterráneo llevado a cabo
por C.L.~Cowan y F.~Reines~\cite{Cowan:1992xc}.

Las primeras características de los neutrinos medidas experimentalmente 
estaban de acuerdo con las predicciones del Modelo Estándar. Por una 
parte, los
neutrinos se introdujeron en el modelo mínimamente, solo para explicar
sus interacciones. Tal y como se detallará en el \cref{chap:SM},
no existe ninguna manera de dar masa a los neutrinos consistentemente con 
las simetrías del modelo sin introducir nuevos grados de libertad. Por 
tanto, el Modelo Estándar predecía que los neutrinos
tienen una masa exactamente cero, tal y como confirmaban cotas
obtenidas mediante medidas cinemáticas~\cite{Aker:2019uuj}.

Por otra parte, la misma estructura matemática que prohíbe que los neutrinos tengan masa también predice que su sabor (definido como el tipo de leptón cargado generado junto con la producción o detección del
neutrino) debe conservarse. Los primeros experimentos con neutrinos confirmaron asimismo esta predicción~\cite{Danby:1962nd}.

Pero los paradigmas en física fundamental parecen incapaces de
sobrevivir más de cinco décadas. A finales del siglo \textsc{xx}, una serie de  experimentos
que estudiaban neutrinos provenientes del Sol o de rayos cósmicos que
colisionaban contra la atmósfera terrestre mostraron que estas partículas pueden
cambiar su sabor~\cite{Ahmad:2002jz, Ahmad:2002ka, Ahmed:2003kj,
  Aharmim:2005gt, BeckerSzendy:1992hq, Fukuda:1994mc, Fukuda:1998mi},
comportándose de una manera que el Modelo Estándar prohíbe
explícitamente. El camino hacia la comprensión de las propiedades de
los neutrinos, que podría revelar la siguiente estructura subyacente
de la naturaleza, ha guiado desde entonces a miles de científicos. El
escenario al inicio del trabajo de esta tesis estaba impulsado por la
última sorpresa experimental de los neutrinos: los primeros indicios que 
apuntan hacia su fuerte violación de la simetría materia-antimateria~\cite{t2k:ichep2016,
t2k:susy2016,nova:nu2016}.
  
En este sentido, uno de los misterios más profundos de la ciencia
moderna, que podría guiarnos hacia una descripción más precisa de la
naturaleza, es la asimetría materia-antimateria del
Universo~\cite{Cohen:1997mt, Kinney:1997ic}. La antimateria aparece de
forma natural debido a que las interacciones en el Modelo
Estándar están mediadas por partículas de espín 1 con una simetría
\emph{gauge} asociada a ellas. Las partículas cargadas bajo
estas interacciones han de venir descritas por campos de números complejos, ya que
las transformaciones \emph{gauge} modifican su fase. Por tanto, la
cantidad de grados de libertad se duplica. Una
consecuencia inmediata es la existencia de antimateria, es decir,
partículas con propiedades cinemáticas idénticas pero con toda «carga» (asociada tanto a simetrías globales como \emph{gauge}) opuesta. Esto permite que
las partículas y antipartículas se puedan aniquilar eficientemente dando lugar a los mediadores de la interacción.

De acuerdo con el modelo del \emph{Big Bang}, toda la materia en el
Universo estuvo en algún momento en equilibrio térmico, es decir, su
abundancia solamente dependía de su masa. Por consiguiente, en el Universo primitivo debería haber
iguales cantidades de materia que de antimateria, que se
habrían aniquilado dejando un Universo que contendría solamente
radiación. Nuestra propia existencia, por tanto, exige una interacción
que distinga entre materia y antimateria, creando más de la primera
fuera del equilibrio térmico~\cite{Sakharov:1967dj}.

Ingenuamente, distinguir entre materia y antimateria simplemente
requeriría que dicha interacción violara la simetría de conjugación de
carga C. Pero esto no es suficiente, porque las partículas
tienen un grado de libertad interno adicional: la helicidad. Por lo
que si una interacción viola C pero conserva CP --- la acción combinada
de C y la inversión espacial (o paridad) P ---, creará la misma cantidad
de partículas con una helicidad que de antipartículas con la helicidad
opuesta. Como, por lo que sabemos, las partículas elementales no
tienen más grados de libertad internos, distinguir entre materia y
antimateria solamente requiere una interacción que viole C y
CP~\cite{Sakharov:1967dj}. Si bien es relativamente sencillo construir
una teoría que viole C (la interacción débil, por ejemplo, lo hace de manera máxima), 
es algo más elaborado formular una interacción que viole también CP.

En el Modelo Estándar, dicha interacción existe para los quarks
siempre que haya tres generaciones de partículas. No obstante,
su intensidad no es lo suficientemente fuerte como para generar la
asimetría materia-antimateria del Universo~\cite{Lesgourgues:2018ncw}.
Sin embargo, la nueva física que induce las transiciones de sabor de
los neutrinos está mostrando indicios de violar fuertemente CP~\cite{t2k:ichep2016,
t2k:susy2016,nova:nu2016}.
Caracterizar su significación estadística, robustez y origen físico es
el principal objetivo que persigue esta tesis.

Para poner este problema en contexto, en el \cref{chap:SM} se
resumen las propiedades básicas del Modelo Estándar relevantes para
las interacciones de neutrinos y para la violación de CP. Además, se
explora cómo el modelo se puede extender para incluir masas de
neutrinos y mezcla leptónica, y cómo esto explica de forma natural las
transiciones de sabor observadas en los neutrinos. Lo que es más, la
mezcla leptónica entre tres generaciones abre la posibilidad de que
exista violación de CP. Por tanto, las masas de los neutrinos serían
no solo nuestra primera prueba en el laboratorio de que existe física
más allá del Modelo Estándar, sino también una fuente potencialmente
grande de asimetría materia-antimateria.

Pero la física, como toda ciencia, ha de estar fundamentada en datos
experimentales. Para establecer y parametrizar empíricamente el
mecanismo responsable de las transiciones de sabor de los neutrinos,
durante las últimas tres décadas se viene llevando a cabo un vasto
programa experimental (resumido en el \cref{chap:3nufit_theor}). Este comenzó detectando con precisión las transiciones de sabor en neutrinos
solares~\cite{Cleveland:1998nv, Abdurashitov:2002nt, Hampel:1998xg,
  Altmann:2005ix, Hirata:1991ub, Hosaka:2005um, Aharmim:2011vm,
  Gando:2014wjd, Bellini:2011rx} y atmosféricos~\cite{Fukuda:1998mi,
  Ambrosio:2001je, Sanchez:2003rb}.  Estas medidas fueron
posteriormente confirmadas con haces de neutrinos artificiales
provenientes de reactores nucleares~\cite{Abe:2008aa} y aceleradores
de partículas~\cite{Ahn:2001cq, Adamson:2013whj}. Después de que una
serie de experimentos con neutrinos producidos en reactores nucleares
midieran el último ángulo de mezcla
leptónico~\cite{DoubleChooz:2019qbj, Adey:2018zwh, Bak:2018ydk}, había
aún tres preguntas sin respuesta experimental: el octante del ángulo
de mezcla responsable de las transiciones de sabor de neutrinos
atmosféricos, el orden de los autoestados de masa de los neutrinos, y
la existencia y magnitud de la violación de CP leptónica.

Para desentrañar estas incógnitas, particularmente la violación de CP, hay
un amplio programa de experimentos de neutrinos con aceleradores a
largas distancias, tanto presentes~\cite{Ayres:2004js, Abe:2011ks}
como futuros~\cite{Acciarri:2016ooe, Abe:2015zbg}. En estos
experimentos, un haz de protones acelerado choca contra un blanco
fijo, produciendo piones que, tras decaer, dan lugar a un haz de
neutrinos. Este haz de neutrinos, originalmente de sabor
principalmente muónico, se detecta en sabores muónico y electrónico
después de viajar centenares de kilómetros. En particular, el canal de
aparición de neutrinos electrónicos aborda las tres cuestiones
abiertas arriba mencionadas.

Al principio del desarrollo de esta tesis, el experimento de neutrinos
con acelerador a larga distancia \NOvA/ publicó sus primeros
resultados. Para obtener una visión global, el \cref{chap:3nufit_fit}
los combina con los resultados de otros experimentos de neutrinos
relevantes. Con ello, se puede evaluar cuantitativamente el estatus de la mezcla leptónica y de la violación de CP. Asimismo, se comprueba que el límite
gaussiano, que normalmente se da por sentado en análisis estadísticos,
es una buena aproximación para evaluar la significación de la
violación de CP leptónica. Posteriormente, se resume la evolución de
dicha significación conforme los experimentos de neutrinos con
aceleradores han publicado resultados. El capítulo finaliza resumiendo
el estatus global de la mezcla leptónica al finalizar esta tesis. En
general, los resultados de este capítulo están basados en los trabajos
publicados~\cite{Esteban:2016qun} y~\cite{Esteban:2018azc}, así como
en las actualizaciones en las Refs.~\cite{nufit-3.1, nufit-3.2,
  nufit-4.1}.

Según las incógnitas empezaban a clarificarse, los datos apuntaban
hacia una violación de CP máxima. Este indicio, en ligera tensión con los
datos de \NOvA/, está dominado por un exceso de neutrinos electrónicos
en el experimento de neutrinos con acelerador a larga distancia T2K.
Dentro del paradigma de tres neutrinos masivos, y con el resto de
parámetros de mezcla leptónica medidos con precisión en varios
experimentos, el exceso solamente se puede acomodar mediante una
violación de CP grande.

A pesar de esto, tres neutrinos masivos es solamente una extensión mínima del
Modelo Estándar: podría haber otra nueva física enmascarando los
resultados, ya que la violación de CP leptónica aún no se ha medido de
manera directa y concluyente. Por ello, el \cref{chap:NSItheor},
basado parcialmente en la publicación~\cite{Esteban:2019lfo}, reseña
la nueva física que podría afectar a los experimentos de transiciones
de sabor de neutrinos. El escenario menos acotado por otros
experimentos, las interacciones no estándar de corriente neutra entre
neutrinos y materia (\emph{NSI}, de sus siglas en inglés), afecta a
las transiciones de sabor de los neutrinos modificando su dispersión
coherente con el medio atravesado. Las \emph{NSI} podrían introducir
nuevas degeneraciones y fuentes de violación de CP, cuyo formalismo se
resume en ese capítulo. Físicamente, su origen son nuevas
interacciones entre neutrinos y materia mediadas por partículas
potencialmente ligeras. En cualquier caso, este trabajo adopta un
punto de vista agnóstico y estudia las consecuencias fenomenológicas
de los modelos de \emph{NSI} independientemente de su origen.

En el \cref{chap:NSIfit}, basado en las
publicaciones~\cite{Esteban:2018ppq} y~\cite{Esteban:2019lfo}, estos
modelos se confrontan con datos experimentales. Debido a la gran dimensión del
espacio de parámetros, primero se exploran \emph{NSI} que conservan CP
(es decir, sus módulos). Se evalúan las cotas actuales, así como las
sinergias y complementariedad entre diferentes experimentos. Gracias a 
que estos detectan neutrinos con diferentes energías y que han recorrido diferentes distancias, la determinación de los parámetros de mezcla leptónica resulta ser bastante robusta. Por ello, es posible
avanzar un paso más y evaluar la sensibilidad actual a la violación de
CP leptónica suponiendo que existen las \emph{NSI} más genéricas que
violan CP. La violación de CP inducida por las masas de los neutrinos
y por la mezcla leptónica resulta ser bastante robusta, debido a la
gran cantidad de datos de transiciones de sabor de neutrinos
recopilados a lo largo de tres décadas.

En cualquier caso, las \emph{NSI} introducen una degeneración exacta a
nivel de probabilidad de transición de sabor, que reduce
la sensibilidad a la violación de CP leptónica. Aunque hoy en día el
efecto no es muy drástico debido a la limitada sensibilidad
experimental, la siguiente generación de experimentos con aceleradores
a largas distancias~\cite{Acciarri:2016ooe, Abe:2015zbg} pretende
llevar a cabo medidas de precisión que podrían verse severamente afectadas. Debido a ello, sería altamente beneficioso constreñir
independientemente las \emph{NSI}.

En principio, se podrían obtener restricciones fuertes con
experimentos de dispersión de corriente neutra o con medidas de
precisión electrodébiles con leptones cargados. Sin embargo, las
\emph{NSI} afectan a la propagación de los neutrinos modificando la
dispersión coherente con el medio atravesado, un proceso con
transferencia de momento nula; mientras que las medidas de dispersión
electrodébil típicamente se realizan a transferencias de momento
 $\mathcal{O}$(GeV).
Si la partícula mediadora de las \emph{NSI} es suficientemente ligera,
sus efectos estarían suprimidos para transferencias de momento altas,
y se evadirían cotas de otros experimentos~\cite{Farzan:2015doa, Farzan:2015hkd, Babu:2017olk,
  Farzan:2017xzy, Denton:2018xmq, Miranda:2015dra, Heeck:2018nzc}.

Afortunadamente, durante los últimos años el experimento
COHERENT~\cite{Akimov:2015nza} ha proporcionado cotas independientes
sobre las \emph{NSI}. Este experimento emplea el flujo de neutrinos, abundante y que se 
comprende bien, producido mediante decaimiento de piones en reposo en una fuente de espalación de neutrones. COHERENT mide la
dispersión coherente de corriente neutra entre neutrinos y núcleos, un
proceso en el que un neutrino interactúa coherentemente con todo un
núcleo atómico. Esto ocurre cuando el momento intercambiado es del
orden del inverso del tamaño nuclear, $\mathcal{O}(\si{MeV})$. Debido
a las bajas transferencias de momento, este proceso es bastante
sensible a \emph{NSI} inducidas por mediadores
ligeros.
En el \cref{chap:coh},
basado en las publicaciones~\cite{Esteban:2019lfo}
y~\cite{Coloma:2019mbs}, los datos del experimento COHERENT se
analizan e integran en los análisis globales del \cref{chap:NSIfit}.
Esto requiere comprender rigurosamente el procedimiento para analizar
los datos de COHERENT. Se presta particular atención a cómo los
resultados dependen de las suposiciones sobre la señal de fondo en el
experimento, la estructura nuclear, y la respuesta del detector. La
combinación de los datos de COHERENT con los de experimentos de
transición de sabor desvela su papel incipiente en aumentar la
robustez de la interpretación global de los datos experimentales de neutrinos.

Estos primeros resultados podrían mejorar ampliamente si se
incrementase la estadística de la señal y/o si las medidas se llevasen
a cabo con diferentes núcleos, sensibles a diferentes modelos de
\emph{NSI}. Para ello, una instalación futura idónea es la Fuente
Europea de Espalación de Neutrones. Producirá un haz de neutrinos cuya 
intensidad será un orden de magnitud mayor que la de 
COHERENT, y como aún
se encuentra en construcción se podría habilitar espacio para diversos
detectores modernos. Sus perspectivas para acotar \emph{NSI}, basadas
en el trabajo publicado~\cite{Baxter:2019mcx}, también se exploran en
el \cref{chap:NSIfit}.

En resumen, esta tesis estudia el indicio en los datos actuales de
una violación sustancial de CP en el sector leptónico. Primero cuantifica su
significación global, y después procede a verificar su robustez
respecto del marco teórico en el que se interpretan los datos
experimentales. Para ello, los experimentos complementarios de
interacción coherente entre neutrinos y núcleos juegan, y continuarán
jugando en el futuro, un papel importante. Por ello, se aborda el
problema desde una perspectiva global para evaluar rigurosamente si
las medidas punteras de física de sabor leptónico están apuntando
hacia una nueva violación fuerte de una simetría de la naturaleza.
\end{otherlanguage}
\cleardoublepage

\tableofcontents

\cleardoublepage

\pagestyle{fancy}
\fancyhead{} 
\fancyhead[LE]{\nouppercase{\leftmark}}
\fancyhead[RE]{\thepage}
\fancyhead[LO]{\thepage}
\fancyhead[RO]{\Author. \Title}
\cfoot{} 
\pagenumbering{arabic} 

\cleardoublepage
\chapter*{List of publications}
\addcontentsline{toc}{chapter}{List of publications}

The original contents of this thesis are based on the following publications:
\begin{enumerate}
\item I.~Esteban, M.~C. Gonzalez-Garcia, M.~Maltoni, I.~Martinez-Soler, and
  T.~Schwetz, ``{Updated fit to three neutrino mixing: exploring the
  accelerator-reactor complementarity}'',
  \href{http://dx.doi.org/10.1007/JHEP01(2017)087}{{\em JHEP} {\bfseries 01}
  (2017) 087},
\href{http://arxiv.org/abs/1611.01514}{{\ttfamily arXiv:1611.01514 [hep-ph]}}.

\item I.~Esteban, M.~C. Gonzalez-Garcia, M.~Maltoni, I.~Martinez-Soler, and
  J.~Salvado, ``{Updated Constraints on Non-Standard Interactions from Global
  Analysis of Oscillation Data}'',
  \href{http://dx.doi.org/10.1007/JHEP08(2018)180}{{\em JHEP} {\bfseries 08}
  (2018) 180},
\href{http://arxiv.org/abs/1805.04530}{{\ttfamily arXiv:1805.04530 [hep-ph]}}.

\item I.~Esteban, M.~C. Gonzalez-Garcia, A.~Hernandez-Cabezudo, M.~Maltoni, and
  T.~Schwetz, ``{Global analysis of three-flavour neutrino oscillations:
  synergies and tensions in the determination of $\theta_{23}$, $\delta_{CP}$,
  and the mass ordering}'',
  \href{http://dx.doi.org/10.1007/JHEP01(2019)106}{{\em JHEP} {\bfseries 01}
  (2019) 106}, \href{http://arxiv.org/abs/1811.05487}{{\ttfamily
  arXiv:1811.05487 [hep-ph]}}.
  
\item I.~Esteban, M.~C. Gonzalez-Garcia, and M.~Maltoni, ``{On the Determination of
  Leptonic CP Violation and Neutrino Mass Ordering in Presence of Non-Standard
  Interactions: Present Status}'',
  \href{http://dx.doi.org/10.1007/JHEP06(2019)055}{{\em JHEP} {\bfseries 06}
  (2019) 055},
\href{http://arxiv.org/abs/1905.05203}{{\ttfamily arXiv:1905.05203 [hep-ph]}}.

\item D.~Baxter {\em et~al.}, ``{Coherent Elastic Neutrino-Nucleus Scattering at the
  European Spallation Source}'',
  \href{http://dx.doi.org/10.1007/JHEP02(2020)123}{{\em JHEP} {\bfseries 02}
  (2020) 123}, \href{http://arxiv.org/abs/1911.00762}{{\ttfamily
  arXiv:1911.00762 [physics.ins-det]}}.

\item P.~Coloma, I.~Esteban, M.~C. Gonzalez-Garcia, and M.~Maltoni, ``{Improved
  global fit to Non-Standard neutrino Interactions using COHERENT energy and
  timing data}'', \href{http://dx.doi.org/10.1007/JHEP02(2020)023}{{\em JHEP}
  {\bfseries 02} (2020) 023},
\href{http://arxiv.org/abs/1911.09109}{{\ttfamily arXiv:1911.09109 [hep-ph]}}.

\item I.~Esteban, M.~Gonzalez-Garcia and M.~Maltoni,
``{On the effect of NSI in the present determination of the mass ordering}'', \href{http://arxiv.org/abs/2004.04745}{{\ttfamily arXiv:2004.04745 [hep-ph]}}.
\end{enumerate}
 
The following publications have been developed in parallel to the content of this thesis, although they have not been included in this dissertation:

\begin{enumerate}
\item I.~Esteban, J.~Lopez-Pavon, I.~Martinez-Soler and J.~Salvado,
 ``{Looking at the axionic dark sector with ANITA}'',
 \href{http://dx.doi.org/10.1140/epjc/s10052-020-7816-y}{\emph{Eur. Phys. J.} \textbf{C80} no.~3, (2020) 259}, \href{http://arxiv.org/abs/1905.10372}{{\ttfamily arXiv:1905.10372 [hep-ph]}}.
\item M.~Dentler, I.~Esteban, J.~Kopp and P.~Machado, ``{Decaying Sterile Neutrinos and the Short Baseline Oscillation Anomalies}'', \href{http://dx.doi.org/10.1103/PhysRevD.101.115013}{\emph{Phys. Rev. D} \textbf{101} no.~11, (2020) 115013}, \href{http://arxiv.org/abs/1911.01427}{{\ttfamily arXiv:1911.01427 [hep-ph]}}.
\end{enumerate}

\cleardoublepage
\chapter*{List of abbreviations}
\addcontentsline{toc}{chapter}{List of abbreviations}

\begin{table}[htb!]
\centering
\begin{tabular}{llll}
\textbf{BSM} & Beyond the Standard Model &  & \\
\textbf{CE$\boldsymbol{\nu}$NS} & Coherent elastic neutrino-nucleus scattering &  & \\
\textbf{CP} & Charge Parity &  & \\
\textbf{ESS} & European Spallation Source &  & \\
\textbf{IO} & Inverted ordering &  & \\
\textbf{LBL} & Long baseline &  & \\
\textbf{NO} & Normal ordering&  & \\
\textbf{NSI} & Non-Standard Interactions &  & \\
\textbf{NSI-CC} & Charged Current Non-Standard Interactions &  & \\
\textbf{NSI-NC} & Neutral Current Non-Standard Interactions &  & \\
\textbf{QF} & Quenching Factor &  & \\
\textbf{SM} & Standard Model &  & \\
\textbf{SNS} & Spallation Neutron Source &  & \\
\end{tabular}
\end{table}
\cleardoublepage
\chapter{Introduction}
\label{chap:intro}

\epigraph{\emph{I wish to establish some sort of system, not guided by
    chance but by some sort of definite and exact principle.}}{ ---
  Dmitri Ivanovich Mendeleev}

\epigraph{\emph{El microcosmos de la física moderna\\ ---después de
    muerto me basta ser electrón---}}{ --- Blas de Otero}
    
Throughout the centuries, the most ambitious quest of fundamental
physics, and thus of fundamental science, has been to understand
Nature at its most basic level. The crusade began in the 19th century,
interpreting the mass ratios in chemical reactions as indicatives of
an underlying structure --- atoms~\cite{Dalton:1808}. The early atomic
theory was then complemented by scattering experiments that explored
the electromagnetic properties of atoms in the early 20th
century~\cite{Geiger:1909, Geiger:1910, Rutherford:1911}.
Startlingly, these experiments showed the existence of even more
fundamental particles: electrons and a nucleus that, eventually, was
observed to be composed of protons and neutrons.

And thus the picture seemed complete, but the exploration of cosmic
rays led to discovering an unexpected zoo of new particles and
interactions. Once again, their properties were understood in terms of
various underlying structures that were further explored and
eventually checked at particle colliders. And so a consistent picture
emerged during the second half of the 20th century: the Standard Model
of Particle Physics (SM)~\cite{Glashow:1961tr, Weinberg:1967tq, Salam1196,
Fritzsch:1973pi, Gross:1973ju, Weinberg:1973un}.

Neutrinos play a special role in the construction of the SM~\cite{Danby:1962nd,
Hasert:1973cr, Eichten:1973cs, Benvenuti:1975ru}.
As a start, they are the most elusive elementary particles we have 
detected. First proposed
by Wolfgang Pauli to account for energy conservation in nuclear beta
decays~\cite{Pauli:83282}, they were back then deemed as impossible to
detect. The reason was their low interaction cross section: the
penetrating power in solid matter of neutrinos emitted in beta decay
is around $10^4$ light years~\cite{Bethe:1934qn}.  Therefore, it was
26 years after their existence was proposed that neutrinos were first
directly observed in an underground experiment carried out by
C.L.~Cowan and F.~Reines~\cite{Cowan:1992xc}.

The observed properties of neutrinos were in accordance with the
predictions of the SM. On the one hand, neutrinos were
introduced in the model minimally, just to explain their
interactions. As will be discussed in \cref{chap:SM}, there is no way to 
provide neutrinos with a mass consistent with the symmetries of the 
model without
introducing new degrees of freedom. Thus, the SM predicted neutrinos to be exactly massless, as confirmed by
bounds from direct kinematic measurements~\cite{Aker:2019uuj}.

On the other hand, the same mathematical structure forbidding neutrino
masses also predicts that their flavour (identified with the charged
lepton generated along with neutrino production or detection) must be
conserved. Early neutrino experiments confirmed this picture,
too~\cite{Danby:1962nd}.

But paradigms in fundamental physics seem to hardly survive more than
five decades. At the end of the 20th century, experiments studying
neutrinos coming from the Sun and cosmic rays hitting the atmosphere
revealed that these particles can change their
flavour~\cite{Ahmad:2002jz, Ahmad:2002ka, Ahmed:2003kj,
  Aharmim:2005gt, BeckerSzendy:1992hq, Fukuda:1994mc, Fukuda:1998mi},
behaving in a way that the SM explicitly forbids. The
quest for understanding the properties of neutrinos, which may reveal
the next underlying structure of Nature, has since then led thousands
of scientists. The scenario when this thesis was initiated was driven
by the latest experimental surprise that neutrinos provided: the initial 
hints towards their strong violation of the particle-antiparticle 
symmetry~\cite{t2k:ichep2016,
t2k:susy2016,nova:nu2016}.

\subsubsection*{CP violation}

One of the deepest mysteries of current science, which may lead us to
a more precise theory of Nature, is the matter-antimatter asymmetry of
the Universe~\cite{Cohen:1997mt, Kinney:1997ic}.  Antimatter arises
naturally as follows: interactions in the SM are mediated
by spin-1 particles with an associated gauge symmetry.  Describing
particles charged under these interactions requires complex-valued
fields, because gauge transformations modify their phase. Thus, the
amount of degrees of freedom doubles. An immediate consequence is the
existence of antimatter, that is, particles with identical kinematic
properties, but with opposite charges under any gauge interaction. As
a result, particles and antiparticles can efficiently annihilate into
the interaction mediators.

Assuming the Hot Big Bang model, all the matter in the Universe was at
some point in thermal equilibrium, i.e., its abundance only depended
on its mass. Thus, equal amounts of matter and antimatter should have
been present, which by today would have annihilated, leaving a
Universe only filled with radiation. Our own existence therefore calls for
an interaction that distinguishes between matter and antimatter,
creating more of the former out of thermal
equilibrium~\cite{Sakharov:1967dj}.

Naively, distinguishing between matter and antimatter would just
require that the interaction violates the charge conjugation symmetry
C. But this is not enough, because particles have an additional
internal degree of freedom: helicity. And so if an interaction
violates C but conserves CP --- the combined action of C and space
inversion, or parity, P --- it will create the same amount of
particles with a given helicity as of antiparticles with the opposite
helicity. Since, as far as we know, elementary particles have no more
internal degrees of freedom, distinguishing between matter and
antimatter only requires that there must exist an interaction that
violates C and CP~\cite{Sakharov:1967dj}. The weak interaction
maximally violates C, but an interaction violating CP as well is more
involved.

In the SM, such an interaction exists in the quark sector
as long as there are three particle generations. Unfortunately, its
magnitude is not big enough to generate the particle-antiparticle
asymmetry of the Universe~\cite{Lesgourgues:2018ncw}.  However, the
new physics inducing neutrino flavour transitions is currently showing
a hint for rather large CP violation~\cite{t2k:ichep2016,
t2k:susy2016,nova:nu2016}.  Characterising the statistical
significance, robustness and physical origin of that hint is the main
goal that this thesis pursuits.

\subsubsection*{Outline of this thesis}

To put the aforementioned problem in context, \cref{chap:SM}
overviews the basic properties of the SM relevant for
neutrino interactions and CP violation. In addition, it explores how
the model can be extended to include neutrino masses and leptonic
mixing, and how these naturally explain the observed neutrino flavour
transitions.  Interestingly, leptonic mixing among three generations
opens the possibility for CP violation. Thus, neutrino masses would
not only be our first laboratory evidence for physics beyond the
SM (BSM), but also a potentially large source of
particle-antiparticle asymmetry.

But physics, as any science, is to be founded upon experimental data.
In order to empirically pin down and parametrise the mechanism
responsible for neutrino flavour transitions, a vast experimental
programme (overviewed in \cref{chap:3nufit_theor}) has been ongoing for the
last three decades. It began with the precise detection of flavour
transitions in solar~\cite{Cleveland:1998nv, Abdurashitov:2002nt,
  Hampel:1998xg, Altmann:2005ix, Hirata:1991ub, Hosaka:2005um,
  Aharmim:2011vm, Gando:2014wjd, Bellini:2011rx} and
atmospheric~\cite{Fukuda:1998mi, Ambrosio:2001je, Sanchez:2003rb}
neutrinos. These were later confirmed with artificial neutrino beams
coming from nuclear reactors~\cite{Abe:2008aa} and particle
accelerators~\cite{Ahn:2001cq, Adamson:2013whj}. After several nuclear
reactor neutrino experiments measured the last leptonic mixing
angle~\cite{DoubleChooz:2019qbj, Adey:2018zwh, Bak:2018ydk}, three
questions remained experimentally unanswered: the octant of the mixing
angle driving atmospheric neutrino flavour transitions, the ordering
of the neutrino mass eigenstates, and the existence and magnitude of
leptonic CP violation.

To unveil these unknowns, particularly CP violation, there is a rich
programme of present~\cite{Ayres:2004js, Abe:2011ks} and
future~\cite{Acciarri:2016ooe, Abe:2015zbg} long baseline (LBL) 
accelerator
neutrino experiments. In these experiments, a beam of accelerated
protons hits a fixed target, generating pions that, after decaying,
give rise to a neutrino beam. This neutrino beam, originally mostly of
muon flavour, is detected in muon and electron flavours after
travelling for several hundred kilometres. In particular, the electron
neutrino appearance channel addresses the three open questions
mentioned above.

At the beginning of the development of this thesis, the LBL
accelerator neutrino experiment \NOvA/ released its first data. To
obtain a global picture, \cref{chap:3nufit_fit} combines it with the
results of other relevant neutrino experiments. The status of leptonic
mixing and CP violation is quantitatively assessed. In addition, it is
also checked that the Gaussian limit usually assumed in statistical
analyses is a good approximation to evaluate the significance of
leptonic CP violation. Then, the evolution of this significance as
LBL accelerator neutrino experiments kept releasing data is
summarised. The chapter finishes with the global status of leptonic
mixing as of the completion of this thesis. Overall, the results in
this chapter are based on the published works~\cite{Esteban:2016qun}
and~\cite{Esteban:2018azc}, as well as on the updates in
Refs.~\cite{nufit-3.1, nufit-3.2, nufit-4.1}.

As the unknowns start getting clarified, the data points towards
maximal CP violation. This hint, though in slight tension with \NOvA/
data, is driven by an excess of electron neutrino appearance events in
the LBL accelerator experiment T2K. In the three massive
neutrino paradigm and with the other leptonic mixing parameters
accurately measured by different experiments, the excess can only be
accommodated by large CP violation.

Nevertheless, three massive neutrinos is just a minimal extension of
the SM: other new physics could be present, masking the
results as direct leptonic CP violation has not yet been conclusively
observed. Because of that, \cref{chap:NSItheor}, partly based on the
published work~\cite{Esteban:2019lfo}, overviews the possible new
physics entering neutrino flavour transition experiments. The scenario
that is less bounded by other experiments, neutral current
Non-Standard Interactions (NSI) among neutrinos and matter, affects
neutrino flavour transitions by modifying neutrino coherent scattering
with the traversed medium. NSI could introduce new degeneracies and
sources of CP violation, whose formalism is overviewed in that
chapter. Physically, they are generated by new interactions among
neutrinos and matter mediated by potentially light particles.
Nevertheless, this work adopts an agnostic point of view and
studies the phenomenological consequences of NSI models regardless of
their origin.

In \cref{chap:NSIfit}, based on the published
works~\cite{Esteban:2018ppq} and~\cite{Esteban:2019lfo}, these models
are confronted with data. Due to the large parameter space involved, first
just CP-conserving NSI (i.e., their moduli) are explored. Current
bounds are evaluated, as well as the synergies and complementarity
among different experiments. Thanks to the experiments working at
various neutrino energies and travelled distances, the determination
of leptonic mixing parameters is found to be quite robust. Thus, it is
possible to move on and evaluate the current sensitivity to leptonic
CP violation assuming the most generic CP-violating NSI are present.
CP violation induced by neutrino masses and leptonic mixing is found
to be quite robust, due to the large amount of neutrino flavour
transition data collected along three decades.

Nevertheless, NSI introduce a degeneracy that is exact at the flavour
transition probability level. This degeneracy reduces the sensitivity
to leptonic CP violation. Although currently the effect is not
dramatic due to the rather limited experimental sensitivity, the next
generation LBL accelerator
experiments~\cite{Acciarri:2016ooe, Abe:2015zbg} are aimed at
precision measurements that could be severely affected. Because of
that, it would be highly beneficial to independently constraint NSI.

In principle, tight constraints could come from neutral current
scattering experiments or from precise electroweak measurements with
charged leptons.  However, NSI affect neutrino propagation by
modifying coherent scattering with the traversed medium, a zero
momentum transfer process to be compared with the typical
$\mathcal{O}$(GeV) momentum transfers in other experiments. If the
particle mediating NSI is quite light, its effects would be
suppressed for high momentum transfers, and bounds from other
experiments would be avoided~\cite{Farzan:2015doa, Farzan:2015hkd, Babu:2017olk,
  Farzan:2017xzy, Denton:2018xmq, Miranda:2015dra, Heeck:2018nzc}.

Luckily, in the last years the COHERENT
experiment~\cite{Akimov:2015nza} has provided independent constraints
on NSI. This experiment makes use of the well-understood neutrino
flux copiously produced by pion decay at rest in a neutron spallation
source. COHERENT measures neutral current coherent neutrino-nucleus
elastic scattering, a process in which a neutrino interacts coherently
with an entire atomic nucleus.  This happens when the exchanged
momentum is of the order of the inverse nuclear size,
$\mathcal{O}(\si{MeV})$. Due to the low momentum transfers, the
process is quite sensitive to NSI induced by light
mediators.
In \cref{chap:coh}, based on
the published works~\cite{Esteban:2019lfo} and~\cite{Coloma:2019mbs},
the data from the COHERENT experiment is analysed and integrated into
the global analyses from \cref{chap:NSIfit}. This requires a rigorous
understanding of the procedure for analysing COHERENT data. Particular
attention is paid to how the results depend on the assumptions about
the experimental background, nuclear structure, and detector
response. Combining COHERENT data with flavour transition experiments
unveils its incipient role in increasing the robustness of their
interpretation.

These first results could be greatly improved by increasing the
statistics of the signal and/or by performing the measurements with
different nuclei sensitive to different NSI models. For that, the
European Spallation Source is an ideal future facility. It will
produce a neutrino beam one order of magnitude more intense than the
one used at COHERENT, and as it is still under construction there is
potential space for various modern detectors. Its prospects for
bounding NSI, based on the published work~\cite{Baxter:2019mcx}, are
also explored in \cref{chap:coh}.

In summary, this thesis deals with the current experimental hint for
large CP violation in the leptonic sector. It first quantifies its
global significance, and then moves on to checking its robustness
against the framework in which the experimental data is
interpreted. For that, complementary experiments on neutrino-nucleus
coherent scattering play, and will keep on playing in the future, a
significant role.  Thus, a global approach is taken to rigorously
assess whether cutting-edge leptonic flavour measurements are pointing
towards a new strong violation of a symmetry of Nature.

\chapter{The Standard Model and the neutrino path beyond it}
\label{chap:SM}

The current paradigm for understanding the basic constituents of 
Nature, and the laws through which they interact, is the SM of Particle Physics. In this chapter, we will overview the 
formalism of the SM and how it implements CP violation, in order to 
set conventions and unify notation. In addition, as mentioned in the 
Introduction, neutrino flavour transitions constitute the first laboratory 
evidence for BSM physics. We will explore why this is so, 
and how neutrino masses can explain the violation of leptonic flavour, 
which until the late 20th century was considered to be a good symmetry 
of Nature.

\section{The Standard Model of Particle Physics: formalism}
\epigraph{\emph{Just because things get a little dingy at the
    subatomic level doesn't mean all bets are off.}}{ --- Murray
  Gell-Mann} 
  
The mathematical formalism of the SM, Quantum Field Theory, can be 
considered an abstraction of the de Broglie hypothesis: all particles 
have an associated wave. This wave is represented by a field that, as 
any quantum observable, is actually an operator acting on a Hilbert 
space.

The matter content of the SM consists of 45 chiral fermions with 
spin $\frac{1}{2}$. These
fields represent all elementary fermions that have ever been observed:
\begin{itemize}
\item The left-handed ($L$) and right-handed ($R$) electron, muon and
  tau: $e_{L/R}$, $\mu_{L/R}$ and $\tau_{L/R}$. These are the
  \emph{charged leptons}.
\item The left-handed electron, muon, and tau neutrinos: $\nu_{e, L}$,
  $\nu_{\mu, L}$ and $\nu_{\tau, L}$. These are the \emph{neutral
  leptons}.
\item The left-handed and right-handed quarks, each of them in three
  different colours $^i$: the up quark, $u^i_{L/R}$; the down quark,
  $d^i_{L/R}$; the charm quark, $c^i_{L/R}$; the strange quark,
  $s^i_{L/R}$; the top quark, $t^i_{L/R}$; and the bottom quark,
  $b^i_{L/R}$.
\end{itemize}
The interactions among these particles can be constructed by first
noticing that they can be arranged in 15
multiplets of the symmetry group $G_\text{SM} = SU(3)_C
\times SU(2)_L \times U(1)_Y$~\cite{Glashow:1961tr, Weinberg:1967tq, 
  Salam1196, Fritzsch:1973pi, Gross:1973ju, Weinberg:1973un}, where the
subindexes $C,L,Y$ refer to colour, left-handedness, and hypercharge
respectively. The 
particular multiplets and irreducible representations they fall in are 
summarised in \cref{tab:SMgaugegroup}.

\begin{table}[hbtp]
\centering
\begin{tabular}{
@{\extracolsep{0.1cm}}c@{\extracolsep{0.1cm}}
@{\extracolsep{0.1cm}}c@{\extracolsep{0.1cm}} |
@{\extracolsep{0.1cm}}c@{\extracolsep{0.1cm}}
@{\extracolsep{0.1cm}}c@{\extracolsep{0.1cm}}
@{\extracolsep{0.1cm}}c@{\extracolsep{0.1cm}}} \toprule
  \multicolumn{5}{c}{$({\color{myGreen} SU(3)_C},{\color{myRed}
      SU(2)_L})_{\color{myBlue} U(1)_Y}$} \\ \midrule $({\color{myGreen}
    1},{\color{myRed} 2})_{\color{myBlue} -\frac{1}{2}}$ &
  $({\color{myGreen} 3},{\color{myRed} 2})_{\color{myBlue}
    \frac{1}{6}}$ & $({\color{myGreen} 1},{\color{myRed}
    1})_{\color{myBlue} -1}$ & $({\color{myGreen} 3},{\color{myRed}
    1})_{\color{myBlue} \frac{2}{3}}$ & $({\color{myGreen}
    3},{\color{myRed} 1})_{\color{myBlue} -\frac{1}{3}}$ \\ \midrule
  $L_L^e=  \Big(\begin{array}{c} \nu_e \\ e
\end{array}\Big)_L$ &
$Q_L^{1,i}=\Big(\begin{array}{c} u^i\\ d^i\end{array}\Big)_L$ & $e_R$ & $u^i_R$&
    $d^i_R$ \\ $L_L^\mu=\Big(\begin{array}{c} \nu_\mu\\\mu
\end{array}\Big)_L$ &
$Q_L^{2,i}=\Big(\begin{array}{c} c^i\\ s^i\end{array}\Big)_L$ & $\mu_R$ &$
      c^i_R$& $s^i_R$ \\ $L_L^\tau=\Big(\begin{array}{c} \nu_\tau\\\tau
\end{array}\Big)_L$ &
$Q_L^{3,i}=\Big(\begin{array}{c} t^i\\ b^i\end{array}\Big)_L$ & $\tau_R$ &$
        t^i_R$& $b^i_R$ \\ \bottomrule
\end{tabular}
\caption{Multiplets and irreducible representations in which SM
  fermions fall. For the $SU(3)_C$ and $SU(2)_L$ groups, $1$, $2$ and $3$
  denote the singlet, doublet and triplet representations. For $U(1)_Y$,
  a subindex $Y$, known as the hypercharge, indicates that the fermion
  $f$ transforms as $f \rightarrow e^{i Y \theta} f$, where $\theta$
  is a real number. The different fermions are also known as
  \emph{flavours}.}
\label{tab:SMgaugegroup}
\end{table}

When the symmetry transformations in $G_\text{SM}$ are allowed to be
position-dependent, additional Lorentz-vector fields (known as gauge
fields) have to be introduced for each of the subgroups in
$G_\text{SM}$. These fields will then mediate interactions among the
fermions. Under the assumption of renormalisability, i.e., that the
theory should be predictive at any energy scale, the most general
Lagrangian --- from which the Heisenberg equations of motion for the
fields follow --- one can write is
\begin{equation}
  \mathcal{L}_\mathrm{fermion}+\mathcal{L}_\mathrm{gauge}
  = \sum_f \bar f(i\gamma^\mu D_\mu) f -
\frac{1}{4} \sum_i (F_{\mu \nu}^i)^2 \, ,
\label{eq:gaugeL}
\end{equation}
where the sum runs over the 15 fermion multiplets $f$ in
\cref{tab:SMgaugegroup} and $\gamma^\mu$ are the Dirac $\gamma$
matrices. $D_\mu = \displaystyle\partial_\mu - i g_s \sum_i G^i_\mu t^i - i g
\sum_i W^i_\mu \tau^i - i g' Y B_\mu$ is the so-called covariant
derivative, where
\begin{itemize}
\item $g_s$, $g$ and $g'$ are the coupling constants of the
  interactions associated with the groups $SU(3)_C$, $SU(2)_L$ and
  $U(1)_Y$, respectively. The former corresponds to the strong interaction
  among quarks, whereas the others correspond to the electroweak
  interactions.
\item $G^i$, with $i \in \{1, \ldots, 8\}$; $W^i$, with $i \in \{1, 2, 3\}$; and $B$ are,
  correspondingly, the $SU(3)$, $SU(2)$, and $U(1)$  gauge fields that
  mediate the  corresponding interactions.
\item $t^i$, with $i \in \{1, \ldots, 8\}$; and $\tau^i$, with $i \in \{1, 2, 3\}$; are the $SU(3)$ and $SU(2)$
  generators in the representation to which $f$ belongs, and $Y$ is its
  hypercharge. 
\end{itemize}
With these definitions, the electric charge operator $Q$ is given by
\begin{equation}
  Q=\tau_3+Y  \; .
\label{eq:Qdef}  
\end{equation}  
Finally, the last term in the Lagrangian~\eqref{eq:gaugeL} is the square of the gauge field tensor, which we have
compactly denoted as $\displaystyle\sum_i (F_{\mu \nu}^i)^2$, given by
\begin{equation}
\begin{split}
  \sum_i (F_{\mu \nu}^i)^2
  \equiv & \sum_{i=1}^8 (\partial_\mu G_\nu^i -
\partial_\nu G_\mu^i + g_s f^{ijk} G_\mu^j G_\nu^k)^2 + \\ & \sum_{i=1}^3
(\partial_\mu W_\nu^i - \partial_\nu W_\mu^i + g \varepsilon^{ijk}
W_\mu^j W_\nu^k)^2 + (\partial_\mu B_\nu - \partial_\nu B_\mu)^2 \,.
\end{split}
\end{equation} 
$f^{ijk}$ are the structure constants of the $SU(3)$ group, and
in each term on the right-hand side the square of a tensor $T_{\mu \nu}$ is to be understood as the
contraction $T_{\mu \nu} T^{\mu \nu}$.

The SM also contains a scalar field, the Higgs field $\Phi$, a singlet
under $SU(3)$ that transforms as a doublet under $SU(2)$ and has
hypercharge $\frac{1}{2}$. Its gauge invariant Lagrangian reads
\begin{equation}
\mathcal{L}_\mathrm{Higgs} = D_\mu \Phi D^\mu \Phi - \mu^2 |\Phi|^2 -
\lambda |\Phi|^4 \, ,
\label{eq:higgsL}
\end{equation}
where $\lambda$ and $\mu$ are constants.

Finally, the most general renormalisable Lagrangian also contains
Yukawa interactions among the Higgs field and the fermions
\begin{equation}
\mathcal{L}_\mathrm{Yukawa} = - \sum_{\alpha \beta} 
Y^u_{\alpha \beta} \overline{Q_L^\alpha} \tilde{\Phi} u_R^\beta - 
\sum_{\alpha \beta} Y^d_{\alpha \beta} \overline{Q_L^\alpha} 
\Phi d_R^\beta - \sum_{\alpha \beta} Y^e_{\alpha \beta} 
\overline{L_L^\alpha} \Phi e_R^\beta + \mathrm{h.c.}\, ,
\label{eq:yukawaL}
\end{equation}
where $\tilde{\Phi} = i \tau_2 \Phi^*$ and $\tau_2$ is the second 
Pauli matrix, $Q^\alpha_L$ is a quark $SU(2)$ doublet ($\alpha \in 
\{1, 2, 3\}$), $u_R^\beta$ is 
a right-handed up-type quark ($\beta \in\{1,2,3\}\equiv \{u, c, t\}$), $d_R^\beta$ is 
a right-handed down-type quark ($\beta \in\{1,2,3\}\equiv \{d, s, b\}$),
  $L^\alpha_L$ 
is a lepton $SU(2)$ doublet ($\alpha \in \{e, \mu,\tau\}$), and 
$e_R^\beta$ is a right-handed charged 
lepton ($\beta \in \{e, \mu,\tau\}$). $Y^u$, $Y^d$ and $Y^e$ are 
$3 \times 3$ matrices, and $\mathrm{h.c.}$ refers to the Hermitian 
conjugate.

The model presented above, built throughout more than 50 years as a
myriad of experimental data suggested and confirmed its different
pieces, is currently able to explain almost all experimental results
in the laboratory. As we shall see, though, data from neutrino
experiments challenge it as the final description of Nature.

\subsection{Spontaneous symmetry breaking: fermion masses and mixing}
\label{sec:ssb}

Before immersing in the experimental data that defies the validity of
the SM, it is important to understand the consequences of its scalar
sector~\cite{Higgs:1964pj, Englert:1964et, Guralnik:1964eu}. The 
Lagrangian in \cref{eq:higgsL} corresponds to a scalar potential 
energy density
\begin{equation}
V(\phi) = \mu^2 |\Phi|^2 + \lambda |\Phi|^4 \, ,
\end{equation}
that, for $\mu^2 < 0$, does not have its minimum at $\Phi = 0$. On the
contrary, there exist a continuous of degenerate minima fulfilling
\begin{equation}
|\Phi_\mathrm{min}|^2 = - \frac{\mu^2}{2 \lambda} \equiv \frac{v^2}{2}
\, .
\end{equation}
All these minima are related by a $G_{SM}$ transformation, so when choosing a
particular one above  which the theory is quantised, the $G_{SM}$ symmetry is
\emph{spontaneously}  broken.

As seen in Eq.~\eqref{eq:Qdef}, the electromagnetic symmetry group, $U(1)_{em}$,
is part of $G_{SM}$ and, since electromagnetism is a good symmetry of
Nature to all tested energies, the vacuum of the theory must leave that
subgroup unbroken. With the chosen representation for $\Phi$ this
is accomplished for
$\Phi_\mathrm{min} = \frac{1} {\sqrt{2}}\begin{pmatrix}0 & v\end{pmatrix}^T$.
So the quantum field for the scalar doublet can be written as 
\begin{equation}
  \Phi(x) = \frac{1}{\sqrt{2}}
\begin{pmatrix} 
0 \\ v + h(x)
\end{pmatrix}\, {\scalebox{2}{e}}^{\displaystyle -i\sum_{i=1}^{3}\,\tau_i\, \xi_i(x)} \, ,
\end{equation}
where the three real fields $\xi_i$ are the would-be Goldstone bosons
associated with the generators of the broken subgroup. Since they can be
gauged away by an $SU(2)_L$ gauge transformation, 
the only quantum field operator whose excitations can be
detected as particles is $h(x)$, the field of the Higgs boson particle
discovered in  2012~\cite{Chatrchyan:2012xdj, Aad:2012tfa}.
In what follows, we choose to work in the so-called unitary gauge~\cite{Weinberg:1973ew},
where the doublet scalar field contains only the physical particle degrees
of freedom and can be written as $\Phi(x)=\frac{1} {\sqrt{2}}\begin{pmatrix}0 \\ v+h(x)\end{pmatrix}$.

Rewriting the entire Lagrangian in terms of $v$ and $h$ yields a very
rich phenomenology. Regarding \cref{eq:higgsL}, the covariant
derivative term generates masses for three of the four gauge bosons in
the $SU(2)_L \times U(1)_Y$ part of $G_\mathrm{SM}$: the $W$ bosons,
$W^\pm_\mu \equiv \frac{1}{\sqrt{2}} (W^1_\mu \mp i W^2_\mu)$, which
acquire a mass $m_W = \frac{1}{2} g v$; and the $Z$ boson, $Z_\mu \equiv
\cos \theta_W W^3_\mu - \sin \theta_W B_\mu$, which accquires a mass $m_Z =
\frac{1}{2} \frac{g v}{\cos \theta_W}$. Here, $\theta_W = \arctan
\frac{g'}{g}$ is the so-called Weinberg angle. A fourth linear
combination, $A_\mu \equiv \sin \theta_W W^3_\mu - \cos \theta_W B_\mu$,
remains massless and corresponds to the photon.

Once the gauge bosons are written in the mass basis, \cref{eq:gaugeL}
gives their interactions with fermions. The interactions involving $W$
and $Z$ bosons, the only mediators that interact with neutrinos, are
given by
\begin{equation}
\begin{split}
\mathcal{L}^\mathrm{weak}_\mathrm{int} = & - \frac{g}{\sqrt{2}}
W_\mu^+ \begin{pmatrix} \bar{\nu}_{e, L} & \bar{\nu}_{\mu, L}
  & \bar{\nu}_{\tau, L}
\end{pmatrix} \gamma^\mu \begin{pmatrix} e_L \\ \mu_L \\ \tau_L
\end{pmatrix} - \frac{g}{\sqrt{2}} W_\mu^+ \begin{pmatrix}
\bar{u}_L & \bar{c}_L & \bar{t}_L
\end{pmatrix} \gamma^\mu \begin{pmatrix}
d_L \\ s_L \\ b_L
\end{pmatrix} + \mathrm{h.c.} \\
& - \frac{g}{\cos \theta_W} Z_\mu \sum_f \left[ \left(T_3 - Q_f \sin^2
  \theta_W \right) \bar{f}_L \gamma^\mu f_L - Q_f \sin^2 \theta_W
  \bar{f}_R \gamma^\mu f_R \right] \, ,
\end{split}
\label{eq:CCandNC}
\end{equation}
where $T_3$ is the 3 component of the \emph{weak isospin} of the
fermion $f$ (i.e., its eigenvalue for $\tau_3$ of $SU(2)_L$)
and $Q_f = T_3 + Y$ its
electric charge. The first two terms are the \emph{charged current}
interactions, and the last ones the \emph{neutral current}
interactions.

Finally, \cref{eq:yukawaL} generates, after spontaneous symmetry
breaking, masses for all charged fermions. Defining the following
matrices
\par\vspace{\abovedisplayskip}\noindent
\begin{minipage}{.33333\textwidth}
    \begin{equation}
    M_u \equiv \frac{v}{\sqrt{2}} Y^u \, ,
    \end{equation}
\end{minipage}%
\begin{minipage}{.33333\textwidth}
    \begin{equation}
    M_d \equiv \frac{v}{\sqrt{2}} Y^d \, ,
    \end{equation}
\end{minipage}%
\begin{minipage}{.33333\textwidth}
    \begin{equation}
    M_e \equiv \frac{v}{\sqrt{2}} Y^e \, ,
    \end{equation}
\end{minipage}
\par\vspace{\belowdisplayskip}\noindent
the terms in \cref{eq:yukawaL} that do not involve the Higgs $h$ read
\begin{equation}
\begin{split}
\mathcal{L}_\mathrm{Yukawa}^0 = & - \begin{pmatrix}
\bar{u}_L & \bar{c}_L & \bar{t}_L
\end{pmatrix} M_u \begin{pmatrix}
u_R 	\\ c_R \\ t_R
\end{pmatrix} - \begin{pmatrix} \bar{d}_L
  & \bar{s}_L & \bar{b}_L
\end{pmatrix} M_d \begin{pmatrix}
d_R \\ s_R \\ b_R
\end{pmatrix} \\
& - \begin{pmatrix} \bar{e}_L & \bar{\mu}_L & \bar{\tau}_L
\end{pmatrix} M_e \begin{pmatrix}
e_R \\ \mu_R \\ \tau_R
\end{pmatrix} + \mathrm{h.c.} \, .
\end{split}
\end{equation}
After unitary diagonalisation, these terms lead to fermion
masses. Since $M$ are not in general Hermitian matrices, two
distinct unitary matrices $V_L$ and $V_R$ are needed to diagonalise them,
\begingroup
\allowdisplaybreaks
\begin{align}
V_L^u M_u V_R^{u \dagger} & = \begin{pmatrix} m_u & 0 & 0 
\\ 0 & m_c & 0 \\ 0 & 0 & m_t
\end{pmatrix} \, , \\
V_L^d M_d V_R^{d \dagger} & = \begin{pmatrix} m_d & 0 & 0 
\\ 0 & m_s & 0 \\ 0 & 0 & m_b
\end{pmatrix} \, , \\
V_L^e M_e V_R^{e \dagger} & = \begin{pmatrix} m_e & 0 & 0 
\\ 0 & m_\mu & 0 \\ 0 & 0 & m_\tau
\end{pmatrix} \, .
\end{align}
\endgroup
After these transformations, the mass eigenstates are given by\footnote{In
 order to make the expressions more compact, we
  often write the vector contaning three fields as a \emph{transposed} 3-vector.
  Notice that the transpose operation acts on the 3-vector space but not
  on the field components, this is we denote
  $\begin{pmatrix}\psi_1&\psi_2&\psi_3\end{pmatrix}^T
    \equiv\begin{pmatrix}\psi_1\\\psi_2\\\psi_3\end{pmatrix}$.}
\begin{align}
\begin{pmatrix}
u & c & t
\end{pmatrix}_\mathrm{mass}^T &= V_L^u \begin{pmatrix}
u_L & c_L & t_L
\end{pmatrix}^T + V_R^u \begin{pmatrix}
u_R & c_R & t_R
\end{pmatrix}^T \, ,\\
\begin{pmatrix}
d & s & b
\end{pmatrix}_\mathrm{mass}^T &= V_L^d \begin{pmatrix}
d_L & s_L & b_L
\end{pmatrix}^T + V_R^d \begin{pmatrix}
d_R & s_R & b_R
\end{pmatrix}^T \, ,\\
\begin{pmatrix}
e & \mu & \tau
\end{pmatrix}_\mathrm{mass}^T &= V_L^e \begin{pmatrix}
e_L & \mu_L & \tau_L
\end{pmatrix}^T + V_R^e \begin{pmatrix}
e_R & \mu_R & \tau_R
\end{pmatrix}^T \, .
\end{align}
Thus, the fermion interaction eigenstates, i.e., the fields that
through the terms in \cref{eq:gaugeL} interact with the gauge bosons;
and the mass eigenstates, i.e., the fields that propagate; are in
principle different. However, for interactions that do not change the
fermion flavour, the interaction term in the mass basis will only get
a factor $V_LV_L^\dagger$ or $V_R V_R^\dagger$, equal to
$\mathbbm{1}_{3 \times 3}$ because of unitarity. Differences
among interaction and mass eigenstates will exclusively appear in
flavour-changing interactions, which in the SM are just the charged
current interactions with the $W$ boson. Indeed, for the quark sector
these interactions read, in the mass basis,
\begin{equation}
\mathcal{L}_{CC}^\mathrm{quark} = - \frac{g}{\sqrt{2}}
W_\mu^+ \begin{pmatrix} \bar{u} & \bar{c} & \bar{t}
\end{pmatrix} V^\mathrm{CKM} \gamma^\mu P_L \begin{pmatrix}
d \\ s \\ b
\end{pmatrix} + \mathrm{h.c.} \, ,
\end{equation}
where $V^\mathrm{CKM} = V_L^u V_L^{d \dagger}$ is the CKM mixing 
matrix~\cite{Cabibbo:1963yz, Kobayashi:1973fv}, and
$P_L = \frac{1 - \gamma_5}{2}$ is a chiral projector.

In the lepton sector, however, there is nothing forbidding the
unphysical rotation
\begin{equation}
\begin{pmatrix}
\nu_{e, L} \\ \nu_{\mu, L} \\ \nu_{\tau, L}
\end{pmatrix} \rightarrow V_L^{e \dagger} \begin{pmatrix}
\nu_{e, L} \\ \nu_{\mu, L} \\ \nu_{\tau, L}
\end{pmatrix} \, ,
\label{eq:neutrinoSymmetry}
\end{equation}
under which the entire interaction Lagrangian~\eqref{eq:gaugeL}
remains invariant even after switching to the mass basis. This
transformation removes any potential mixing in leptonic charged current
interactions~\eqref{eq:CCandNC}. Notice also that the symmetry above is
incompatible with neutrino masses, as mass terms mix left-handed and
right-handed components of fermions.

As we will see,
the fact that the most general renormalisable Lagrangian compatible
with the symmetry $G_\mathrm{SM}$ turns out to be invariant under the
transformation~\eqref{eq:neutrinoSymmetry} is intimately related not
only with neutrino masses, but also with the sources of
particle-antiparticle asymmetry in the leptonic sector.

\subsection{CP violation and global symmetries}
\label{sec:CPviol_SM}
As discussed in the Introduction, the matter-antimatter asymmetry of
the universe calls for an interaction that violates CP, the combined 
action of charged conjugation C and parity P.

Mathematically, the CP conjugate of a Lagrangian changes all
parameters into their complex conjugates. And so to have CP violation
there must be complex parameters in the Lagrangian of the theory,
which in the SM can only be the Yukawa matrices in
\cref{eq:yukawaL}. However, their observability is not immediate
because, as has been noticed for the neutrinos, there exist global
symmetry transformations that can remove these parameters from the
theory.

This problem is better formulated as follows~\cite{Bernabeu:1986fc, 
Gronau:1986xb, Jarlskog:1985cw, Jarlskog:1985ht}: the SM Lagrangian, 
i.e., the most general renormalisable Lagrangian that can be built 
with the SM fermion content and gauge group, is accidentally invariant 
under a
\begin{equation}
SU(3)_{Q_L} \times SU(3)_{u_R} \times SU(3)_{d_R} \times
SU(3)_{L_L} \times SU(3)_{e_R} \times U(1)_{B-L}
\label{eq:SMglobal}
\end{equation} 
global symmetry~\cite{Chivukula:1987py, DAmbrosio:2002vsn}. This 
symmetry, exact both at the classical and quantum levels~\cite{tHooft:1976rip}, acts as 
follows over the fermion fields
\begin{alignat}{3} 
& SU(3)_{Q_L} : \nonumber \\ & \qquad \begin{pmatrix} Q^1_L
    & Q^2_L & Q^3_L
\end{pmatrix}^T && \rightarrow P_{Q_L} \begin{pmatrix}
Q^1_L & Q^2_L & Q^3_L
\end{pmatrix}^T \quad && P_{Q_L} \in SU(3) \, , 
\label{eq:symFerm1}\\
& SU(3)_{u_R} : \nonumber \\ & \qquad \begin{pmatrix} u_R & c_R & t_R
\end{pmatrix}^T && \rightarrow P_{u_R} \begin{pmatrix}
u_R & c_R & t_R
\end{pmatrix}^T \; && P_{u_R} \in SU(3) \, , \\
& SU(3)_{d_R} : \nonumber \\ & \qquad \begin{pmatrix} d_R & s_R & b_R
\end{pmatrix}^T && \rightarrow P_{d_R} \begin{pmatrix}
d_R & s_R & b_R
\end{pmatrix}^T \; && P_{d_R} \in SU(3) \, , \\
& SU(3)_{L_L} : \nonumber \\ & \qquad \begin{pmatrix}
    L^e_L & L^\mu_L & L^\tau_L
\end{pmatrix}^T && \rightarrow P_{L_L} \begin{pmatrix}
L^e_L & L^\mu_L & L^\tau_L
\end{pmatrix}^T \quad && P_{L_L} \in SU(3) \, , \label{eq:symLep1} \\
& SU(3)_{e_R} : \nonumber \\ & \qquad \begin{pmatrix} e_R & \mu_R &
    \tau_R
\end{pmatrix}^T && \rightarrow P_{e_R} \begin{pmatrix}
e_R & \mu_R & \tau_R
\end{pmatrix}^T \; && P_{e_R} \in SU(3) \, , \label{eq:symLep2}\\
& U(1)_{B-L} : \nonumber \\ & \qquad \qquad \qquad \qquad f &&
  \rightarrow e^{i \delta (B-L)} f \; && \delta \in \mathbb{R} \,
  , \label{eq:symFerm3}
\end{alignat}
where $f$ is any fermion, $B$ its baryon number ($\frac{1}{3}$ for
quarks and $0$ for leptons) and $L$ its lepton number ($1$ for leptons
and $0$ for quarks). The symmetry also modifies the Yukawa matrices as
\begin{align}
Y^u & \rightarrow P_{Q_L}^\dagger Y^u P_{u_R} \,
, \label{eq:symYuk1}\\ Y^d & \rightarrow P_{Q_L}^\dagger
Y^d P_{d_R} \, ,\\ Y^e & \rightarrow P_{L_L}^\dagger Y^e
P_{e_R} \, .\label{eq:symYuk3}
\end{align}

Besides, CP acts on the Yukawa matrices as
\begin{equation}
  Y \overset{\text{CP}}{\rightarrow} Y^* \, .
  \label{eq:CPYukawa}
\end{equation} 
And so CP conjugation is physical if and only if the transformation
\eqref{eq:CPYukawa} is not equivalent to a symmetry in \cref{eq:SMglobal}. I.e., CP
conjugation is physical if and only if there is no set of unitary matrices
$\{P_{Q_L}, P_{u_R}, P_{d_R}, P_{L_L}, P_{e_R} \}$
satisfying
\begin{align}
P_{Q_L}^\dagger Y^u P_{u_R} & = Y^{u *}\, ,\\ 
P_{Q_L}^\dagger Y^d P_{d_R} & = Y^{d *}\, ,\\ 
P_{L_L}^\dagger Y^e P_{e_R} & = Y^{e *} \, .
\end{align}
Now, given a matrix $A$, its ``square'' $A A^\dagger$ determines $A$
up to unitary rotations $A \rightarrow A U$. Thus, one can work with
the ``squares'' of the Yukawa matrices, and there is CP conservation
if and only if there exist unitary matrices $\{P_{Q_L},
P_{L_L} \}$ such that
\begin{align}
P_{Q_L}^\dagger \left( Y^d Y^{d \dagger} \right) P_{Q_L} & = 
\left( Y^d Y^{d \dagger}\right)^*\, ,\\ 
P_{Q_L}^\dagger \left( Y^u Y^{u \dagger} \right) P_{Q_L} & = 
\left( Y^u Y^{u \dagger}\right)^*\, ,\\ 
P_{L_L}^\dagger \left( Y^e Y^{e \dagger} \right) P_{L_L} & = 
\left( Y^e Y^{e \dagger}\right)^*\,
. \label{eq:CPviolationSMlepton}
\end{align}

The condition \eqref{eq:CPviolationSMlepton} is trivially satisfied as
any Hermitian matrix can be made real by a unitary rotation, simply
choosing the unitary rotation that diagonalises it. Therefore, in the
SM there is no CP violation in the leptonic sector. This happens 
because in the lepton sector there is a single flavour matrix, $Y^e$,
whereas in the quark sector there are two, $Y^u$ and $Y^d$. Indeed, 
the flavour transformation in \cref{eq:neutrinoSymmetry} is unphysical
because there is no ``neutrino Yukawa matrix'' $Y^\nu$.

Regarding the other two conditions, they can be shown to be equivalent
to~\cite{Bernabeu:1986fc, Gronau:1986xb, Jarlskog:1985cw, 
Jarlskog:1985ht}
\begin{equation}
\Im \Tr \left[\left( Y^d Y^{d \dagger}\right)^2
  \left( Y^u Y^{u \dagger} \right)^2 \left( Y^d Y^{d \dagger}\right)
  \left( Y^u Y^{u \dagger} \right) \right] = 0 \, .
\end{equation}
Since this condition is invariant under the symmetry transformation
\eqref{eq:SMglobal}, we can use it to diagonalise $Y^u$. The condition
then reads
\begin{equation}
(m_b^2 - m_s^2)(m_b^2 - m_d^2)(m_s^2-m_d^2) (m_t^2 - m_c^2)(m_t^2 -
  m_u^2)(m_c^2-m_u^2) J = 0
\label{eq:jarlskogQuark}
\end{equation}
where $J$ is the so-called Jarlskog invariant
\begin{equation}
J = \Im [V^\mathrm{CKM}_{us} V^\mathrm{CKM}_{cb} V^\mathrm{CKM
    *}_{ub} V^\mathrm{CKM *}_{cs}] \, .
\label{eq:jarlskogMatrix}
\end{equation}
And so in the SM there is only CP violation in quark charged currents,
and all CP-violating phenomena are proportional to
$J$. Its experimentally measured value ($J = (3.18 \pm 0.15) \times
10^{-5}$~\cite{Tanabashi:2018oca}), though, is too small to explain
the matter-antimatter asymmetry of the 
Universe~\cite{Lesgourgues:2018ncw}. The SM is thus 
craving for an additional source of CP violation. As we shall see
at the end of this chapter, neutrinos might provide it.

\subsubsection{Particle number conservation laws}

The symmetry group introduced in \cref{eq:SMglobal} should, because of
Noether's theorem, lead to conservation laws~\cite{Noether:1918zz}. 
Notice, though, that some of the symmetry transformations require 
modifying the parameters in the Lagrangian as in 
\crefrange{eq:symYuk1}{eq:symYuk3}. Thus, some of the symmetry 
transformations do not leave the Lagrangian invariant, and therefore 
Noether's theorem does not apply to them.

To obtain the conservation laws stemming from the symmetry group
\eqref{eq:SMglobal}, we have to disentangle the symmetry
transformations that leave the Lagrangian invariant. For that, we
first switch to the fermion mass basis with no
leptonic mixing. In there, flavour-mixing symmetry transformations are
forbidden because of the mass terms, and so the remaining global symmetries
are
\begin{align}
q_\alpha & \rightarrow e^{i \delta^q_\alpha} q_\alpha \, ,\\ l_\alpha
& \rightarrow e^{i \delta^l_\alpha} l_\alpha \, ,\\ \nu_\alpha &
\rightarrow e^{i \delta^l_\alpha} \nu_\alpha \, ,
\end{align}
where $l_\alpha$ is a charged lepton. Now, the quark rephasings
require modifying the CKM matrix
\begin{equation}
V^\mathrm{CKM}_{\alpha \beta} \rightarrow e^{-i (\delta^q_\alpha -
  \delta^q_\beta)} \, ,
\end{equation}
and so the only global symmetries that leave the Lagrangian invariant are
\begin{align}
q_\alpha & \rightarrow e^{i \delta^q} q_\alpha \, \forall \alpha \,
,\\ l_\alpha & \rightarrow e^{i \delta^l_\alpha} l_\alpha \,
, \label{eq:symmetryLepton1}\\ \nu_\alpha & \rightarrow e^{i
  \delta^l_\alpha} \nu_\alpha \,
. \label{eq:symmetryLepton2}\end{align}
Invariance at the quantum level requires $\delta^q + \sum_\alpha
\delta^l_\alpha = 0$~\cite{tHooft:1976rip}, and so we are left with 3
independent conservation laws:
\begin{itemize}
\item The conservation of baryon minus lepton number, $B-L$.
\item The conservation of the difference among any pair of lepton
  flavour numbers. E.g., the conservation of $L_\mu - L_\tau$ and $L_e
  - L_\mu$.
\end{itemize}
Nevertheless, as mentioned in the Introduction, paradigms in fundamental
physics seem to fail after about five decades. And, as we shall see,
at the end of the 20th century some of these conservation laws were
observed to be explicitly violated in laboratory experiments.

\section{Neutrino phenomenology}

\epigraph{\emph{I have done a terrible thing, I have postulated a
    particle that cannot be detected.}}{ --- Wolfgang Pauli}

\epigraph{\emph{In the old days, Pauli considered terrible a single
    undetectable particle. And now, we are not even ashamed by the
    entire supersymmetric spectrum!}}{ --- M.~C.~Gonzalez-García}
    
As we have seen, the structure of the SM dictates several conservation
laws that can be checked in experiments with leptons and, in
particular, with neutrinos. 


Some years after their first detection, it was realised that the
nuclear processes that fuel the Sun would emit copious amounts of
neutrinos that could be detected at Earth~\cite{Fowler:1958zz,
  Cameron:1958vx, Bahcall:1997ha}. Since the energies of solar
processes are not high enough to produce charged leptons other than
electrons, all solar neutrinos should be of electron flavour. Their
observation thus constitutes a direct check of lepton flavour
conservation. When such neutrinos were detected, though, a deficit
with respect to the theoretical predictions was sistematically
observed~\cite{Davis:1968cp, Cleveland:1998nv, Abdurashitov:2002nt,
  Hampel:1998xg, Altmann:2005ix, Fukuda:1996sz, Smy:2003jf,
  Ahmad:2001an, Ahmad:2002jz, Ahmad:2002ka, Ahmed:2003kj,
  Aharmim:2005gt}. Particularly enlightening were the results from the
SNO observatory~\cite{Ahmad:2002jz, Ahmad:2002ka, Ahmed:2003kj,
  Aharmim:2005gt}: when detecting \emph{electron} neutrinos through
charged current interactions a deficit was found. But if neutral
current interactions were used to detect \emph{any} neutrino flavour,
there was no deficit. In other words, the electron neutrinos were
undoubtely changing their flavour as they propagated between their
production point
in the Sun and their detection point on the 
Earth. The leptonic flavours were seen to be violated, a clear sign of
new physics.

Neutrino flavour change had also been observed in experiments
detecting neutrinos
generated when a cosmic ray hits the
atmosphere~\cite{BeckerSzendy:1992hq, Fukuda:1994mc,
  Fukuda:1998mi}. The ratio among muon and electron neutrinos in these
experiments significantly deviated from the expectations. Furthermore,
the deviation was seen to be more important for neutrinos that
travelled for longer distances.

Since the original discoveries, the effect has been independently
confirmed with large statistical significance, both with natural and
artificial neutrino sources~\cite{Cleveland:1998nv, Kaether:2010ag,
  Abdurashitov:2009tn, Hosaka:2005um, Cravens:2008aa, Abe:2010hy,
  Nakano:th, ikeda_motoyasu_2018_1286858, Aharmim:2011vm,
  Bellini:2011rx, Bellini:2008mr, Bellini:2014uqa, Aartsen:2014yll,
  Abe:2017aap, Gando:2013nba, An:2016srz, dc:cabrera2016,
  Adey:2018zwh, Bak:2018ydk, Adamson:2013whj, Adamson:2013ue,
  t2k:vietnam2016, t2k:kek2019, Acero:2019ksn}. The relevant
experiments will be described in \cref{chap:3nufit_theor}.

\subsection{Massive neutrinos}

The experimental results above are in direct contradiction
with the SM. To understand which kind of new physics could induce
them, we recall that lepton flavours are conserved due to the symmetry
in \cref{eq:symmetryLepton1,eq:symmetryLepton2}, that allows to
independently rephase each leptonic flavour. This is not possible in
the quark sector, because both up-type and down-type quarks have
masses, flavour mixing is physical, and each quark flavour cannot be
independently rephased without modifying the Lagrangian.

Correspondingly, if neutrinos are massive leptonic flavours can also
mix, each of the leptonic flavours will no longer be conserved, and
the experiments above may be explained. The mass, however, must be
very tiny, $\lesssim \mathcal{O}\left(\si{eV}\right)$, to avoid
entering in conflict with direct searches~\cite{Aker:2019uuj}.

Intriguingly, though, the field content and gauge symmetry of the SM may also
provide a mechanism for generating neutrino masses provided that
we assume that some new physics must exist above a certain energy
scale $\Lambda$. The Lagrangian introduced in the previous section
is based on the assumption of renormalisability, i.e., the theory
should give definite predictions up to arbitrarily large energy
scales. If we relax this assumption, higher-dimensional operators
made of the SM fields and respecting the SM gauge symmetry will 
enter the Lagrangian suppressed by the scale $\Lambda$ at which the
theory stops being predictive. Interestingly, the lowest order (least
suppressed) effective operator that only contains SM fields and is
consistent with gauge symmetry is~\cite{Weinberg:1979sa}
\begin{equation}
\mathcal{O} = \frac{Z_{\alpha \beta}}{\Lambda} \left(
\overline{L_L^\alpha} \tilde{\Phi} \right) \left( \tilde{\Phi}^T L_L^{\beta C} \right) + \text{h.c.} \, ,
\label{eq:WeinbergOperator}
\end{equation}
where $L_L^{\beta C} \equiv C \overline{L_L^{\beta}}^T = - i
\gamma^2 \gamma^0 \overline{L_L^{\beta}}^T$ is the charge conjugated
field of the lepton doublet $L_L^\beta$ (the transpose affects the
Dirac indices of each spinor in the doublet). This operator violates
total lepton number (and $B-L$) by two units but, being this an
accidental symmetry of the SM, there is no reason why a more
fundamental theory should conserve it.

After spontaneous symmetry breaking, \cref{eq:WeinbergOperator} leads
to
\begin{equation}
\mathcal{O} = \frac{Z_{\alpha \beta}}{2}\frac{v^2}{\Lambda}
\bar{\nu}_{L \alpha} \nu_{L \beta}^c + \text{h.c.} \, ,
\label{eq:WeinbergOperatorAfterSSB}
\end{equation}
a term that induces neutrino masses and leptonic mixing. Furthermore,
the mass eigenstates are Majorana fermions, i.e., $\nu^c_\mathrm{mass} =
\nu_\mathrm{mass}$. Their mass is suppressed by $\Lambda$, that is, the lightness of
neutrinos is naturally explained by the large value of $\Lambda$.

Even though there are a variety of BSM theories that can
generate the operator in \cref{eq:WeinbergOperator}, there is a
particularly simple example. In principle, one can add to the theory
any number of singlets under its gauge group $G_\mathrm{SM}$: since
they do not interact through the Lagrangian in \cref{eq:gaugeL}, they
may have easily evaded detection. These particles would be
right-handed Weyl fermions with no weak hypercharge, commonly known as
sterile neutrinos $\nu_s$. The most general gauge invariant
renormalisable Lagrangian one can add to the SM is 
then~\cite{GonzalezGarcia:2002dz, GonzalezGarcia:2007ib}
\begin{equation}
-\mathcal{L}_{M_\nu} = \sum_{\substack{i=1, \, \ldots, \, m \\ \alpha
    \in \{e,\mu,\tau\}}}Y^\nu_{i\alpha} \bar\nu_{s i}
\tilde{\Phi}^\dagger L_L^\alpha + \frac{1}{2} \sum_{i, j=1}^m M_{N ij}
\bar\nu_{s i} \nu_{s j}^c + \text{h.c.}  \, , \label{eq:SMLbeforeSSB}
\end{equation}
where $Y^\nu$ and $M_N = M_N^T$ are complex $m \times 3$ and $m \times
m$ matrices, respectively.

After electroweak spontaneous symmetry breaking, the first term in
\cref{eq:SMLbeforeSSB} generates at low energy a standard, or Dirac,
mass term,
\begin{equation}
M_{D i\alpha} \bar\nu_{s i} \nu_{\alpha, L} \, ,
\end{equation}
with $M_{D i\alpha} = Y^\nu_{i\alpha} \frac{v}{\sqrt{2}}$. This term
is similar to the one present for the charged fermions, and with no
additional operators gives neutrinos a mass $\sim M_D$.

The second term in \cref{eq:SMLbeforeSSB}, however, is more
interesting. It has the structure of a Majorana mass term, violating
any $U(1)$ charge carried by $\nu_s$; in particular it breaks lepton
number L by two units if L is assigned to $\nu_s$ so as to make the Dirac
mass term L conserving.

The diagonalisation of the whole Lagrangian~\eqref{eq:SMLbeforeSSB}
leads to $3+m$ mass eigenstates $\nu_M$ that are Majorana fermions,
i.e., $\nu^c_M = \nu_M$. In what refers to the mass eigenvalues, two
interesting cases can be distinguished:
\begin{itemize}
\item $M_N = 0$: this option is equivalent to imposing \emph{by hand}
  lepton number conservation in any theory embedding the SM
  (otherwise, even if $M_N = 0$ at tree level, loop corrections from
  new physics could induce $M_N \neq 0$~\cite{Kniehl:1996bd}). It
  allows to rearrange, for $m=3$, the 6 Majorana eigenstates in 3
  Dirac fermions. In this case, there is no natural explanation for
  the lightness of neutrinos, which would require $Y^\nu \lesssim
  10^{-11}$.
\item $M_N \gg M_D$: this is expected in SM extensions such as SO(10)
  GUTs~\cite{Ramond:1979py, GellMann:1980vs, Yanagida:1979as} or
  left-right symmetric models~\cite{Mohapatra:1979ia}, where the new  
  physics scale $\sim M_N$ is much larger than the electroweak scale.
  In this case, the diagonalisation leads to 3 light eigenstates of 
  masses $\sim M_D^2/M_N$, which are mostly left-handed; and $m$ heavy
  eigenstates of masses $\sim M_N$, which are mostly
  right-handed. This naturally explains the lightness of neutrinos
  through what is known as the \emph{see-saw mechanism}: the heavier
  are the heavy states, the lighter are the light ones.
  
  Indeed, if the Lagrangian~\eqref{eq:SMLbeforeSSB} with $M_N \gg M_D$
  is considered, integrating out the heavy eigenstates leads to a
  Lagrangian of the form \eqref{eq:WeinbergOperator} with
  $\Lambda \sim M_N$.
\end{itemize}

\subsection{Charged current interaction Lagrangian}

Besides the mass Lagrangian~\eqref{eq:SMLbeforeSSB}, the charged
current Lagrangian also gets modified by introducing neutrino masses,
leading to flavour mixing as in the quark sector. For $3+m \equiv n$
neutrino mass eigenstates, in the mass basis this Lagrangian reads
\begin{equation}
-\mathcal{L}_{CC} = \frac{g}{\sqrt{2}} \begin{pmatrix} \bar{e}_L &
  \bar{\mu}_L & \bar{\tau}_L \end{pmatrix}
U^\text{lep} \begin{pmatrix} \nu_{1} \\ \nu_{2} \\ \nu_{3}
  \\ \vdots \\ \nu_{n}
\end{pmatrix} W_\mu^+ - \text{h.c.} \, ,
\end{equation}
where $\nu_i \, , \, i \in \{1,\ldots,n\}$, are the neutrino mass
eigenstates and $U^\text{lep}$ is a $3 \times n$ matrix verifying
\begin{equation}
U^\text{lep} U^{\text{lep} \dagger} = \mathbbm{1}_{3 \times 3} \, .
\end{equation}
If there is no new interactions for the charged leptons,
$U^\text{lep}$ can be identified as a $3 \times n$ submatrix of the $n
\times n$ matrix $V^\nu$ relating neutrino flavour and mass
eigenstates. In particular, for the
Lagrangian~\eqref{eq:SMLbeforeSSB}
\begin{equation}
\begin{pmatrix}
\nu_{L 1} & \nu_{L 2} & \nu_{L 3} & \nu_{s 1}^c & \ldots & \nu_{s
  m}^c
\end{pmatrix}^T = V^\nu P_L 
 \begin{pmatrix}
\nu_{1} & \nu_{2} & \nu_{3} & \ldots & \nu_{n}
\end{pmatrix}^T \,  ,
\end{equation} 
with $P_L$ the left-handed projector.

If there is only three Majorana neutrinos, $U^\text{lep}$ is a
$3 \times 3$ unitary matrix usually referred to as the PMNS
matrix~\cite{Pontecorvo:1967fh, Maki:1962mu, Kobayashi:1973fv}.  Using three angles $\theta_{12} ,
\theta_{13} , \theta_{23} \in [0,90^\circ]$ and three phases
$\delta_\text{CP}, \eta_1, \eta_2 \in [0,2\pi)$, it can be
conveniently parametrised as~\cite{Coloma:2016gei}\footnote{This 
expression differs from the usual one (see, e.g., Ref.~\cite{Tanabashi:2018oca}) 
by an overall unphysical rephasing. Its advantage is that vacuum CPT 
transformations, which will be relevant in 
\cref{chap:NSItheor,chap:NSIfit,chap:coh}, are more transparent to 
implement.}

\begin{equation}
\begin{split}
U^\text{lep} & = \begin{pmatrix} 1 & 0 & 0 \\ 0 & c_{23} & s_{23} \\ 0
  & -s_{23} & c_{23}
\end{pmatrix} \begin{pmatrix}
c_{13} & 0 & s_{13}  \\ 0 & 1 & 0 \\ -s_{13}
 & 0 & c_{13}
\end{pmatrix} \begin{pmatrix}
c_{12} & s_{12} e^{-i \delta_{\text{CP}}} & 0 \\ -s_{12} e^{i \delta_\text{CP}} & c_{12} & 0 \\ 0 & 0 & 1
\end{pmatrix} \begin{pmatrix}[c@{\extracolsep{\arraycolsep}}c@{\extracolsep{\arraycolsep}}c] 
e^{i \eta_1} & 0 & 0\\ 0 & e^{i \eta_2} & 0 \\ 0 & 0 & 1
\end{pmatrix} \\[10pt]
&
= \begin{pmatrix}[c@{\extracolsep{1.2\arraycolsep}}c@{\extracolsep{1.2\arraycolsep}}c]
  c_{12} c_{13} & s_{12} c_{13} e^{i \delta_\text{CP}} & s_{13} 
  \\ -s_{12} c_{23}e^{-i \delta_\text{CP}} - c_{12} s_{13} s_{23} &
  c_{12} c_{23} - s_{12} s_{13} s_{23}e^{i \delta_\text{CP}} & c_{13}
  s_{23} \\ s_{12} s_{23}e^{-i \delta_\text{CP}} - c_{12} s_{13} c_{23} & -c_{12} s_{23} - s_{12} s_{13} c_{23} e^{i \delta_\text{CP}} & c_{13} c_{23}
\end{pmatrix} \begin{pmatrix}[c@{\extracolsep{\arraycolsep}}c@{\extracolsep{\arraycolsep}}c] 
e^{i \eta_1} & 0 & 0\\ 0 & e^{i \eta_2} & 0 \\ 0 & 0 & 1
\end{pmatrix} \, ,
\label{eq:PMNS}
\end{split}
\end{equation}
where $c_{i j} \equiv \cos \theta_{ij}$ and $s_{i j} \equiv \sin
\theta_{ij}$. Note that, unlike in the parametrisation of the quark
CKM mixing matrix, there are two new phases $\eta_1$ and
$\eta_2$. These phases appear due to the Majorana mass
term~\eqref{eq:WeinbergOperatorAfterSSB} $\propto \bar{\nu} \nu^*$: because of
it, some phases in $U^\mathrm{lep}$ cannot be absorbed in the neutrino fields.

For three Dirac neutrinos, the parametrisation in
\cref{eq:PMNS} still holds with $\eta_1 = \eta_2 = 0$. Finally, for
three light and $m$ heavy neutrinos stemming from the
Lagrangian~\eqref{eq:SMLbeforeSSB}, $U^\text{lep}$ has the form
\begin{equation}
U^\text{lep} \simeq \begin{pmatrix} \left( 1 - \frac{1}{2}
  M_D^\dagger M_N^{* -1} M_N^{-1} M_D \right) V_l &
  M_D^\dagger M_N^{* -1} V_h
\end{pmatrix} \, 
,
\end{equation}
where $V_l$ and $V_h$ are $3 \times 3$ and $m \times m$ unitary
matrices, respectively. Therefore, if only light neutrinos are
considered a $3 \times 3$ non-unitary mixing matrix is obtained. The
unitarity violation, however, is suppressed by a factor $\sim \left(
M_D/M_N \right)^2$ and can be safely ignored in what follows.

\subsection{Neutrino flavour oscillations}
\label{sec:vacuumOsc}

An immediate phenomenological consequence of introducing neutrino
masses, and consequently lepton flavour mixing, is that the flavour of
neutrinos \emph{oscillates} during their propagation. That is, since
flavour eigenstates are not propagation eigenstates, a neutrino
produced with a given flavour could, after travelling, be detected as
a neutrino of a different flavour.

Getting into more detail, a neutrino flavour eigenstate produced in a
weak interaction process, $\ket{\nu_\alpha}$, will in general be a
superposition of mass eigenstates
\begin{equation}
\ket{\nu_\alpha} = \sum_{i=1}^n U^{\text{lep} *}_{\alpha i}
\ket{\nu_i} \, ,
\end{equation}
where the sum runs over light neutrino mass eigenstates. The origin of
the complex conjugation is that $\ket{\nu_\alpha} \propto
\bar{\nu}_\alpha \ket{0} = \sum_i U^{\text{lep} *}_{\alpha i}
\bar{\nu}_i \ket{0} \propto \sum_i U^{\text{lep} *}_{\alpha i}
\ket{\nu_i}$, where $\nu_\alpha$ is the neutrino field and $\ket{0}$ the 
vacuum.

After travelling for a time $t$, the state of the neutrino will be
\begin{equation}
e^{-i \hat{H} t} \ket{\nu_\alpha} \, ,
\end{equation}
where $\hat{H}$ is the Hamiltonian operator. It is diagonal in the 
mass eigenstate basis, and for ultra-relativistic neutrinos,
\begin{equation}
e^{-i \hat{H} t} \ket{\nu_\alpha} =
e^{i E L} \sum_{i=1}^n U^{\text{lep} *}_{\alpha i} e^{-i \frac{m_i^2 L}{2E}}
\ket{\nu_i} \, ,
\label{eq:vacuumHamiltonian}
\end{equation}
to lowest order in $\frac{m_i^2}{E}$. $m_i$ is the mass of the $i$-th eigenstate,
$L = ct$ is the travelled distance, and E is the average energy of the 
neutrino wave packet~\cite{Akhmedov:2009rb, Cohen:2008qb}.

Therefore, the probability $P_{\alpha \beta}$ for the neutrino to
be detected in a charged current process associated to a flavour
$\ket{\nu_\beta}$ is given by
\begin{equation}
\begin{split}
P_{\alpha \beta} &= \left| \braket{\nu_\beta | e^{-i \hat{H} t} |
 \nu_\alpha} \right|^2 = \left| \sum_{i=1}^n
U^{\text{lep} *}_{\alpha i} U^{\text{lep}}_{\beta i} e^{-i
  \frac{m_i^2 L}{2E}} \right|^2 \\ &= \delta_{\alpha \beta} - 4
\sum_{i<j}^n \Re \left[U^{\text{lep} *}_{\alpha i} U^{\text{lep}}_{\beta i} U^{\text{lep}}_{\alpha j} U^{\text{lep} *}_{\beta j}
  \right] \sin^2 \frac{\Delta m_{ij}^2 L}{4E} \\ & \hspace*{10mm} + 2
\sum_{i<j}^n \Im \left[U^{\text{lep} *}_{\alpha i} U^{\text{lep}}_{\beta i} U^{\text{lep}}_{\alpha j} U^{\text{lep} *}_{\beta j}
  \right] \sin \frac{\Delta m_{ij}^2 L}{2E} \, ,
\end{split}
\label{eq:oscillationVacuum}
\end{equation}
where $\Delta m_{ij}^2 \equiv m_i^2 - m_j^2$. Because of the trigonometric functions, these transitions are also known as \emph{neutrino flavour oscillations}. The transition
probability for antineutrinos is obtained exchanging $U^\text{lep}
\rightarrow U^{\text{lep} *}$, thus modifying the sign of the last
term. Because of that, this probability can be different for neutrinos
and antineutrinos, and CP violation in the leptonic sector can be
detected studying this phenomenon.

Several aspects of \cref{eq:oscillationVacuum} are to be noticed. In
general, to have mass-induced flavour transitions neutrinos must have
different masses ($\Delta m_{ij}^2 \neq 0$) and must mix
($U^{\text{lep} *}_{\alpha i} U^{\text{lep}}_{\beta i} \neq
\delta_{\alpha \beta}$). Besides, the particular functional form makes
the Majorana phases in \cref{eq:PMNS} cancel out when multiplying
$U^{\text{lep} *}_{ \alpha i} U^{\text{lep}}_{\beta i}$, so they are
not observable. This is expected: the transition does not depend on
the Dirac or Majorana nature of the neutrinos. Lastly,
expression~\eqref{eq:oscillationVacuum} has an oscillatory behaviour
with characteristic oscillation lengths

\begin{equation}
L^{\text{osc}}_{i j} = \frac{4 \pi E}{\left|\Delta m_{ij}^2\right|}
\simeq \SI{2.48}{km} \frac{E/\si{GeV}}{\left|\Delta m_{ij}^2\right|/\si{eV^2}} \, .
\end{equation}
Since in real experiments neutrino beams are not monoenergetic but an
incoherent superposition of different energy states and detectors have
finite energy resolution, experiments do not measure $P_{\alpha
  \beta}$ but an average of it over some energy range.  Thus,
depending on the length $L$ that neutrinos travel in an experiment,
three different cases can be distinguished:
\begin{itemize}
\item $L \ll L^{\text{osc}}_{i j}$: in this case, oscillations do not
  have enough time to develop, the sines in
  \cref{eq:oscillationVacuum} are small and neither $\Delta m_{ij}^2$
  nor the leptonic mixing matrix elements $U^{\text{lep}}_{\alpha i}$
  are measurable.
\item $L \sim L^{\text{osc}}_{i j}$: in this, case a well-designed
  experiment is sensitive to both $\Delta m_{ij}^2$ and the leptonic
  mixing matrix elements.
\item $L \gg L^{\text{osc}}_{i j}$: in this case, the oscillation
  phase goes through many cycles when averaging over the energy and
  $\sin^2 \frac{\Delta m_{ij}^2 L}{4E}$ is averaged to
  $\frac{1}{2}$. The experiment can be sensitive to the leptonic
  mixing matrix elements but not to $\Delta m_{ij}^2$.
\end{itemize}

This behaviour can also be understood graphically. \Cref{fig:oscProbs}
shows the $\nu_e \rightarrow \nu_e$ and
$\nu_\mu \rightarrow \nu_\mu$ transition
probabilities as a function of the energy to distance ratio E/L for 
three light neutrino mass eigenstates with the oscillation parameters given in
Ref.~\cite{Esteban:NuFIT41}. In this scenario, there are three
non-independent oscillation lengths,
$\frac{E}{L^\text{osc}_{21}} \simeq \SI{2.98e-5}{GeV/km}$, $\frac{E}
{L^\text{osc}_{32}} \simeq \SI{1.95e-4}{GeV/km}$, and $\frac{E}{L^
\text{osc}_{31}} = \frac{E}{L^\text{osc}_{32}} + \frac{E}{L^\text{osc}
_{21}} \simeq \SI{1.02e-3}{GeV/km}$.
\begin{pagefigure}

\begin{subfigure}{\textwidth}
\centering \includegraphics[width=.96\textwidth]{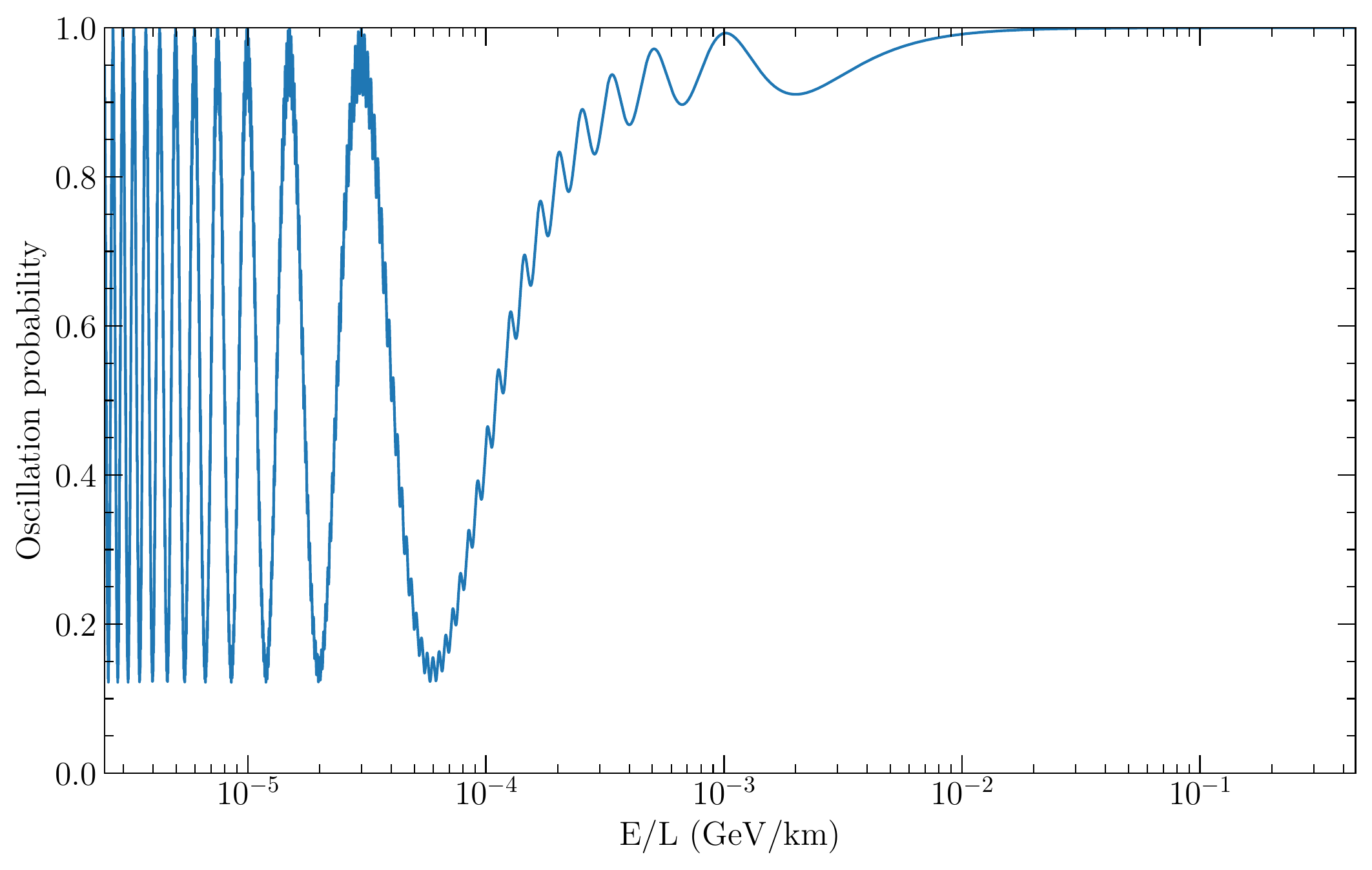}
\caption{$\nu_e \rightarrow \nu_e$ transition probability.}
\label{fig:eToeProb}
\end{subfigure}

\begin{subfigure}{\textwidth}
\centering
\includegraphics[width=.96\textwidth]{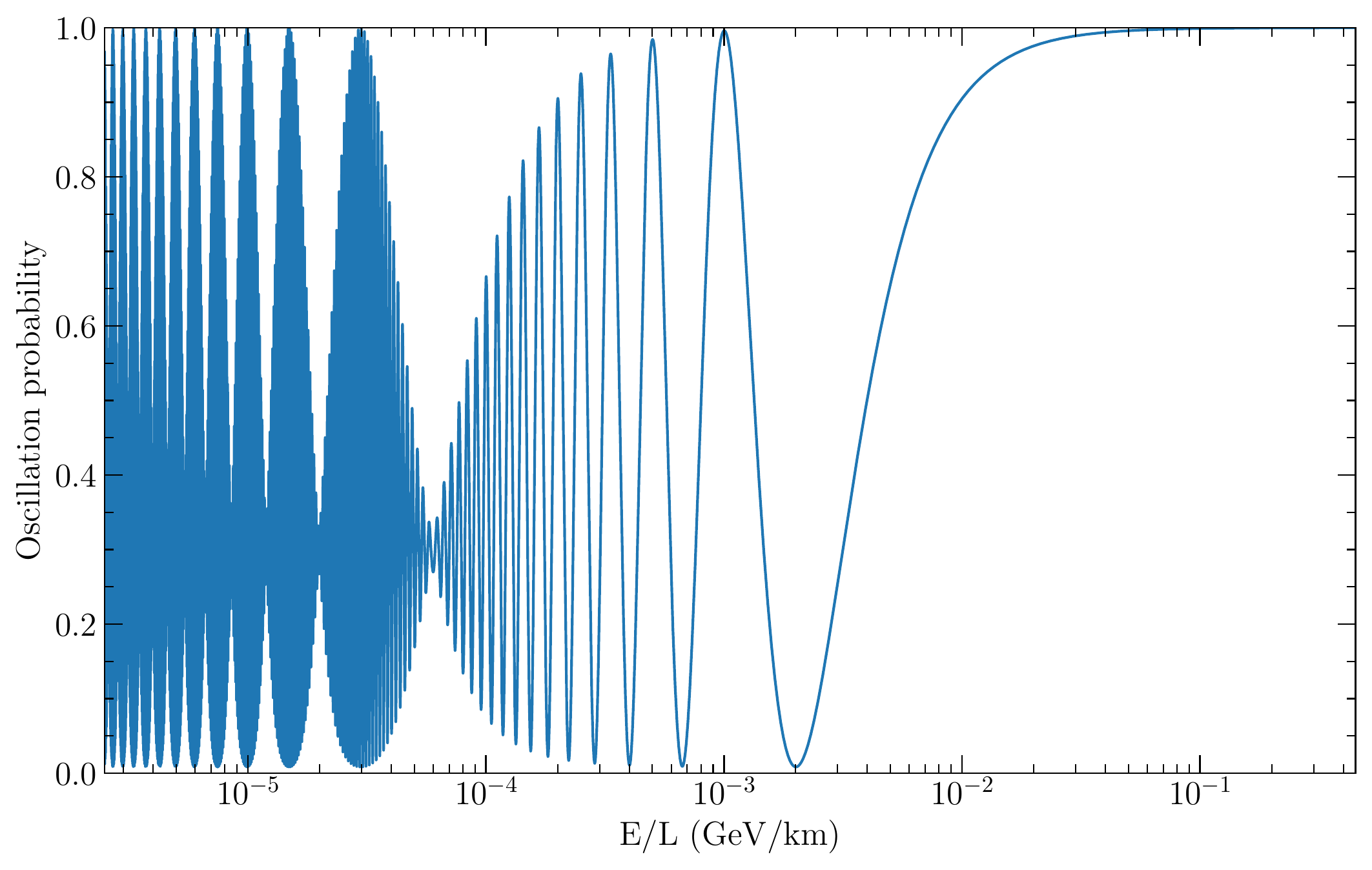}
\caption{$\nu_\mu \rightarrow \nu_\mu$ transition probability.}
\label{fig:muTomuProb}
\end{subfigure}

\caption{Some neutrino flavour transition probabilities. 
  The results have
  been obtained using \cref{eq:oscillationVacuum} under the assumption
  of 3 light neutrinos and the values for the oscillation parameters
  in Ref.~\cite{Esteban:NuFIT41}.}
\label{fig:oscProbs}
\end{pagefigure}

\Cref{fig:eToeProb} shows the transition between the regions where
$L^{\text{osc}}_{21}$, $L^{\text{osc}}_{23}$, or none contribute:
oscillations driven by $L^{\text{osc}}_{13}$ are suppressed with respect to the ones driven by
$L^{\text{osc}}_{23}$ because of their similar frequencies
and the smallness of $U^\text{lep}_{13} <
U^\text{lep}_{12}$. Also, since $U^\text{lep}_{23} \sim
U^\text{lep}_{22}$, \cref{fig:muTomuProb} shows the beats that
typically appear when two oscillations of similar frequencies
superpose.

Due to the three different oscillation regimes described above, some
experimental results can be roughly understood in terms of an
approximate two neutrino mixing. In that case, there is an angle
dominantly controlling the observable oscillation amplitude and a mass
difference dominantly controlling the observable frequency.
Technically, for 2 light neutrino flavours $U^\text{lep}$ is
parametrised by a single angle $\theta$ and the transition formula is
then quite simple:
\begin{equation}
P_{\alpha \beta} = \delta_{\alpha \beta} - \left( 2 \delta_{\alpha
  \beta} - 1\right) \sin^2 2 \theta \sin^2 \frac{\Delta m^2 L}{4E} \,
,
\label{eq:2gen}
\end{equation}
where $\Delta m^2$ is the squared mass difference between the
considered mass eigenstates. In this case, the oscillation probability
is symmetric under the exchange $\theta \leftrightarrow
\frac{\pi}{2}-\theta$ and/or $\Delta m^2\leftrightarrow - \Delta
m^2$. Furthermore, no physical CP violating phase is left
here. More-than-two neutrino mixing as well as matter effects (see
below) break all these symmetries.

\subsection{Flavour transitions in matter}
\label{sec:matterEffects}

The transition probability computed in \cref{eq:oscillationVacuum}
assumed that neutrinos travelled in vacuum. Even though their
inelastic scattering cross section is very small and can be safely neglected in most cases, coherent forward
elastic scattering on dense matter gets accumulated over long 
distances, as scattered and unscattered waves interfere. It is therefore important to take it into account.

In more detail, the matrix elements of the Hamiltonian operator in
\cref{eq:vacuumHamiltonian} have to be computed among states 
containing all particles present in the medium. Thus, the interaction 
operators among neutrinos and fermions in the Lagrangian~
\eqref{eq:CCandNC} will contribute to these matrix elements,
generating an effective potential $V$.

As an example, in a medium containing electrons, the charged current interaction Hamiltonian will 
contribute to the $ee$ matrix element in flavour space. If the electrons
have spin $s$ and momentum $\vec{p}_e$ distributed as $f(\vec{p}_e, s,
 \vec{x})$, this matrix element is computed 
 as~\cite{GonzalezGarcia:2002dz}
\begin{equation}
\begin{split}
V_{ee} & = \sum_s \int \mathrm{d}^3\vec{p}_e \, f(\vec{p}_e, s,
 \vec{x}) \braket{\nu_e, \, e(\vec{p}_e, s)| \hat{H}_\mathrm{CC} | 
 \nu_e, \, e(\vec{p}_e, s)} \\
& = 2\sqrt{2} G_F \sum_s \int \mathrm{d}^3\vec{p}_e \mathrm{d}^3x \, 
f(\vec{p}_e, s, \vec{x})\braket{\nu_e, \, e(\vec{p}_e, s)| \bar{e}(x) 
\gamma^\mu P_L \nu_e(x) \bar{\nu}_e(x) \gamma_\mu 
P_L e(x) | \nu_e, \, e(\vec{p}_e, s)} \, ,
\end{split}
\end{equation}
where $G_F = \frac{\sqrt{2}}{8} \frac{g^2}{M_W^2} \simeq \SI{1.166e-5}
{GeV^{-2}}$ is the Fermi constant. Notice that coherence implies that the spins and 
momenta of all initial and final particles are equal. As a 
consequence, the momentum transfer among neutrinos and electrons is 
zero,\footnote{Strictly, if neutrino mass eigenstates have different 
masses the momentum transfer cannot be zero. Nevertheless, coherence
holds as long as the momentum transfer is much smaller than the typical electron 
quantum mechanical momentum spread, $\mathcal{O}(\si{eV})$.} much 
smaller than the $W$ boson mass, and the effective Fermi Lagrangian ${\mathcal{L}_\mathrm{eff} = - 2 \sqrt{2} G_F \left(\bar{\nu}_e \gamma^\mu L e\right) \left(\bar{e} \gamma_\mu L \nu_e\right)}$
can be used instead of the Lagrangian~\eqref{eq:CCandNC} to compute 
the Hamiltonian. After Fierz rearrangement,
\begin{equation}
V_{ee} = 2 \sqrt{2} G_F \sum_s \int \mathrm{d}^3\vec{p}_e 
\mathrm{d}^3x \, \braket{\nu_e | \bar{\nu}_e(x) \gamma^\mu P_L \nu_e(x)
 | \nu_e} f(\vec{p}_e, s, \vec{x}) \braket{e(\vec{p}_e, s)| \bar{e}(x) 
 \gamma_\mu P_L e(x) | e(\vec{p}_e, s)} \, .
\end{equation}
Expanding the electron fields in plane waves, standard Dirac algebra
gives
\begin{equation}
\braket{e(\vec{p}_e, s)| \bar{e}(x) \gamma_\mu P_L 
e(x) | e(\vec{p}_e, s)} = \frac{1}{2 E_e} \bar{u}_s(p_e) \gamma_\mu P_L u_s(p_e) \, ,
\end{equation}
with $E_e$ the electron energy. The one-particle states have been 
normalised as $\ket{e(\vec{p}, s)} = a_{\vec{p}}^{s \dagger} \ket{0}$, 
with $a_{\vec{p}}^s$ an annihilation operator with momentum $\vec{p}$
and spin $s$, and $\ket{0}$ the vacuum.

For an isotropic and unpolarised medium, $f(\vec{p}_e, s, \vec{x}) = 
\frac{1}{2} f(p_e, \vec{x})$ and thus
\begin{equation}
\begin{split}
& \sum_s \int  \mathrm{d}^3\vec{p}_e \, f(\vec{p}_e, s, \vec{x}) 
\braket{e(\vec{p}_e, s)| \bar{e}(x) \gamma_\mu P_L e(x) | e(\vec{p}_e,
 s)} \\
= & \frac{1}{2} \int \mathrm{d}^3\vec{p}_e \, f(p_e, \vec{x}) \sum_s 
\frac{1}{2 E_e}\bar{u}_s(p_e) \gamma_\mu P_L u_s(p_e) = \frac{1}{2} \int 
\mathrm{d}^3\vec{p}_e \, f(p_e, \vec{x}) \mathrm{Tr}\left[\frac{m_e +
 \slashed{p}}{2 E_e} \gamma_\mu P_L \right] \\
= & \frac{1}{2} \int \mathrm{d}^3\vec{p}_e \, f(p_e, \vec{x}) 
\frac{p_\mu}{E_e} \, ,
\end{split}
\label{eq:matterEffectsDerivation}
\end{equation}
where $m_e$ is the electron mass, and the Dirac equation $\sum_s 
u_s(p) \bar{u}_s(p) = \slashed{p} + m$ has been used. Because of 
isotropy, $\int \mathrm{d}^3\vec{p}_e \, \vec{p}_e = 0$, and only the 
$\mu=0$ term survives in the last integral. Furthermore, $\int \mathrm{d}^3\vec{p}_e \,
 f(p_e, \vec{x}) = n_e(x)$, the electron number density in the medium.
 Thus,
\begin{equation}
\sum_s \int  \mathrm{d}^3\vec{p}_e \, f(\vec{p}_e, s, \vec{x}) 
\braket{e(\vec{p}_e, s)| \bar{e}(x) \gamma_\mu P_L e(x) | e(\vec{p}_e,
 s)} = \frac{1}{2} n_e(x) \delta_\mu^0 \, .
\end{equation} 

Finally, the neutrino matrix element
\begin{equation}
\braket{\nu_e| \bar{\nu}_e(x) \gamma^\mu P_L \nu_e(x) | \nu_e}
\end{equation}
gives, for a narrow neutrino wavepacket centered at a position $\vec{x}_0$,
a term $\sim \delta^3(\vec{x}-\vec{x}_0)$. Putting 
everything together,
\begin{equation}
V_{ee}(\vec{x}_0) = \sqrt{2} G_F n_e(\vec{x}_0) \, .
\end{equation}
For antineutrinos, there is one additional anticommutation when 
computing the matrix element, and the result changes sign.

Finally, neutral current interactions will also contribute to the 
matrix elements. A similar calculation gives for the effective potential matrix elements in flavour space 
\begin{equation}
V_{\alpha \beta}(\vec{x}_0) = - \delta_{\alpha \beta} \frac{G_F}{\sqrt{2}}\left[
(n_e(\vec{x}_0) - n_p(\vec{x}_0))(1-4\sin^2 \theta_W) + n_n(\vec{x}_0)
\right] \, ,
\end{equation}
where $n_p$ and $n_n$ are the proton and neutron number densities.
Notice that this contribution to the Hamiltonian matrix is 
proportional to the identity, and thus unobservable in this scenario.

All in all, we have seen that matter effects lead to an effective
potential difference between electron neutrinos and other flavours
\begin{equation}
V_{\text{eff}} = \pm \sqrt{2} G_F n_e \, ,
\label{eq:matterPotential}
\end{equation}
where the + (-) sign refers to neutrinos (antineutrinos). Adding it to
the free Hamiltonian, the $\nu_\alpha \rightarrow \nu_\beta$ transition probability for 3 light
neutrinos travelling through matter of constant density is given by
\begin{equation}
P_{\alpha \beta} = |\mathcal{M}_{\beta \alpha}|^2 \, ,
\end{equation}
with
\begin{equation}
\mathcal{M} = \exp \left\lbrace -i L\left[ \begin{pmatrix} \sqrt{2}
    G_F n_e & 0 & 0 \\ 0 & 0 & 0 \\ 0 & 0 & 0
\end{pmatrix} + \frac{1}{2E} U^\text{lep} \begin{pmatrix}
0 & 0 & 0 \\ 0 & \Delta m_{21}^2 & 0 \\ 0 & 0 & \Delta m_{31}^2
\end{pmatrix} U^{\text{lep} \dagger} \right] \right\rbrace \, ,
\end{equation}
where $U^\text{lep}$ is the leptonic mixing matrix~\eqref{eq:PMNS}. In
general there is no compact exact expression for this probability.

As a consequence, a two neutrino scenario gets significantly modified
by matter effects (take for example $\nu_\mu \leftrightarrow \nu_e$
oscillations by setting $\theta_{13} = \theta_{23} = 0$). There, the
presence of the matter potential allows to distinguish the sign of
$\Delta m^2_{12}$ (or, equivalently, whether $\theta_{12}$ is below or above
$45^\circ$). It also induces a different oscillation probability between
neutrinos and antineutrinos, this is, matter breaks CP because it
contains only electrons and not positrons.

\subsubsection{Non-uniform density: the MSW effect}

If the matter density cannot be considered constant along neutrino
propagation, which is the case for instance for solar neutrinos, the derivation of
the relevant transition probability is more involved. As rigorously 
derived in Refs.~\cite{Halprin:1986pn, Baltz:1988sv, Mannheim:1987ef} 
from field theory first principles, the evolution equation for n light 
neutrino mass eigenstates is given by
\begin{equation}
i \frac{\mathrm{d}}{\mathrm{d}x} \begin{pmatrix}
\nu_1 \\
\vdots \\
\nu_n
\end{pmatrix} = H \begin{pmatrix}
\nu_1 \\
\vdots \\
\nu_n
\end{pmatrix} \, ,
\label{eq:eom}
\end{equation}
where $x$ is the coordinate along the neutrino trajectory and $H$ is
the Hamiltonian matrix
\begin{equation}
H = \frac{1}{2E} \diag(m_1^2, \ldots, m_n^2) + U^{\mathrm{lep} 
\dagger} V U^\mathrm{lep} \, ,
\label{eq:matterH}
\end{equation}
with $V$ the effective potential matrix in the flavour basis.  For 
the particular case of three SM neutrinos with just SM interactions in 
a medium made of electrons, protons and neutrons,
\begin{equation}
V = \begin{pmatrix}
2 \sqrt{2} G_F n_e(x) & 0 & 0 \\
0 & 0 & 0 \\
0 & 0 & 0
\end{pmatrix} \, .
\end{equation}
For antineutrinos, the leptonic mixing matrix has to be replaced by 
its complex conjugate, and the SM matter potential flips sign.

To solve this equation,\footnote{The derivation here closely follows 
Refs.~\cite{GonzalezGarcia:2002dz, Tanabashi:2018oca}.} we first switch 
to the basis of instantaneous mass eigenstates in matter, i.e., the 
eigenstates of the full Hamiltonian~\eqref{eq:matterH} for a fixed $x$,
\begin{equation}
H(x) \begin{pmatrix}
\nu^m_1 \\
\vdots \\
\nu^m_n
\end{pmatrix} = \frac{1}{2E} \diag(\mu_1^2(x), \ldots, \mu_n^2(x)) 
\begin{pmatrix}
\nu^m_1 \\
\vdots \\
\nu^m_n
\end{pmatrix} \, ,
\end{equation}
where the relationship between instantaneous mass eigenstates $\begin{pmatrix}\nu_1^m & \ldots & \nu_n^m \end{pmatrix}^T$  and flavour
eigenstates $\begin{pmatrix}\nu_\alpha & \ldots &\end{pmatrix}^T$ 
can be written as 
\begin{equation}
\begin{pmatrix}
\nu_\alpha \\
\vdots \\
\phantom{\nu_\alpha}
\end{pmatrix} = \tilde{U}(x) \begin{pmatrix}
\nu^m_1 \\
\vdots \\
\nu^m_n
\end{pmatrix} \, ,
\label{eq:MSW_massBasis}
\end{equation}
with $\tilde{U}(x)$ an effective mixing matrix in matter.

For the particular case of two neutrino flavours $\alpha$ and $\beta$, 
\begin{equation}
\mu^2_{1, 2}(x) = \frac{m_1^2 + m_2^2}{2} + E(V_\alpha(x) + V_\beta(x)) \mp
\frac{1}{2} \sqrt{(\Delta m^2 \cos 2 \theta -
A(x))^2 + (\Delta m^2 \sin 2 \theta)^2} \, ,
\end{equation}
where $A(x) \equiv 2E(V_\alpha(x) - V_\beta(x))$ and $V_\alpha$ and $V_\beta$ 
are the matter potentials of $\nu_\alpha$ and $\nu_\beta$. $\theta$ is 
the angle parametrising the ${2 \times 2}$ leptonic mixing matrix
\begin{equation}
\begin{pmatrix}
\nu_\alpha \\
\nu_\beta
\end{pmatrix} = \begin{pmatrix}
\cos \theta & \sin \theta \\
-\sin \theta & \cos \theta
\end{pmatrix} \begin{pmatrix}
\nu_1 \\
\nu_2
\end{pmatrix} \, .
\end{equation}
On top of that, the effective mixing angle $\theta_m$
parametrising $\tilde{U}(x)$ is given by
\begin{equation}
\tan 2 \theta_m(x) = \frac{\Delta m^2 \sin 2 \theta}{\Delta m^2 \cos 2 
\theta - A(x)} \, .
\label{eq:resonantCondition}
\end{equation}
From here, we notice that for the \emph{resonance condition}
\begin{equation}
V_\alpha - V_\beta = \frac{\Delta m^2}{2E} \cos 2 \theta \, ,
\end{equation}
the effective mixing angle is maximal. 

We can now write the evolution equation in the instantaneous mass 
basis. Taking the derivative of \cref{eq:MSW_massBasis} and using 
\cref{eq:eom},
\begin{equation}
i \frac{\mathrm{d}}{\mathrm{d}x} \begin{pmatrix}
\nu^m_1 \\
\vdots \\
\nu^m_n
\end{pmatrix} = \left[\frac{1}{2E} \diag(\mu_1^2,\ldots,\mu_n^2) - i \tilde{U}^\dagger (x) \frac{\mathrm{d}\tilde{U}(x)}
{\mathrm{d}x}\right] \begin{pmatrix}
\nu^m_1 \\
\vdots \\
\nu^m_n
\end{pmatrix} \, .
\end{equation}
If the potential varies slowly enough, the second term can be 
neglected. Then, the instantaneous mass eigenstates do not mix in the 
evolution. Instead, they just pick a phase $\propto \frac{\mu_i^2}{2E}$ 
and the transition probability takes a simple form, similar to the 
vacuum one~\cite{Tanabashi:2018oca},
\begin{equation}
P_{\alpha \beta} = \left| \sum_i \tilde{U}^*_{\alpha i}(0) 
\tilde{U}_{\beta i}(L) \exp\left(-i \frac{1}{2E} \int_0^L \mu_i^2(x') 
\, \mathrm{d}x'\right) \right|^2 \, .
\end{equation}
This is known as the \emph{adiabatic approximation}. Physically, it 
means that the potential varies slowly enough so that the effective 
mixing in matter changes over scales much larger than the oscillation 
length. If, on the contrary, there are regions where the potential has 
strong variations, the adiabatic approximation breaks down and 
there can be transitions among instantaneous mass eigenstates. In these 
regions, one has to resort for instance to the WKB approximation to 
solve the evolution equation analytically~\cite{Kim:1994dy}. 

For the two-neutrino case, the adiabaticity condition 
reads~\cite{GonzalezGarcia:2002dz}
\begin{equation}
\frac{2 E A(x) \Delta m^2 
\sin 2 \theta}{\Delta(x)^3} \left|\frac{1}{A} 
\frac{\mathrm{d}A}{\mathrm{d}x}\right| \ll 1 \, ,
\label{eq:adiabaticApprox}
\end{equation}
with $\Delta(x) \equiv \mu_1^2(x) - \mu_2^2(x)$. For small mixing 
angles $\theta$, the left-hand side is largest at the resonance. If the 
adiabaticity condition is always satisfied, for very large $L$
\begin{equation}
P_{\alpha \alpha} = 1 - P_{\alpha \beta} = \frac{1}{2}(1+\cos 2 
\theta_{m, \, 0} \cos 2 \theta_{m, \, L}) \, ,
\label{eq:MSWtransitions}
\end{equation}
where $\theta_{m, \, 0}$ and $\theta_{m, \, L}$ are the effective 
mixing angles at the production and detection points, respectively. Notice that
with constant matter effects $\theta_{m, \, 0}$ and $\theta_{m, \, L}$ have
the same sign, and so $P_{\alpha \alpha} \geq \frac{1}{2}$. However, as
can be seen from \cref{eq:resonantCondition}, if neutrinos cross the 
resonance along their path, $\theta_m$ changes its octant. 
Thus, $\cos 2 \theta_{m, \, 0} \cos 2 \theta_{m, \, L} < 0$ and 
$P_{\alpha \alpha} < \frac{1}{2}$. 
This is referred to as the Mihheev-Smirnov-Wolfenstein (MSW)
effect~\cite{Wolfenstein:1977ue, Mikheev:1986gs}, and plays a 
fundamental role in explaining the observed deficit of solar neutrinos. 
This will be further explored in \cref{sec:solarExperiments}.

\subsection{Leptonic CP violation as a consequence of neutrino masses}
\label{sec:leptonicCPviol}

Besides flavour oscillations, another important consequence of
neutrino masses and mixing is the possibility of having CP violation
in the leptonic sector.  As in the quark sector, flavour mixing
between 3 particle flavours opens the door to breaking this symmetry. 
In particular, a non zero value for any of the phases in the 
parametrisation~\eqref{eq:PMNS} will introduce a phase in the
Lagrangian that violates CP. For the rest of this work, the Majorana
phases will be ignored: they depend on the Dirac or Majorana nature of
neutrino mass eigenstates and, as discussed above, they are irrelevant
for neutrino flavour oscillations.

The fact that a non zero value of $\delta_\text{CP}$ can lead to CP
violation is directly present in the oscillation
formula~\eqref{eq:oscillationVacuum}: due to the chiral structure of
the SM, antineutrinos (right-handed) are the CP conjugates of
neutrinos (left-handed); any difference in $P_{\alpha \beta}$ between
neutrinos and antineutrinos is a sign of CP violation. Since
\cref{eq:oscillationVacuum} accounts for antineutrino flavour 
oscillations just by substituting $U^\text{lep} \rightarrow 
U^{\text{lep} *}$, only the terms in the second sum
\begin{equation}
\Im\left[U^{\text{lep} *}_{\alpha i} U^\text{lep}_{\beta i}
  U^{\text{lep} *}_{\alpha j} U^\text{lep}_{\beta j} \right] \, , \,
i<j \, , \, \alpha \neq \beta
\label{eq:CPviolProd}
\end{equation}
violate CP. It is immediate to see that if $\delta_\text{CP} \neq 0$
these elements are different from zero. Furthermore, for three light
neutrinos the matrix element products~\eqref{eq:CPviolProd} can be
shown to be all equal, up to signs, to
\begin{equation}
J_\mathrm{CP} \equiv J_\mathrm{CP}^\mathrm{max} \sin \delta_\text{CP} = c_{12} c_{23} c_{13}^2 s_{12} s_{23} s_{13} \sin \delta_\text{CP} \, .
\label{eq:jarlskogNeutrino}
\end{equation} 
A quantity completely equivalent to the Jarlskog invariant in
\cref{eq:jarlskogMatrix}. Indeed, the same procedure followed in \cref{sec:CPviol_SM} and leading to \cref{eq:jarlskogQuark,eq:jarlskogMatrix}
could be applied here: the presence of neutrino masses makes the
transformation~\eqref{eq:neutrinoSymmetry} physical as there is an
additional flavour matrix as in the quark sector. In the end, the same
condition for CP conservation would be obtained: for
Dirac neutrinos, CP violation requires all three neutrino masses to be
different (so that the mixing angles are physical), all three charged
lepton masses to be different (so that the concept of neutrino flavour
is meaningful), and the invariant in \cref{eq:jarlskogNeutrino} not to
vanish. This will be explicitly shown for the more general case in
which there is new physics in addition to neutrino masses in
\cref{chap:NSItheor}.

\section{Summary}

The paradigmatic fundamental theory of Nature until the end of the
20th century was the SM of Particle Physics. Although it
was ultimately confirmed with the discovery of the Higgs boson, it
predicts leptonic flavours to be exactly conserved, and as a
consequence neutrinos to be strictly massless. The conservation of leptonic flavour has been experimentally checked to fail, constituting our first laboratory evidence for BSM physics. Intriguingly, assuming the SM to be a theory valid up to some high
energy scale, the first observable effect of new physics is actually
\emph{predicted} to be leptonic flavour mixing driven by neutrino masses.

An immediate experimental consequence is that neutrinos periodically change their
flavour as they travel, a phenomenon known as neutrino
oscillations. The experimental scrutiny of this process allows to
measure the first properties of a theory more fundamental than the
SM. To this end, there is a strong experimental programme that
will be overviewed in the next chapter.

In addition, mixing among three neutrino flavours induces a new source
of CP violation. This could be linked to the matter-antimatter
asymmetry of the Universe, and so it is interesting on its
own. Current experiments are slowly starting to show a hint that CP
could be violated in the leptonic sector~\cite{t2k:ichep2016,
t2k:susy2016,nova:nu2016}. Rigorously assessing the
significance of this signal and its robustness is the main goal of
this thesis.

\chapter{Three-neutrino fit to oscillation experiments: framework}
\label{chap:3nufit_theor}

\epigraph{\emph{In that simple statement is the key to science. It
does not make any difference how beautiful your guess is. It does not
make any difference how smart you are, who made the guess, or what his
name is --- if it disagrees with experiment it is wrong.}}{ --- Richard
P.~Feynman}

\epigraph{\emph{Mientras en el mar o en el cielo haya un abismo, \\
que al cálculo resista; \\ Mientras la humanidad siempre avanzando, \\
no sepa a dó camina; \\ Mientras haya un misterio para el hombre, \\
¡habrá poesía! }}{ --- Gustavo A.~Bécquer}

As mentioned in the previous chapter, data from neutrino experiments
conclusively show that leptonic flavours are not conserved, and
thus that there is BSM physics. In
principle, neutrino masses could provide the source for leptonic
flavour violation. However, any other mechanism introducing a
distinction between neutrino interaction and propagation eigenstates could
explain the same phenomenon. Some alternatives, that generically
introduce a dependence of the flavour transition probability on the
energy different from $\sim \sin
\frac{L}{E}$~\cite{GonzalezGarcia:2004wg}, include neutrino
decay~\cite{Barger:1998xk}, quantum decoherence~\cite{Lisi:2000zt}, or
Lorentz invariance violation~\cite{Coleman:1997xq,
Glashow:1997gx}. Nevertheless, the precise analysis of neutrino
spectra conclusively proved that
the observed phenomena were mass-induced
flavour-oscillations~\cite{GonzalezGarcia:2008ru, Lisi:2000zt,
GonzalezGarcia:2004wg}.

Once mass-induced neutrino oscillations are experimentally
established, the next step is their quantitative characterisation. 
Currently, there 
exist some tensions in the experimental data: the so-called short 
baseline reactor anomaly~\cite{Mueller:2011nm, Huber:2011wv,
 Mention:2011rk, Hayes:2013wra, Fang:2015cma, Hayes:2016qnu}, the 
gallium anomaly~\cite{Acero:2007su, Giunti:2010zu}, and the 
LSND~\cite{Aguilar:2001ty} and MiniBooNE~\cite{Aguilar-Arevalo:2018gpe} 
short baseline accelerator results. Since they have low statistical 
significance and they are not fully consistent among 
themselves~\cite{Dentler:2018sju, Boser:2019rta}, they will be ignored 
in what follows. Apart from these anomalies, all the existing oscillation
data can be explained by a three light neutrino paradigm parametrised by 
the mixing matrix in \cref{eq:PMNS}.

\begin{figure}[hbtp]
\centering \includegraphics[width=0.8\textwidth]{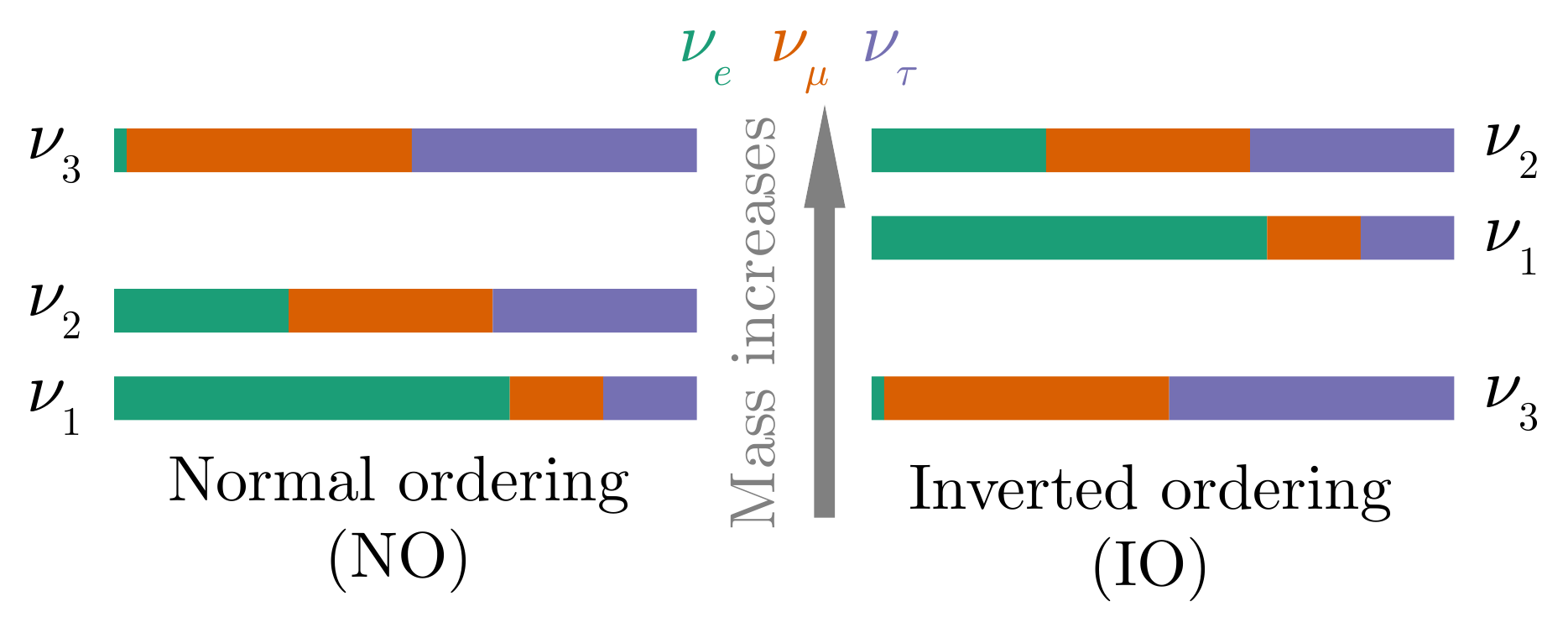}
\caption{Convention for the numbering of mass eigenstates and possible
orderings (NO in left, IO in right). The colours indicate the amount
of mixing between mass and flavour eigenstates.}
\label{fig:ordering}
\end{figure}

The convention regarding the numbering of mass eigenstates,
diagrammatically shown in \cref{fig:ordering}, makes use of the
experimental fact (see description of the data below) that two of them
have masses close to each other ($|\Delta m^2| \sim \SI{e-5}{eV^2} $)
whereas the third one is further away ($|\Delta m^2| \sim
\SI{e-3}{eV^2}$). Based on this, one can always chose a convention in
which the close eigenstates are denoted as $\nu_1$ and $\nu_2$, with
$m_1 < m_2$, keeping the ranges of angles and phases as $\theta_{12} ,
\theta_{13} , \theta_{23} \in [0,90^\circ]$ and $\delta_\text{CP}\in
[0,2\pi)$. The other mass eigenstate is denoted as $\nu_3$. It is currently
unknown whether $m_3 > m_1$ (known as Normal Ordering, NO) or $m_3 <
m_1$ (known as Inverted Ordering, IO).

The current picture of neutrino flavour transitions has been built by
combining a variety of experimental results. Indeed, due to the
different oscillation regimes described in \cref{sec:vacuumOsc} and
explicitly seen in \cref{fig:oscProbs}, there are qualitatively
different experiments that look for different sectors of the leptonic
mixing matrix. They will be explored in the rest of this chapter.

\section{Solar neutrinos}
\label{sec:solarExperiments} 
\begin{figure}[hbtp] \centering
\includegraphics[width=\textwidth]{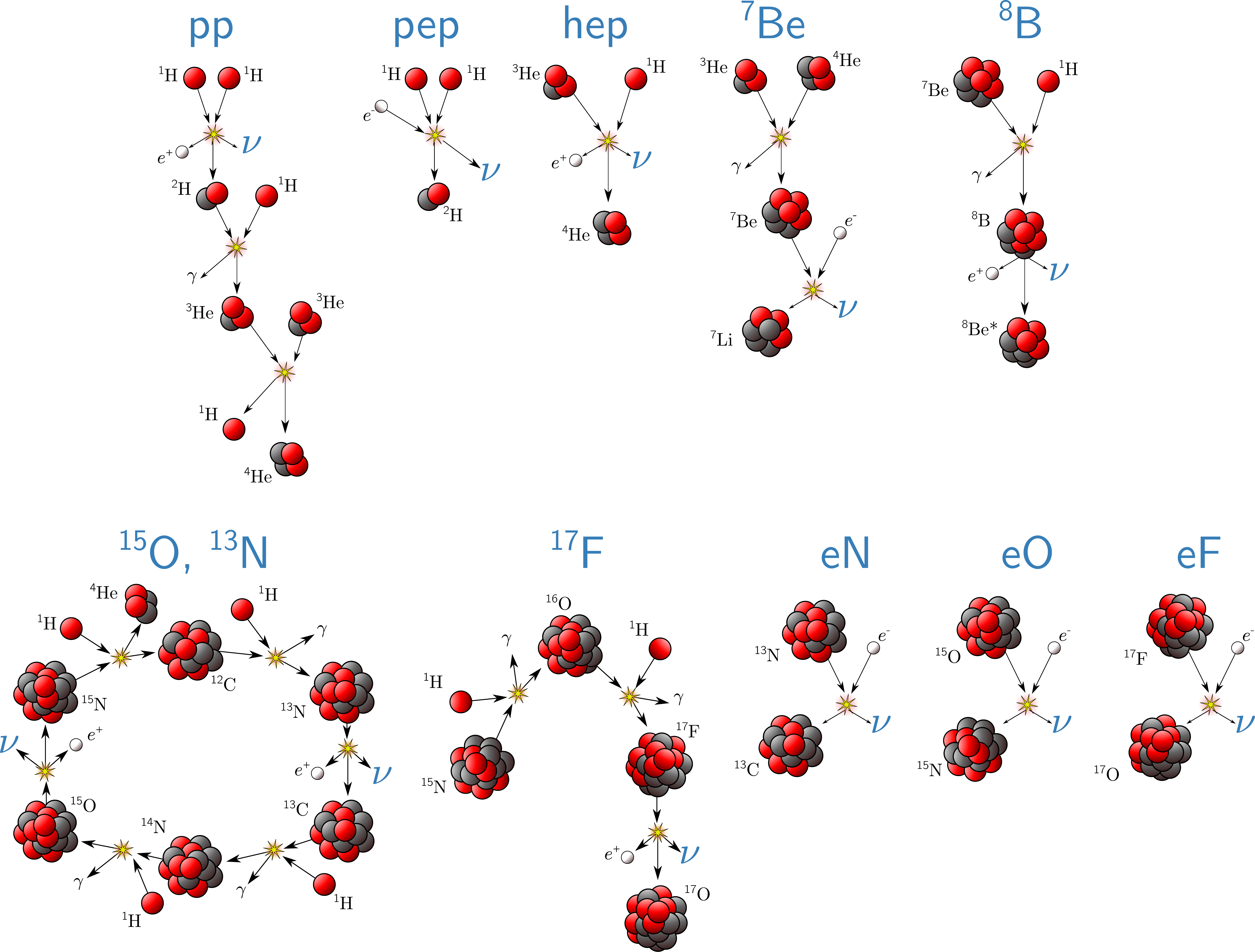}
\caption{Nuclear reactions in the Sun that produce neutrinos. The
top row corresponds to the pp cycle, and the bottom row to the CNO
cycle. Above each reaction, the name with which the neutrinos produced
there are denoted is indicated. Adapted from Refs.~\cite{wiki:pp,
wiki:CNO}.}
\label{fig:sunReactions}
\end{figure}

\begin{figure}[hbtp] \centering
\includegraphics[width=0.75\textwidth] {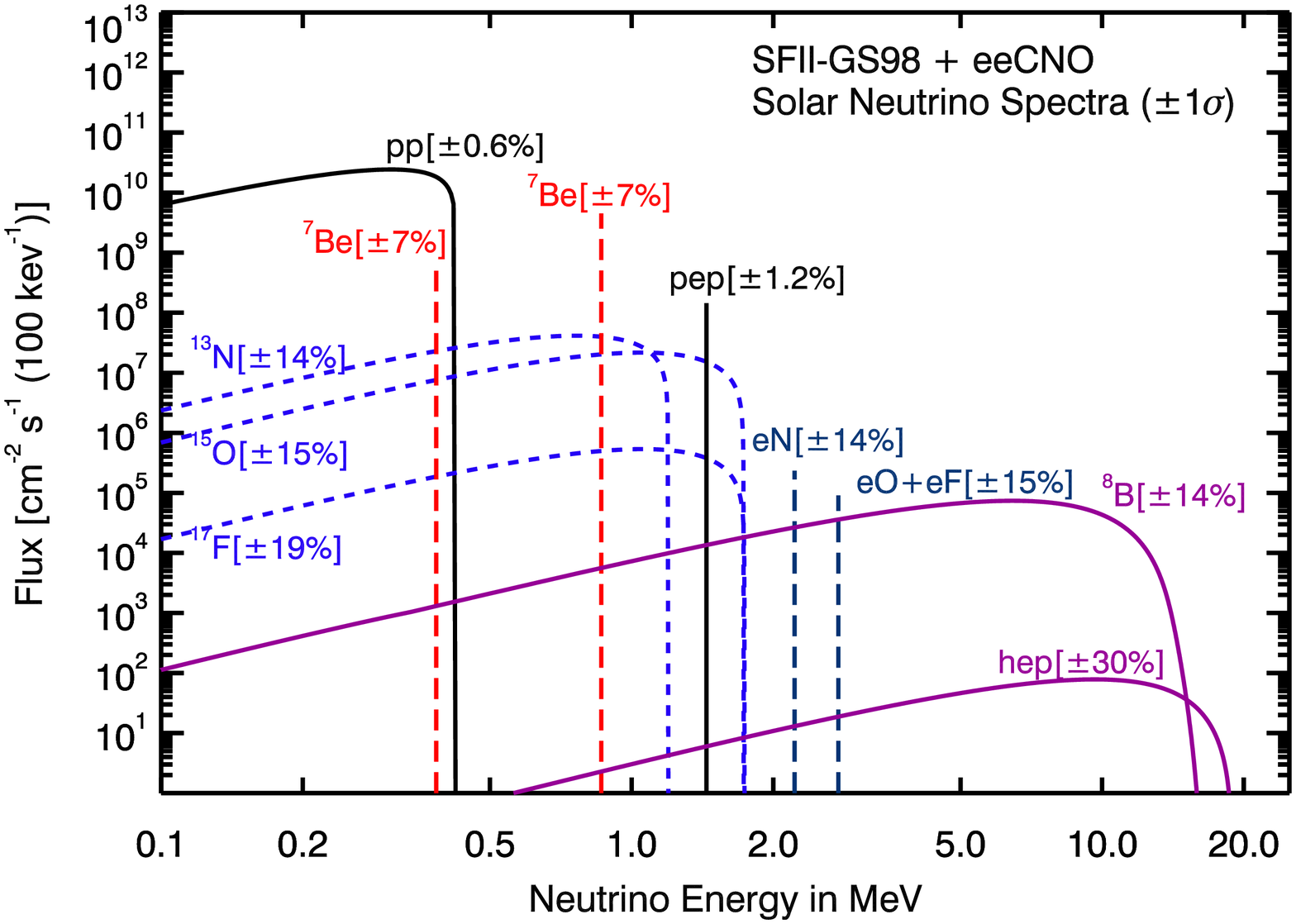}
\caption{Solar neutrino spectrum. Above each line, the reaction in
which neutrinos are generated is indicated (see
\cref{fig:sunReactions}). Extracted from
Ref.~\cite{Serenelli:2016dgz}.  The theoretical computations and
uncertainties on flux normalisations are given in
Refs.~\cite{Serenelli:2011py, Villante:2014txa}.  Monochromatic
fluxes, corresponding to electron capture processes, are given in
units of \si{cm^{-2}.s^{-1}}.}
\label{fig:sunSpectrum}
\end{figure}

Neutrinos coming from the Sun provided
the first experimental hint for neutrino flavour transitions. Solar
neutrinos are copiously generated in the thermonuclear reactions that
fuel our star, which occur through two main chains, the pp
chain and the CNO cycle: the relevant processes producing neutrinos
are shown in \cref{fig:sunReactions}. The resulting spectrum, with
energies $\mathcal{O}(\si{MeV})$ as is typical in nuclear reactions,
is shown in \cref{fig:sunSpectrum}. Its precise computation requires a
detailed knowledge of the Sun and its evolution, and the Solar Models
rely on observational parameters that in turn give the normalisation
uncertainties shown in the figure~\cite{Serenelli:2011py,
Villante:2014txa}.

\begin{figure}[hbtp] \centering
\includegraphics[width=0.75\textwidth]{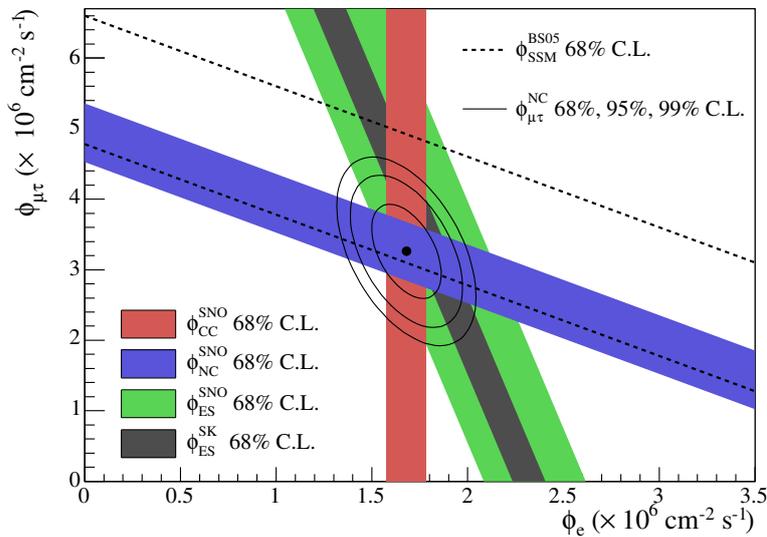}
\caption{$\nu_\mu + \nu_\tau$ flux vs $\nu_e$ flux, as measured by the
SNO neutral current (purple), charged current (red), and elastic
scattering (green) data. The region between the dashed lines
corresponds to the prediction in Ref.~\cite{Bahcall:2004pz} assuming
neutrino flavour oscillations. The contours correspond to the joint
confidence regions. Also shown in grey is the Super-Kamiokande
measurement~\cite{Fukuda:2002pe}. The hypothesis of no flavour
transitions ($\phi_{\mu \tau}=0$) is clearly excluded. Figure from
Ref.~\cite{Aharmim:2005gt}.}
\label{fig:SNO}
\end{figure}

Solar neutrinos were first detected by the Chlorine experiment in
1968~\cite{Davis:1968cp}. Since then, they have been detected in many
experiments that can be generically classified
as~\cite{Tanabashi:2018oca}
\begin{itemize}
\item \emph{Radiochemical experiments}: these experiments detect solar
neutrinos through inverse beta decay
\begin{equation} n + \nu_e \rightarrow e^- + p \, ,
\end{equation} which modifies the chemical composition of the
detector. The amount of generated protons is measured after a certain
period of time, which allows to extract the solar neutrino flux.

The detectors either exploit the
\begin{equation} {}^{37}\mathrm{Cl} + \nu_e \rightarrow {}^{37}\mathrm{Ar} + e^-
\end{equation} reaction~\cite{Cleveland:1998nv}, that with a threshold
of \SI{0.814}{MeV} is mostly sensitive to $^7\mathrm{Be}$ and $^8\mathrm{B}$ neutrinos;
or the
\begin{equation} {}^{71}\mathrm{Ga} + \nu_e \rightarrow {}^{71}\mathrm{Ge} + e^-
\end{equation} reaction~\cite{Abdurashitov:2002nt, Hampel:1998xg,
Altmann:2005ix}, that has a lower threshold (\SI{0.233}{MeV}) and a
larger capture cross section, thus detecting as well pp neutrinos. The
former set of experiments detected a $\sim 70\%$ deficit of solar
neutrinos, whereas the latter observed a $\sim 45\%$ deficit,
indicating that the solar neutrino deficit is energy dependent.

\item \emph{Real time experiments}: unlike radiochemical experiments,
these are capable of detecting the solar neutrino interaction in real
time. In addition, they can measure the energy and incoming direction 
of each event. Therefore, they are sensitive to the energy dependence 
of flavour transitions, as well as to transitions induced by the Earth
matter as discussed in \cref{sec:matterEffects}.

The real time experiments that have detected solar neutrinos are
\begin{itemize}
\item Kamiokande~\cite{Hirata:1991ub} and
Super-Kamiokande~\cite{Hosaka:2005um}, water tanks that detect the
Cherenkov light emitted by electrons elastically scattered in the
$\nu_\alpha + e^- \rightarrow \nu_\alpha + e^-$ process. Requiring the
electrons to emit enough Cherenkov light sets a threshold $E_\nu
\gtrsim \SI{5}{MeV}$, and so the experiments are mostly sensitive to
${}^8\mathrm{B}$ neutrinos. A deficit $\sim 60\%$ was observed in these
experiments.
\item SNO~\cite{Aharmim:2011vm}, a heavy water ($\mathrm{D}_2\mathrm{O}$) Cherenkov
detector. Interestingly, in addition to elastic scattering $\nu_\alpha
+ e^- \rightarrow \nu_\alpha + e^-$, SNO could measure both charged
current interactions of electron neutrinos, $\nu_e + {}^2\mathrm{H} \rightarrow
p + p + e^-$, and neutral current interactions of all interacting
neutrinos, $\nu_\alpha + {}^2\mathrm{H} \rightarrow n + p + \nu_\alpha$.
Comparing the reaction rates (see \cref{fig:SNO}), SNO checked that
the solar neutrino deficit was due to electron neutrinos transitioning
to other flavours.
\item KamLAND~\cite{Gando:2014wjd} and Borexino~\cite{Bellini:2011rx}, detectors filled with liquid scintillator that detect neutrinos
through elastic scattering with electrons. Unlike with Cherenkov
detectors, there is no physical barrier for detecting low-energy
neutrinos other than the scintillator being sensitive to low-energy
electrons. This allowed to detect ${}^7\mathrm{Be}$ and pp neutrinos.
\end{itemize}
\end{itemize}

\subsection{Analysis and interpretation of solar neutrino data}

Since solar neutrinos traverse large non-uniform densities before
leaving the Sun, the results of the experiments above must be analysed
by properly taking into account the effects of non-uniform matter 
discussed in \cref{sec:matterEffects}. As experiments with a higher 
energy threshold observed $P_{ee} < \frac{1}{2}$, the MSW effect 
overviewed there is expected to play a significant role in explaining the 
data.

\begin{figure}[hbtp] \centering
\includegraphics[width=0.75\textwidth]{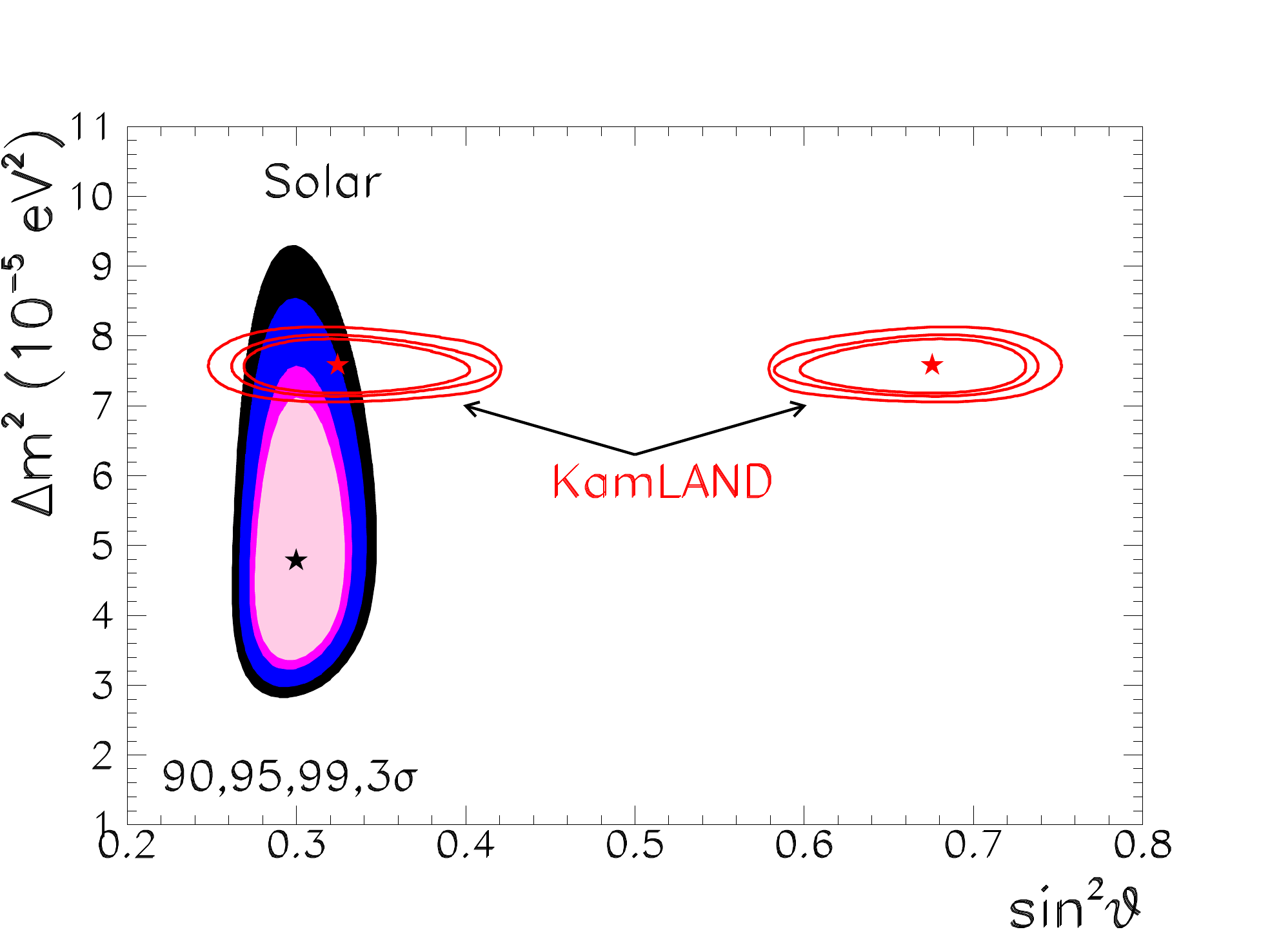}
\caption{Allowed regions of $\Delta m^2$ and $\sin^2 \theta$ that best
describe solar data. The red regions correspond to the KamLAND reactor
experiment, to be discussed later. Figure courtesy of
M.~C.~Gonzalez-García and M.~Maltoni.}
\label{fig:solRegions}
\end{figure}

Solar neutrino experiments measure the $\nu_e \rightarrow \nu_e$
transition probability and matter affects identically $\nu_\mu$ and
$\nu_\tau$, and so we can consider to a good approximation transitions
between $\nu_e$ and $\nu_X$, where the latter is some linear
combination of $\nu_\mu$ and $\nu_\tau$. These transitions are
parametrised in terms of a single squared mass difference $\Delta m^2$
and a single mixing angle $\theta$. When analysing all solar neutrino
data, the region of masses and mixings that best describes the
experiments is shown in \cref{fig:solRegions}. As can be seen, the
squared mass difference is $\sim \SI{e-5}{eV^2}$, and the mixing angle
is $\sim 32^\circ$.

Given the allowed parameters, the amplitude and energy dependence of
the $\nu_e \rightarrow \nu_e$ survival probability can be easily
understood from the discussion in \cref{sec:matterEffects}, as the
adiabatic approximation~\eqref{eq:adiabaticApprox} always
holds~\cite{GonzalezGarcia:2002dz}. Recalling
\cref{eq:MSWtransitions},
\begin{equation} P_{ee} = \frac{1}{2}(1+ \cos 2 \theta_{m, \, 0} \cos
2 \theta) \, ,
\end{equation} where
\begin{equation} \tan 2 \theta_m = \frac{\tan 2 \theta}{1 - A/(\Delta
m^2 \cos 2 \theta)} \, .
\end{equation} For the best fit parameters in \cref{fig:solRegions},
\begin{equation} \frac{A}{\Delta m^2 \cos 2 \theta} \simeq 0.72
\left(\frac{E}{\si{MeV}}\right) \left(\frac{\SI{4.8e-5}{eV^2}} {\Delta
m^2}\right)\, ,
\end{equation} where $E$ is the neutrino energy, and we have taken a
solar density $\sim \SI{100}{gm/cm^3}$, representative of its core.

Thus, for $E \ll \si{MeV}$, matter effects are not important and
$P_{ee} \simeq \frac{1}{2} (1 + \cos^2 2 \theta) \simeq 0.6$. For $E
\gg \si{MeV}$, however, $P_{ee} \simeq \frac{1}{2} (1 - \cos
2 \theta) \simeq 0.3$. At $E \sim \si{MeV}$, the transition between
both regimes takes place. In addition, matter effects allow to
determine the octant of $\theta$,
as the MSW effect only happens when $1-\frac{A}{\Delta m^2 \cos 2
\theta}$ changes sign along neutrino propagation. That is, for $\theta
< 45^\circ$.\footnote{In our convention, we have set $\Delta
m^2 > 0$, as one can always change the sign of $\Delta m^2$ by
simultaneously changing the octant of $\theta$. The whole procedure is
equivalent to relabelling the mass eigenstates.}

These results have been discussed in a two-neutrino
approximation. When embedded in a three-neutrino framework
parametrised by the mixing matrix~\eqref{eq:PMNS}, they
correspond to the limit $\frac{|\Delta m^2_{32}|}{G_F n_e E}
\rightarrow \infty$, $\theta_{13} \rightarrow 0$~\cite{Kuo:1986sk}
(see description of the atmospheric and reactor data below). In this
limit, the determined parameters correspond to $\Delta m^2 = \Delta
m^2_{21}$ and $\theta = \theta_{12}$.

\begin{figure}[hbtp] \centering
\includegraphics[width=0.75\textwidth]{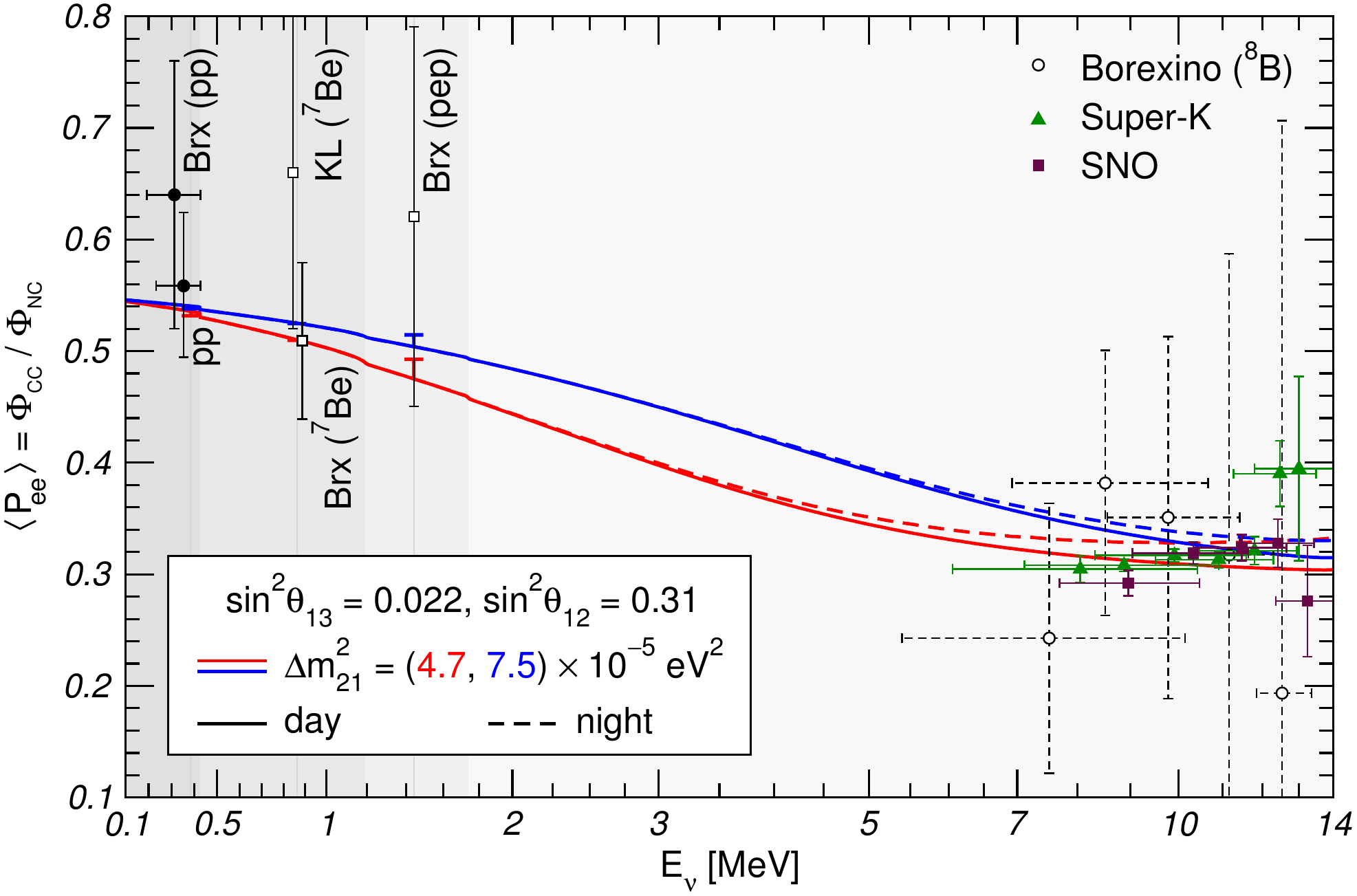}
\caption{Solar $\nu_e$ survival probability as a function of the
neutrino energy, as well as some experimental data. Brx means Borexino
and KL KamLAND. Earth matter effects, which induce a small day-night
asymmetry, are also included. The transition between the
vacuum-dominated regime, at lower energies and with $P_{ee} >
\frac{1}{2}$; and the MSW-dominated regime, at higher energies and
with $P_{ee} < \frac{1}{2}$, is clearly visible.  Extracted from
Ref.~\cite{Maltoni:2015kca}.}
\label{fig:solarNuProb}
\end{figure}

For illustration, the $\nu_e \rightarrow \nu_e$ survival probability as a
function of the neutrino energy, as well as the experimental data, is
shown in \cref{fig:solarNuProb}, where subleading $\theta_{13}$
effects have been included. The transition from $P_{ee} < 0.5$ to
$P_{ee} > 0.5$ is clearly visible. As we see, Nature was kind to
provide solar neutrinos with energies around the transition between
vacuum-dominated and MSW-dominated oscillations, $\sim \si{MeV}$, so
that the squared mass splitting could be determined.

\section{Atmospheric neutrinos} Besides solar neutrinos, the other
experiments that historically first established neutrino flavour
oscillations were atmospheric neutrino experiments.

\begin{figure}[hbtp] \centering
\includegraphics[width=0.75\textwidth]{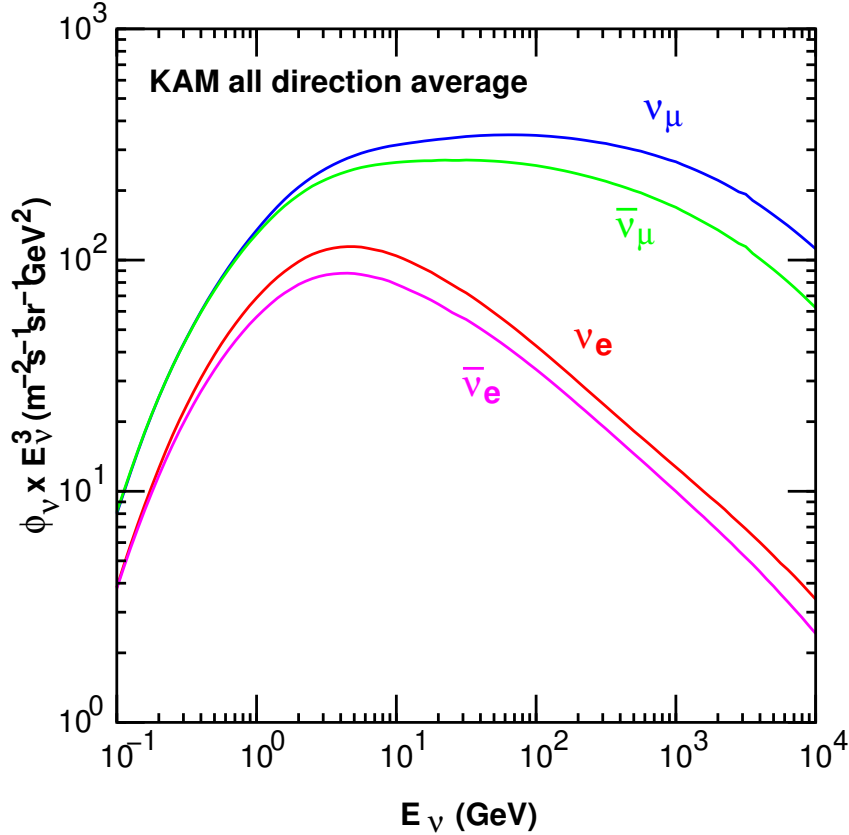}
\caption{Atmospheric neutrino flux at the Super-Kamiokande site,
averaged over all directions and over one year. Extracted from
Ref.~\cite{Honda:2015fha}.}
\label{fig:atmosFlux}
\end{figure}

Muon and electron neutrinos and antineutrinos are abundantly produced
in particle cascades created when cosmic rays hit the atmosphere at
altitudes $\sim \SI{10}{km}$. These cascades involve charged mesons,
mainly pions, that decay to charged leptons and neutrinos as
\begin{equation} \pi^\pm \rightarrow \mu^\pm + \parenbar{\nu}_\mu \, ,
\end{equation} followed by muon decay
\begin{equation} \mu^\pm \rightarrow e^\pm + \parenbar{\nu}_\mu
+ \parenbar{\nu}_e\, .
\end{equation} Realistic calculations introduce decays of other
subdominant mesons, mostly kaons, producing a neutrino flux depicted
in \cref{fig:atmosFlux}.

Atmospheric neutrinos were first detected in underground experiments,
where plastic scintillators detected the muons produced by atmospheric
$\parenbar{\nu}_\mu$ in charged current interactions~\cite{Achar:1965ova,
Reines:1965qk}. In the 1980s, large underground experiments searching
for nucleon decay started observing atmospheric neutrinos. There are
two main kind of such experiments
\begin{itemize}
\item \emph{Water Cherenkov detectors}: these experiments detect the
Cherenkov light emitted by muons and electrons generated in charged
current interactions of neutrinos with nuclei. They can estimate the
neutrino energy and incoming direction, as well as discern between
electrons and muons, as the former generate an electromagnetic shower
that blurs the Cherenkov ring.  The main experiments are
IMB~\cite{Casper:1990ac}, Kamiokande~\cite{Fukuda:1994mc} and
Super-Kamiokande~\cite{Fukuda:1998mi}. On top of them, in the last
years the IceCube experiment~\cite{Aartsen:2014yll} has also detected
atmospheric neutrino oscillations using the Antarctic ice as a Cherenkov detector.

Both IMB and Kamiokande reported a 60--70\% deficit in muon neutrinos,
and no significant deficit in electron neutrinos. The Super-Kamiokande
experiment explored this anomaly as a function of the energy and
arrival direction of the neutrinos. The results, which will be
discussed below, confirmed the deficit with large significance.

\item \emph{Iron calorimeters}: these detectors are sensitive to the
energy deposition of neutrino-generated electrons and muons as they
traverse a large iron volume. The Frejus~\cite{Daum:1994bf} and
NUSEX~\cite{Aglietta:1988be} experiment did not observe any
atmospheric neutrino deficit (although the results were compatible
with IMB and Kamiokande within $\sim 2\sigma$). Nevertheless, some
years later the MACRO~\cite{Ambrosio:2001je} and
Soudan-2~\cite{Sanchez:2003rb} calorimetric experiments independently
confirmed the Super-Kamiokande results on a significant $\parenbar{\nu}_\mu$
deficit.
\end{itemize}

Since atmospheric neutrinos have energies $\mathcal{O}(\si{GeV})$, the
oscillation length induced by the ``solar'' squared mass splitting,
$\mathcal{O}(\SI{e-5}{eV^2})$, is $\sim \SI{2e4}{km}$, very large even
for neutrinos that traverse the entire Earth. Thus, it can be safely
ignored when analysing the data. Furthermore, since the data shows no
evidence for $\parenbar{\nu}_e$ appearance, the process can be understood in
terms of $\parenbar{\nu}_\mu \rightarrow \parenbar{\nu}_\tau$ oscillations. As both flavours
have the same interactions with matter, matter effects are subdominant
and the $\parenbar{\nu}_\mu$ survival probability is simply given by
\begin{equation} P_{\mu \mu} = 1 - \sin^2 2 \theta\sin^2
\frac{\Delta m^2 L}{4E} \, ,
\label{eq:Pmumuatm}
\end{equation} parametrised by a single mixing angle
$\theta$ and a single squared mass splitting $\Delta
m^2$. Notice that here, since matter effects are
irrelevant, there is no sensitivity to the octant of
$\theta$ or to the sign of $\Delta m^2$.

\begin{figure}[hbtp] \centering
\includegraphics[width=\textwidth]{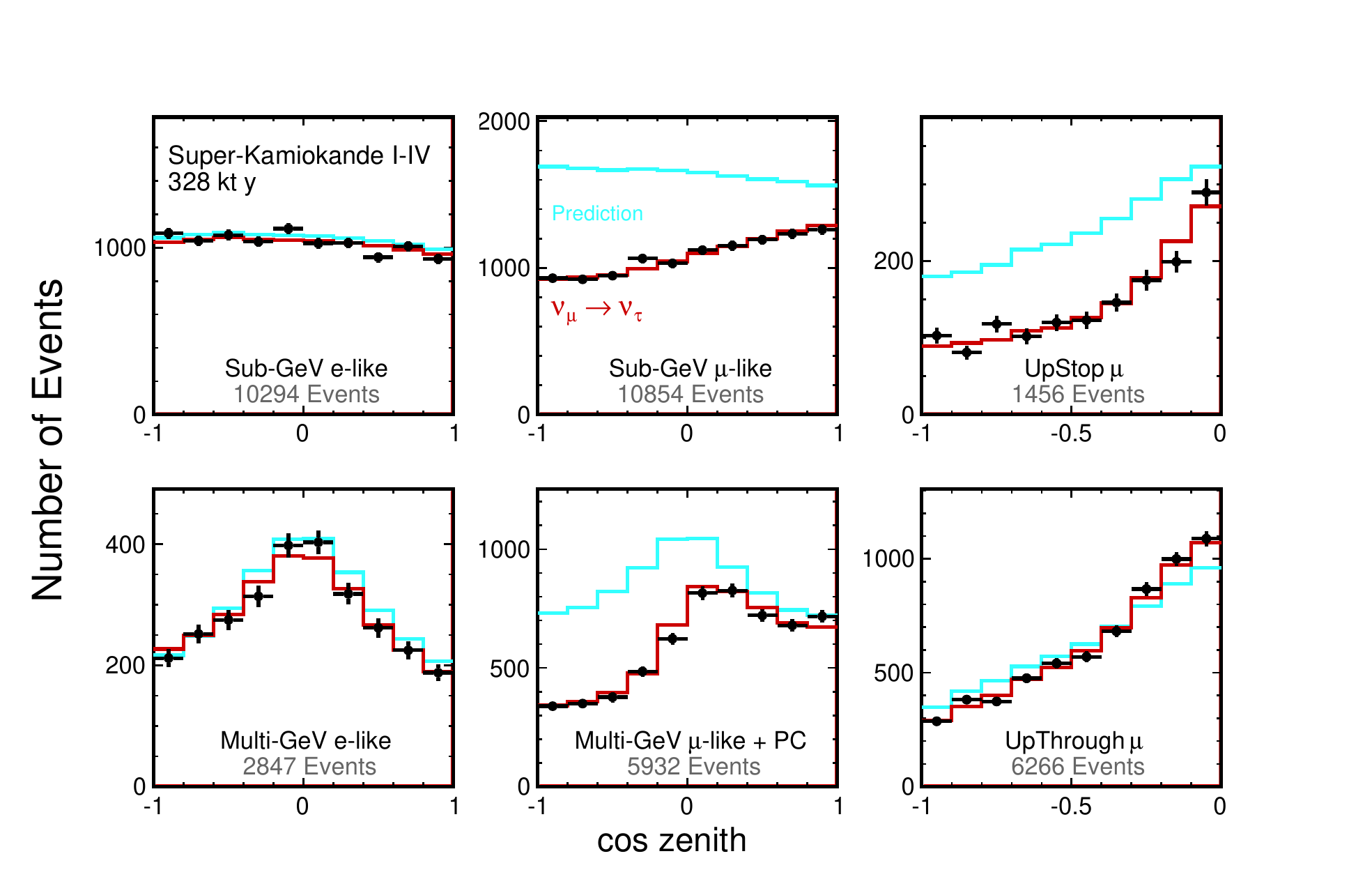}
\caption{Super-Kamiokande neutrino events as a function of the zenith
angle. The blue (red) lines shown the expectations without (with) neutrino oscillations. The data points
show a clear preference for muon neutrino disappearance.  The events are
classified as ``Sub-GeV'' (``Multi-GeV'') if their deposited energy
was $<\SI{1.33}{GeV}$ ($>\SI{1.33}{GeV}$).  The third column
corresponds to up-going muons, generated by neutrino interactions with
the surrounding rock, and classified as ``UpStop'' or ``UpThrough''
depending on whether the muons stopped inside the detector or not,
respectively. The latter are generated by much more energetic
neutrinos. Figure from Ref.~\cite{Tanabashi:2018oca}.}
\label{fig:SK-zenith}
\end{figure}

Interestingly, atmospheric neutrinos allow to directly test the 
oscillatory behaviour of \cref{eq:Pmumuatm} by exploiting that neutrinos 
with different
incoming directions have travelled different distances before reaching
the detector. This is parametrised in terms of the zenith angle, i.e.,
the angle between the neutrino direction and the vertical. The amount
of $\parenbar{\nu}_e$ and $\parenbar{\nu}_\mu$ events observed by Super-Kamiokande as a
function of this angle is shown in \cref{fig:SK-zenith}.

\begin{figure}[hbtp] \centering
\includegraphics[width=0.45\textwidth]{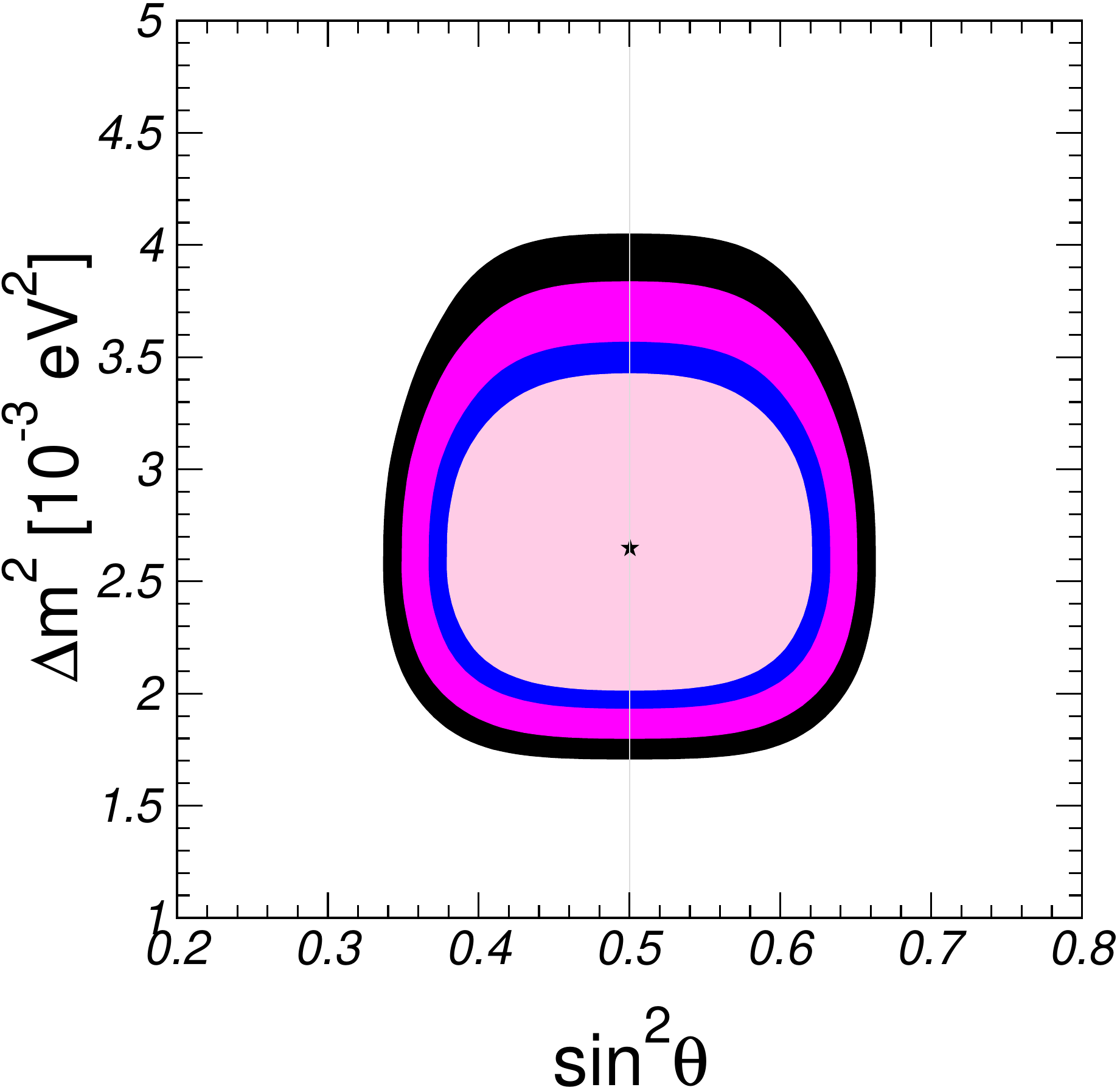}
\caption{Allowed regions of $|\Delta m^2|$ and $\sin^2 \theta$ that
best describe atmospheric data at 90\% CL, 95\% CL, 99\% CL and
$3\sigma$. Figure courtesy of M.~C.~Gonzalez-García and M.~Maltoni.}
\label{fig:2nu-atmos-sk}
\end{figure}

The data shows a clear preference for $\parenbar{\nu}_\mu$ disappearance, and no
significant $\parenbar{\nu}_e$ appearance or disappearance. Furthermore, the
deficit grows for larger zenith angles, i.e., muon neutrinos
travelling for larger distances are more likely to disappear. Finally,
neutrinos with larger energies are less likely to disappear: this is
particularly visible in the ``UpThrough'' sample, which shows no
significant $\parenbar{\nu}_\mu$ depletion; and in the ``Multi-GeV'' sample at
smaller zenith angles.

These properties are exactly what is expected from \cref{eq:Pmumuatm}.
When interpreted in terms of $\parenbar{\nu}_\mu \rightarrow \parenbar{\nu}_\tau$
oscillations, the allowed regions for the squared mass splitting and
the effective mixing angle are shown in \cref{fig:2nu-atmos-sk}.

When discussing these results, only mixing among two neutrinos has been
considered. In a three-neutrino framework parametrised by the mixing
matrix~\eqref{eq:PMNS}, this corresponds to the
limit $\frac{\Delta m^2_{21} L}{E} \rightarrow 0$, $\theta_{13} \rightarrow 0$
(see description of solar and reactor neutrino data). In this limit,
the determined parameters correspond to $|\Delta m^2| = |\Delta
m^2_{32}| = |\Delta m^2_{31}|$ and $\theta = \theta_{23}$.

\section{Reactor neutrinos}

\begin{figure}[hbtp] \centering
\includegraphics[width=0.75\textwidth]{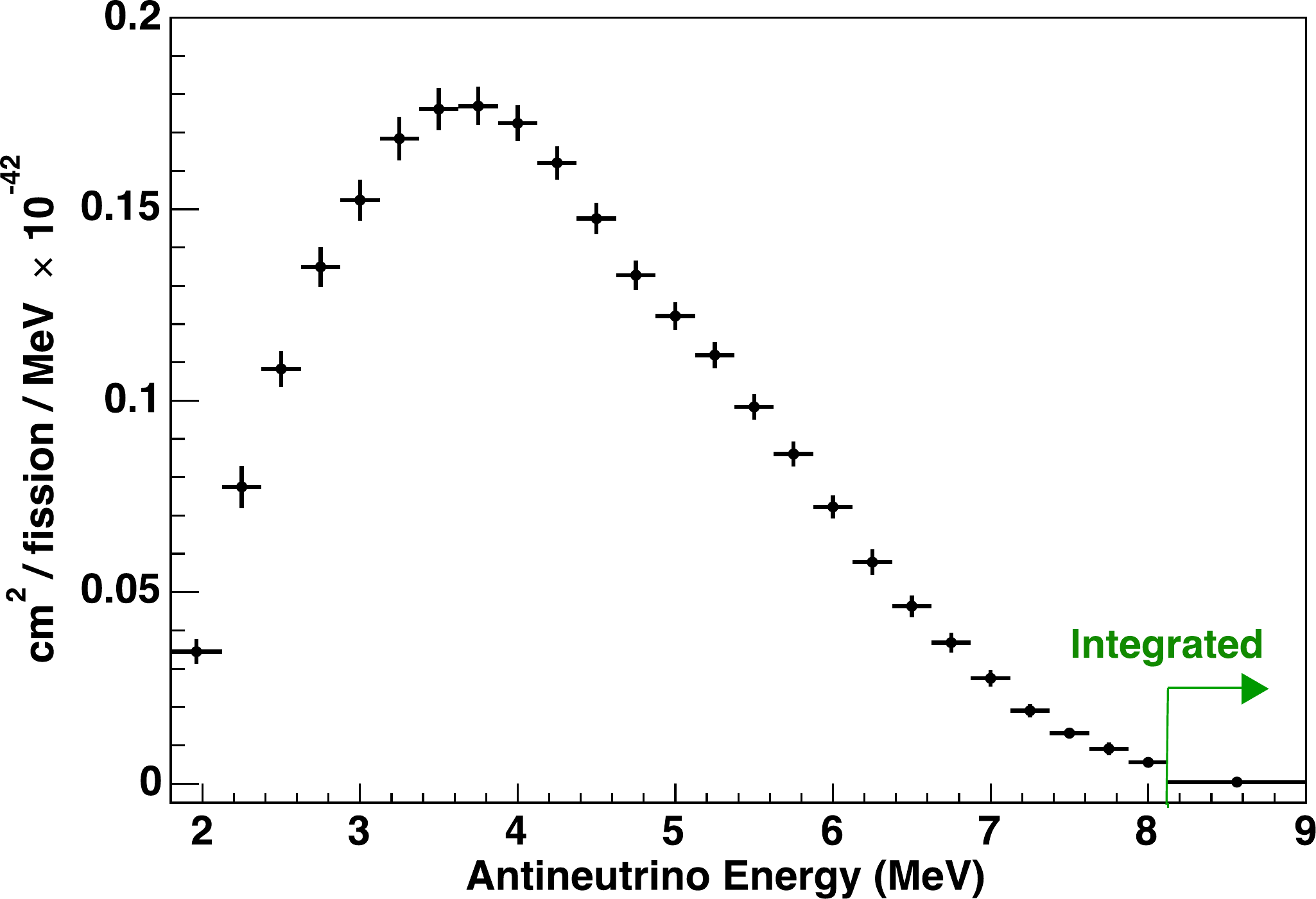}
\caption{Reactor $\bar{\nu}_e$ spectrum, as measured by the Daya Bay
collaboration~\cite{An:2015nua}.}
\label{fig:DB-spectrum}
\end{figure}

Nuclear reactors produce copious amounts of electron anti-neutrinos in
beta decay processes. Indeed, the first experimentally detected
neutrinos came from a reactor~\cite{Reines:1965qk}. Their typical
energies are $\mathcal{O}(\si{MeV})$, as is characteristic in nuclear
processes, and their spectrum is shown in \cref{fig:DB-spectrum}. They
are relatively easy to detect, as low-energy $\bar{\nu}_e$ induce
inverse beta decay
\begin{equation} \bar{\nu}_e + p \rightarrow e^+ + n \, ,
\end{equation} a process with a very characteristic signature: $e^+$
annihilation into two photons with a total energy $\sim E_\nu + m_p -
m_n + m_e$, where $E_\nu$ is the antineutrino energy and $m_p$, $m_n$
and $m_e$ are the proton, neutron and electron masses; as well as
neutron capture if the detector material is sensitive to it. $e^+$
annihilation also allows to cleanly measure the antineutrino energy.

As reactor antineutrinos have low energies, they are sensitive to
rather low squared mass splittings. In this sense, we can distinguish
two types of reactor neutrino experiments
\begin{itemize}
\item \emph{Long baseline reactor experiments}: these experiments look
for reactor $\bar{\nu}_e$ disappearance at baselines
$\mathcal{O}(\SI{100}{km})$. I.e., they are sensitive to squared mass
splittings $\mathcal{O}(\SI{e-5}{eV^2})$, and so they can
independently confirm the neutrino oscillation solution to the solar
neutrino deficit.

With this idea, the KamLAND liquid scintillator
detector~\cite{Gando:2013nba} was built in the Kamioka mine in Japan,
about \SI{100}{km} away from several Japanese nuclear power
plants. The ratio between the observed number of events and the
expectation without $\bar{\nu}_e$ disappearance is shown in
\cref{fig:kamlandProb}. As can be seen, the data clearly shows the
oscillatory pattern expected from mass-induced neutrino flavour
transitions.

The resulting allowed region in parameter space is shown
in \cref{fig:solRegions} (where the octant degeneracy associated with
vacuum experiments is also visible). The compatibility with solar neutrino data casts no doubt
on the interpretation of the solar neutrino deficit. As in the solar
neutrino analysis, in a three-neutrino framework parametrised by
\cref{eq:PMNS} the measured parameters correspond to
$\Delta m^2 = \Delta m^2_{21}$ and $\theta = \theta_{12}$, as long as
subleading $\theta_{13}$ effects are neglected.

\begin{figure}[hbtp] \centering
\includegraphics[width=0.85\textwidth]{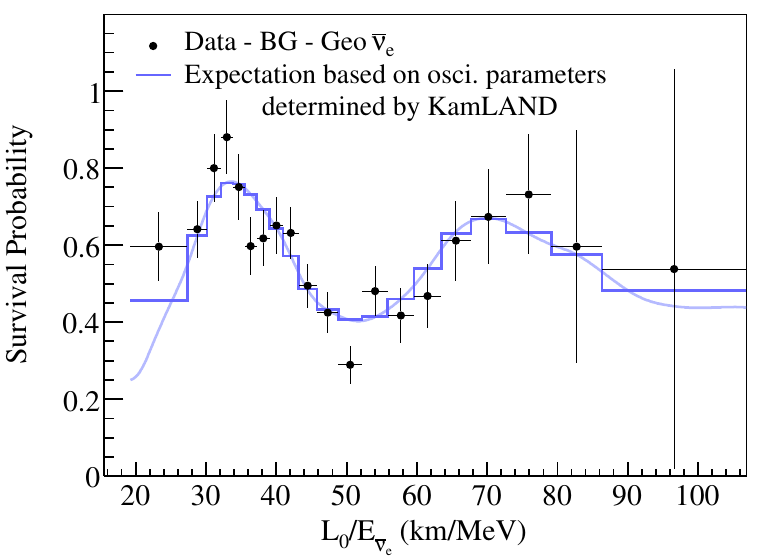}
\caption{Ratio between the observed number of $\bar{\nu}_e$ events and
the expectation without oscillations, as a function of
$L_0/E_{\bar{\nu}_e}$. $L_0$ is the flux-weighted average distance to
nuclear reactors, and $E_{\bar{\nu}_e}$ the antineutrino energy. The
blue line corresponds to the expectation from oscillations. Figure
from Ref.~\cite{Abe:2008aa}.}
\label{fig:kamlandProb}
\end{figure}

\item \emph{Medium baseline reactor experiments}: reactor antineutrino
experiments can also look for $\bar{\nu}_e$ disappearance induced by
the ``atmospheric'' squared mass splitting,
$\Delta m^2_{32} \sim \mathcal{O}(\SI{e-3}{eV^2})$. The corresponding baseline has to be
$\mathcal{O}(\SI{1}{km})$, and so \emph{a priori} the experiments look
relatively easy to carry out. However, the atmospheric neutrino
results showed no evidence for $\nu_e$ appearance nor disappearance,
and so the mixing angle between $\nu_e$ and the mass eigenstates
involved in atmospheric $\nu_\mu$ disappearance ($\nu_2$ and $\nu_3$
in our convention) must be tiny. In the parametrisation in
\cref{eq:PMNS}, this angle is $\theta_{13}$.

Originally, reactor experiments looking for a nonzero $\theta_{13}$,
such as CHOOZ~\cite{Apollonio:2002gd} or Palo
Verde~\cite{Boehm:2001ik}, could only set limits $\sin^2 \theta_{13}
\lesssim 0.03$. Being sensitive to smaller mixing angles required
reducing the flux uncertainty below the $3\%$ level. To this end, a
new generation of experiments was built: Double
Chooz~\cite{DoubleChooz:2019qbj}, Daya Bay~\cite{Adey:2018zwh} and
RENO~\cite{Bak:2018ydk}. These experiments had an additional detector near the
reactor, that minimised systematics and flux uncertainties.

All these experiments have reported an energy-dependent $\bar{\nu}_e$
deficit shown in \cref{fig:reactorProb}, pointing towards $\sin^2
\theta_{13} \sim 0.02$. When analysed in a two-neutrino framework, the
allowed regions for $\sin^2 \theta$
and $\Delta m^2$, which in the parametrisation~\eqref{eq:PMNS}
correspond to $\sin^2 \theta_{13}$ and $\Delta
m^2_{32}$ up to subleading $\Delta m^2_{21}$ and
$\theta_{12}$ effects, is shown in \cref{fig:2nu-reac}.

\begin{figure}[hbtp] \centering
\begin{subfigure}[b]{0.49\textwidth}
\includegraphics[width=\textwidth]{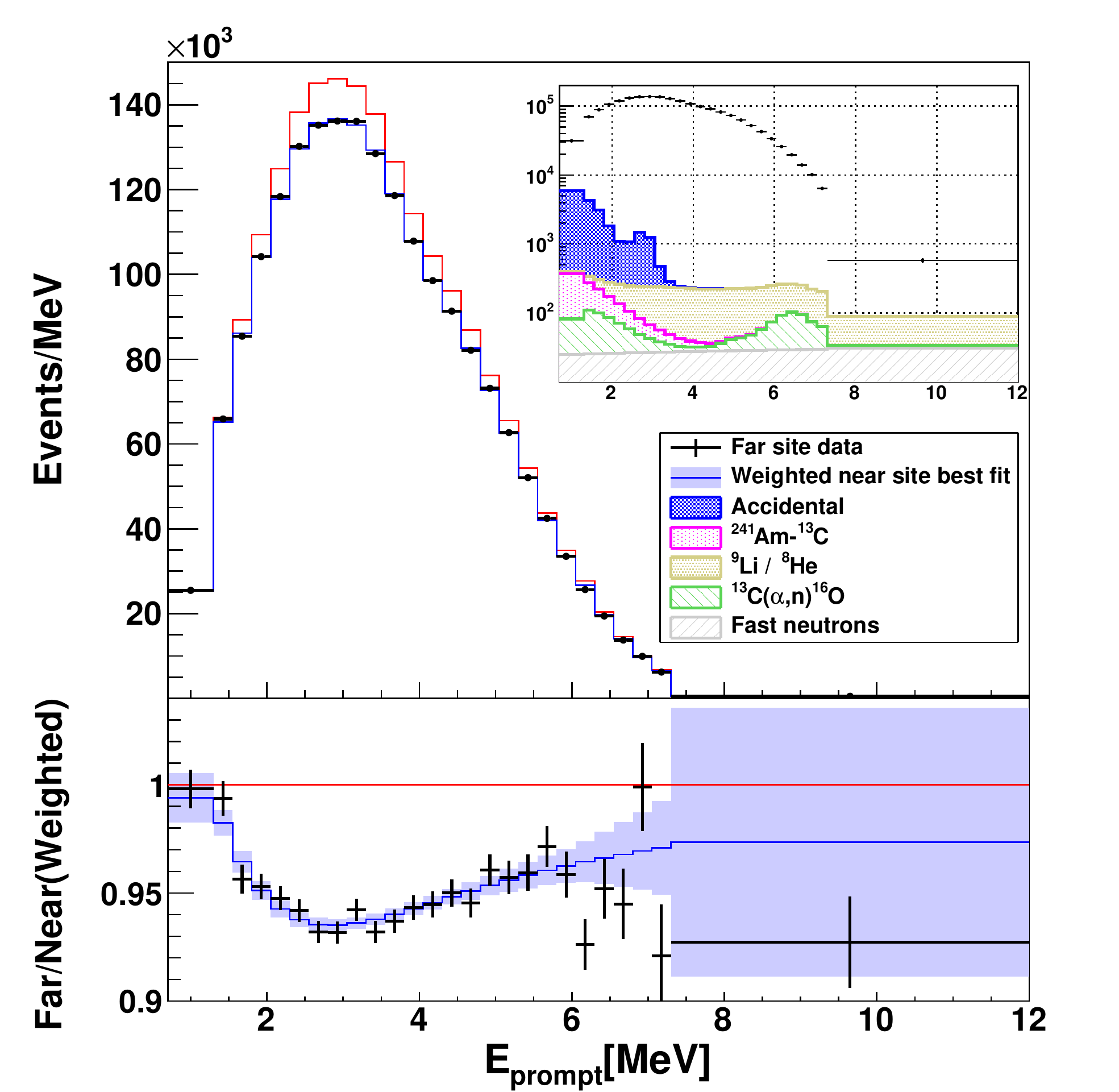}
\caption{Daya Bay results. Figure from Ref.~\cite{Adey:2018zwh}.}
\end{subfigure}
\begin{subfigure}[b]{0.49\textwidth}
\includegraphics[width=\textwidth]{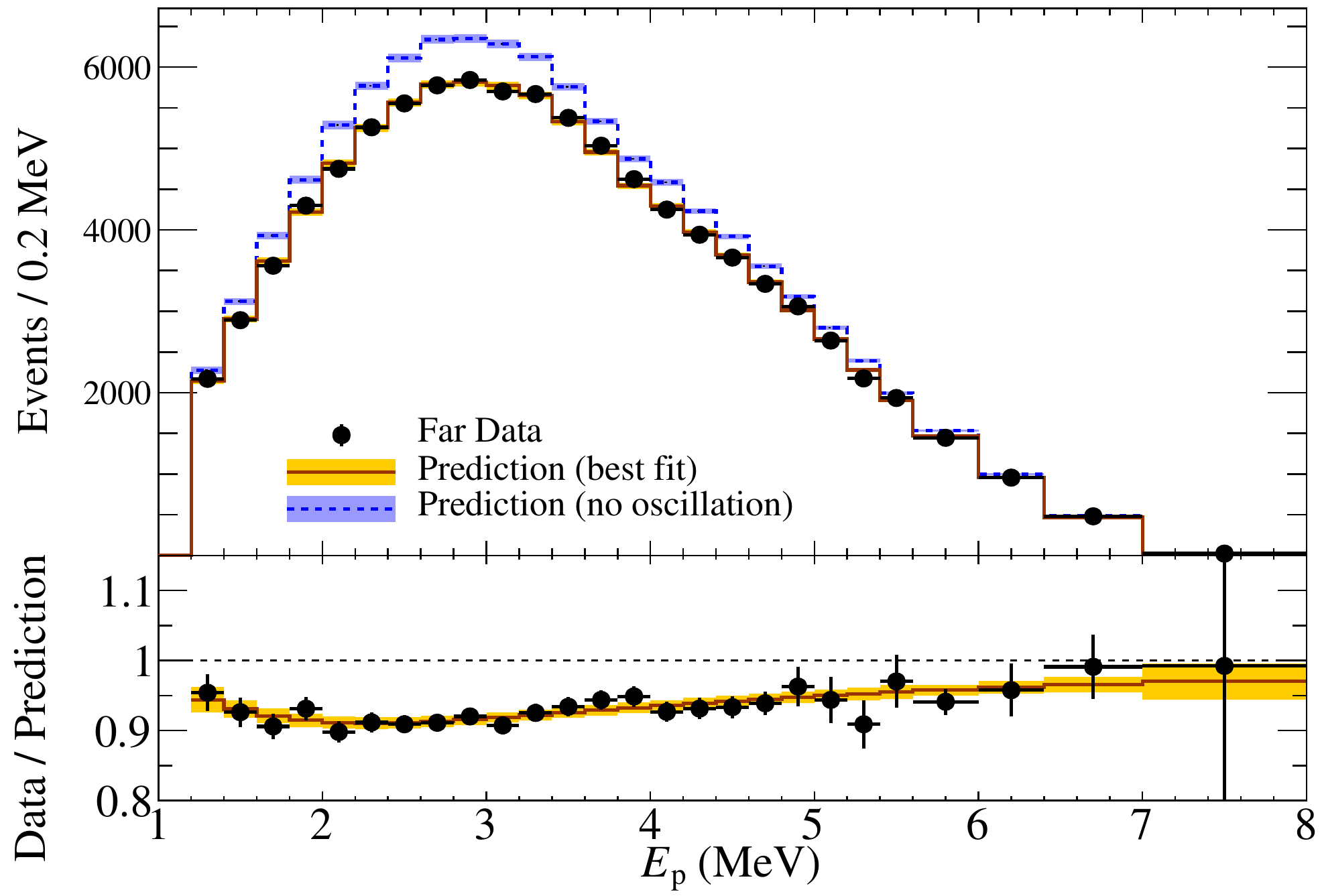}
\caption{RENO results. Figure from Ref.~\cite{Bak:2018ydk}.}
\end{subfigure}
\caption{Energy spectra of events at the RENO and Daya Bay far
detectors. The lower panels show the ratio among observed and
predicted number of events, as well as the prediction from neutrino
oscillations.  The energy is given in terms of the prompt event energy
$\simeq E_{\bar{\nu}_e} - \SI{0.78}{MeV}$, with $E_{\bar{\nu}_e}$ the
antineutrino energy.}
\label{fig:reactorProb}
\end{figure}

\begin{figure}[hbtp] \centering
\includegraphics[width=0.5\textwidth]{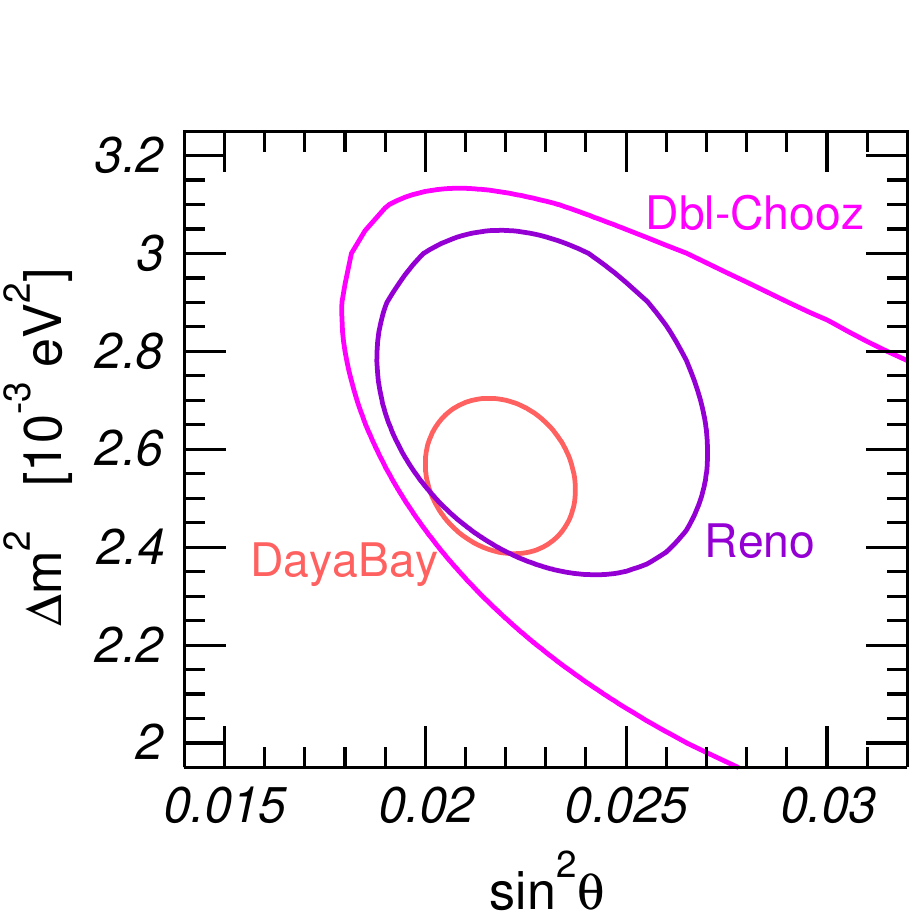}
\caption{Allowed region of $|\Delta m^2|$ and $\sin^2 \theta$ at 95\%
CL for each relevant medium baseline reactor experiment. Figure
courtesy of M.~C.~Gonzalez-García and M.~Maltoni.}
\label{fig:2nu-reac}
\end{figure}
\end{itemize}

\section{Accelerator neutrinos}
\label{sec:3nutheor_accel}

Neutrino beams produced at proton accelerators played a significant
role in the early development of the SM~\cite{Danby:1962nd,
Hasert:1973cr, Eichten:1973cs, Benvenuti:1975ru}. The idea, which
dates back to 1960~\cite{Pontecorvo:1959sn, Schwartz:1960hg} (see
Ref.~\cite{Dore:2018ldz} for a review), is to collide high-energy
protons onto a fixed target, generating a large amount of mesons
(mostly pions) that decay as
\begin{equation} \setlength{\arraycolsep}{2pt}
\begin{array}{lll} p + \mathrm{target} \rightarrow &
\multicolumn{2}{l}{\pi^\pm + X} \\ & \pi^\pm \rightarrow & \mu^\pm
+ \parenbar{\nu}_\mu \\ & & \mu^\pm \rightarrow e^\pm +
 \parenbar{\nu}_\mu + \parenbar{\nu_e} \, .
\end{array}
\end{equation} If the muons are stopped before they decay, a beam of
muon neutrinos and antineutrinos will be generated. Furthermore, a
magnetic field can be used to discard negative or positive pions,
generating a beam of muon neutrinos or antineutrinos,
respectively. This allows exploring matter effects and CP violation, 
that affect differently neutrinos and antineutrinos.

\begin{figure}[hbtp] \centering
\begin{subfigure}[b]{0.49\textwidth}
\includegraphics[width=\textwidth]{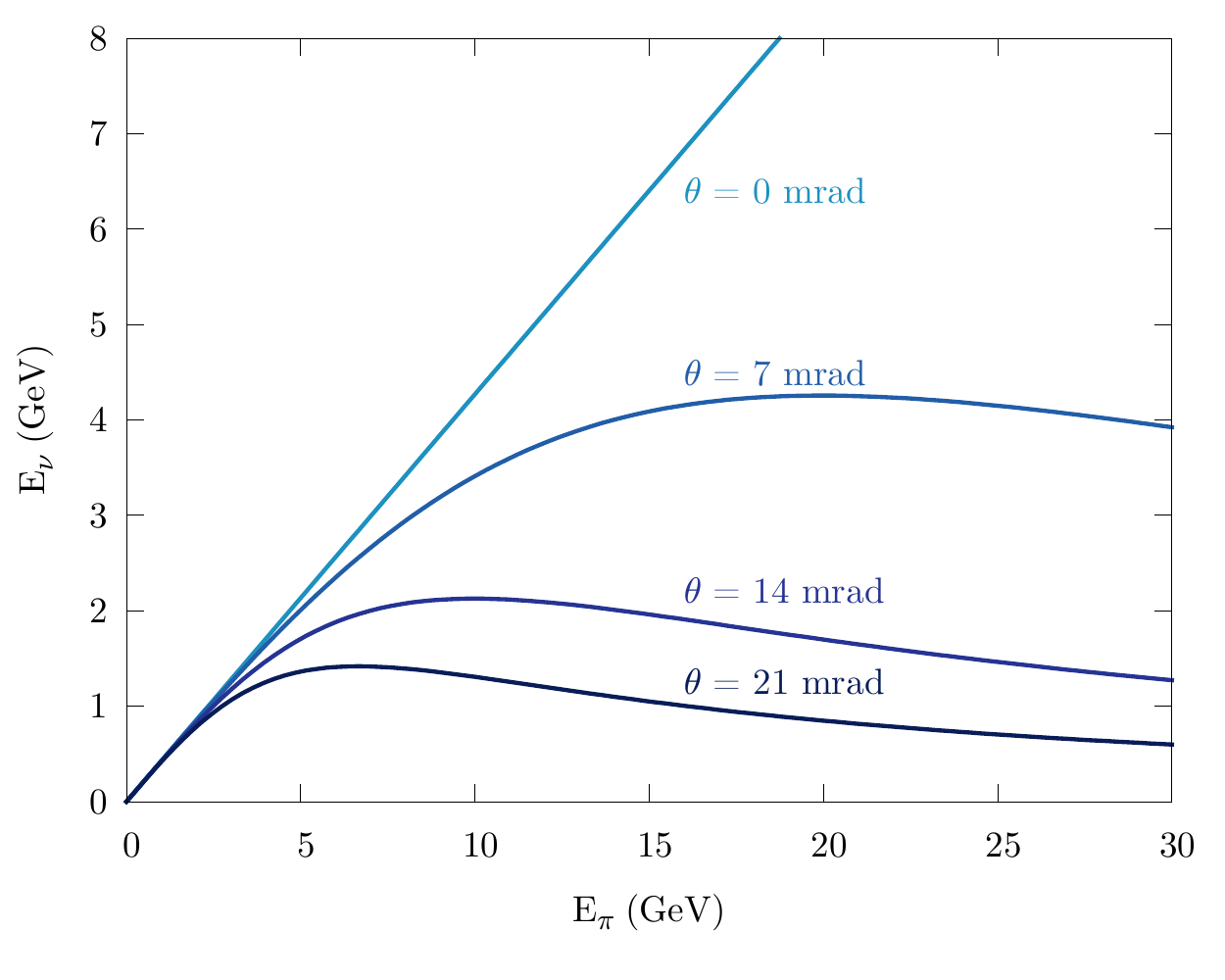}
\end{subfigure}
\begin{subfigure}[b]{0.49\textwidth}
\includegraphics[width=\textwidth]{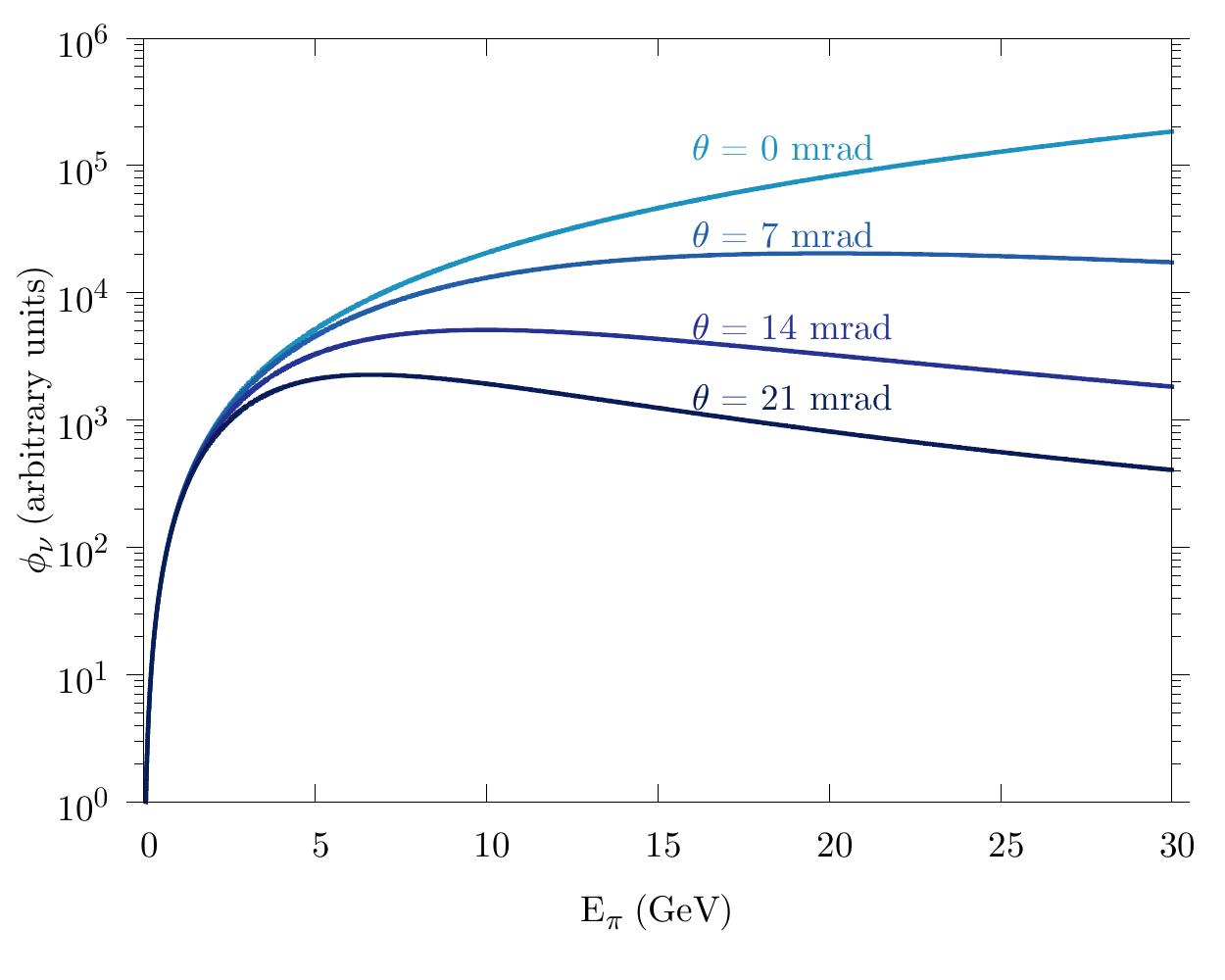}
\end{subfigure}
\caption{Neutrino energy (left) and flux (left) as a function of the
parent pion energy at different angles $\theta$ with respect to the
parent beam. Placing the neutrino detector off-axis ($\theta \neq 0$)
significantly reduces the neutrino energy spread for a non-monochromatic 
parent beam. The neutrino flux is also reduced, though.}
\label{fig:offAxis}
\end{figure}

In addition, the orientation of the detector with respect to the beam
can be exploited to precisely characterise neutrino oscillations. As pions are
spinless particles and their muonic decays two-body decays, they emit
in their rest frame neutrinos isotropically and with a fixed
energy. Considering Lorentz boosts, however, the picture changes and
isotropy is lost. In particular, in the laboratory frame the neutrino
energy $E_\nu$ and flux $\phi_\nu$ as a function of the angle $\theta$
with respect to the meson beam are given by
\begin{equation} E_\nu (\theta) = \left(1 - \frac{m_\mu^2}{m_\pi^2}
\right) \frac{E_\pi} {1+ \theta^2 \gamma^2} \, ,
\end{equation}
\begin{equation} \phi_\nu (\theta) \propto \left(\frac{2 \gamma}{1 +
\theta^2 \gamma^2}\right)^2 \, ,
\end{equation} to lowest order in $\theta$. Here, $m_\mu$ is the muon mass, $m_\pi$ and $E_\pi$ are the
mass and energy of the parent pion respectively, and $\gamma$ is its Lorentz factor. The values of
$E_\nu$ and $\phi_\nu$ as a function of the parent pion energy are
shown in \cref{fig:offAxis}. As can be seen there, an off-axis beam
($\theta \neq 0$) produces a quite monochromatic neutrino spectrum,
allowing for a more precise exploration of neutrino oscillations, at
the expense of reducing the flux.

\begin{figure}[hbtp] \centering
\includegraphics[width=0.5\textwidth]{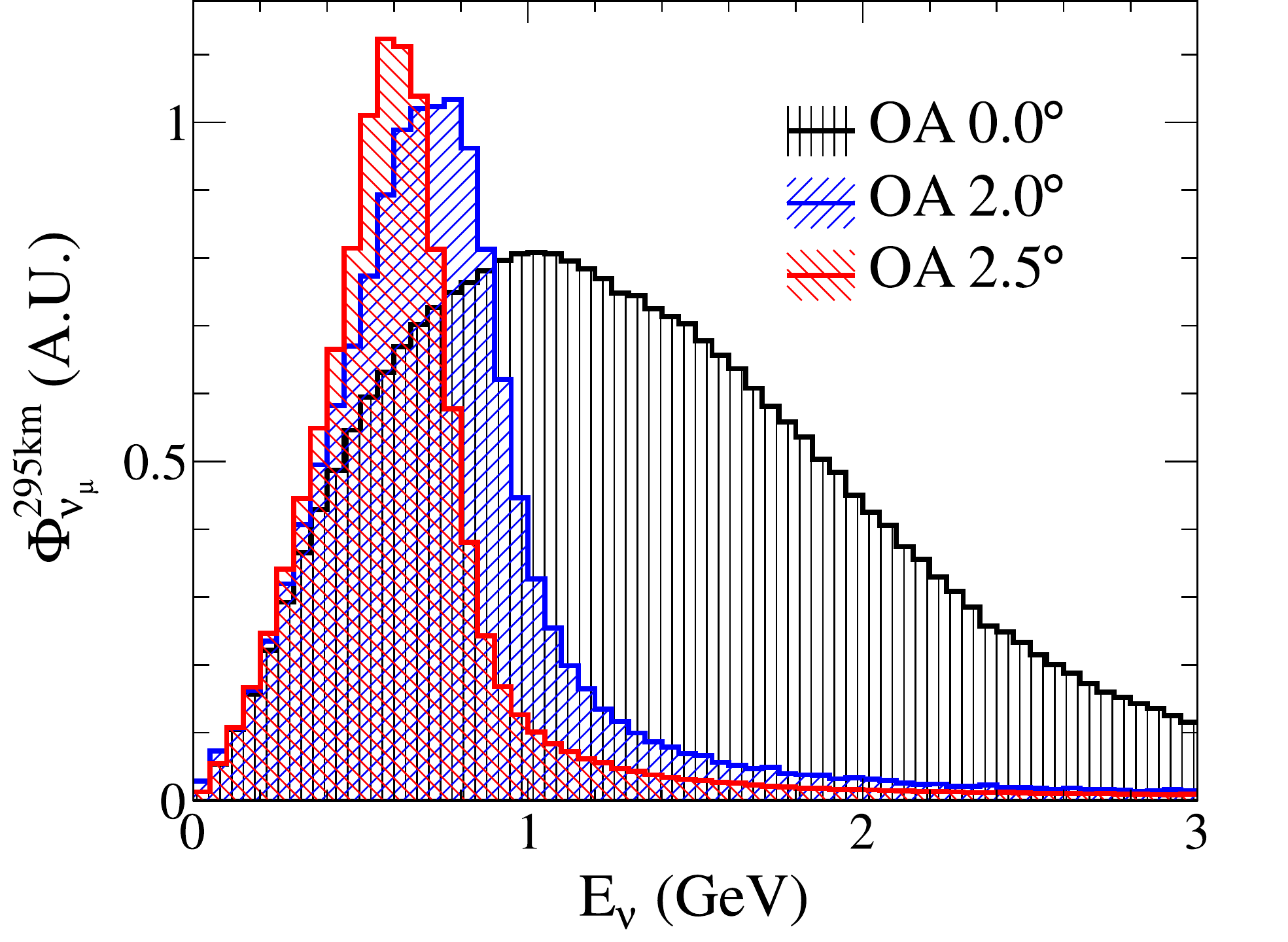}
\caption{Muon neutrino flux at the T2K experiment. OA refers to the
off-axis angle $\theta$. Extracted from Ref.~\cite{Abe:2012av}.}
\label{fig:T2Kflux}
\end{figure}

Accelerator-produced neutrinos have typical energies $\mathcal{O}
(\si{GeV})$: a standard flux is shown in \cref{fig:T2Kflux}. Thus, by
placing large detectors at distances $\mathcal{O}(\SI{100}{km})$,
the neutrino oscillation interpretation of the atmospheric $\nu_\mu$
deficit can be checked with a controlled beam. Due to the large
baselines involved, these are usually referred to as long baseline
(LBL) accelerator neutrino experiments.

\begin{figure}[hbtp] \centering
\begin{subfigure}[b]{0.49\textwidth}
\includegraphics[width=\textwidth]{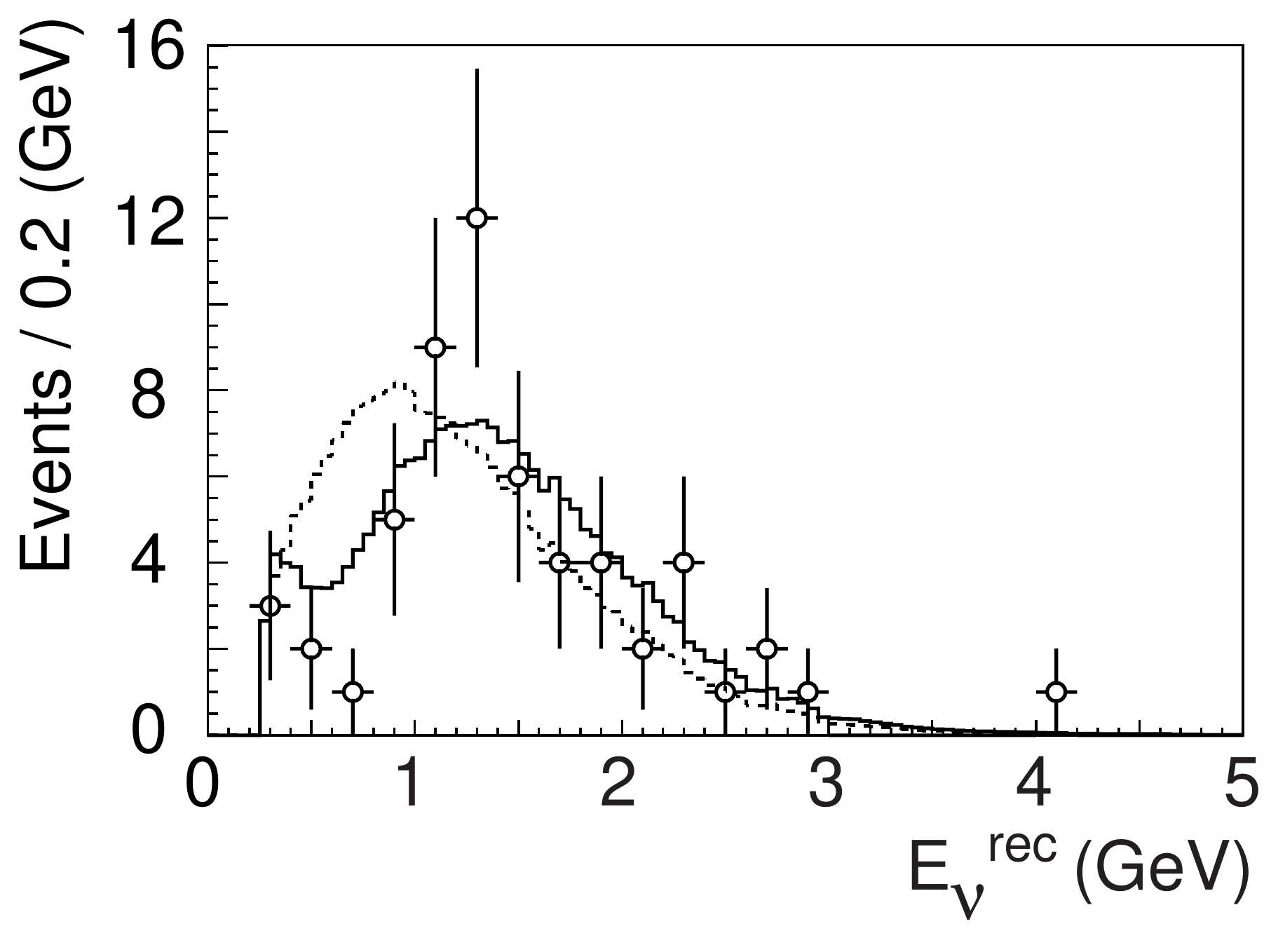}
\caption{K2K results. Figure from Ref.~\cite{Aliu:2004sq}.}
\end{subfigure} \begin{subfigure}[b]{0.49\textwidth}
\includegraphics[width=\textwidth]{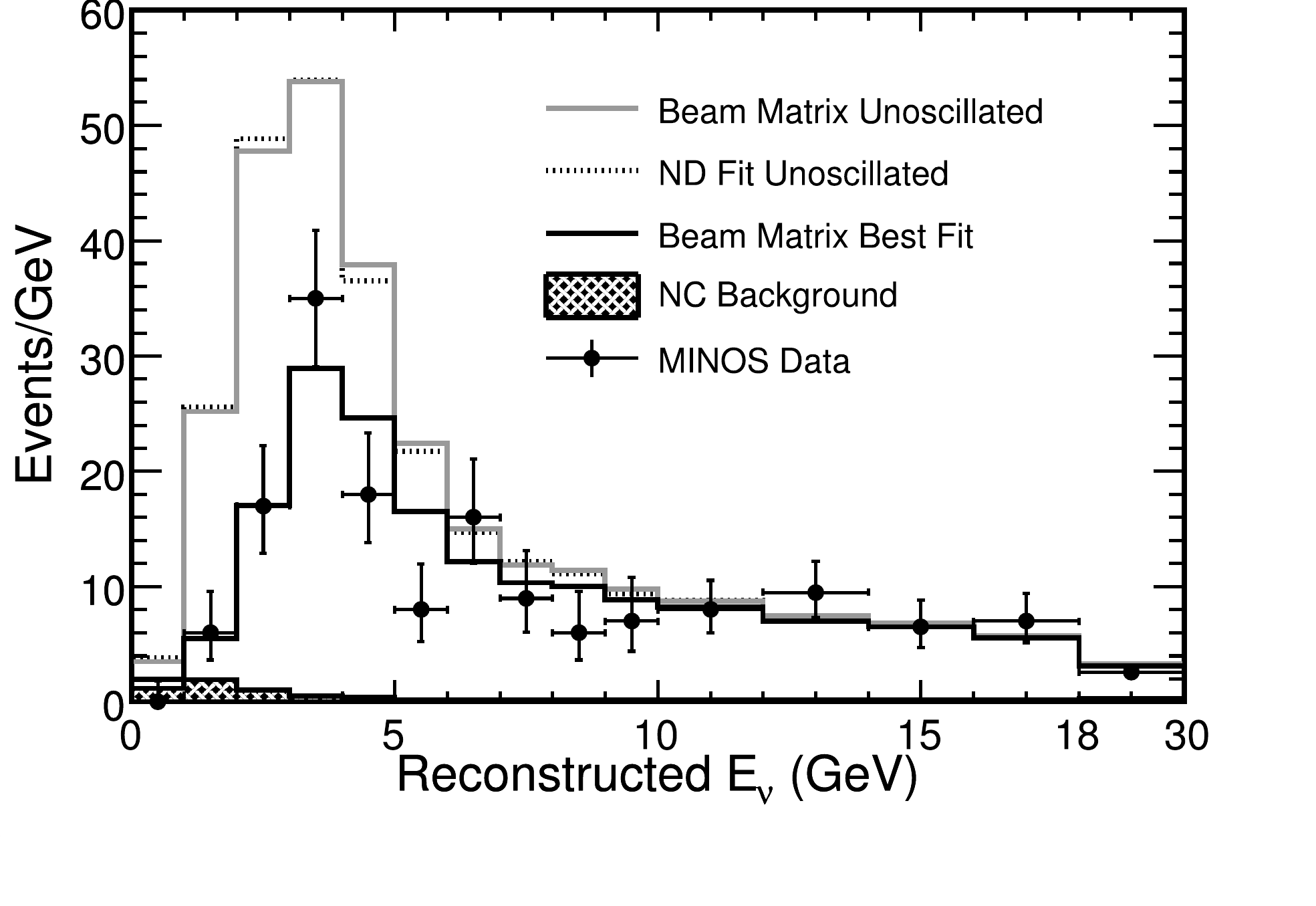}
\caption{MINOS results. Figure from Ref.~\cite{Michael:2006rx}.}
\end{subfigure}
\caption{Energy spectra of events at the K2K and MINOS detectors,
along with the expectations. For K2K,
the dashed line is the expected spectrum without oscillations
\emph{normalised to the observed number of events}. For MINOS, the
clear line shows the expected number of events without oscillations. In both cases, the dark solid line is the expected number of events with neutrino oscillations.}
\label{fig:LBL1prob}
\end{figure}

\begin{figure}[hbtp] \centering
\begin{subfigure}[b]{0.49\textwidth}
\includegraphics[width=\textwidth]{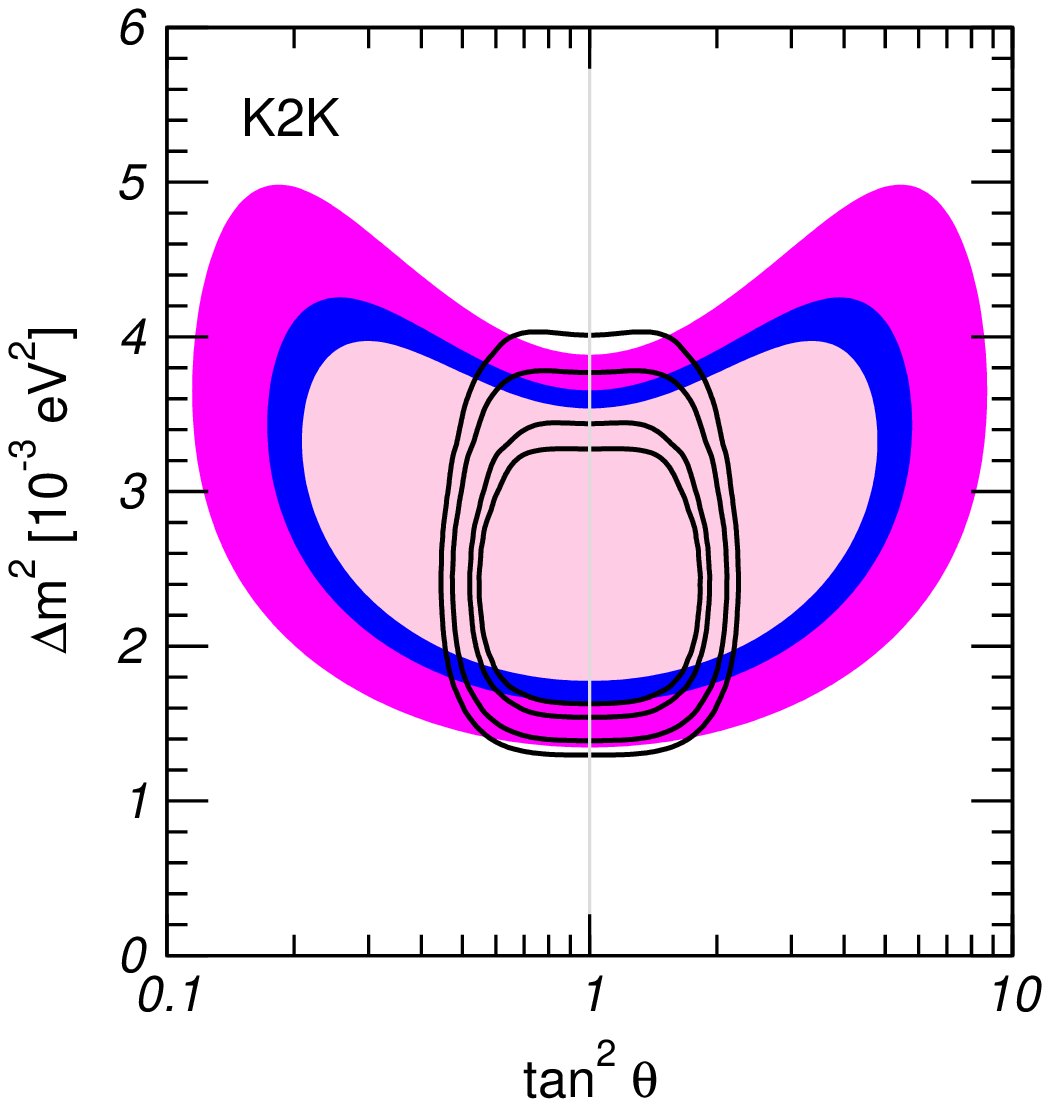}
\end{subfigure}
\begin{subfigure}[b]{0.49\textwidth}
\includegraphics[width=\textwidth]{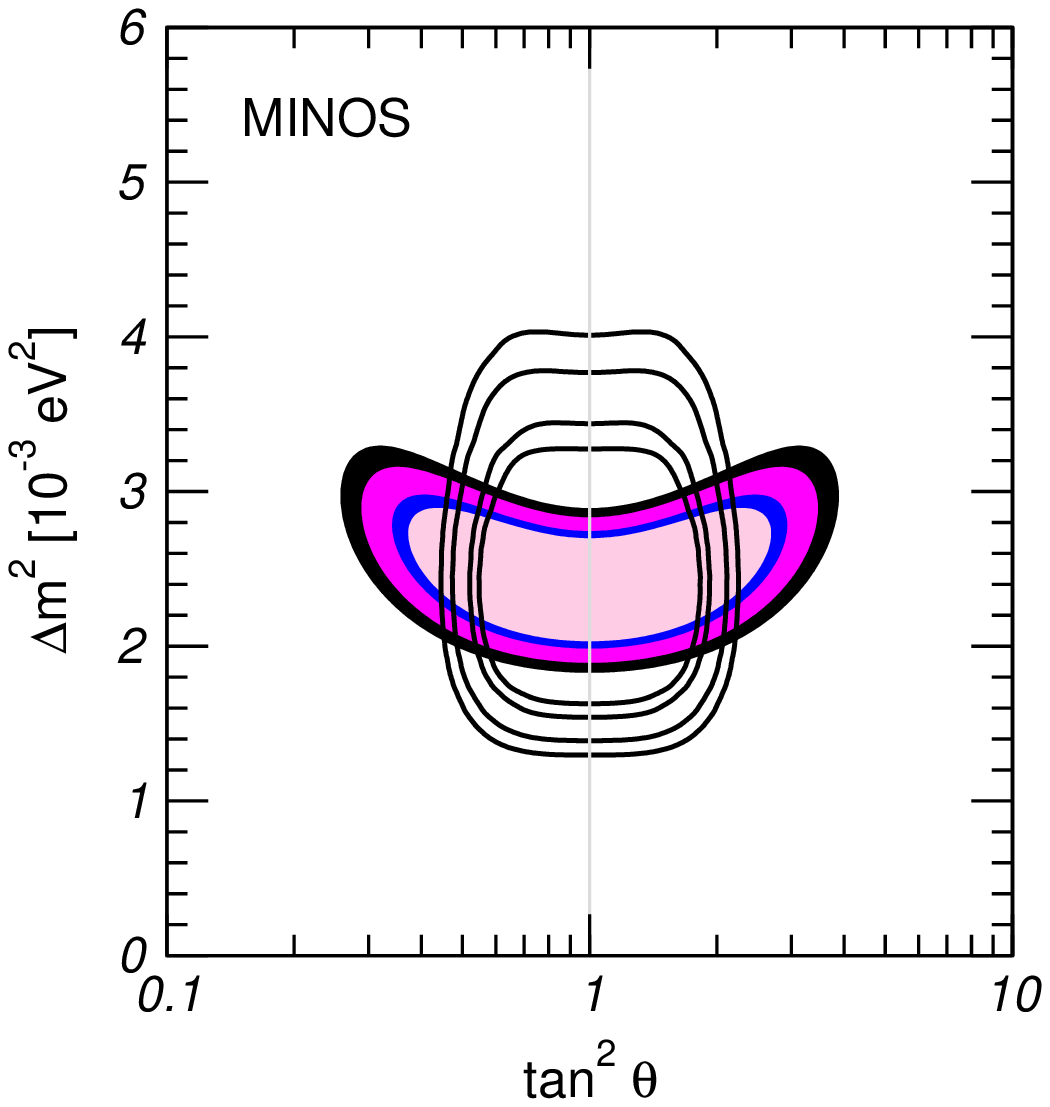}
\end{subfigure}
\caption{Allowed regions, in colour, of $|\Delta m^2|$ and $\tan^2
\theta$ that best describe K2K (left) and MINOS (right) data at 90\%
CL, 95\% CL, 99\% CL and $3\sigma$. For comparison, the corresponding
regions from atmospheric neutrino data are also shown in lines at the
same CL. Figure from Ref.~\cite{GonzalezGarcia:2007ib}.}
\label{fig:LBL1Regions}
\end{figure}

The first experiments of this kind were K2K~\cite{Ahn:2001cq} and
MINOS~\cite{Adamson:2013whj}. The former, with a baseline of about
$\SI{235}{km}$, sent neutrinos from the KEK facility to Super-Kamiokande. The latter, with a larger baseline of about $\SI{735}{km}$,
sent neutrinos from Fermilab to a detector in Soudan mine. The spectra
they observed along with the expectations with and without neutrino
oscillations are shown in \cref{fig:LBL1prob}. As the figure shows,
they both confirmed muon neutrino disappearance with a controlled
beam. The allowed parameter region is shown in \cref{fig:LBL1Regions},
which also displays the compatibility with the atmospheric $\parenbar{\nu}_\mu$
disappearance results. As in the atmospheric
neutrino analysis, in a three-neutrino framework parametrised by
\cref{eq:PMNS} the measured parameters correspond to
$|\Delta m^2| = |\Delta m^2_{32}| = |\Delta m^2_{31}|$ and $\theta =
\theta_{23}$, as long as subleading $\Delta m^2_{21}$ and
$\theta_{13}$ effects are neglected. 

All the data presented up to now robustly established the
three-neutrino mixing paradigm. For convenience and clarity, we
summarise in \cref{tab:neutrinoExperiments} the different experiments
which dominantly contribute to the present determination of the
different parameters in the chosen convention
\cref{eq:PMNS}.

\begin{table}[hbtp] \centering
  \begin{tabular}{ccc} 
    Experiment & Dominant & Important \\
    \toprule 
    Solar experiments & $\theta_{12}$ & $\Delta m^2_{21}, \,
    \theta_{13}$ \\
    \midrule 
    \begin{tabular}{@{}c@{}}Long baseline reactor experiments \\ (KamLAND)\end{tabular} & $\Delta m^2_{21}$ & $\theta_{12}, \,
    \theta_{13}$ \\
    \midrule
    \begin{tabular}{@{}c@{}}Medium baseline reactor experiments \\ (Daya Bay, RENO, Double Chooz)\end{tabular}
    & $\theta_{13}, \, |\Delta m^2_{31, 32}|$ & \\
    \midrule
    \begin{tabular}{@{}c@{}}Atmospheric experiments \\ (Super-Kamiokande, IceCube-DeepCore)\end{tabular}
    & & $\theta_{23}, \, |\Delta m^2_{31, 32}|, \, \theta_{13}, \,
    \delta_\mathrm{CP}$ \\
    \midrule
    \begin{tabular}{@{}c@{}}Accelerator LBL, $\parenbar{\nu}_\mu$ disappearance \\ (K2K, MINOS,
    T2K, \NOvA/)\end{tabular}
    & $|\Delta m^2_{31, 32}|, \, \theta_{23} $ &  \\
    \midrule
    \begin{tabular}{@{}c@{}}Accelerator LBL, $\parenbar{\nu}_e$ appearance \\ (MINOS,
    T2K, \NOvA/)\end{tabular}
    & $\delta_\mathrm{CP}$ & $\theta_{13},\, \theta_{23}$  \\
\bottomrule
\end{tabular}
\caption{Experiments contributing to the present determination of the
  oscillation parameters. For each experiment set, we indicate as
  ``Dominant'' the parameter(s) whose global determination is dominated
  by that set of experiments, whereas ``Important'' denotes parameters
  about which the experiments add some information.}
\label{tab:neutrinoExperiments}
\end{table}

Nevertheless, there were still some open questions at
the beginning of the 2010s:
\begin{itemize}
\item The ``atmospheric'' parameters were known with a rather low
precision. In particular, it was not clear whether the corresponding mixing 
angle, $\theta_{23}$ in the parametrisation~\eqref{eq:PMNS}, is
maximal ($\theta_{23} = 45^\circ$) or not.
\item The octant of $\theta_{23}$ was unknown.
\item The sign of the ``atmospheric'' squared mass splitting, $\Delta
m^2_{32}$ in our convention, was unknown. In other words, it was not
known whether the mass ordering was normal or inverted.
\item Perhaps the physically most relevant question was whether CP is
violated in the leptonic sector. This question became more significant
when a non-zero $\theta_{13}$ was measured: CP violation in the quark
sector is rather small because the mixing angles entering the
invariant~\eqref{eq:jarlskogMatrix} are small. However, leptonic
mixing angles are quite large, and so the Jarlskog invariant could be
up to three orders of magnitude larger in the leptonic sector than in the quark sector.
\end{itemize}

The first question is difficult to assess with atmospheric
neutrino data, as \cref{eq:Pmumuatm} has a minimum at $\theta_{23} =
45^\circ$. Thus, determining the maximality of this angle accounts to
measuring whether the $\parenbar{\nu}_\mu$ disappearance probability is exactly
zero for $E = \frac{L}{4 |\Delta m^2|}$ or not. Such measurement can be easily spoiled by a not well-understood neutrino spectrum and by the unavoidable experimental energy resolution. The octant of
$\theta_{23}$ and the sign of $|\Delta m^2|$ cannot be measured
without three-neutrino mixing effects and/or matter effects. And,
finally, CP violation is a three-flavour effect that requires
sensitivity to three-neutrino mixing to be detectable.

Luckily, LBL accelerator experiments can overcome these issues. By
placing the detector off-axis, the neutrino beam can be quite
monochromatic and relatively well-understood, as seen in \cref{fig:T2Kflux}. A near detector can also
reduce flux uncertainties to improve the measurements. In addition, as
oscillations driven by $\Delta m^2_{32}$ are mostly $\nu_\mu
\rightarrow \nu_\tau$, detecting $\nu_\mu \rightarrow \nu_e$
transitions driven by this squared mass splitting directly tests
three-neutrino effects. That is, $\nu_e$ appearance in LBL
experiments is sensitive all to the $\theta_{23}$ octant, the mass
ordering, and CP violation. Sensitivity to matter effects and CP
violation can be enhanced by switching between a $\nu_\mu$ and a $
\bar\nu_\mu$ beam, as these phenomena affect differently neutrinos and
antineutrinos.

\begin{figure}[hbtp] \centering
\begin{subfigure}{0.49\textwidth} \centering
\includegraphics[width=\textwidth]{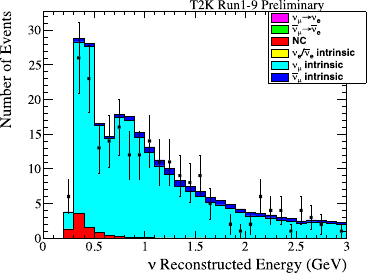}
\caption{$\nu_\mu$ disappearance.}
\end{subfigure} \begin{subfigure}{0.49\textwidth} \centering
\includegraphics[width=\textwidth]{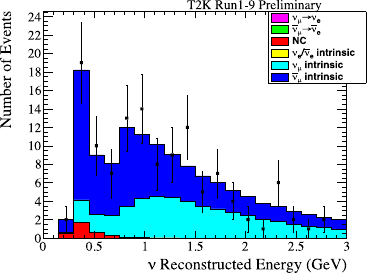}
\caption{$\bar\nu_\mu$ disappearance.}
\end{subfigure}

\begin{subfigure}{0.49\textwidth}
\includegraphics[width=\textwidth]{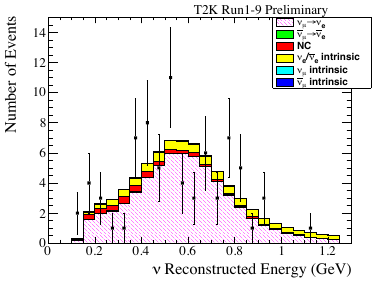}
\caption{$\nu_e$ appearance, CCQE.}
\end{subfigure} \begin{subfigure}{0.49\textwidth}
\includegraphics[width=\textwidth]{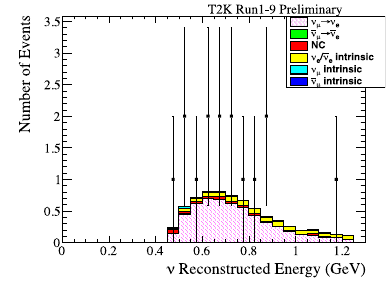}
\caption{$\nu_e$ appearance, CC1$\pi$.}
\end{subfigure} \begin{subfigure}{0.49\textwidth}
\includegraphics[width=\textwidth]{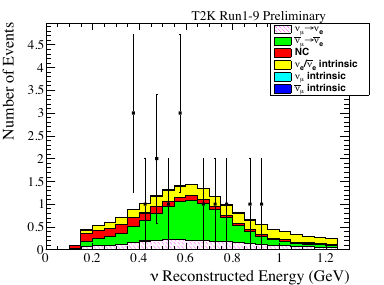}
\caption{$\bar\nu_e$ appearance, CCQE.}
\end{subfigure}

\caption{T2K results as of January 2019. The $\nu_e$ appearance
results are separated as CCQE or CC1$\pi$ depending on whether the
final state contains no visible pions or a single visible
pion. Extracted from Ref.~\cite{t2k:kek2019}.}
\label{fig:T2Kresults}
\end{figure}

\begin{figure}[hbtp] \centering
\begin{subfigure}{0.49\textwidth} \centering
\includegraphics[width=\textwidth]{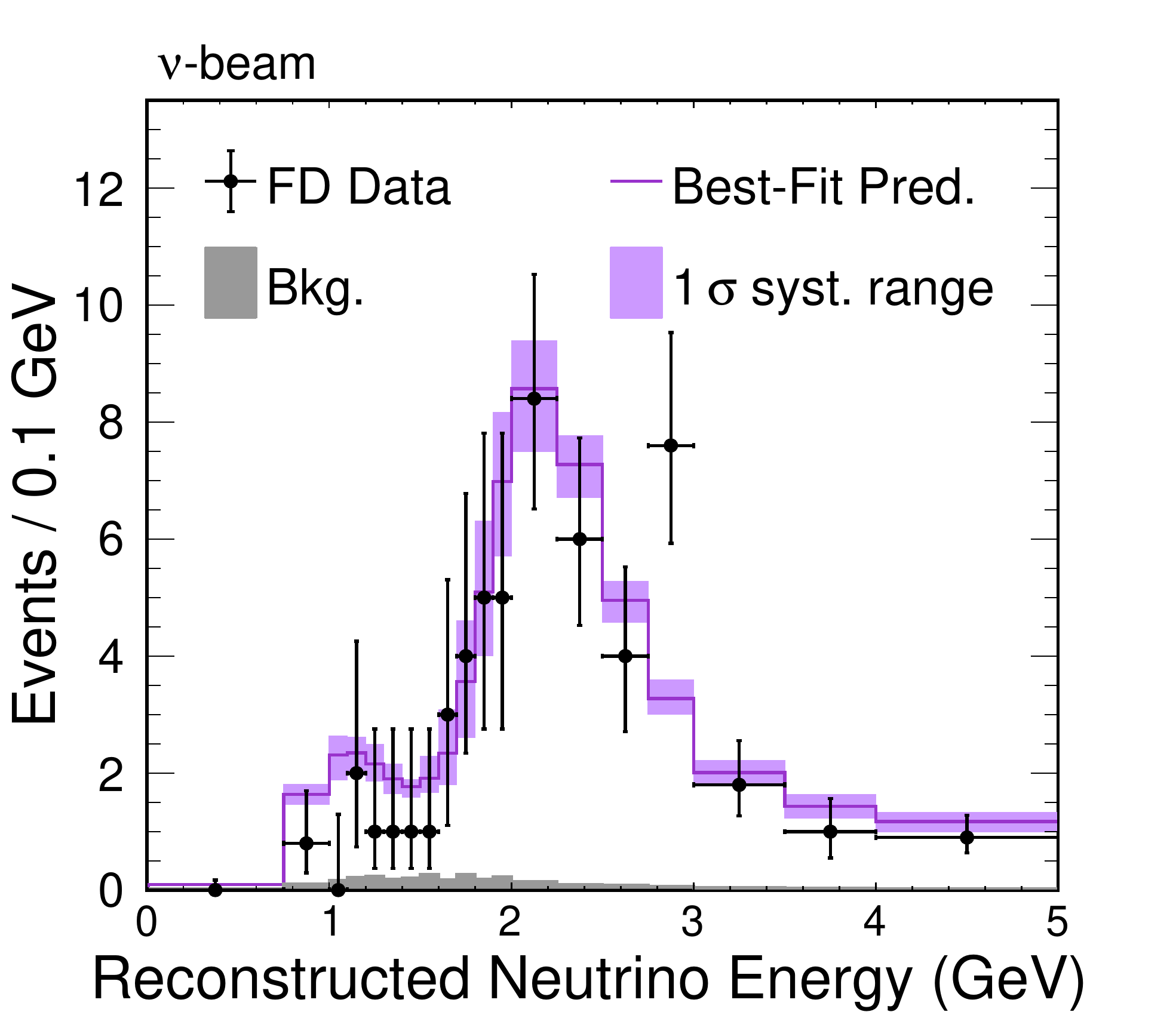}
\caption{$\nu_\mu$ disappearance.}
\end{subfigure} \begin{subfigure}{0.49\textwidth} \centering
\includegraphics[width=\textwidth]{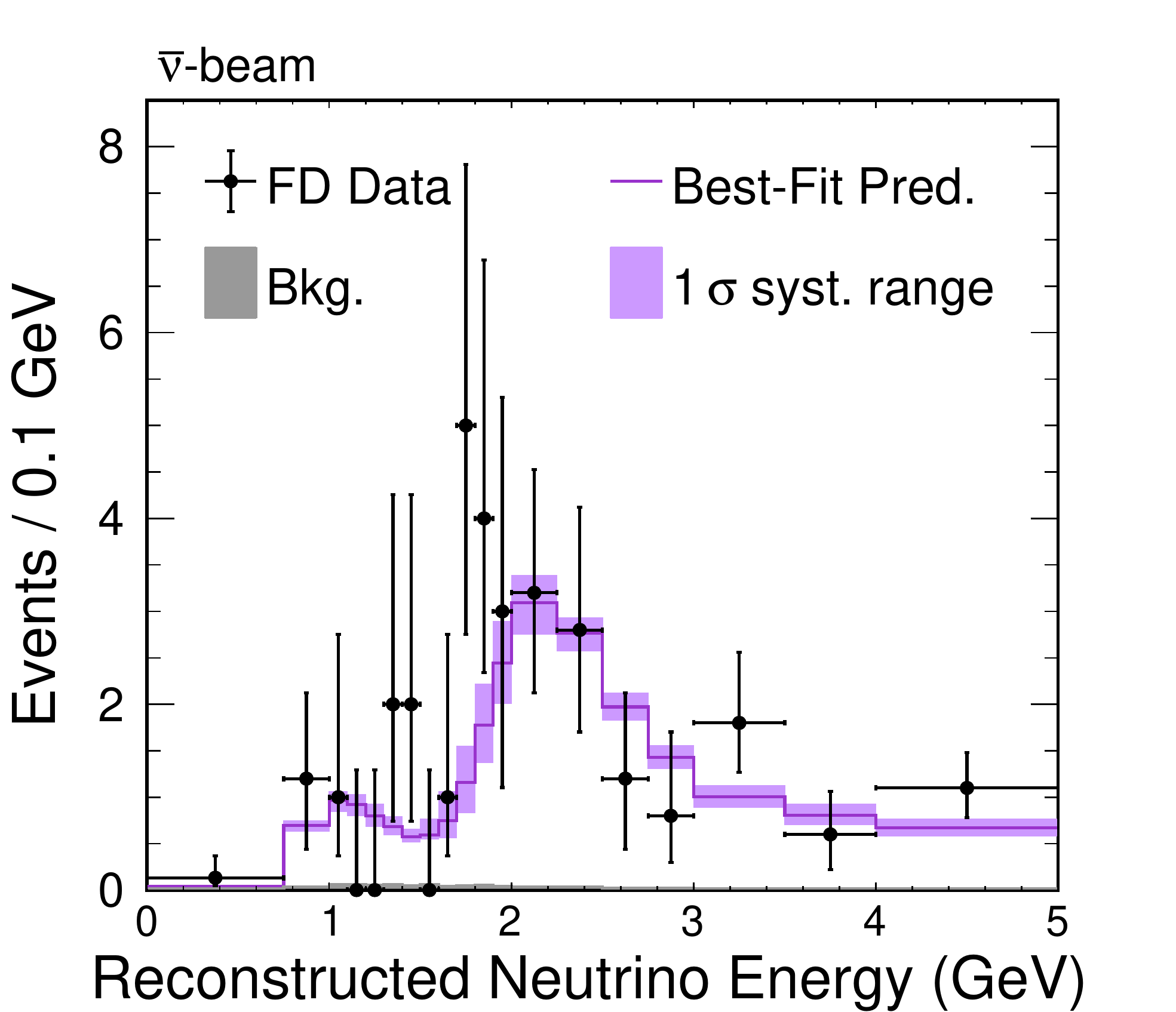}
\caption{$\bar\nu_\mu$ disappearance.}
\end{subfigure}

\begin{subfigure}{0.49\textwidth} \centering
\includegraphics[width=\textwidth]{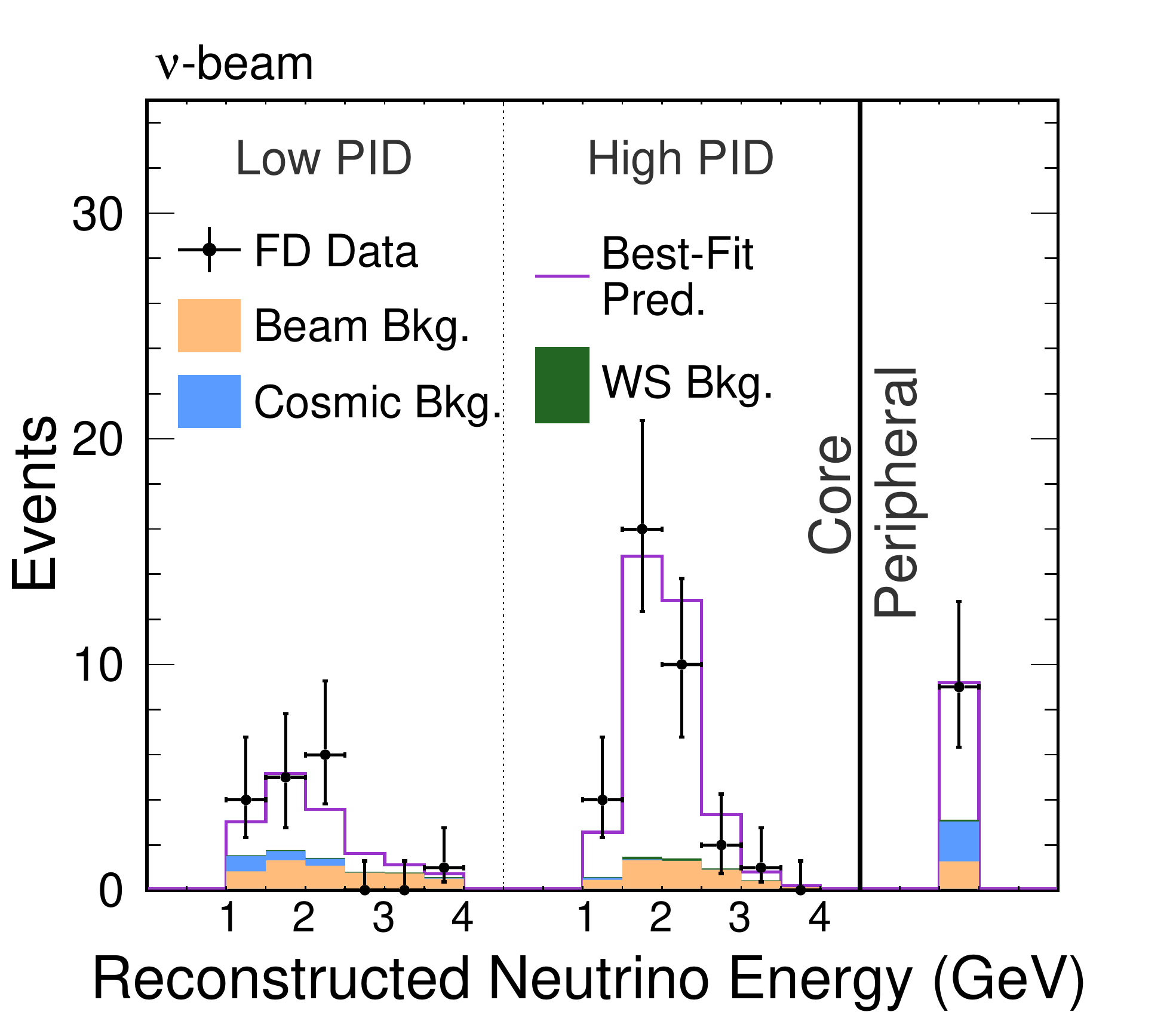}
\caption{$\nu_e$ appearance.}
\end{subfigure} \begin{subfigure}{0.49\textwidth} \centering
\includegraphics[width=\textwidth]{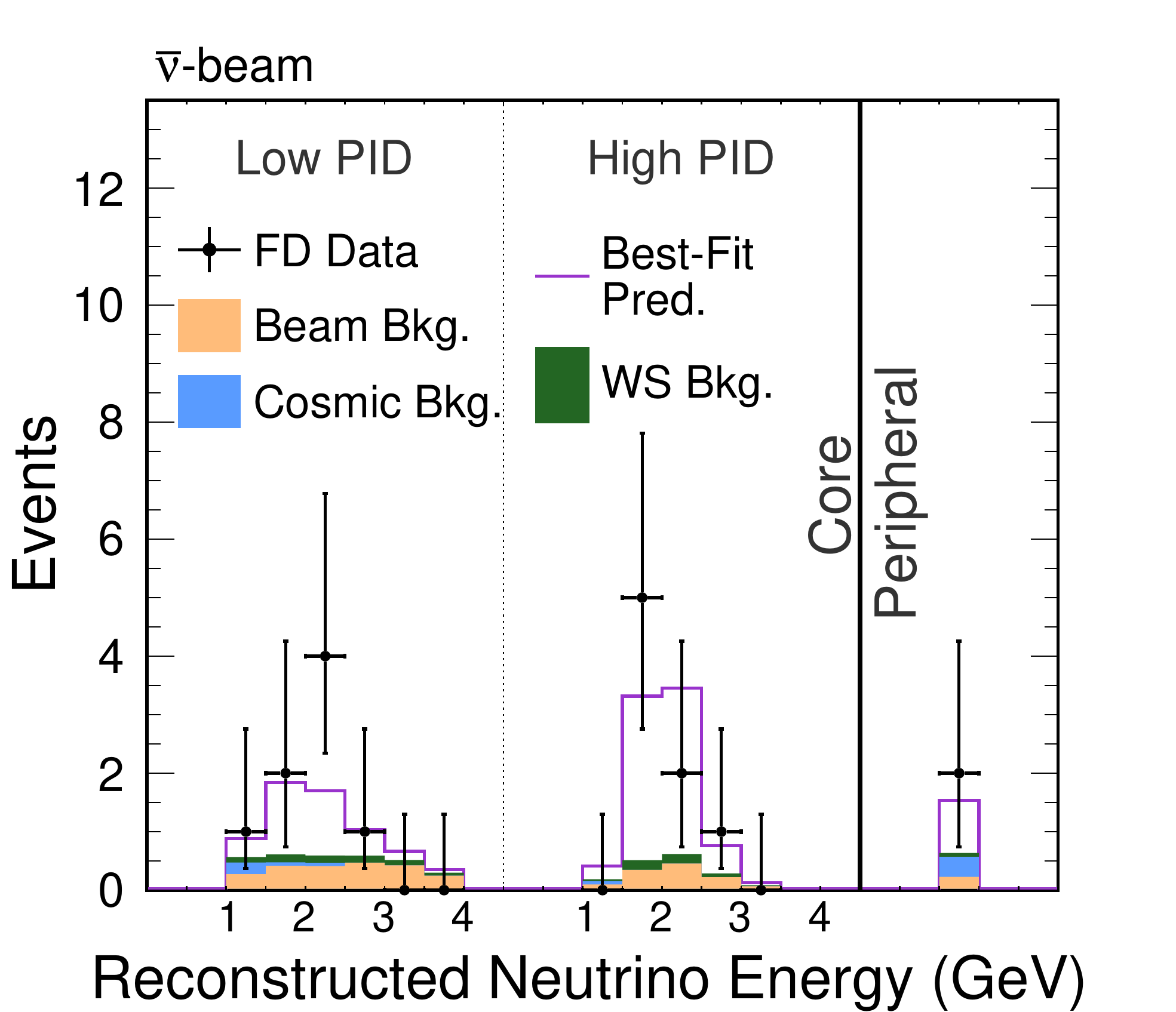}
\caption{$\bar\nu_e$ appearance.}
\end{subfigure}

\caption{\NOvA/ results as of June 2019. The appearance events are
classified in three bins from lowest to highest purity:
``Peripheral'', ``Low PID'', and ``High PID''. Extracted from
Ref.~\cite{Acero:2019ksn}.}
\label{fig:NOvAresults}
\end{figure}

With this in mind, the T2K and \NOvA/ experiments were built. T2K
sends a neutrino beam from J-PARC to Super-Kamiokande, at
\SI{295}{km}; and \NOvA/ sends a beam from Fermilab to a liquid scintillator far detector
in Minnesota, at \SI{810}{km}. The latter experiment has a higher
energy and baseline, and thus it is more sensitive to matter
effects. The latest results from both experiments are shown in
\cref{fig:T2Kresults,fig:NOvAresults}. They clearly show
$\parenbar{\nu}_\mu$ disappearance and $\parenbar{\nu}_e$ appearance.
Assessing the significance, compatibility, robustness and physical
consequences of the combined signal in a three-neutrino paradigm
including all the neutrino experiments discussed in this chapter is the main
goal that this thesis pursuits.

\subsection{Simulation of long baseline accelerator experiments}
\label{sec:LBL_simulation}

As explained above, the current unknowns in leptonic flavour mixing are
being explored with LBL accelerator experiments. However, since
each of them has limited statistical significance, a definite answer
only comes from combining the data. Furthermore, as most of the
unknowns are assessed through 3-neutrino effects, it is essential to
combine these experiments with solar, atmospheric, and reactor
neutrino results in order to get the most out of the data and assess
the true significance of the signals. Because of that,
phenomenological simulations of the \NOvA/ and T2K experiments have
been developed as part of this thesis. These simulations allow to
analyse the data and approximately reproduce the results from each
experiments, so that they can be embedded in a global framework.

\subsubsection{Analysis framework}

As in any statistical data analysis, the compatibility between data
and different parameter values is determined through a log-likelihood
test.  In particular, for an experiment with data
$\text{data}_{\text{exp}}$ and for a set of parameters $\Theta$, the
log-likelihood for that experiment $\chi_\text{exp}^2$ is given by
\begin{equation} \chi^2_\text{exp} (\Theta) = - \ln
\mathcal{P}(\text{data}_{\text{exp}} | \Theta) \, ,
\label{eq:chisqDefinition}
\end{equation} where $\mathcal{P}(\text{data}_{\text{exp}} | \Theta)$
is the probability of obtaining the data points
$\text{data}_{\text{exp}}$ assuming the parameters to be $\Theta$. The
probabilistic nature stems both from systematic uncertainties in the
experiment and from quantum mechanical fluctuations. The test
statistic~\eqref{eq:chisqDefinition} is particularly useful because,
for independent experiments, $\chi^2 = {\displaystyle
\sum_{\text{exp}}} \chi^2_\text{exp}$. For the particular case of
binned data $n_i$ following a Poisson distribution with averages
$\mu_i$~\cite{Tanabashi:2018oca},
\begin{equation} \chi^2_\text{exp} = 2 \sum_i \mu_i - n_i + n_i \log
\frac{n_i}{\mu_i} \, ,
\end{equation} whereas if the data follow a Gaussian distribution with
means $\mu_i$ and standard deviations $\sigma_i$,
\begin{equation} \chi^2_\text{exp} = \sum_i \frac{(n_i -
\mu_i)^2}{\sigma_i^2} \, .
\end{equation}

The best fit parameters are chosen to be the ones that minimise
$\chi^2 (\Theta)$, whereas confidence intervals are obtained by
evaluating
\begin{equation} \Delta \chi^2 (\Theta) = \chi^2 (\Theta) -
\min_{\Theta} \chi^2(\Theta) \, .
\label{eq:deltaChisqDefinition}
\end{equation} According to Wilks' theorem~\cite{Wilks:1938dza}, in the
large sample limit (what is usually known as the \emph{Gaussian
limit}) the test statistic~\eqref{eq:deltaChisqDefinition} is
distributed following a $\chi^2$ distribution whose number of degrees
of freedom is the number of parameters in $\Theta$. Therefore, in the
Gaussian limit, a $\Delta \chi^2$ confidence interval $[0, \Delta
\chi^2_\lambda]$ with confidence level (CL) $\lambda$ fulfils
\begin{equation} \int_0^{\Delta \chi^2_\lambda}
\chi^2_{\dimension \Theta} (x) \, \mathrm{d}x = \lambda \,
,
\label{eq:confidenceLevels}
\end{equation} where $\chi^2_{\dimension \Theta} (x)$ is a
$\chi^2$ distribution whose number of degrees of freedom is the number
of parameters in $\Theta$. Since $\chi^2 = \chi^2
\left(\Theta\right)$, confidence intervals on $\chi^2$ can be
translated into confidence regions on $\Theta$ in a straightforward
manner. \Cref{tab:chisqLimits} shows numerically obtained values of
$\Delta \chi^2_\lambda$ for different confidence levels and
degrees of freedom.

Finally, if $\Theta$ contains additional parameters apart from the
ones on which we want to obtain confidence intervals, $\chi^2$ is
minimised over them and the number of degrees of freedom is accordingly reduced. Examples of such parameters include variables
parametrising systematic uncertainties or additional model parameters.

\begin{table}[hbtp] \centering
\begin{tabular}{cccccc} \toprule \multirow{2}{*}{CL (\%)} &
\multicolumn{5}{c}{$\Delta \chi^2_\lambda$} \\ \cmidrule(l){2-6} & 1
d.o.f. & 2 d.o.f. & 3 d.o.f. & 4 d.o.f. & 5 d.o.f.\\ \midrule 68.27 &
1.00 & 2.30 & 3.53 & 4.72 & 5.89 \\ 90.00 & 2.71 & 4.61 & 6.25 & 7.78
& 9.24 \\ 95.00 & 3.84 & 5.99 & 7.81 & 9.49 & 11.07 \\ 95.45 & 4.00 &
6.18 & 8.02 & 9.72 & 11.31 \\ 99.00 & 6.63 & 9.21 & 11.34 & 13.28 &
15.09 \\ 99.73 & 9.00 & 11.83 & 14.16 & 16.25 & 18.21 \\ \bottomrule
\end{tabular}
\caption{Value of $\Delta \chi^2_\lambda$ corresponding to a given
confidence level (CL) for different degrees of freedom (d.o.f.) of the
underlying $\chi^2$ distribution.}
\label{tab:chisqLimits}
\end{table}

\subsubsection{Prediction of the number of events} The only ingredient
left for simulating LBL accelerator experiments is the expected number
of events in a given energy bin $i$ and for a given channel $\alpha$
($\alpha \in \{\nu_\mu, \nu_e, \bar\nu_\mu, \bar\nu_e\}$). This can
generically be calculated as
\begin{equation} N_i^\alpha = N_\text{bkg,i} + \int^{E_{i+1}}_{E_i} \,
\mathrm{d}E_\text{rec} \int_0^\infty \, \mathrm{d}E_\nu
R(E_\text{rec}, E_\nu) \varepsilon (E_\nu) \sum_\beta \frac{\mathrm{d}
\Phi^\beta}{\mathrm{d} E_\nu } P_{\nu_\beta
\rightarrow \nu_\alpha}(E_\nu) \sigma_\alpha (E_\nu) \, ,
\end{equation} where
\begin{itemize}
\item $N_\text{bkg,i}$ is the number of background events in that bin.
If there is a neutrino component in the background, its oscillation
has to be consistently included.
\item $[E_i, E_{i+1}]$ are the bin limits.
\item $E_\text{rec}$ is the reconstructed neutrino energy.
\item $E_\nu$ is the true neutrino energy.
\item $R(E_\text{rec}, E_\nu)$ is the energy reconstruction function:
the probability to observe a reconstructed energy $E_\text{rec}$ if
the true neutrino energy is $E_\nu$. We usually take it to be Gaussian, i.e.,
\begin{equation} R(E_\text{rec}, E_\nu) = \frac{1}{\sqrt{2 \pi}
\sigma_E E_\nu} \exp\left[-\frac{1}{2 \sigma_E^2}\left(\frac{E_\nu -
E_\text{rec}} {E_\nu}\right)^2\right] \, .
\end{equation} That is, $\frac{E_\nu - E_\text{rec}}{E_\nu}$ is
Gaussian-distributed around zero with standard deviation
$\sigma_E$. The $\frac{1}{E_\nu}$ prefactor is for normalisation
purposes.
\item $\varepsilon$ is the detection efficiency.
\item $\frac{\mathrm{d} \Phi^\beta}{\mathrm{d} E_\nu }$ is the
incident neutrino flux with flavour $\beta$.
\item $P_{\nu_\beta \rightarrow \nu_\alpha} (E_\nu)$ is the $\nu_\beta
  \rightarrow \nu_\alpha$ transition probability.
\item $\sigma_\alpha$ is the $\nu_\alpha$ detector cross section.
\end{itemize} The antineutrino channels are obtained switching $\nu$ by
$\bar{\nu}$.

The number of background events, neutrino fluxes, cross sections,
and energy resolutions can be usually obtained from the public
information released by the experimental collaborations. The detection
efficiencies can then be adjusted to reproduce the official predicted
spectra. Finally, one can add systematic uncertainties on the
global normalisation and/or on the reconstructed energy scale (i.e., 
the bin limits in reconstructed energy are multiplied by an additional variable 
over which $\chi^2$ is minimised).  These can be sometimes obtained from the 
collaborations,
or otherwise they are adjusted to reproduce their confidence
intervals.

\section{Summary}

In the last decades, the study of neutrino flavour transitions has 
tremendously progressed. What started as odd anomalies in solar and 
atmospheric neutrino fluxes has evolved into a precision science, where 
tiny distortions in well-controlled neutrino beams are detected and 
explored. Remarkably, all data from many different experiments, 
detection techniques, neutrino energies, baselines, traversed matter 
densities... fits within a simple 3 massive neutrino framework. 

This effort of parametrising our first laboratory evidence for BSM physics is 
currently facing the challenge of determining its 
last unknowns. These are the maximality and octant of the mixing angle 
involved in atmospheric muon neutrino disappearance, the sign of the 
largest squared mass splitting, and the possible presence of CP 
violation. All these unknowns are to be assessed by LBL accelerator 
experiments which, at the same time, will serve as a validation of the 3 
massive neutrino paradigm.

Due to the importance of these experiments, in particular to assess 
leptonic CP violation, detailed simulations to describe their results 
have been developed. This will allow to combine their data with other 
experiments in a consistent framework in \cref{chap:3nufit_fit}. Thus,
a global picture will emerge that will precisely quantify our 
current experimental knowledge about neutrino flavour transitions and 
leptonic CP violation.

\chapter{Three-neutrino fit to oscillation data: results}
\label{chap:3nufit_fit}

\epigraph{\emph{Shut up and calculate.}}{ --- David Mermin}

\epigraph{\emph{La vocación del arma es el blanco.}}{ --- Manuel Machado}

In \cref{chap:3nufit_theor}, we have summarised the results from the
relevant neutrino experiments that have conclusively observed leptonic 
flavour transitions. Before the \NOvA/ and T2K experiments released 
data, each of them could be analysed to a good approximation in an 
effective two-neutrino framework. Because of that, there were three 
unknowns that cannot be assessed in this approximation: the octant of 
the mixing angle involved in atmospheric muon neutrino disappearance, 
the sign of the largest squared mass splitting, and the possible 
presence of leptonic CP violation. 

LBL accelerator experiment address all these three questions. Once their 
data is analysed as described in \cref{sec:LBL_simulation}, it can be 
combined in a three-neutrino framework\footnote{As described in the
  introduction of \cref{chap:3nufit_theor}, this three-neutrino
  framework is parametrised by three non-independent squared mass
  splittings ($\Delta m^2_{32}$, $\Delta m^2_{21}$ and $\Delta
  m^2_{31} = \Delta m^2_{32} + \Delta m^2_{21}$) and the leptonic
  mixing matrix~\eqref{eq:PMNS}. Since Majorana phases $\{\eta_1,
  \eta_2\}$ do not enter neutrino oscillations, they will be ignored.}
with data from solar, atmospheric and reactor neutrinos as analysed by
the NuFIT group. In this chapter, we will discuss the results,
synergies and tensions that arise from that combination. We will begin
with the results obtained at the beginning of this thesis, as the
\NOvA/ experiment released its first data.

\section{Global fit as of November 2016}

\subsection{Global analysis: determination of oscillation parameters}
\label{sec:nufit3_global}

\subsubsection{Data samples analysed}
\label{subsec:nufit3_data}

In the analysis of solar neutrino data we consider the total rates
from the radiochemical experiments Chlorine~\cite{Cleveland:1998nv},
Gallex/GNO~\cite{Kaether:2010ag} and SAGE~\cite{Abdurashitov:2009tn},
the results for the four phases of
Super-Kamiokande~\cite{Hosaka:2005um, Cravens:2008aa, Abe:2010hy,
sksol:nakano2016, sksol:ichep2016}, the data of the three phases of
SNO included in the form of the parametrisation presented
in~\cite{Aharmim:2011vm}, and the results of both Phase-I and Phase-II
of Borexino~\cite{Bellini:2011rx, Bellini:2008mr, Bellini:2014uqa}.

Results from LBL accelerator experiments as of November 2016 include the
final energy distribution of events from MINOS~\cite{Adamson:2013whj,
Adamson:2013ue} in $\nu_\mu$ and $\bar\nu_\mu$ disappearance and
$\nu_e$ and $\bar\nu_e$ appearance channels, as well as the
energy spectrum for T2K in the same four channels~\cite{t2k:ichep2016,
t2k:susy2016} and for NO$\nu$A on the $\nu_\mu$ disappearance and
$\nu_e$ appearance neutrino modes~\cite{nova:nu2016}.

Data samples on $\bar\nu_e$ disappearance from reactor include the
full results of the long baseline reactor data in
KamLAND~\cite{Gando:2010aa}, as well as the results from medium
baseline reactor experiments from CHOOZ~\cite{Apollonio:1999ae} and
Palo Verde~\cite{Piepke:2002ju}. Concerning running experiments we
include spectral data from
Double-Chooz~\cite{dc:moriond2016} and Daya-Bay~\cite{db:nu2016},
while for RENO we use the total rates obtained with their largest data
sample corresponding to 800 days of data-taking~\cite{reno:nu2014}.

In the analysis of the reactor data, the unoscillated reactor flux is
determined as described in~\cite{Kopp:2013vaa} by including in the fit
the results from short baseline reactor data from
ILL~\cite{Kwon:1981ua}, G\"osgen~\cite{Zacek:1986cu},
Krasnoyarsk~\cite{Vidyakin:1987ue, Vidyakin:1994ut},
ROVNO88~\cite{Afonin:1988gx}, ROVNO4~\cite{Kuvshinnikov:1990ry},
Bugey3~\cite{Declais:1994su}, Bugey4~\cite{Declais:1994ma}, and
SRP~\cite{Greenwood:1996pb}.

For the analysis of atmospheric neutrinos we include the results from
IceCube/DeepCore 3-year data~\cite{Aartsen:2014yll}.

The above data sets constitute the samples included in our NuFIT 3.0
analysis. For Super-Kamiokande atmospheric neutrino data from phases
SK1--4 we will comment on our strategy in Sec.~\ref{subsec:nufit3_SK}.

\subsubsection{Results: oscillation parameters}
\label{subsec:nufit3_oscparam}

The results of our standard analysis are presented in
Figs.~\ref{fig:nufit3_region-glob} and~\ref{fig:nufit3_chisq-glob}
where we show projections of the allowed six-dimensional parameter
space.\footnote{$\Delta\chi^2$ tables from the global analysis
corresponding to all 1-dimensional and 2-dimensional projections are
available for download at the NuFIT website~\cite{nufit}.}  In all
cases when including reactor experiments we leave the normalisation of
reactor fluxes free and include data from short baseline (less than
100 m) reactor experiments. A previous
analysis~\cite{Gonzalez-Garcia:2014bfa, GonzalezGarcia:2012sz} studied
the impact of this choice versus that of fixing the reactor fluxes to
the prediction of the latest calculations~\cite{Mueller:2011nm,
Huber:2011wv, Mention:2011rk}.  As expected, the overall description
is better when the flux normalisation $f_\text{flux}$ is fitted
against the data.  We find $\chi^2(f_\text{flux}~\text{fix}) -
\chi^2(f_\text{flux}~\text{fit}) \simeq 6$ which is just another way
to quantify the well-known short baseline reactor anomaly to be $\sim
2.5\sigma$.  However, the difference in the resulting parameter
determination (in particular for $\theta_{13}$) between these two
reactor flux normalisation choices has become marginal, since data
from the reactor experiments with near detectors such as Daya-Bay,
RENO and Double-Chooz (for which the near-far comparison allows for
flux-normalisation independent analysis) is now dominant.
Consequently, in what follows we show only the $\Delta\chi^2$
projections for our standard choice with fitted reactor flux
normalisation.

\begin{pagefigure}\centering
  \includegraphics[width=0.79\textwidth]{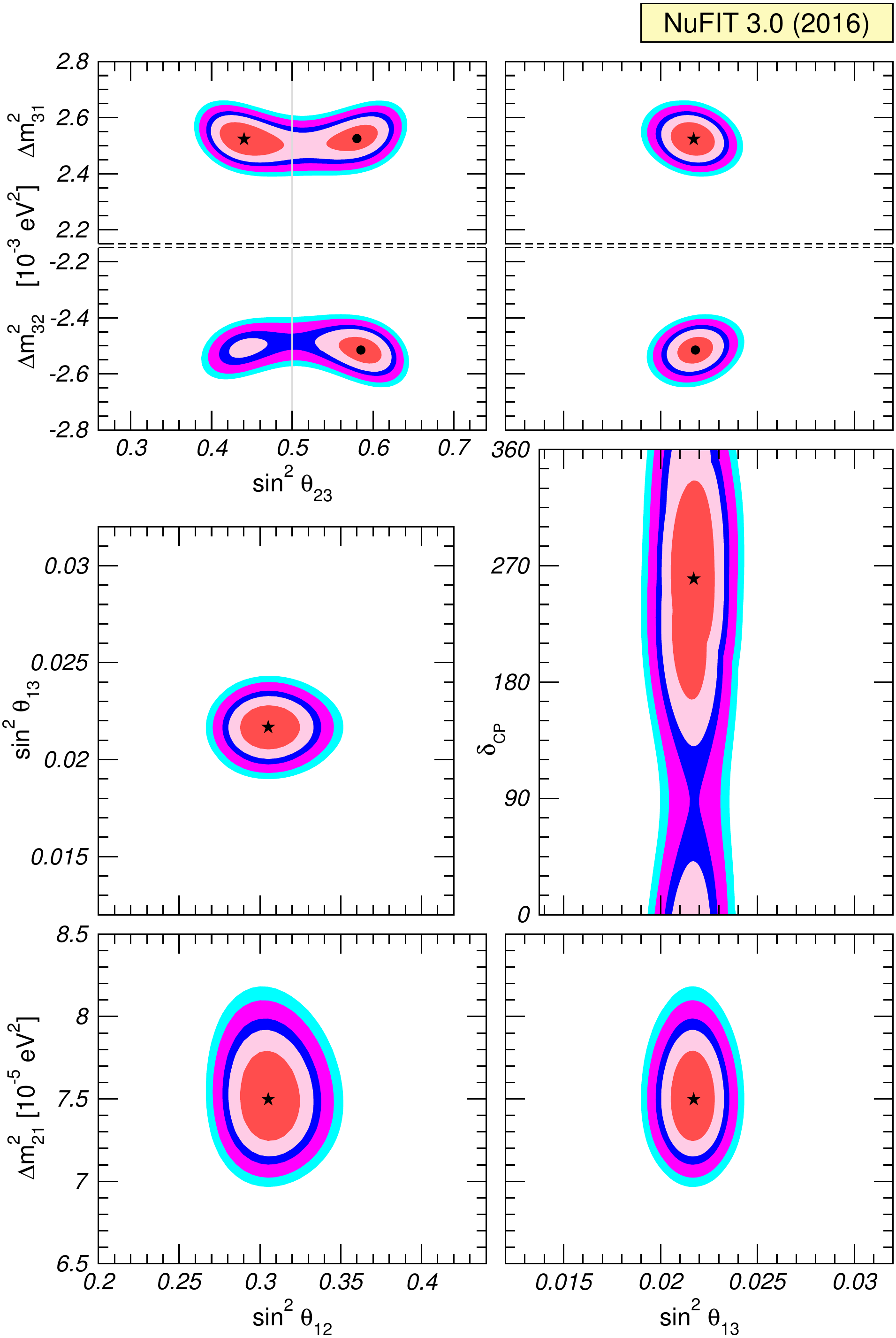}
  \caption{Global $3\nu$ oscillation analysis. Each panel shows the
two-dimensional projection of the allowed six-dimensional region after
marginalisation with respect to the undisplayed parameters. The
different contours correspond to $1\sigma$, 90\%, $2\sigma$, 99\%,
$3\sigma$ CL (2 dof).  The normalisation of reactor fluxes is left
free and data from short baseline reactor experiments are included as
explained in the text. Note that as atmospheric mass-squared splitting
we use $\Delta m^2_{31}$ for NO and $\Delta m^2_{32}$ for IO. The
regions in the four lower panels are obtained from $\Delta\chi^2$
minimised with respect to the mass ordering.}
  \label{fig:nufit3_region-glob}
\end{pagefigure}

\begin{pagefigure}\centering
\includegraphics[width=0.84\textwidth]{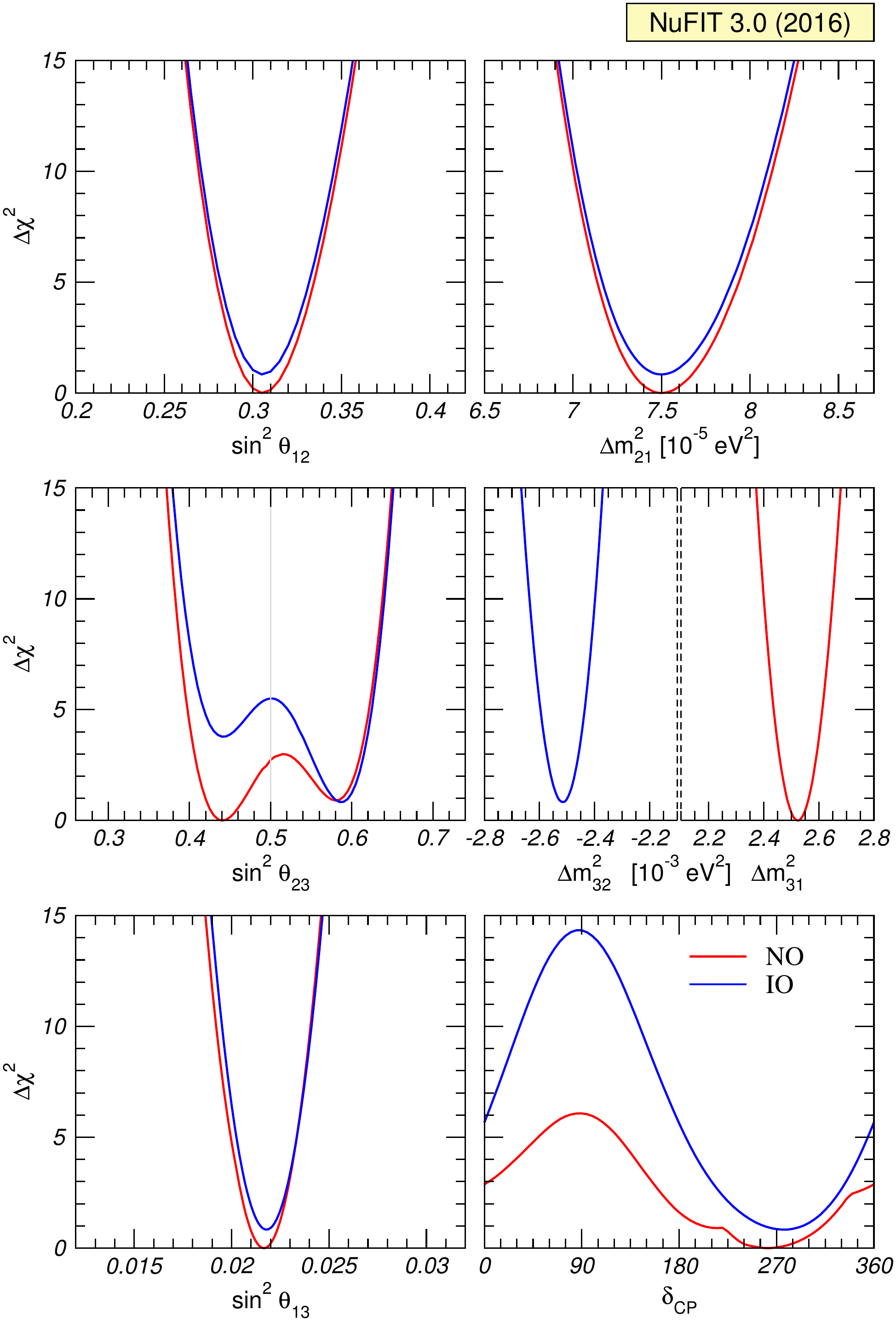}
  \caption{Global $3\nu$ oscillation analysis. The red (blue) curves
correspond to Normal (Inverted) Ordering.  The normalisation of
reactor fluxes is left free and data from short baseline reactor
experiments are included.  Note that as atmospheric mass-squared
splitting we use $\Delta m^2_{31}$ for NO and $\Delta m^2_{32}$ for
IO.}
  \label{fig:nufit3_chisq-glob}
\end{pagefigure}

The best fit values and the derived ranges for the six parameters at
the $1\sigma$ ($3\sigma$) level are given in
Tab.~\ref{tab:nufit3_bfranges}.  For each parameter $x$ the ranges are
obtained after marginalising with respect to the other
parameters\footnote{We use the term
``marginalisation'' over a given parameter as synonym for minimising
the $\chi^2$ function with respect to that parameter.}  and under the
assumption that $\Delta\chi_\text{marg}^2(x)$ follows a $\chi^2$
distribution. Hence the $1\sigma$ ($3\sigma$) ranges are given by the
condition $\Delta\chi_\text{marg}^2(x)=1$ (9). It is known that
because of its periodic nature and the presence of parameter
degeneracies the statistical distribution of the marginalised
$\Delta\chi^2$ for $\delta_\text{CP}$ and $\theta_{23}$ (and
consequently the corresponding CL intervals) may be
modified~\cite{Schwetz:2006md, Blennow:2014sja}.  In
Sec.~\ref{sec:nufit3_MC} we will discuss and quantify these effects.

In Tab.~\ref{tab:nufit3_bfranges} we list the results for three
scenarios. In the first and second columns we assume that the ordering
of the neutrino mass states is known \emph{a priori} to be Normal or
Inverted, respectively, so the ranges of all parameters are defined
with respect to the minimum in the given scenario.  In the third
column we make no assumptions on the ordering, so in this case the
ranges of the parameters are defined with respect to the global
minimum (which corresponds to Normal Ordering) and are obtained
marginalising also over the ordering. For this third case we only give
the $3\sigma$ ranges. In this case the range of $\Delta m^2_{3\ell}$
is composed of two disconnected intervals, one containing the absolute
minimum (NO) and the other the secondary local minimum (IO).

\begin{table}\centering
  \begin{footnotesize} \makebox[\textwidth][c]{
    \begin{tabular}{l|cc|cc|c} \toprule &
\multicolumn{2}{c|}{Normal Ordering (best fit)} &
\multicolumn{2}{c|}{Inverted Ordering ($\Delta\chi^2=0.83$)} & Any
Ordering \\  & bfp $\pm 1\sigma$ & $3\sigma$ range & bfp $\pm
1\sigma$ & $3\sigma$ range & $3\sigma$ range \\ \cmidrule(l){2-6}
\rule{0pt}{4mm}\ignorespaces $\sin^2\theta_{12}$ &
$0.306_{-0.012}^{+0.012}$ & $0.271 \to 0.345$ &
$0.306_{-0.012}^{+0.012}$ & $0.271 \to 0.345$ & $0.271 \to 0.345$
\\[1mm] $\theta_{12}/^\circ$ & $33.56_{-0.75}^{+0.77}$ & $31.38 \to
35.99$ & $33.56_{-0.75}^{+0.77}$ & $31.38 \to 35.99$ & $31.38 \to
35.99$ \\[3mm] $\sin^2\theta_{23}$ & $0.441_{-0.021}^{+0.027}$ &
$0.385 \to 0.635$ & $0.587_{-0.024}^{+0.020}$ & $0.393 \to 0.640$ &
$0.385 \to 0.638$ \\[1mm] $\theta_{23}/^\circ$ & $41.6_{-1.2}^{+1.5}$
& $38.4 \to 52.8$ & $50.0_{-1.4}^{+1.1}$ & $38.8 \to 53.1$ & $38.4 \to
53.0$ \\[3mm] $\sin^2\theta_{13}$ & $0.02166_{-0.00075}^{+0.00075}$ &
$0.01934 \to 0.02392$ & $0.02179_{-0.00076}^{+0.00076}$ & $0.01953 \to
0.02408$ & $0.01934 \to 0.02397$ \\[1mm] $\theta_{13}/^\circ$ &
$8.46_{-0.15}^{+0.15}$ & $7.99 \to 8.90$ & $8.49_{-0.15}^{+0.15}$ &
$8.03 \to 8.93$ & $7.99 \to 8.91$ \\[3mm] $\delta_\text{CP}/^\circ$ &
$261_{-59}^{+51}$ & $\hphantom{00}0 \to 360$ & $277_{-46}^{+40}$ &
$145 \to 391$ & $\hphantom{00}0 \to 360$ \\[3mm] $\dfrac{\Delta
m^2_{21}}{10^{-5}~\ensuremath{\text{eV}^2}}$ & $7.50_{-0.17}^{+0.19}$
& $7.03 \to 8.09$ & $7.50_{-0.17}^{+0.19}$ & $7.03 \to 8.09$ & $7.03
\to 8.09$ \\[3mm] $\dfrac{\Delta
m^2_{3\ell}}{10^{-3}~\ensuremath{\text{eV}^2}}$ &
$+2.524_{-0.040}^{+0.039}$ & $+2.407 \to +2.643$ &
$-2.514_{-0.041}^{+0.038}$ & $-2.635 \to -2.399$ & $\begin{bmatrix}
+2.407 \to +2.643\\[-2pt] -2.629 \to -2.405
      \end{bmatrix}$ \\[3mm] \bottomrule
    \end{tabular} }
  \end{footnotesize}
  \caption{Three-flavour oscillation parameters from our fit to global
data after the NOW~2016 and ICHEP-2016 conferences.  The numbers in the
1st (2nd) column are obtained assuming NO (IO), i.e.,
relative to the respective local minimum, whereas in the 3rd column we
minimise also with respect to the ordering. Note that $\Delta
m^2_{3\ell} \equiv \Delta m^2_{31} > 0$ for NO and $\Delta m^2_{3\ell}
\equiv \Delta m^2_{32} < 0$ for IO.}
  \label{tab:nufit3_bfranges}
\end{table}

Defining the $3\sigma$ relative precision of a parameter by
$2(x^\text{up} - x^\text{low}) / (x^\text{up} + x^\text{low})$, where
$x^\text{up}$ ($x^\text{low}$) is the upper (lower) bound on a
parameter $x$ at the $3\sigma$ level, we read $3\sigma$ relative
precision of 14\% ($\theta_{12}$), 32\% ($\theta_{23}$), 11\%
($\theta_{13}$), 14\% ($\Delta m^2_{21}$) and 9\% ($|\Delta
m^2_{3\ell}|$) for the various oscillation parameters.

\subsubsection{Results: leptonic mixing matrix and CP violation}
\label{subsec:nufit3_CP}

From the global $\chi^2$ analysis described in the previous section
and following the procedure outlined in
Ref.~\cite{GonzalezGarcia:2003qf} one can derive the $3\sigma$ ranges
on the magnitude of the elements of the leptonic mixing matrix:
\begin{equation}
  \label{eq:nufit3_umatrix} |U^\mathrm{lep}| = \begin{pmatrix} 0.800
\to 0.844 &\qquad 0.515 \to 0.581 &\qquad 0.139 \to 0.155 \\ 0.229 \to
0.516 &\qquad 0.438 \to 0.699 &\qquad 0.614 \to 0.790 \\ 0.249 \to
0.528 &\qquad 0.462 \to 0.715 &\qquad 0.595 \to 0.776
  \end{pmatrix} .
\end{equation}
Note that there are strong correlations between the elements due to
the unitary constraint.

\begin{figure}\centering
  \includegraphics[width=0.9\textwidth]{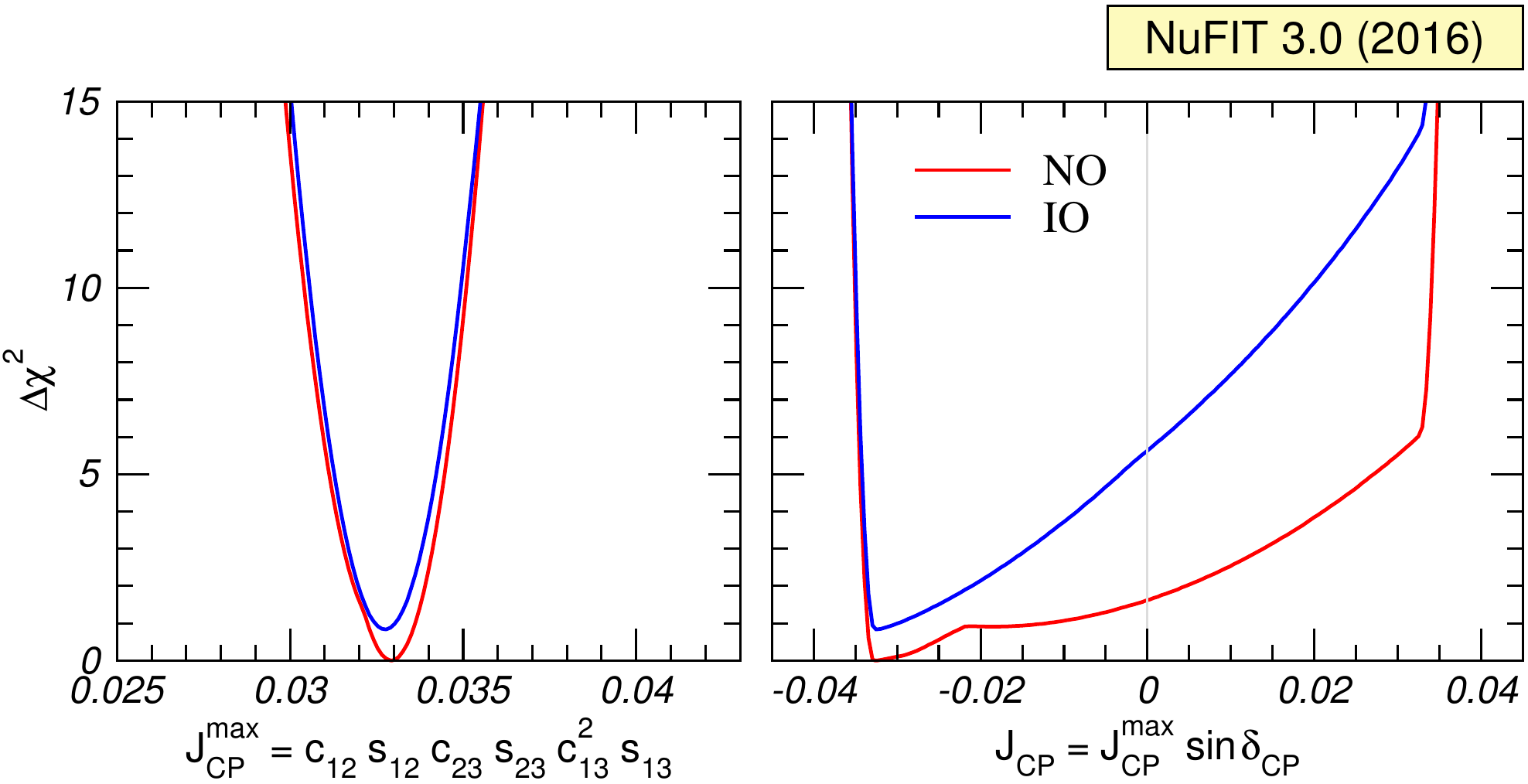}
  \caption{Dependence of the global $\Delta\chi^2$ function on the
Jarlskog invariant. The red (blue) curves are for NO (IO).}
  \label{fig:nufit3_chisq-viola}
\end{figure}

The significance of leptonic CP violation is
illustrated in Fig.~\ref{fig:nufit3_chisq-viola}. In the left panel we
show the dependence of $\Delta\chi^2$ of the global analysis on the
Jarlskog invariant, defined in
Eq.~\eqref{eq:jarlskogNeutrino}, which gives a convention-independent 
measure of CP
violation~\cite{Jarlskog:1985ht}.  Thus the determination of the mixing
angles yields a maximum allowed CP violation
\begin{equation}
  \label{eq:nufit3_jmax} J_\text{CP}^\text{max} = 0.0329 \pm 0.0007 \,
(^{+0.0021}_{-0.0024})
\end{equation}
at $1\sigma$ ($3\sigma$) for both orderings.  The preference of the
data for non-zero $\delta_\text{CP}$ implies a best fit value
$J_\text{CP}^\text{best} = -0.033$, which is favoured over CP
conservation with $\Delta\chi^2 = 1.7$.  These numbers can be compared
with the size of the Jarlskog invariant in the quark sector, which is
determined to be $J_\text{CP}^\text{quarks} = (3.04^{+0.21}_{-0.20})
\times 10^{-5}$~\cite{PDB2016}.

\begin{figure}\centering
\includegraphics[width=0.9\textwidth]{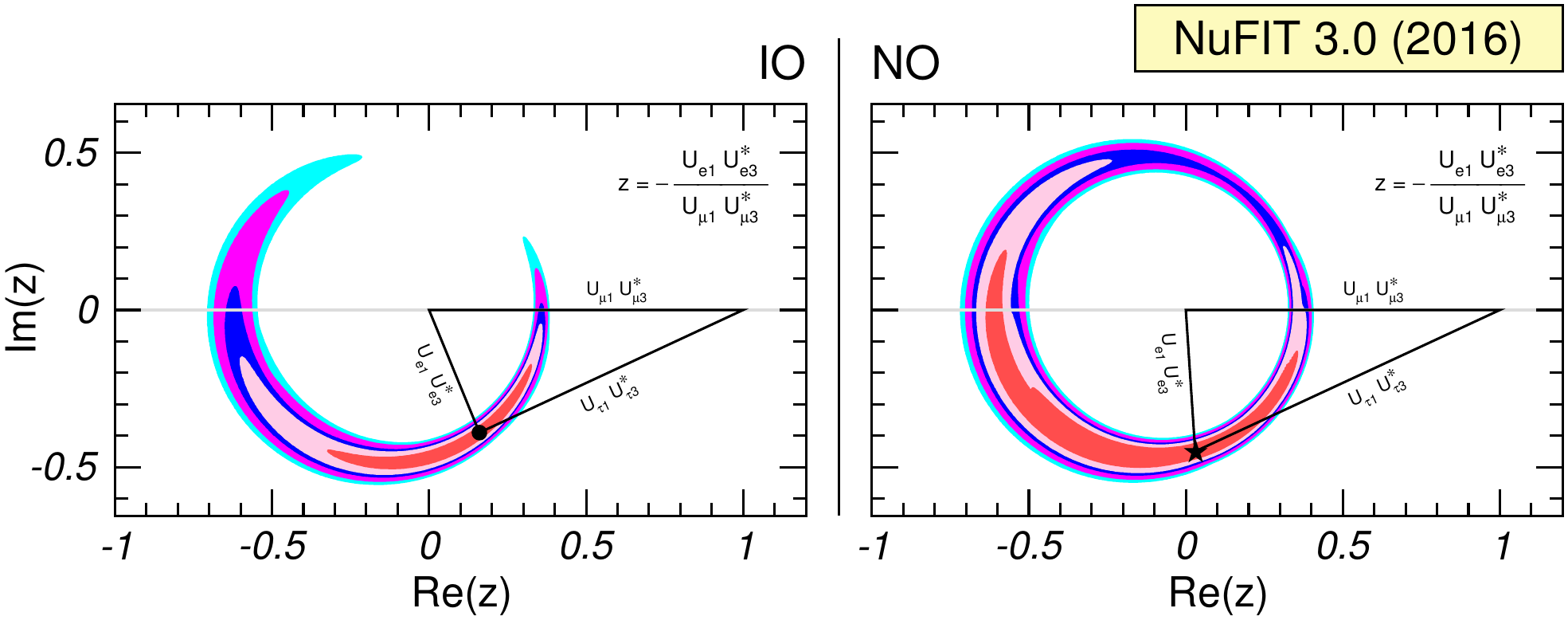}
  \caption{Leptonic unitarity triangle for the first and third columns
of the mixing matrix.  After scaling and rotating the triangle so that
two of its vertices always coincide with $(0,0)$ and $(1,0)$ we plot
the $1\sigma$, 90\%, $2\sigma$, 99\%, $3\sigma$ CL (2~dof) allowed
regions of the third vertex. Note that in the construction of the
triangle the unitarity of the $U^\mathrm{lep}$ matrix is always
explicitly imposed. The regions for both orderings are defined with
respect to the common global minimum which is in NO.}
  \label{fig:nufit3_region-viola}
\end{figure}

In Fig.~\ref{fig:nufit3_region-viola} we recast the allowed regions
for the leptonic mixing matrix in terms of one leptonic unitarity
triangle. Since in the analysis $U^\mathrm{lep}$ is unitary by
construction, any given pair of rows or columns can be used to define
a triangle in the complex plane. In the figure we show the triangle
corresponding to the unitarity conditions on the first and third
columns which is the equivalent to the one usually shown for the quark
sector.  In this figure the absence of CP violation implies a flat
triangle, i.e., $\Im(z) = 0$. As can be seen, for NO the
horizontal axis crosses the $1\sigma$ allowed region, which for 2~dof
corresponds to $\Delta\chi^2 \leq 2.3$. This is consistent with the
preference for CP violation, $\chi^2(J_\text{CP} = 0) -
\chi^2(J_\text{CP}~\text{free}) = 1.7$, mentioned above. We will
comment on the statistical interpretation of this number in
Sec.~\ref{sec:nufit3_MC}.

\subsection{Issues in the analysis}
\label{sec:nufit3_issues}

The $3\nu$ fit results in the previous section provide a statistically
satisfactory description of all the neutrino oscillation data
considered.  There are however some issues in the determination of
some of the parameters which, although not of statistical significance 
yet, deserve some attention.

\subsubsection{Status of \texorpdfstring{$\Delta m^2_{21}$}{dmq21} in solar experiments versus KamLAND}
\label{subsec:nufit3_dm12}

\begin{figure}\centering
\includegraphics[width=0.9\textwidth]{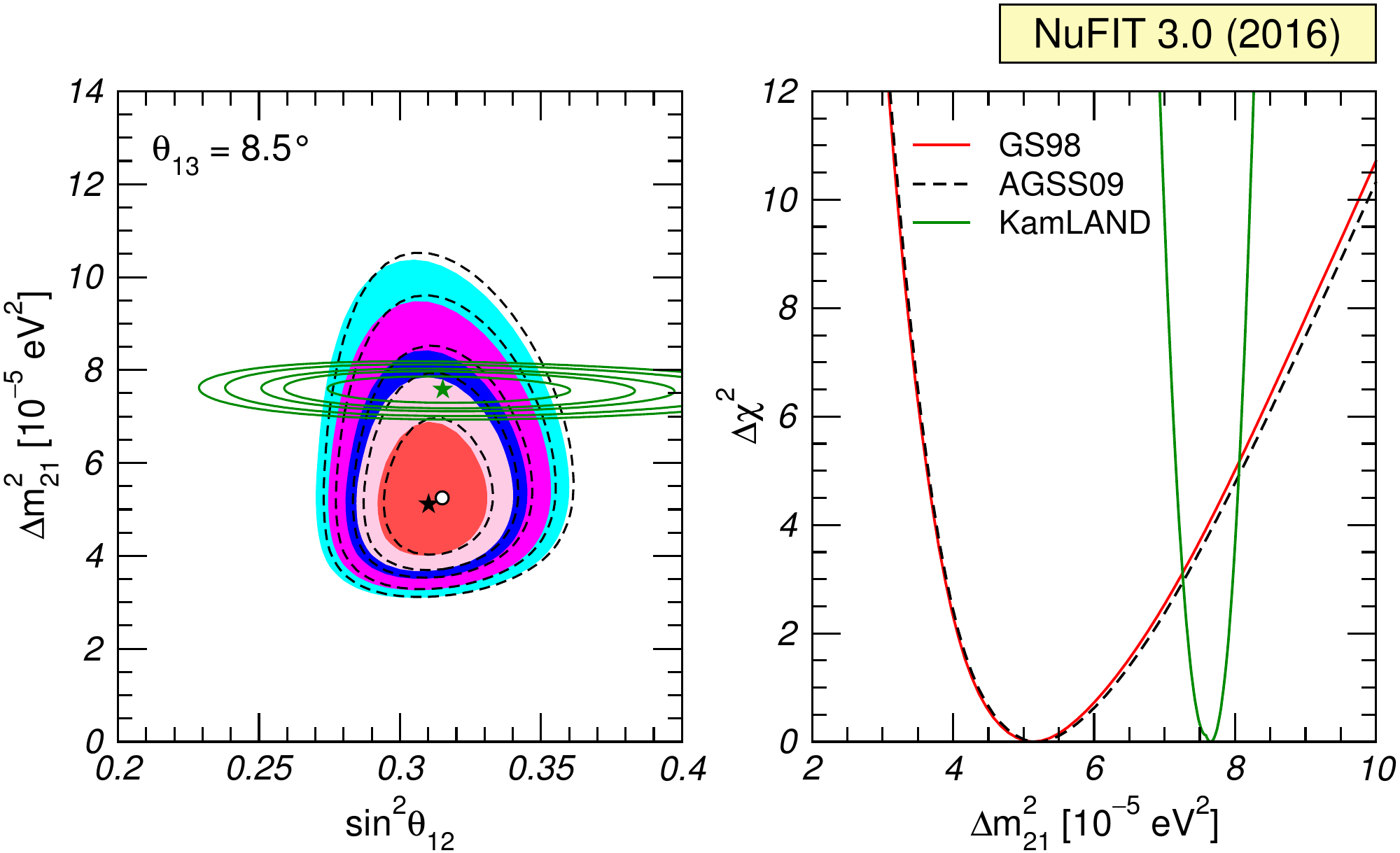}
  \caption{Left: Allowed parameter regions (at $1\sigma$, 90\%,
$2\sigma$, 99\% and $3\sigma$ CL for 2~dof) from the combined analysis
of solar data for GS98 model (full regions with best fit marked by
black star) and AGSS09 model (dashed void contours with best fit
marked by a white dot), and for the analysis of KamLAND data (solid
green contours with best fit marked by a green star) for fixed
$\theta_{13}=8.5^\circ$.  Right: $\Delta\chi^2$ dependence on $\Delta
m^2_{21}$ for the same three analyses after marginalising over
$\theta_{12}$.}
  \label{fig:nufit3_sun-tension}
\end{figure}

The analyses of the solar experiments and of KamLAND give the dominant
contribution to the determination of $\Delta m^2_{21}$ and
$\theta_{12}$.  It has been a result of global analyses for several
years already, that the value of $\Delta m^2_{21}$ preferred by
KamLAND is somewhat higher than the one from solar experiments. This
tension arises from a combination of two effects which have not
changed significantly over the last years (see also \cref{fig:solarNuProb}): a) the well-known fact
that none of the $^8$B measurements performed by SNO, Super-Kamiokande and Borexino
shows any evidence of the low energy spectrum turn-up expected in the
standard LMA-MSW~\cite{Wolfenstein:1977ue, Mikheev:1986gs} solution
for the value of $\Delta m^2_{21}$ favoured by KamLAND; b) the
observation of a non-vanishing day-night asymmetry in Super-Kamiokande, whose size
is larger than the one predicted for the $\Delta m^2_{21}$ value
indicated of KamLAND (for which Earth matter effects are very small).
Ref.~\cite{Gonzalez-Garcia:2014bfa} discussed the differences in the
physics entering in the analyses of solar and KamLAND data which are
relevant to this tension, and to which we refer the reader for
details. Here for sake of completeness we show in
Fig.~\ref{fig:nufit3_sun-tension} the quantification of this tension
in this global analysis. As seen in the figure, the best fit
value of $\Delta m^2_{21}$ of KamLAND lays at the boundary of the
$2\sigma$ allowed range of the solar neutrino analysis.

Also for illustration of the independence of these results with
respect to the solar modelling, the solar neutrino regions are shown
for two latest versions of the Standard Solar Model, namely the GS98
and the AGSS09 models~\cite{Bergstrom:2016cbh} obtained with two
different determinations of the solar
abundances~\cite{Vinyoles:2016djt}.

\subsubsection{\texorpdfstring{$\Delta m^2_{3\ell}$}{dmq3l} determination in LBL accelerator experiments versus reactors}
\label{subsec:nufit3_dm32}

\begin{figure}\centering
\includegraphics[width=0.9\textwidth]{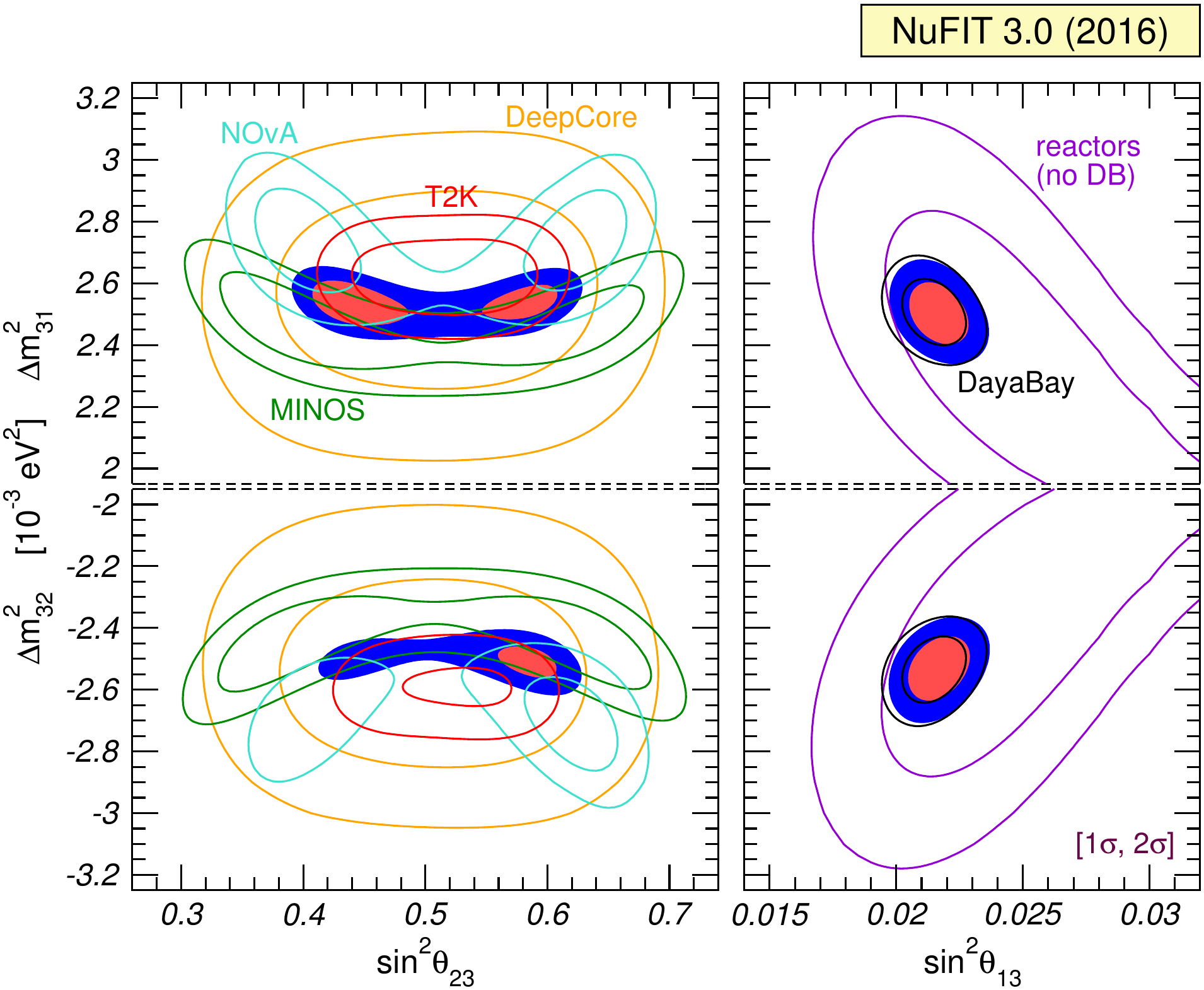}
  \caption{Determination of $\Delta m^2_{3\ell}$ at $1\sigma$ and
$2\sigma$ (2~dof), where $\ell=1$ for NO (upper panels) and $\ell=2$
for IO (lower panels). The left panels show regions in the
$(\theta_{23}, \Delta m^2_{3\ell})$ plane using both appearance and
disappearance data from MINOS (green line), T2K (red lines), NO$\nu$A
(light blue lines), as well as IceCube/DeepCore (orange lines) and the
combination of them (colored regions). In these panels the constraint
on $\theta_{13}$ from the global fit (which is dominated by the
reactor data) is imposed as a Gaussian bias.  The right panels show
regions in the $(\theta_{13}, \Delta m^2_{3\ell})$ plane using only
Daya-Bay (black lines), reactor data without Daya-Bay (violet lines),
and their combination (colored regions). In all panels solar and
KamLAND data are included to constrain $\Delta m^2_{21}$ and
$\theta_{12}$. Contours are defined with respect to the global minimum
of the two orderings.}
  \label{fig:nufit3_region-sample}
\end{figure}

Figure~\ref{fig:nufit3_region-sample} illustrates the contribution to
the determination of $\Delta m^2_{3\ell}$ from the different
data sets.  In the left panels we focus on the determination from LBL
 experiments, which is mainly from $\nu_\mu$ disappearance
data. We plot the $1\sigma$ and $2\sigma$ allowed regions (2~dof) in
the dominant parameters $\Delta m^2_{3\ell}$ and $\theta_{23}$. As
seen in the figure, although the agreement between the different
experiments is reasonable, some ``tension'' starts to appear in the
determination of both parameters among the LBL accelerator
experiments.  In particular we see that the results from
NO$\nu$A, unlike those from T2K, favour a non-maximal value of
$\theta_{23}$. It is important to notice that in the context of $3\nu$
mixing the relevant oscillation probabilities for the LBL accelerator
experiments also depend on $\theta_{13}$ (and on the $\theta_{12}$ and
$\Delta m^2_{21}$ parameters which are independently well constrained
by solar and KamLAND data).  To construct the regions plotted in the
left panels of Fig.~\ref{fig:nufit3_region-sample}, we adopt the
procedure currently followed by the LBL accelerator experiments: we
marginalise with respect to $\theta_{13}$, taking into account the
information from reactor data by adding a Gaussian penalty term to the
corresponding $\chi^2_\text{LBL}$. This is not the same as making a
combined analysis of LBL and reactor data as we will quantify in
Sec.~\ref{subsec:nufit3_t23ordcp}.

Concerning $\nu_e$ disappearance data, the total rates observed in
reactor experiments at different baselines can provide an independent
determination of $\Delta m^2_{3\ell}$~\cite{Bezerra:2012at,
GonzalezGarcia:2012sz}.  On top of this, the observation of the
energy-dependent oscillation effect due to $\theta_{13}$ allows to
further strengthen such measurement.  In the right panels of
Fig.~\ref{fig:nufit3_region-sample} we show therefore the allowed
regions in the $(\theta_{13}, \Delta m^2_{3\ell})$ plane based on
global data on $\nu_e$ disappearance. The violet contours are obtained
from all the medium baselines reactor experiments with the exception
of Daya-Bay; these regions emerge from the baseline effect mentioned
above plus spectral information from Double-Chooz.\footnote{RENO also
presented a spectral analysis based on an exposure of 500
days~\cite{Seo:2016uom}. Here we prefer to include from RENO only the
total rate measurement, based on the larger exposure of 800
days~\cite{reno:nu2014}.}  The black contours are based on the energy
spectrum in Daya-Bay, whereas the colored regions show the
combination.

\begin{figure}\centering
\includegraphics[width=0.8\textwidth]{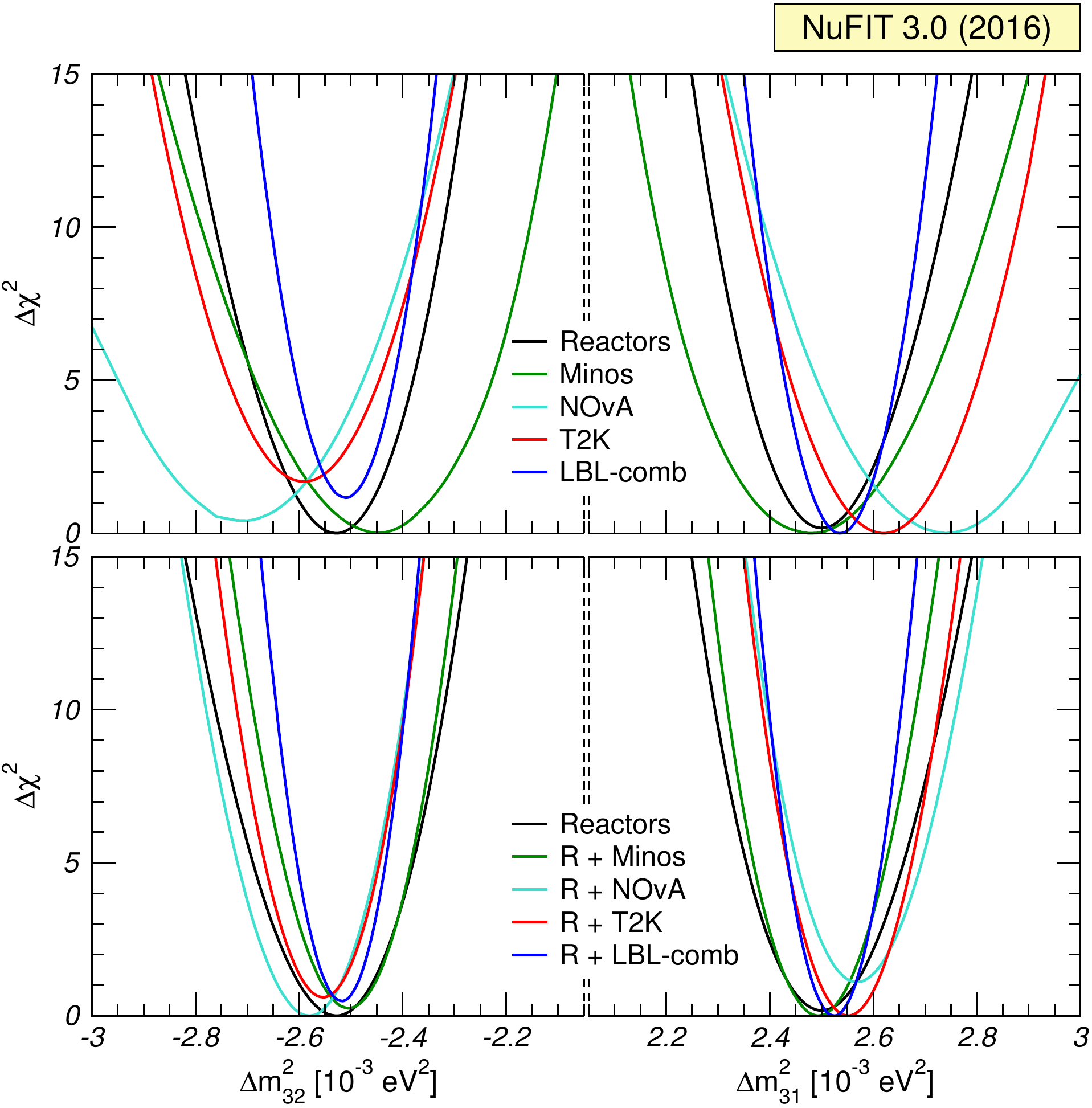}
\caption{$\Delta m^2_{3\ell}$ determination from LBL accelerator
experiments, reactor experiments and their combination. Left (right)
panels are for IO (NO). The upper panels show the 1-dim $\Delta\chi^2$
from LBL accelerator experiments after constraining \emph{only}
$\theta_{13}$ from reactor experiments (this is, marginalising
Eq.~\eqref{eq:nufit3_lblt13r} with respect to $\theta_{23}$ and
$\delta_\text{CP}$). For each experiment $\Delta\chi^2$ is defined
with respect to the global minimum of the two orderings.  The lower
panels show the corresponding determination when the full information
of LBL and reactor experiments is used in the combination (this is,
marginalising Eq.~\eqref{eq:nufit3_lblreac} with respect to
$\theta_{23}$ and $\delta_\text{CP}$).}
 \label{fig:nufit3_chisq-dma}
\end{figure}

\begin{figure}\centering
  \includegraphics[width=0.8\textwidth]{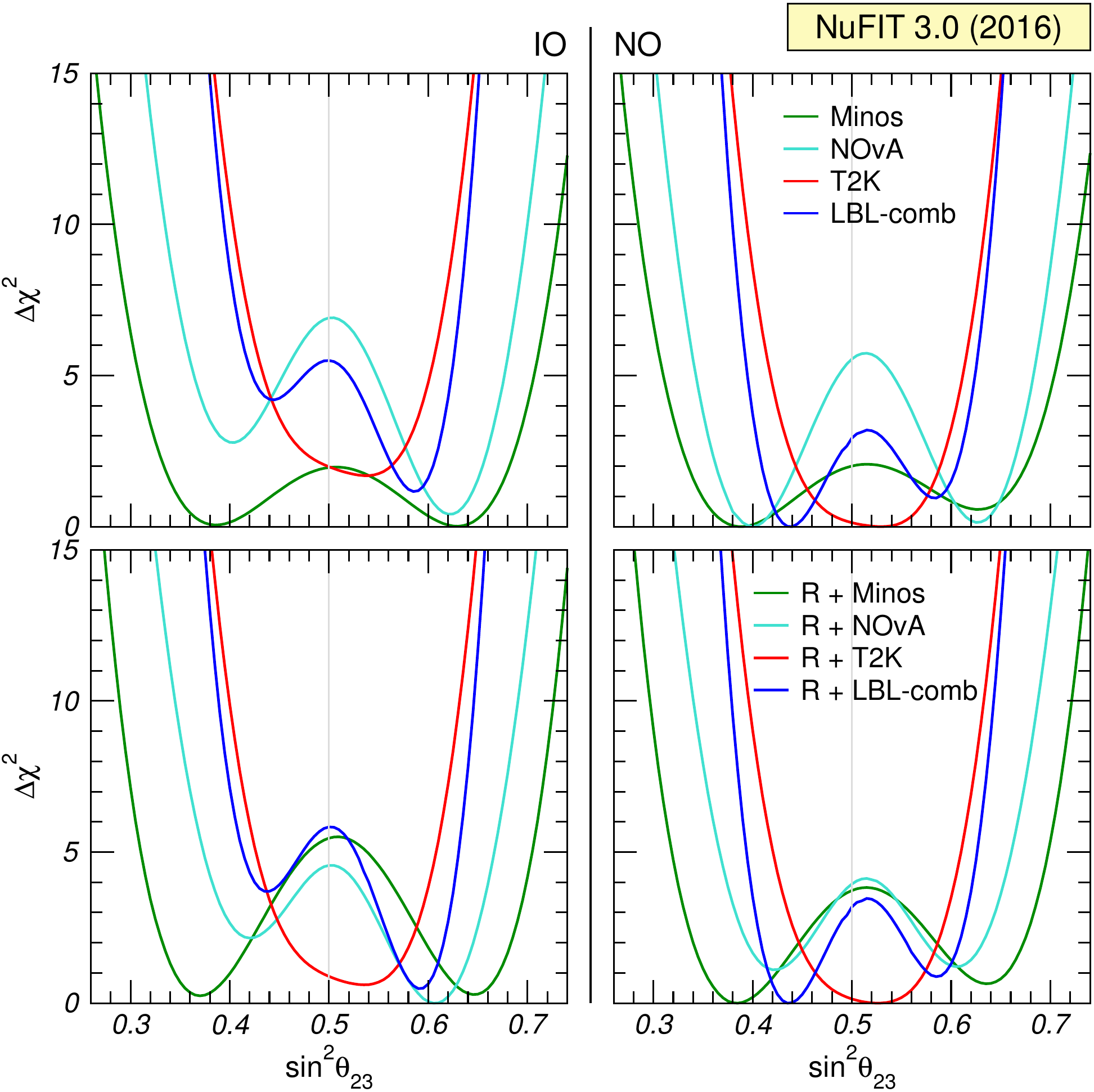}
  \caption{$\theta_{23}$ determination from LBL, reactor and their
combination. Left (right) panels are for IO (NO). The upper panels
show the 1-dim $\Delta\chi^2$ from LBL experiments after constraining
\emph{only} $\theta_{13}$ from reactor experiments (this is,
marginalising Eq.~\eqref{eq:nufit3_lblt13r} with respect to $\Delta
m^2_{3\ell}$ and $\delta_\text{CP}$). For each experiment
$\Delta\chi^2$ is defined with respect to the global minimum of the
two orderings.  The lower panels show the corresponding determination
when the full information of LBL accelerator and reactor experiments
is used in the combination (this is, marginalising
Eq.~\eqref{eq:nufit3_lblreac} with respect to $\Delta m^2_{3\ell}$ and
$\delta_\text{CP}$).}
  \label{fig:nufit3_chisq-t23}
\end{figure}

\begin{figure}\centering
  \includegraphics[width=0.8\textwidth]{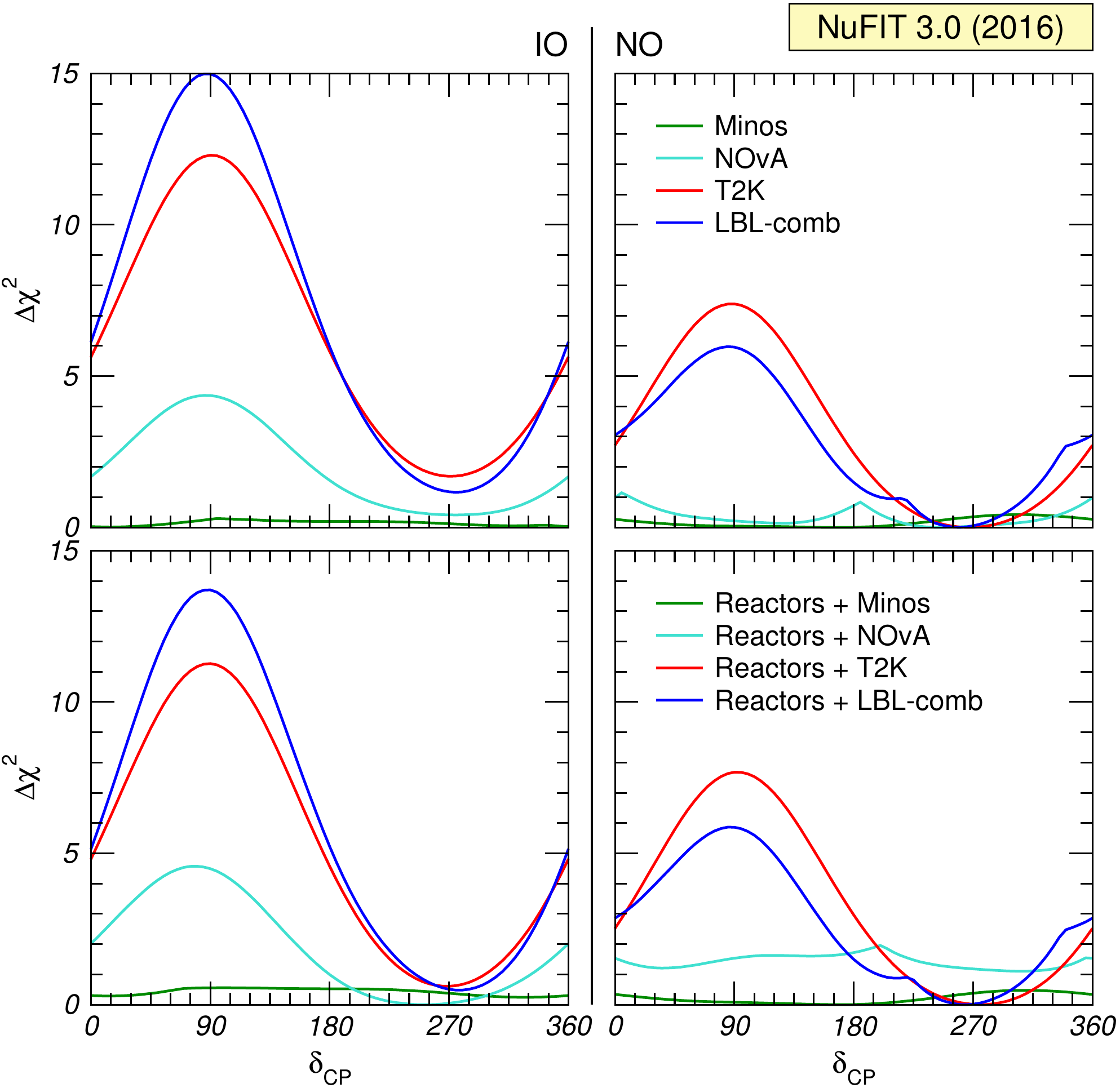}
  \caption{$\delta_\text{CP}$ determination from LBL, reactor and
their combination. Left (right) panels are for IO (NO). The upper
panels show the 1-dim $\Delta\chi^2$ from LBL experiments after
constraining \emph{only} $\theta_{13}$ from reactor experiments (this
is, marginalising Eq.~\eqref{eq:nufit3_lblt13r} with respect to
$\Delta m^2_{3\ell}$ and $\theta_{23}$). For each experiment
$\Delta\chi^2$ is defined with respect to the global minimum of the
two orderings.  The lower panels show the corresponding determination
when the full information of LBL accelerator and reactor experiments
is used in the combination (this is, marginalising
Eq.~\eqref{eq:nufit3_lblreac} with respect to $\Delta m^2_{3\ell}$ and
$\theta_{23}$).}
  \label{fig:nufit3_chisq-dcp}
\end{figure}

By comparing the left and right panels of
Fig.~\ref{fig:nufit3_region-sample} we observe that the combined
$\nu_\mu$ and $\nu_e$ disappearance experiments provide a consistent
determination of $|\Delta m^2_{3\ell}|$ with similar precision.
However when comparing the region for each LBL experiment with that of
the reactor experiments we find some dispersion in the best fit values
and allowed ranges.
This is more clearly illustrated in the upper panels of
Fig.~\ref{fig:nufit3_chisq-dma}, where we plot the one dimensional
projection of the regions in Fig.~\ref{fig:nufit3_region-sample} as a
function of $\Delta m^2_{3\ell}$ after marginalisation over
$\theta_{23}$ for each of the LBL experiments and for their
combination, together with that from reactor data after
marginalisation over $\theta_{13}$.  The projections are shown for
NO(right) and IO(left). Let us stress that the curves corresponding to
LBL experiments in the upper panels of Fig.~\ref{fig:nufit3_chisq-dma}
(as well as those in the upper panels of
Figs.~\ref{fig:nufit3_chisq-t23} and~\ref{fig:nufit3_chisq-dcp}) have
been obtained by a partial combination of the information on the shown
parameter ($\Delta m^2_{3\ell}$ or $\theta_{23}$ or
$\delta_\text{CP}$) from LBL with that of $\theta_{13}$ from reactors,
because in these plots only the $\theta_{13}$ constraint from reactors
is imposed while the dependence on $\Delta m^2_{3\ell}$ is
neglected. This corresponds to the 1-dim projections of the function:
\begin{multline} \Delta\chi^2_\text{LBL+$\theta_{13}^\text{REA}$}
(\theta_{23}, \delta_\text{CP}, \Delta m^2_{3\ell}) \\ =
\min_{\theta_{13}} \Big[ \chi^2_\text{LBL}(\theta_{13}, \theta_{23},
\delta_\text{CP}, \Delta m^2_{3\ell}) + \min_{\Delta
m^2_{3\ell}}\chi^2_\text{REA}(\theta_{13}, \Delta m^2_{3\ell}) \Big] -
\chi^2_\text{min} \,.
  \label{eq:nufit3_lblt13r}
\end{multline}

However, since reactor data also depends on $\Delta m^2_{3\ell}$ the
full combination of reactor and LBL results implies that one must add
consistently the $\chi^2$ functions of the LBL experiment with that of
reactors evaluated the same value of $\Delta m^2_{3\ell}$, this is
\begin{multline} \Delta\chi^2_\text{LBL+REA} (\theta_{23},
\delta_\text{CP}, \Delta m^2_{3\ell}) \\ = \min_{\theta_{13}} \Big[
\chi^2_\text{LBL}(\theta_{13}, \theta_{23}, \delta_\text{CP}, \Delta
m^2_{3\ell}) + \chi^2_\text{REA}(\theta_{13}, \Delta m^2_{3\ell})
\Big] - \chi^2_\text{min} \,.
  \label{eq:nufit3_lblreac}
\end{multline}
We discuss next the effect of combining consistently the information
from LBL and reactor experiments in the determination of
$\theta_{23}$, $\delta_\text{CP}$ and the ordering.

\paragraph{Impact on the determination of
\texorpdfstring{$\theta_{23}$}{t23}, mass ordering, and
\texorpdfstring{$\delta_\text{CP}$}{dCP}}
\label{subsec:nufit3_t23ordcp}

We plot in the lower panels of
Figs.~\ref{fig:nufit3_chisq-dma}--\ref{fig:nufit3_chisq-dcp} the one
dimensional projections of $\Delta\chi^2_\text{LBL+REA}$ for each of
the parameters $\theta_{23}$, $\delta_\text{CP}$, $\Delta m^2_{3\ell}$
(marginalised with respect to the two undisplayed parameters) for the
consistent LBL+REA combinations with both the information on
$\theta_{13}$ and $\Delta m^2_{3\ell}$ from reactors included,
Eq.~\eqref{eq:nufit3_lblreac}.  As mentioned before, the curves in the
upper panels for these figures show the corresponding 1-dimensional
projections for the partial combination, in which only the
$\theta_{13}$ constraint from reactors is used,
Eq.~\eqref{eq:nufit3_lblt13r}.  For each experiment the curves in
these figures are defined with respect to the global minimum of the
two orderings, so the relative height of the minimum in one ordering
vs the other gives a measure of the ordering favoured by each of the
experiments.

Comparing the upper and lower panels in
Figs.~\ref{fig:nufit3_chisq-dma}, \ref{fig:nufit3_chisq-t23}
and~\ref{fig:nufit3_chisq-dcp} one sees how the contribution to the
determination of the mass ordering, the octant and non-maximality of
$\theta_{23}$, and the presence of leptonic CP violation of each LBL
experiment in the full LBL+REA combination
(Eq.~\ref{eq:nufit3_lblreac}) can differ from those derived from the
LBL results imposing only the $\theta_{13}$ constraint from reactors
(Eq.~\ref{eq:nufit3_lblt13r}). This is due to the additional
information on $\Delta m^2_{3\ell}$ from reactors, which is missing in
this last case.  In particular:
\begin{itemize}
\item When only combining the results of the accelerator LBL
experiments with the reactor bound of $\theta_{13}$, both NO$\nu$A and
T2K favour NO by $\chi^2_\text{LBL+$\theta_{13}^\text{REA}$}(\text{IO})
-\chi^2_\text{LBL+$\theta_{13}^\text{REA}$}(\text{NO}) \simeq 0.4$
($1.7$) for $\text{LBL} = \text{NO$\nu$A}$ (T2K). This is in agreement
with the analyses shown by the collaborations for example in
Refs.~\cite{nova:nu2016, t2k:ichep2016}.  However, when consistently
combining with the reactor data, we find that the preference for NO by
T2K+REA is reduced, and NO$\nu$A+REA actually favours IO. This is due
to the slightly lower value of $|\Delta m^2_{3\ell}|$ favoured by the
reactor data, in particular in comparison with NO$\nu$A for both
orderings, and also with T2K for NO.  Altogether we find that for the
full combination of LBL accelerator experiments with reactors the
``hint'' towards NO is below $1\sigma$.

\item Figure~\ref{fig:nufit3_chisq-t23} illustrates how both NO$\nu$A
and MINOS favour non-maximal $\theta_{23}$.  From this figure we see
that while the significance of non-maximality in NO$\nu$A seems more
evident than in MINOS when only the information of $\theta_{13}$ is
included (upper panels), the opposite holds for the full combination
with the reactor data (lower panels). In particular,
  \begin{equation}
    \begin{aligned}
\chi^2_\text{LBL+$\theta_{13}^\text{REA}$}(\theta_{23}=45^\circ,\text{NO})
- \min_{\theta_{23}}
\chi^2_\text{LBL+$\theta_{13}^\text{REA}$}(\theta_{23},\text{NO}) &=
5.5~(2.0) \,, \\
\chi^2_\text{LBL+$\theta_{13}^\text{REA}$}(\theta_{23}=45^\circ,\text{IO})
- \min_{\theta_{23}}
\chi^2_\text{LBL+$\theta_{13}^\text{REA}$}(\theta_{23},\text{IO}) &=
6.5~(1.9) \,, \\ \chi^2_\text{LBL+REA}(\theta_{23}=45^\circ,\text{NO})
- \min_{\theta_{23}} \chi^2_\text{LBL+REA}(\theta_{23},\text{NO}) &=
2.8~(3.7) \,, \\ \chi^2_\text{LBL+REA}(\theta_{23}=45^\circ,\text{IO})
- \min_{\theta_{23}} \chi^2_\text{LBL+REA}(\theta_{23},\text{IO}) &=
4.6~(5.2) \,,
    \end{aligned}
  \end{equation}
  for LBL = NO$\nu$A (MINOS).  The T2K results, though, are
compatible with $\theta_{23} = 45^\circ$ for any ordering. Altogether
we find that for NO the full combination of LBL accelerator
experiments and reactors disfavour maximal $\theta_{23}$ mixing by
$\Delta\chi^2 = 3.2$.

\item Regarding the octant of $\theta_{23}$, for IO all LBL
accelerator experiments are better described with $\theta_{23} >
45^\circ$, adding up to a $\sim 1.8\sigma$ preference for that
octant. Conversely, for NO $\theta_{23} < 45^\circ$ is favoured at
$\sim 1\sigma$.

\item From Fig.~\ref{fig:nufit3_chisq-dcp} we see that the ``hint''
for a CP phase around $270^\circ$ is mostly driven by T2K data, with
some extra contribution from NO$\nu$A in the case of IO.  Within the 
precision of the data samples in this section, the favoured ranges of 
$\delta_\text{CP}$ in each
ordering by the combination of LBL accelerator experiments are pretty
independent on the inclusion of the $\Delta m^2_{3\ell}$ information
from reactors.
\end{itemize}

\subsubsection{Analysis of Super-Kamiokande atmospheric data}
\label{subsec:nufit3_SK}

\begin{figure}\centering
  \includegraphics[width=0.9\textwidth]{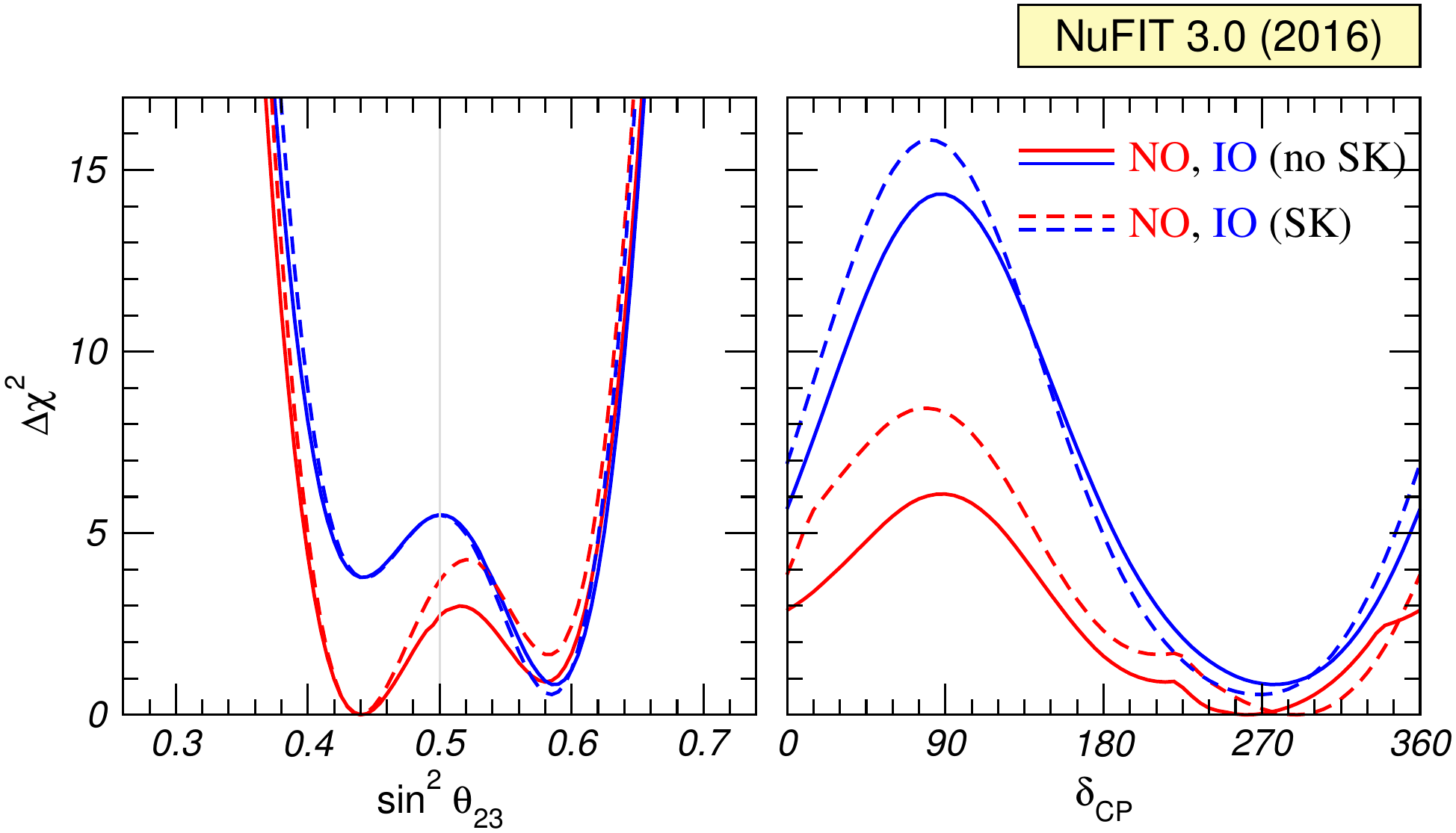}
  \caption{Impact of our re-analysis of Super-Kamiokande atmospheric neutrino
data~\cite{Wendell:2014dka} (70 bins in energy and zenith angle) on
the determination of $\sin^2\theta_{23}$, $\delta_\text{CP}$, and the
mass ordering. The impact on all other parameters is negligible.}
  \label{fig:nufit3_chisq-atmos}
\end{figure}

In all the results discussed so far we have not included information
from Super-Kamiokande atmospheric data. The reason is that our
oscillation analysis could not reproduce that of the collaboration
presented in their talks (see for example
Ref.~\cite{skatm:nufact2016}).

\begin{table}\centering
  \begin{footnotesize}
    \begin{tabular}{l|cc|cc|c} \toprule &
\multicolumn{2}{c|}{Normal Ordering (best fit)} &
\multicolumn{2}{c|}{Inverted Ordering ($\Delta\chi^2=0.56$)} & Any
Ordering \\  & bfp $\pm 1\sigma$ & $3\sigma$ range & bfp $\pm
1\sigma$ & $3\sigma$ range & $3\sigma$ range \\ \cmidrule(l){2-6}
\ignorespaces $\sin^2\theta_{23}$ &
$0.440_{-0.019}^{+0.024}$ & $0.388 \to 0.630$ &
$0.584_{-0.022}^{+0.019}$ & $0.398 \to 0.634$ & $0.388 \to 0.632$
\\[1mm] $\theta_{23}/^\circ$ & $41.5_{-1.1}^{+1.4}$ & $38.6 \to 52.5$
& $49.9_{-1.3}^{+1.1}$ & $39.1 \to 52.8$ & $38.6 \to 52.7$ \\[3mm]
$\delta_\text{CP}/^\circ$ & $289_{-51}^{+38}$ & $\hphantom{00}0 \to
360$ & $269_{-45}^{+40}$ & $146 \to 377$ & $\hphantom{00}0 \to 360$
\\[1mm] \bottomrule
    \end{tabular}
  \end{footnotesize}
  \caption{Three-flavour oscillation parameters from our fit to global
data, including also our re-analysis of SK1--4 (4581 days) atmospheric
data.  The numbers in the 1st (2nd) column are obtained assuming NO
(IO), i.e., relative to the respective local minimum, whereas
in the 3rd column we minimise also with respect to the ordering.  The
omitted parameters are identical to Tab.~\ref{tab:nufit3_bfranges}.}
  \label{tab:nufit3_skranges}
\end{table}

Already since SK2 the Super-Kamiokande collaboration has been
presenting its experimental results in terms of a growing number of
data samples.  The rates for some of those samples cannot be predicted
(and therefore included in a statistical analysis) without a detailed
simulation of the detector, which can only be made by the experimental
collaboration itself. The NuFIT analysis of Super-Kamiokande data has
been always based on the ``classical'' set of samples for which the
simulations were reliable enough: sub-GeV and multi-GeV $e$-like and
$\mu$-like fully contained events, as well as partially contained,
stopping and through-going muon data, each divided into 10 angular
bins for a total of 70 energy and zenith angle bins (details on the
simulation of the data samples and the statistical analysis are given
in the Appendix of Ref.~\cite{GonzalezGarcia:2007ib}).  Despite the
limitations, until recently these results represented the most
up-to-date analysis of the atmospheric neutrino data which could be
performed outside the collaboration, and they were able to reproduce
with reasonable precision the oscillation results of the full analysis
presented by Super-Kamiokande~--- both for what concerns the determination of the
dominant parameters $\Delta m^2_{3\ell}$ and $\theta_{23}$, as well as
their rather marginal sensitivity to the subdominant $\nu_e$
appearance effects driven by $\theta_{13}$ (and consequently to
$\delta_\text{CP}$ and the ordering). Thus the NuFIT collaboration
confidently included their own implementation of the Super-Kamiokande
$\chi^2$ in the global fit.

However, in the last years Super-Kamiokande has developed a new
analysis method in which a set of neural network based selections are
introduced, some of them with the aim of constructing $\nu_e +
\bar\nu_e$ enriched samples which are then further classified into
$\nu_e$-like and $\bar\nu_e$-like subsamples, thus increasing the
sensitivity to subleading parameters such as the mass ordering and
$\delta_\text{CP}$~\cite{Wendell:2014dka, skatm:thesis}.  The
selection criteria are constructed to exploit the expected differences
in the number of charged pions and transverse momentum in the
interaction of $\nu_e$ versus $\bar\nu_e$.  With this new analysis
method Super-Kamiokande has been reporting in talks an increasing
sensitivity to the ordering and to $\delta_\text{CP}$: for example,
the preliminary results of the analysis of SK1--4 (including 2520 days
of SK4)~\cite{skatm:nufact2016} in combination with the reactor
constraint of $\theta_{13}$ show a preference for NO with a
$\Delta\chi^2(\text{IO}) = 4.3$ and variation of
$\chi^2(\delta_\text{CP})$ with the CP phase at the level of $\sim
1.7\sigma$.

Unfortunately, with publicly available information this analysis is
not reproducible outside the collaboration.  Conversely the
``traditional'' analysis based on their reproducible data samples
continues to show only marginal dependence on these effects. This is
illustrated in Fig.~\ref{fig:nufit3_chisq-atmos} and
Tab.~\ref{tab:nufit3_skranges} where we show the impact of inclusion
of our last re-analysis of Super-Kamiokande atmospheric data using the above
mentioned 70 bins in energy and zenith angle.\footnote{We use the same
data and statistical treatment as in the previous global fit NuFIT
2.0~\cite{Gonzalez-Garcia:2014bfa} as well as in versions 2.1 and
2.2~\cite{nufit} which is based on 4581 days of data from
SK1--4~\cite{Wendell:2014dka} (corresponding to 1775 days of SK4).}
We only show the impact on the determination of $\sin^2\theta_{23}$,
$\delta_\text{CP}$, and the mass ordering as the effect on all other
parameters is negligible. We observe that $\Delta\chi^2$ for maximal
mixing and the second $\theta_{23}$ octant receive an additional
contribution of about 1 unit in the case of NO, whereas the
$\theta_{23}$ result for IO is practically unchanged. Values of
$\delta_\text{CP} \simeq 90^\circ$ are slightly more disfavoured,
whereas there is basically no effect on the mass ordering
discrimination.

In summary, with the information at hand we are not able to reproduce
the elements driving the main dependence on the subdominant effects of
the official Super-Kamiokande
results, while the dominant parameters are currently well determined
by LBL experiments. For these reasons we have decided not to include
our re-analysis of Super-Kamiokande data in our preferred global fit
presented in the previous section.  Needless to say that when enough
quantitative information becomes available to allow a reliable
simulation of the subdominant $\nu_e$-driven effects, we will proceed
to include it in our global analysis.

\subsection{Monte Carlo evaluation of confidence levels for \texorpdfstring{$\theta_{23}$}{t23}, \texorpdfstring{$\delta_\text{CP}$}{dCP} and ordering}
\label{sec:nufit3_MC}

From the analysis presented in \cref{chap:3nufit_theor} we see that the three least known neutrino oscillation parameters are
the Dirac CP violating phase $\delta_\text{CP}$, the octant of
$\theta_{23}$ and the mass ordering (which in what follows we will
denote by ``O'').  In order to study the information from data on
these parameters one can use two $\Delta\chi^2$ test
statistics~\cite{Elevant:2015ska, Blennow:2014sja}:
\begin{align}
  \label{eq:nufit3_testStatistics1} \Delta\chi^2
\left(\delta_\text{CP}, \text{O} \right) &= \min_{x_1}
\chi^2\left(\delta_\text{CP}, \text{O}, x_1\right) -\chi^2_\text{min}
\,, \\
  \label{eq:nufit3_testStatistics2} \Delta \chi^2\left(\theta_{23},
\text{O}\right) &= \min_{x_2} \chi^2\left(\theta_{23}, \text{O},
x_2\right) -\chi^2_\text{min} \,,
\end{align}
where the minimisation in the first equation is performed with respect
to all oscillation parameters except $\delta_\text{CP}$ and the
ordering ($x_1 = \lbrace \theta_{12}, \theta_{13}, \theta_{23}, \Delta
m^2_{21}, |\Delta m^2_{3\ell}| \rbrace$), while in the second equation
the minimisation is over all oscillation parameters except
$\theta_{23}$ and the ordering ($x_2 = \lbrace \theta_{12},
\theta_{13}, \delta_\text{CP}, \Delta m^2_{21}, |\Delta m^2_{3\ell}|
\rbrace$). Here $\chi^2_\text{min}$ indicates the $\chi^2$ minimum
with respect to all oscillation parameters including the mass
ordering.

We have plotted the values of these test statistics in the lower right
and central left panels in Fig.~\ref{fig:nufit3_chisq-glob}. We can
use them not only for the determination of $\delta_\text{CP}$ and
$\theta_{23}$, respectively, but also of the mass ordering. For
instance, using Eq.~\eqref{eq:nufit3_testStatistics1} we can determine
a confidence interval for $\delta_\text{CP}$ at a given CL for both
orderings. However, below a certain CL no interval will appear for the
less favoured ordering. In this sense we can exclude that ordering at
the CL at which the corresponding interval for $\delta_\text{CP}$
disappears. Note that a similar prescription to test the mass ordering
can be built for any other parameter as well, e.g., for
$\theta_{23}$ using
Eq.~\eqref{eq:nufit3_testStatistics2}.\footnote{Let us mention that
this method to determine the mass ordering is different from the one
based on the test statistic $T$ discussed in
Ref.~\cite{Blennow:2013oma}.}

\begin{figure}\centering
  \includegraphics[width=0.8\textwidth]{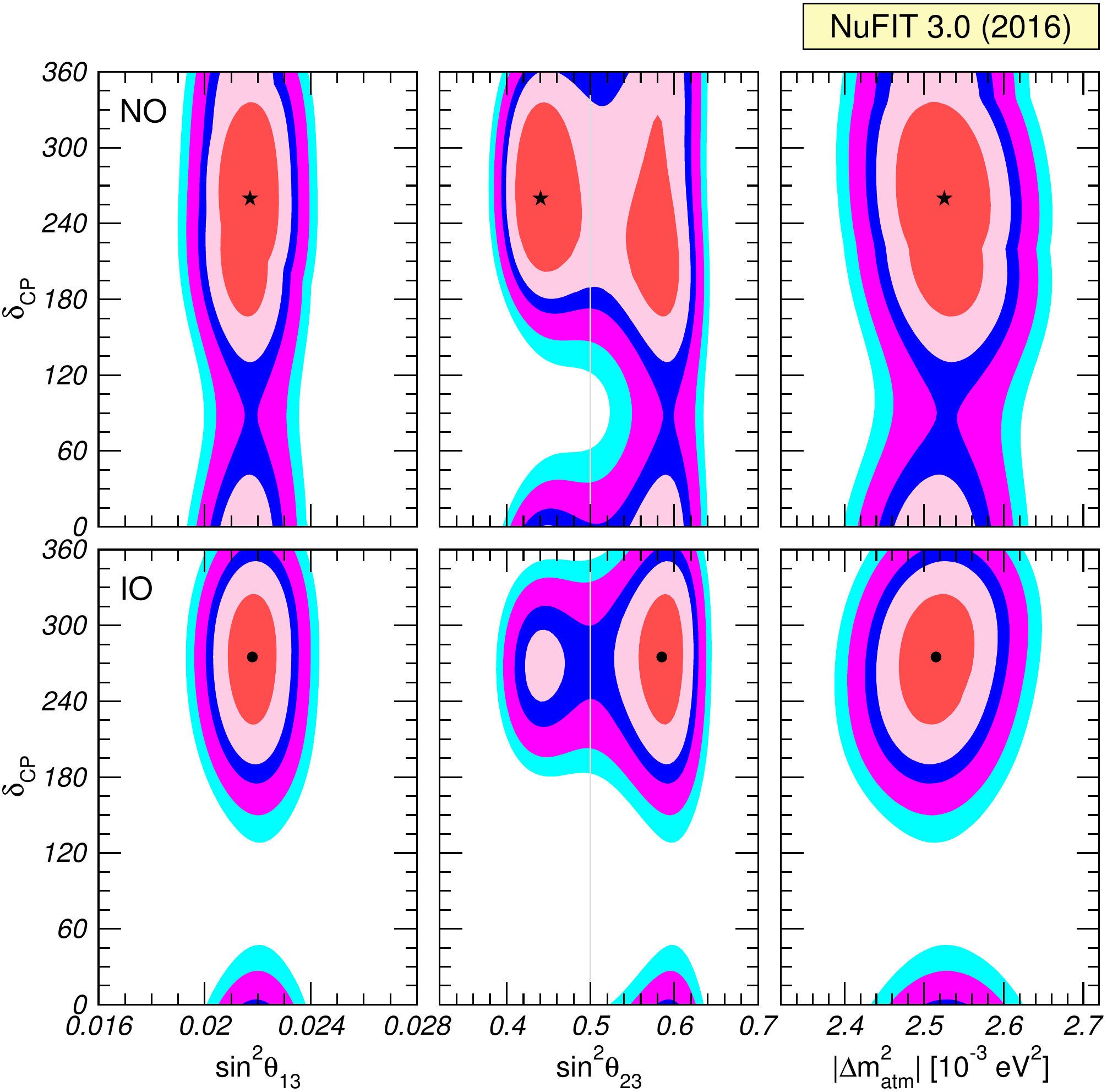}
  \caption{Allowed regions from the global data at $1\sigma$, 90\%,
$2\sigma$, 99\% and $3\sigma$ CL (2~dof). We show projections onto
different planes with $\delta_\text{CP}$ on the vertical axis after
minimising with respect to all undisplayed parameters. The lower
(upper) panels correspond to IO (NO).  Contour regions are derived
with respect to the global minimum which occurs for NO and is
indicated by a star. The local minimum for IO is shown by a black
dot.}
  \label{fig:nufit3_region-hier}
\end{figure}

In Sec.~\ref{sec:nufit3_global} we have presented confidence intervals
assuming that the test statistics follow a $\chi^2$-distribution with
1~dof, relying on Wilks' theorem to hold~\cite{Wilks:1938dza} (this is
what we call the Gaussian limit). However, the test statistics in
Eqs.~\eqref{eq:nufit3_testStatistics1}
and~\eqref{eq:nufit3_testStatistics2} are expected not to follow
Wilks' theorem because of several reasons~\cite{Elevant:2015ska}:
\begin{itemize}
\item Sensitivity of the data presented in this section to $\delta_\text{CP}$ is still
limited, as can be seen in Fig.~\ref{fig:nufit3_chisq-glob}: all
values of $\delta_\text{CP}$ have $\Delta \chi^2 < 14$, and for NO not
even $\Delta \chi^2 = 6$ is attained.

\item Regarding $\theta_{23}$, its precision is dominated by $\nu_\mu$
disappearance experiments. Since the relevant survival probability
depends dominantly on $\sin^2 2\theta_{23}$, there is both a physical
boundary of their parameter space at $\theta_{23} = 45^\circ$ (because
$\sin 2\theta_{23}<1$), as well as a degeneracy related to the octant.

\item The mass ordering is a discrete parameter.

\item The dependence of the theoretical predictions on
$\delta_\text{CP}$ is significantly non-linear, even more considering
the periodic nature of this parameter. Furthermore, there are
complicated correlations and degeneracies between $\delta_\text{CP}$,
$\theta_{23}$, and the mass ordering (see
Fig.~\ref{fig:nufit3_region-hier} for illustration).
\end{itemize}
Therefore, one may expect deviations from the Gaussian limit of the
$\Delta\chi^2$ distributions, and confidence levels for these
parameters should be cross checked through a Monte Carlo simulation of
the relevant experiments. We consider in the following the combination
of the T2K, NO$\nu$A, MINOS and Daya-Bay experiments, which are most
relevant for the parameters we are interested in this section. For a
given point of assumed true values for the parameters we generate a
large number ($10^4$) of pseudo-data samples for each of the
experiments. For each pseudo-data sample we compute the two statistics
given in Eqs.~\eqref{eq:nufit3_testStatistics1}
and~\eqref{eq:nufit3_testStatistics2} to determine their distributions
numerically. In Ref.~\cite{Elevant:2015ska} it has been shown that the
distribution of test statistics for 2-dimensional parameter region
(such as for instance the middle panels of
Fig.~\ref{fig:nufit3_region-hier}) are more close to Gaussianity than
1-dimensional ones such as Eqs.~\eqref{eq:nufit3_testStatistics1}
and~\eqref{eq:nufit3_testStatistics2}. Therefore we focus here on the
1-dimensional cases.

First, let us note that in order to keep calculation time manageable
one can fix all parameters which are known to be uncorrelated with the
three we are interested in (i.e., $\theta_{23}$,
$\delta_\text{CP}$, O).  This is certainly the case for $\Delta
m^2_{21}$ and $\theta_{12}$ which are determined independently by
solar and KamLAND data. As for $\theta_{13}$, the most
precise information arises from reactor data whose results are
insensitive to $\delta_\text{CP}$ and $\theta_{23}$.  Consequently,
marginalising over $\theta_{13}$ within reactor uncertainties or
fixing it to the best fit value gives a negligible difference in the
simulations. Concerning $|\Delta m^2_{3\ell}|$ we observe that there
are no strong correlations or degeneracies with $\delta_\text{CP}$
(see Fig.~\ref{fig:nufit3_region-hier}), and we assume that the
distributions of the test statistics do not significantly depend on
the assumed true value. Therefore we consider only the global best fit
values for each ordering as true values for $|\Delta m^2_{3\ell}|$ to
generate pseudo-data. However, since the relevant observables
\emph{do} depend non-trivially on its value, it is important to keep
$|\Delta m^2_{3\ell}|$ as a free parameter in the fit and to minimise
the $\chi^2$ for each pseudo-data sample with respect to it.  Hence,
we approximate the test statistics in
Eqs.~\eqref{eq:nufit3_testStatistics1}
and~\eqref{eq:nufit3_testStatistics2} by using
\begin{align} \chi^2 \left(\delta_\text{CP}, \text{O}, x_1\right)
&\equiv \min_{\theta_{23}, |\Delta m^2_{3\ell}|} \chi^2
\left(\theta_{23}, \delta_\text{CP}, \text{O}, |\Delta
m^2_{3\ell}|\right) \,, \\ \chi^2 \left(\theta_{23}, \text{O}, x_2
\right) &\equiv \min_{\delta_\text{CP}, |\Delta m^2_{3\ell}|} \chi^2
\left(\theta_{23}, \delta_\text{CP}, \text{O}, |\Delta m^2_{3\ell}|
\right) \,,
\end{align}
with the other oscillation parameters kept fixed at their best fit
points: $\Delta m^2_{21} = 7.5 \times 10^{-5}~\text{eV}^2$,
$\sin^2\theta_{12} = 0.31$, and $\sin^2\theta_{13} = 0.022$.

\subsubsection{\texorpdfstring{$\delta_\text{CP}$}{dCP} and the mass
ordering}

\begin{figure}[hbtp]\centering
  \includegraphics[width=\textwidth]{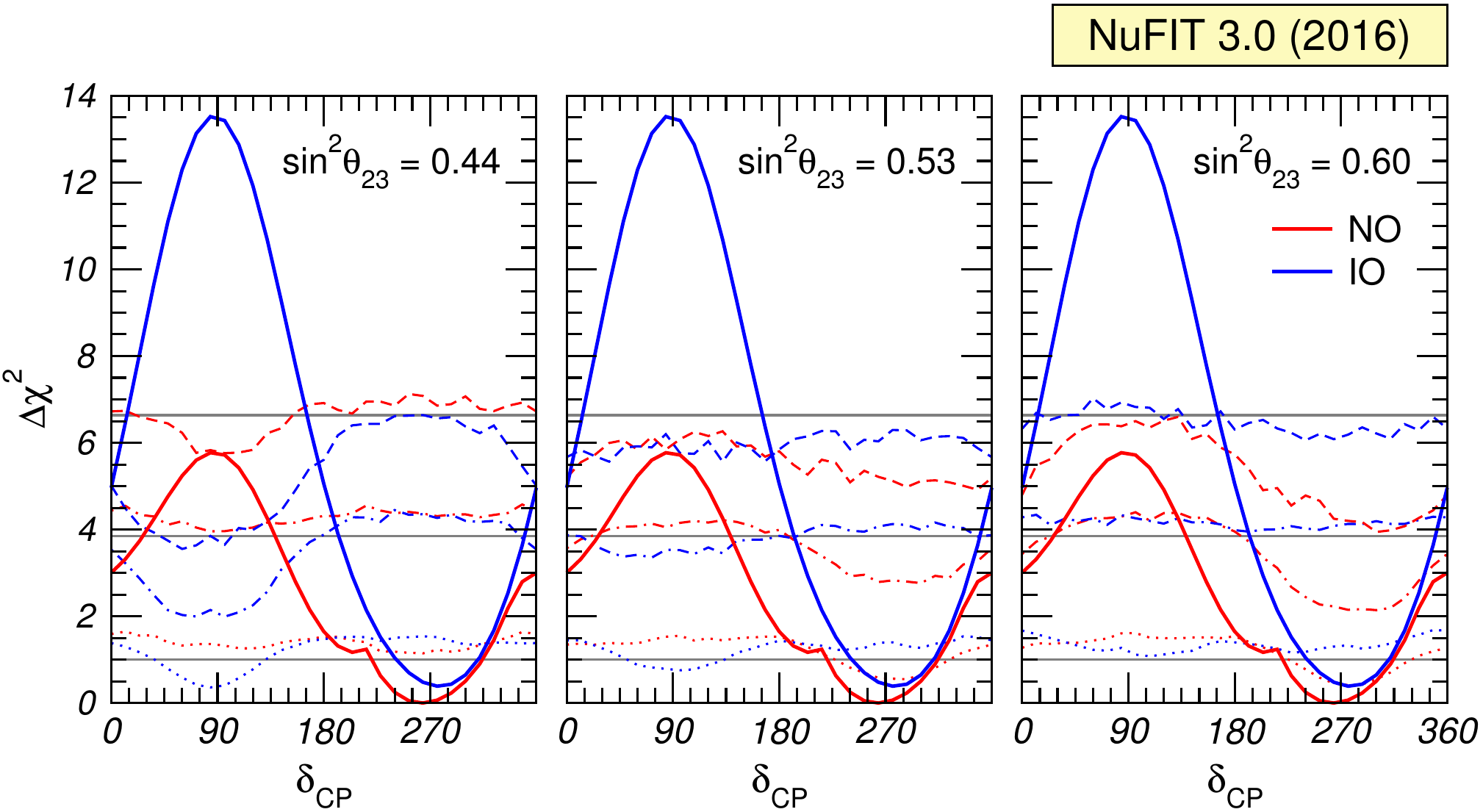}
  \caption{68\%, 95\% and 99\% confidence levels (broken curves) for
the test statistics~\eqref{eq:nufit3_testStatistics1} along with its
value (solid curves) for the combination of T2K, NO$\nu$A, MINOS and
reactor data.  The value of $\sin^2\theta_{23}$ given in each panel
corresponds to the assumed true value chosen to generate the
pseudo-experiments and for all panels we take $\Delta
m^2_{3\ell,\text{true}} = -2.53 \times 10^{-3}~\text{eV}^2$ for IO and
$+2.54 \times 10^{-3}~\text{eV}^2$ for NO.  The solid horizontal lines
represent the 68\%, 95\% and 99\% CL predictions from Wilks' theorem.}
  \label{fig:nufit3_probab-dcp}
\end{figure}

The value of the test statistics~\eqref{eq:nufit3_testStatistics1} is
shown in Fig.~\ref{fig:nufit3_probab-dcp} for the combination of T2K,
NO$\nu$A, MINOS and Daya-Bay as a function of $\delta_\text{CP}$ for
both mass orderings. In the generation of the pseudo-data we have
assumed three representative values of $\theta_{23,\text{true}}$ as
shown in the plots.  The broken curves show, for each set of true
values, the values of $\Delta\chi^2(\delta_\text{CP}, \text{O})$ which
are larger than 68\%, 95\%, and 99\% of all generated data samples.

From the figure we read that if the $\Delta\chi^2$ from real data
(solid curve, identical in the three panels) for a given ordering is
above the $x\%$ CL lines for that ordering for a given value of
$\delta_\text{CP}$, that value of $\delta_\text{CP}$ and the mass
ordering can be rejected with $x\%$ confidence. So if the minimum of
the $\Delta\chi^2$ curve for one of the orderings (in this case IO is
the one with non-zero minimum) is above the $x\%$ CL line one infers
that that ordering is rejected at that CL.

For the sake of comparison we also show in
Fig.~\ref{fig:nufit3_probab-dcp} the corresponding 68\%, 95\% and 99\%
Gaussian confidence levels as horizontal lines.  There are some
qualitative deviations from Gaussianity that have already been
reported~\cite{Elevant:2015ska}:
\begin{itemize}
\item For $\theta_{23}<45^\circ$, $\delta_\text{CP} = 90^\circ$, and
IO as well as for $\theta_{23} > 45^\circ$, $\delta_\text{CP} =
270^\circ$ and NO, the confidence levels decrease. This effect arises
because at those points in parameter space the $\nu_\mu \to \nu_e$
oscillation probability has a minimum or a maximum,
respectively. Therefore, statistical fluctuations leading to less (or
more) events than predicted cannot be accommodated by adjusting the
parameters.  $\Delta \chi^2$ is small more often and the confidence
levels decrease. This is an effect always present at boundaries in
parameter space, usually referred to as an effective decrease in the
number of degrees of freedom in the model.

\item Conversely for $\delta_\text{CP} \sim 90^\circ$ for
$\theta_{23}>45^\circ$, and $\delta_\text{CP} \sim 270^\circ$ for
$\theta_{23}<45^\circ$, the confidence levels increase. This is
associated with the prominent presence of the octant
degeneracy. Degeneracies imply that statistical fluctuations can drive
you away from the true value, $\Delta \chi^2$ increases, and the
confidence levels increase. This is usually referred to as an
effective increase in the number of degrees of freedom in the model
due to degeneracies.

\item Overall we find that with the data presented in this section confidence levels are
clearly closer to Gaussianity than found in
Refs.~\cite{Gonzalez-Garcia:2014bfa, Elevant:2015ska}, where similar
simulations have been performed with less data available. For those
data sets confidence levels were consistently below their Gaussian
limit. This was mainly a consequence of the limited statistics and the
cyclic nature of $\delta_\text{CP}$ which lead to an effective
decrease in the number of degrees of freedom.  We now find that when
the full combination of data currently available is included this
effect is reduced, as expected if experiments become more sensitive.

\item For all true values considered, IO is not rejected even at
$1\sigma$.  In particular we find IO disfavoured at $30\% - 40\%$ for
$\sin^2\theta_{23}=0.44 - 0.60$.
\end{itemize}

\begin{table}\centering
  \begin{tabular}{cc|ccc} \toprule $\sin^2\theta_{23,\text{true}}$
& Ordering & CP cons.  & 90\% CL range & 95\% CL range \\ \midrule 0.44
& NO & 70\% & $[0^\circ, 14^\circ] \cup [151^\circ, 360^\circ]$ &
$[0^\circ, 37^\circ] \cup [133^\circ, 360^\circ]$ \\ & IO & 98\% &
$[200^\circ, 341^\circ]$ & $[190^\circ, 350^\circ]$ \\ 0.53 & NO &
70\% & $[150^\circ, 342^\circ]$ & $[0^\circ, 28^\circ] \cup
[133^\circ, 360^\circ]$ \\ & IO & 98\% & $[203^\circ, 342^\circ]$ &
$[193^\circ, 350^\circ]$ \\ 0.60 & NO & 70\% & $[148^\circ,
336^\circ]$ & $[0^\circ, 28^\circ] \cup [130^\circ, 360^\circ]$ \\ &
IO & 97\% & $[205^\circ, 345^\circ]$ & $[191^\circ, 350^\circ]$ \\
\midrule Gaussian & NO & 80\% & $[158^\circ, 346^\circ]$ & $[0^\circ,
26^\circ] \cup [139^\circ, 360^\circ]$ \\ & IO & 97\% & $[208^\circ,
332^\circ]$ & $[193^\circ, 350^\circ]$ \\ \bottomrule
  \end{tabular}
  \caption{Confidence level with which CP conservation
($\delta_\text{CP} = 0, 180^\circ$) is rejected (third column) and
90\% and 95\% confidence intervals for $\delta_\text{CP}$ (fourth and
fifth column) for different sets of true values of the parameters and
in the Gaussian approximation.  Confidence intervals for
$\delta_\text{CP}$ as well as the CL for CP conservation are defined
for both orderings with respect to the global minimum (which happens
for NO).}
  \label{tab:nufit3_CPCL}
\end{table}

Quantitatively we show in Tab.~\ref{tab:nufit3_CPCL} the CL at which
CP conservation ($\delta_\text{CP}=0, 180^\circ$) is disfavoured as
well as the 90\% and 95\% confidence intervals for $\delta_\text{CP}$.
We find that the CL of rejection of CP conservation as well as the
allowed ranges do not depend very significantly on
$\theta_{23,\text{true}}$. This can be understood from
Fig.~\ref{fig:nufit3_probab-dcp}: the dependence on
$\theta_{23,\text{true}}$ occur mostly for $\delta_\text{CP} \sim
90^\circ$ and IO, a region discarded with a large CL, and for
$\delta_\text{CP} \sim 270^\circ$ and NO, a region around the best
fit.

Note that in the table the intervals for $\delta_\text{CP}$ are
defined for both orderings with respect to the global minimum (which
happens for NO). Hence the intervals for IO include the effect that IO
is slightly disfavoured with respect to NO. They cannot be directly
compared to the intervals given in Tab.~\ref{tab:nufit3_bfranges},
where we defined intervals relative to the local best fit point for
each ordering.

A similar comment applies also to the CL quoted in the table to reject
CP conservation. For IO this is defined relative to the best fit point
in NO. We find that for NO, CP conservation is allowed at 70\% CL,
i.e., slightly above $1\sigma$ (with some deviations from the
Gaussian result of 80\%~CL), while for IO the CL for CP conservation
is above $2\sigma$. Note that values of $\delta_\text{CP} \simeq
90^\circ$ are disfavoured at around 99\%~CL for NO, while for IO the
rejection is at even higher CL: the $\Delta\chi^2$ with respect to the
global minimum is around 14, which would correspond to $3.7\sigma$ in
the Gaussian limit. Our Monte Carlo sample of $10^4$ pseudo-data sets
is not large enough to confirm such a high confidence level.

\subsubsection{\texorpdfstring{$\theta_{23}$}{t23} and the mass
ordering}

\begin{figure}\centering
  \includegraphics[width=\textwidth]{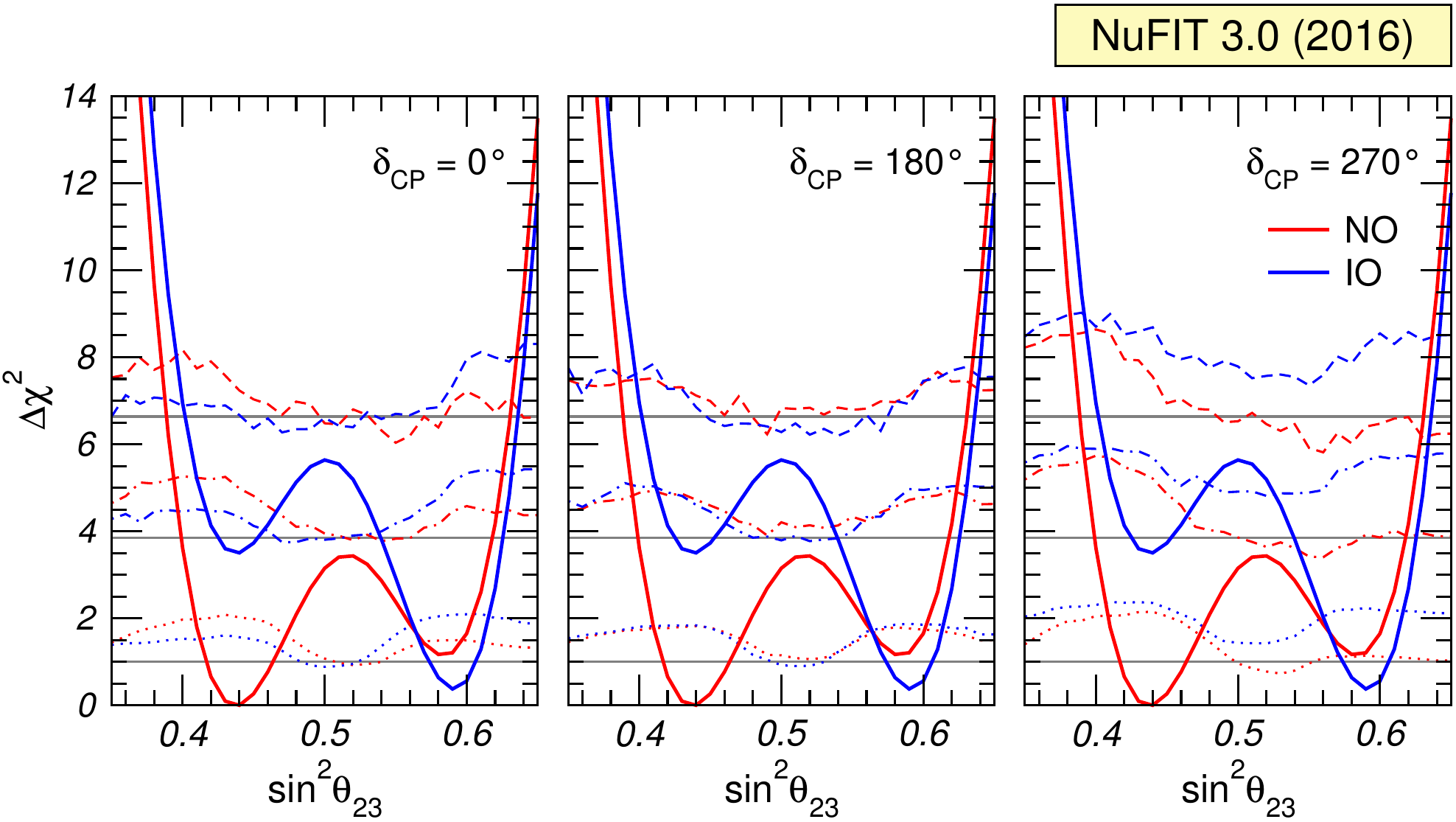}
  \caption{68\%, 95\% and 99\% confidence levels (broken curves) for
the test statistics~\eqref{eq:nufit3_testStatistics2} along with its
value (solid curves) for the combination of T2K, NO$\nu$A, MINOS and
reactor data.  The value of $\delta_\text{CP}$ above each plot
corresponds to the assumed true value chosen to generate the
pseudo-experiments and for all panels we take $\Delta
m^2_{3\ell,\text{true}} = -2.53 \times 10^{-3}~\text{eV}^2$ for IO and
$+2.54 \times 10^{-3}~\text{eV}^2$ for NO.  The solid horizontal lines
represent the 68\%, 95\% and 99\% CL predictions from Wilks' theorem.}
  \label{fig:nufit3_probab-t23}
\end{figure}

Moving now to the discussion of $\theta_{23}$, we show the value of
the test statistics~\eqref{eq:nufit3_testStatistics2} in
Fig.~\ref{fig:nufit3_probab-t23} for the combination of T2K, NO$\nu$A,
MINOS and Daya-Bay experiments as a function of $\theta_{23}$, for
both mass orderings.  For the generation of the pseudo-data we have
assumed three example values $\delta_\text{CP,true} = 0, 180^\circ,
270^\circ$. We do not show results for $\delta_\text{CP,true}
=90^\circ$, since this value is quite disfavoured by data,
especially for IO.\footnote{We are aware of the fact that this choice
is somewhat arbitrary and implicitly resembles Bayesian reasoning. In
the strict frequentist sense we cannot a priori exclude any true value
of the parameters.}  The broken curves show for each set of true
values, the values of $\Delta\chi^2(\theta_{23}, \text{O})$ which are
larger than 68\%, 95\%, and 99\% of all generated data samples. From
the figure we see that the deviations from Gaussianity are not very
prominent and can be understood as follows:
\begin{itemize}
\item The confidence levels decrease around maximal mixing because of
the boundary on the parameter space present at maximal mixing for
disappearance data.

\item There is some increase and decrease in the confidence levels for
$\delta_\text{CP} = 270^\circ$, in the same parameter region as the
corresponding ones in Fig.~\ref{fig:nufit3_probab-dcp}.
\end{itemize}

\begin{table}\centering \catcode`?=\active\def?{\hphantom{0}}
  \begin{tabular}{cc|ccc} \toprule $\delta_\text{CP,true}$ &
Ordering & $\theta_{23} = 45^\circ$ & 90\% CL range & 95\% CL range \\
\midrule $??0^\circ$ & NO & 92\% & $[0.40, 0.49] \cup [0.55, 0.61]$ &
$[0.39, 0.62]$ \\ & IO & 98\% & $[0.55, 0.62]$ & $[0.42, 0.46] \cup
[0.54, 0.63]$ \\ $180^\circ$ & NO & 91\% & $[0.40, 0.50] \cup [0.54,
0.61]$ & $[0.40, 0.62]$ \\ & IO & 98\% & $[0.43, 0.44] \cup [0.55,
0.62]$ & $[0.41, 0.46] \cup [0.54, 0.63]$ \\ $270^\circ$ & NO & 92\% &
$[0.40, 0.49] \cup [0.55, 0.61]$ & $[0.39, 0.62]$ \\ & IO & 97\% &
$[0.42, 0.45] \cup [0.55, 0.62]$ & $[0.41, 0.48] \cup [0.53, 0.63]$ \\
\midrule Gaussian & NO & 92\% & $[0.41, 0.49] \cup [0.55, 0.61]$ &
$[0.40, 0.62]$ \\ & IO & 98\% & $[0.56, 0.62]$ & $[0.43, 0.45] \cup
[0.54, 0.63]$ \\ \bottomrule
  \end{tabular}
  \caption{CL for the rejection of maximal $\theta_{23}$ mixing (third
column), and 90\% and 95\% CL intervals for $\sin^2\theta_{23}$ for
different sets of true parameter values and in the Gaussian
approximation (last row).}
  \label{tab:nufit3_t23CL}
\end{table}

In Tab.~\ref{tab:nufit3_t23CL} we show the CL at which the combination
of LBL and reactor experiments can disfavour maximal $\theta_{23}$
mixing ($\theta_{23} = 45^\circ$) as well as the 90\% and 95\%
confidence intervals for $\sin^2\theta_{23}$ for both orderings with
respect to the global best fit. We observe from the table that the
Gaussian approximation is quite good for both, the CL of maximal
mixing as well as for the confidence intervals. We conclude that
the data presented in this section excludes maximal mixing at slightly more than
90\%~CL. Again we note that the intervals for $\sin^2\theta_{23}$ for
IO cannot be directly compared with the ones from
Tab.~\ref{tab:nufit3_bfranges}, where they are defined with respect to
the local minimum in each ordering.

\begin{table}\centering \catcode`?=\active\def?{\hphantom{0}}
  \begin{tabular}{c|ccc} \toprule $\delta_\text{CP,true}$ & NO/2nd
Oct. & IO/1st Oct. & IO/2nd Oct.  \\ \midrule $??0^\circ$ & 62\% & 91\%
& 28\% \\ $180^\circ$ & 56\% & 89\% & 32\% \\ $270^\circ$ & 70\% &
83\% & 27\% \\ \midrule Gaussian & 72\% & 94\% & 46\% \\ \bottomrule
  \end{tabular}
  \caption{CL for the rejection of various combinations of mass
ordering and $\theta_{23}$ octant with respect to the global best fit
(which happens for NO and 1st octant). We quote the CL of the local
minima for each ordering/octant combination, assuming three example
values for the true value of $\delta_\text{CP}$ as well as for the
Gaussian approximation (last row).}
  \label{tab:nufit3_octantCL}
\end{table}

In Tab.~\ref{tab:nufit3_octantCL} we show the CL at which a certain
combination of mass ordering and $\theta_{23}$ octant can be excluded
with respect to the global minimum in the NO and 1st $\theta_{23}$
octant.  We observe that the CL of the second octant for NO shows
relatively large deviations from Gaussianity and dependence on the
true value of $\delta_\text{CP}$. In any case, the sensitivity is very
low and the 2nd octant can be reject at most at 70\% CL ($1\sigma$)
for all values of $\delta_\text{CP}$. The first octant for IO can be
excluded at between 83\% and 91\%~CL, depending on $\delta_\text{CP}$.
As discussed above, the exclusion of the IO/2nd octant case
corresponds also to the exclusion of the IO, since at that point the
confidence interval in IO would vanish. Also in this case we observe
deviations from the Gaussian approximation and the CL of at best 32\%
is clearly less than $1\sigma$ (consistent with the results discussed
in the previous subsection), showing that the considered data set has
essentially no sensitivity to the mass ordering.

\subsection{Conclusions}
\label{sec:nufit3_summary}

In this section we have presented the results of the analysis as of fall 2016
of relevant neutrino data in the framework of mixing among
three massive neutrinos.  Quantitatively the determination of
the two mass differences, three mixing angles and the relevant CP
violating phase obtained under the assumption that their
log-likelihood follows a $\chi^2$ distribution is listed in
Tab.~\ref{tab:nufit3_bfranges}, and the corresponding leptonic mixing
matrix is given in Eq.~\eqref{eq:nufit3_umatrix}.  We have found that
the maximum allowed CP violation in the leptonic sector parametrised
by the Jarlskog determinant is $J_\text{CP}^\text{max} = 0.0329 \pm
0.0007 \, (^{+0.0021}_{-0.0024}))$ at $1\sigma$ ($3\sigma$).

We have studied in detail how the sensitivity to the least-determined
parameters $\theta_{23}$, $\delta_\text{CP}$ and the mass ordering
depends on the proper combination of the different data samples
(Sec.~\ref{subsec:nufit3_dm32}).  Furthermore we have quantified
deviations from the Gaussian approximation in the evaluation of the
confidence intervals for $\theta_{23}$ and $\delta_\text{CP}$ by
performing a Monte Carlo study of the LBL accelerator and
reactor results (Sec.~\ref{sec:nufit3_MC}).  We can summarise the main
conclusions in these sections as follows:
\begin{itemize}
\item The precision on the determination of $|\Delta
m^2_{3\ell}|$ from $\nu_\mu$ disappearance in LBL accelerator
experiments NO$\nu$A, T2K and MINOS is comparable to that from $\nu_e$
disappearance in reactor experiments, in particular with the spectral
information from Daya-Bay. When comparing the region for each LBL
experiment with that of the reactor experiments we find some
dispersion in the best fit values and allowed ranges.

\item The interpretation of the data from accelerator LBL experiments
in the framework of $3\nu$ mixing requires using information from the
reactor experiments, in particular about the mixing angle
$\theta_{13}$.  But since, as mentioned above, reactor data also
constrain $|\Delta m^2_{3\ell}|$, the resulting CL of low
confidence effects (in particular the non-maximality of $\theta_{23}$
and the mass ordering) is affected by the inclusion of this
information in the combination.

\item We find that the mass ordering favoured by NO$\nu$A changes from
NO to IO when the information on $\Delta m^2_{3\ell}$ from reactor
experiments is correctly included in the LBL+REA combination, and the
$\Delta\chi^2$ of NO in T2K is reduced from around 2 to 0.5 (see
Fig.~\ref{fig:nufit3_chisq-dma}).  Our MC study of the combination of
LBL and reactor data shows that for all cases generated, NO is favoured
but with a CL of less than $1\sigma$.

\item About the non-maximality of $\theta_{23}$, we find that when the
information on $\Delta m^2_{3\ell}$ from reactor experiments is
correctly included in the LBL+REA combination, it is not NO$\nu$A but
actually MINOS which contributes most to the preference for
non-maximal $\theta_{23}$ (see Fig.~\ref{fig:nufit3_chisq-t23}).
Quantitatively our MC study of the combination of LBL and reactor data
shows that for all the cases generated the CL for rejection of maximal
$\theta_{23}$ is about 92\% for NO.  As seen in
Fig.~\ref{fig:nufit3_probab-t23} and Tab.~\ref{tab:nufit3_t23CL}, the
CL of maximal mixing as well as confidence intervals for
$\sin^2\theta_{23}$ derived with MC simulations are not very different
from the corresponding Gaussian approximation.

\item The same study shows that for NO (IO) the favoured octant is
$\theta_{23}<45^\circ$ ($\theta_{23}>45^\circ$).  The CL for rejection
of the disfavoured octant depends on the true value of
$\delta_\text{CP}$ assumed in the MC study and it is generically lower
than the one obtained in the Gaussian limit (see
Tab.~\ref{tab:nufit3_octantCL}). For example, for NO the second octant
is disfavoured at a confidence level between $0.9\sigma$ and
$1.3\sigma$ depending on the assumed true value of $\delta_\text{CP}$.

\item The sensitivity to $\delta_\text{CP}$ is driven by T2K
with a minor contribution from NO$\nu$A for IO (see
Fig.~\ref{fig:nufit3_chisq-dcp}). The dependence of the combined CL of
the ``hint'' towards leptonic CP violation and in particular for
$\delta_\text{CP} \simeq 270^\circ$ on the true value of $\theta_{23}$
is shown in Fig.~\ref{fig:nufit3_probab-dcp}, from which we read that
for all cases generated CP conservation is disfavoured only at 70\%
($1.05\sigma$) for NO.  Values of $\delta_\text{CP} \simeq 90^\circ$
are disfavoured at around 99\%~CL for NO, while for IO the rejection is
at higher CL ($\Delta\chi^2 \simeq 14$ with respect to the global
minimum).
\end{itemize}
Finally we comment that the increased statistics in SK4 and Borexino
has had no major impact in the long-standing tension between the best
fit values of $\Delta m^2_{21}$ as determined from the analysis of
KamLAND and solar data, which remains an unresolved $\sim 2\sigma$
effect.

\section{Results on \texorpdfstring{$\delta_\text{CP}$}{dCP}: from 2016 to present}
\label{sec:dCPevolution}

The results presented above correspond to the status just after the first \NOvA/ data
release. Along the following years, the LBL accelerator experiments
(and also some reactor experiments) have continued releasing data. In
this section, we will overview how this has affected the status of the
leptonic CP phase $\delta_\text{CP}$.  As has been shown in \cref{sec:nufit3_MC}, evaluating
confidence intervals with a Monte Carlo simulation gives a similar
result to using Wilks' theorem, and so the latter will be assumed in
what follows.

\subsection{November 2017 update} 
About one year after the results
presented in the previous section, the T2K experiment released new
data with twice as statistics in the neutrino mode and a $\sim 2\%$
increase in the antineutrino mode data~\cite{Abe:2018wpn}.  The
$\parenbar{\nu}_\mu$ disappearance spectrum pointed towards $P_{\mu
\mu}=0$ at energies $\sim \SI{0.6}{GeV}$, increasing the significance
for maximal $\theta_{23}$. In addition, the RENO experiment published
additional spectral data~\cite{reno:eps2017}, which slightly improved the precision on $\theta_{13}$
and $|\Delta m^2_{3\ell}|$.

The T2K $\parenbar{\nu}_e$ appearance results, though, significantly
impacted the determination of $\delta_\mathrm{CP}$. This can be understood in terms of the
approximate $\parenbar{\nu}_\mu \rightarrow \parenbar{\nu}_e$
transition probability~\cite{Cervera:2000kp, Freund:2001pn,
Akhmedov:2004ny}
\begin{equation}
\begin{split} P_{\parenbar{\nu}_\mu \rightarrow \parenbar{\nu}_e}
\simeq & 4 \sin^2 \theta_{13} \sin^2 \theta_{23} \frac{\sin^2
\Delta}{(1-A)^2} + \left(\frac{\Delta m^2_{21}}{\Delta
m^2_{31}}\right)^2 \sin^2 2 \theta_{12} \cos^2 \theta_{23}
\frac{\sin^2 A \Delta}{A^2} \\ & + 8 \frac{\Delta m^2_{21}}{\Delta
m^2_{31}} J^\mathrm{max}_\mathrm{CP} \cos (\Delta \pm
\delta_\text{CP}) \frac{\sin \Delta A}{A} \frac{\sin \Delta
(1-A)}{1-A} \, ,
\end{split}
\label{eq:appProb}
\end{equation} where $J^\mathrm{max}_\mathrm{CP}$ is the maximum value
of the leptonic Jarlskog invariant as defined in
\cref{eq:jarlskogNeutrino}, $\Delta \equiv \frac{\Delta m^2_{31} L}{4 E}$
and $A \equiv \frac{2E V}{\Delta m^2_{31}}$ with $V$ the matter potential
in \cref{eq:matterPotential}. The $+$($-$) sign applies to neutrinos
(antineutrinos). The probability has been expanded to second order in
the small parameters $\sin \theta_{13} \sim 0.15$ and $\frac{\Delta
m^2_{21}}{\Delta m^2_{31}} \sim 0.03$. 

The last term gives the
sensitivity to $\delta_\text{CP}$ and is the most relevant for our
discussion.  Since T2K has a narrow neutrino spectrum peaked around
maximal $\nu_\mu$ disappearance, $\Delta \sim
\frac{\pi}{2}$. Therefore, the $\nu_e$ appearance probability is
maximised for $\delta_\text{CP} \sim \frac{3\pi}{2}$. At the same
time, the $\bar{\nu}_e$ appearance probability is minimised for that
same value of $\delta_\text{CP}$. The T2K result presented in the
previous section, favouring maximal CP violation ($\delta_\text{CP}
\sim \frac{3\pi}{2} $), was indeed driven by an excess of $\nu_e$ events and
a deficit of $\bar\nu_e$ events with respect to the expectations, as
\cref{tab:T2K2016app} shows. As a result, the significance for maximal
CP violation was larger than the expected sensitivity.

\begin{table}[hbtp] \centering
\begin{tabular}{cccccc} \toprule Channel & $\delta_\text{CP} = 0$ &
$\delta_\text{CP} = \frac{\pi}{2}$ & $\delta_\text{CP} = \pi$ &
$\delta_\text{CP} = \frac{3\pi}{2}$ & Observed \\ \midrule $\nu_e$ &
24.2 & 19.6 & 24.1 & 28.7 & 32\\ $\bar\nu_e$ & 6.9 & 7.7 & 6.8 & 6.0 &
4 \\ \bottomrule
\end{tabular}
\caption{Expected and observed number of $\nu_e$ and $\bar\nu_e$
events in T2K, as of late 2016, for different values of
$\delta_\text{CP}$. The expectations are shown for NO, which increases the
amount of $\nu_e$ events and decreases the amount of $\bar\nu_e$
events as suggested by the data. The other mixing parameters are set to
$\Delta m^2_{21} = \SI{7.53e-5}{eV^2}$, $\sin^2 2 \theta_{12} =
0.846$, $\sin^2 2 \theta_{13} = 0.085$, $\Delta m^2_{32} =
\SI{2.509e-3}{eV^2}$ and $\sin^2 2 \theta_{23} = 0.528$.  Table
adapted from Ref.~\cite{Abe:2017uxa}.}
\label{tab:T2K2016app}
\end{table}

The 2017 T2K results in the $\parenbar{\nu}_e$ appearance channels
also present in the 2016 analysis, which correspond to events in
which only one electron Cherenkov ring was reconstructed in the final
state, were closer to the expected values for $\delta_\text{CP} \sim
\frac{3\pi}{2}$.  However, the new T2K results included data from an
additional $\nu_e$ channel where a pion was also reconstructed in the final
state. This channel (see \cref{tab:T2K2017app}) presented a $\sim 2
\sigma$ upper fluctuation that increased the significance for
$\delta_\text{CP} \sim \frac{3\pi}{2}$ above the expectations.

\begin{table}[hbtp] \centering
\begin{tabular}{cccccc} \toprule Channel & $\delta_\text{CP} = 0$ &
$\delta_\text{CP} = \frac{\pi}{2}$ & $\delta_\text{CP} = \pi$ &
$\delta_\text{CP} = \frac{3\pi}{2}$& Observed \\ \midrule $\nu_e$ &
61.4 & 49.9 & 61.9 & 73.5 & 74\\ $\nu_e$ (CC1$\pi$) & 6.0 & 4.9 & 5.8
& 6.9 & 15 \\ $\bar\nu_e$ & 9.8 & 10.9 & 9.7 & 8.5 & 7\\ \bottomrule
\end{tabular}
\caption{Expected and observed number of $\nu_e$ and $\bar\nu_e$
events in T2K, as of late 2017, for different values of
$\delta_\text{CP}$. The expectations are shown for NO, which increases the
amount of $\nu_e$ events and decreases the amount of $\bar\nu_e$
events as suggested by the data.  $\nu_e$ (CC1$\pi$) refers to the
channel where an electron and a single pion are detected in the final
state. The other mixing parameters are set to $\Delta m^2_{21} =
\SI{7.53e-5}{eV^2}$, $\sin^2 \theta_{12} = 0.304$, $\sin^2 \theta_{13}
= 0.0219$, $\Delta m^2_{32} = \SI{2.509e-3}{eV^2}$ and $\sin^2 2
\theta_{23} = 0.528$. Table adapted from Ref.~\cite{Abe:2018wpn}.}
\label{tab:T2K2017app}
\end{table}

\begin{figure}[hbtp] \centering
\includegraphics[width=0.8\textwidth]{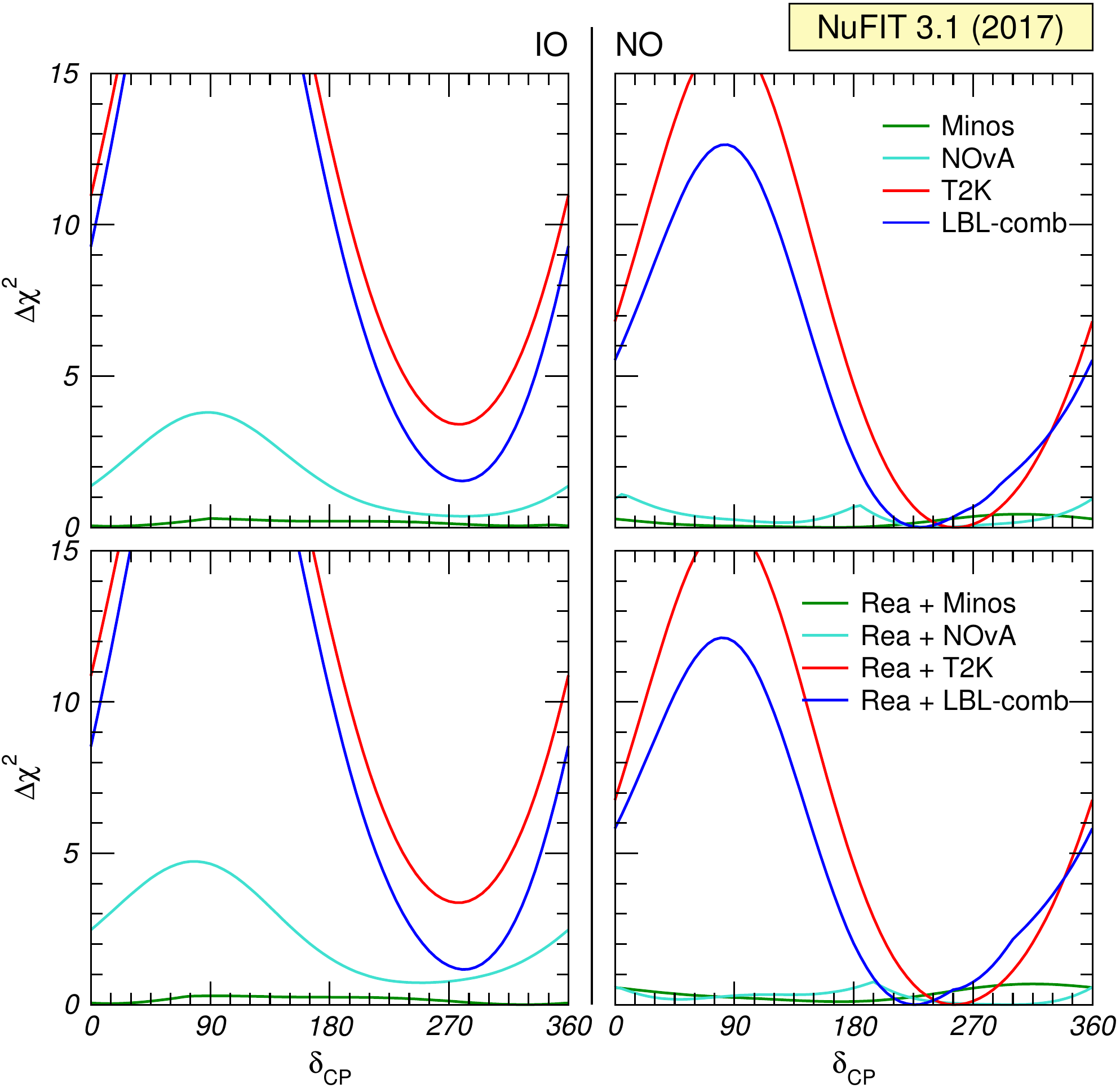}
\caption{$\delta_\text{CP}$ determination as of November 2017. Left
(right) panels are for IO (NO). The upper panels constrain only
$\theta_{13}$ from reactor experiments, whereas the lower panels
include the full information from them: see
\cref{fig:nufit3_chisq-dcp} and its description in the text for the
difference among both procedures.}
\label{fig:dCP31}
\end{figure}

The combined $\Delta \chi^2$ as a function of $\delta_\text{CP}$ is shown in \cref{fig:dCP31}.\footnote{See
\url{http://www.nu-fit.org/?q=node/150} for all the results in the
global fit.} Comparing with \cref{fig:nufit3_chisq-dcp}, we notice
that the significance for $\delta_\text{CP} \sim \frac{3\pi}{2}$
increased noticeably: the new results rejected $\delta_\text{CP} \sim
\frac{\pi}{2}$ with more than 3$\sigma$. In the global combination,
though, CP conservation was still allowed within $1 \sigma$, less stringently than in
the T2K result alone. This was due to \NOvA/ and MINOS pushing
$\theta_{23}$ away from $45^\circ$ (see \cref{fig:nufit3_chisq-t23}).
As a consequence, for $\theta_{23} > 45^\circ$, the first term in
\cref{eq:appProb} increases and the T2K excess could be explained
without resorting to very large CP violation. This result emphasises that
the hint towards maximal CP violation is \emph{not} driven by
directly observing CP violation, but by a $\nu_e$ excess requiring
a very large $\nu_e$ appearance probability that maximal CP
violation provides. The T2K hint for NO also has the same origin, and
because of that the 2017 update increased its significance.  These
hints, being indirect and coming from statistical fluctuations, are
expected to be sensitive to the theoretical model under which the data
is analysed. This will be explored in the next chapters.

\subsection{January 2018 update} 
\begin{figure}[hbtp]
\centering
\begin{subfigure}{\textwidth}
\centering
\includegraphics[width=0.55\textwidth]{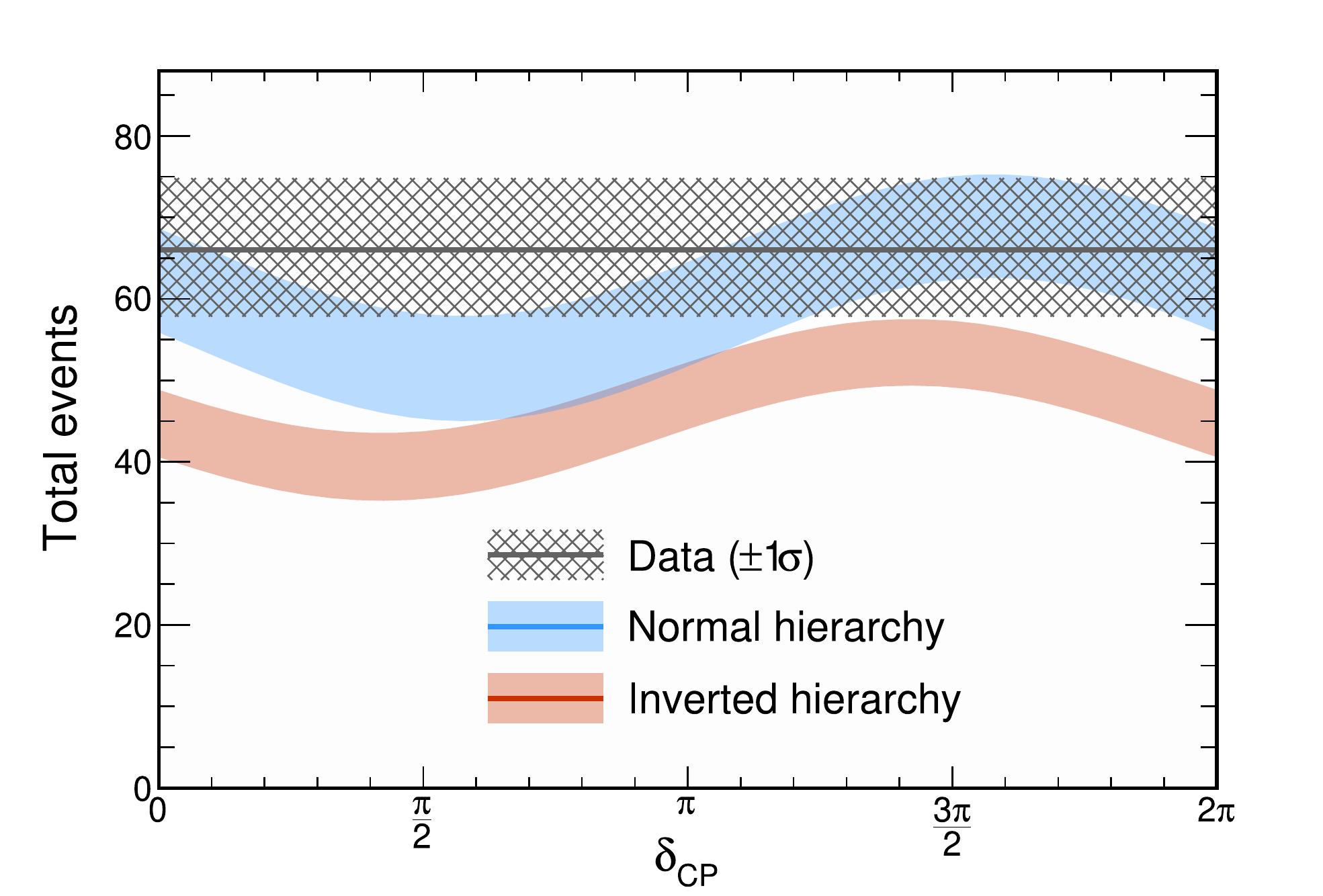}
\caption{Total number of observed $\nu_e$ events at \NOvA/ (gray). The prediction (colour) is shown as a function of
$\delta_text{CP}$ for both mass orderings (here named as
``hierarchy''). The bands represent the variation as $\sin^2
\theta_{23}$ changes from 0.43 to 0.60. Adapted from
Ref.~\cite{NOvA:2018gge}.}
\label{fig:NOvA2018jan}
\end{subfigure}

\begin{subfigure}{\textwidth}
\centering
\includegraphics[width=0.8\textwidth]{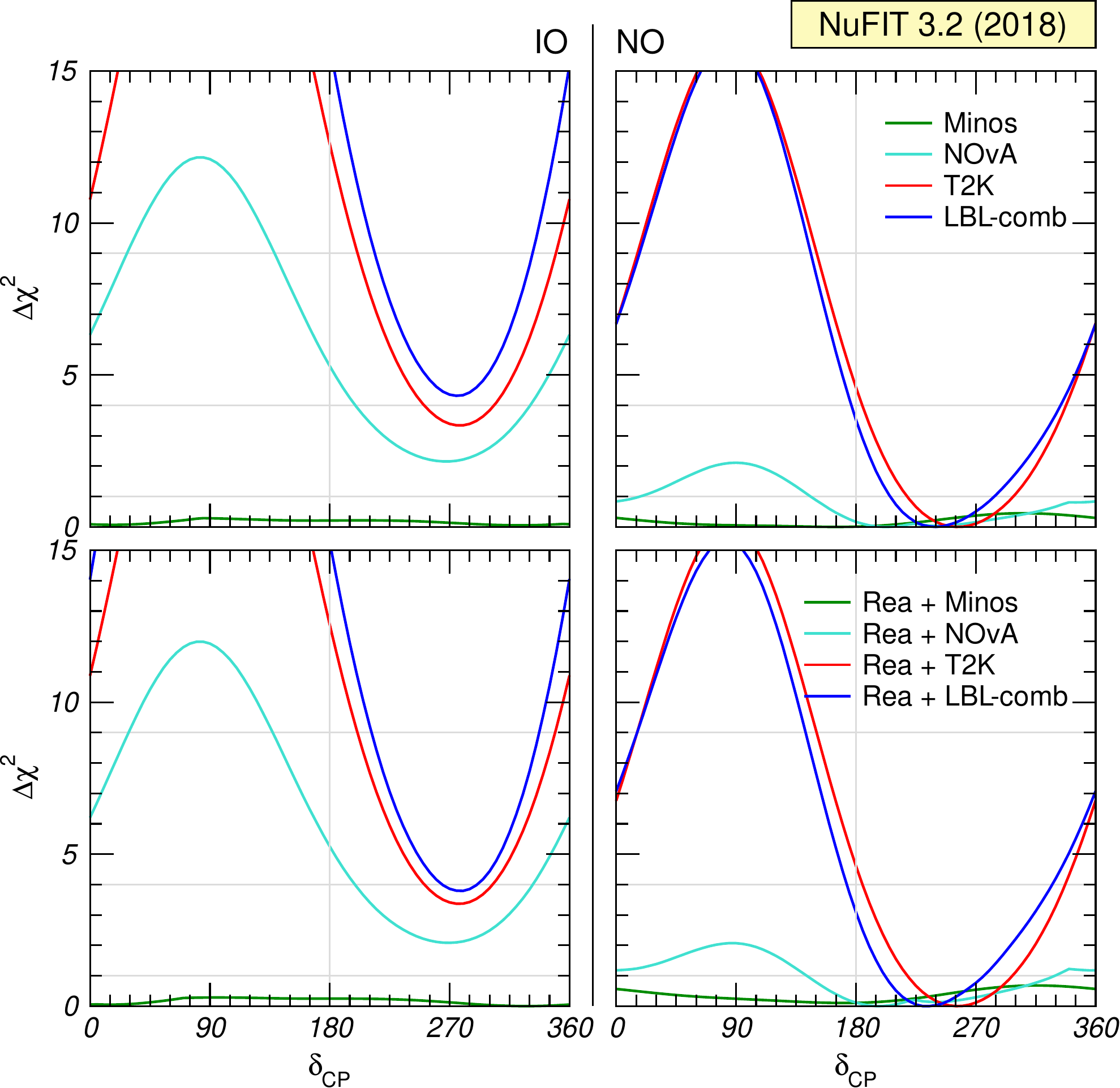}
\caption{$\delta_\text{CP}$ determination. Left
(right) panels are for IO (NO). The upper panels constrain only
$\theta_{13}$ from reactor experiments, whereas the lower panels
include the full information from them: see
\cref{fig:nufit3_chisq-dcp} and its description in the text for the
difference among both procedures.}
\label{fig:dCP32}
\end{subfigure}
\caption{Updated \NOvA/ results on $\delta_\text{CP}$ and global combination as of January 2018.}
\end{figure}

Shortly after the update mentioned
above, \NOvA/ released new data with $\sim 50\%$ more
statistics~\cite{NOvA:2018gge}. They also observed a large amount of
events in the $\nu_e$ appearance channel, increasing the significance for maximal 
CP violation and NO. Their results are shown in \cref{fig:NOvA2018jan}: unlike for T2K, they were well within the expectations.

In addition, they also improved the simulation of the
energy response of their detector by including propagation of Cherenkov light inside it. As has
been discussed in \cref{sec:3nutheor_accel}, the issue of whether $\theta_{23} = 45^\circ$ or not is
related to observing a minimum in the $\nu_\mu$ spectrum. Thus, it is
rather sensitive to energy mismodelling. Indeed, after this
improvement the \NOvA/ results became perfectly compatible with $\theta_{23} =
45^\circ$. As a consequence, the $\nu_e$ excess in T2K and \NOvA/ could no longer be explained with a large
$\theta_{23}$, and so the significance of CP violation in the combined
analysis grew up to $\sim 2 \sigma$. This effect, along with the
\NOvA/ preference for maximal CP violation, can be seen in
\cref{fig:dCP32}.\footnote{See \url{http://www.nu-fit.org/?q=node/166} for all
the results of the global fit.}

\subsection{November 2018 update}

At the end of 2018, Daya Bay and RENO published new results with
increased statistics~\cite{Adey:2018zwh,Bak:2018ydk}, T2K increased the statistics of their
antineutrino sample~\cite{t2k:nu2018}, and \NOvA/ released their first
antineutrino data~\cite{nova:nu2018}.  This is particularly relevant,
as comparing neutrino and antineutrino data constitutes a direct check
of CP violation. The whole global fit is discussed in detail in
Ref.~\cite{Esteban:2018azc}, and here we will explore the results
affecting $\delta_\text{CP}$.

\begin{figure}[hbtp] \centering
\begin{subfigure}[b]{\textwidth}
\centering
\includegraphics[width=0.35\textwidth]{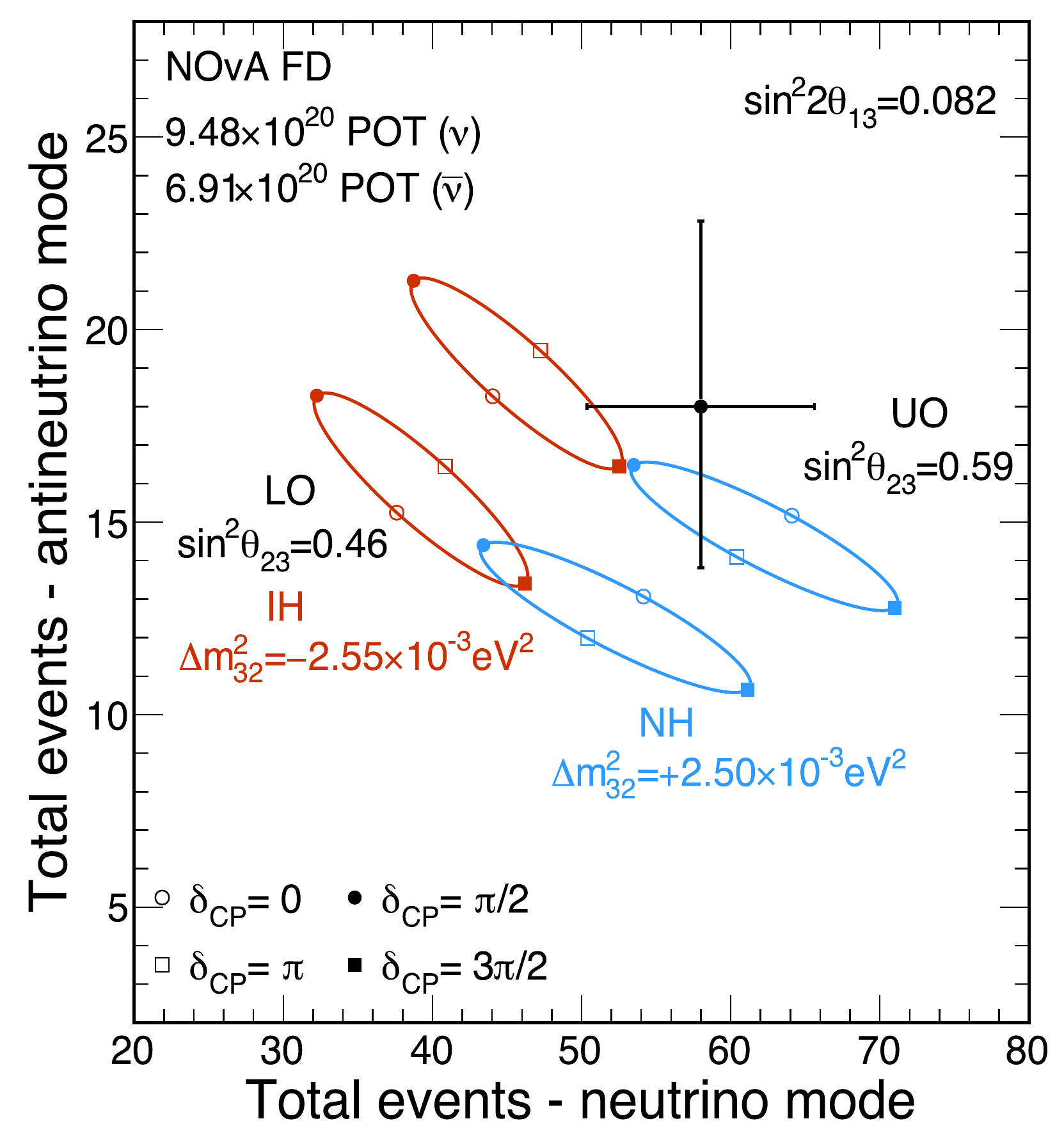} \includegraphics[width=0.44\textwidth]{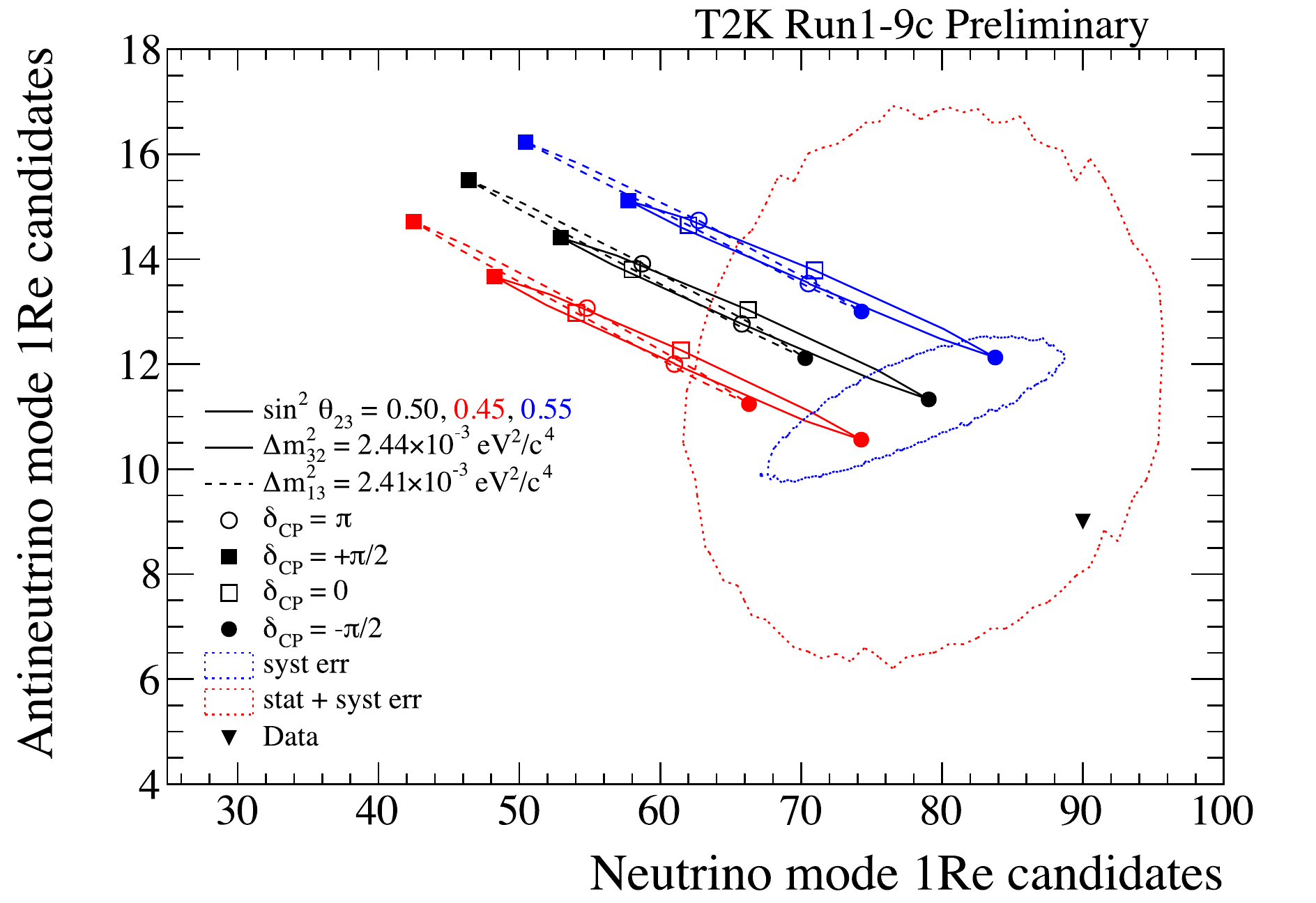}
\caption{Observed number of $\nu_e$ and $\bar\nu_e$ events at \NOvA/ 
(left) and T2K (right), extracted from Refs.~\cite{nova:nu2018,t2k:nu2018}. The expectations are shown 
for different values of $\delta_\text{CP}$, $\theta_{23}$ and the mass 
ordering. In the left panel, ``UO'' means Upper Octant ($\theta_{23} > 
45^\circ$), ``LO'' Lower Octant ($\theta_{23} < 45^\circ$), ``NH'' 
Normal Ordering and ``IH'' Inverted Ordering. Notice that the \NOvA/ 
ellipses are more open and separated, giving more sensitivity to the 
$\theta_{23}$ octant, the mass ordering and $\delta_\text{CP}$. This 
is due to the larger baseline of this experiment, that increases matter 
effects.}
\label{fig:neutrino2018biprob}
\end{subfigure}

\begin{subfigure}{\textwidth}
\centering
\includegraphics[width=0.688\textwidth]{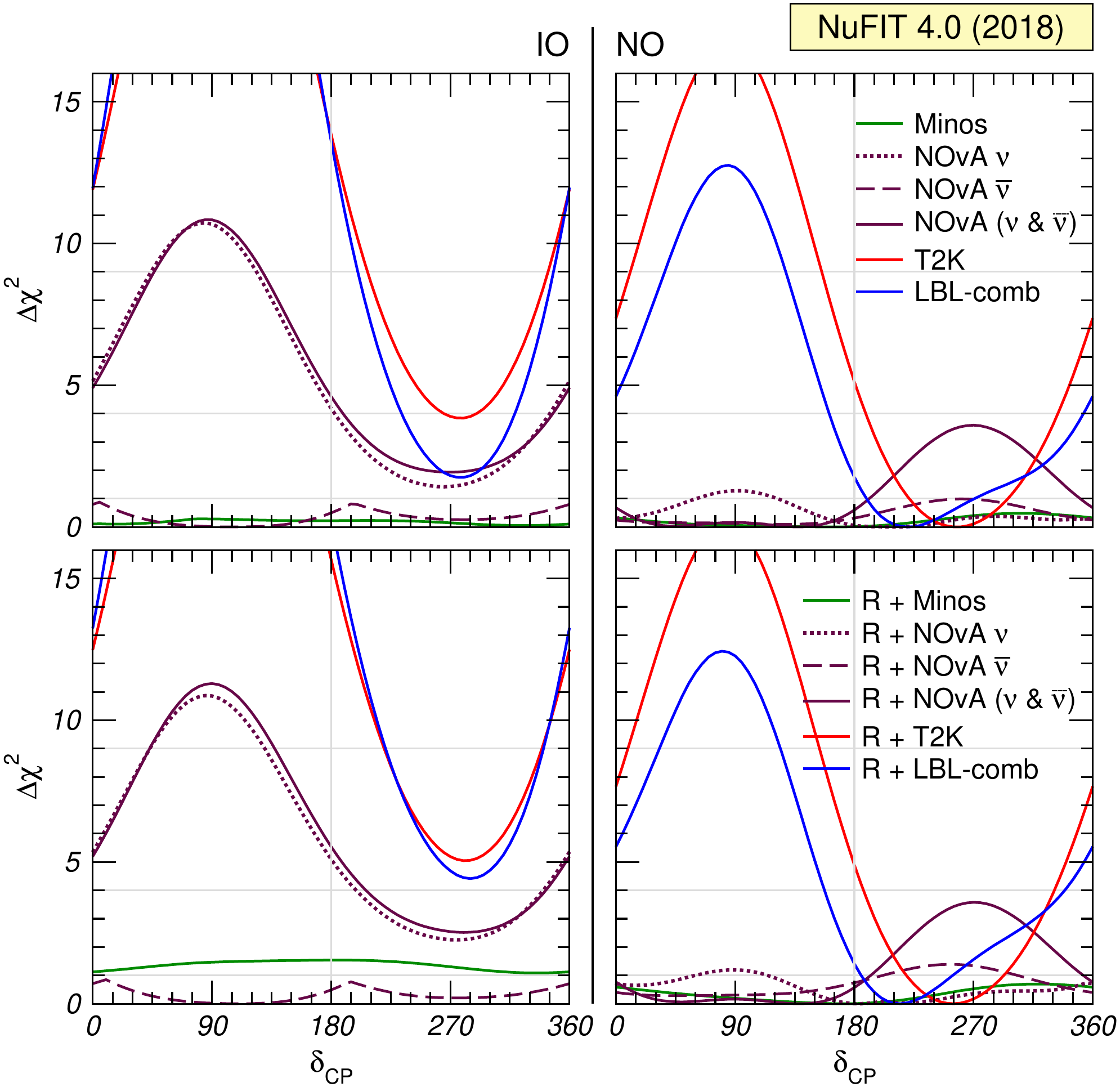}
\caption{$\delta_\text{CP}$ determination. Left
(right) panels are for IO (NO). The upper panels constrain only
$\theta_{13}$ from reactor experiments, whereas the lower panels
include the full information from them: see
\cref{fig:nufit3_chisq-dcp} and its description in the text for the
difference among both procedures. The \NOvA/ results are separated into 
neutrino and antineutrino samples.}
\label{fig:dCP40}
\end{subfigure}
\caption{Updated \NOvA/ and T2K $\protect\parenbar{\nu}_e$ appearance results and global combination projection on $\delta_\text{CP}$ as of November 2018.}
\end{figure}

This information is mostly driven by the $\parenbar{\nu}_e$ appearance signals, depicted
in \cref{fig:neutrino2018biprob}. As shown there, the T2K $\bar\nu_e$ 
sample still showed a slight downward fluctuation with respect to the expectations, driving $\delta_\text{CP}$ 
towards $\frac{3\pi}{2}$. The \NOvA/ data, however, did 
\emph{not} show a deficit of electron antineutrinos. As a consequence, 
when combining the \NOvA/ neutrino and antineutrino samples, they did 
not point towards $\delta_\text{CP} \sim \frac{3\pi}{2}$.

Furthermore, the \NOvA/ $\bar\nu_\mu$ sample rejected maximal 
$\bar\nu_\mu$ disappearance with $\sim 3 \sigma$. As has been 
discussed earlier in this section, this reduces the T2K significance 
for CP violation. The global combination is shown in \cref{fig:dCP40}. There, we also see that only with the combination of neutrino and 
antineutrino samples could \NOvA/ reject $\delta_\text{CP} \sim 
\frac{3\pi}{2}$ with $\sim 2 \sigma$. In the global fit, CP conservation was just rejected with $\sim 1 \sigma$.

\subsection{July 2019 update}

\begin{figure}[hbtp] \centering
\begin{subfigure}[b]{\textwidth}
\centering
\includegraphics[width=0.38\textwidth]{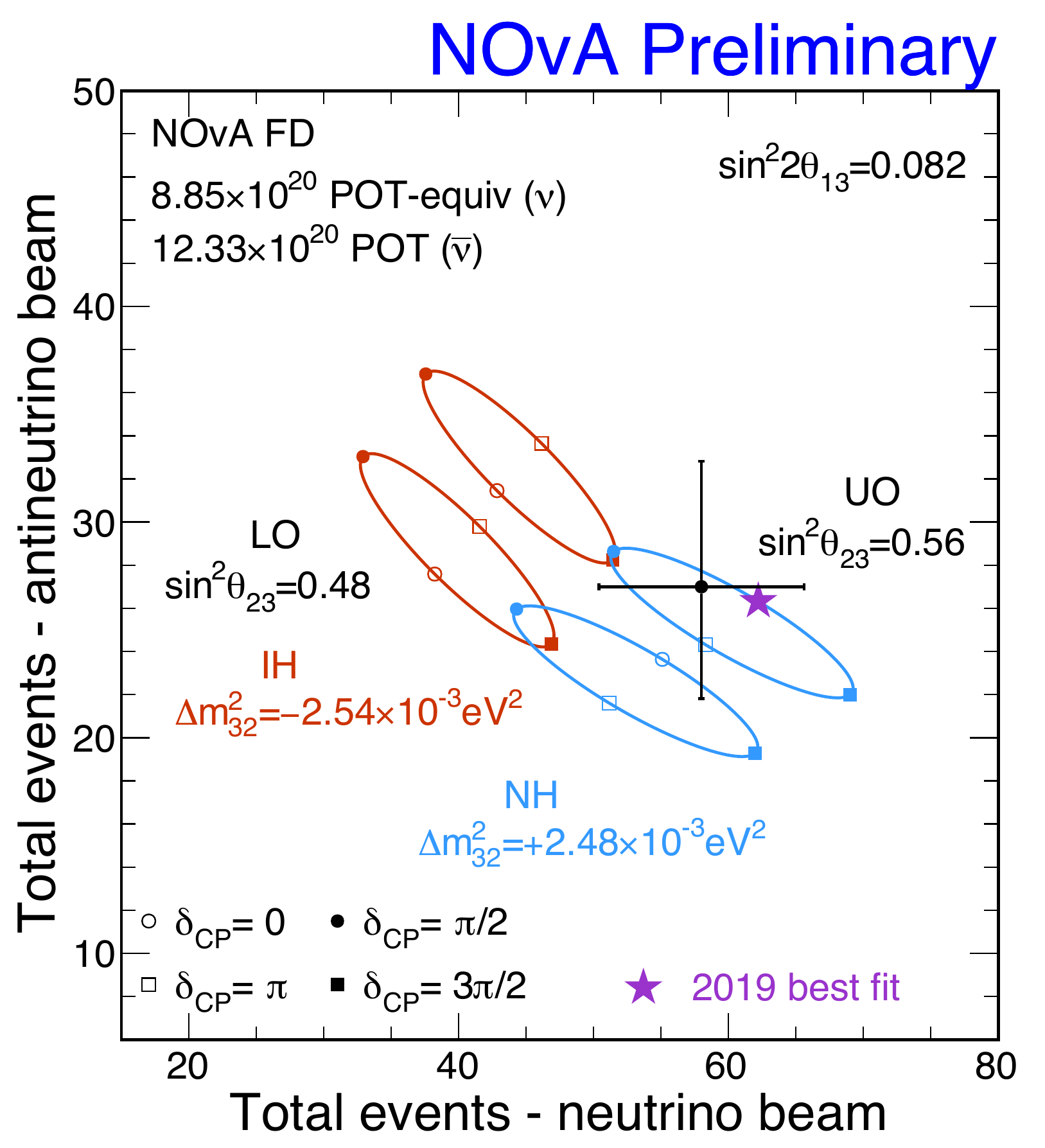} \includegraphics[width=0.45\textwidth]{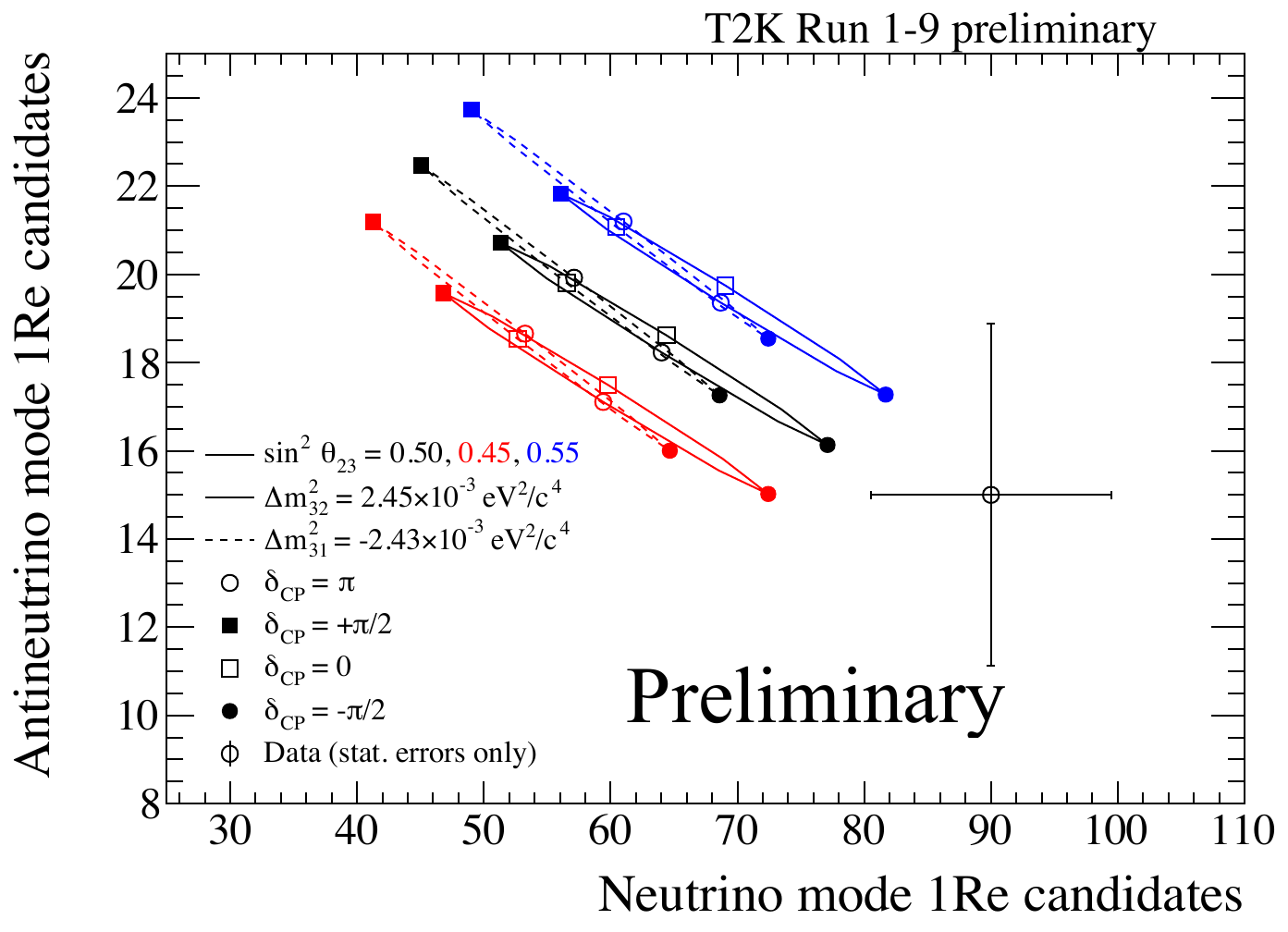}
\caption{Observed number of $\nu_e$ and $\bar\nu_e$ events at \NOvA/ 
(left) and T2K (right), extracted from Refs.~\cite{nova:nuphys2019,t2k:kek2019}. The expectations are shown 
for different values of $\delta_\text{CP}$, $\theta_{23}$ and the mass 
ordering. In the left panel, ``UO'' means Upper Octant ($\theta_{23} > 
45^\circ$), ``LO'' Lower Octant ($\theta_{23} < 45^\circ$), ``NH'' 
Normal Ordering and ``IH'' Inverted Ordering.}
\label{fig:2019biprob}
\end{subfigure}

\begin{subfigure}{\textwidth}
\centering
\includegraphics[width=0.688\textwidth]{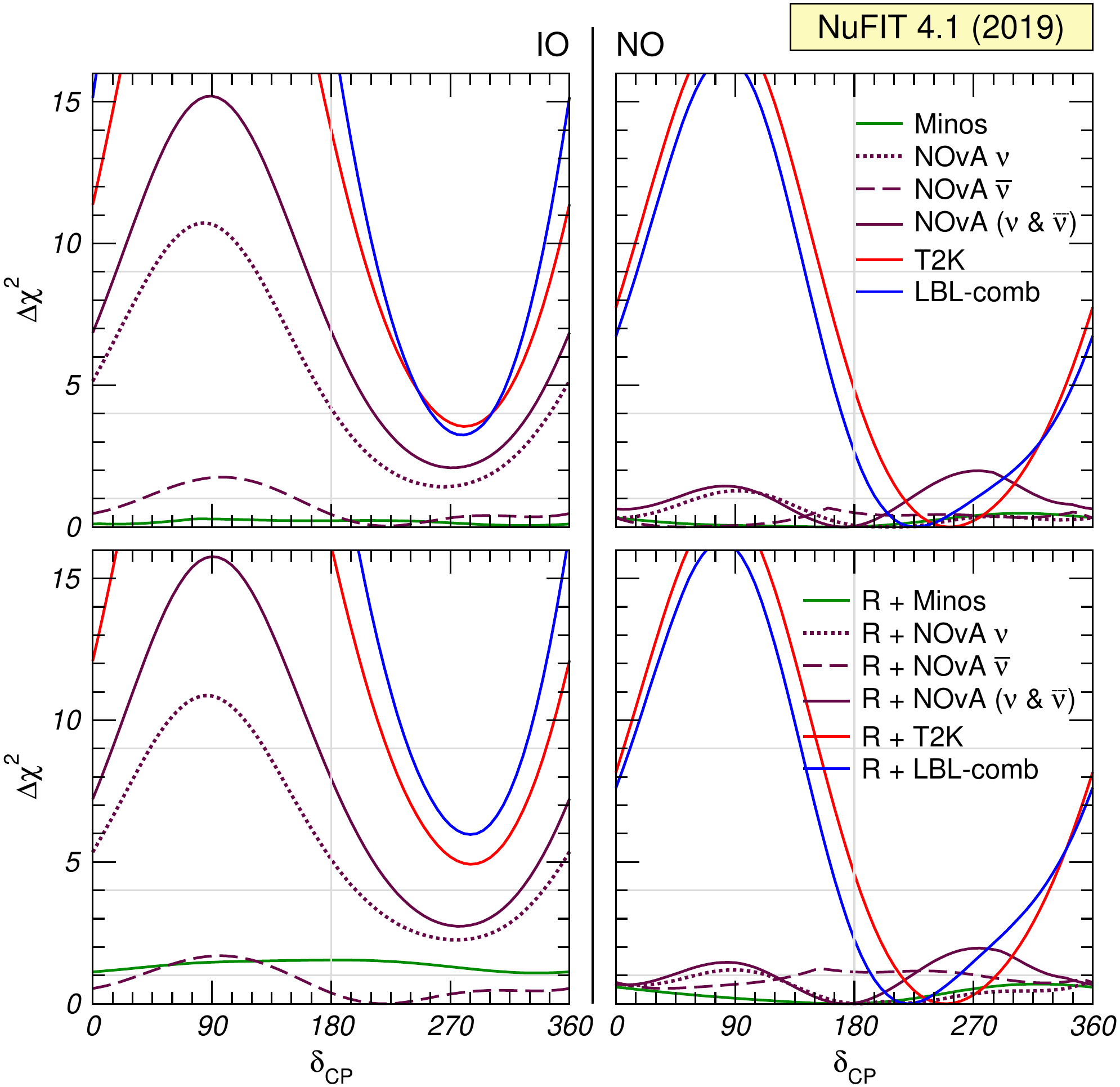}
\caption{$\delta_\text{CP}$ determination. Left
(right) panels are for IO (NO). The upper panels constrain only
$\theta_{13}$ from reactor experiments, whereas the lower panels
include the full information from them: see
\cref{fig:nufit3_chisq-dcp} and its description in the text for the
difference among both procedures. The \NOvA/ results are separated into 
neutrino and antineutrino samples.}
\label{fig:dCP41}
\end{subfigure}
\caption{Updated \NOvA/ and T2K $\protect\parenbar{\nu}_e$ appearance results and global combination projection on $\delta_\text{CP}$ as of July 2019.}
\end{figure}

The last update before this thesis was completed took place in summer 
2019, as the T2K~\cite{t2k:kek2019} and \NOvA/~\cite{Acero:2019ksn} 
experiments released new antineutrino data. The observed and expected 
number of events in the $\parenbar{\nu}_e$ appearance channels are shown in 
\cref{fig:2019biprob}. As can be seen, the T2K $\bar\nu_e$ appearance 
signal has moved towards the expectation. The \NOvA/ $\bar\nu_e$ data, 
when combined with the $\nu_e$ data, also keeps pointing towards 
$\delta_\text{CP} \neq \frac{3\pi}{2}$ and normal mass ordering.

In addition, the \NOvA/ $\bar\nu_\mu$ spectrum now allows $\theta_{23}
= 45^\circ$ within less than $2 \sigma$: the previous indication for
non-maximal $\theta_{23}$ was probably a statistical fluctuation due
to the limited statistics. This has slightly increased the global significance for CP violation.

The combined $\Delta \chi^2$ as a function of $\delta_\text{CP}$ is shown in \cref{fig:dCP41}. 
As can be seen, there is still a slight tension among T2K, that prefers
$\delta_\text{CP} \sim \frac{3\pi}{2}$ due to the large amount of 
observed $\nu_e$ events; and \NOvA/, that does not observe a 
significant $\nu_e$ excess and $\bar\nu_e$ deficit. As a consequence, 
the T2K $\sim 2 \sigma$ hint for leptonic CP violation gets diluted to 
$\sim 1.5 \sigma$ when all the data is combined. 

Finally, as a summary of the current status of three-neutrino mixing, 
in \cref{fig:nufit41_region-glob,fig:nufit41_chisq-glob} we show the 
current projections of the allowed six-dimensional parameter 
space.\footnote{More results are 
available in \url{http://www.nu-fit.org/?q=node/211}.} The results are
shown with and without the Super-Kamiokande atmospheric neutrino data:
even though, as explained in \cref{subsec:nufit3_SK}, we cannot 
reproduce their results, they have provided a $\Delta \chi^2$ table that 
allows to include them in a global fit~\cite{Abe:2017aap}.

These figures are to be compared with 
\cref{fig:nufit3_region-glob,fig:nufit3_chisq-glob}
, the first global combination produced as part of this thesis. 
Because of all the data released mostly by LBL accelerator experiments, there is no 
longer an octant degeneracy in $\theta_{23}$, although whether 
$\theta_{23} = 45^\circ$ or not is still unknown. The mass ordering, 
about which there was no clue, is now favoured to be normal at 
2--3$\sigma$. And, finally, we have gained a lot of information 
regarding the CP phase $\delta_\text{CP}$. Values around $\frac{3\pi}
{2}$ are still favoured, and now $\delta_\text{CP} \sim \frac{\pi}{2}$ is disfavoured with 
$\sim 4 \sigma$. The issue of CP conservation, though, remains unclear as $\delta_\text{CP} = \pi$ is still allowed within $\sim 1.5 \sigma$.

\begin{pagefigure}\centering
  \includegraphics[width=0.79\textwidth]{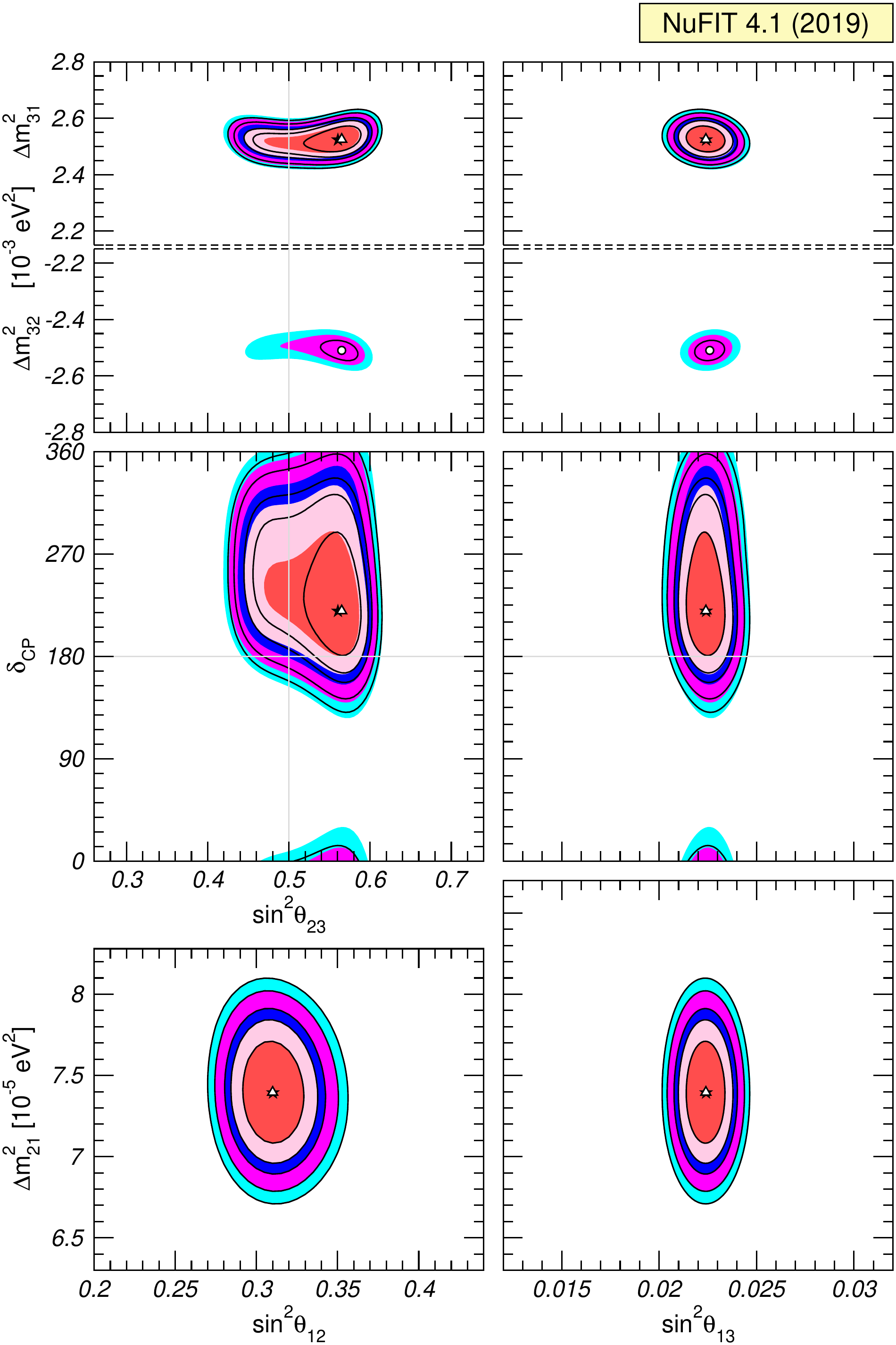}
  \caption{Global $3\nu$ oscillation analysis. Each panel shows the
  two-dimensional projection of the allowed six-dimensional region
  after minimisation with respect to the undisplayed parameters. The
  regions in the four lower panels are obtained from $\Delta\chi^2$
  minimised with respect to the mass ordering. The different contours
  correspond to $1\sigma$, 90\%, $2\sigma$, 99\%, $3\sigma$ CL (2
  dof). Coloured regions (black contour curves) are without (with)
  adding the Super-Kamiokande 
  atmospheric results, provided as a $\Delta \chi^2$ table by the 
  collaboration. Note that as atmospheric mass-squared splitting we
  use $\Delta m^2_{31}$ for NO and $\Delta m^2_{32}$ for IO.}
  \label{fig:nufit41_region-glob}
\end{pagefigure}

\begin{pagefigure}\centering
\includegraphics[width=0.84\textwidth]{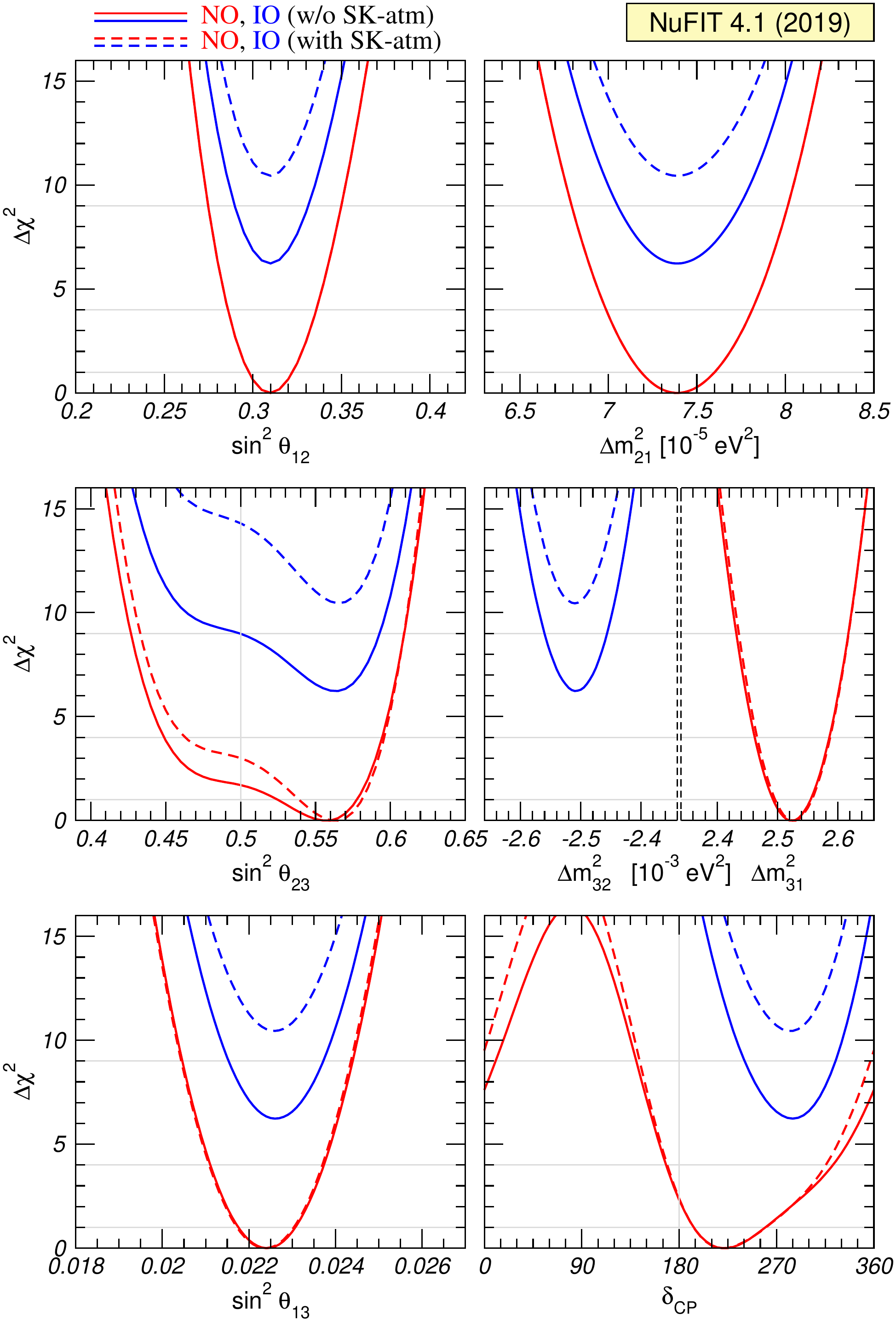}
  \caption{Global $3\nu$ oscillation analysis. We show $\Delta \chi^2$
  profiles minimised with respect to all undisplayed parameters. The
  red (blue) curves correspond to Normal (Inverted) Ordering. Solid
  (dashed) curves are without (with) adding the Super-Kamiokande 
  atmospheric results, provided as a $\Delta \chi^2$ table by the 
  collaboration.  Note that as
  atmospheric mass-squared splitting we use $\Delta m^2_{31}$ for NO and
  $\Delta m^2_{32}$ for IO.}
  \label{fig:nufit41_chisq-glob}
\end{pagefigure}

\section{Summary and conclusions}

The experimental programme for exploring neutrino flavour transitions 
successfully established the three massive neutrino framework as 
summarised in \cref{chap:3nufit_theor}. Currently, it is determining
its last unknowns, including leptonic CP violation, with LBL accelerator 
experiments.

As a result of the interplay in these experiments
between $\parenbar{\nu}_\mu$ disappearance, $\nu_e$ appearance and 
$\bar\nu_e$ appearance data, the unknowns start to clarify. There is no 
longer a strong degeneracy in the $\theta_{23}$ octant, normal mass 
ordering is currently favoured by the data, and there is a hint for 
$\delta_\text{CP} \sim \frac{3\pi}{2}$, i.e., maximal CP violation. 
$\delta_\text{CP} \sim \frac{\pi}{2}$ is disfavoured with 
$\sim 4\sigma$, but CP conservation is still allowed within $\sim 1.5 
\sigma$. We have also checked that Wilks' theorem can be safely applied
to extract the corresponding confidence levels, as no significant 
deviation from gaussianity is expected in the currently favoured 
parameter regions with the present statistics.

The hint for maximal CP violation, which would imply that the 
strongest measured source of CP violation is provided by the lepton sector, is 
mostly driven by a $\nu_e$ excess in T2K. This excess has been 
consistently present for the last years, and within the 3-neutrino 
framework it can only be accommodated by large CP violation once 
information from other experiments is consistently included. 
Nevertheless, as neutrino masses already constitute BSM physics, it is legitimate to ask whether we are detecting large leptonic CP 
violation or, on the contrary, whether other new physics could also explain 
the data. This is even more 
relevant considering that, in the near future, a new generation of LBL 
accelerator experiments will precisely explore the current hints coming 
from \NOvA/ and T2K~\cite{Abi:2020wmh, Abi:2020evt, Abi:2020oxb,
 Abi:2020loh, Abe:2018uyc} . Answering that question will be the main 
goal of the rest of this thesis.

\chapter{Beyond the three-neutrino paradigm: framework}
\label{chap:NSItheor}

\epigraph{\emph{Non-renormalizable interactions may also be detected. 
I doubt that they would not.}}{ ---  Steven Weinberg}

\epigraph{\emph{Sed tengo, y sal se vuelven tus arenas}}{ --- Blas de Otero}

As discussed in the previous chapter, there is a hint for maximal 
leptonic CP violation mostly coming from the T2K experiment. If their 
large $\nu_\mu \rightarrow \nu_e$ appearance signal is interpreted in 
the three light neutrino paradigm, it can only be accommodated by 
asumming large leptonic CP violation. Nevertheless, three light 
neutrinos is just the consequence of extending the SM with the 
operator~\eqref{eq:WeinbergOperator}. Other operators of higher 
dimension are also expected if the SM is a theory valid up to certain 
energy scale $\Lambda$. If they are present, they could be masking the 
results and even introducing new degeneracies.

In this chapter, we will present the formalism for parametrising 
leptonic CP violation in these extended scenarios, in particular when 
including Non-Standard neutrino Interactions (NSI). We will end by describing 
their other relevant effect in the existing oscillation experiments, which 
is the appearance of intrinsic parameter degeneracies.

\section{Formalism}

The lowest order effective operator that only contains SM fields and
is consistent with gauge symmetry is the dimension 5
operator~\eqref{eq:WeinbergOperator}, whose low-energy effects have
been extensively studied in the previous chapters. The next operators
with observable consequences at low energies come at dimension 6, and
the ones affecting neutrinos include
\begin{itemize}
\item Operators modifying the neutrino kinetic term,
\begin{equation}
\left(\overline{L_L^\alpha} \tilde \Phi \right) i \slashed{\partial} 
\left( \tilde \Phi^\dagger L_L^\beta \right) + \mathrm{h.c.} \, .
\end{equation}
After spontaneous symmetry breaking, these operators generate 
non-unitary corrections to the leptonic mixing 
matrix~\cite{Broncano:2002rw, Broncano:2003fq, Abada:2007ux}. 
Generically, they appear as a low-energy consequence of SM neutrinos
mixing with heavy mass eigenstates. Nevertheless, leptonic
non-unitarity is strongly
constrained by precision electroweak data~\cite{Fernandez-Martinez:2016lgt}.
Thus, its quantitative effect in present neutrino oscillation 
experiments is suppressed enough to be safely ignored in what 
follows~\cite{Blennow:2016jkn}.
\item Four-fermion operators leading to so-called NSI~\cite{Wolfenstein:1977ue, Valle:1987gv,
 Guzzo:1991hi} between neutrinos and matter (for recent reviews, see
 Refs.~\cite{Ohlsson:2012kf, Miranda:2015dra}), both in charged current 
interactions (NSI-CC)
\begin{equation}
  \label{eq:NSI-cc}
  (\bar\nu_\alpha \gamma_\mu P_L \ell_\beta) (\bar f' \gamma^\mu P f) \, ,
\end{equation}
and in neutral current interactions (NSI-NC)
\begin{equation}
  \label{eq:NSI-nc}
  (\bar\nu_\alpha \gamma_\mu P_L \nu_\beta) (\bar f \gamma^\mu P f) \,.
\end{equation}
Here $\alpha,\beta$ are lepton flavour indices, $l_\alpha$ is a charged 
lepton, $f$ and $f'$ are SM charged
fermions, and the chiral projector $P$ can be either $P_L$ or $P_R$. These operators are expected
to arise generically from the exchange of some mediator state assumed
to be heavier than the characteristic momentum transfer in the $\nu$
interaction process. They have all been written after spontaneous 
electroweak symmetry breaking: see Ref.~\cite{Gavela:2008ra} for a 
discussion of gauge invariant operators leading to them.
\end{itemize}

Since operators in both Eqs.~\eqref{eq:NSI-cc} and~\eqref{eq:NSI-nc}
modify the inelastic neutrino scattering cross sections with other SM
fermions they can be bounded by precision electroweak data (see for
example Refs.~\cite{Davidson:2003ha, Biggio:2009nt, Biggio:2009kv}).
In general these ``scattering'' bounds on NSI-CC operators are rather
stringent, whereas the bounds on NSI-NC tend to be weaker.
In turn, the operators in Eq.~\eqref{eq:NSI-nc} can also
modify the forward-coherent scattering (i.e., at zero
momentum transfer) of neutrinos as they propagate through matter, as 
discussed in \cref{sec:matterEffects}. Consequently
their effect can be significantly enhanced in oscillation experiments. 
Indeed, a global analysis of
data from oscillation experiments in the framework of mass induced
oscillations in presence of NSI provided some of the
strongest constraints on the size of the NSI affecting neutrino
propagation~\cite{GonzalezGarcia:2011my, Gonzalez-Garcia:2013usa}.

Of course, for models with a high energy new physics scale,
electroweak gauge invariance generically implies that the NSI-NC
parameters are still expected to be subject to tight constraints from
charged lepton observables~\cite{Gavela:2008ra, Antusch:2008tz},
leading to no visible effect in oscillations. However, more recently
it has been argued that viable gauge models with light mediators
(i.e., below the electroweak scale) may lead to observable
effects in oscillations without entering in conflict with other
bounds~\cite{Farzan:2015doa, Farzan:2015hkd, Babu:2017olk,
  Farzan:2017xzy, Denton:2018xmq} (see also
Refs.~\cite{Miranda:2015dra, Heeck:2018nzc} for discussions). In particular, for
light mediators bounds from high-energy neutrino scattering
experiments such as CHARM~\cite{Dorenbosch:1986tb} and
NuTeV~\cite{Zeller:2001hh} do not apply.

Because of that, in this work we will consider generic NSI affecting neutral current processes
relevant to neutrino propagation in matter. The coefficients
accompanying the new operators are usually parametrised in the form:
\begin{equation}
  \label{eq:NSILagrangian}
  \mathcal L_\text{NSI} = -2\sqrt2 G_F
  \sum_{f,P,\alpha,\beta} \varepsilon_{\alpha\beta}^{f,P}
  (\bar\nu_\alpha\gamma^\mu P_L\nu_\beta)
  (\bar f\gamma_\mu P f) + \mathrm{h.c.} \,,
\end{equation}
where $G_F$ is the Fermi constant. In this notation,
$\varepsilon_{\alpha\beta}^{f,P}$ parametrises the strength of the new
interaction with respect to the Fermi constant,
$\varepsilon_{\alpha\beta}^{f,P} \sim \mathcal{O}(G_X/G_F)$.
If we now assume that the neutrino flavour structure of the
interactions is independent of the charged fermion type, we can
factorise $\varepsilon_{\alpha\beta}^{f,P}$ as the product of two terms:
\begin{equation}
  \label{eq:eps-fact}
  \varepsilon_{\alpha\beta}^{f,P} \equiv \varepsilon_{\alpha\beta} \, \xi^{f,P}
\end{equation}
where the matrix $\varepsilon_{\alpha\beta}$ describes the neutrino part
and the coefficients $\xi^{f,P}$ parametrise the coupling to the
charged fermions. Under this assumption the Lagrangian in
Eq.~\eqref{eq:NSILagrangian} takes the form:
\begin{equation}
  \mathcal L_\text{NSI} = -2\sqrt2 G_F
  \bigg[ \sum_{\alpha,\beta} \varepsilon_{\alpha\beta}
  (\bar\nu_\alpha\gamma^\mu P_L\nu_\beta) \bigg]
  \bigg[ \sum_{f,P} \xi^{f,P} (\bar f\gamma_\mu P f) \bigg]  + \mathrm{h.c.}  \,.
  \label{eq:NSILagrangian2}
\end{equation}

If we follow the derivation of matter effects in 
\cref{sec:matterEffects}, we immediately see that only vector NSI 
contribute to the matter potential in neutrino oscillations, as any 
$\gamma_5$ factor does not contribute to the trace in 
\cref{eq:matterEffectsDerivation}. It is therefore convenient to define:
\begin{equation}
  \label{eq:eps-xi}
  \varepsilon_{\alpha\beta}^f
  \equiv \varepsilon_{\alpha\beta}^{f,L} + \varepsilon_{\alpha\beta}^{f,R} 
  = \varepsilon_{\alpha\beta} \, \xi^f
  \quad\text{with}\quad
  \xi^f \equiv \xi^{f,L} + \xi^{f,R} \,.
\end{equation}

\subsection{Neutrino oscillations in the presence of NSI}

In general, the evolution of the neutrino and antineutrino flavour
state during propagation is governed by \cref{eq:eom}, with the 
Hamiltonian:
\begin{equation}
  H^\nu = H_\text{vac} + H_\text{mat}
  \quad\text{and}\quad
  H^{\bar\nu} = ( H_\text{vac} - H_\text{mat} )^* \,,
\end{equation}
where $H_\text{vac}$ is the vacuum part which in the flavour basis
$(\nu_e, \nu_\mu, \nu_\tau)$ reads
\begin{equation}
  \label{eq:Hvac}
  H_\text{vac} = U^\text{lep} D_\text{vac} U^{\text{lep} \dagger}
  \quad\text{with}\quad
  D_\text{vac} = \frac{1}{2E_\nu} \diag(0, \Delta m^2_{21}, \Delta m^2_{31}) \,.
\end{equation}
Here $U^\text{lep}$ denotes the three-lepton mixing matrix in
vacuum~\eqref{eq:PMNS}.

Concerning the matter part $H_\text{mat}$ of the Hamiltonian which
governs neutrino oscillations, if all possible operators in
Eq.~\eqref{eq:NSILagrangian} are added to the SM Lagrangian we get:
\begin{equation}
  \label{eq:Hmat}
  H_\text{mat} = \sqrt{2} G_F N_e(x)
  \begin{pmatrix}
    1+\mathcal{E}_{ee}(x) & \mathcal{E}_{e\mu}(x) & \mathcal{E}_{e\tau}(x) \\
    \mathcal{E}_{e\mu}^*(x) & \mathcal{E}_{\mu\mu}(x) & \mathcal{E}_{\mu\tau}(x) \\
    \mathcal{E}_{e\tau}^*(x) & \mathcal{E}_{\mu\tau}^*(x) & \mathcal{E}_{\tau\tau}(x)
  \end{pmatrix}
\end{equation}
where the ``$+1$'' term in the $ee$ entry accounts for the standard
contribution, and
\begin{equation}
  \label{eq:epx-nsi}
  \mathcal{E}_{\alpha\beta}(x) \equiv \sum_{f=e,u,d}
  \frac{N_f(x)}{N_e(x)} \varepsilon_{\alpha\beta}^f
\end{equation}
describes the non-standard part. Here $N_f(x)$ is the number density
of fermion $f$ as a function of the distance traveled by the neutrino
along its trajectory.  In Eq.~\eqref{eq:epx-nsi} we have limited the
sum to the charged fermions present in ordinary matter,
$f=e,u,d$. Since quarks are always confined inside protons ($p$) and
neutrons ($n$), it is convenient to define:
\begin{equation}
  \label{eq:eps-nucleon}
  \varepsilon_{\alpha\beta}^p = 2\varepsilon_{\alpha\beta}^u + \varepsilon_{\alpha\beta}^d \,,
  \qquad
  \varepsilon_{\alpha\beta}^n = 2\varepsilon_{\alpha\beta}^d + \varepsilon_{\alpha\beta}^u \,.
\end{equation}
Taking into account that $N_u(x) = 2N_p(x) + N_n(x)$, that $N_d(x) =
N_p(x) + 2N_n(x)$, and that matter neutrality implies $N_p(x) =
N_e(x)$, Eq.~\eqref{eq:epx-nsi} becomes:
\begin{equation}
  \label{eq:epx-nuc}
  \mathcal{E}_{\alpha\beta}(x) =
  \big( \varepsilon_{\alpha\beta}^e + \varepsilon_{\alpha\beta}^p \big)
  + Y_n(x) \varepsilon_{\alpha\beta}^n
  \quad\text{with}\quad
  Y_n(x) \equiv \frac{N_n(x)}{N_e(x)} \, ,
\end{equation}
Since this matter term can be determined by oscillation experiments
only up to an overall multiple of the identity, each
$\varepsilon_{\alpha\beta}^f$ matrix introduces 8 new parameters: two
differences of the three diagonal real parameters (e.g.,
$\varepsilon_{ee}^f - \varepsilon_{\mu\mu}^f$ and $\varepsilon_{\tau\tau}^f -
\varepsilon_{\mu\mu}^f$) and three off-diagonal complex parameters
(i.e., three additional moduli and three complex phases). If, on top
of that, we assume the factorisation in \cref{eq:eps-xi} to hold,
there are only 8 parameters describing $\varepsilon_{\alpha \beta}$ and
two angles characterising the relative strength of couplings with 
electrons, up quarks, and down quarks.

In summary, NSI-NC operators parametrised by 
\cref{eq:NSILagrangian2,eq:eps-xi} could be present and affect
neutrino oscillation experiments. In the following, we will discuss the 
new sources of CP violation and parameter degeneracies that they 
introduce.
\section{Leptonic CP violation beyond the three-neutrino paradigm}
\label{sec:BSMinvariants}

As discussed in \cref{sec:CPviol_SM}, CP violation arises whenever there 
are physical complex phases in the Lagrangian. In the SM and its minimal 
extension to include neutrino masses, flavour transformations
can remove these phases and so determining if a theory violates CP is
not straightforward. As 
described in \cref{sec:leptonicCPviol}, in the framework of three 
massive neutrinos all leptonic CP violating observables in neutrino 
oscillations depend on a unique physical parameter, which can be written 
in a basis independent form as the so-called leptonic Jarslokg invariant 
in \cref{eq:jarlskogNeutrino}.

In this section, we repeat the procedure to derive a set of flavour 
basis invariants that characterise leptonic CP violation in the 
presence of NSI factorisable as in \cref{eq:eps-xi}. We follow the 
methodology introduced in Refs.~\cite{Bernabeu:1986fc, Gronau:1986xb} 
for generalising the construction of such invariants in the quark sector,
first introduced for three generations in~\cite{Jarlskog:1985cw,
  Jarlskog:1985ht}. We will work with Dirac neutrinos,
which is all it is needed when interested in CP violation in neutrino
oscillations (see Ref.~\cite{Branco:1998bw} for the invariants relevant 
for Majorana neutrinos).

The relevant parts of the Lagrangian that can contain complex phases 
and affect neutrino oscillations are:
\begin{equation}
  \begin{split}
    -\mathcal{L} = & \begin{pmatrix}
    \bar\nu_{e, L} & \bar\nu_{\mu, L} & \bar\nu_{\tau, L}
    \end{pmatrix}  M_D \begin{pmatrix}
    \nu_{e, R} \\ \nu_{\mu, R} \\ \nu_{\tau, R}
     \end{pmatrix} 
    + \begin{pmatrix}
    \bar\nu_{e, L} & \bar\nu_{\mu, L} & \bar\nu_{\tau, L}
    \end{pmatrix}  M_e \begin{pmatrix}
    \nu_{e, R} \\ \nu_{\mu, R} \\ \nu_{\tau, R}
     \end{pmatrix}  \\
                 & + 2 \sqrt{2} G_F \begin{pmatrix}
    \bar\nu_{e, L} & \bar\nu_{\mu, L} & \bar\nu_{\tau, L}
    \end{pmatrix}  \varepsilon \gamma^\mu \begin{pmatrix}
    \nu_{e, L} \\ \nu_{\mu, L} \\ \nu_{\tau, L}
     \end{pmatrix} 
    \bigg[\sum_{f} \xi^f \left(\bar{f}\gamma_\mu f\right) \bigg] - \mathrm{h.c.} \, .
  \end{split}
\end{equation}
Unphysical flavour basis rotations leaving the fermion kinetic and 
gauge Lagrangian~\eqref{eq:gaugeL} invariant are given by the following 
field transformations (see \cref{eq:symLep1,eq:symLep2})
\begin{align}
  \begin{pmatrix} e_L & \mu_L & \tau_L \end{pmatrix}^T   & \stackrel{\text{flavour}}{\longrightarrow} P_{L_L} \, \begin{pmatrix} e_L & \mu_L & \tau_L \end{pmatrix}^T \,,
  \\
  \begin{pmatrix}
    \nu_{e, L} & \nu_{\mu, L} & \nu_{\tau, L}
    \end{pmatrix}^T & \stackrel{\text{flavour}}{\longrightarrow} P_{L_L} \,   \begin{pmatrix}
    \nu_{e, L} & \nu_{\mu, L} & \nu_{\tau, L}
    \end{pmatrix}^T \,,
  \\
  \begin{pmatrix} e_R & \mu_R & \tau_R \end{pmatrix}^T & \stackrel{\text{flavour}}{\longrightarrow} P_{e_R} \, \begin{pmatrix} e_R & \mu_R & \tau_R \end{pmatrix}^T
  \,,
  \\
  \begin{pmatrix} \nu_{e, R} & \nu_{\mu, R} & \nu_{\tau, R} \end{pmatrix}^T & \stackrel{\text{flavour}}{\longrightarrow} P_{\nu_R} \, \begin{pmatrix} \nu_{e, R} & \nu_{\mu, R} & \nu_{\tau, R} \end{pmatrix}^T \,,
\end{align}
with all the $P \in SU(3)$. Correspondingly the matrices
with flavour indices transform as
\begin{equation}
  M_D \stackrel{\text{flavour}}{\longrightarrow}
  P^{\dagger}_{L_L} M_D P_{\nu_R} \,,
  \qquad
  M_e \stackrel{\text{flavour}}{\longrightarrow}
  P^{\dagger}_{L_L} M_e P_{e_R} \,,
  \qquad
  \varepsilon \stackrel{\text{flavour}}{\longrightarrow}
  P^{\dagger}_{L_L} \varepsilon P_{L_L} \,.
\end{equation}
So clearly the CP transformation, that changes these matrices into their 
complex conjugates, will be unphysical if (and only if)
it is equivalent to some flavour rotation. That is, there is CP
conservation if and only if there exists a set of matrices
$\{P_{L_L}, P_{\nu_R}, P_{e_R}\} \in SU(3)$ such that
\begin{equation}
  P^{\dagger}_{L_L} \varepsilon P_{L_L}  = \varepsilon^* \,,
  \qquad
  P^{\dagger}_{L_L} M_D P_{\nu_R}  = M_D^* \,,
  \qquad
  P^{\dagger}_{L_L} M_e P_{e_R}  = M_e^* \,.
\end{equation}
Since given a matrix $A$, $A A^\dagger$ determines $A$ up to unitary
rotations we can work with the ``squares'' of the mass matrices
instead, and we find that there is CP conservation if and only if there
exists a matrix $P \in SU(3)$ such that
\begin{equation}
  \label{eq:CPconsCondition}
  P^\dagger \varepsilon P  = \varepsilon^* \,,
  \qquad
  P^\dagger S_\nu P  = S_\nu^* \,,
  \qquad
  P^\dagger S_e P  = S_e^* \,, 
\end{equation}
with $S_e = M_e M_e^\dagger$ and $S_\nu = M_D M_D^\dagger$. In the 
charged lepton mass basis (i.e., where $P_{L_L}$ and $P_{e_R}$ are 
chosen to diagonalise the charged lepton mass matrix as in 
\cref{sec:ssb,sec:CPviol_SM}), these matrices read
\begin{equation}
  S_e =
  \begin{pmatrix}
    m_e^2 & 0 & 0 \\
    0 & m_\mu^2 & 0 \\
    0 & 0 & m_\tau^2
  \end{pmatrix} \,,
  \qquad
  S_\nu  = U^\mathrm{lep}
  \begin{pmatrix}
    m_1^2 & 0 & 0 \\
    0 & m_2^2 & 0 \\
    0 & 0 & m_3^2
  \end{pmatrix}
  U^{\mathrm{lep} \dagger} \,,
\end{equation}
as $U^\mathrm{lep}$ is the product of the matrices that diagonalise 
$S_e$ and $S_\nu$. $\{m_1, m_2, m_3\}$ are the neutrino masses.

This basis is particularly convenient, as there the last condition in 
\cref{eq:CPconsCondition} states that the matrices $P$ and $S_e$ 
commute. Therefore, $P$ is also diagonal, and being an element 
of $SU(3)$ we can write it as $P = \diag
(e^{i \delta_1}, e^{i \delta_2}, e^{- i (\delta_1 + \delta_2)})$. Thus, 
writing the other conditions and using the hermiticity of $\varepsilon$ 
and $S_\nu$, we find that there is CP conservation if and only if there exist 
$\{\delta_1, \delta_2\} \in [0, 2\pi)$ such that
\begin{alignat}{3}
\varepsilon_{\mu e} e^{i(\delta_1 - \delta_2)} & = \varepsilon_{\mu e}^* \, ,
\qquad \qquad 
S_{\nu \mu e} e^{i(\delta_1 - \delta_2)} && = S_{\nu \mu e}^* \, , \\
\varepsilon_{\tau e} e^{i(2\delta_1 + \delta_2)} & = \varepsilon_{\tau e}^* \, ,
\qquad \qquad
S_{\nu \tau e} e^{i(2\delta_1 + \delta_2)} && = S_{\nu \tau e}^* \, , \\
\varepsilon_{\tau \mu} e^{i(2\delta_2 + \delta_1)} & = \varepsilon_{\tau \mu}^* \, ,
\qquad \qquad
S_{\nu \tau \mu} e^{i(2\delta_2 + \delta_1)} && = S_{\nu \tau \mu}^* \, .
\end{alignat}
If we write each complex matrix element in polar form, 
we arrive to a linear system of six equations with two unkowns 
$\{\delta_1, \delta_2\}$. Imposing the existence of a solution, we
conclude that there is CP 
conservation if and only if
\begin{align}
\Ph (\varepsilon_{\mu e}) - \Ph (\varepsilon_{\tau e}) 
+ \Ph(\varepsilon_{\tau \mu}) & = 0 \, , \\
\Ph (S_{\nu \mu e}) + \Ph (S_{\nu \tau e}) 
- \Ph (S_{\nu \tau \mu}) & = 0 \, , \\
\Ph (\varepsilon_{\mu e}) - \Ph (S_{\nu \mu e}) & = 0 \, , \\
\Ph (\varepsilon_{\mu \tau}) - \Ph (S_{\nu \mu \tau}) & = 0 \, , \\
\Ph (\varepsilon_{e \tau}) - \Ph (S_{\nu e \tau}) & = 0 \, ,
\end{align}
where $\Ph(z)$ refers to the phase of the complex number $z$. 
These conditions can equivalently be written as
\begin{align}
\Im \left(\varepsilon_{\mu e} \, \varepsilon_{e \tau} \, \varepsilon_{\tau \mu}\right) & = 0 \, ,\\
\Im \left(S_{\nu \mu e} \, S_{\nu e \tau} \, S_{\nu \tau \mu}\right) & = 0 \, ,\\
\Im \left(\varepsilon_{\alpha \beta} \, S_{\nu \beta \alpha}\right) & = 0 \, .
\end{align}
where we have used the hermiticity of the $\varepsilon$ and $S_\nu$ 
matrices, and the last condition has to be fulfilled for $\{\alpha,\beta\}
= \{e,\mu\}, \{e,\tau\}, \{\mu,\tau\}$. However, since
\begin{equation}
    \label{eq:nonIndependentCondition}
  \varepsilon_{\mu\tau}\, S_{\nu \tau\mu} =
  \frac{\big(\varepsilon_{e\mu}\, \varepsilon_{\mu\tau}\, \varepsilon_{\tau e}\big)
    \big(S_{\nu e\mu}\, S_{\nu \mu\tau}\, S_{\nu \tau e}\big)^*
    \big(\varepsilon_{e\tau} \,S_{\nu \tau e}\big)
    \big(\varepsilon_{e\mu}\, S_{\nu \mu e}\big)^*}
    {\big|\varepsilon_{e\mu}\big|^2 \big|\varepsilon_{e\tau}\big|^2
      \big|S_{\nu e\mu}\big|^2 \big|S_{\nu e\tau}\big|^2} \, ,
\end{equation}
there are only four independent conditions.

Using the projector technique~\cite{Jarlskog:1987zd} the four
conditions can be expressed in a basis-invariant form. For example as
\begin{align}
  \Im\Tr\left[ S_e^2\, S_\nu^2\, S_e\, S_\nu \right]
  &= \frac{2}{i} \Det [S_e, S_\nu] = 0 \,,
  \\
  \Im\Tr\left[ S_e^2\, \varepsilon^2\, S_e\, \varepsilon \right]
  &= \frac{2}{i} \Det [S_e, \varepsilon]=0\,,
  \\
  \Im\Tr\left[ S_\nu\, S_e\,\varepsilon \right] &= 0\,,
  \\
  \Im\Tr\left[ S_e\, S_\nu\, S_e^2\, \varepsilon \right] &= 0 \,.
\end{align}
In the basis where the lepton mass matrix is diagonal these invariants read
\begin{align}
  \label{eq:appinv1}
  \Im\Tr\left[ S_e^2\, S_\nu^2\, S_e\, S_\nu \right]
  & = v(m_e, m_\mu,m_\tau)
  \Im\left[ S_{\nu e\mu}\, S_{\nu \mu\tau}\, S_{\nu \tau e} \right] \, ,
  \\
  \label{eq:appinv2}
  \Im\Tr\left[ S_e^2\, \varepsilon^2\, S_e\, \varepsilon \right]
  & = v(m_e, m_\mu,m_\tau)
  \Im\left[ \varepsilon_{e\mu}\, \varepsilon_{\mu\tau}\, \varepsilon_{\tau e} \right] \,,
  \\
  \label{eq:appinv3}
  \begin{split}
    \Im \Tr\left[ S_\nu\, S_e\, \varepsilon \right]
    &= (m_\mu^2-m_e^2) \Im\left( S_{\nu e\mu} \varepsilon_{\mu e} \right)
    + (m_\tau^2-m_e^2) \Im\left( S_{\nu e\tau} \varepsilon_{\tau e} \right)
    \\
    & \quad + (m_\tau^2-m_\mu^2)
    \Im \left( S_{\nu \mu\tau}\varepsilon_{\tau\mu} \right) \, ,
  \end{split} 
  \\
  \label{eq:appinv4}
  \begin{split}
    \Im \Tr\left[ S_e \, S_\nu \, S_e^2 \, \varepsilon \right]
    &= m_e m_\mu (m_\mu^2-m_e^2)
    \Im\left( S_{\nu e\mu} \varepsilon_{\mu e} \right)
    + m_e m_\tau (m_\tau^2-m_e^2)
    \Im\left( S_{\nu e\tau} \varepsilon_{\tau e} \right)
    \\
    & \quad + m_\mu m_\tau(m_\tau^2-m_\mu^2)
    \Im \left( S_{\nu \mu\tau} \varepsilon_{\tau\mu} \right)\, ,
  \end{split} 
\end{align}
with $v(m_e,m_\mu,m_\tau) = (m_\tau^2-m_\mu^2) (m_\tau^2-m_e^2)
(m_\mu^2-m_e^2)$.

Written in this form, the conditions for which the four independent
phases are physically realisable becomes explicit, in particular the
requirement of the non-zero difference between all or some of the
charged lepton masses. We thus identify four invariants characterising
CP violation in neutrino oscillations with NSI. In the charged 
lepton mass basis with the leptonic mixing matrix parametrised as in 
\cref{eq:PMNS}, they can be chosen as
\begin{itemize}
\item The standard Jarlskog invariant~\eqref{eq:jarlskogNeutrino}, 
\begin{multline}
  \label{eq:Jvac}
  \Im\Tr\Big( S_e^2\, S_\nu^2\, S_e\, S_\nu \Big)
  = \frac{2}{i} \Det [S_e, S_\nu]
  = \frac{1}{4} v(m_e,m_\mu,m_\tau)
  \Delta m^2_{21}\, \Delta m^2_{31}\, \Delta m^2_{23}
  \\
  \times \sin(2\theta_{23}) \sin(2\theta_{12})
  \sin(\theta_{13}) \cos^2(\theta_{13})
  \sin \delta_\text{CP} \, .
\end{multline}
It parametrises CP violation in vacuum (e.g., in the T2K experiment to 
a good approximation), and as is well known it 
requires three-flavour effects to be non-zero. That is, all three charged 
leptons must have different masses (to sensibly define neutrino 
flavour), all three neutrinos must have different masses (to sensibly 
define mixing angles), and all three mixing angles must be non-zero.

\item An invariant characterising CP violation in neutrino propagation 
in matter in the $E_\nu\rightarrow \infty $ limit. That is, the source 
of CP violation induced solely by the NSI,
\begin{multline}
  \label{eq:Jmat}
  \Im\Tr\Big( S_e^2\, \varepsilon^2\, S_e\, \varepsilon \Big) =
  \frac{2}{i} \Det [S_e, \varepsilon]= v(m_e,m_\mu,m_\tau)\,
  \Im(\varepsilon_{e\mu}\, \varepsilon_{\mu\tau}\, \varepsilon_{\tau e})
  \\
  = v(m_e,m_\mu,m_\tau)\,
  |\varepsilon_{e\mu}|\, |\varepsilon_{e\tau}|\, |\varepsilon_{\mu\tau}|\,
  \sin(\phi_{e\mu} - \phi_{e\tau} + \phi_{\mu\tau}) \, ,
\end{multline}
where $\phi_{\alpha \beta}$ is the phase of $\varepsilon_{\alpha \beta}$.  For it to be non-zero, all 
three charged leptons must have different masses (to sensibly define 
neutrino flavour), and all three flavour-violating NSI must be non-zero. 
Thus, it is also a three-flavour CP violating effect.

\item The other two basis invariants involve both $\varepsilon$ and $S_\nu$ and can
be formed by two combinations of the rephasing invariants $\Im\big(
\varepsilon_{\alpha\beta} \,S_{\nu \beta\alpha} \big)$ for
${\alpha\beta}={e\mu},\,{e\tau},\,{\mu\tau}$ as shown, for example,
in Eqs.~\eqref{eq:appinv3} and~\eqref{eq:appinv4}.  In the charged
lepton mass basis they read:
\begin{align}
  \label{eq:inv3}
  \begin{split}
    \Im\big(\varepsilon_{e\mu} S_{\nu \mu e} \big)
    &= \frac{1}{2} \cos \theta_{13} \varepsilon_{e\mu} \big[
      \Delta m^2_{21} \cos\theta_{23} \sin 2\theta_{12}
      \sin(\delta_\text{CP} + \phi_{e\mu})
      \\
      &\quad + (2 \Delta m^2_{31} - \Delta m^2_{21} + \Delta m^2_{21} \cos 2 \theta_{12})
      \sin\theta_{13} \sin\theta_{23}
      \sin \phi_{e\mu} \big] \,,
  \end{split}
  \\
  \label{eq:inv4}
  \begin{split}
    \Im\big(\varepsilon_{e\tau} S_{\nu \tau e} \big)
    &= \frac{1}{2} \cos\theta_{13} \varepsilon_{e\tau} \big[
      -\Delta m^2_{21} \sin\theta_{23} \sin 2\theta_{12}
      \sin(\delta_\text{CP} + \phi_{e\tau})
      \\
      &\quad + (\Delta m^2_{31} + \Delta m^2_{32} + \Delta m^2_{21}\cos 2\theta_{12})
      \sin\theta_{13} \cos\theta_{23} \sin \phi_{e\tau} \big] \,,
  \end{split}
\end{align}
and $\Im\big(\varepsilon_{\mu\tau} S_{\nu \tau\mu} \big)$ can be written in
terms of the two above using the equality in
Eq.~\eqref{eq:nonIndependentCondition}.

Unlike for the case of the invariants in Eqs.~\eqref{eq:Jvac}
and~\eqref{eq:Jmat}, there is not a clear physical setup which could
single out the contribution from Eq.~\eqref{eq:inv3} and
Eq.~\eqref{eq:inv4} (or any combination of those) to a leptonic CP
violating observable. Nevertheless, they are ``interference'' effects 
between vacuum CP-violation induced by lepton mixing and matter 
CP-violation induced by NSI. They are present whenever matter and 
vacuum oscillations are both relevant (for instance, in the \NOvA/ 
experiment), and they require only two-neutrino flavour mixing.
\end{itemize}

Admittedly the discussion above is only academic for the
quantification of the effects induced by the NSI matter potential on
neutrino propagation, because the relevant probabilities cannot be
expressed in any practical form in terms of these basis invariants and
one is forced to work in some specific parametrisation. What these
basis invariants clearly illustrate is that in order to study the
possible effects (in experiments performed in matter) of NSI on the
determination of the phase which parametrises CP violation in vacuum
\emph{without introducing an artificial basis dependence}, one needs
to include in the analysis the most general complex NSI matter
potential containing \emph{all} the three additional arbitrary phases.

Furthermore, it also illustrates the origin of the four sources of CP 
violation: three-neutrino vacuum effects, three-neutrino matter 
effects, and two-neutrino interference effects among vacuum and matter.
An experiment sensitive only to vacuum or only to matter could be 
analysed in terms of one single CP violation source. But experiments 
with relevant vacuum \emph{and} matter effects, as LBL accelerator 
experiments, are sensitive to all sources of CP violation. Moreover,
the interference CP violation sources are \emph{not} suppressed by
three-neutrino mixing, and so they could in principle be comparable to
vacuum CP violation.

\section{The generalised mass ordering degeneracy}
\label{sec:LMAD_theor}

Apart from explicitly introducing new sources of CP violation, NSI 
also introduce a degeneracy that affects the determination of 
$\delta_\text{CP}$ and the mass ordering~\cite{Gonzalez-Garcia:2013usa,
 Coloma:2016gei, Bakhti:2014pva, Miranda:2004nb}.

Neutrino evolution is invariant if the Hamiltonian $H^\nu =
H_\text{vac} + H_\text{mat}$ is transformed as $H^\nu \to -(H^\nu)^*$.
This transformation, that for the vacuum Hamiltonian stems from CPT 
invariance, requires a 
simultaneous change of both the vacuum and the
matter terms. The transformation of $H_\text{vac}$ is implemented 
exactly (up to an irrelevant multiple of the identity) by
the following transformation of the parameters:
\begin{equation}
  \label{eq:osc-deg}
  \begin{aligned}
    \Delta m^2_{31} & \to -\Delta m^2_{31} + \Delta m^2_{21} = -\Delta m^2_{32} \,,
    \\
    \theta_{12} & \to \pi/2 - \theta_{12} \,,
    \\
    \delta_\text{CP} & \to \pi - \delta_\text{CP} \,,
  \end{aligned}
\end{equation}
which does not spoil the commonly assumed restrictions on the range of
the vacuum parameters ($\Delta m^2_{21} > 0$ and $0 \leq \theta_{ij} 
\leq \pi/2$). It involves a change in the octant of
$\theta_{12}$ as well as a change in the neutrino mass ordering
(i.e., the sign of $\Delta m^2_{31}$), which is why it has been
called ``generalised mass ordering degeneracy'' in
Ref.~\cite{Coloma:2016gei}. This degeneracy was first explored for 
solar neutrino oscillation data, where it was called the LMA-D 
solution, standing for ``Large Mixing Angle - Dark'' as $\theta_{12}$
is quite large and bigger than $45^\circ$ (a parameter region named ``the dark side'' in 
Ref.~\cite{deGouvea:2000pqg}).

As for $H_\text{mat}$ we need:
\begin{equation}
  \label{eq:NSI-deg}
  \begin{aligned}
    \big[ \mathcal{E}_{ee}(x) - \mathcal{E}_{\mu\mu}(x) \big]
    &\to - \big[ \mathcal{E}_{ee}(x) - \mathcal{E}_{\mu\mu}(x) \big] - 2  \,,
    \\
    \big[ \mathcal{E}_{\tau\tau}(x) - \mathcal{E}_{\mu\mu}(x) \big]
    &\to -\big[ \mathcal{E}_{\tau\tau}(x) - \mathcal{E}_{\mu\mu}(x) \big] \,,
    \\
    \mathcal{E}_{\alpha\beta}(x)
    &\to - \mathcal{E}_{\alpha\beta}^*(x) \qquad (\alpha \neq \beta) \,,
  \end{aligned}
\end{equation}
see Refs.~\cite{Gonzalez-Garcia:2013usa, Bakhti:2014pva,
  Coloma:2016gei}. As seen in Eqs.~\eqref{eq:epx-nsi} 
and~\eqref{eq:epx-nuc} the matrix
$\mathcal{E}_{\alpha\beta}(x)$ depends on the chemical composition of the
medium, which may vary along the neutrino trajectory, so that in
general the condition in Eq.~\eqref{eq:NSI-deg} is fulfilled only in
an approximate way. The degeneracy becomes exact in the following two
cases:\footnote{Strictly speaking, Eq.~\eqref{eq:NSI-deg} can be
  satisfied exactly for \emph{any} matter chemical profile $Y_n(x)$ if
  $\varepsilon_{\alpha\beta}^u$, $\varepsilon_{\alpha\beta}^e$ and $\varepsilon_{\alpha\beta}^d$ are allowed to
  transform independently of each other. This possibility, however, is
  incompatible with the factorisation constraint of
  Eq.~\eqref{eq:eps-fact}, so it will not be discussed here.}
\begin{itemize}
\item If the effective NSI coupling to neutrons vanishes, so that
  $\varepsilon_{\alpha\beta}^n = 0$ in Eq.~\eqref{eq:epx-nuc}. In terms of
  fundamental quantities this occurs when $\varepsilon_{\alpha\beta}^u = -2
  \varepsilon_{\alpha\beta}^d$, i.e., the NSI couplings are
  proportional to the electric charge of quarks.

\item If the neutron/proton ratio $Y_n(x)$ is constant along the
  entire neutrino propagation path. This is certainly the case for
  reactor and LBL experiments, where only the Earth's mantle
  is involved, and to a good approximation also for atmospheric
  neutrinos, since the differences in chemical composition between
  mantle and core can safely be neglected in the context of
  NSI~\cite{GonzalezGarcia:2011my}. In this case the matrix
  $\mathcal{E}_{\alpha\beta}(x)$ becomes independent of $x$ and can be
  regarded as a new phenomenological parameter, as we will describe in
  \cref{sec:nsifit1_formalism}.
\end{itemize}
Further details on the implications of this degeneracy for different
classes of neutrino experiments (solar, atmospheric, etc.)
will be provided later in \cref{chap:NSIfit}.

Nevertheless, we already foresee that unless enough data from neutrino
experiments with a non-constant chemical composition along the 
trajectory (essentially solar neutrino experiments) is available, this
degeneracy is exact and thus completely spoils the sensitivity to the 
mass ordering.

Furthermore, the sensitivity to $\delta_\text{CP}$ may also get 
spoiled. Even though the transformation 
$\delta_\text{CP} \to \pi - \delta_\text{CP}$ does 
not change the Jarlskog invariant~\eqref{eq:jarlskogNeutrino}, 
it introduces a degenerate solution when determining $\delta_\text{CP}$. 
A precise measurement of this parameter would thus be compromised.

\section{Summary}

Neutrino flavour transition data is usually analysed assuming that the 
only BSM physics affecting it are three light neutrino mass 
eigenstates. Nevertheless, there could be additional new physics in the 
form of higher-dimensional operators masking the results. Some of these
operators can be constrained with electroweak precision observables, 
but others involve a two-neutrino vertex and are harder to explore. 
These include what are usually called NSI-NC.

Furthermore, these operators directly affect neutrino propagation 
in matter, and so they are expected to noticeably impact neutrino 
oscillation experiments. As we have seen, they introduce new sources of
leptonic CP violation and an intrinsic degeneracy involving both 
$\delta_\text{CP}$ and the mass ordering. Thus, their impact should be
assessed to assure the robustness of leptonic CP violation 
measurements. This will be the goal of the following chapter.

\chapter{Beyond the three-neutrino paradigm: fit to oscillation data}
\label{chap:NSIfit}

\epigraph{\emph{We are boys playing on the sea-shore, and diverting 
ourselves in now and then finding a smoother pebble or a prettier shell 
than ordinary, whilst the great ocean of truth lay all undiscovered 
before us.}}{ --- Isaac Newton}

\epigraph{\emph{Despacito y buena letra, \\
que el hacer las cosas bien, importa más que el hacerlas}}
{ --- Antonio Machado}
    
By analysing data from neutrino oscillation experiments in 
\cref{chap:3nufit_fit}, we have obtained a hint for maximal CP violation in 
the leptonic sector. The hint, though, is indirect, and might be masked 
if other BSM operators apart from the  
operator \eqref{eq:WeinbergOperator} are present. Among them, NSI-NC 
induced by rather light mediators are difficult to 
constraint but directly affect neutrino propagation in matter. 
Furthermore, they introduce new sources of CP violation and a 
degenerate solution known as LMA-D.

In this chapter, we confront these models with the bulk of data from
neutrino oscillation experiments. Due to the large 
parameter space and variety of experiments involved, we will first 
assess the sensitivity of neutrino oscillation experiments to CP 
conserving NSI-NC. 
We will evaluate current bounds, the complementarity among different
experiments, and the level at which the LMA-D 
solution can be tested. We will also evaluate how robustly are neutrino 
masses and mixing angles determined in the presence of the maximally 
allowed values for the NSI-NC.

Afterwards, we will assess whether experimentally allowed NSI-NC 
can spoil the sensitivity to CP violation in LBL 
accelerator experiments. As detailed in the previous chapter, this 
requires introducing all possible CP-violating phases in the analysis. 
Therefore, we will explore the entire parameter space in the presence
of NSI-NC.

\section{CP-conserving analysis: bounds on NSI moduli}
\label{sec:nsifit1}
In this section we revisit our current knowledge of the size and flavour
structure of NSI-NC which affect the matter background in the
evolution of solar, atmospheric, reactor and LBL
accelerator neutrinos as determined by a global analysis of
oscillation data. This updates and extends the analysis in
Ref.~\cite{Gonzalez-Garcia:2013usa} where NSI-NC with either up or
down quarks were considered. Here we extend that analysis to
account for the possibility of NSI with up \emph{and} down quarks
simultaneously, under the simplifying assumption that they carry the
same lepton flavour structure. To this aim, in 
Sec.~\ref{sec:nsifit1_formalism}
we briefly summarise the framework of our study and discuss the
simplifications used in the analysis of the atmospheric and LBL data
on one side and of the solar and KamLAND sector on the other side. In
Sec.~\ref{sec:nsifit1_solar} we present the results of the updated analysis of
solar and KamLAND data and quantify the impact of the modified matter
potential on the data description, as well as the status of the LMA-D
solution~\cite{Miranda:2004nb} in presence of the most general NSI
scenario considered here.  In Sec.~\ref{sec:nsifit1_globalosc} we describe the
constraints implied by the analysis of atmospheric, LBL and reactor
experiments, and combine them with those arising from the
solar+KamLAND data.  We show how the complementarity and synergy of
the different data sets is important for a robust determination of
neutrino masses and mixing in the presence of these general NSI, and
we derive allowed ranges on NSI couplings. 

\subsection{Formalism}
\label{sec:nsifit1_formalism}

We will consider NSI-NC mediated by the 
Lagrangian~\eqref{eq:NSILagrangian2}. Furthermore, we restrict 
ourselves to NSI with quarks, so that
only $\xi^u$ and $\xi^d$ (see \cref{eq:eps-xi}) are relevant for 
neutrino propagation. The reason is that the presence of NSI with 
electrons would affect not only neutrino propagation in matter, but 
also the neutrino-electron cross section in experiments such as Super-Kamiokande and Borexino. Since here we are only 
interested in studying the propagation bounds, we limit ourselves to NSI with quarks.
Also, \cref{eq:epx-nuc} shows that from the phenomenological point of 
view the propagation effects of NSI with electrons can be mimicked by 
NSI with quarks by means of a suitable combination of up-quark and 
down-quark contributions. Our choice of neglecting 
$\varepsilon_{\alpha\beta}^e$ in this work does not therefore imply a loss of 
generality.

In what respects the parametrisation of the NSI couplings to quarks, from \cref{eq:eps-fact} it is clear that a global rescaling of both $
\xi^u$ and $\xi^d$ by a common
factor can be reabsorbed into a rescaling of
$\varepsilon_{\alpha\beta}$, so that only the direction in the $(\xi^u,
\xi^d)$ plane is phenomenologically non-trivial. We parametrise such
direction in terms of an angle $\eta$, which for later convenience we
have related to the NSI couplings of protons and neutrons 
(see Eqs.~\eqref{eq:eps-nucleon}
and~\eqref{eq:nsifit1_epx-eta} for a formal definition). In terms of 
the ``quark'' couplings introduced in Eq.~\eqref{eq:eps-xi} we have:
\begin{equation}
  \label{eq:nsifit1_xi-eta}
  \xi^u = \frac{\sqrt{5}}{3} (2 \cos\eta - \sin\eta) \,,
  \qquad
  \xi^d = \frac{\sqrt{5}}{3} (2 \sin\eta - \cos\eta)
\end{equation}
where we have chosen the normalisation so that $\eta = \arctan(1/2)
\approx 26.6^\circ$ corresponds to NSI with up quarks ($\xi^u=1$,
$\xi^d=0$) while $\eta = \arctan(2) \approx 63.4^\circ$ corresponds to
NSI with down quarks ($\xi^u=0$, $\xi^d=1$). Note that the
transformation $\eta \to \eta + \pi$ simply results in a sign flip of
$\xi^u$ and $\xi^d$, hence it is sufficient to consider $-\pi/2 \leq
\eta \leq \pi/2$.

In terms of neutron and proton NSI, $\varepsilon_{\alpha\beta}^p =
\varepsilon_{\alpha\beta} \, \xi^p$ and $\varepsilon_{\alpha\beta}^n =
\varepsilon_{\alpha\beta}\, \xi^n$, which leads to:
\begin{equation}
  \label{eq:nsifit1_epx-eta}
  \mathcal{E}_{\alpha\beta}(x) =
  \varepsilon_{\alpha\beta} \big[ \xi^p + Y_n(x) \xi^n \big]
  \quad\text{with}\quad
  \xi^p = \sqrt{5} \cos\eta
  \quad\text{and}\quad
  \xi^n = \sqrt{5} \sin\eta
\end{equation}
so that the phenomenological framework adopted here is characterised
by 9 matter parameters: eight related to the matrix
$\varepsilon_{\alpha\beta}$ plus the direction $\eta$ in the
$(\xi^p,\xi^n)$ plane.

\subsubsection{Matter potential in atmospheric and long baseline neutrinos}
\label{sec:nsifit1_formalism-earth}
  
As discussed in Ref.~\cite{GonzalezGarcia:2011my}, in the Earth the
neutron/proton ratio $Y_n(x)$ which characterise the matter chemical
composition can be taken to be constant to very good approximation.
The PREM model~\cite{Dziewonski:1981xy} fixes $Y_n = 1.012$ in the
Mantle and $Y_n = 1.137$ in the Core, with an average value
$Y_n^\oplus = 1.051$ all over the Earth. Setting therefore $Y_n(x)
\equiv Y_n^\oplus$ in Eqs.~\eqref{eq:epx-nsi} and~\eqref{eq:epx-nuc}
we get $\mathcal{E}_{\alpha\beta}(x) \equiv \varepsilon_{\alpha\beta}^\oplus$ with:
\begin{equation}
  \varepsilon_{\alpha\beta}^\oplus
  = \varepsilon_{\alpha\beta}^e + \big( 2 + Y_n^\oplus \big) \varepsilon_{\alpha\beta}^u
  + \big( 1 + 2Y_n^\oplus \big) \varepsilon_{\alpha\beta}^d
  = \big( \varepsilon_{\alpha\beta}^e + \varepsilon_{\alpha\beta}^p \big)
  + Y_n^\oplus \varepsilon_{\alpha\beta}^n \,.
\end{equation}
If we drop $\varepsilon_{\alpha\beta}^e$ and impose quark-lepton
factorisation as in Eq.~\eqref{eq:nsifit1_epx-eta} we get:
\begin{equation}
  \label{eq:nsifit1_eps-earth}
  \varepsilon_{\alpha\beta}^\oplus
  = \varepsilon_{\alpha\beta} \big( \xi^p + Y_n^\oplus \xi^n \big)
  = \sqrt{5} \left( \cos\eta + Y_n^\oplus \sin\eta \right)
  \varepsilon_{\alpha\beta} \,.
\end{equation}
In other words, within this approximation the analysis of atmospheric
and LBL neutrinos holds for any combination of NSI with up quarks, down quarks or
electrons and it can be performed in terms of the effective NSI
couplings $\varepsilon_{\alpha\beta}^\oplus$, which play the role of
phenomenological parameters. In particular, the best-fit value and
allowed ranges of $\varepsilon_{\alpha\beta}^\oplus$ are independent of
$\eta$, while the bounds on the physical quantities
$\varepsilon_{\alpha\beta}$ simply scale as $(\cos\eta + Y_n^\oplus
\sin\eta)$. Moreover, it is immediate to see that for $\eta =
\arctan(-1/Y_n^\oplus) \approx -43.6^\circ$ the contribution of NSI to
the matter potential vanishes, so that no bound on
$\varepsilon_{\alpha\beta}$ can be derived from atmospheric and LBL data
in such case.

Following the approach of Ref.~\cite{GonzalezGarcia:2011my}, the
matter Hamiltonian $H_\text{mat}$, given in Eq.~\eqref{eq:Hmat} after
setting $\mathcal{E}_{\alpha\beta}(x) \equiv \varepsilon_{\alpha\beta}^\oplus$, can
be parametrised in a way that mimics the structure of the vacuum
term~\eqref{eq:Hvac}:
\begin{equation}
  \label{eq:nsifit1_HmatGen}
  H_\text{mat} = Q_\text{rel} U_\text{mat} D_\text{mat}
  U_\text{mat}^\dagger Q_\text{rel}^\dagger
  \text{~~with~~}
  \left\lbrace
  \begin{aligned}
    Q_\text{rel} &= \diag\left(
    e^{i\alpha_1}, e^{i\alpha_2}, e^{-i\alpha_1 -i\alpha_2} \right),
    \\
    U_\text{mat} &= R_{12}(\varphi_{12}) R_{13}(\varphi_{13})
    \tilde{R}_{23}(\varphi_{23}, \delta_\text{NS}) \,,
    \\
    D_\text{mat} &= \sqrt{2} G_F N_e(x) \diag(\varepsilon_\oplus, \varepsilon_\oplus', 0)
  \end{aligned}\right.
\end{equation}
where $R_{ij}(\varphi_{ij})$ is a rotation of angle $\varphi_{ij}$ in
the $ij$ plane and $\tilde{R}_{23}(\varphi_{23},\delta_\text{NS})$ is
a complex rotation by angle $\varphi_{23}$ and phase
$\delta_\text{NS}$.
Note that the two phases $\alpha_1$ and $\alpha_2$ included in
$Q_\text{rel}$ are not a feature of neutrino-matter interactions, but
rather a relative feature of the vacuum and matter terms. This is in 
accordance with the analysis in \cref{sec:BSMinvariants}, where two CP 
violating invariants arose as ``interference'' among vacuum and matter 
terms. There is a single invariant for matter-only CP violation, 
parametrised here in terms of the phase $\delta_\text{NS}$.

In order to simplify the analysis we neglect $\Delta m^2_{21}$ and also
impose that two eigenvalues of $H_\text{mat}$ are equal
($\varepsilon_\oplus'=0$).  The latter assumption is justified since, as
shown in Ref.~\cite{Friedland:2004ah}, strong cancellations in the
oscillation of atmospheric neutrinos occur when two eigenvalues of
$H_\text{mat}$ are equal, and it is precisely in this situation that
the weakest constraints can be placed.
Setting $\Delta m^2_{21} \to 0$ implies that the $\theta_{12}$ angle and the
$\delta_\text{CP}$ phase disappear from the expressions of the
oscillation probabilities, and the same happens to the $\varphi_{23}$
angle and the $\delta_\text{NS}$ phase in the limit $\varepsilon_\oplus' \to
0$.
Under these approximations the effective NSI couplings
$\varepsilon_{\alpha\beta}^\oplus$ can be parametrised as:
\begin{equation}
  \label{eq:nsifit1_eps_atm}
  \begin{aligned}
    \varepsilon_{ee}^\oplus - \varepsilon_{\mu\mu}^\oplus
    &= \hphantom{-} \varepsilon_\oplus \, (\cos^2\varphi_{12} - \sin^2\varphi_{12})
    \cos^2\varphi_{13} - 1\,,
    \\
    \varepsilon_{\tau\tau}^\oplus - \varepsilon_{\mu\mu}^\oplus
    &= \hphantom{-} \varepsilon_\oplus \, (\sin^2\varphi_{13}
    - \sin^2\varphi_{12} \, \cos^2\varphi_{13}) \,,
    \\
    \varepsilon_{e\mu}^\oplus
    &= -\varepsilon_\oplus \, \cos\varphi_{12} \, \sin\varphi_{12} \,
    \cos^2\varphi_{13} \, e^{i(\alpha_1 - \alpha_2)} \,,
    \\
    \varepsilon_{e\tau}^\oplus
    &= -\varepsilon_\oplus \, \cos\varphi_{12} \, \cos\varphi_{13} \,
    \sin\varphi_{13} \, e^{i(2\alpha_1 + \alpha_2)} \,,
    \\
    \varepsilon_{\mu\tau}^\oplus
    &= \hphantom{-} \varepsilon_\oplus \, \sin\varphi_{12} \, \cos\varphi_{13} \,
    \sin\varphi_{13} \, e^{i(\alpha_1 + 2\alpha_2)} \,.
  \end{aligned}
\end{equation}
With all this the relevant flavour transition probabilities for
atmospheric and LBL experiments depend on eight parameters:
($\Delta m^2_{31}$, $\theta_{13}$, $\theta_{23}$) for the vacuum part,
($\varepsilon_\oplus$, $\varphi_{12}$, $\varphi_{13}$) for the matter part,
and ($\alpha_1$, $\alpha_2$) as relative phases.  Notice that in this
case only the relative sign of $\Delta m^2_{31}$ and $\varepsilon_\oplus$ is
relevant for atmospheric and LBL neutrino oscillations: this is just a
manifestation of the generalised mass ordering degeneracy described in
\cref{sec:LMAD_theor} once $\Delta m^2_{21}$ and
$\varepsilon_\oplus'$ are set to zero~\cite{GonzalezGarcia:2011my}.

As further simplification, in order to keep the fit manageable we
assume real NSI, which we implement by choosing $\alpha_1 = \alpha_2 =
0$ with $\varphi_{ij}$ range $-\pi/2 \leq \varphi_{ij} \leq \pi/2$.
It is important to note that with these approximations the formalism
for atmospheric and LBL data is CP-conserving. We will go back to this
point when discussing the experimental results included in the
analysis.

In addition to atmospheric and LBL experiments, important information
on neutrino oscillation parameters is provided also by reactor
experiments with a baseline of about 1~km. Due to the very small
amount of matter crossed, both standard and non-standard matter
effects are completely irrelevant for these experiments, and the
corresponding $P_{ee}$ survival probability depends only on the vacuum
parameters. However, in view of the high precision recently attained
by both reactor and LBL experiments in the determination of the
atmospheric mass-squared difference (see \cref{fig:nufit3_chisq-dma}
and its discussion in \cref{subsec:nufit3_dm32}), combining them without adopting a
full $3\nu$ oscillation scheme requires a special care. In
Ref.~\cite{Nunokawa:2005nx} it was shown that, in the limit $\Delta m^2_{21}
\ll \Delta m^2_{31}$ as indicated by the data, the $P_{\mu\mu}$ probability
relevant for LBL-disappearance experiments can be accurately described
in terms of a single effective mass parameter $\Delta m^2_{\mu\mu} =
\Delta m^2_{31} - r_2 \Delta m^2_{21}$ with $r_2 = |U_{\mu2}^\text{lep}|^2 \big/
(|U_{\mu1}^\text{lep}|^2 + |U_{\mu2}^\text{lep}|^2)$. In the rest of
this section we will therefore make use of $\Delta m^2_{\mu\mu}$ as the
fundamental quantity parametrising the atmospheric mass-squared
difference. For each choice of the vacuum mixing parameters in
$U_\text{lep}$, the calculations for the various data sets are then
performed as follows:
\begin{itemize}
\item for atmospheric and LBL data we assume $\Delta m^2_{21}=0$ and set
  $\Delta m^2_{31} = \Delta m^2_{\mu\mu}$;
  
\item for reactor neutrinos we keep $\Delta m^2_{21}$ finite and set
  $\Delta m^2_{31} = \Delta m^2_{\mu\mu} + r_2 \Delta m^2_{21}$.
\end{itemize}
In this way the information provided by reactor and LBL data
on the atmospheric mass scale is consistently combined in spite of the
approximation $\Delta m^2_{21} \to 0$ discussed above. Note that the
correlations between solar and reactor neutrinos are properly taken
into account in our fit, in particular for what concerns the octant of
$\theta_{12}$.

\subsubsection{Matter potential for solar and KamLAND neutrinos}

For the study of propagation of solar and KamLAND neutrinos one can
work in the one mass dominance approximation, $\Delta m^2_{31} \to \infty$
(which effectively means that $G_F \sum_f N_f(x) \varepsilon_{\alpha\beta}^f
\ll \Delta m^2_{31} / E_\nu$). In this approximation the survival
probability $P_{ee}$ can be written as~\cite{Kuo:1986sk, Guzzo:2000kx}
\begin{equation}
  \label{eq:nsifit1_peesun}
  P_{ee} = c_{13}^4 P_\text{eff} + s_{13}^4
\end{equation}
The probability $P_\text{eff}$ can be calculated in an effective
$2\times 2$ model described by the Hamiltonian $H_\text{eff} =
H_\text{vac}^\text{eff} + H_\text{mat}^\text{eff}$, with:
\begin{align}
  \label{eq:nsifit1_HvacSol}
  H_\text{vac}^\text{eff}
  &= \frac{\Delta m^2_{21}}{4 E_\nu}
  \begin{pmatrix}
    -\cos2\theta_{12} \, \hphantom{e^{-i\delta_\text{CP}}}
    & ~\sin2\theta_{12} \, e^{i\delta_\text{CP}}
    \\
    \hphantom{-}\sin2\theta_{12} \, e^{-i\delta_\text{CP}}
    & ~\cos2\theta_{12} \, \hphantom{e^{i\delta_\text{CP}}}
  \end{pmatrix} ,
  \\
  \label{eq:nsifit1_HmatSol}
  H_\text{mat}^\text{eff}
  &= \sqrt{2} G_F N_e(x)
  \left[
    \begin{pmatrix}
      c_{13}^2 & 0 \\
      0 & 0
    \end{pmatrix}
    + \big[ \xi^p + Y_n(x) \xi^n \big]
    \begin{pmatrix}
      -\varepsilon_D^{\hphantom{*}} & \varepsilon_N \\
      \hphantom{+} \varepsilon_N^{*} & \varepsilon_D
    \end{pmatrix}
    \right],
\end{align}
where we have imposed the quark-lepton factorisation of
Eq.~\eqref{eq:nsifit1_epx-eta} and used the parametrisation convention of
Eq.~\eqref{eq:PMNS} for $U^\text{lep}$.  The coefficients
$\varepsilon_D$ and $\varepsilon_N$ are related to the original parameters
$\varepsilon_{\alpha\beta}$ by the following relations:
\begin{align}
  \label{eq:nsifit1_eps_D}
  \begin{split}
    \varepsilon_D
    &= c_{13} s_{13}\, \Re\!\big( s_{23} \, \varepsilon_{e\mu}
    + c_{23} \, \varepsilon_{e\tau} \big)
    - \big( 1 + s_{13}^2 \big)\, c_{23} s_{23}\,
    \Re\!\big( \varepsilon_{\mu\tau} \big)
    \\
    & \hphantom{={}}
    -\frac{c_{13}^2}{2} \big( \varepsilon_{ee} - \varepsilon_{\mu\mu} \big)
    + \frac{s_{23}^2 - s_{13}^2 c_{23}^2}{2}
    \big( \varepsilon_{\tau\tau} - \varepsilon_{\mu\mu} \big) \,,
  \end{split}
  \\[2mm]
  \label{eq:nsifit1_eps_N}
  \varepsilon_N &=
  c_{13} \big( c_{23} \, \varepsilon_{e\mu} - s_{23} \, \varepsilon_{e\tau} \big)
  + s_{13} \left[
    s_{23}^2 \, \varepsilon_{\mu\tau} - c_{23}^2 \, \varepsilon_{\mu\tau}^{*}
    + c_{23} s_{23} \big( \varepsilon_{\tau\tau} - \varepsilon_{\mu\mu} \big)
    \right].
\end{align}
Note that the $\delta_\text{CP}$ phase appearing in
Eq.~\eqref{eq:nsifit1_HvacSol} could be transferred to Eq.~\eqref{eq:nsifit1_HmatSol}
without observable consequences by means of a global rephasing. Hence,
for each fixed value of $\eta$ the relevant probabilities for solar
and KamLAND neutrinos depend effectively on six quantities: the three
real oscillation parameters $\Delta m^2_{21}$, $\theta_{12}$ and
$\theta_{13}$, one real matter parameter $\varepsilon_D$, and one
complex vacuum-matter combination $\varepsilon_N
e^{-i\delta_\text{CP}}$.
As stated before, in this section we will
assume real NSI, implemented here by setting $\delta_\text{CP} = 0$
and considering only real (both positive and negative) values for
$\varepsilon_N$.

Unlike in the Earth, the matter chemical composition of the Sun varies
substantially along the neutrino trajectory, and consequently the
potential depends non-trivially on the specific combinations of
couplings with up and down quarks --- i.e., on the value of
$\eta$. This implies that the generalised mass-ordering degeneracy is
not exact, except for $\eta=0$ (in which case the NSI potential is
proportional to the standard MSW potential and an exact inversion of
the matter sign is possible). However, as we will see in
Sec.~\ref{sec:nsifit1_solar}, the generalised mass ordering 
transformation described in
Eqs.~\eqref{eq:osc-deg} and~\eqref{eq:NSI-deg} still results in a good
fit to the global analysis of oscillation data for a wide range of
values of $\eta$, and non-oscillation data are needed to break this
degeneracy~\cite{Coloma:2017egw, Coloma:2017ncl}, as will be discussed
in \cref{chap:coh}. Because of the
change in the $\theta_{12}$ octant implied by Eq.~\eqref{eq:osc-deg}
and given that the standard LMA solution clearly favours $\theta_{12} <
45^\circ$, this alternative ``LMA-D'' solution is characterised by a value of
$\theta_{12} > 45^\circ$.

\subsection{Analysis of solar and KamLAND data}
\label{sec:nsifit1_solar}

Let us start by presenting the results of the updated analysis of
solar and KamLAND experiments in the context of oscillations with the
generalised matter potential in Eq.~\eqref{eq:nsifit1_HmatSol}. We 
include the same data as the latest 3$\nu$ analysis in \cref{chap:3nufit_fit}.

\afterpage{\clearpage\begin{figure}[p!]\centering
  \includegraphics[width=0.89\textwidth]{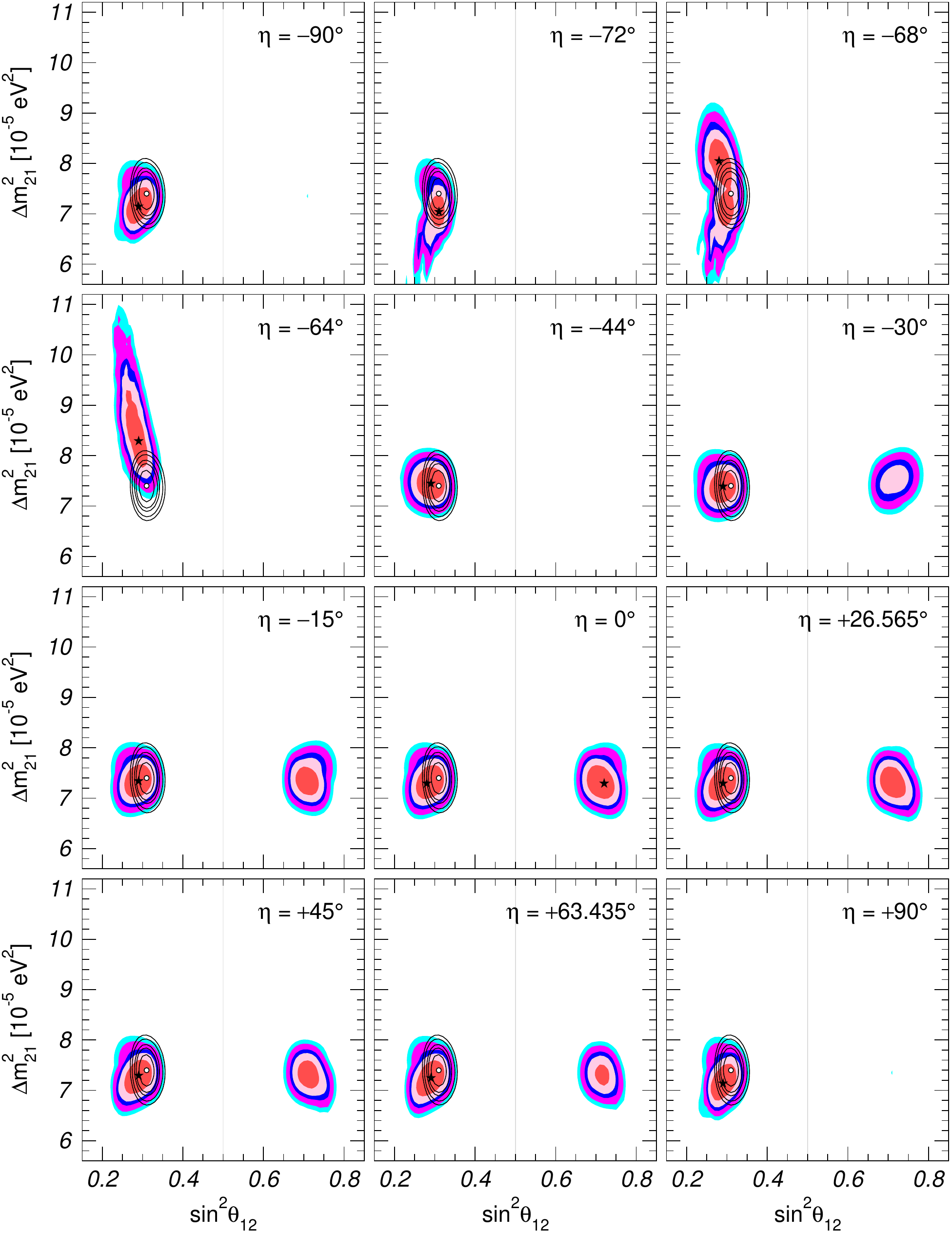}
  \caption{Two-dimensional projections of the $1\sigma$, 90\%,
    $2\sigma$, 99\% and $3\sigma$ CL (2~dof) allowed regions from the
    analysis of solar and KamLAND data in the presence of non-standard
    matter potential for the oscillation parameters $(\theta_{12},
    \Delta m^2_{21})$ after marginalising over the NSI parameters and for
    $\theta_{13}$ fixed to $\sin^2\theta_{13} = 0.022$.  The best-fit
    point is marked with a star. The results are shown for fixed
    values of the NSI quark coupling parameter $\eta$. For comparison
    the corresponding allowed regions for the analysis in terms of
    $3\nu$ oscillations without NSI are shown as black void
    contours. Note that, as a consequence of the periodicity of
    $\eta$, the regions in the first ($\eta = -90^\circ$) and last
    ($\eta = +90^\circ$) panels are identical.}
  \label{fig:nsifit1_sun-oscil}
\end{figure}\clearpage}

We present different projections of the allowed parameter space in
Figs.~\ref{fig:nsifit1_sun-oscil}--\ref{fig:nsifit1_sun-range}. In the analysis we
have fixed $\sin^2\theta_{13} = 0.022$ which is the best-fit value
from the global analysis of $3\nu$ oscillations~\cite{Esteban:2016qun,
  nufit-3.2}.\footnote{Note that the determination of $\theta_{13}$ is
  presently dominated by reactor experiments, which have negligible
  matter effects and are therefore unaffected by the presence of NSI.
  Allowing for variations of $\theta_{13}$ within its current
  well-determined range has no quantitative impact on our results.} So
for each value of $\eta$ there are four relevant parameters:
$\Delta m^2_{21}$, $\sin^2\theta_{12}$, $\varepsilon_D$, and $\varepsilon_N$.
As mentioned above, for simplicity the results are shown for real
$\varepsilon_N$. Also strictly speaking the sign of $\varepsilon_N$ is not
physically observable in oscillation experiments, as it can be
reabsorbed into a redefinition of the sign of $\theta_{12}$. However,
for definiteness we have chosen to present our results in the
convention $\theta_{12} \geq 0$, and therefore we consider both
positive and negative values of $\varepsilon_N$.
Fig.~\ref{fig:nsifit1_sun-oscil} shows the two-dimensional projections on the
oscillation parameters $(\theta_{12}, \Delta m^2_{21})$ for different values
of $\eta$ after marginalising over the NSI parameters, while
Fig.~\ref{fig:nsifit1_sun-epses} shows the corresponding two-dimensional
projections on the matter potential parameters $(\varepsilon_D,
\varepsilon_N)$ after marginalising over the oscillation
parameters. The one-dimensional ranges for the four parameters as a
function of $\eta$ are shown in Fig.~\ref{fig:nsifit1_sun-range}.

\afterpage{\clearpage\begin{figure}\centering
  \includegraphics[width=0.82\textwidth]{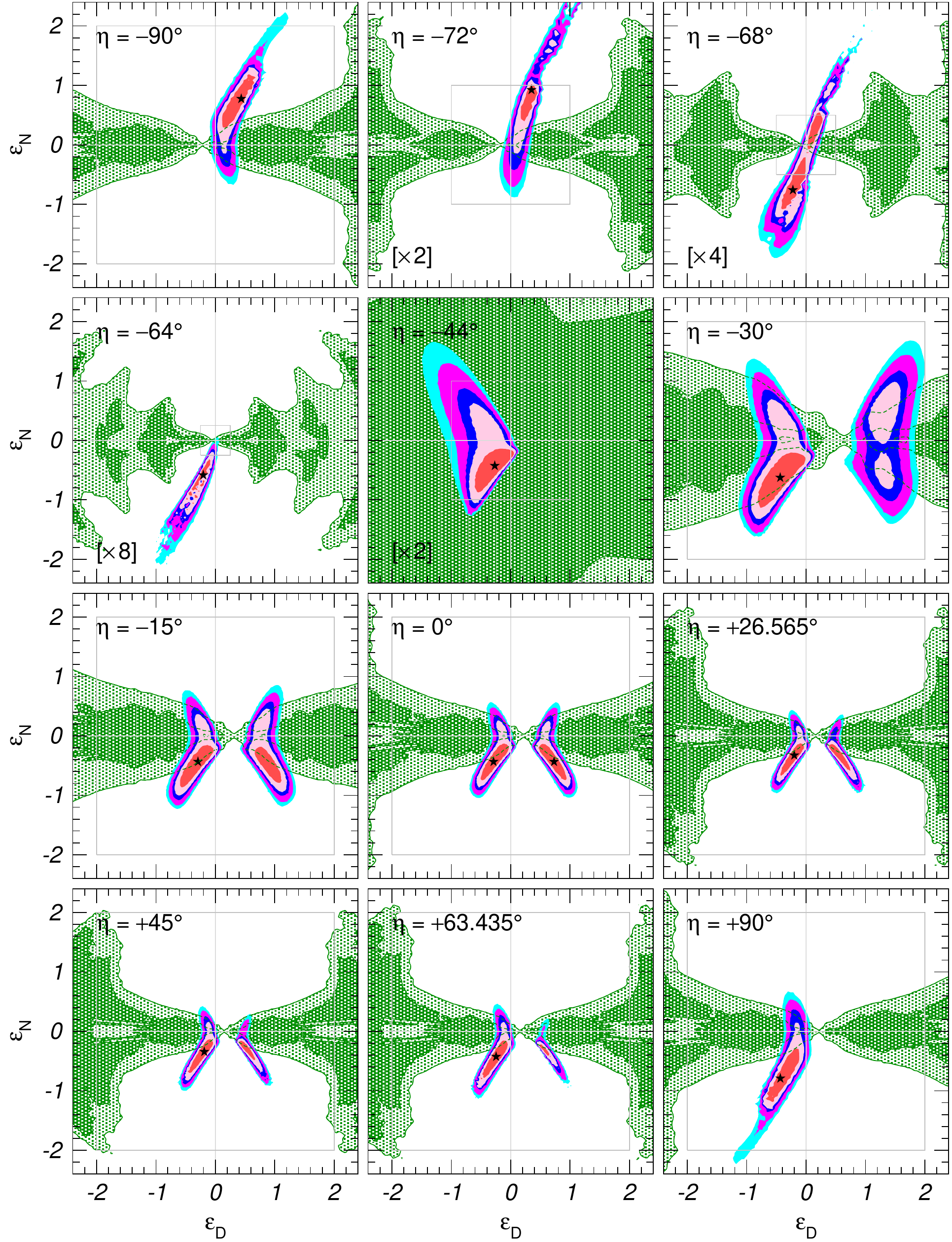}
  \caption{Two-dimensional projections of the $1\sigma$, 90\%,
    $2\sigma$, 99\% and $3\sigma$ CL (2~dof) allowed regions from the
    analysis of solar and KamLAND data in the presence of non-standard
    matter potential for the matter potential parameters
    $(\varepsilon_D, \varepsilon_N)$, for $\sin^2\theta_{13} = 0.022$ and
    after marginalising over the oscillation parameters.  The best-fit
    point is marked with a star. The results are shown for fixed
    values of the NSI quark coupling parameter $\eta$.  The panels
    with a scale factor ``$[\times N]$'' in their lower-left corner
    have been ``zoomed-out'' by such factor with respect to the
    standard axis ranges, hence the grey square drawn in each panel
    always corresponds to $\max\big( |\varepsilon_D|, |\varepsilon_N|
    \big) = 2$ and has the same size in all the panels.  For
    illustration we also show as shaded green areas the 90\% and
    $3\sigma$ CL allowed regions from the analysis of the atmospheric
    and LBL data. Note that, as a consequence of the periodicity of
    $\eta$, the regions in the first ($\eta = -90^\circ$) and last
    ($\eta = +90^\circ$) panels are identical up to an overall sign
    flip.}
  \label{fig:nsifit1_sun-epses}
\end{figure}\clearpage}

The first thing to notice in the figures is the presence of the LMA-D
solution for a wide range of values of $\eta$. This is a consequence
of the approximate degeneracy discussed in the previous section. In
particular, as expected, for $\eta=0$ the degeneracy is exact and the
LMA-D region in Fig.~\ref{fig:nsifit1_sun-oscil} is perfectly symmetric to the
LMA one with respect to maximal $\theta_{12}$.  Looking at the
corresponding panels of Fig.~\ref{fig:nsifit1_sun-epses} we note that the
allowed area in the NSI parameter space is composed by two
disconnected regions, one containing the SM case (i.e., the
point $\varepsilon_D = \varepsilon_N = 0$) which corresponds to the
``standard'' LMA solution in the presence of the modified matter
potential, and another which does not include such point and
corresponds to the LMA-D solution.  Although the appearance of the
LMA-D region is a common feature, there is also a range of values of
$\eta$ for which such solution is strongly disfavoured and does not
appear at the displayed CLs.

In order to further illustrate the $\eta$ dependence of the results,
it is convenient to introduce the functions $\chi^2_\text{LMA}(\eta)$
and $\chi^2_\text{LMA-D}(\eta)$ which are obtained by marginalising
the $\chi^2$ for a given value of $\eta$ over both the oscillation and
the matter potential parameters with the constraint $\theta_{12} <
45^\circ$ and $\theta_{12} > 45^\circ$, respectively. With this, in
the left panel of Fig.~\ref{fig:nsifit1_chisq-eta} we plot the differences
$\chi^2_\text{LMA}(\eta) - \chi^2_\text{no-NSI}$ (full lines) and
$\chi^2_\text{LMA-D}(\eta) - \chi^2_\text{no-NSI}$ (dashed lines),
where $\chi^2_\text{no-NSI}$ is the minimum $\chi^2$ for standard
$3\nu$ oscillations (i.e., without NSI), while in the right
panel we plot $\chi^2_\text{LMA-D}(\eta) - \chi^2_\text{LMA}(\eta)$
which quantifies the relative quality of the LMA and LMA-D solutions.
From this plot we can see that even for the analysis of solar and
KamLAND data alone (red lines) the LMA-D solution is disfavoured at
more than $3\sigma$ when $\eta \lesssim -40^\circ$ or $\eta \gtrsim
86^\circ$. Generically for such range of $\eta$ the modified matter
potential in the Sun, which in the presence of NSI is determined not
only by the density profile but also by the chemical composition, does
not allow for a degenerate solution compatible with KamLAND data. In
particular, as discussed below, for a fraction of those $\eta$ values
the NSI contribution to the matter potential in the Sun becomes very
suppressed and therefore the degeneracy between NSI and octant of
$\theta_{12}$ cannot be realised. In what respects the LMA solution,
we notice that it always provides a better fit (or equivalent for
$\eta=0$) than the LMA-D solution to solar and KamLAND data, for any
value of $\eta$. This does not have to be the case in general, and
indeed it is no longer so when atmospheric data are also included in
the analysis. We will go back to this point in the next section.

\begin{figure}\centering
  \includegraphics[width=0.8\textwidth]{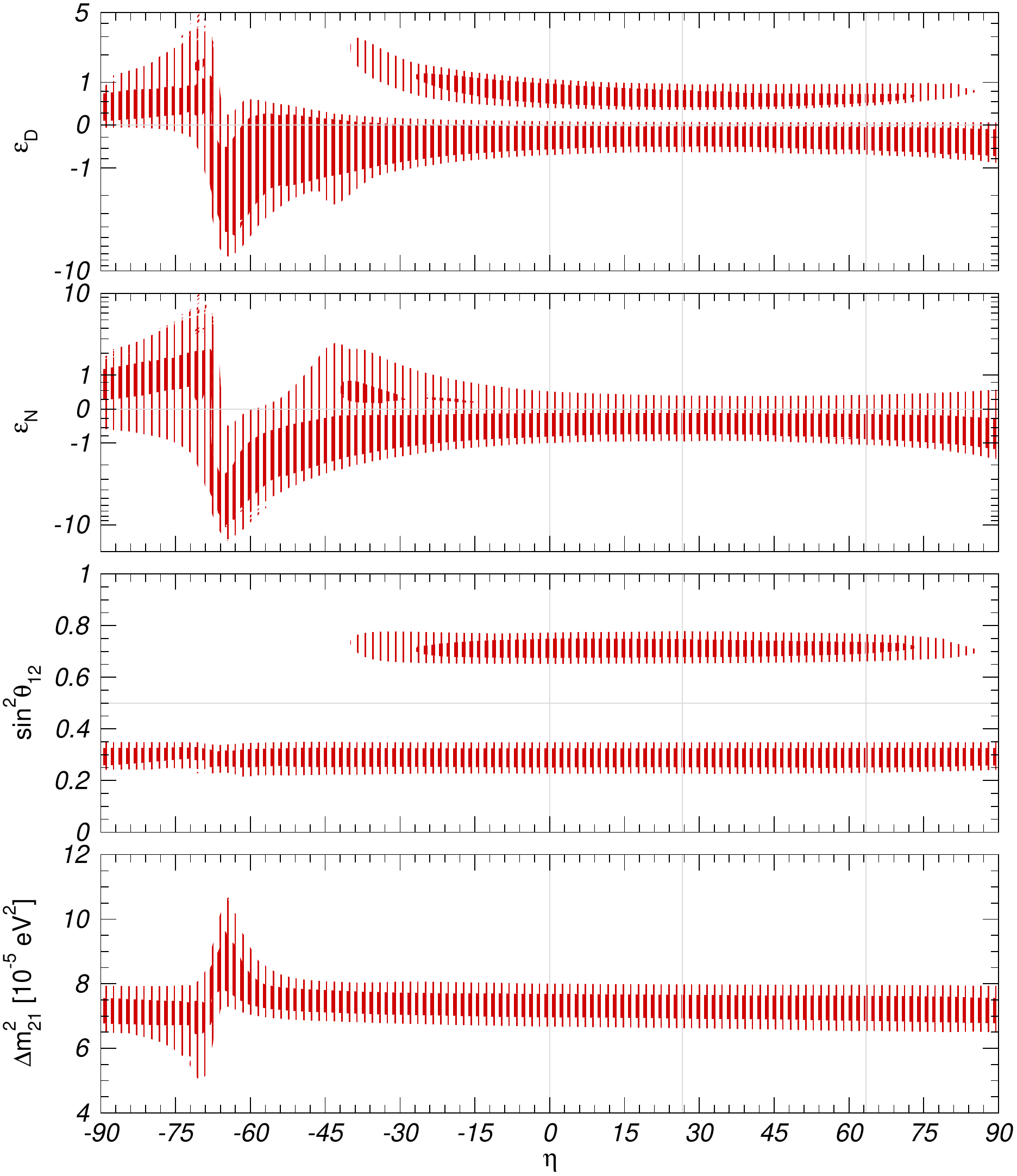}
  \caption{90\% and $3\sigma$ CL (1~dof) allowed ranges from the
    analysis of solar and KamLAND data in the presence of NSI, for the four relevant parameters
    (the matter potential parameters $\varepsilon_D$ and $\varepsilon_N$
    as well as the oscillation parameters $\Delta m^2_{21}$ and
    $\sin^2\theta_{12}$) as a function of the NSI quark coupling
    parameter $\eta$, for $\sin^2\theta_{13}=0.022$.  In each panel
    the three undisplayed parameters have been marginalised.}
    \label{fig:nsifit1_sun-range}
\end{figure}

From the left panel in Fig.~\ref{fig:nsifit1_chisq-eta} we see that the
introduction of NSI can lead to a substantial improvement in the
analysis of solar and KamLAND data, resulting in a sizeable decrease
of the minimum $\chi^2$ with respect to the standard oscillation
scenario.  The maximum gain occur for $\eta \simeq -64^\circ$ and is
about $11.2$ units in $\chi^2$ (i.e., a $3.3\sigma$ effect),
although for most of the values of $\eta$ the inclusion of NSI
improves the combined fit to solar and KamLAND by about $2.5\sigma$.
This is mainly driven by the well known $2.7\sigma$ tension between solar and
KamLAND data in the determination of $\Delta m^2_{21}$ described in 
\cref{subsec:nufit3_dm12}.
Such tension can be alleviated in presence of a non-standard matter
potential, thus leading to the corresponding decrease in the minimum
$\chi^2$ for most values of $\eta$ --- with the exception of the range
$-70^\circ \lesssim \eta \lesssim -60^\circ$.  Furthermore, as seen in
the lower panel in Fig.~\ref{fig:nsifit1_sun-range} the allowed range of
$\Delta m^2_{21}$ implied by the combined solar and KamLAND data is pretty
much independent of the specific value of $\eta$, except again for
$-70^\circ \lesssim \eta\lesssim -60^\circ$ in which case it can
extend well beyond the standard oscillation LMA values.

The special behaviour of the likelihood of solar and KamLAND in the
range $-70^\circ \lesssim \eta\lesssim -60^\circ$ is a consequence of
the fact that for such values the NSI contributions to the matter
potential in the Sun approximately cancel. As mentioned in the
previous section, the matter chemical composition of the Sun varies
substantially along the neutrino production region, with $Y_n(x)$
dropping from about $1/2$ in the center to about $1/6$ at the border
of the solar core. Thus for $-70^\circ \lesssim \eta \lesssim
-60^\circ$ (corresponding to $-2.75 \lesssim \tan\eta \lesssim -1.75$)
the effective NSI couplings $\mathcal{E}_{\alpha\beta}(x) =
\varepsilon_{\alpha\beta}^p + Y_n(x) \varepsilon_{\alpha\beta}^n \propto 1 + Y_n(x)
\tan\eta$ vanish at some point inside the neutrino production
region.  This means that for such values of $\eta$ the constraints on
the NSI couplings from solar data become very weak, being prevented
from disappearing completely only by the \emph{gradient} of
$Y_n(x)$. This is visible in the two upper panels in
Fig.~\ref{fig:nsifit1_sun-range} and in the panels of Fig.~\ref{fig:nsifit1_sun-epses}
with $\eta$ in such range, where a multiplicative factor 2--8 has to
be included to make the regions fit in the same axis range.  Indeed
for those values of $\eta$ the allowed NSI couplings can be so large
that their effect in the propagation of long baseline reactor
neutrinos through the Earth becomes sizable, and can therefore lead
to spectral distortions in KamLAND which affect the determination of
$\Delta m^2_{21}$ --- hence the ``migration'' and distortion of the LMA
region observed in the corresponding panels in
Fig.~\ref{fig:nsifit1_sun-oscil}.  In particular, it is precisely for $\eta =
-64^\circ$ for which the ``migration'' of the KamLAND region leads to
the best agreement with the solar determination of $\Delta m^2_{12}$,
whereas for $\eta = -68^\circ$ we find the worst agreement.  In any
case, looking at the shaded green regions in the corresponding panels
of Fig.~\ref{fig:nsifit1_sun-epses} we can anticipate that the inclusion of
atmospheric and LBL oscillation experiments will rule out almost
completely such very large NSI values.

As for $\theta_{12}$, looking at the relevant panel in
Fig.~\ref{fig:nsifit1_sun-range} we can see that its determination is pretty
much independent of the value of $\eta$, however a comparison between
coloured and void regions in Fig.~\ref{fig:nsifit1_sun-oscil} shows that its
allowed range always extends to lower values than in the standard
$3\nu$ case without NSI. This is expected since the presence of
non-diagonal NSI parametrised by $\varepsilon_N$ provides another source
of flavour transition, thus leading to a weakening of the lower bound
on $\theta_{12}$.

We finish this section by noticing that two of the panels in
Figs.~\ref{fig:nsifit1_sun-oscil} and~\ref{fig:nsifit1_sun-epses} correspond to the
values of NSI only with $f=u$ ($\eta \approx 26.6^\circ$) and only
with $f=d$ ($\eta \approx 63.4^\circ$) and can be directly compared
with the results of the previous global OSC+NSI analysis in
Ref.~\cite{Gonzalez-Garcia:2013usa}.  For illustration we also show in
one of the panels the results for $\eta = -44^\circ$ which is close to
the value for which NSI effects in the Earth matter cancel.

\begin{figure}\centering
  \includegraphics[width=0.9\textwidth]{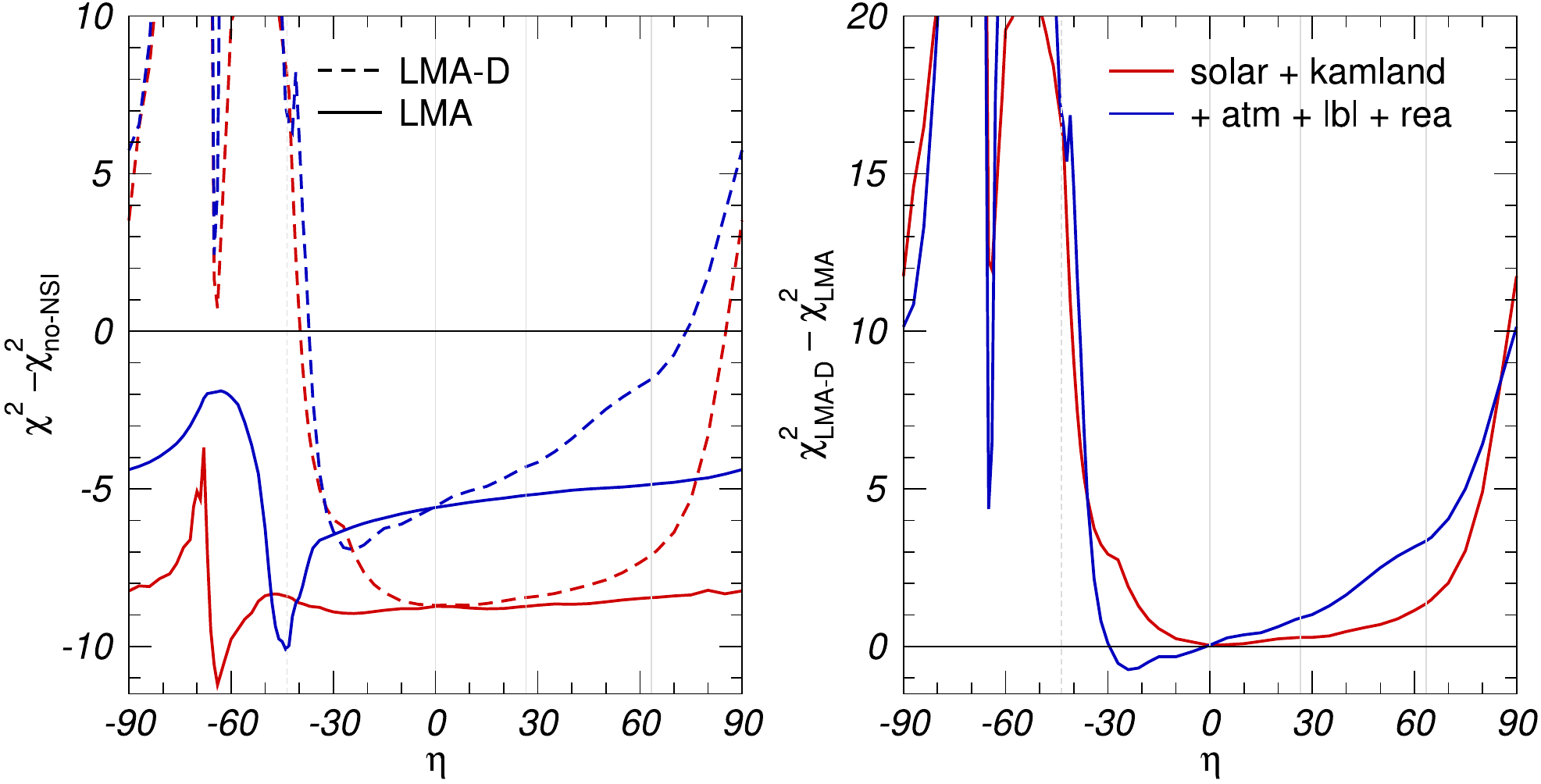}
  \caption{Left: $\chi^2_\text{LMA}(\eta) - \chi^2_\text{no-NSI}$
    (full lines) and $\chi^2_\text{LMA-D}(\eta) -
    \chi^2_\text{no-NSI}$ (dashed lines) for the analysis of different
    data combinations (as labeled in the figure) as a function of the
    NSI quark coupling parameter $\eta$.  Right:
    $\chi^2_\text{LMA-D}(\eta) - \chi^2_\text{LMA}(\eta)$ as a
    function of $\eta$. See text for details.}
  \label{fig:nsifit1_chisq-eta}
\end{figure}

\subsection{Results of the global oscillation analysis}
\label{sec:nsifit1_globalosc}

In addition to the solar and KamLAND data discussed so far, in our
global analysis we also consider the following data sets:
\begin{itemize}
\item atmospheric neutrino data: this sample includes the four phases
  of Super-Kamiokande (up to 1775 days of SK4~\cite{Wendell:2014dka})
  in the form of the ``classical'' samples of $e$-like and $\mu$-like
  events (70 energy and zenith angle bins), together with the complete
  set of DeepCore 3-year $\mu$-like events (64 data points) presented
  in Ref.~\cite{Aartsen:2014yll} and publicly released in
  Ref.~\cite{deepcore:2016}. The calculations of the event rates for
  both detectors are based on the atmospheric neutrino flux
  calculations described in Ref.~\cite{Honda:2015fha}. In addition, we
  also include the results on $\nu_\mu$-induced upgoing muons reported
  by IceCube~\cite{Jones:2015, Arguelles:2015, TheIceCube:2016oqi},
  based on one year of data taking;
  
\item LBL experiments: we include here the $\nu_\mu$ and
  $\bar\nu_\mu$ disappearance as well as the $\nu_e$ and $\bar\nu_e$
  appearance data in MINOS~\cite{Adamson:2013whj} (39, 14, 5, and 5
  data points, respectively); the $\nu_\mu$ and $\bar\nu_\mu$
  disappearance data in T2K~\cite{t2k:vietnam2016} (39 and 55 data
  points, respectively), the latter as of January 2018; and the $\nu_\mu$ disappearance data in
  NO$\nu$A~\cite{nova:fnal2018} (72 data points).\footnote{We do not 
  include the \NOvA/ antineutrino data or the latest T2K $\bar{\nu}_\mu$  results, because they were not available when the fit was performed. 
  As the purpose of this section is just to understand the sensitivity 
  to NSI of different experiments, these datasets will be included in 
  the more complete fits in 
  \cref{sec:nsifit2,sec:coh_plusOsc}.}  As mentioned in
  Sec.~\ref{sec:nsifit1_formalism}, in order to keep the fit manageable we
  restrict ourselves to the CP-conserving scenario. At present, the
  results of the full $3\nu$ oscillation analysis with standard matter
  potential show a hint of CP violation~\cite{Esteban:2016qun,
    nufit-3.2}, which is mainly driven by the LBL $\nu_e$ and
  $\bar\nu_e$ appearance data at T2K~\cite{t2k:vietnam2016} and
  NO$\nu$A~\cite{nova:fnal2018}.  Conversely, allowing for CP
  violation has negligible impact on the determination of the
  CP-conserving parameters in the analysis of MINOS appearance data
  and of any LBL disappearance data samples, as well as in our
  analysis of atmospheric events mentioned above.  Hence, to ensure
  full consistency with our CP-conserving parametrisation we have
  chosen \emph{not} to include in the present study the data from the
  $\nu_e$ and $\bar\nu_e$ appearance channels in NO$\nu$A and
  T2K. This also renders our fit only marginally sensitive to the
  neutrino mass ordering. In what follows we will refer to the
  LBL data included here as LBL-CPC.  Note that for
  simplicity we have omitted from our analysis the MINOS+ results on
  $\nu_\mu$ disappearance, despite the fact that they probe higher
  neutrino energies than the other LBL experiments and are therefore,
  at least in principle, more sensitive to the NSI parameters than,
  e.g., MINOS~\cite{Graf:2015egk}.  The rationale behind this
  choice is that the LBL experiments which we include are crucial to
  determine the oscillation parameters in an energy range where NSI
  effects are subdominant, whereas at present MINOS+ data lack this
  capability. As for the NSI parameters involved in $\nu_\mu$
  disappearance, they are more strongly constrained by the atmospheric
  neutrino data of Super-Kamiokande and IceCube, which extends to energies well
  beyond those of MINOS+;
  
\item medium baseline reactor experiments: since these
  experiments are largely insensitive to matter effects (either
  standard or non-standard), the results included here coincide with
  those of the standard $3\nu$ analysis presented in
  Ref.~\cite{nufit-3.2} and illustrated in the black lines of the plot
  tagged ``Synergies:~determination of $\Delta m^2_{3\ell}$''. Such analysis
  is based on a reactor-flux-independent approach as described in
  Ref.~\cite{Dentler:2017tkw}, and includes the Double-Chooz FD-I/ND
  and FD-II/ND spectral ratios with 455-day (FD-I), 363-day (FD-II),
  and 258-day (ND) exposures~\cite{dc:cabrera2016} (56 data points),
  the Daya-Bay 1230-day EH2/EH1 and EH3/EH1 spectral
  ratios~\cite{An:2016ses} (70 data points), and the Reno 1500-day
  FD/ND spectral ratios~\cite{reno:eps2017} (26 data points).
\end{itemize}

\begin{figure}\centering
  \includegraphics[width=\textwidth]{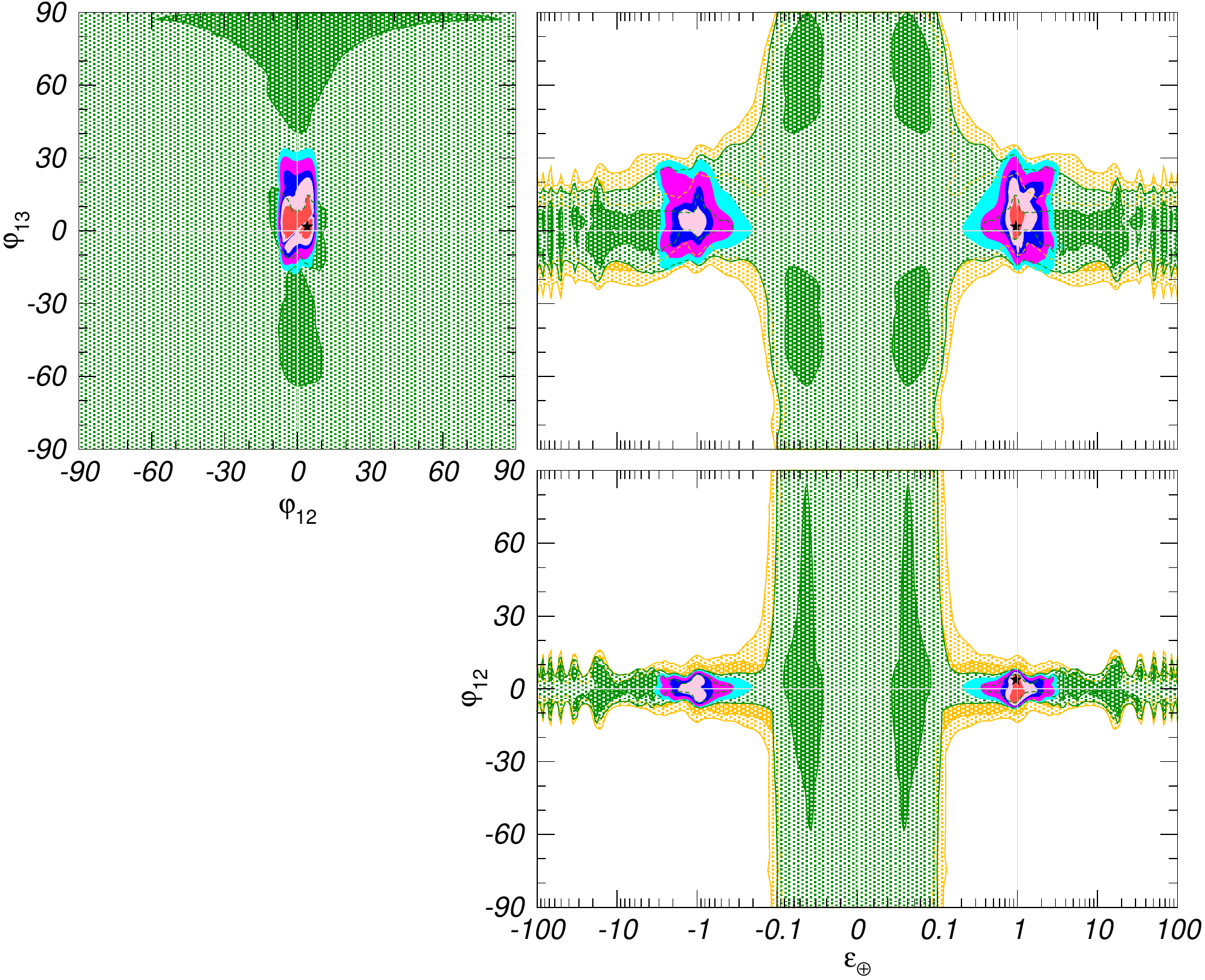}
  \caption{Two-dimensional projections of the allowed regions onto the
    matter potential parameters $\varepsilon_\oplus$, $\varphi_{12}$, and
    $\varphi_{13}$ after marginalisation with respect to the
    undisplayed parameters. The large green regions correspond to the
    analysis of atmospheric, LBL-CPC, and medium baseline reactor data at 90\% and
    $3\sigma$ CL. For comparison we show in yellow the corresponding
    results when omitting IceCube and reactor data. The solid coloured
    regions show the $1\sigma$, 90\%, $2\sigma$, 99\% and $3\sigma$ CL
    allowed regions once solar and KamLAND data are included. The
    best-fit point is marked with a star.}
  \label{fig:nsifit1_glb-epsil}
\end{figure}

Let us begin by showing in Figure~\ref{fig:nsifit1_glb-epsil} the
two-dimensional projections of the allowed regions in the Earth's
matter potential parameters $\varepsilon_\oplus$, $\varphi_{12}$ and
$\varphi_{13}$ (i.e., in the parametrisation of
Eq.~\eqref{eq:nsifit1_eps_atm} with $\alpha_i = 0$) after marginalising over
the oscillation parameters.  The green regions show the 90\% and
$3\sigma$ confidence regions (2~dof) from the analysis of atmospheric,
LBL-CPC and medium baseline reactor experiments.  Besides the increase in
statistics on low-energy atmospheric events provided by the updated
Super-Kamiokande and the new DeepCore data samples, the main
difference with respect to the analysis in
Refs.~\cite{GonzalezGarcia:2011my, Gonzalez-Garcia:2013usa} is the
inclusion of the bounds on NSI-induced $\nu_\mu$ disappearance
provided by IceCube high-energy data as well as the precise
information on $\theta_{13}$ and $|\Delta m^2_{31}|$ from medium baseline reactor
experiments.  To illustrate their impact we show as yellow regions the
results obtained when IceCube and reactor data are omitted. For what
concerns the projection over the matter potential parameters shown
here, we have verified that the difference between the yellow and
green regions is mostly driven by IceCube, which restricts the allowed
values of the $\varphi_{12}$ for $|\varepsilon_\oplus| \sim 0.1$--$1$.  This
can be understood since, for neutrinos with energies above
$\mathcal{O}(100~\text{GeV})$, the vacuum oscillation is very
suppressed and the survival probability of atmospheric $\nu_\mu$
arriving at zenith angle $\Theta_\nu$ is dominated by the matter
induced transitions
\begin{equation}
  P_{\mu\mu} \simeq 1 - \sin^2( 2\varphi_{\mu\mu} )
  \sin^2\left( \frac{d_e(\Theta_\nu) \varepsilon_\oplus}{2} \right)
  \quad\text{with}\quad
  \sin^2\varphi_{\mu\mu} = \sin^2\varphi_{12} \cos^2\varphi_{13}
\end{equation}
where $d_e(\Theta_\nu) = \sqrt{2} G_F X_e(\Theta_\nu)$ and the column
density $X_e(\Theta_\nu)$ is the integral of $N_e(x)$ along the
neutrino path in the Earth~\cite{Gonzalez-Garcia:2016gpq}. Since
$0.2\lesssim d_e(\Theta_\nu) \lesssim 20$ for $-1 \leq \cos\Theta_\nu
\leq -0.2$, the range $0.1 \lesssim |\varepsilon_\oplus| \lesssim 1$
corresponds to the first oscillation maximum for some of the
trajectories.  Also, the effective parameter $\varphi_{\mu\mu}$
entering in the expression of $P_{\mu\mu}$ depends linearly on
$\varphi_{12}$ and only quadratically on $\varphi_{13}$, which
explains why the bounds on the mixings are stronger for $\varphi_{12}$
than for $\varphi_{13}$.

As can be seen in Fig.~\ref{fig:nsifit1_glb-epsil}, even with the inclusion of
IceCube neither upper nor lower bounds on the overall strength of the
Earth's matter effects, $\varepsilon_\oplus$, can be derived from the
analysis of atmospheric, LBL-CPC and medium baseline reactor
experiments~\cite{Friedland:2004ah, Friedland:2005vy,
  GonzalezGarcia:2011my}.\footnote{See Refs.~\cite{Esmaili:2013fva,
    Salvado:2016uqu} for constraints in more restricted NSI
  scenarios.}  This happens because the considered data sample is
mainly sensitive to NSI through $\nu_\mu$ disappearance, and lacks
robust constraints on matter effects in the $\nu_e$ sector.  As a
consequence, when marginalising over $\varepsilon_\oplus$ (as well as over
the oscillation parameters) the full flavour projection $(\varphi_{12},
\varphi_{13})$ plane is allowed.  On the other hand, once the results
of solar and KamLAND experiments (which are sensitive to $\nu_e$) are
included in the analysis a bound on $\varepsilon_\oplus$ is obtained and the
flavour structure of the matter potential in the Earth is significantly
constrained.

\begin{figure}\centering
  \includegraphics[width=\textwidth]{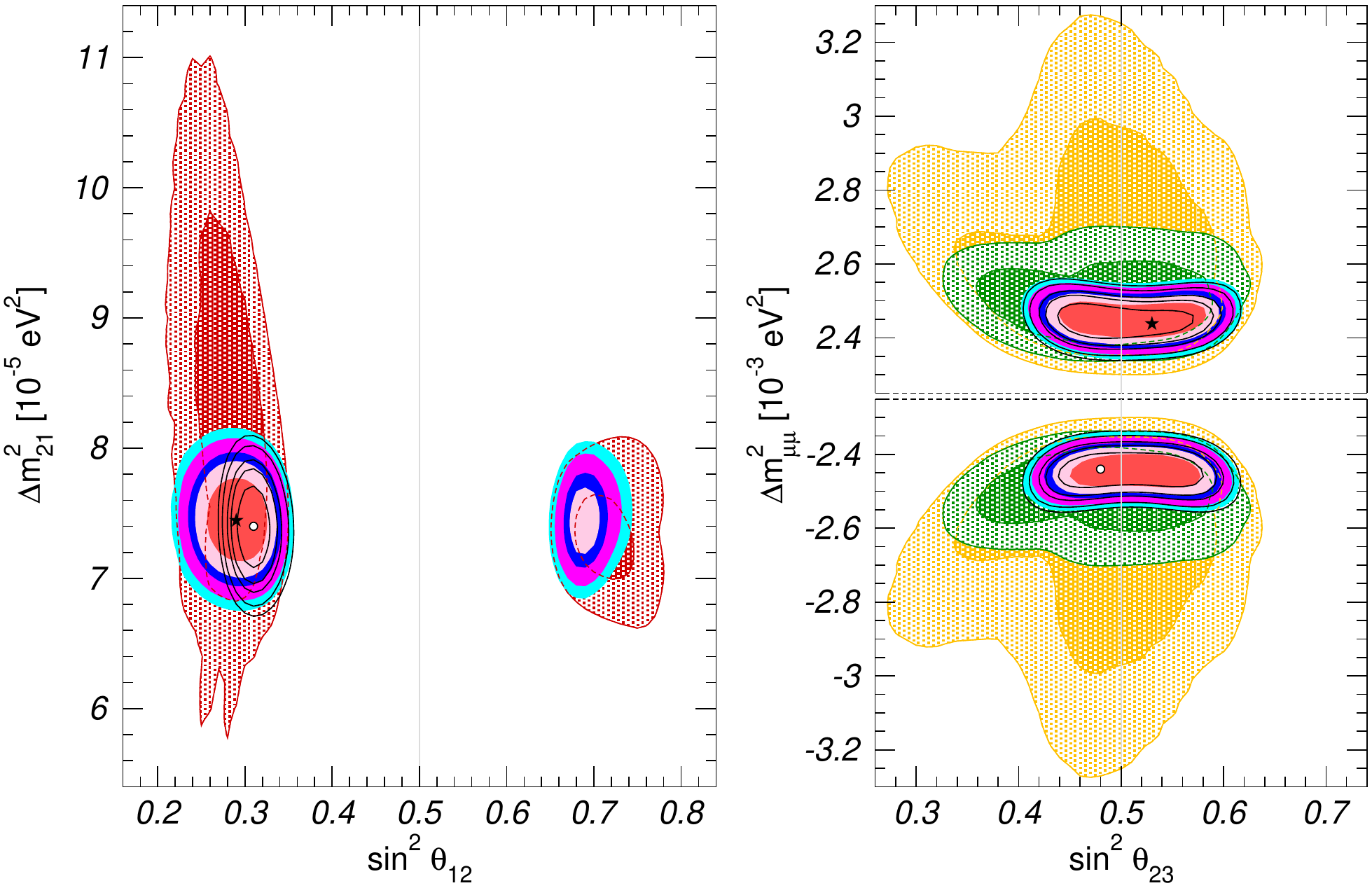}
  \caption{Two-dimensional projections of the allowed regions onto
    different vacuum parameters after marginalising over the matter
    potential parameters (including $\eta$) and the undisplayed
    oscillation parameters.  The solid coloured regions correspond to
    the global analysis of all oscillation data, and show the
    $1\sigma$, 90\%, $2\sigma$, 99\% and $3\sigma$ CL allowed regions;
    the best-fit point is marked with a star.  The black void regions
    correspond to the analysis with the standard matter potential
    (i.e., without NSI) and its best-fit point is marked with
    an empty dot.  For comparison, in the left panel we show in red
    the 90\% and $3\sigma$ allowed regions including only solar and
    KamLAND results, while in the right panels we show in green the
    90\% and $3\sigma$ allowed regions excluding solar and KamLAND
    data, and in yellow the corresponding ones excluding also IceCube
    and reactor data.}
  \label{fig:nsifit1_glb-oscil}
\end{figure}

In Fig.~\ref{fig:nsifit1_glb-oscil} we show the two-dimensional projections of
the allowed regions from the global analysis onto different sets of
oscillation parameters.  These regions are obtained after
marginalising over the undisplayed vacuum parameters as well as the
NSI couplings.  For comparison we also show as black-contour void
regions the corresponding results with the standard matter potential,
i.e., in the absence of NSI.  As discussed in
Sec.~\ref{sec:nsifit1_formalism-earth}, in the right panels we have chosen to
plot the regions in terms of the effective mass-squared difference
relevant for $\nu_\mu$ disappearance experiments,
$\Delta m^2_{\mu\mu}$. Notice that, having omitted NO$\nu$A and T2K
appearance data and also set $\Delta m^2_{21} = 0$ in atmospheric and
LBL-CPC experiments, the impact of the mass ordering on the results of
the fit is greatly reduced.

This figure clearly shows the robustness of the determination of the
$\Delta m^2_{21}$, $|\Delta m^2_{\mu\mu}|$ and $\theta_{23}$ vacuum oscillation
parameters even in the presence of the generalised NSI. This result relies on the complementarity and synergies
between the different data sets, which allows to constrain those
regions of the parameter space where cancellations between standard
and non-standard effects occur in a particular data set.  To
illustrate this we show as shaded regions the results obtained when
some of the data are removed.  For example, comparing the solid
coloured regions with the shaded red ones in the left panel we see how,
in the presence of NSI with arbitrary values of $\eta$, the precise
determination of $\Delta m^2_{21}$ requires the inclusion of atmospheric,
LBL-CPC and medium baseline reactor data: if these sets are omitted, the huge
values of the NSI couplings allowed by solar data for $-70^\circ
\lesssim \eta \lesssim -60^\circ$ destabilise KamLAND's determination
of $\Delta m^2_{21}$, as discussed in Sec.~\ref{sec:nsifit1_solar}.  The inclusion
of these sets also limits the margins for NSI to alleviate the tension
between solar and KamLAND data on the preferred $\Delta m^2_{21}$ value, as
can be seen by comparing the full dark-blue and red lines in the left
panel of Fig.~\ref{fig:nsifit1_chisq-eta}: indeed, in the global analysis the
best-fit is achieved for $\eta \simeq -44^\circ$, which is precisely
when the NSI effects in the Earth matter cancel so that no restriction
on NSI contributions to solar and KamLAND data is imposed.

In the same way we see on the right panels that, if the solar and
KamLAND data are removed from the fit, the determination of
$\Delta m^2_{\mu\mu}$ and $\theta_{23}$ degrades because of the possible
cancellations between NSI and mass oscillation effects in the relevant
atmospheric and LBL-CPC probabilities. As NSI lead to
energy-independent contributions to the oscillation phase, such
cancellations allow for larger values of $|\Delta m^2_{\mu\mu}|$.  Comparing
the yellow and green regions we see the inclusion medium baseline reactor
experiments, for which NSI effects are irrelevant due to the short
baselines involved, is crucial to reduce the degeneracies and provide
a NSI-independent measurement of $|\Delta m^2_{\mu\mu}|$. Even so, only the
inclusion of solar and KamLAND allows to recover the full sensitivity
of atmospheric and LBL-CPC experiments and derive limits on
$\Delta m^2_{\mu\mu}$ and $\theta_{23}$ as robust as the standard ones.

The most dramatic implications of NSI for what concerns the
determination of the oscillation parameters affect $\theta_{12}$.  In
particular, for generic NSI with arbitrary $\eta$ the LMA-D solution
is still perfectly allowed by the global oscillation analysis, as
indicated by the presence of the corresponding region in the left
panel in Fig.~\ref{fig:nsifit1_glb-oscil}.  Turning to
Fig.~\ref{fig:nsifit1_chisq-eta} we see that even after including all the
oscillation data (dark-blue lines) the LMA-D solution is allowed at
$3\sigma$ for $-38^\circ \lesssim \eta \lesssim 87^\circ$ (as well as
in a narrow window around $\eta \simeq -65^\circ$), and indeed for
$-28^\circ \lesssim \eta\lesssim 0^\circ$ it provides a slightly
better global fit than LMA.
From Fig.~\ref{fig:nsifit1_glb-oscil} we also see that the lower bound on
$\theta_{12}$ in the presence of NSI is substantially weaker than the
standard $3\nu$ case.  We had already noticed such reduction in the
analysis of solar and KamLAND data for any value of $\eta$; here we
point out that the cancellation of matter effects in the Earth for
$\eta \approx -43.6^\circ$ prevents any improvement of that limit from
the addition of Earth-based oscillation experiments.

\begin{table}[hbtp]\centering
  \definecolor{grey}{gray}{0.75}
  \newcommand{\grsep}{~\color{grey}\vrule}
  \begin{tabular}{lrr}
    \toprule
    & {LMA\hfil} & $\text{LMA}\oplus\text{LMA-D}$     \\
    \cmidrule(l){2-3}
    \begin{tabular}{@{}l@{}}
      $\varepsilon_{ee}^u - \varepsilon_{\mu\mu}^u$ \\
      $\varepsilon_{\tau\tau}^u - \varepsilon_{\mu\mu}^u$
    \end{tabular}
    &
    \begin{tabular}{@{}r@{}}
      $[-0.020, +0.456]$ \\
      $[-0.005, +0.130]$
    \end{tabular}
    &
    \begin{tabular}{@{}r@{}}
      $\oplus [-1.192, -0.802]$ \\
      $[-0.152, +0.130]$
    \end{tabular}
\\
    $\varepsilon_{e\mu}^u$ & $[-0.060, +0.049]$ & $[-0.060, +0.067]$    \\
    $\varepsilon_{e\tau}^u$ & $[-0.292, +0.119]$ & $[-0.292, +0.336]$    \\
    $\varepsilon_{\mu\tau}^u$ & $[-0.013, +0.010]$ & $[-0.013, +0.014]$     \\
    \midrule
    \begin{tabular}{@{}l@{}}
      $\varepsilon_{ee}^d - \varepsilon_{\mu\mu}^d$ \\
      $\varepsilon_{\tau\tau}^d - \varepsilon_{\mu\mu}^d$
    \end{tabular}
    &
    \begin{tabular}{@{}r@{}}
      $[-0.027, +0.474]$ \\
      $[-0.005, +0.095]$
    \end{tabular}
    &
    \begin{tabular}{@{}r@{}}
      $\oplus [-1.232, -1.111]$ \\
      $[-0.013, +0.095]$
    \end{tabular} \\
    $\varepsilon_{e\mu}^d$ & $[-0.061, +0.049]$ & $[-0.061, +0.073]$    \\
    $\varepsilon_{e\tau}^d$ & $[-0.247, +0.119]$ & $[-0.247, +0.119]$    \\
    $\varepsilon_{\mu\tau}^d$ & $[-0.012, +0.009]$ & $[-0.012, +0.009]$     \\
    \midrule
    \begin{tabular}{@{}l@{}}
      $\varepsilon_{ee}^p - \varepsilon_{\mu\mu}^p$ \\
      $\varepsilon_{\tau\tau}^p - \varepsilon_{\mu\mu}^p$
    \end{tabular}
    &
    \begin{tabular}{@{}r@{}}
      $[-0.041, +1.312]$ \\
      $[-0.015, +0.426]$
    \end{tabular}
    &
    \begin{tabular}{@{}r@{}}
      $\oplus [-3.327, -1.958]$ \\
      $[-0.424, +0.426]$
    \end{tabular}
    \\
    $\varepsilon_{e\mu}^p$ & $[-0.178, +0.147]$ & $[-0.178, +0.178]$    \\
    $\varepsilon_{e\tau}^p$ & $[-0.954, +0.356]$ & $[-0.954, +0.949]$    \\
    $\varepsilon_{\mu\tau}^p$ & $[-0.035, +0.027]$ & $[-0.035, +0.035]$     \\
    \bottomrule
  \end{tabular}
  \caption{$2\sigma$ allowed ranges for the NSI couplings
    $\varepsilon_{\alpha\beta}^u$, $\varepsilon_{\alpha\beta}^d$ and
    $\varepsilon_{\alpha\beta}^p$ as obtained from the global analysis of
    oscillation data.  The results are obtained after marginalising over
    oscillation and the other matter potential parameters either
    within the LMA only and within both LMA and LMA-D subspaces
    respectively (this second case is denoted as $\text{LMA} \oplus
    \text{LMA-D}$).}
  \label{tab:nsifit1_ranges}
\end{table}

\begin{figure}\centering
  \includegraphics[width=0.8\textwidth]{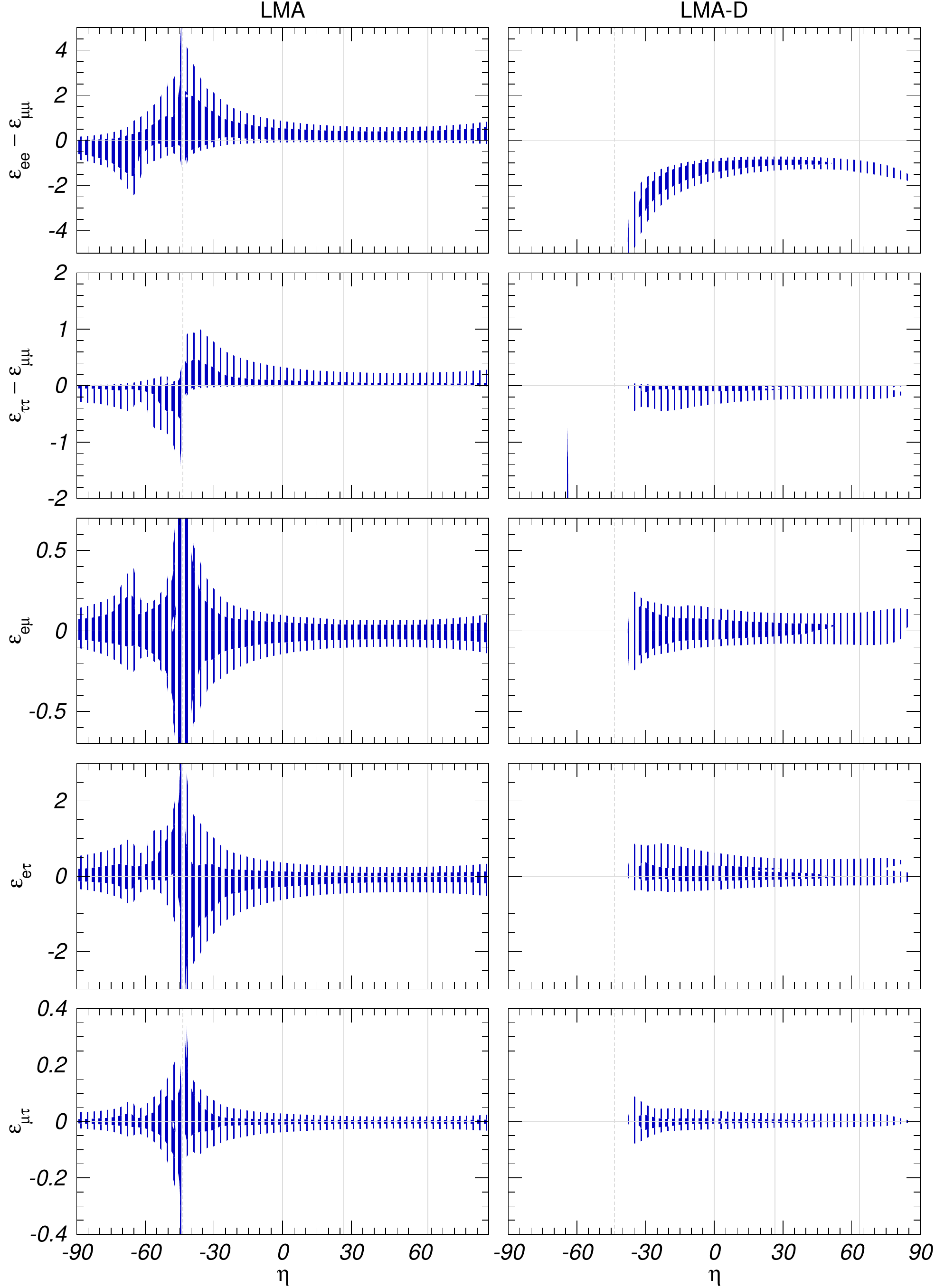}
  \caption{90\%, and $3\sigma$ CL (1~dof) allowed ranges for the NSI
    couplings from the global oscillation analysis in the presence of
    non-standard matter potential as a function of the NSI quark
    coupling parameter $\eta$.  In each panel the undisplayed
    parameters have been marginalised. On the left panels the
    oscillation parameters have been marginalised within the LMA
    region while the right panels corresponds to LMA-D solutions.  The
    ranges are defined with respect to the minimum for each $\eta$.}
  \label{fig:nsifit1_glb-range}
\end{figure}

The bounds on the five relevant NSI couplings (two diagonal
differences and three non-diagonal entries) from the global
oscillation analysis are displayed in Fig.~\ref{fig:nsifit1_glb-range} as a
function of $\eta$. Concretely, for each value of $\eta$ we plot as
vertical bars the 90\% and $3\sigma$ allowed ranges (1~dof) after
marginalising with respect to the undisplayed parameters. The left and
right panels correspond to the limits for $\theta_{12}$ within the LMA
and LMA-D solution, respectively, both defined with respect to the
same common minimum for each given $\eta$.  For the sake of
convenience and comparison with previous results we list in the first
columns in Table~\ref{tab:nsifit1_ranges} the 95\% CL ranges for NSI with
up-quarks only ($\eta \approx 26.6^\circ$), down-quarks only ($\eta
\approx 63.4^\circ$) and couplings proportional to the electric charge
($\eta=0^\circ$); in this last case we have introduced an extra
$\sqrt{5}$ normalisation factor so that the quoted bounds can be
directly interpreted in terms of $\varepsilon_{\alpha\beta}^p$.  Let us point
out that the sign of each non-diagonal $\varepsilon_{\alpha\beta}$ can
be flipped away by a suitable change of signs in some of the mixing
angles; it is therefore not an intrinsic property of NSI, but rather a
relative feature of the vacuum and matter Hamiltonians.  Thus,
strictly speaking, once the results are marginalised with respect to
all the other parameters in the most general parameter space, the
oscillation analysis can only provide bounds on
$|\varepsilon_{\alpha\neq\beta}|$.  However, for definiteness we have
chosen to restrict the range of the mixing angles to $0 \leq
\theta_{ij} \leq \pi/2$ and to ascribe the relative vacuum-matter
signs to the NSI couplings, so that the ranges of the non-diagonal
$\varepsilon_{\alpha\beta}$ in Fig.~\ref{fig:nsifit1_glb-range} as well as in Table~\ref{tab:nsifit1_ranges} are given
for both signs.

From Fig.~\ref{fig:nsifit1_glb-range} and Tab.~\ref{tab:nsifit1_ranges} we see that
the allowed range for all the couplings (except $\varepsilon_{ee} -
\varepsilon_{\mu\mu}$) obtained marginalising over both $\theta_{12}$
octants, which we denote in the table as $\text{LMA} \oplus
\text{LMA-D}$, is only slighter wider than what obtained considering
only the LMA solution. Conversely, for $\varepsilon_{ee} -
\varepsilon_{\mu\mu}$ the allowed range is composed by two disjoint
intervals, each one corresponding to a different $\theta_{12}$
octant. Note that for this coupling the interval associated with the
LMA solution is not centered at zero due to the tension between the
value of $\Delta m^2_{21}$ preferred by KamLAND and solar experiments, even
after including the bounds from atmospheric and LBL data.
In general, we find that the allowed ranges for all the couplings do
not depend strongly on the value of $\eta$ as long as $\eta$ differs
enough from the critical value $\eta \approx -43.6^\circ$. As already
explained, at this point NSI in the Earth cancel
out, so that no bound on the NSI parameters can be derived from any
Earth-based experiment. This leads to a breakdown of the limits on
$\varepsilon_{\alpha\beta}$, since solar data are only sensitive to the
$\varepsilon_D$ and $\varepsilon_N$ combinations and cannot constrain the
five NSI couplings simultaneously.  In addition to the region around
$\eta \approx -43.6^\circ$, there is also some mild weakening of the
bounds on NSI couplings involving $\nu_e$ for $-70^\circ \lesssim \eta
\lesssim -60^\circ$, corresponding to the window where NSI effects in
the Sun are suppressed.  Apart from these special cases, the bounds
quoted in Table~\ref{tab:nsifit1_ranges} are representative of the
characteristic sensitivity to the NSI coefficients from present
oscillation experiments, which at 95\% CL ranges from
$\mathcal{O}(1\%)$ for $|\varepsilon_{\mu\tau}|$ to $\mathcal{O}(30\%)$
for $|\varepsilon_{e\tau}|$ ---~the exception being, of course,
$\varepsilon_{ee} - \varepsilon_{\mu\mu}$.

\begin{figure}[hbtp]\centering
  \includegraphics[width=\textwidth]{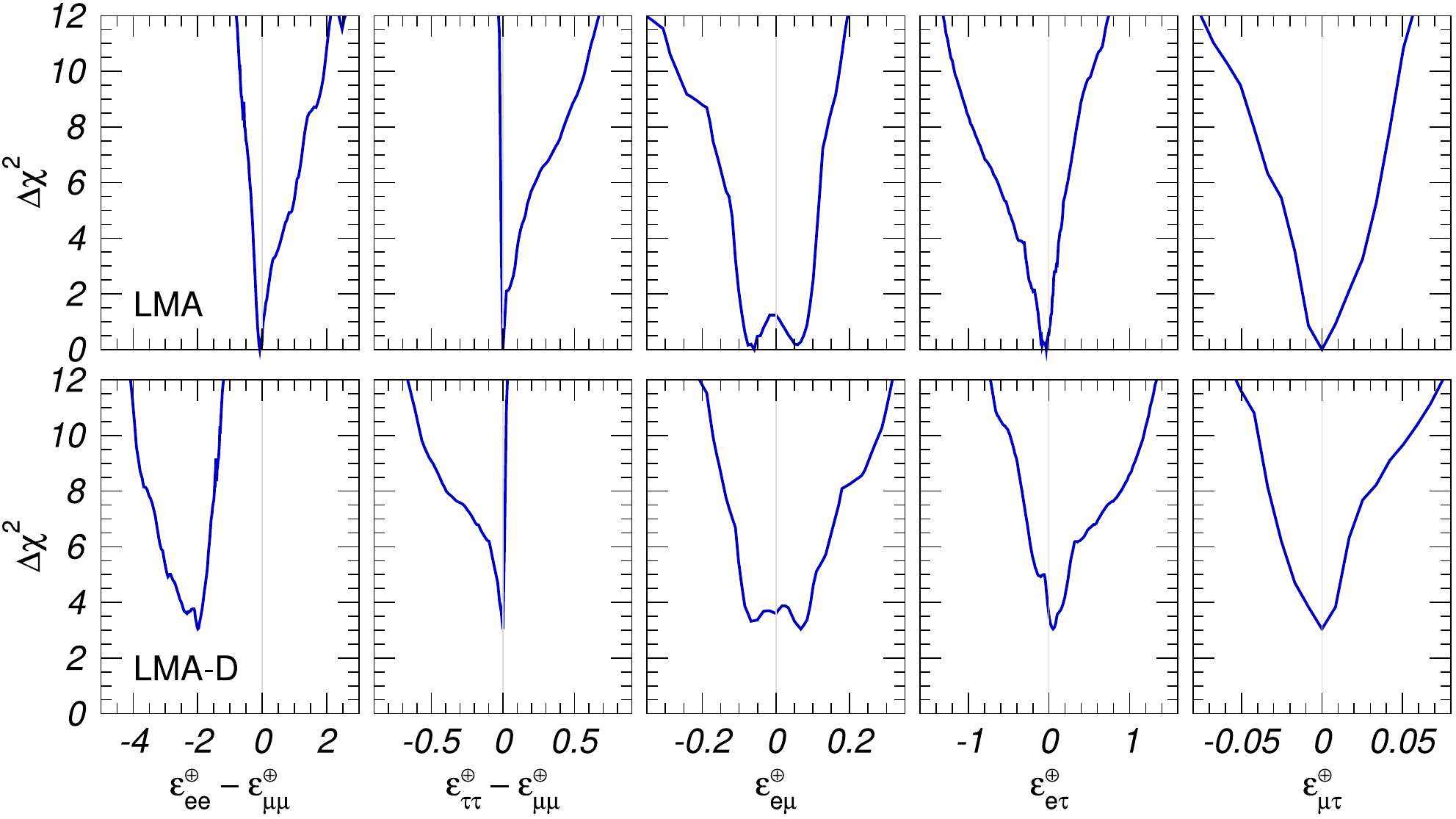}
  \caption{Dependence of the $\Delta\chi^2$ function on the effective
    NSI parameters relevant for matter effects in LBL experiments with
    arbitrary values of $\eta$, from the global analysis of solar,
    atmospheric, LBL-CPC and reactor data.  The upper (lower) panels correspond
    to solutions within the LMA (LMA-D) subset of parameter space.}
  \label{fig:nsifit1_chisq-rng}
\end{figure}

We finish by quantifying the results of our analysis in terms of the
effective NSI parameters which describe the generalised Earth matter
potential and are, therefore, the relevant quantities for the study of
LBL experiments.  The results are shown in
Fig.~\ref{fig:nsifit1_chisq-rng} where we plot the dependence of the global
$\chi^2$ on each NSI effective couplings after marginalisation over
all other parameters.\footnote{Notice that the correlations among the
  allowed values for these parameters are important and they are
  required for reconstruction of the allowed potential at given CL.}
Let us point out that, if only the results from Earth-based
experiments such as atmospheric, LBL and reactor data were
included in the analysis, the curves would be independent of
$\eta$. However, when solar experiments are also
considered the global $\chi^2$ becomes sensitive to the value of
$\eta$.  Given that, what we quantify in Fig.~\ref{fig:nsifit1_chisq-rng} is
our present knowledge of the matter potential for neutrino propagation
in the Earth for \emph{any unknown value} of $\eta$. Technically this
is obtained by marginalising the results of the global $\chi^2$ with
respect to $\eta$ as well, so that the $\Delta\chi^2$ functions
plotted in the figure are defined with respect to the absolute minimum
for any $\eta$ (which, as discussed above and shown in
Fig.~\ref{fig:nsifit1_chisq-eta}, lies close to $\eta \sim -45^\circ$).  In
the upper panels the oscillation parameters have been marginalised
within the LMA solution and in the lower ones within the LMA-D
solution.

\subsection{Summary}
\label{sec:nsifit1_summary}

In this section we have presented an updated analysis of neutrino
oscillation results with the aim of establishing how well we can
presently determine the size and flavour structure of NSI-NC which
affect the evolution of neutrinos in a matter background. In
particular we have extended previous studies by considering NSI with
an arbitrary ratio of couplings to up and down quarks (parametrised by
an angle $\eta$) and a lepton-flavour structure independent of the
quark type (parametrised by a matrix $\varepsilon_{\alpha\beta}$). We
have included in our fit all the solar, atmospheric, reactor and
accelerator data commonly used for the standard $3\nu$ oscillation
analysis, with the only exception of T2K and NO$\nu$A appearance data
whose recent hints in favour of CP violation are not easily
accommodated within the CP-conserving approximation assumed in this
fit.  We have found that:
\begin{itemize}
\item classes of experiments which are sensitive to NSI only through
  matter characterised by a limited range of proton/neutron ratios
  $Y_n$ unavoidably exhibit suppression of NSI effects for specific
  values of $\eta$. This is the case for solar data at $-70^\circ
  \lesssim \eta \lesssim -60^\circ$, and for Earth-based (atmospheric,
  LBL, reactor) experiments at $\eta \approx -44^\circ$. Such
  cancellations limit the sensitivity to the NSI couplings;

\item moreover, the interplay between vacuum and matter contributions
  to the flavour transition probabilities in classes of experiments
  with limited energy range and/or sensitive only to a specific
  oscillation channel spoils the accurate determination of the
  oscillation parameters achieved in the standard $3\nu$
  scenario. This is particularly visible in $\Delta m^2_{21}$ and
  $\theta_{12}$ as determined by solar and KamLAND data, as well as in
  $\Delta m^2_{31}$ and $\theta_{23}$ as determined by atmospheric, LBL-CPC
  and medium baseline reactor data;

\item however, both problems can be efficiently resolved by combining
  together different classes of experiments, so to ensure maximal
  variety of matter properties, energy ranges, and oscillation
  channels.  In particular, our calculations show that the precise
  determination of the vacuum parameters is fully recovered (except
  for $\theta_{12}$) in a joint analysis of solar and Earth-based
  oscillation experiments, even when arbitrary values of $\eta$ are
  considered;

\item the well-known LMA-D solution, which arises in the presence of
  of NSI as a consequence of the generalised mass ordering degeneracy, is allowed at $3\sigma$
  for $-38^\circ \lesssim \eta \lesssim 87^\circ$ from the global
  analysis of oscillation data.
\end{itemize}
In addition, we have determined the allowed range of the NSI couplings
$\varepsilon_{\alpha\beta}$ as a function of the up-to-down coupling
$\eta$, showing that such constraints are generically robust except
for a few specific values of $\eta$ where cancellations
occurs. Finally, in view of the possible implications that generic
NSI-NC may have for future Earth-based facilities, we have recast the
results of our analysis in terms of the effective NSI parameters
$\varepsilon_{\alpha\beta}^\oplus$ which describe the generalised matter
potential in the Earth, and are therefore the relevant quantities for
the study of atmospheric and LBL experiments.

\section{General analysis: robustness of LBL accelerator experiments under NSI}
\label{sec:nsifit2}
In the previous section, we have derived bounds on CP conserving
NSI-NC from a global analysis of oscillation data.
The obtained constraints are strong in general, but in some
flavour channels $\mathcal{O}(0.1)$ NSI are still
allowed. Being a fit to CP conserving effects, only $\nu_\mu$
disappearance data from the T2K and \NOvA/ accelerator LBL experiments 
was included. 
But with the results we found, the allowed NSI could have 
an impact on the $\parenbar{\nu}_\mu \rightarrow
\parenbar{\nu}_e$ appearance probability as well.
This is mostly so because in appearance channels the standard
contribution is suppressed by three-flavour mixing whereas the new 
physics contributions may not be~\cite{Kopp:2007ne} (see also
\cref{eq:inv3,eq:inv4}).

We have seen that the $\parenbar{\nu}_e$ appearance channel is 
particularly relevant
in assessing the questions still open in the three-neutrino analysis:
the maximality and octant of $\theta_{23}$, the ordering of the mass
eigenstates, and leptonic CP violation.
The clarification of these unknowns is the main
focus of the running LBL experiments and its precise determination is
at the center of the physics programme of the upcoming LBL facilities,
in particular the Deep Underground Neutrino Experiment
(DUNE)~\cite{Acciarri:2016ooe} and the Tokai to HyperKamiokande (T2HK)
experiment~\cite{Abe:2015zbg}.

In the presence of NSI, however, the task of exploring leptonic
CP violation in LBL experiments becomes enriched (to the point of
confusion) by new sources of CP violation and an intrinsic degeneracy 
in the Hamiltonian describing neutrino evolution, as explored in
\cref{chap:NSItheor}. This has resulted in an intense phenomenological 
activity to quantify
these issues and to devise strategies to clarify them, first in
proposed facilities like the Neutrino
Factory~\cite{Bandyopadhyay:2007kx, Gago:2009ij, Coloma:2011rq} and
most recently in the context of the upcoming
experiments~\cite{Coloma:2015kiu, Masud:2015xva, deGouvea:2015ndi,
  Liao:2016hsa, Huitu:2016bmb, Bakhti:2016prn, Masud:2016bvp,
  C.:2017yqh, Rashed:2016rda, Masud:2016gcl, Blennow:2016etl,
  Ge:2016dlx, Forero:2016ghr, Blennow:2016jkn, Fukasawa:2016lew,
  Liao:2016orc, Deepthi:2016erc, Deepthi:2017gxg, Meloni:2018xnk,
  Flores:2018kwk, Verma:2018gwi, Chatterjee:2018dyd, Masud:2018pig}.

Also very interestingly, it has been argued that NSI can already play a
role in the significance of the ``hints'' of CP violation and of
NO~\cite{Forero:2016cmb, Liao:2016bgf}. In particular, in
Ref.~\cite{Forero:2016cmb} it was pointed out the discomforting
possibility of confusing CP conserving NSI with a non-zero value of
$\delta_\text{CP}$ in the analysis of $\nu_e$ and $\bar\nu_e$
appearance results at T2K and \NOvA/.  Clearly such confusion could lead
to an incorrect claim of the observation of leptonic CP violation in a
theory which is CP conserving.

As the analysis presented in the previous section only constrained the CP
conserving part of the Hamiltonian --- and for consistency the observables
most sensitive to CP violating effects, i.e., $\nu_e$ and $\bar\nu_e$
appearance at LBL experiments, were not included in the fit ---, 
the issue of the possible confusion between real NSI and
leptonic CP violation could not be addressed.  Furthermore, under the
simplifying assumptions employed, the analysis could not either yield any
conclusion on the status of the mass ordering determination in 
the presence of NSI.

So, in order to address these questions, in this section we extend the
analysis to account for the effect of complex NSI in the observables sensitive to leptonic CP violation. We also include the effects of $\Delta m^2_{21}$
required to determine the sensitivity to the mass ordering.
Our goal is to quantify the robustness of the present ``hints'' for
these effects in the presence of NSI \emph{which are consistent with the
bounds imposed by the
  CP-conserving observables}.
  
\subsection{Analysis framework}

This section builds upon the results of the comprehensive global fit 
in \cref{sec:nsifit1}, performed in the framework of
three-flavour oscillations plus NSI with quarks. In addition, as 
discussed in \cref{chap:NSItheor}, in order to study the
possible effects (in experiments performed in matter) of NSI on the
determination of the phase which parametrises CP violation in vacuum
\emph{without introducing an artificial basis dependence}, one needs
to include in the analysis the most general complex NSI matter
potential containing \emph{all} the three additional arbitrary phases.

Thus, in principle, for each choice of the $(\xi^p, \xi^n)$
couplings the analysis depends on six oscillations parameters plus
eight NSI parameters, of which five are real and three are phases. To
keep the fit manageable, in the previous section only real NSI
were considered and $\Delta m^2_{21}$ effects were neglected in the fit 
of atmospheric and LBL experiments. This rendered such 
analysis independent of the CP phase in the leptonic mixing matrix and 
of the ordering of the states.

In this section we extend the previous analysis to include the effect of
the four CP-phases in the Hamiltonian as well as the $\Delta m^2_{21}$
effects, in particular where they are most relevant which is the fit
of the LBL experiments.  In order to do so while still maintaining the
running time under control, we split the global $\chi^2$ in a part
containing the latest data from the LBL experiments MINOS, T2K and
\NOvA/ (accounting for both appearance and disappearance data in
neutrino and antineutrino modes, see \cref{sec:dCPevolution}), for which both the extra phases and
the interference between $\Delta m^2_{21}$ and $\Delta m^2_{31}$ are properly
included, and a part containing CP-conserving observables where the
complex phases can be safely neglected and are therefore implemented
as described in the previous section. In what follows we label
as ``OTH'' (short for ``others'') these non-LBL observables which
include the results from solar neutrino
experiments~\cite{Cleveland:1998nv, Kaether:2010ag,
  Abdurashitov:2009tn, Hosaka:2005um, Cravens:2008aa, Abe:2010hy,
  sksol:nakano2016, Aharmim:2011vm, Bellini:2011rx, Bellini:2008mr,
  Bellini:2014uqa}, from the KamLAND reactor
experiment~\cite{Gando:2013nba}, from medium baseline reactor
experiments~\cite{dc:cabrera2016, An:2016ses, reno:eps2017}, from
atmospheric neutrinos collected by IceCube~\cite{TheIceCube:2016oqi}
and its sub-detector DeepCore~\cite{Aartsen:2014yll}, and from our
analysis of Super-Kamiokande atmospheric
data~\cite{Wendell:2014dka}.\footnote{As in
  the previous section, we include here our analysis of Super-Kamiokande
  atmospheric data in the form of the ``classical'' samples of
  $e$-like and $\mu$-like events (70 energy and zenith angle bins). As
  discussed in \cref{subsec:nufit3_SK} with such analysis in the
  framework of standard 3-nu oscillations we cannot reproduce the
  sensitivity to subdominant effects associated with the mass ordering
  and $\delta_\text{CP}$ found by Super-Kamiokande in their analysis of more
  dedicated samples~\cite{Abe:2017aap}. For that reason we include Super-Kamiokande
  atmospheric data but only as part of the ``OTH'' group.}. These 
correspond to the same data as the latest analysis in \cref{sec:dCPevolution}.
  
Schematically, if we denote by $\vec{w}$ the five real oscillation
parameters (i.e., the two mass differences and the three
mixing angles), by $\eta$ the direction in the $(\xi^p, \xi^n)$ plane,
by $|\varepsilon_{\alpha\beta}|$ the five real components of the neutrino
part of the NSI parameters (two differences of the three diagonal
entries of $\varepsilon_{\alpha\beta}$, as well as the modulus of the three
non-diagonal entries\footnote{More precisely, in our analysis of OTH
  experiments we consider both the modulus and the sign of the
  non-diagonal $\varepsilon_{\alpha\beta}$, i.e., we explicitly
  account for the all the CP-conserving values of the three phases:
  $\phi_{\alpha\beta} = 0,\pi$. However, in the construction of
  $\chi^2_\text{OTH}$ these signs are marginalised, so that only the
  modulus $|\varepsilon_{\alpha\beta}|$ is correlated with
  $\chi^2_\text{LBL}$.}), and by $\phi_{\alpha\beta}$ the three phases
of the non-diagonal entries of $\varepsilon_{\alpha\beta}$, we split the
global $\chi^2$ for the analysis as
\begin{equation}
  \chi^2_\text{GLOB}(\vec{w}, \delta_\text{CP},
  |\varepsilon_{\alpha\beta}|, \phi_{\alpha\beta}, \eta)
  = \chi^2_\text{OTH}(\vec{w}, |\varepsilon_{\alpha\beta}|, \eta)
  + \chi^2_\text{LBL}(\vec{w}, \delta_\text{CP},
  |\varepsilon_{\alpha\beta}|, \phi_{\alpha\beta}, \eta)
\end{equation}
so $\chi^2_\text{OTH}$ and $\chi^2_\text{LBL}$ depend on $5+5+1=11$
and $5+1+5+3+1 = 15$ parameters, respectively.

To make the analysis in such large parameter space treatable, we
introduce a series of simplifications. First, we notice that in medium baseline
reactor experiments the baseline is short enough to safely neglect the
effects of the matter potential, so that we have:
\begin{equation}
  \chi^2_\text{OTH}(\vec{w}, |\varepsilon_{\alpha\beta}|, \eta)
  = \chi^2_\text{SOLAR+KAMLAND+ATM}(\vec{w}, |\varepsilon_{\alpha\beta}|, \eta)
  + \chi^2_\text{MBL-REA}(\vec{w}) \,.
\end{equation}
Next, we notice that in LBL experiments the sensitivity to
$\theta_{12}$, $\Delta m^2_{21}$ and $\theta_{13}$ is marginal compared to
solar and reactor experiments; hence, in $\chi^2_\text{LBL}$ we can
safely fix $\theta_{12}$, $\theta_{13}$ and $\Delta m^2_{21}$ to their best
fit value as determined by the experiments included in
$\chi^2_\text{OTH}$. However, in doing so we must notice that, within
the approximations used in the construction of $\chi^2_\text{OTH}
(\vec{w}, |\varepsilon_{\alpha\beta}|, \eta)$, there still remains the effect
associated to the NSI/mass-ordering degeneracy which leads to the
appearance of a new solution in the solar sector with a mixing angle
$\theta_{12}$ in the second octant, the so-called LMA-Dark
(LMA-D)~\cite{Miranda:2004nb} solution. Although LMA-D is not totally
degenerate with LMA, due to the variation of the matter chemical
composition along the path travelled by solar neutrinos, it still
provides a good fit to the data for a wide range of quark couplings,
as found in the previous section.  Concretely, after
marginalisation over $\eta$ we get that the parameter region
containing the LMA-D solution lies at
\begin{equation}
  \label{eq:dclmad}
  \chi^2_\text{OTH,LMA-D} - \chi^2_\text{OTH,LMA} = 3.15 \,.
\end{equation}

Therefore, when marginalising over $\theta_{12}$ we consider two
distinct parts of the parameter space, labelled by the tag ``REG'':
one with $\theta_{12}<45^\circ$, which we denote as $\text{REG} =
\text{LIGHT}$, and one with $\theta_{12}>45^\circ$, which we denote by
$\text{REG} = \text{DARK}$.  Correspondingly, the fixed value of
$\theta_{12}$ used in the construction of $\chi^2_\text{LBL,REG}$ is
the best fit value within either the LMA or the LMA-D region:
$\sin^2\bar\theta_{12}^\textsc{light} = 0.31$ or
$\sin^2\bar\theta_{12}^\textsc{dark} = 0.69$, respectively. The best
fit values for the other two oscillation parameters fixed in
$\chi^2_\text{LBL,REG}$ are the same for LMA and LMA-D:
$\Delta\bar{m}^2_{21} = \SI{7.4e-5}{eV^2}$ and
$\sin^2\bar\theta_{13} = 0.0225$.

Further simplification arises from the fact that for LBL experiments
the dependence on the NSI neutrino and quark couplings enters only via
the effective Earth-matter NSI combinations
$\varepsilon_{\alpha\beta}^\oplus$ defined in Eq.~\eqref{eq:nsifit1_eps-earth}.  It
is therefore convenient to project also $\chi^2_\text{OTH}$ over these
combinations, and to marginalise it with respect to $\eta$. In
addition we neglect the small correlations introduced by the common
dependence of the atmospheric experiments in $\chi^2_\text{OTH}$ and
the LBL experiments on $\Delta m^2_{31}$ and $\theta_{23}$, and we also
marginalise the atmospheric part of $\chi^2_\text{OTH}$ over these two
parameters.  This means that in our results we do not account for the
information on $\Delta m^2_{31}$ and $\theta_{23}$ arising from atmospheric
experiments, however we keep the information on $\Delta m^2_{31}$ from medium baseline
reactor experiments. With all this, we can define a function
$\chi^2_\text{OTH,REG}$ depending on six parameters:
\begin{equation}
  \chi^2_\text{OTH,REG} \big( \Delta m^2_{31}, |\varepsilon^\oplus_{\alpha\beta}| \big)
  \equiv
  \min_{\substack{\eta,\, \theta_{12} \in \,\textsc{reg}\\
      \theta_{13}, \theta_{23}, \Delta m^2_{21}}}
  \chi^2_\text{OTH}\big( \vec{w},\,
  |\varepsilon_{\alpha\beta}^\oplus| \big/ [\xi^p + Y_n^\oplus \xi^n],\, \eta
  \big)
\end{equation}
while our final global $\chi^2_\text{GLOB,REG}$ is a function of
eleven parameters which takes the form
\begin{multline}
\label{eq:chi2global}
  \chi^2_\text{GLOB,REG} \big( \theta_{23}, \Delta m^2_{31},
  \delta_\text{CP}, |\varepsilon_{\alpha\beta}^\oplus|, \phi_{\alpha\beta} \big)
  = \chi^2_\text{OTH,REG} \big( \Delta m^2_{31}, |\varepsilon^\oplus_{\alpha\beta}| \big)
  \\
  + \chi^2_\text{LBL,REG} \big( \theta_{23}, \Delta m^2_{31}, \delta_\text{CP},
  |\varepsilon_{\alpha\beta}^\oplus|, \phi_{\alpha\beta}
  \,\big\Vert\,
  \bar\theta_{12}^\textsc{reg}, \bar\theta_{13}, \Delta\bar{m}^2_{21} \big)
\end{multline}
with $\text{REG} = \text{LIGHT}$ or DARK.

\subsection{Results}
\label{sec:nsifit1_results}

In order to quantify the effect of the matter NSI on the present
oscillation parameter determination we have performed a set of 12
different analyses in the eleven-dimensional parameter space. Each
analysis corresponds to a different combination of observables. The
results of the LBL experiment MINOS are always included in
all the cases, so for convenience in what follows we define
$\chi^2_\text{OTHM} \equiv \chi^2_\text{OTH} + \chi^2_\text{MINOS}$.
To this we add $\chi^2_\text{LBL}$ with $\text{LBL} = \text{T2K}$,
\NOvA/, and T2K+\NOvA/. In
addition, we perform the analysis in four distinctive parts of the
parameter space: the solar octant ``REG'' being LIGHT or DARK, and the
mass ordering being normal (NO) or inverted (IO).

To efficiently explore such a large dimensional parameter space, with potential 
flat directions and quasidegenerate minima, we have employed 
\verb+MultiNest+~\cite{Feroz:2007kg, Feroz:2008xx, Feroz:2013hea} and the 
\verb+GNU GSL+ simplex minimiser~\cite{gough2009gnu}.

For illustration we show in Figs.~\ref{fig:nsifit2_triangle-light} and
Fig.~\ref{fig:nsifit2_triangle-dark} all the possible one-dimensional and
two-dimensional projections of the eleven-dimensional parameter space
accounting for the new CP violating phases, parametrised as
$\phi_{\alpha\beta}^\oplus \equiv \arg(\varepsilon_{\alpha\beta}^\oplus)$. In both figures we show the regions
for the GLOBAL analysis including both T2K and \NOvA/ results. In Fig.~\ref{fig:nsifit2_triangle-light}
we present the results for the LIGHT sector and Normal Ordering, while
in Fig.~\ref{fig:nsifit2_triangle-dark} we give the regions corresponding to
the DARK sector and Inverted Ordering; in both cases the allowed
regions are defined with respect to the local minimum of each
solution. From these figures we can see that, with the exception of
the required large value of $\varepsilon_{ee}^\oplus - \varepsilon_{\mu\mu}^\oplus$
in the DARK solution, there is no statistically significant feature
for the $\varepsilon_{\alpha\beta}^\oplus$ parameters other than their
bounded absolute values, nor there is any meaningful information on
the $\phi_{\alpha\beta}$ phases.  The most prominent non-trivial
feature is the preference for a non-zero value of $\varepsilon_{e\mu}^\oplus$
at a $\Delta\chi^2 \sim 3$ level, associated with a $\phi_{e\mu}$
phase centered at the CP-conserving values $\pi$ ($0$) for the LIGHT
(DARK) solution.  More on this below.

\begin{pagefigure}\centering
  \includegraphics[width=0.99\textwidth]{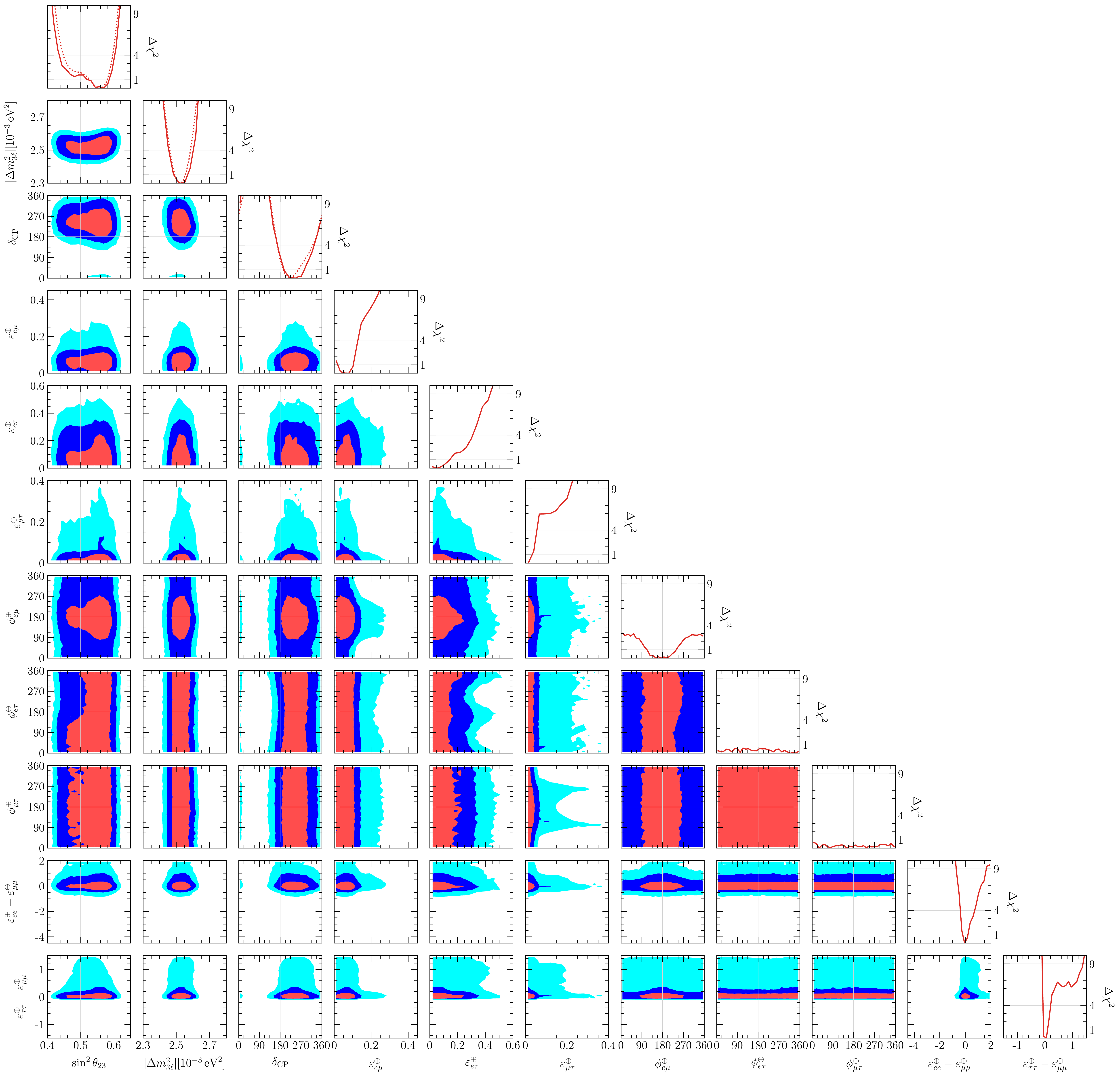}
  \caption{Global analysis of solar, atmospheric, reactor and
    accelerator oscillation experiments, in the LIGHT side of the
    parameter space and for Normal Ordering of the neutrino states.
    The panels show the two-dimensional projections of the allowed
    parameter space after marginalisation with respect to the
    undisplayed parameters.  The different contours correspond to the
    allowed regions at $1\sigma$, $2\sigma$ and $3\sigma$ for
    2~degrees of freedom.  Note that as atmospheric mass-squared
    splitting we use $\Delta m^2_{3\ell} = \Delta m^2_{31}$ for NO. Also shown are
    the one-dimensional projections as a function of each
    parameter. For comparison we show as dotted lines the
    corresponding one-dimensional dependence for the same analysis
    assuming only standard $3\nu$ oscillation (i.e., setting
    all the NSI parameters to zero).}
  \label{fig:nsifit2_triangle-light}
\end{pagefigure}

\begin{pagefigure}\centering
  \includegraphics[width=0.99\textwidth]{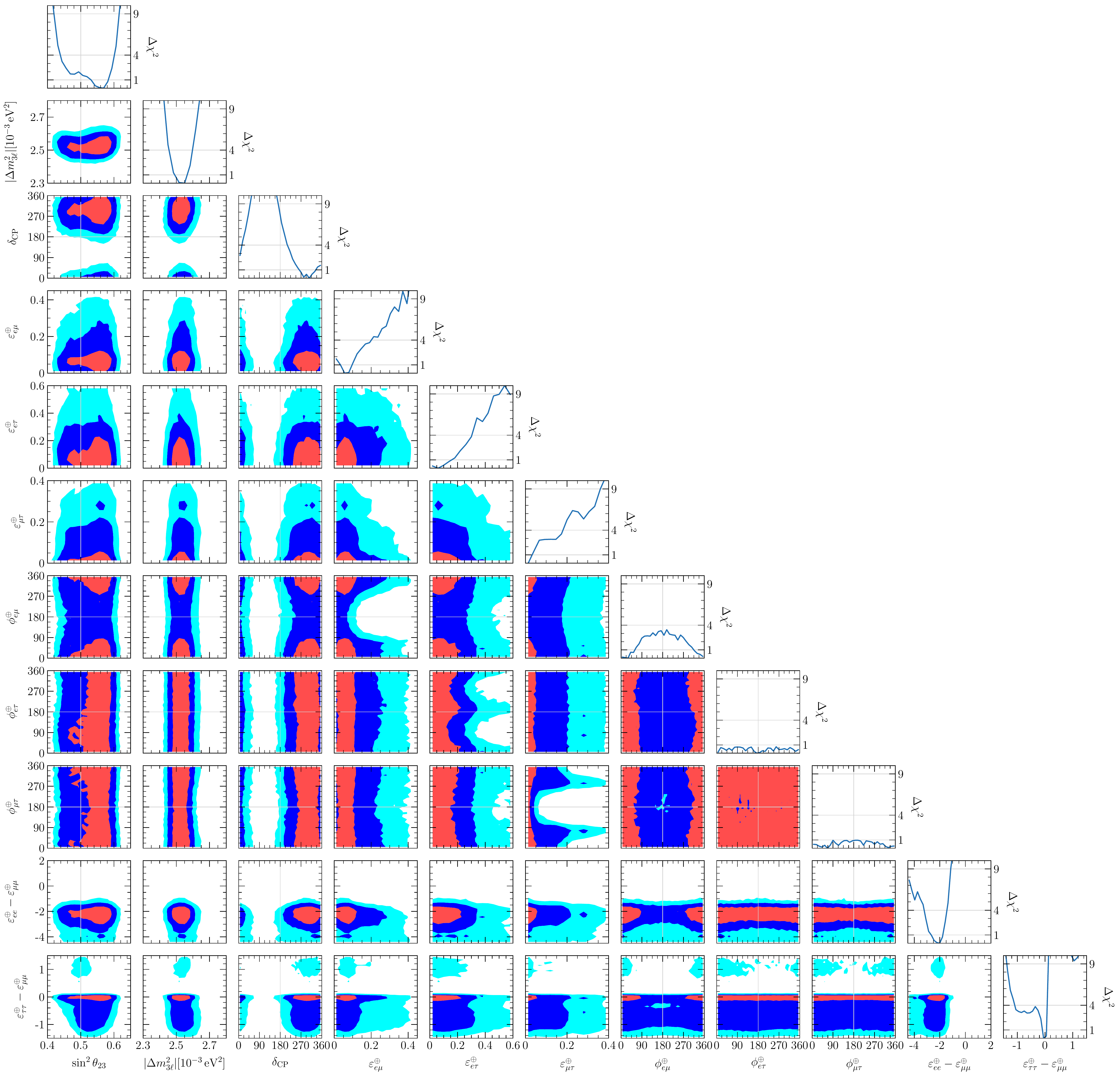}
  \caption{Same as Fig.~\ref{fig:nsifit2_triangle-dark} but for DARK-IO
    solution.  In this case $\Delta m^2_{3\ell} = \Delta m^2_{32}<0$ and we plot
    its absolute value. The regions and one-dimensional projections
    are defined with respect to the \emph{local} minimum in this
    sector of the parameter space.}
  \label{fig:nsifit2_triangle-dark}
\end{pagefigure}

\begin{figure}\centering
  \includegraphics[width=\textwidth]{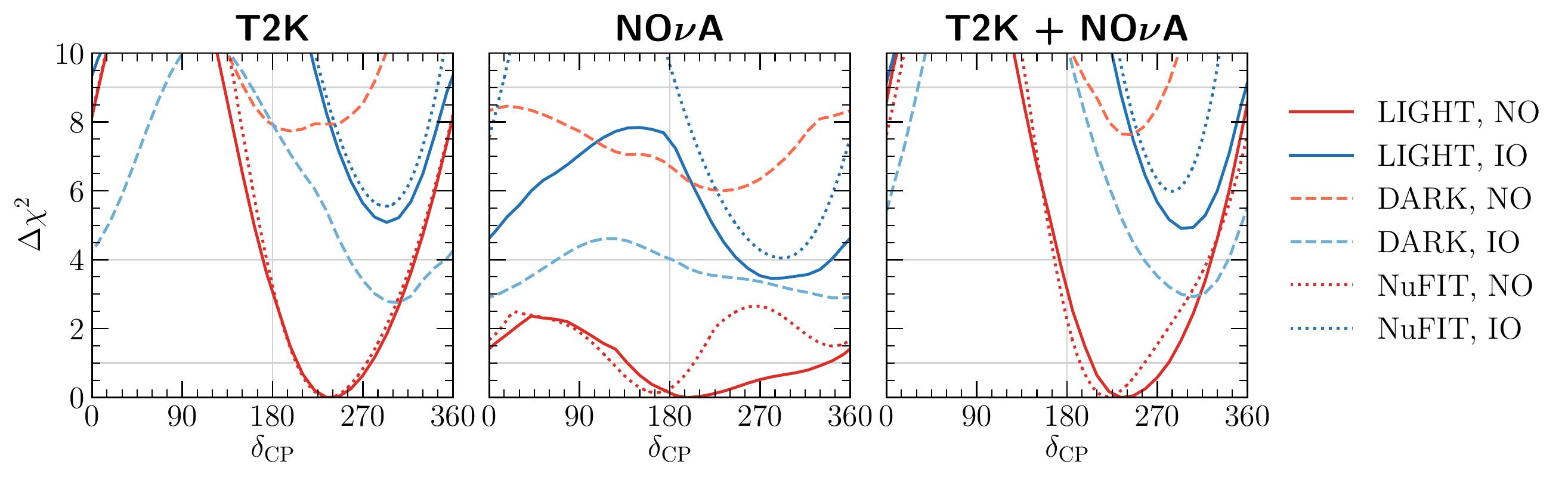}
  \caption{$\Delta\chi^2_\text{GLOB}$ as a function of
    $\delta_\text{CP}$ after marginalising over all the undisplayed
    parameters, for different combination of experiments. We include $\text{SOLAR} + \text{KamLAND} + \text{ATM} + \text{MBL-REA}
    + \text{MINOS}$ to which we add T2K (left), \NOvA/ (center) and
    $\text{T2K} + \text{NO}\nu\text{A}$ (right). The different curves
    are obtained by marginalising within different regions of the
    parameter space, as detailed in the legend. See text for details.}
  \label{fig:nsifit2_chi2dcpnsi}
\end{figure}

In order to quantify the effect of the matter NSI on the present
determination of $\delta_\text{CP}$ and the mass ordering we plot in
Fig.~\ref{fig:nsifit2_chi2dcpnsi} the one-dimensional
$\chi^2(\delta_\text{CP})$ function obtained from the above
$\chi^2_\text{GLOB,REG}$ after marginalising over the ten undisplayed
parameters. In the left, central and right panels we focus on the
GLOBAL analysis including T2K, \NOvA/, and T2K+\NOvA/ respectively.   
In each panel we plot the
curves obtained marginalising separately in NO (red curves) and IO
(blue curves) and within the $\text{REG} = \text{LIGHT}$ (full lines)
and $\text{REG} = \text{DARK}$ (dashed) regions. For the sake of
comparison we also plot the corresponding $\chi^2(\delta_\text{CP})$
from the $3\nu$ oscillation analysis with the SM matter potential
(labeled ``NuFIT'' in the figure).

For what concerns the analysis which includes T2K but not \NOvA/,
i.e., the left panels in Fig.~\ref{fig:nsifit2_chi2dcpnsi}, we find
that:
\begin{itemize}
\item The statistical significance of the hint for a non-zero
  $\delta_\text{CP}$ in T2K is robust under the inclusion of the
  NSI-induced matter potential for the most favoured solution
  (i.e., LIGHT-NO), as well as for LIGHT-IO.

\item The $\Delta\chi^2$ for DARK solutions exhibits the expected
  inversion of the ordering as well as the $\delta_\text{CP} \to \pi -
  \delta_\text{CP}$ transformation when compared with the LIGHT
  ones. This is a consequence of the NSI-mass-ordering degeneracy
  discussed in \cref{sec:LMAD_theor}.

\item We notice that $\Delta\chi^2_\text{min,OTHM+T2K,DARK,IO} \neq
  \Delta\chi^2_\text{min,OTHM+T2K,LIGHT,NO}$ because of the breaking
  of the NSI-mass-ordering degeneracy in the analysis of solar
  experiments as a consequence of the sizeable variation of the
  chemical composition of the matter crossed by solar neutrinos along
  their path. However as we see in the left panel
  \begin{equation}
    \chi^2_\text{min,OTHM+T2K,DARK,IO}
    - \chi^2_\text{min,OTHM+T2K,LIGHT,NO} \simeq 2.75 < 3.15
  \end{equation}
  so the DARK solutions become less disfavoured when T2K is included
  (see \cref{eq:dclmad}).  This suggests that the DARK-IO
  solution can provide a perfect fit to T2K data, i.e., there
  is an almost total loss of sensitivity to the ordering in T2K.
  
  Indeed what the inequality above shows is that within the allowed
  DARK parameter space it is possible to find areas where the fit to
  T2K-only data for IO are slightly better than the fit for NO in the
  LIGHT sector (and than NO oscillations without NSI).  
  For example, in
  DARK-NO we find that the best fit value for $\Delta m^2_{31}$ can be
  slightly larger than the best-fit $|\Delta m^2_{32}|$ in LIGHT-IO, which
  leads to a slightly better agreement with the results on $\Delta m^2_{31}$
  from medium baseline reactors.  These solutions, however, involve large NSI
  parameters, in particular $\varepsilon_{e\mu}$ and $\varepsilon_{e\tau}$.

\item For the same reason, the statistical
  significance of the hint of CP violation in T2K is reduced for the
  DARK solutions with respect to the LIGHT ones. We find that CP
  conservation (CPC), that is, a fit with all phases either zero or
s  $\pi$, lies at
  \begin{equation}
    \begin{aligned}
      \chi^2_\text{CPC,OTHM+T2K,DARK,IO}
      - \chi^2_\text{min,OTHM+T2K,DARK,IO} &\simeq 1.5 
      \\
      \chi^2_\text{CPC,OTHM+T2K,LIGHT,NO}
      - \chi^2_\text{min,OTHM+T2K,LIGHT,NO} &\simeq 3.2
    \end{aligned}
    \Bigg\rbrace \Rightarrow 1.5 < 3.2 \,.
  \end{equation}
  
\item However we still find that
  \begin{equation}
    \begin{aligned}
      \chi^2_\text{CPC,OTHM+T2K,LIGHT,NO} &\simeq
      \chi^2_\text{OTHM+T2K,LIGHT,NO}(\delta_\text{CP}=\pi) \,,
      \\
      \chi^2_\text{CPC,OTHM+T2K,DARK,IO} &\simeq
      \chi^2_\text{OTHM+T2K,DARK,IO}(\delta_\text{CP}=0) \,.
    \end{aligned}
  \end{equation}
  So the CL for CPC as naively read from the curves of
  $\delta_\text{CP}$ still holds, or what is the same, there is no
  leptonic CP violation ``hidden'' when there is no CP violation from
  $\delta_\text{CP}$.
\end{itemize}

For the global combination including \NOvA/ without T2K (central panels
in Fig.~\ref{fig:nsifit2_chi2dcpnsi}) we notice that:
\begin{itemize}
\item The sensitivity to $\delta_\text{CP}$ diminishes with respect to
  that of the oscillation only analysis both in the LIGHT and DARK
  sectors.

\item Within the DARK sector, IO is the best solution as expected from
  the NSI/mass-ordering degeneracy, but it is still disfavoured at
  $\Delta\chi^2_\text{min,OTHM+NOvA,DARK,IO}\sim 3$ because of
  SOLAR + KamLAND, Eq.~\eqref{eq:dclmad}. By chance, this happens to be
  of the same order of the disfavouring of IO in the pure oscillation
  analysis, $\Delta\chi^2_\text{min,OTHM+NOvA,OSC}\sim 3.5$ (although
  the physical effect responsible for this is totally different).
\end{itemize}

In the global analysis including both T2K and \NOvA/ (right panels in
Fig.~\ref{fig:nsifit2_chi2dcpnsi}) we find qualitatively similar conclusions
than for the analysis without \NOvA/, albeit with a slight washout of
the statistical significance for $\delta_\text{CP}$ due to the tensions between T2K and \NOvA/ discussed in \cref{sec:dCPevolution}. Such
washout is already present in the oscillation-only analysis and within
the LIGHT sector it is only mildly affected by the inclusion of
NSI. For the same reason, the significance for the disfavouring of IO is also reduced by $\sim 0.2 \sigma$, although this requires relatively large values of $\varepsilon_{e\tau}$.

However one also observes that in the favoured solution, LIGHT-NO,
maximal $\delta_\text{CP} = 3\pi/2$ is more allowed than without
NSI. This happens because, as mentioned above, in \NOvA/ the presence of
NSI induces a loss of sensitivity on $\delta_\text{CP}$, so in the
global analysis with both T2K and \NOvA/ the behaviour observed in T2K
dominates.

In the global analysis there remains, still, the DARK-IO solution at
\begin{equation}
  \chi^2_\text{min,GLOB,DARK,IO}
  - \chi^2_\text{min,GLOB,LIGHT,NO} = 3 \, .
\end{equation}

\begin{figure}\centering
  \includegraphics[width=0.75\textwidth]{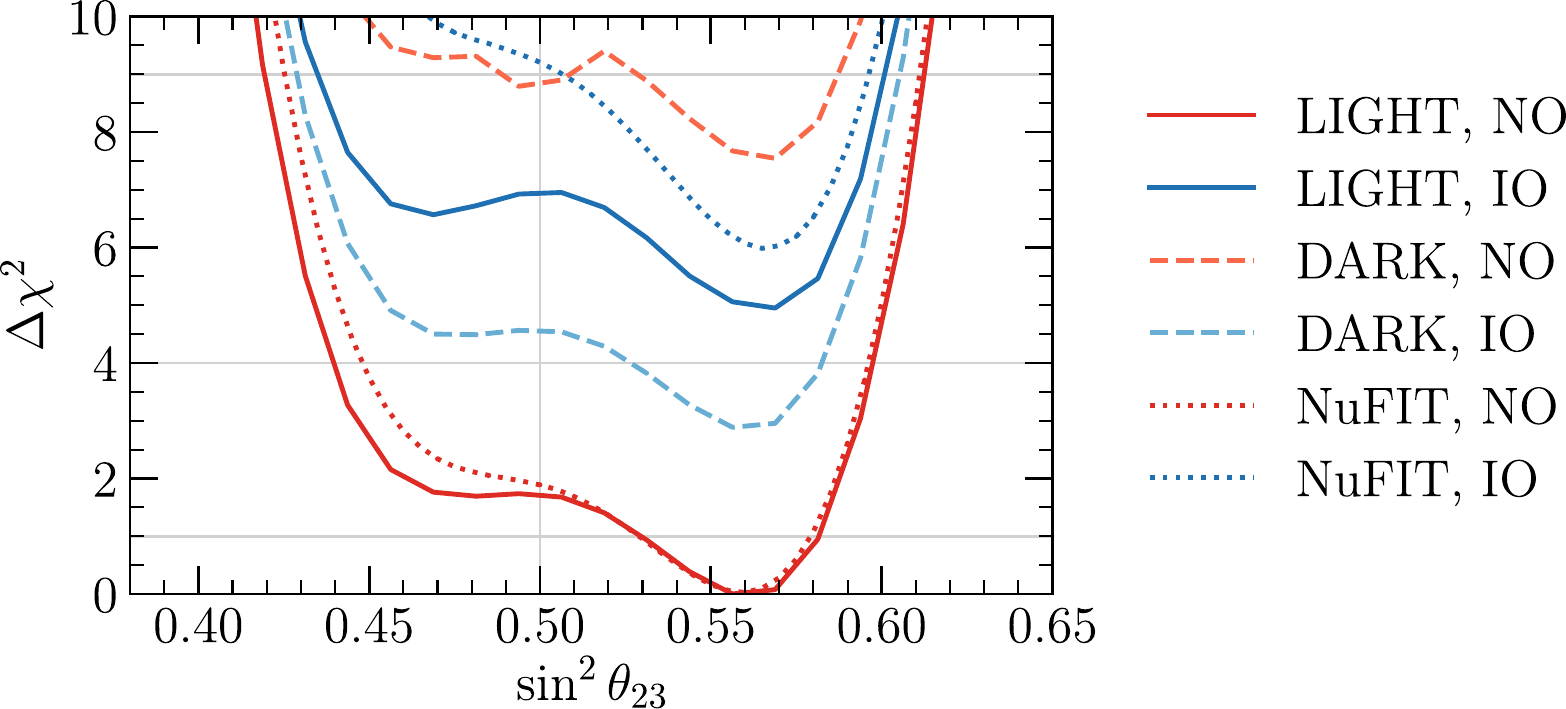}
  \caption{$\Delta\chi^2_\text{GLOB}$ as a function of
    $\sin^2\theta_{23}$ after marginalising over all other parameters
    for the GLOBAL combination of oscillation experiments.  The different curves
    correspond to marginalisation within the different regions of the
    parameter space, as detailed in the legend. See text for details.}
  \label{fig:nsifit2_t23}
\end{figure}

The status of the non-maximality and octant determination for
$\theta_{23}$ is displayed in Fig.~\ref{fig:nsifit2_t23} where we show the
one-dimensional $\chi^2(\sin^2\theta_{23})$ obtained from
$\chi^2_\text{GLOB,REG}$ including both T2K and \NOvA/ and after
marginalising over all the undisplayed parameters (so these are the
corresponding projections to the left panels in
Fig.~\ref{fig:nsifit2_chi2dcpnsi}).
As
seen in the figure the global analysis including NSI for both
orderings, for both LIGHT and DARK sectors, still disfavours the maximal $\theta_{23}=\pi/4$
at a CL $\sim 1.5\sigma$.  The main effect of the
generalised NSI matter potential on $\theta_{23}$ is the mild
improvement of the CL for the first octant.
This is so because the
disfavouring of the first octant in the global oscillation-only
analysis is driven by the excess of appearance events in \NOvA/.  These
events can now be fitted better with $\theta_{23}$ in the first octant
when including a non-zero $\varepsilon_{e\mu}$ to enhance the $\nu_\mu \to
\nu_e$ flavour transition probability.

\subsection{Summary}
\label{sec:nsifit2_summary}

In this section we have extended the analysis in
\cref{sec:nsifit1} to account for the effect of NSI affecting
neutrino propagation in matter on the observables sensitive to
leptonic CP violation and to the mass ordering. We have quantified the
robustness of the present hints for these effects in the presence of
NSI as large as allowed by the global oscillation analysis itself. We
conclude that the CL for the preference for a CKM-like CP phase close
to $3\pi/2$ in T2K, which is the one that drives the preference in the
global analysis, remains valid even when including all other phases in
the extended scenario. On the contrary the preference for NO in LBL
experiments is totally lost when including NSI as large as allowed
by the global analysis because of the intrinsic NSI/mass-ordering
degeneracy in the Hamiltonian which implies the existence of an
equally good fit to LBL results with IO and reversed octant of
$\theta_{12}$ and $\delta_\text{CP}\rightarrow \pi - \delta_\text{CP}$
(so in this solution the favoured $\delta_{CP}$ is also close to
$3\pi/2$).  In the global analysis the only relevant breaking of this
degeneracy comes from the composition dependence of the matter
potential in the Sun which disfavours the associated LMA-D with CL
below $2\sigma$. Finally, we have also studied the effect of NSI in
the status of the non-maximality and octant determination for
$\theta_{23}$ and find that for both
orderings, for both LIGHT and DARK sectors, maximal $\theta_{23}=\pi/4$ is still disfavoured 
in the global fit at a CL $\sim 1.5\sigma$.

\section{Summary and conclusions}

In this chapter, we have performed a fit to neutrino oscillation data
assuming there is BSM physics other than 
neutrino masses. In particular, we have introduced NSI-NC, very 
difficult to bound with other experiments but crucial for neutrino 
propagation in matter. 

We have first obtained bounds on the moduli of
the new parameters, i.e., their CP conserving part. We have found that 
individual experiments allow very large NSI, particularly when they 
are adjusted to be suppressed for the particular matter composition 
traversed by the neutrino beam. Consequently, the bounds on the 
oscillation parameters get weakened. However, different experiments are sensitive to different matter 
profiles, energy ranges and oscillation channels. When combining all 
data, we find that the oscillation parameters are robustly determined 
(except slightly for $\theta_{12}$) and that $\mathcal{O}(1)$ NSI are 
disfavoured. Having observed neutrino oscillations in a large variety 
of environments is crucial for this.

Experimentally allowed NSI, though, could still spoil the sensitivity of LBL accelerator 
experiments to CP violation. To check whether this is the case, we have performed a 
generic fit to all experiments where CP-violating NSI are also 
allowed. We have found that, within the LMA solution,
 the results from the T2K experiment are 
quite robust: it has a relatively short baseline, and so NSI strong 
enough to affect it would have been detected by other 
experiments more sensitive to matter effects. Nevertheless, the 
\NOvA/ baseline is larger, and NSI within experimental bounds can 
significantly affect the interpretation of its data. Since the 
information on $\delta_\text{CP}$ is dominated by T2K, the global 
result within the LMA solution is robust, and the hint for large CP
violation driven by the Jarlskog invariant \eqref{eq:Jvac} persists.

Nevertheless, the generalised mass ordering degeneracy discussed in 
\cref{sec:LMAD_theor} implies a good fit at $\sim 2 \sigma$ that 
spoils the global sensitivity to the mass ordering. Furthermore, it 
also has the potential of worsening the sensitivity to 
$\delta_\text{CP}$.

These results are particularly worrisome in the context of future 
experiments. The Deep Underground Neutrino Experiment, for instance, is 
expected to
determine CP violation with a very large statistical significance. This 
experiment is dominated by matter effects, and so it is 
potentially affected by the same sensitivity loss as \NOvA/ (see
\cref{fig:nsifit2_t23}). The advent
of more neutrino oscillation experiments will not alleviate the 
situation, as part of the sensitivity loss comes from the generalised 
mass ordering degeneracy, exact for neutrino oscillations.

The only hope for lifting this degeneracy is bounding NSI with non-oscillation 
experiments. Traditional neutrino 
scattering experiments have large momentum transfers, and so bounds 
from them could be evaded if the mediator inducing NSI is light. 
Fortunately, in the recent years coherent neutrino-nucleus elastic 
scattering has been detected~\cite{Akimov:2017ade}. In this process,  
neutrinos interact coherently with an entire atomic nucleus by 
exchanging very low momenta, $\mathcal{O}(\SI{10}{MeV})$. Therefore,  
light mediator-induced NSI could be bounded with these experiments. 
This will be the subject of the next chapter.

\chapter{COHERENT constraint of beyond three-neutrino scenarios}
\label{chap:coh}

\epigraph{\emph{Our suggestion may be an act of hubris, because the
    inevitable constraints on interaction rate, resolution, and
    background pose grave experimental difficulties for elastic
    neutrino-nucleus scattering.}}{ --- Daniel Z.~Freedman (1974)}

\epigraph{\emph{We shall not cease from exploration. And the end of
    all our exploring will be to arrive where we started and know the
    place for the first time.}}{ --- T.~S.~Eliot}

As explored in the previous chapters, there could be new physics in
the form of NSI-NC affecting neutrino propagation in
matter. Under its presence, the measurement of leptonic CP violation
can get compromised. This is not only because of the additional
parameters in the model, but also because NSI introduce a degeneracy
exact at the oscillation probability level.

Therefore, robustly determining leptonic CP violation in present and
next generation experiments would highly benefit from independent
constraints on NSI-NC. In principle, these could come
from neutral current neutrino scattering experiments such as
CHARM~\cite{Dorenbosch:1986tb} and NuTeV~\cite{Zeller:2001hh}. These
experiments, however, probe typical momentum exchanges 
$\mathcal{O}(\SI{10}{GeV})$, whereas neutrino matter effects are a
coherent, zero momentum exchange process.

More explicitly, if NSI are mediated by a particle with mass $M_X$
and coupling $g$, the coupling times propagator entering neutrino
scattering will generically be
\begin{equation}
 \sim \frac{g^2}{q^2 - M_X^2} \, ,
\end{equation}
with $q$ the momentum transfer. Thus, neutrino matter effects are
sensitive to $g^2/M_X^2$, whereas for light mediators ($M_X \ll q$)
scattering experiments can only bound $g^2/q^2$. If the mediator is
light enough, scattering bounds can consequently be
avoided~\cite{Farzan:2015doa, Farzan:2015hkd, Babu:2017olk,
  Farzan:2017xzy, Denton:2018xmq, Miranda:2015dra,
  Heeck:2018nzc}. 
Thus, bounding NSI that
could spoil neutrino oscillation experiments requires a neutrino
scattering experiment with very low momentum transfers.

\begin{figure}[hbtp]
\centering \includegraphics[width=0.5\textwidth]{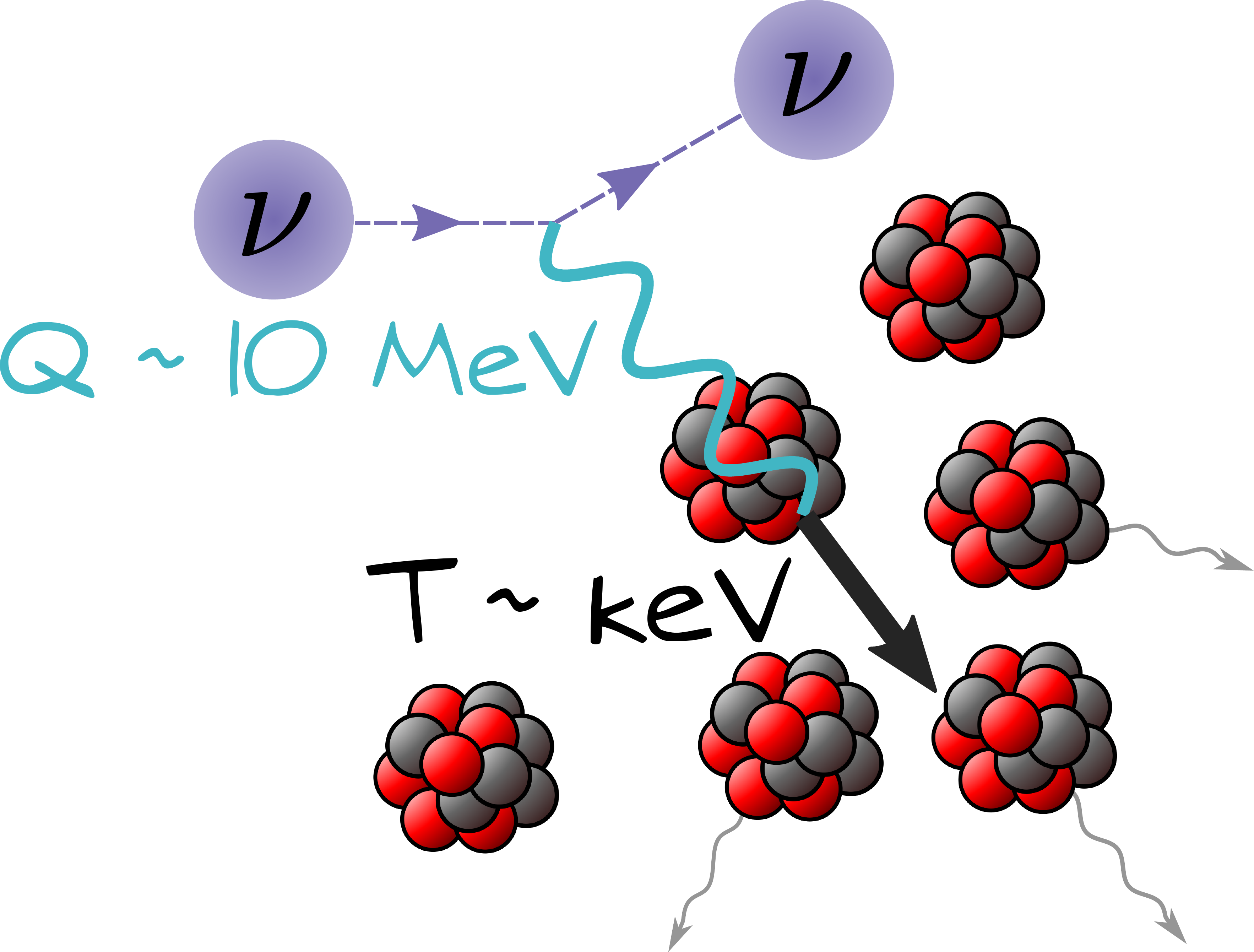}
\caption{Schematic representation of coherent neutrino-nucleus elastic
  scattering. Here, Q is the momentum transfer, and T the recoil
  energy of the hit nucleus.}
\label{fig:CEvNS}
\end{figure}

Fortunately, such an experiment exists and has released data during
the completion of this thesis. The COHERENT
experiment~\cite{Akimov:2015nza}, which will be further explained
below, measures coherent elastic neutrino-nucleus scattering
(CE$\nu$NS), a process in which a neutrino interacts coherently with
an entire atomic nucleus, with a very low momentum exchange that
enables exploring NSI mediated by light particles. Due to coherence, scattering
amplitudes get enhanced by
roughly the number of neutrons in each target nucleus,\footnote{Neutral current
  interactions with protons are suppressed by {$1-4 \sin^2 \theta_W
  \simeq 0.05$}~\cite{Tanabashi:2018oca} see \cref{eq:CCandNC}.} $N$,
and cross sections thus increase
by a factor $N^2$.  Despite the large cross sections, coherent
interaction with an entire nucleus only happens when the de Broglie
wavelength of the exchanged mediator is of the order of the nuclear
size, $\mathcal{O}(\si{MeV^{-1}})$. The only signal that the
interaction leaves is thus a nucleus recoiling with tiny energies,
$\mathcal{O}(\si{keV})$, very challenging to detect. The process is
depicted in \cref{fig:CEvNS}.

Nevertheless, modern low-threshold detectors are sensitive to such
tiny recoils and in 2017 this process was
detected~\cite{Akimov:2017ade}. In this chapter, we will develop a
comprehensive analysis of data from the COHERENT experiment in the
framework of NSI-NC. We
will combine the results with the oscillation analyses from the previous
chapter, to assess the importance of coherent scattering experiments
in interpreting neutrino oscillation data.

Finally, we will also explore the prospects of a near future facility,
the European Spallation Source, that can accumulate data one order of
magnitude faster than the COHERENT experiment.

\section{Analysis of COHERENT data}
\label{sec:coh_fit}

\subsection{Coherent elastic neutrino-nucleus scattering}
\label{sec:coh_coh1}

The COHERENT experiment uses neutrinos produced at the Spallation
Neutron Source (SNS) sited at the Oak Ridge National
Laboratory. There, an abundant flux of both $\pi^+$ and $\pi^-$ is
produced in proton-nucleus collisions in a mercury target. While the
$\pi^-$ are absorbed by nuclei before they can decay, the $\pi^+$ lose
energy as they propagate and eventually decay at rest into $\pi^+ \to
\mu^+ \nu_\mu$, followed by $\mu^+ \to e^+ \nu_e \bar\nu_\mu$. Since
the muon lifetime is much longer than that of the pion, the $\nu_\mu$
component is usually referred to as the prompt contribution to the
flux, as opposed to the delayed contributions from $\mu^+$ decay
($\bar\nu_\mu$ and $\nu_e$).

\begin{figure}[hbtp]
\centering
\includegraphics[width=0.5\textwidth]{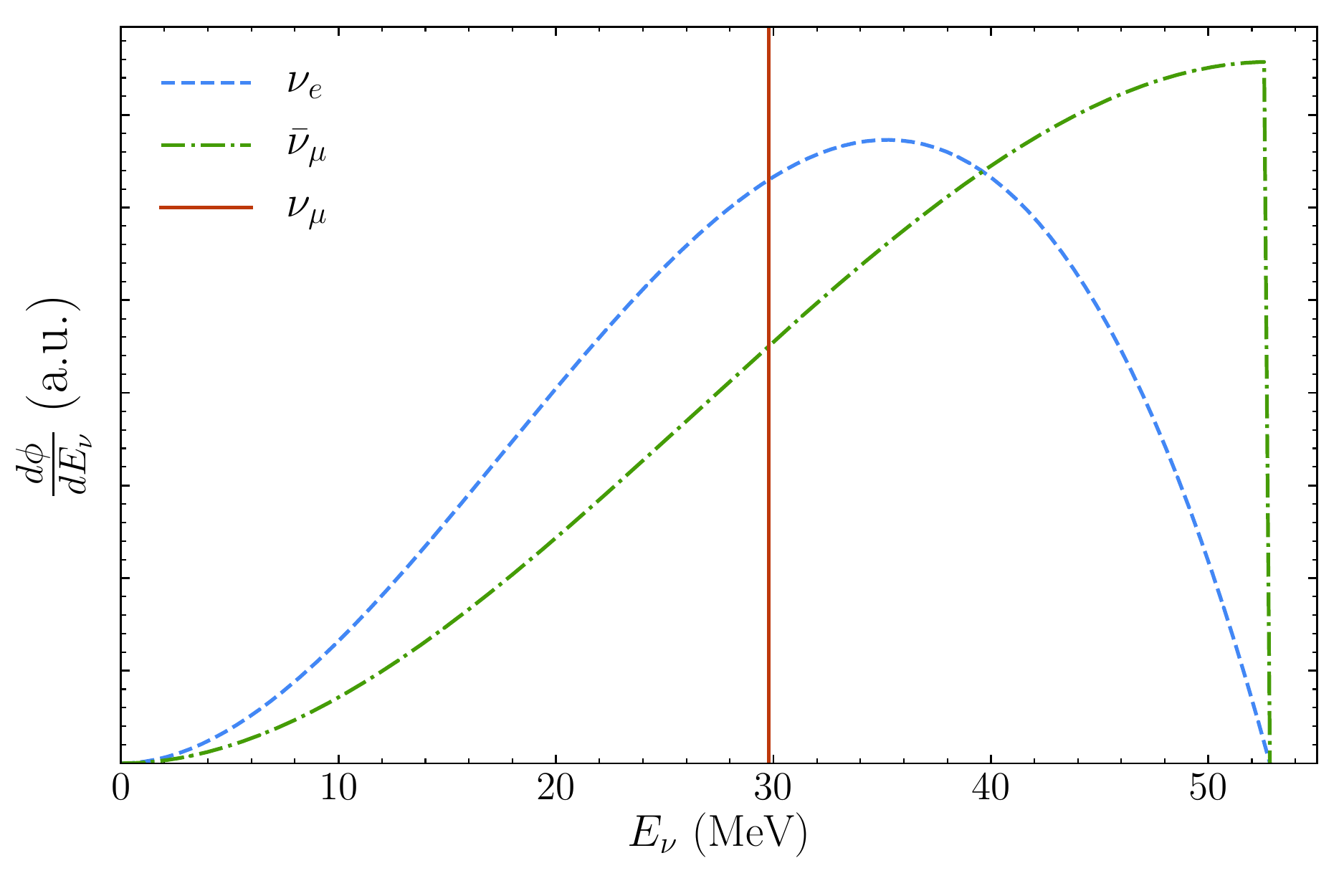}
\caption{Neutrino flux spectra expected from $\pi^+$ decay at rest, in arbitrary units (a.u.), as a function of the neutrino energy in MeV. The three components of the flux are shown separately as indicated by the legend. The distributions have been normalised to one.}
\label{fig:coh_flux}
\end{figure}

Given that the prompt neutrinos are a by-product of two-body decays at
rest, their contribution to the total flux is a monochromatic line at
$E_\text{pr} = (m_\pi^2 - m_\mu^2)/(2 m_\pi) \simeq 29.7$~MeV, where
$m_\pi$ and $m_\mu$ refer to the pion and muon masses,
respectively. Conversely, the delayed neutrino fluxes follow a
continuous spectra at energies $E_{\nu_e, \bar\nu_\mu} < m_\mu/2\simeq
52.8$~MeV. At a distance $\ell$ from the source, they read:
\begin{equation}
  \label{eq:coh_COHflux}
  \begin{aligned}
    \frac{\mathrm{d}\phi_{\nu_\mu}}{\mathrm{d}E_\nu} &= \frac{1}{4\pi \ell^2 }
    \delta(E_\nu - E_\text{pr}) \,,
    \\ \frac{\mathrm{d}\phi_{\bar\nu_\mu}}{\mathrm{d}E_\nu} &= \frac{1}{4\pi \ell^2}
    \frac{64}{m_\mu} \left[ \left( \frac{E_\nu}{m_\mu} \right)^2
      \left( \frac{3}{4} - \frac{E_\nu}{m_\mu} \right) \right],
    \\ \frac{\mathrm{d}\phi_{\nu_e}}{\mathrm{d}E_\nu} &= \frac{1}{4\pi \ell^2}
    \frac{192}{m_\mu} \left[ \left( \frac{E_\nu}{m_\mu} \right)^2
      \left( \frac{1}{2} - \frac{E_\nu}{m_\mu} \right) \right],
  \end{aligned}
\end{equation}
and are normalised to each proton collision on the target. For
reference the distance $\ell$ at COHERENT is 19.3~m. These spectra are 
shown in \cref{fig:coh_flux}.

The differential cross section for coherent elastic neutrino-nucleus
scattering, for a neutrino with incident energy $E_\nu$ interacting
with a nucleus with $Z$ protons and $N$ neutrons,
reads~\cite{Freedman:1973yd}:
\begin{equation}
  \label{eq:coh_xsec-SM}
  \frac{\mathrm{d}\sigma_\text{SM} (T, E_\nu)}{\mathrm{d}T} = \frac{G_F^2}{ 2 \pi}
  \mathcal{Q}^2 (Z, N) F^2(Q^2) M \left(2 - \frac{M T}{E_\nu^2}
  \right)
\end{equation}
where $G_F$ is the Fermi constant and $\mathcal{Q}^2$ is the weak
charge of the nucleus.  In this notation, $T$ is the recoil energy of
the nucleus, $M$ is its mass, and $F$ is its nuclear form factor
evaluated at the squared momentum transfer of the process, $Q^2 = 2 M
T$.  In our calculations we have first used a Helm form factor\footnote{The
  collaboration used a slightly different form factor, taken from
  Ref.~\cite{Klein:1999qj} However, we have checked that the results
  of the fit using their parametrisation gives identical results to
  those obtained using the Helm form factor.}
parametrisation~\cite{Helm:1956zz}:
\begin{equation}
  \label{eq:coh_Helm}
  F(Q^2) = 3 \frac{j_1(Q R_0)} {Q R_0} e^{-Q^2 s^2 / 2}
\end{equation}
where $s = 0.9$~fm~\cite{Lewin:1995rx} and $j_1(x)$ is the order-1
spherical Bessel function of the first kind. The value of $R_0$
relates to the value of $s$ and the neutron radius $R_n$ as
\begin{equation}
  \label{eq:coh_Rn}
  \frac{R_0^2}{5} = \frac{ R_n^2 }{3} - s^2 \,.
\end{equation}
In the absence of an experimental measurement of the neutron radius in
CsI, we tune its value so that the prediction for the total number of
events at COHERENT matches the official one provided in
Refs.~\cite{Akimov:2018vzs, Akimov:2017ade} (173 events). However, by
doing so we obtain $R_n = 4.83$~fm, a value that is unphysical as it
approaches the proton radius~\cite{Fricke:1995zz} which all models
predict to be smaller.  Given that this is a phenomenological
parametrisation, though, and seeing the large differences in the
prediction for the total number of events obtained with different
values of $R_n$, it is worth asking whether this is accurate enough
for CsI, and exploring the impact of the nuclear form factor on the results of
the fit. Therefore, in Sec.~\ref{sec:coh_results} we will also show
the results obtained using a state-of-the-art theoretical calculation
for the nuclear form factor, taken from Refs.~\cite{menendez, Klos:2013rwa}
(calculated using the same methodology as in
Refs.~\cite{Hoferichter:2016nvd, Hoferichter:2018acd}).

In the SM, the weak charge of a nucleus only depends on the SM vector
couplings to protons ($g_p^V$) and neutrons ($g_n^V$) and is
independent of the neutrino flavour (see \cref{eq:CCandNC}):
\begin{equation}
  \mathcal{Q}^2 \equiv \big( Z g_p^V + N g_n^V \big)^2 \,,
\end{equation}
where $g_p^V = 1/2 - 2\sin^2\theta_W$ and $g_n^V = -1/2$, with
$\theta_W$ being the weak mixing angle.  For CsI, we obtain
$\mathcal{Q}^2 \simeq 1352.5$ in the SM.
However, in presence of NSI-NC, this effective charge gets
modified\footnote{In practice, unless the ratio of the new couplings
  to up and down quarks remains the same as in the SM, the form factor of the
  nucleus would also be affected by the NP and should be recomputed
  including the NP terms. However, in the case of vector-vector
  interactions (as in the case of NSI) the modifications to the
  nuclear form factor are expected to be subleading, and the factorisation of
  the NP effects into the weak charge approximately holds.  We warmly
  thank Martin Hoferichter for pointing this out.}  by the new
operators introduced as~\cite{Barranco:2005yy}:
\begin{multline}
  \label{eq:coh_Qalpha-nsi}
  \mathcal{Q}^2_\alpha(\vec\varepsilon) = \left[ Z \big(g_p^V +
    2\varepsilon_{\alpha\alpha}^u + \varepsilon_{\alpha \alpha}^d
    \big) + N \big( g_n^V + \varepsilon_{\alpha\alpha}^u +
    2\varepsilon_{\alpha \alpha}^d \big) \right]^2 \\ + \sum_{\beta
    \neq \alpha} \left[ Z \big( 2\varepsilon_{\alpha \beta}^u +
    \varepsilon_{\alpha \beta}^d) + N \big(
    \varepsilon_{\alpha\beta}^u + 2\varepsilon_{\alpha\beta}^d \big)
    \right]^2 ,
\end{multline}
and in general its value may now depend on the NSI parameters
$\vec\varepsilon \equiv \{ \varepsilon_{\alpha\beta}^f \}$ as well as
the incident neutrino flavour $\alpha$. Since the COHERENT experiment
observes interactions of both electron and muon neutrinos, its results
are sensitive to both $\mathcal{Q}^2_e$ and $\mathcal{Q}^2_\mu$.

As can be seen from Eq.~\eqref{eq:coh_Qalpha-nsi}, the modification of
NSI to the CE$\nu$NS event rate comes in as a normalisation
effect. Therefore, adding nuclear recoil energy information to the
analysis of the data is not expected to have a significant effect on
the results of our fit to NSI.\footnote{In principle, a subleading
  effect can be observed for experiments with large statistics, due to
  the different maximum recoil energies expected for the prompt and
  delayed neutrino components of the beam as will be explored in
  \cref{sec:ESS}. However, we find that the COHERENT experiment is
  insensitive to this effect with the current exposure.} Conversely,
the addition of timing information is crucial as it translates into a
partial discrimination between neutrino flavours, thanks to the
distinct composition of the prompt ($\nu_\mu$) and delayed
($\bar\nu_\mu$ and $\nu_e$) neutrino flux. This translates into an
enhanced sensitivity to NSI, since the fit will now be sensitive to a
change in normalisation affecting neutrino flavours differently.

\subsection{Implementation of the COHERENT experiment}
\label{sec:coh_coh2}

The COHERENT collaboration has released publicly both the energy and
timing information of the events~\cite{Akimov:2018vzs}. In this
section we describe the procedure used to implement in our fit the
information provided in such data release.

\subsubsection{Computation of the signal}

The differential event distribution at COHERENT, as a function of the
nuclear recoil energy $T$, reads
\begin{equation}
  \label{eq:coh_dNdT}
  \frac{\mathrm{d}N}{\mathrm{d}T} = N_\text{pot} f_{\nu/p} N_\text{nuclei} \sum_\alpha
  \int_{E_\nu^\text{min}}^{m_\mu/2} \frac{\mathrm{d}\sigma_\alpha}{\mathrm{d}T}
  \frac{\mathrm{d}\phi_{\nu_\alpha}}{\mathrm{d}E_\nu} \, \mathrm{d}E_\nu \,,
\end{equation}
where $N_\text{nuclei}$ is the total number of nuclei in the detector,
$N_\text{pot} = 1.76 \cdot 10^{23}$ is the total number of protons on
target considered, and $f_{\nu/p} = 0.08$ is the neutrino yield per
proton.  In Eq.~\eqref{eq:coh_dNdT} the sum runs over all neutrino
flux components ($\nu_e$, $\nu_\mu$, $\bar\nu_\mu$), and the upper
limit of the integral is given by the end-point of the spectrum from
pion DAR, while the minimum neutrino energy that can lead to an event
with a nuclear recoil energy $T$ is given by
\begin{equation}
  E_\nu^\text{min} = \sqrt{\frac{M T}{2}} \,.
  \label{eq:coh_T}
\end{equation}

At COHERENT, the observable that is actually measured is the number of
photo-electrons (PE) produced by an event with a certain nuclear
recoil. In fact, the nuclear recoil energy in CE$\nu$NS events is
typically dissipated through a combination of scintillation (that is,
ionisation) and secondary nuclear recoils (that is, heat). While
secondary recoils are \emph{the} characteristic signal of a nuclear
recoil (as opposed to an electron recoil, which favours ionisation
instead), their measurable signal is much smaller than that of
electron recoils. The ratio between the light yields from a nuclear
and an electron recoil of the same energy is referred to as the
Quenching Factor (QF).

Besides being a detector-dependent property, the QF may also depend
non-trivially on the recoil energy of the nucleus. In general, the
relation between PE and nuclear recoil $T$ can be expressed as:
\begin{equation}
  \label{eq:coh_QF}
  \text{PE} = T \cdot \text{LY} \cdot \text{QF}(T) \,,
\end{equation}
where LY is the light yield of the detector (that is, the number of PE
produced by an electron recoil of one keV), and we have explicitly
noted that the QF may depend on the nuclear recoil energy. Therefore,
the expected number of events in a certain bin $i$ in PE space can be
computed as:
\begin{equation}
  \label{eq:coh_Nevents}
  N_i = \int_{T(\text{PE}_i^\text{min})}^{T(\text{PE}_i^\text{max})}
  \frac{\mathrm{d}N}{\mathrm{d}T} \, \mathrm{d}T \,,
\end{equation}
where the limits of the integral correspond to the values of $T$
obtained for the edges of the PE bin ($\text{PE}_i^\text{min}$,
$\text{PE}_i^\text{max}$) from Eq.~\eqref{eq:coh_QF}.

In their analysis, the COHERENT collaboration adopted an
energy-independent QF throughout the whole energy range considered in
the analysis, between 5 and 30~keV~\cite{Akimov:2017ade}. Also, given
the tension observed between the different calibration measurements
available at the time, they assigned large error bars to the assumed
central value $\overline{\text{QF}} = 8.78\%$. Taking a central value
for the light yield $\overline{\text{LY}} = 13.348$~PE per keV of
electron recoil~\cite{Akimov:2018vzs}, this means that approximately
1.17~PE are expected per keV of nuclear recoil energy. Very recently,
however, the authors of Ref.~\cite{Collar:2019ihs} have re-analysed
past calibration data used to derive this result. They concluded that
the tension between previous measurements was partially due to an
unexpected saturation of the photo-multipliers used in the calibration
and, after correcting for this effect, a much better agreement was
found between the different data sets.  This allows for a significant
reduction of the error bars associated to the QF, as well as for the
implementation of an energy-dependent QF.  Nevertheless, the
COHERENT collaboration has not confirmed the claims of the authors of
Ref.~\cite{Collar:2019ihs}. After repeating the calibration
measurements, their new data still shows a good
agreement~\cite{PhilBarbeau} with the original measurements performed
by the Duke (TUNL) group~\cite{Akimov:2018vzs, Akimov:2017ade}.

\begin{figure}\centering
  \includegraphics[scale=0.48]{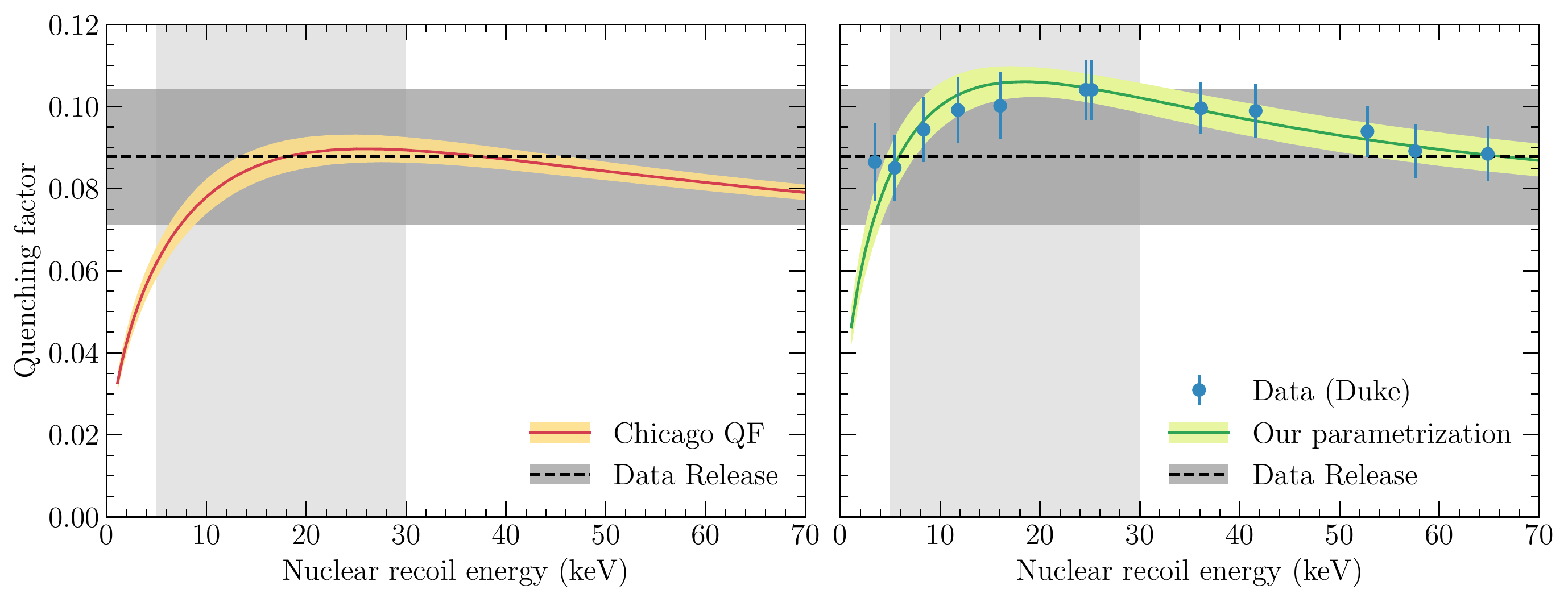}
  \caption{QF parametrisations used in our analysis of
    the COHERENT data.  The left panel shows the curve provided in
    Ref.~\recite{Collar:2019ihs} (solid curve), while the right panel
    shows the corresponding result obtained for our fit to the
    calibration data of the Duke (TUNL) group~\recite{Akimov:2017ade},
    provided as part of the data release~\recite{Akimov:2018vzs}. In
    both panels, the constant QF used in Ref.~\recite{Akimov:2017ade}
    is also shown for comparison. The three parametrisations are shown
    with a shaded band to indicate the values allowed at the $1\sigma$
    CL in each case. For illustration, the vertical shaded area
    indicates approximately the range of nuclear recoil energies that
    enters the signal region used in the fit (the exact range varies
    with the nuisance parameters, and the exact QF parametrisation
    used).}
  \label{fig:coh_QF}
\end{figure}

Given that this issue has not been settled yet, we will study and
quantify the effect of these new measurements in the results of the
fit in Sec.~\ref{sec:coh_results}. We will present our results
obtained for three different QF parametrisations: the original
(constant) parametrisation used in the data
release~\cite{Akimov:2018vzs}; the best-fit obtained by the authors of
Ref.~\cite{Collar:2019ihs}; and the results from our fit to the
calibration measurements performed by the TUNL
group~\cite{Akimov:2017ade, Akimov:2018vzs}. In order to fit the data
of the Duke group, we use the phenomenological parametrisation
proposed in~\cite{Collar:2019ihs}, which is based on a modification of
the semi-empirical approach by Birks~\cite{Birks:1951boa} and depends
only on two parameters, $E_0$ and $\textit{kB}$.\footnote{In brief,
  the fitted functional form of the QF as a function of the nuclear
  recoil energy $T$ is $\text{QF}(T) = [1-\exp(-T/E_0)] /
  [\textit{kB}\, dE/dR(T)]$ where $dE/dR(T)$ is the energy loss per
  unit length of the ions. For simplicity, we take it to be the
  average between that of Cs and I, obtained from
  SRIM-2013~\cite{SRIM}. We have also verified that using a simple
  polynomial parametrisation for $QF(T)$ leads to very similar
  results.}  Following this approach, we obtain a best-fit to the Duke
group data for $E_0 = 9.54 \pm 0.84$ and $\textit{kB} = 3.32 \pm
0.10$, with a correlation $\rho_{\textit{kB},E_0} = -0.69$. The three
QF parametrisations used in our calculations are shown in
Fig.~\ref{fig:coh_QF}, as a function of the recoil energy of the
nucleus.

Once the expected event distribution in PE has been computed following
Eq.~\eqref{eq:coh_Nevents}, the expected number of events in each bin
has to be smeared according to a Poisson distribution, to account for
the probability that a given event yields a different number of PE
than the average. On top of that, signal acceptance efficiencies are
applied to each bin:
\begin{equation}
  \label{eq:coh_acceptance}
  \eta(\text{PE}) = \frac{\eta_0}{1 + e^{-k (\text{PE} -
      \text{PE}_0)}} \, \Theta(\text{PE} - 5) \,,
\end{equation}
where the function $\Theta$ is defined as:
\begin{equation}
  \label{eq:coh_Theta}
  \Theta(\text{PE} - 5) =
  \begin{cases}
    0 & \text{if~} \text{PE} < 5 \,, \\ 0.5 & \text{if~} 5 < \text{PE}
    < 6 \,, \\ 1 & \text{if~} 6 < \text{PE} \,.
  \end{cases}
\end{equation}
Following Ref.~\cite{Akimov:2018vzs}, the central values of the signal
acceptance parameters are set to $\bar\eta_0 = 0.6655$, $\bar{k} =
0.4942$, $\overline{\text{PE}}_0 = 10.8507$.

Finally, once the predicted energy spectrum has been computed, one
should consider the arrival times expected for the different
contributions to the signal. This is implemented using the
distributions provided by the COHERENT collaboration in the data
release~\cite{Akimov:2018vzs}, which are normalised to one.

\begin{figure}\centering
  \includegraphics[width=0.47\textwidth]{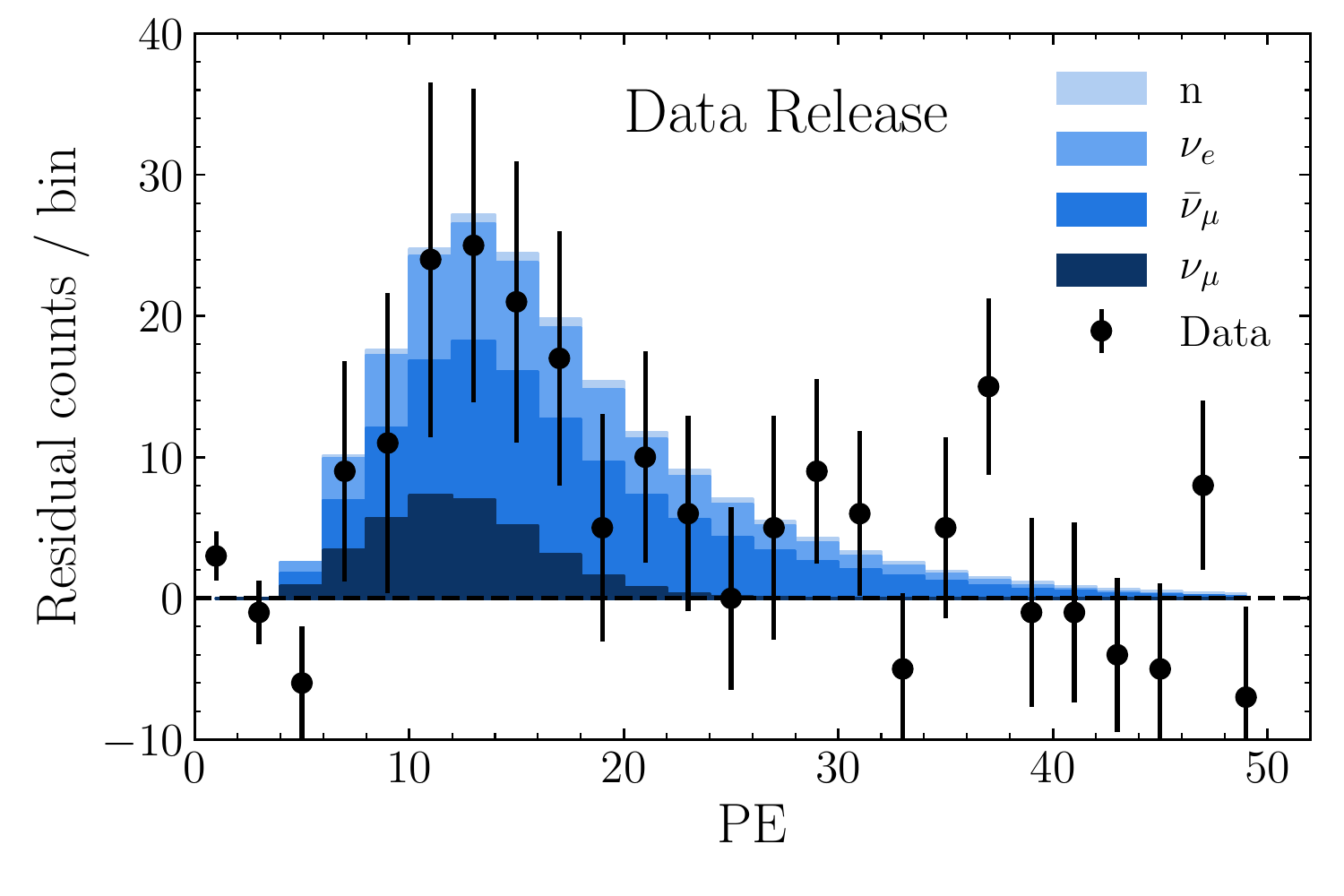}
  \includegraphics[width=0.47\textwidth]{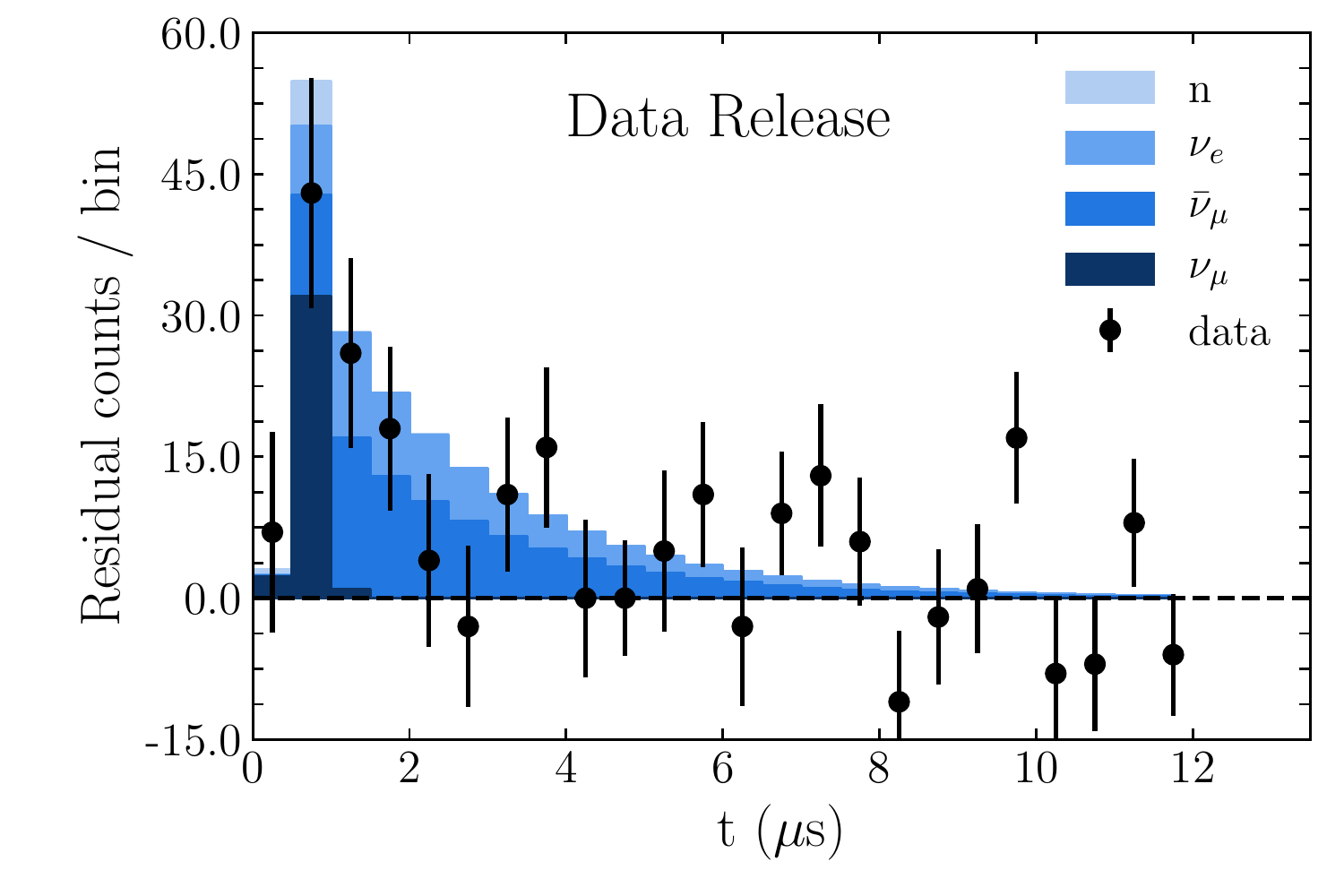}
  \includegraphics[width=0.47\textwidth]{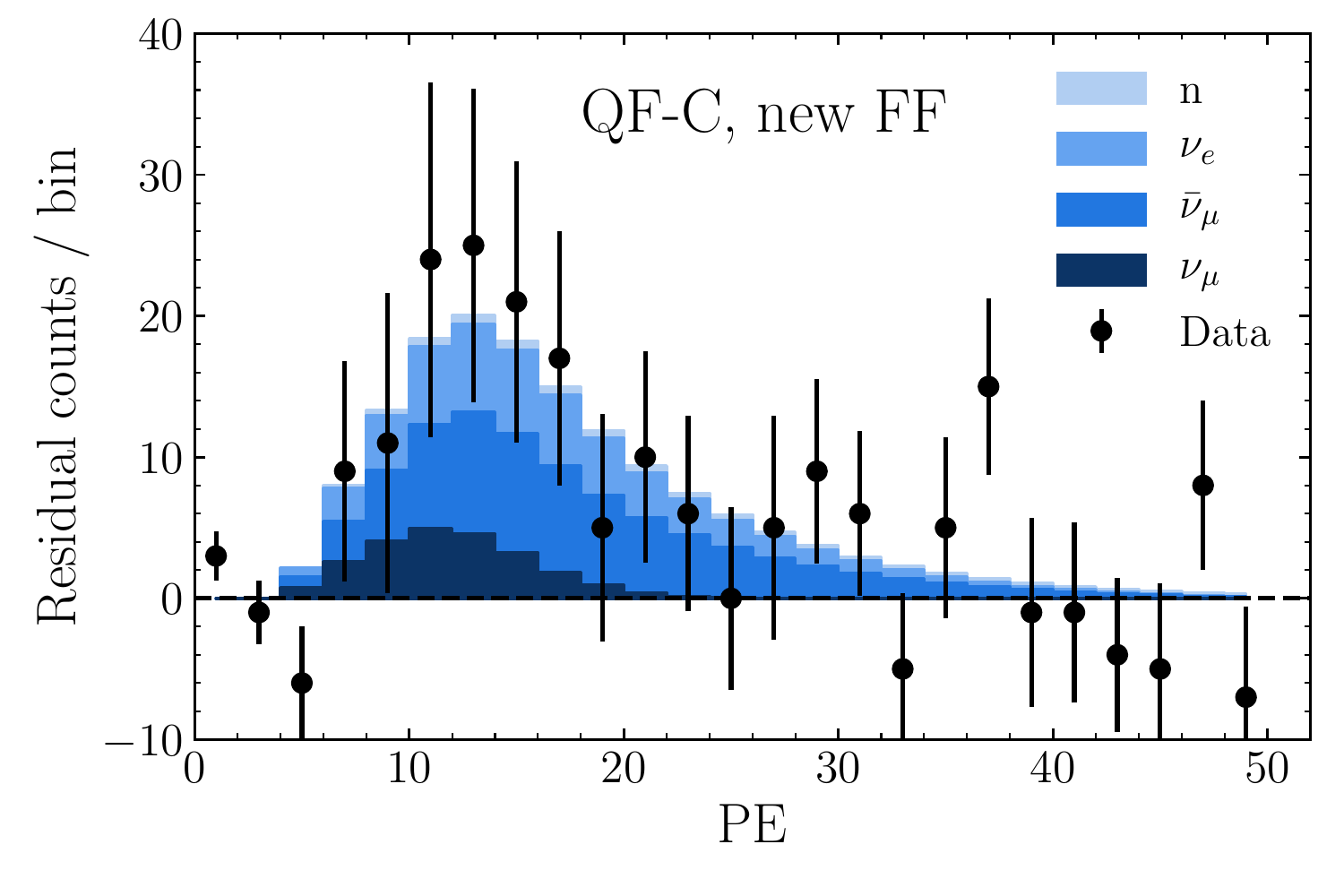}
  \includegraphics[width=0.47\textwidth]{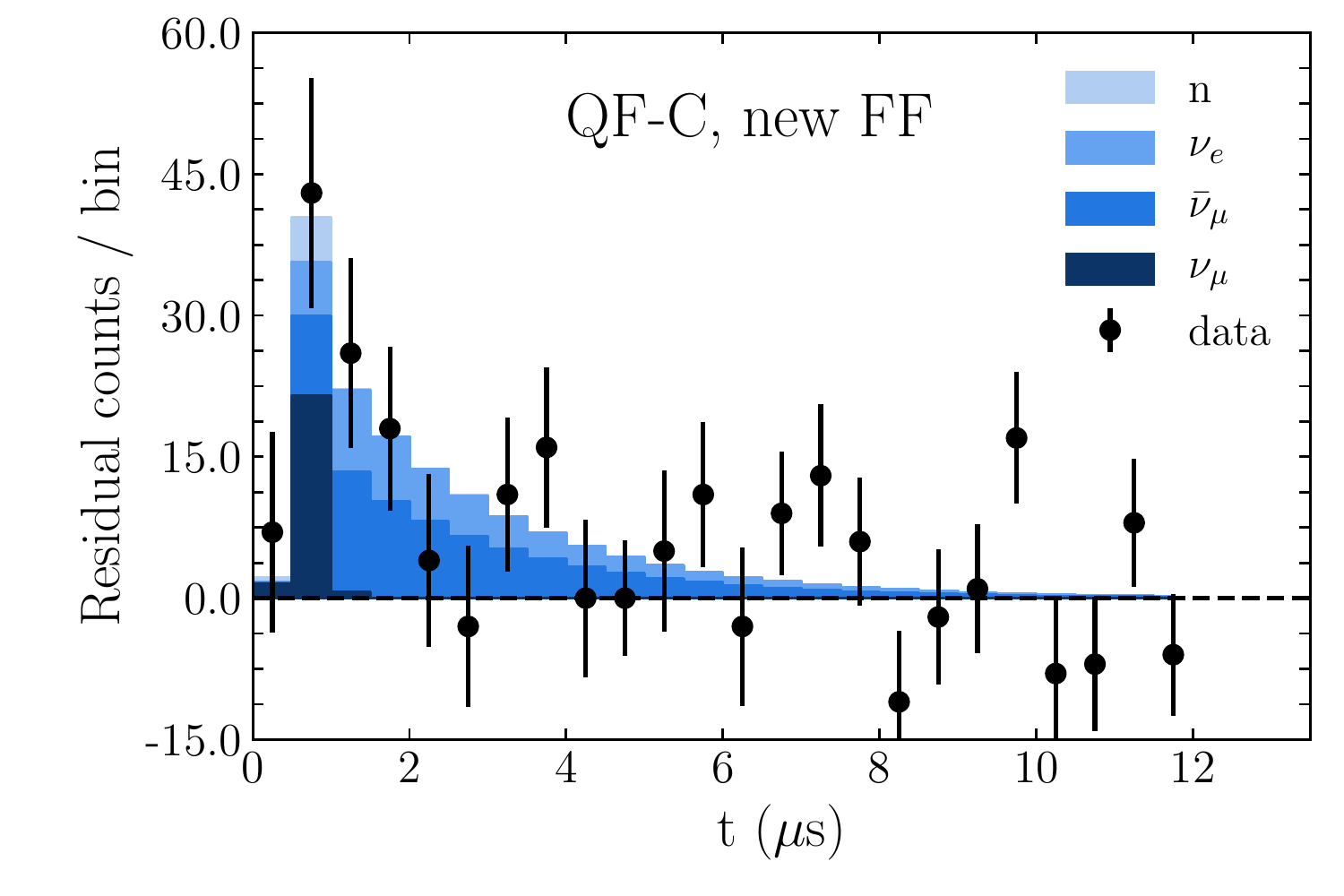}
  \includegraphics[width=0.47\textwidth]{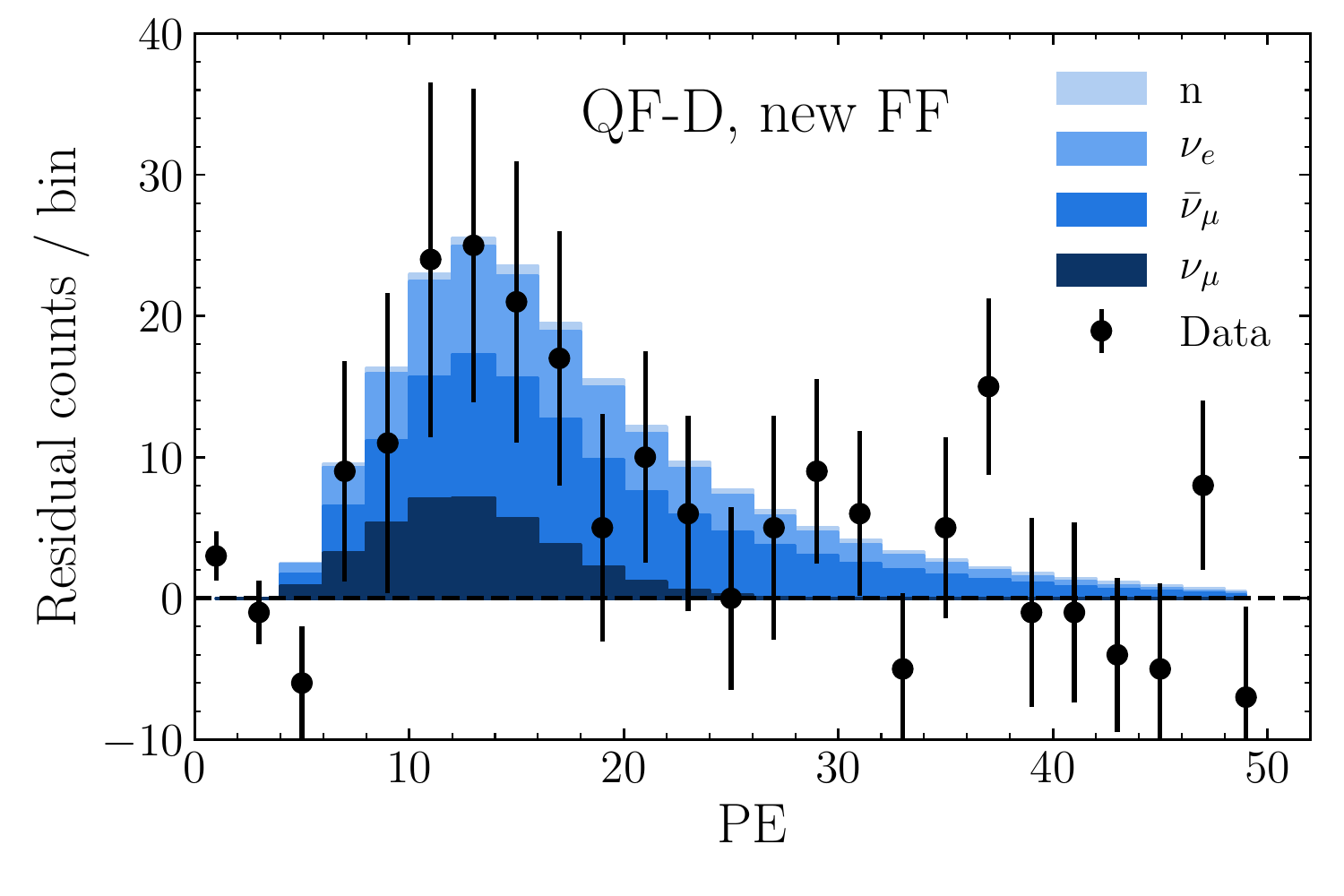}
  \includegraphics[width=0.47\textwidth]{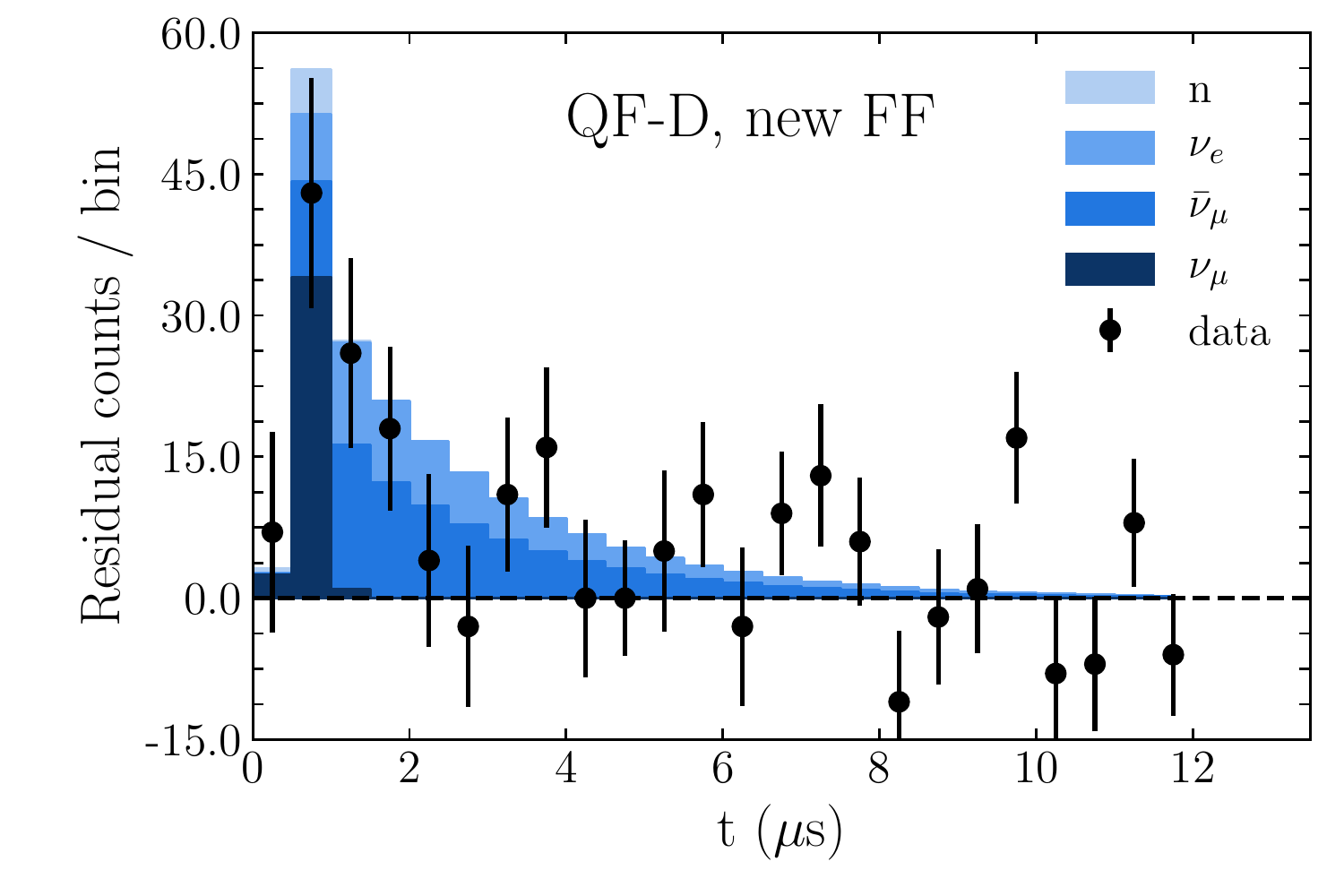}
  \caption{Residual events per bin obtained after subtracting C and AC
    data for the beam-ON sample, after being projected onto the PE
    (left panels) and time (right panels) axes and using the same cuts
    in PE an time as those applied to Fig.~3 in
    Ref.~\cite{Akimov:2017ade}. The observed data points are indicated
    with statistical error bars, as in Ref.~\cite{Akimov:2017ade}. In
    the upper panels the shaded histograms show the predicted event
    rates in the SM using the QF and nuclear form factor from
    Ref.~\cite{Akimov:2018vzs}. In the middle panels they correspond
    to the predictions with the QF from the Chicago group (QF-C) in
    Ref.~\cite{Collar:2019ihs} and the nuclear form factor (FF) from
    Ref.~\cite{menendez,Klos:2013rwa}.  The lower panels have been
    obtained with the same form factor, but changing the QF to match the Duke
    (TUNL) measurements in Ref.~\cite{Akimov:2017ade} (QF-D).  In all
    panels the prompt neutron background prediction is also shown for
    completeness. All the event histograms shown in this figure
    correspond to the SM prediction.}
  \label{fig:coh_histo-res}
\end{figure}

This final prediction can be compared with the published data.  This
is provided in two different time windows for each trigger in the data
acquisition system (that is, for each proton pulse). On the one hand,
the region where signals and beam-induced backgrounds associated with
the SNS beam are expected is referred to as the coincidence (C)
region, which can therefore be considered a ``signal'' region.
Conversely, the region where no contribution from the SNS beam is
expected is referred to as the anti-coincidence (AC) region and could
be considered a ``background'' region. While the collaboration
provides data separately for the beam-ON and beam-OFF data taking
periods, in this chapter we only use the beam-ON samples. The total
exposure considered in this chapter corresponds to 308.1 live-days of
neutrino production, which correspond to 7.48 GW-hr ($\sim 1.76\times
10^{23}$ protons on target). The residual event counts for this
period, i.e., the C data with the AC data subtracted, are shown in
Fig.~\ref{fig:coh_histo-res}, projected onto the time and PE axes, for
different choices of the QF and form factor as indicated by the labels.

As can be seen from the comparison between the upper and lower panels,
the change in QF between a constant approximation (Data Release) and
the energy-dependent result obtained by the TUNL group (QF-D) does not
affect significantly the predicted event distributions. This will lead
to a minor change in the results of the numerical fit to the data in
Sec.~\ref{sec:coh_results}. A larger difference is observed with
respect to the predictions using the QF by the Chicago group (QF-C,
middle panels): in this case, the very different central values at $T
\sim 10$~keV (corresponding to $\text{PE}\sim 10$) lead to a reduced
number of events, which will have a larger impact on the results.

\subsubsection{Computation of the background}
\label{sec:coh_bg}

The COHERENT measurement is affected by three main background sources:
(i) the steady-state background, coming from either cosmic rays or
their by-products entering the detector; (ii) prompt neutrons produced
in the target station and exiting it, and (iii) neutrino-induced
neutrons (NINs) that originate in the shielding surrounding the
detector. While the latter is irreducible, it has been shown to be
negligible at the COHERENT experiment and is therefore ignored here.

The procedure used to compute the expected number of background events
for the steady-state and the prompt neutron components follows the
prescription given in Ref.~\cite{Akimov:2018vzs}. For both
backgrounds, it is assumed that the temporal and energy dependence on
the number of events can be factorised as
\begin{equation}
  \label{eq:coh_bg}
  N_\text{bg}(t, \text{PE}) = f(t) \cdot g(\text{PE}) \,,
\end{equation}
where $f$ contains the temporal dependence of the signal and $g$ its
energy dependence.

For the prompt neutron background, the collaboration provides both its
expected energy distribution before acceptance efficiencies are
applied (that is, $g(\text{PE})$), and the total expected counts as a
function of time (that is, $f(t)$). The expected 2D distribution can
be obtained simply by multiplying the two distributions. After the
number of events in each bin has been computed, the same acceptance
efficiency as for the signal, Eq.~\eqref{eq:coh_acceptance}, is
applied to determine the expected number of events in each bin.

The steady-state is the most significant background source to this
analysis, and it has the largest impact on the fit. In this case, the
functions $g(\text{PE})$ and $f(t)$ are not provided in
Ref.~\cite{Akimov:2018vzs} but inferred from the data, which is
provided per bin in energy and time. In particular, the projected data
onto the PE axis is then used directly as $g(\text{PE})$, while $f(t)$
is assumed to follow an exponential:
\begin{equation}
  \label{eq:coh_expo-fit}
  f_\text{ss}(t) = a_\text{ss} e^{-b_\text{ss} t} \,.
\end{equation}
By taking the AC data and projecting it into the time axis, a best-fit
to the steady-state background is obtained for $a_\text{ss} = 58.5$
and $b_\text{ss} = 0.062$. The value of $f(t)$ is then normalised so
that its integral over the whole range in time is equal to one. Since
in this case the expected background events are inferred from a
measurement, the signal acceptance has already been included into the
calculation and there is no need to apply it here.

\begin{figure}\centering
  \includegraphics[width=0.65\textwidth]{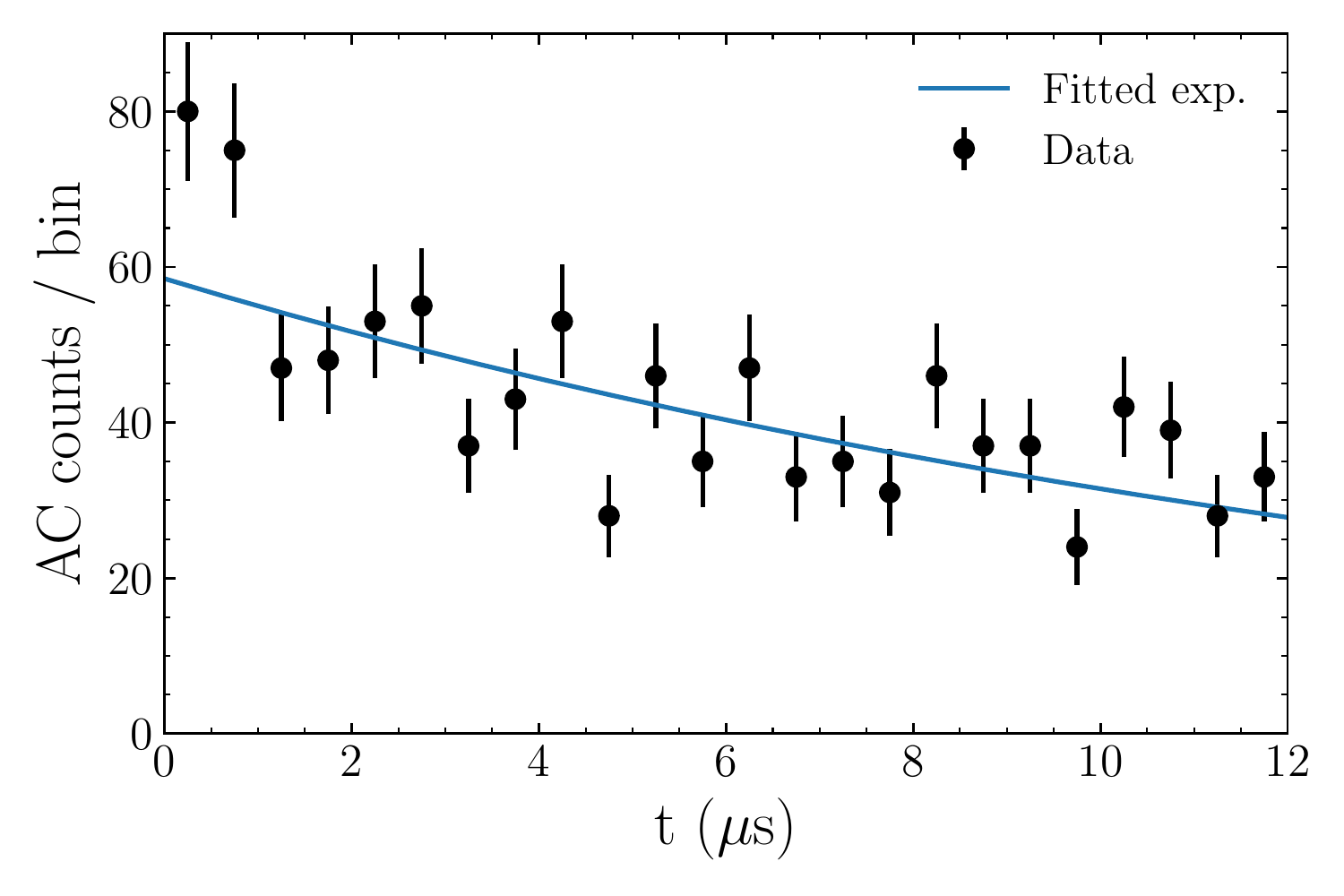}
  \caption{Total AC counts per bin, compared to the results of the
    exponential fit employed in Ref.~\recite{Akimov:2018vzs} and
    described in Sec.~\ref{sec:coh_bg} used to model the steady-state
    background. No cuts on the observed number of PE have been applied
    to this figure.}
  \label{fig:coh_bckg}
\end{figure}

The procedure outlined above for the steady-state component is meant
to eliminate biases in the fit due to the limited statistics of the
data sample used. However, by treating the background in this way the
analysis is rather sensitive to a mismodelling of its temporal
component. In particular, if we plot the separate C and AC event
distribution as a function of time, instead of looking at their
difference, it is easy to see that there is an excess in the first two
bins in the data, which cannot be accommodated by the simplified
exponential fit. This is shown in Fig.~\ref{fig:coh_bckg}, where we
show the total AC counts (which should include only the steady-state
background as measured by the detector) together with the exponential
that gives a best fit to the data. As clearly seen, the first two bins
are not well fitted by a simple exponential model.

\begin{figure}\centering
  \includegraphics[width=0.48\textwidth]{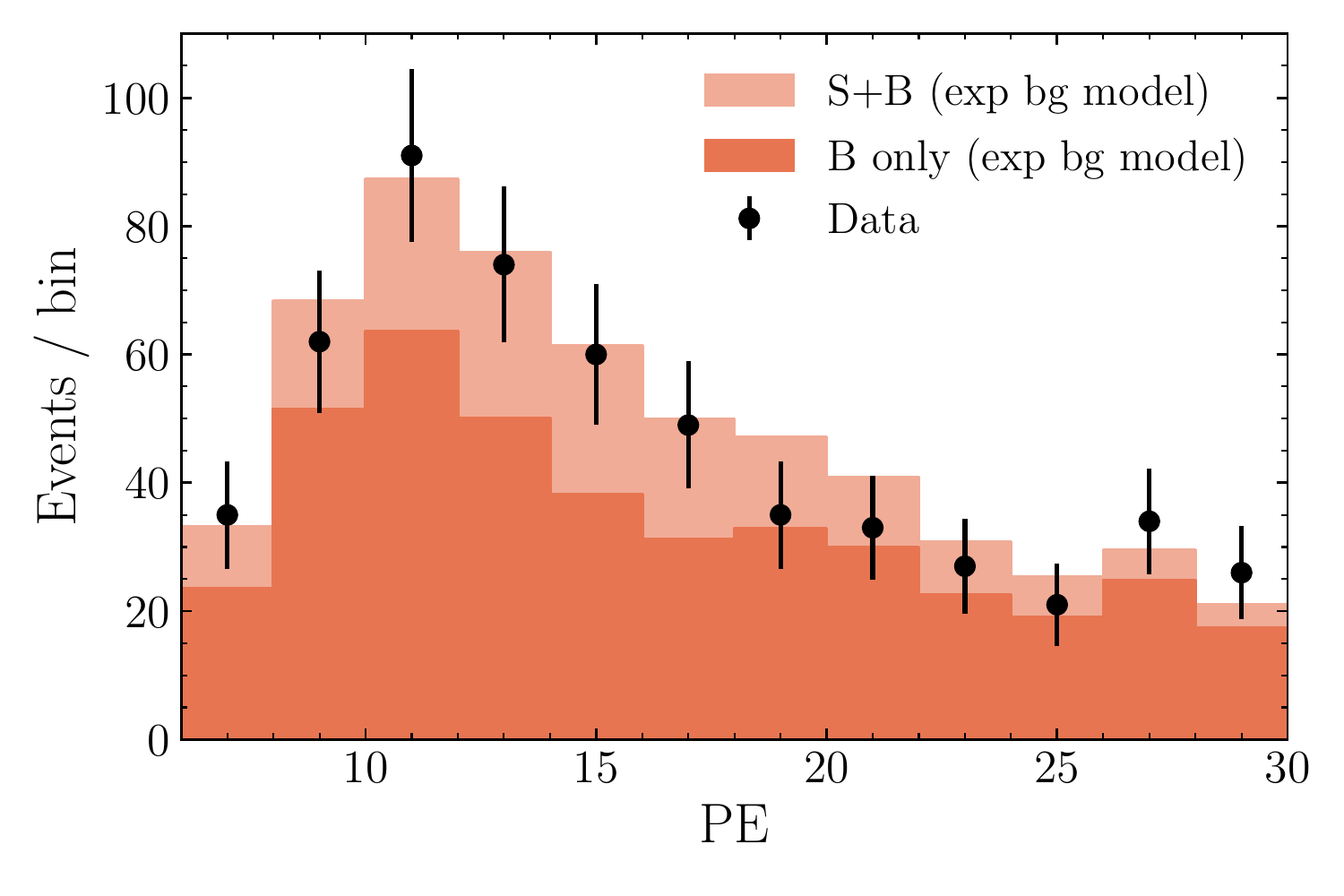}
  \includegraphics[width=0.48\textwidth]{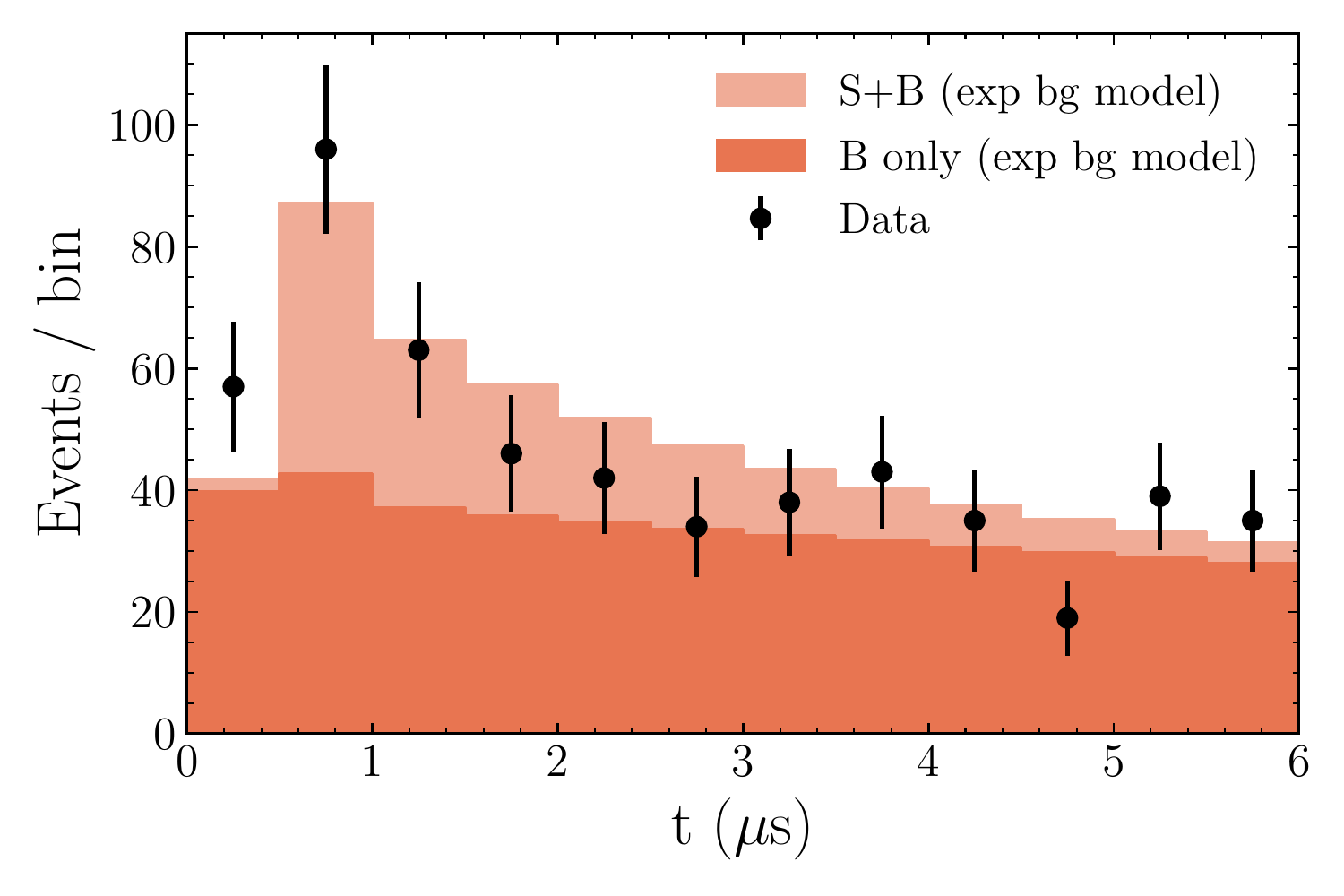}
  \includegraphics[width=0.48\textwidth]{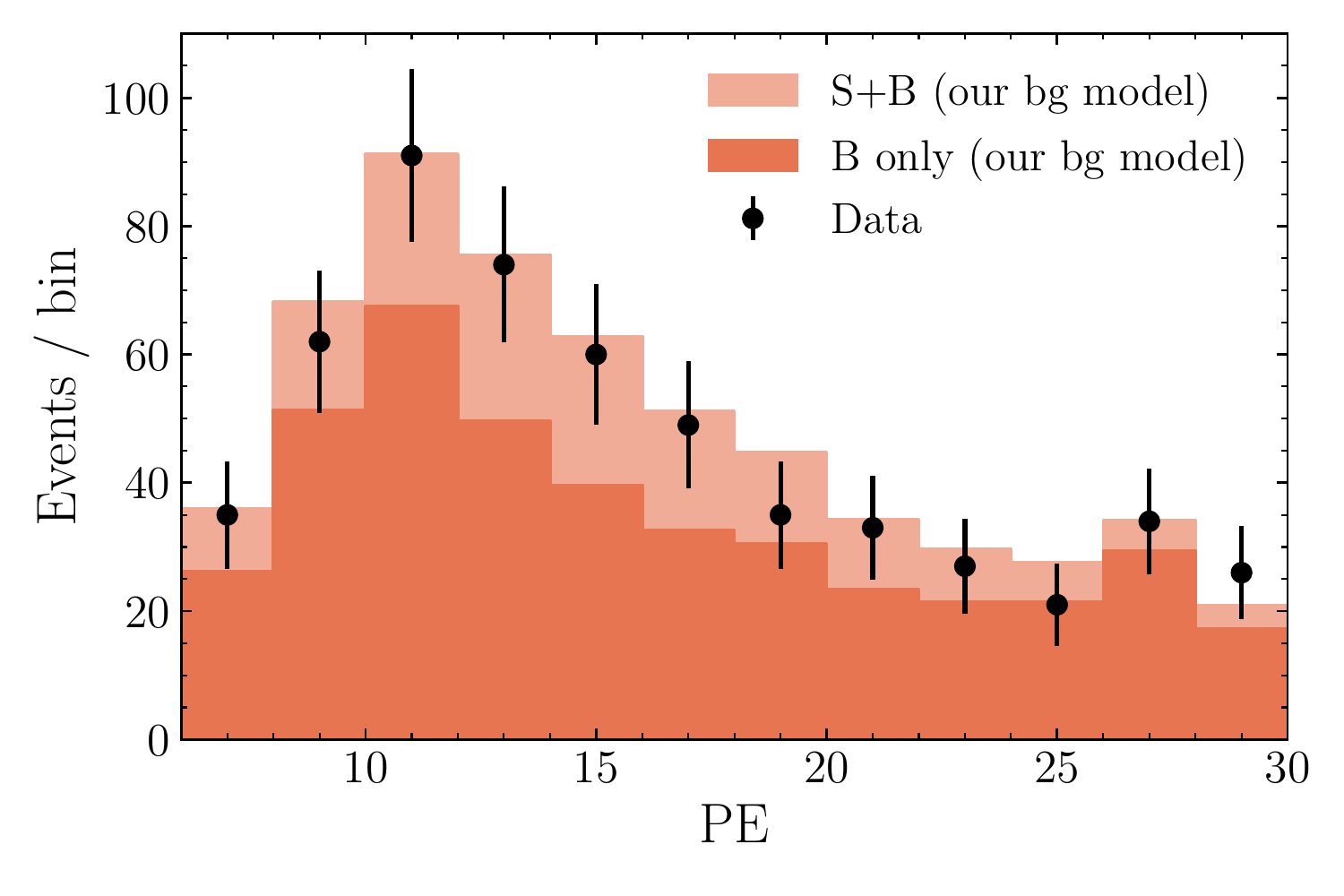}
  \includegraphics[width=0.48\textwidth]{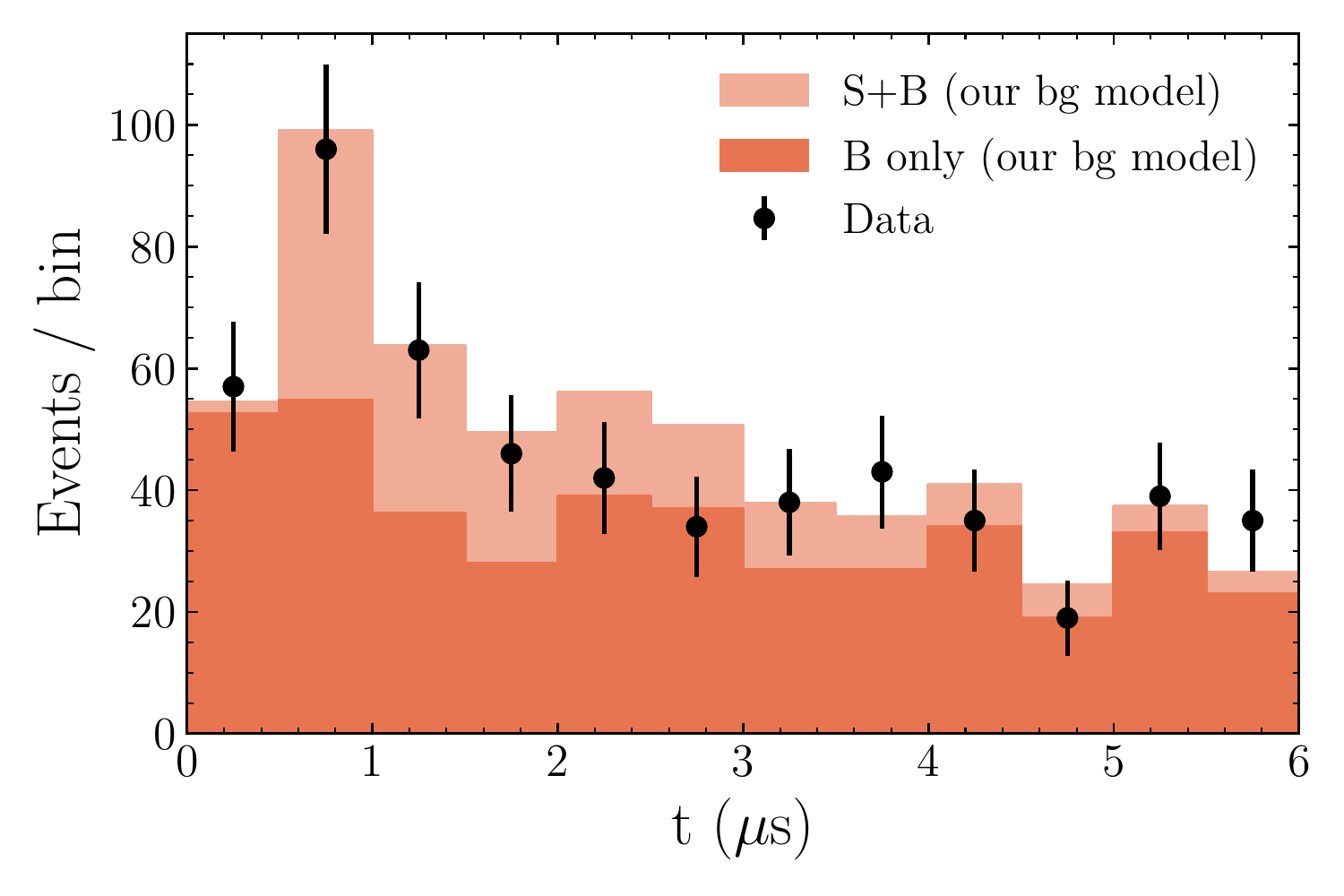}
  \caption{Total events per bin in the beam-ON C sample, after being
    projected onto the time (left) and PE (right) and imposing the
    cuts $5 < \text{PE} \leq 30$ and $t < 6~\mu\text{s}$. The observed
    data points are indicated with statistical error bars as in
    Fig.~\ref{fig:coh_histo-res}.  The dark histograms show the
    expected background events for the steady-state contribution only.
    The upper panels have been obtained assuming that the time
    dependence of the background follows an exponential model, as in
    Ref.~\cite{Akimov:2018vzs}, while the lower panels have been
    obtained using our model (which follows the time dependence of the
    AC events, see text for details). The light histograms show the
    predicted total number of events, after adding all signal and
    background contributions. To ease the comparison among different
    panels, in this figure the signal has been computed in all cases
    using the same form factor and QF as in the data
    release~\cite{Akimov:2018vzs}. All histograms shown correspond to
    the SM predicted event rates.}
  \label{fig:coh_histo-total}
\end{figure}

Interestingly enough, both the C and AC samples seem to observe a
similar excess in the first two temporal bins, which suggests that
this contribution is not related to the neutrino signal but to some
mismodelling of the background. A possible way to correct for this is
to directly use the measured time dependence of the AC sample as a
direct prediction for the expected behaviour of the steady-state
background in the C sample. Doing this on a bin-per-bin basis would
not provide a good predictor for the expected number of events in each
bin, due to the limited statistics. However, the projected data onto
the time axis may still be used, as in the case of the exponential
fit, to get a prediction for the function $f(t)$. In other words, the
prediction for $f(t)$ may be obtained following the same procedure as
was done for $g(\text{PE})$, i.e., projecting the events onto the
corresponding axis.  Figure~\ref{fig:coh_histo-total} shows the
observed total event counts for the beam-ON, C sample (which includes
both signal and background) projected onto PE (left) and time (right),
compared with the predictions using these two different background
models.

As can be seen from this figure, both background models are able to
reproduce the observed spectrum in PE relatively well, and give very
similar results. However, the event rates obtained with this second
method provide a better fit to the data when projected onto the time
axis and, as clearly observed from the figure, the effect is specially
noticeable in the first two bins. Therefore, in
Sec.~\ref{sec:coh_results} we will show two sets of results: with and
without using an exponential model for the background.

\subsubsection{Systematic errors and implementation of the $\chi^2$}

Once the predicted event distributions for the signal and backgrounds
have been computed, a $\chi^2$ function is built as:
\begin{equation}
  \label{eq:coh_chi2}
  \chi^2\big[P_{ij}(\vec\xi)\big] = \sum_{ij} 2 \left[ P_{ij}(\vec\xi)
    - O_{ij} + O_{ij} \ln\left( \frac{O_{ij}}{P_{ij}(\vec\xi)}\right)
    \right],
\end{equation}
where $O_{ij}$ stands for the observed number of events in PE bin $i$
and time bin $j$, while $P_{ij}$ stands for the total number of
predicted events in that bin, including the signal plus all background
contributions. Following Ref.~\cite{Akimov:2018vzs}, we consider only
the events with $5 < \text{PE} \leq 30$ and $t < 6~\mu\text{s}$ in the
analysis. The predicted number of events depends on the nuisance
parameters $\vec\xi \equiv \{ \xi_a \}$ included in the fit, which
account for the systematic uncertainties affecting the QF, signal
acceptance, neutrino production yield, and normalisation of the
backgrounds. These are implemented replacing the original quantity as
$x \to (1 + \sigma_x \xi_x)\bar{x}$, where $\bar{x}$ denotes the
central value assumed for $x$ prior to the experiment and $\sigma_x$
denotes the relative uncertainty for nuisance parameter $\xi_x$
summarised in Tab.~\ref{tab:coh_sys} for convenience.  More
specifically:
\begin{multline}
  \label{eq:coh_pij}
  P_{ij}(\vec\xi) = (1 + \sigma_\text{ss} \xi_\text{ss})
  N_{ij}^\text{ss} \\ + \eta(\text{PE}_i \,|\, \xi_{\eta_0}, \xi_k,
  \xi_{\text{PE}_0}) \big[ (1 + \sigma_\text{n} \xi_\text{n})
    N_{ij}^\text{n} + (1 + \sigma_\text{sig}\xi_\text{sig})
    N_{ij}^\text{sig}(\xi_\text{QF}) \big] ,
\end{multline}
where $N_{ij}^\text{ss}$, $N_{ij}^\text{n}$ and $N_{ij}^\text{sig}$
stand for the predicted number of events for the steady-state
background, the prompt neutron background and the signal. In
Eq.~\eqref{eq:coh_pij} we have generically denoted as $\xi_\text{QF}$
the set of nuisance parameters characterising the uncertainty on the
QF employed. For the constant parametrisation used in the data
release~\cite{Akimov:2018vzs} we introduce a unique nuisance parameter
with constant uncertainty.  For QF-C we introduce also a unique
nuisance parameter, but with an energy-dependent uncertainty inferred
from the uncertainty band in Fig.~1 of Ref.~\cite{Collar:2019ihs}
(also shown in the left panel of Fig.~\ref{fig:coh_QF}), which varies
from 6.5\% to 3.5\% in the range of recoil energies relevant for
COHERENT.  For QF-D we introduce two nuisance parameters
characterising the uncertainty on parameters $E_0$ and $\textit{kB}$
with their corresponding correlation.\footnote{We have verified that,
  in practice, it is equivalent to using a single nuisance parameter
  with an energy-dependent uncertainty ranging between 8\% and 3\%
  (corresponding to the shaded band shown in the right panel in
  Fig.~\ref{fig:coh_QF}).}

Altogether the likelihood for some physics model parameters
$\vec\varepsilon$, leading to a given set of predictions
$P_{ij}^{\vec\varepsilon}(\vec\xi)$ for the events in bin $ij$, is
obtained including the effects of the nuisance parameters as in
Eq.~\eqref{eq:coh_pij} and minimising over those within their assumed
uncertainty. This is ensured by adding a pull term to the $\chi^2$
function in Eq.~\eqref{eq:coh_chi2} for each of the nuisance
parameters introduced:
\begin{equation}
  \label{eq:coh_chi2min}
  \chi^2_\text{COH} (\vec\varepsilon) = \min\limits_{\vec\xi} \bigg\{
  \chi^2\big[ P_{ij}^{\vec\varepsilon}(\vec\xi) \big] + \sum_{ab}
  \xi_a (\rho^{-1})_{ab} \xi_b \bigg\} \,,
\end{equation}
where $\rho$ is the correlation matrix, whose entries are $\rho_{ab} =
\delta_{ab}$ for all parameters except those entering our
parametrisation of the QF of the Duke group (the corresponding
correlation coefficient can be found in Tab.~\ref{tab:coh_sys}).

\begin{table}\centering
  \catcode`!=\active\def!{\hphantom{-}}
  \catcode`?=\active\def?{\hphantom{0}}
  \begin{tabular}{lc}
    \toprule Parameter & Uncertainty (\%) \\ \midrule Steady-state norm. &
    $!?5.0?$ \\ Prompt n norm. & $!25.0?$ \\ Signal norm. & $!11.2?$
    \\ $\eta_0$ & $!?4.5?$ \\ k & $!?4.7?$ \\ $\text{PE}_0$ & $!?2.7?$
    \\ QF (data release) & $!18.9?$ \\ QF (our fit, QF-C)& $!?6.5 -
    3.5?$ \\ $\textit{kB}$ (our fit, QF-D) & $!?3.0?$ \\ $E_0$ (our
    fit, QF-D) & $!?8.8?$ \\ $\rho_{E_0,\textit{kB}}$ (our fit, QF-D)
    & $?{-0.69}$ \\ \bottomrule
  \end{tabular}
  \caption{Systematic uncertainties considered in the fit on
    acceptance efficiency parameters (Eq.~\eqref{eq:coh_acceptance}),
    normalisation of the signal and background contributions, and the
    QF. The steady-state normalisation uncertainty includes the
    statistical error of the sample (AC data). The quoted
    uncertainties on the QF also includes the error on the light yield
    (0.14\%), which is however subdominant. For details on the QF
    parametrisation, see Sec.~\ref{sec:coh_coh2}.}
  \label{tab:coh_sys}
\end{table}

\begin{figure}\centering
  \includegraphics[width=0.7\textwidth]{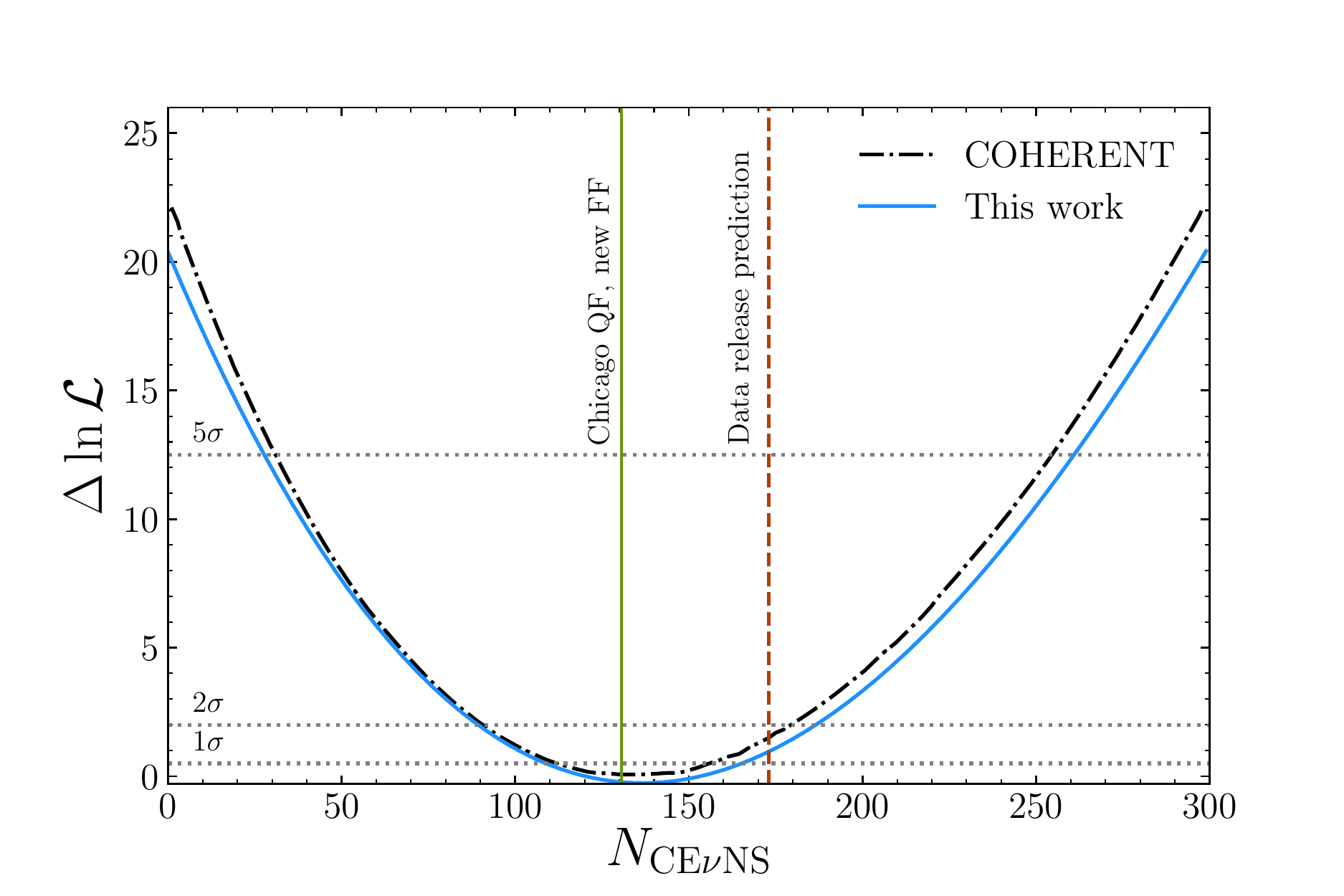}
  \caption{Comparison of our $\chi^2$ for the COHERENT timing and
    energy data as a function of the number of signal CE$\nu$NS events
    under the same assumptions on the background, systematics and
    expected time and energy dependence of signal, compared to that
    provided by COHERENT in figure S13 of
    Ref.~\cite{Akimov:2017ade}. For comparison, the vertical lines
    indicate the predicted event rates in the SM (with no systematic
    uncertainties), for different choices of QF and form factor used: dashed
    red corresponds to the prediction provided in the data release of
    173 events~\cite{Akimov:2018vzs}, while solid green indicates our
    prediction using the QF from Ref.~\recite{Collar:2019ihs} (left
    panel in Fig.~\ref{fig:coh_QF}) and nuclear form factor of~\cite{menendez,
      Klos:2013rwa}.}
  \label{fig:coh_compachi2}
\end{figure}

As validation of our $\chi^2$ construction we have performed a fit to
extract the total number of signal CE$\nu$NS events when using the
same assumptions on the background, systematics and energy and time
dependence of the signal as those employed by COHERENT in their data
release~\cite{Akimov:2018vzs}.  This can be directly compared with
their corresponding likelihood extracted from Figure S13 of
Ref.~\cite{Akimov:2017ade}. The result of this comparison is shown in
Fig.~\ref{fig:coh_compachi2}.  Strictly speaking, the $\chi^2$
function plotted in Fig.~\ref{fig:coh_compachi2} depends on the
assumed energy dependence of the signal. Therefore it is expected to
vary if, instead of using the QF and form factor quoted in the data release, we
employed a different QF parametrisation and nuclear form factor.
Quantitatively, within the systematic uncertainties used in the
construction of the shown $\chi^2(N_\text{CE$\nu$NS}$), we find that
changing the QF and form factor has a negligible effect on this curve.
Conversely, we find a stronger dependence on the systematic
uncertainties introduced, and therefore Fig.~\ref{fig:coh_compachi2}
serves as validation of our implementation for these.  For
illustration, we also indicate the predicted event rates in the SM
predicted by the collaboration (173 events) as well as our result
obtained using the Chicago QF parametrisation~\cite{Collar:2019ihs},
as shown in the left panel of Fig.~\ref{fig:coh_QF} and the new
nuclear form factor from Refs.~\cite{menendez, Klos:2013rwa}. In both cases the
vertical lines correspond to the prediction without accounting for
systematic uncertainties. The predicted result using the Duke QF and
the new form factor from Refs.~\cite{menendez, Klos:2013rwa} is very similar to
the one obtained by the collaboration (168 events) and is therefore
not shown here.

\subsection{Results: Fit to COHERENT data}
\label{sec:coh_results}

In this section we present our results. We discuss in detail the improvements coming from
the inclusion of energy and timing information in the fit, and
investigate the impact of the different choices of QF,
nuclear form factor and background implementation.

\begin{figure}\centering
  \includegraphics[width=\textwidth]{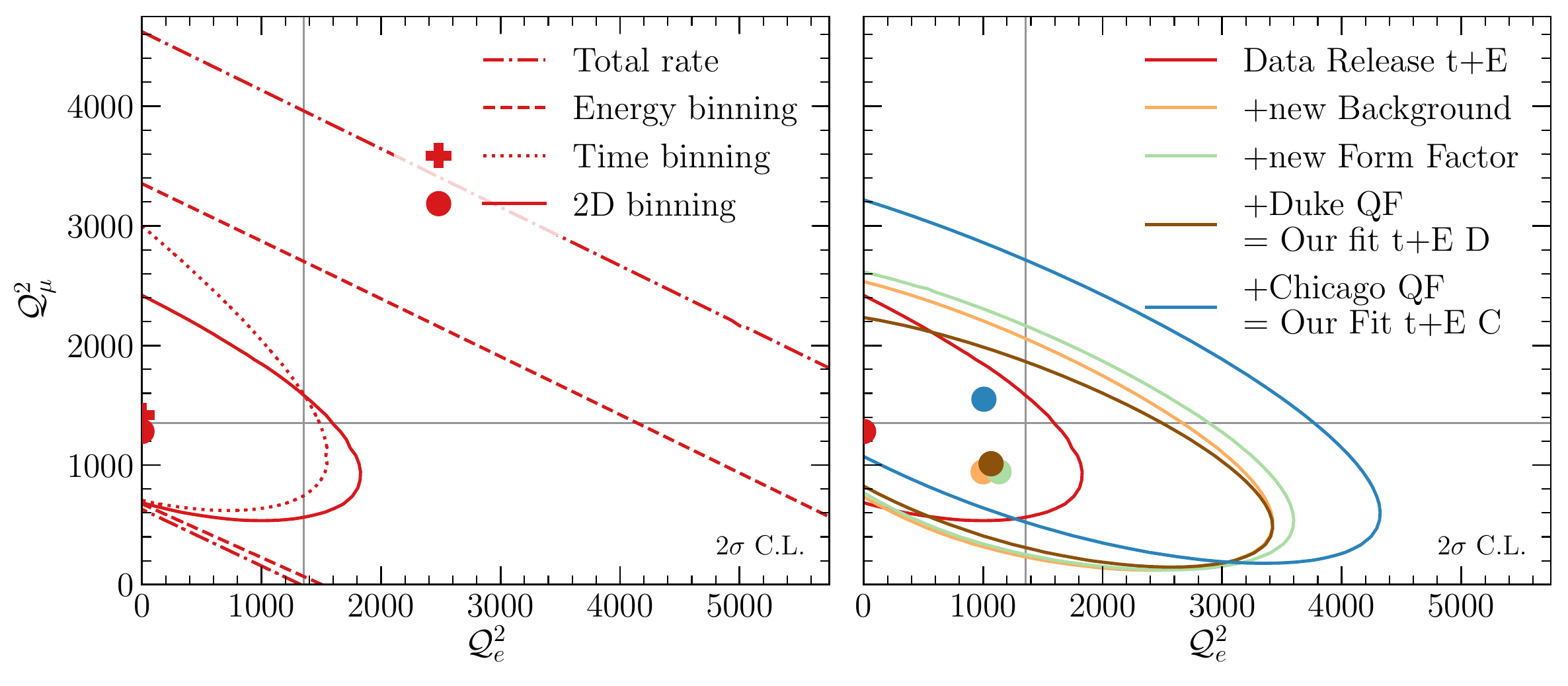}
  \caption{$2\sigma$ allowed regions for the flavour-dependent weak
    charges for a variety of fits to COHERENT data as labeled in the
    figure. In all cases shown in the left panel, the QF, nuclear form factor
    and background assumptions are those employed in the data
    release~\recite{Akimov:2018vzs}.  On the right panel we show the
    dependence on the steady background modelling, nuclear form factor, and
    QF. The vertical lines indicate the SM value $\mathcal{Q}^2_e =
    \mathcal{Q}^2_\mu = 1353.5$. The coloured dots and the red cross
    mark the position of the best-fit for the various cases.}
  \label{fig:coh_qw}
\end{figure}

In order to study the dependence of the results on the different
assumptions, we have performed a set of fits to COHERENT data in terms
of two effective flavour-dependent weak charges $\mathcal{Q}^2_\alpha$
(assumed to be energy-independent). Figure~\ref{fig:coh_qw} shows the
corresponding allowed regions at $2\sigma$ from the fit to COHERENT
data alone.  In the left panel we illustrate the effect of including
the energy and timing information in the fit, by comparing the allowed
values of the weak charges obtained: (i) using only the total rate
information (dot-dashed); (ii) adding only the energy information
(dashed); (iii) using only the event timing information (dotted); and
(iv) fitting the data binned in both timing and energy (solid). In all
cases shown in the left panel, the QF, nuclear form factor and background have
been implemented following closely the prescription given in the data
release~\cite{Akimov:2018vzs}. It is well-known that, when only the
total event rate information is considered, there is a degeneracy in
the determination of the flavour-dependent weak charges, since the
number of predicted events approximately behaves as
\begin{equation}
  \label{eq:coh_Qbands}
  \mathcal{Q}^2_e f_{\nu_e} + \mathcal{Q}^2_\mu\, (f_{\nu_\mu} +
  f_{\bar\nu_\mu}) \approx \frac{1}{3} \mathcal{Q}^2_e + \frac{2}{3}
  \mathcal{Q}^2_\mu
\end{equation}
where $f_\alpha$ indicates the fraction of expected SM events from
interactions of $\nu_\alpha$ in the final event sample and we have
assumed that one neutrino of each species is produced for each pion
DAR.  Under these assumptions, the allowed region in the
$(\mathcal{Q}^2_e, \mathcal{Q}^2_\mu)$ plane is a straight band with a
negative slope, $\arctan(-0.5) \approx -27^\circ$. This behaviour is
also observed from our exact fit to the data, as shown by the
dot-dashed lines in the left panel in Fig.~\ref{fig:coh_qw}.

As expected, the timing information is most relevant in breaking of
this degeneracy: since the prompt component of the beam contains only
$\nu_\mu$, the inclusion of time information allows for a partial
discrimination between $\mathcal{Q}^2_e$ and $\mathcal{Q}^2_\mu$.
Notice, however, that in this case the best fit is obtained at the
edge of the physically allowed region, $\mathcal{Q}^2_e \simeq 0$ (in
fact, it would probably take place for a negative value, but this is
not the case since we are effectively imposing the restriction
$\mathcal{Q}^2_\alpha > 0$ in the fit).  This is driven by the small
excess for the event rates in the first two time bins (with respect to
the SM prediction) when using the exponential fit model for the
steady-state background, as described in Sec.~\ref{sec:coh_bg} (see
Fig.~\ref{fig:coh_histo-total}).  Such excess can be accommodated
thanks to the overall normalisation uncertainty of the signal,
combined with a decrease of the $\nu_e$ contribution as required to
match the distribution observed for the delayed events.  Within the
systematic uncertainties in the analysis, this results into a higher
rate at short times without a major distortion of the PE spectrum.  We
also observe that, including only the PE spectrum in the fit, the
degeneracy still remains but the width of the band in this plane
decreases.  For values of $\mathcal{Q}^2_\alpha$ in the
non-overlapping region, the fit using the event rate information alone
is able to fit the data, albeit at the price of very large nuisance
parameters and, in particular, of the QF-related uncertainties (which
affect the shape of the event distributions in PE space). Therefore,
once the PE information is added the allowed regions are consequently
reduced.

The right panel in Fig.~\ref{fig:coh_qw} shows the dependence of the
allowed region on the assumed background model, nuclear form factor and QF
choice in the fit, for the 2D fit using both time and PE
information. As seen in the figure, if one uses the steady-state
background prediction without the exponential model for its temporal
dependence the region becomes considerably larger, and the best fit moves
closer to the SM. This is expected because, with this background,
there is no excess of events in the first time bins with respect to
the SM prediction (see Fig.~\ref{fig:coh_histo-total}). This also
leads to a better overall fit, with $\chi^2_\text{min} = 145.24$ (for
$12\times 12=144$ data points) compared to $\chi^2_\text{min} = 150.8$
obtained for the exponential model of the steady-state background.
Altogether we observe that modifying the nuclear form factor has a very small
effect on the current results, as can be seen from the comparison
between the orange and green lines in the figure.  Changing the QF
does not have a dominant impact either, once the exponential fit to
the background has been removed.  This can be seen from the comparison
between the green, brown and blue lines in the figure, which all
provide similar results.  Overall, we find a slightly better agreement
with the SM result for the QF-C parametrisation, albeit the effect is
small.

\begin{figure}\centering
  \includegraphics[width=\textwidth]{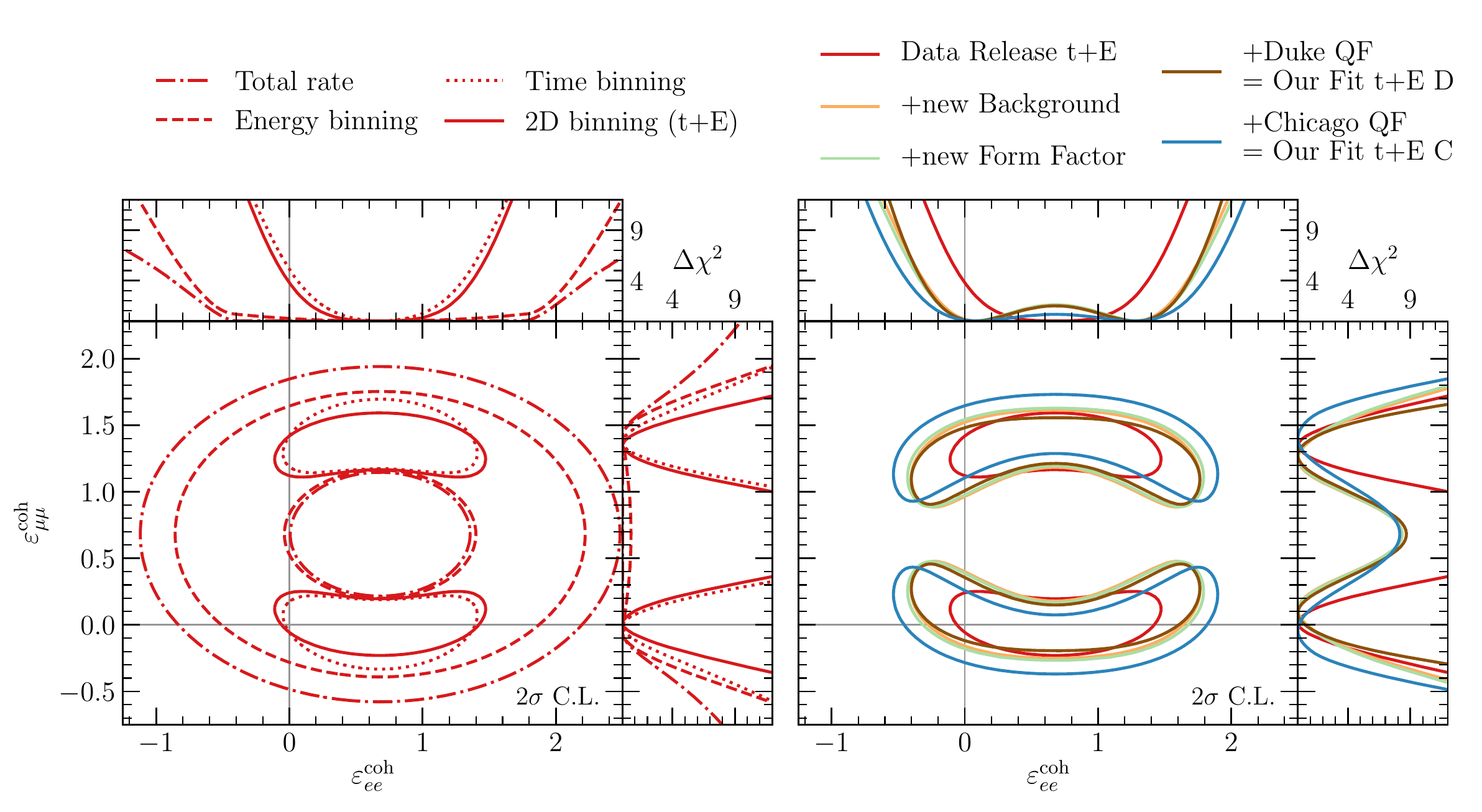}
  \caption{$2\sigma$ allowed regions for the flavour-diagonal NSI
    coefficients $\varepsilon_{\alpha\beta}^\text{coh}$ (assuming zero
    non-diagonal couplings) for a variety of fits to COHERENT data as
    labelled in the figure. In all cases shown in the left panel, the
    QF, nuclear form factor and background assumptions are those employed in
    the data release~\recite{Akimov:2018vzs}.  On the right panel we
    show the dependence on the assumptions for steady background
    modelling, nuclear form factor, and QF. For simplicity, in this figure we
    set all off-diagonal NSI parameters to zero, but it should be kept
    in mind that the results of our global analysis presented in
    Sec.~\ref{sec:coh_resglob} have been obtained allowing all
    operators simultaneously in the fit.}
  \label{fig:coh_epscoh}
\end{figure}

In the framework of NSI, the constraints on the weak charges derived
above can be directly translated into constraints on the
$\varepsilon_{\alpha \beta}^f$. This is shown in Fig.~\ref{fig:coh_epscoh}, where
we plot the allowed regions for the two relevant flavour-diagonal NSI
couplings after setting the flavour-changing ones to zero.  In doing
so we notice that the fact that the neutron/proton ratio in the two
target nuclei is very similar ($N_\text{Cs} \big/ Z_\text{Cs} \simeq
1.419$ for cesium and $N_\text{I} \big/ Z_\text{I} \simeq 1.396$ for
iodine) allows to approximate Eq.~\eqref{eq:coh_Qalpha-nsi} as:
\begin{equation}
  \label{eq:coh_Qalpha-coh}
  \mathcal{Q}^2_\alpha (\vec\varepsilon) \propto \big[ (g_p^V +
    Y_n^\text{coh} g_n^V) + \varepsilon_{\alpha\alpha}^\text{coh}
    \big]^2 + \sum_{\beta\neq\alpha} \big(
  \varepsilon_{\alpha\beta}^\text{coh} \big)^2
\end{equation}
with an average value $Y_n^\text{coh} = 1.407$ and
\begin{equation}
  \label{eq:coh_eps-nucleon}
  \varepsilon_{\alpha\beta}^\text{coh} \equiv
  \varepsilon_{\alpha\beta}^p + Y_n^\text{coh}
  \varepsilon_{\alpha\beta}^n \,, \qquad \varepsilon_{\alpha\beta}^p
  \equiv 2\varepsilon_{\alpha\beta}^u + \varepsilon_{\alpha\beta}^d
  \,, \qquad \varepsilon_{\alpha\beta}^n \equiv
  2\varepsilon_{\alpha\beta}^d + \varepsilon_{\alpha\beta}^u \,.
\end{equation}
From Eq.~\eqref{eq:coh_Qalpha-nsi} it is evident that COHERENT can
only be sensitive to a certain combination of NSI operators
$\varepsilon_{\alpha\beta}^\text{coh}$, which are ultimately
determined by just two factors: (a) the value of $Y_n^\text{coh}$,
which depends on the nuclei in the detector, and (b) the strength of
the coupling of the new interaction to up and down quarks (or,
equivalently, to protons and neutrons). In fact, using the $\eta$
parametrisation in Eqs.~\eqref{eq:nsifit1_xi-eta}
and~\eqref{eq:nsifit1_epx-eta}, $\varepsilon_{\alpha\beta}^\text{coh}$
can be written as:
\begin{equation}
  \label{eq:coh_eps-coh}
  \varepsilon_{\alpha\beta}^\text{coh} = \sqrt{5} \left( \cos\eta +
  Y_n^\text{coh} \sin\eta \right) \varepsilon_{\alpha\beta}^\eta \,.
\end{equation}
It is clear from the expressions above that the best-fit value and
allowed ranges of $\varepsilon_{\alpha\beta}^\text{coh}$ implied by
COHERENT are independent of $\eta$. Once these have been determined,
the corresponding bounds on the associated couplings
$\varepsilon_{\alpha\beta}^\eta$ for a given NSI model (identified by
a particular value of $\eta$) can be obtained in a very simple way, by
just rescaling the values of $\varepsilon_{\alpha\beta}^\text{coh}$ as
$[\sqrt{5} (\cos\eta + Y_n^\text{coh} \sin\eta)]^{-1}$.  For example,
the results in Fig.~\ref{fig:coh_epscoh} can be immediately translated
in the corresponding ranges for NSI models where the new interaction
couples only to $f=u$, $f=d$, or $f=p$ ($\eta \approx 26.6^\circ$,
$63.4^\circ$, and $0$, respectively), after rescaling the bounds on
$\varepsilon_{\alpha\beta}^\text{coh}$ by the corresponding factors of
$0.293$, $0.262$, and $1$ in each case. Furthermore, from
Eq.~\eqref{eq:coh_eps-coh} it becomes evident that, for NSI models
with $\eta = \arctan(-1/Y_n^\text{coh}) \approx -35.4^\circ$, no bound
can be derived from COHERENT data. This could be improved by measuring CE$\nu$NS in nuclei with different neutron/proton ratios, which could be achieved in the future as explored in \cref{sec:ESS}.

The impact of the timing information on the fit can be readily
observed from the left panel in Fig.~\ref{fig:coh_epscoh}.  In the
absence of any timing information and using total rate information
alone, it is straightforward to show that, if the experiment observes
a result compatible with the SM expectation, the allowed confidence
regions in this plane should obey the equation of an ellipse. This
automatically follows from Eqs.~\eqref{eq:coh_Qbands}
and~\eqref{eq:coh_Qalpha-coh}:
\begin{equation}
  \label{eq:coh_ellipse}
   \frac{1}{3} [R + \varepsilon_{ee}^\text{coh}]^2 + \frac{2}{3} [R +
     \varepsilon_{\mu\mu}^\text{coh}]^2 = R^2 \,,
\end{equation}
where $R \equiv g_p^V + Y_n^\text{coh} g_n^V \approx -0.68$. This is
also shown by our numerical results in the left panel of
Fig.~\ref{fig:coh_epscoh} which do not include timing information in
the fit (dashed and dot-dashed contours).

While the inclusion of a non-zero $\varepsilon_{ee}^\text{coh}$ can be
compensated by a change in $\varepsilon_{\mu\mu}^\text{coh}$ that
brings the total number of events in the opposite direction without
significantly affecting the delayed events, this would be noticed in
the prompt event distribution once timing information is added to the
fit. In particular, too large/small values of
$\varepsilon_{ee}^\text{coh}$ would require a consequent modification
of the $\varepsilon_{\mu\mu}^\text{coh}$ to recover the same event
rate in the delayed time bins, which is however not allowed by the
prompt events observed. Thus, once timing information is included in
the fit the ellipse is broken in this plane and two separate minima
are obtained (dotted and solid lines).

It should also be noted that the central region in
Fig.~\ref{fig:coh_epscoh} (around the centre of the ellipse in
Eq.~\eqref{eq:coh_ellipse}, $\varepsilon_{ee}^\text{coh} =
\varepsilon_{\mu\mu}^\text{coh} = -R$) can be excluded at COHERENT
\emph{only} in the case when the off-diagonal NSI operators are not
included in the fit. This is so because in this region the effect of
the diagonal parameters leads to a destructive interference in the
total cross section and therefore to a reduction of the number of
events, in contrast with the experimental observation. Once the
off-diagonal operators are introduced this is no longer the case and
the central region becomes allowed~\cite{Giunti:2019xpr}. However,
since global neutrino oscillation data provide tight constraints on
the off-diagonal NSI operators, in our results the two minima remain
separate even after the off-diagonal operators are allowed in the fit,
as we will show in Sec.~\ref{sec:coh_resglob}.

\subsection{Summary}
\label{sec:coh_summary}

In this section, we analysed the
latest results obtained for coherent neutrino-nucleus scattering data
at the COHERENT experiment, which provide both energy and timing
information. 

We have quantified the dependence of our results for COHERENT with
respect to the choice of QF, nuclear form factor, and
the treatment of the backgrounds. We find that the implementation of
the steady-state background has a strong impact on the results of the
analysis of COHERENT due to a slight background excess in the first
two bins, which is present in both the coincident and anti-coincident
data samples provided by the collaboration. Once this effect has been
accounted for in the modelling of the expected backgrounds, the choice
of QF and nuclear form factor has a minor impact on the
results obtained from the fit.

\section{Combining COHERENT and oscillation data}
\label{sec:coh_plusOsc}

Once the results from the COHERENT experiments are properly understood
and analysed, their bounds on NSI can be combined with the global 
analyses of oscillation data in \cref{chap:NSIfit}. This will allow to
quantitatively assess the current complementarity between CE$\nu$NS and 
neutrino oscillation experiments.

\subsection{Including COHERENT in the CP-conserving analysis of \texorpdfstring{Section~\ref{sec:nsifit1}}{Section~6.1}}
\label{sec:coh_resglob}

As a start, we present the 
results of the global CP-conserving analysis of oscillation plus 
COHERENT data. We will show two
main sets of results, which quantify: (1) the quality of the global
fit once NSI are allowed, compared to the fit obtained under the SM
hypothesis; and (2) the status of the LMA-D solution after the
inclusion of COHERENT energy and timing data in the global fit. Both
sets of results are presented for a wide range of NSI models (that is,
for different values of $\eta$). Finally, we also provide the allowed
ranges obtained for the $\varepsilon_{\alpha \beta}^f$ for three particular NSI
models, assuming that the new mediator couples predominantly to either
up/down quarks or to protons.

To this end we construct a combined $\chi^2$ function
\begin{equation}
  \chi^2_\text{global}(\vec\varepsilon) = \min\limits_{\vec\omega}
  \left[ \chi^2_\text{OSC}(\vec\omega, \vec\varepsilon) +
    \chi^2_\text{COH}(\vec\varepsilon) \right] \,,
\end{equation}
where we denote by $\vec\omega \equiv \{ \theta_{ij},
\delta_\text{CP}, \Delta m^2_{ji} \}$ the ``standard'' $3\nu$
oscillation parameters.  For the detailed description of methodology
and data included in $\chi^2_\text{OSC}$ we refer to the comprehensive
global fit in \cref{sec:nsifit1} performed in the framework of
three-flavour oscillations plus NSI with quarks parametrised as
Eqs.~\eqref{eq:nsifit1_xi-eta} and~\eqref{eq:nsifit1_epx-eta}. In this
section we minimally update the results from \cref{sec:nsifit1} to
account for the latest LBL data samples discussed in \cref{sec:dCPevolution}. To keep
the fit manageable in \cref{sec:nsifit1} only the CP-conserving case
with real NSI and $\delta_\text{CP} \in \{ 0, \pi \}$ was considered,
and consequently the T2K and NO$\nu$A appearance data (which exhibit
substantial dependence on the leptonic CP phase) were not included in
the fit. Here we follow the same approach and consistently update only
the disappearance samples from these experiments.\footnote{For a
  discussion of CP violation in the presence of NSI, see
  \cref{chap:NSItheor,sec:nsifit2,sec:coh_dCP}.}

\begin{figure}[hbtp]\centering
  \includegraphics[width=\textwidth]{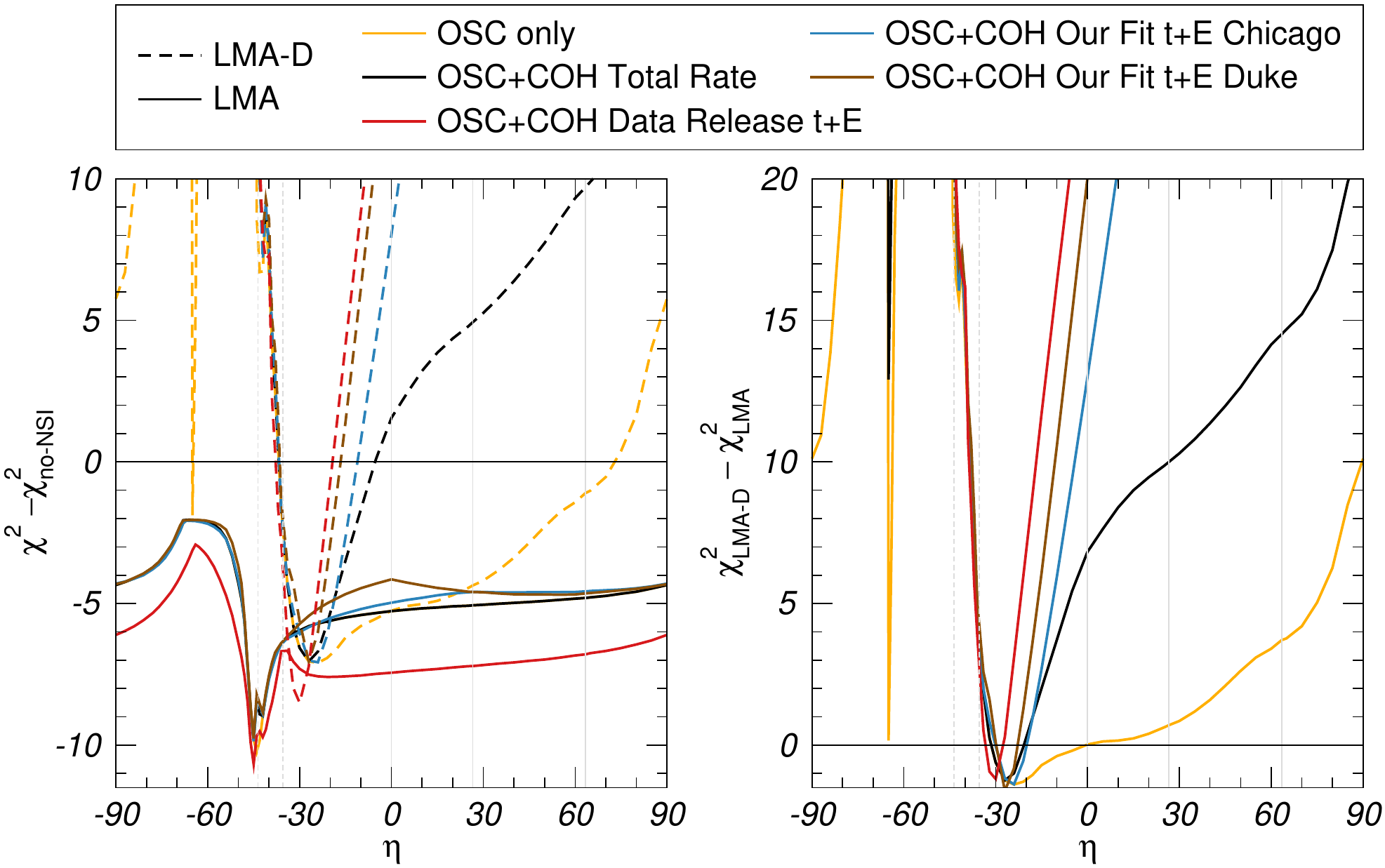}
  \caption{Left: $\chi^2_\text{LMA}(\eta) - \chi^2_\text{no-NSI}$
    (full lines) and $\chi^2_\text{LMA-D}(\eta) -
    \chi^2_\text{no-NSI}$ (dashed lines) for the analysis of different
    data combinations (as labeled in the figure) as a function of the
    NSI quark coupling parameter $\eta$.  Right:
    $\chi^2_\text{dark}-\chi^2_\text{light}\equiv
    \chi^2_\text{LMA-D}(\eta) - \chi^2_\text{LMA}(\eta)$ as a function
    of $\eta$. See text for details.}
  \label{fig:coh_chisq-eta}
\end{figure}

Figure~\ref{fig:coh_chisq-eta} shows the impact of COHERENT on the
global fit (left panel) as well as on the LMA-D degeneracy (right
panel).  In doing so, we have defined the functions
$\chi^2_\text{LMA}(\eta)$ and $\chi^2_\text{LMA-D}(\eta)$, obtained by
marginalising $\chi^2_\text{global}(\vec\omega, \vec\varepsilon)$ over
both $\vec\omega$ and $\vec\varepsilon$ for a given value of $\eta$,
with the constraint $\theta_{12} < 45^\circ$ (in the LMA case) and
$\theta_{12} > 45^\circ$ (for LMA-D).  With these definitions, we
show in the left panel the differences $\chi^2_\text{LMA}(\eta) -
\chi^2_\text{no-NSI}$ (full lines) and $\chi^2_\text{LMA-D}(\eta) -
\chi^2_\text{no-NSI}$ (dashed lines), where $\chi^2_\text{no-NSI}$ is
the minimum $\chi^2$ for standard $3\nu$ oscillations (i.e., setting
all the NSI parameters to zero). Then, in the right panel we show the
values of $\chi^2_\text{LMA-D}(\eta) - \chi^2_\text{LMA}(\eta)$, which
quantifies the relative quality of the LMA and LMA-D solutions as a
function of $\eta$.

First, from the left panel in Fig.~\ref{fig:coh_chisq-eta} we notice
that the introduction of NSI leads to a substantial improvement of the
fit already for the LMA solution (solid lines) with respect to the
oscillation data analysis, resulting in a sizable decrease of the
minimum $\chi^2_\text{LMA}$ with respect to the standard oscillation
scenario.  This is mainly driven by a well-known tension (although
mild, at the level of $\Delta\chi^2\sim 7.4$ in the present analysis)
between solar and KamLAND data in the determination of $\Delta
m^2_{21}$ (see \cref{subsec:nufit3_dm12,sec:nsifit1_solar}).  As seen in the figure, the inclusion of NSI improves the
combined fit by about $2.2\sigma$ over a broad range of values of
$\eta$.  The improvement is maximised for NSI models with values of
$\eta$ for which the effect is largest in the Sun without entering in
conflict with terrestrial experiments.  This occurs for $\eta \simeq
-44^\circ$ (as for this value the NSI in the Earth matter essentially
cancel) and leads to an improvement of about $10$ units in $\chi^2$
(i.e., a $\sim 3.2\sigma$ effect).  From the figure we also conclude
that adding the information from COHERENT on rate only, as well as on
timing and energy (t+E), still allows for this improved fit in the LMA
solution for most values of $\eta$. Indeed, the maximum effect at
$\eta \simeq -44^\circ$ still holds after the combination since it
falls very close to $-35.4^\circ$, for which NSI effects cancel at
COHERENT as seen in Eq.~\eqref{eq:coh_eps-coh}.  Interestingly, the
improvement is slightly larger for the combination with COHERENT t+E
data using the data release assumptions. This is so because, as
described in the previous section, in this case the fit pulls the weak
charge $\mathcal{Q}^2_e$ towards zero (see Fig.~\ref{fig:coh_qw})
while leaving the value of $\mathcal{Q}^2_\mu$ around the SM
expectation. Such situation can be easily accommodated by invoking
diagonal NSI operators and, in particular, favours the non-standard
values $\varepsilon_{ee}^\eta - \varepsilon_{\mu\mu}^\eta \neq 0$,
thus bringing the fit to a better agreement with solar+KamLAND
oscillation data.

Most importantly, Fig.~\ref{fig:coh_chisq-eta} shows that the main
impact of including COHERENT data in the analysis is on the status of
the LMA-D degeneracy. We see in the figure that with oscillation data
alone the LMA-D solution is still allowed at $3\sigma$ for a wide
range of NSI models ($-38^\circ \lesssim \eta \lesssim 87^\circ$, as
well as a narrow window around $\eta \simeq -65^\circ$) and, in fact,
for $-31^\circ \lesssim \eta\lesssim 0^\circ$ it provides a slightly
better global fit than the LMA solution. The addition of COHERENT to
the analysis of oscillation data disfavours the LMA-D degeneracy for
most values of $\eta$, and the inclusion of the timing and energy
information makes this conclusion more robust.  More quantitatively we
find that, when COHERENT results are taken into account, LMA-D is
allowed below $3\sigma$ only for values of $\eta$ in the following
ranges:
\begin{equation}
  \label{eq:coh_LMADcoh}
  \begin{aligned}
    -38^\circ &\lesssim \eta \lesssim \hphantom{+}15^\circ
    &&\text{COHERENT Total Rate,} \\ -38^\circ &\lesssim \eta \lesssim
    -18^\circ && \text{COHERENT t+E Data Release,} \\ -38^\circ
    &\lesssim \eta \lesssim \hphantom{0}{-6^\circ} && \text{COHERENT
      t+E Our Fit Chicago,} \\ -38^\circ &\lesssim \eta \lesssim
    -12^\circ &&\text{COHERENT t+E Our Fit Duke.}
  \end{aligned}
\end{equation}

\begin{figure}\centering
  \includegraphics[width=\textwidth]{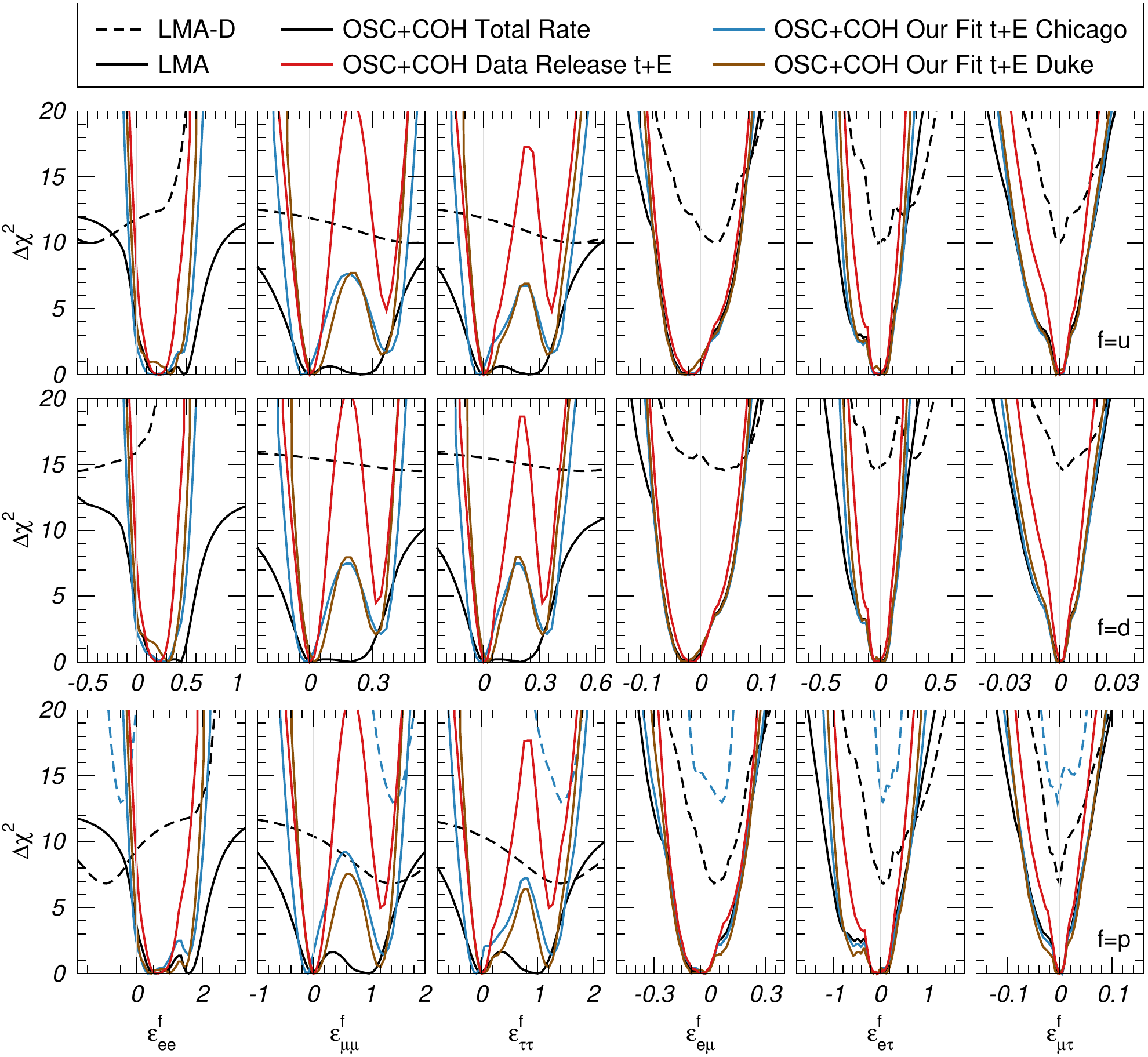}
  \caption{Dependence of the $\Delta\chi^2_\text{global}$ function on
    the NSI couplings with up quarks (upper row), down quark (central
    row) and, protons (lower row) for the global analysis of
    oscillation and COHERENT data. In each panel
    $\chi^2_\text{global}$ is marginalised with respect to the other
    five NSI couplings not shown and with respect to the oscillation
    parameters for the LMA (solid) and LMA-D (dashed) solutions.  The
    different curves correspond to the different variants of the
    COHERENT analysis implemented in this chapter: total rate (black),
    t+E Data Release (red), t+E with QF-C (blue), and t+E with QF-D
    (brown); see text for details.}
  \label{fig:coh_chisq-qrk}
\end{figure}

Finally, we provide in Fig.~\ref{fig:coh_chisq-qrk} the $\chi^2$
profiles for each of the six NSI coefficients after marginalisation
over the undisplayed oscillation parameters and the other five NSI
coefficients not shown in a given panel. We show these results for
three representative cases of NSI models including couplings to up
quarks only, down quarks only and to protons. The corresponding
$2\sigma$ ranges are also provided in Tab.~\ref{tab:coh_ranges} for
convenience.  This figure shows that the LMA-D solution for NSI models
that couple only to protons ($\eta = 0$) can only be excluded beyond
$3\sigma$ if both energy and timing information are included for
COHERENT, in agreement with Eq.~\eqref{eq:coh_LMADcoh}. From this
figure we also see that for the LMA solution the allowed ranges for
the off-diagonal NSI couplings are only moderately reduced by the
addition of the COHERENT results and, moreover, the impact of the
energy and timing information is small in these cases.  This is
because they are already very well constrained by oscillation data
alone.

\begin{sidewaystable}[!p]
  \centering
  \begin{tabular}{lcccc}
    \toprule & Total Rate & Data Release t+E & Our Fit t+E Chicago & Our
    Fit t+E Duke \\ \cmidrule(l){2-5} $\varepsilon_{ee}^u$ & $[-0.012, +0.621]$ &
    $[+0.043, +0.384]$ & $[-0.032, +0.533]$\hfill~ & $[-0.004,
      +0.496]$\hfill~ \\ $\varepsilon_{\mu\mu}^u$ & $[-0.115, +0.405]$
    & $[-0.050, +0.062]$ & $[-0.094, +0.071] \oplus [+0.302, +0.429]$
    & $[-0.045, +0.108] \oplus [+0.290, +0.399]$
    \\ $\varepsilon_{\tau\tau}^u$ & $[-0.116, +0.406]$ & $[-0.050,
      +0.065]$ & $[-0.095, +0.125] \oplus [+0.302, +0.428]$ &
    $[-0.045, +0.141] \oplus [+0.290, +0.399]$
    \\ $\varepsilon_{e\mu}^u$ & $[-0.059, +0.033]$ & $[-0.055,
      +0.027]$ & $[-0.060, +0.036]$\hfill~ & $[-0.060, +0.034]$\hfill~
    \\ $\varepsilon_{e\tau}^u$ & $[-0.250, +0.110]$ & $[-0.141,
      +0.090]$ & $[-0.243, +0.118]$\hfill~ & $[-0.222, +0.113]$\hfill~
    \\ $\varepsilon_{\mu\tau}^u$ & $[-0.012, +0.008]$ & $[-0.006,
      +0.006]$ & $[-0.013, +0.009]$\hfill~ & $[-0.012, +0.009]$\hfill~
    \\ \hline $\varepsilon_{ee}^d$ & $[-0.015, +0.566]$ & $[+0.036,
      +0.354]$ & $[-0.030, +0.468]$\hfill~ & $[-0.006, +0.434]$\hfill~
    \\ $\varepsilon_{\mu\mu}^d$ & $[-0.104, +0.363]$ & $[-0.046,
      +0.057]$ & $[-0.083, +0.077] \oplus [+0.278, +0.384]$ &
    $[-0.037, +0.099] \oplus [+0.267, +0.356]$
    \\ $\varepsilon_{\tau\tau}^d$ & $[-0.104, +0.363]$ & $[-0.046,
      +0.059]$ & $[-0.083, +0.083] \oplus [+0.279, +0.383]$ &
    $[-0.038, +0.104] \oplus [+0.268, +0.354]$
    \\ $\varepsilon_{e\mu}^d$ & $[-0.058, +0.032]$ & $[-0.052,
      +0.024]$ & $[-0.059, +0.034]$\hfill~ & $[-0.058, +0.034]$\hfill~
    \\ $\varepsilon_{e\tau}^d$ & $[-0.198, +0.103]$ & $[-0.106,
      +0.082]$ & $[-0.196, +0.107]$\hfill~ & $[-0.181, +0.101]$\hfill~
    \\ $\varepsilon_{\mu\tau}^d$ & $[-0.008, +0.008]$ & $[-0.005,
      +0.005]$ & $[-0.008, +0.008]$\hfill~ & $[-0.007, +0.008]$\hfill~
    \\ \hline $\varepsilon_{ee}^p$ & $[-0.035, +2.056]$ & $[+0.142,
      +1.239]$ & $[-0.095, +1.812]$\hfill~ & $[-0.024, +1.723]$\hfill~
    \\ $\varepsilon_{\mu\mu}^p$ & $[-0.379, +1.402]$ & $[-0.166,
      +0.204]$ & $[-0.312, +0.138] \oplus [+1.036, +1.456]$ &
    $[-0.166, +0.337] \oplus [+0.952, +1.374]$
    \\ $\varepsilon_{\tau\tau}^p$ & $[-0.379, +1.409]$ & $[-0.168,
      +0.257]$ & $[-0.313, +0.478] \oplus [+1.038, +1.453]$ &
    $[-0.167, +0.582] \oplus [+0.950, +1.382]$
    \\ $\varepsilon_{e\mu}^p$ & $[-0.179, +0.112]$ & $[-0.174,
      +0.086]$ & $[-0.179, +0.120]$\hfill~ & $[-0.187, +0.131]$\hfill~
    \\ $\varepsilon_{e\tau}^p$ & $[-0.877, +0.340]$ & $[-0.503,
      +0.295]$ & $[-0.841, +0.355]$\hfill~ & $[-0.817, +0.386]$\hfill~
    \\ $\varepsilon_{\mu\tau}^p$ & $[-0.041, +0.025]$ & $[-0.020,
      +0.019]$ & $[-0.044, +0.026]$\hfill~ & $[-0.048, +0.030]$\hfill~
    \\ \hline
  \end{tabular}
  \caption{$2\sigma$ allowed ranges for the NSI couplings
    $\varepsilon_{\alpha\beta}^u$, $\varepsilon_{\alpha\beta}^d$ and
    $\varepsilon_{\alpha\beta}^p$ as obtained from the global analysis
    of oscillation plus COHERENT data. See text for details.}
  \label{tab:coh_ranges}
  \afterpage\clearpage
\end{sidewaystable}

More interestingly, the addition of COHERENT data allows to derive
constraints on each of the diagonal parameters separately and, for
those, the timing (and to less degree energy) information has a
quantitative impact. In particular we see in the figure the appearance
of the two minima corresponding to the degenerate solutions for
$\varepsilon_{\mu\mu}^\text{coh}$ in Fig.~\ref{fig:coh_epscoh},
obtained after the inclusion of timing information for
COHERENT. Noticeably, now the non-standard solution (obtained for
$\varepsilon_{\mu\mu}^f \neq 0$) is partially lifted by the
combination with oscillation data, but it remains well allowed around
$\sim 2\sigma$ depending on the assumptions for the COHERENT analysis.
Figure~\ref{fig:coh_chisq-qrk} also shows the corresponding two minima
for $\varepsilon_{\tau\tau}^f$ arising from the combination of the
information on $\varepsilon_{\tau\tau}^f - \varepsilon_{\mu\mu}^f$ and
$\varepsilon_{ee}^f - \varepsilon_{\mu\mu}^f$ from the oscillation
experiments with the constraints on $\varepsilon_{\mu\mu}^f$ from the
COHERENT t+E data.  In particular, in all the three cases $f=u$, $d$,
$p$ shown in the figure the bound on $\varepsilon_{\tau\tau}^f$
becomes about two orders of magnitude stronger than previous indirect
(loop-induced) limits~\cite{Davidson:2003ha} when the t+E analysis
with the data release assumptions is used. Indeed this conclusion
holds for most $\eta$ values, with exception of $\eta \sim -45^\circ$
to $-35^\circ$ for which NSI effects are suppressed in either the
Earth matter or in COHERENT. In particular, for $\eta=-35.4^\circ$ the
NSI effects in COHERENT totally cancel, as described above, and
consequently no separate determination of the three diagonal
parameters is possible around such value.

\subsubsection{Summary}
In this section, we have combined neutrino oscillation data with the 
latest results from the COHERENT experiment. The results of our analysis 
are used to constrain the
whole set of neutral current operators
leading to CP-conserving NSI involving up and down quarks simultaneously.

We have found that the inclusion of COHERENT timing information affects the
global fit significantly and, most notably, has a large impact on the
constraints that can be derived for the flavour-diagonal NSI
operators, for which separate constraints can only be derived after
combination of COHERENT and oscillation data.

Furthermore, the presence of NSI is known to introduce a degeneracy in
the oscillation probabilities for neutrinos propagating in matter,
leading in particular to the appearance of the LMA-D
solution.  We find that the inclusion of COHERENT to the analysis of
oscillation data disfavours the LMA-D degeneracy for NSI models over a
wide range of $\eta$, and the addition of the timing and energy
information makes this conclusion more robust, see
Eq.~\eqref{eq:coh_LMADcoh} and Fig.~\ref{fig:coh_chisq-eta}. In
particular, the LMA-D solution for NSI models that couple only to
protons ($\eta = 0$) can only be excluded beyond $3\sigma$ once both
energy and timing information are included for the COHERENT data.

Finally, the introduction of NSI is known to alleviate the well-known
(albeit mild) tension between solar and KamLAND data in the
determination of the $\Delta m^2_{21}$, thus leading to an overall
improvement in the quality of the global fit to oscillation data.  We
find that this still remains the case after the inclusion of COHERENT
results.

\subsection{Including COHERENT in the CP violation and mass ordering analysis of \texorpdfstring{Section~\ref{sec:nsifit2}}{Section~6.2}}
\label{sec:coh_dCP}

The results from the COHERENT experiment impose relevant constraints
on NSI that could improve the robust determination of leptonic CP violation
in LBL accelerator experiments. In this section, we will add the
bounds from COHERENT to the CP-violating NSI analysis in
\cref{sec:nsifit2} (see that section for the details of the fit). We will use the time and energy information; and for the background, nuclear form factor and QF we use the procedure named as ``Our fit t+E Duke'' in the previous section. Nevertheless, the results of the fit do not significantly depend on the nuclear form factor and QF assumed.

To begin with, in
\cref{fig:coh2_triangle-light,fig:coh2_triangle-dark} we show all the
possible one-dimensional and two-dimensional projections of the
eleven-dimensional parameter space after including COHERENT
data. Comparing with
\cref{fig:nsifit2_triangle-light,fig:nsifit2_triangle-dark}, we see
that COHERENT improves the constraints on large NSI moduli. This will
be relevant for CP violation.

\begin{pagefigure}\centering
  \includegraphics[width=0.99\textwidth]{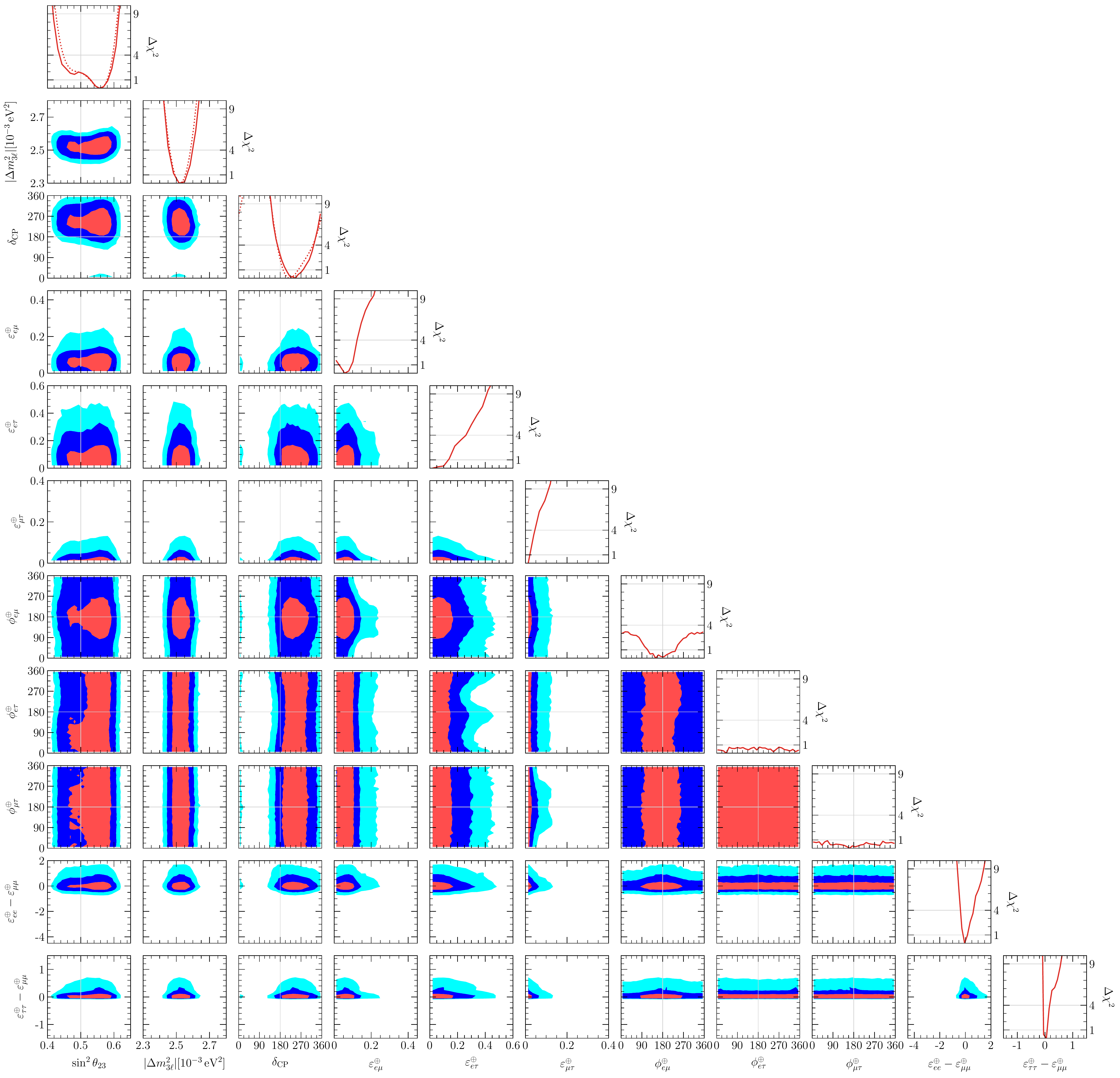}
  \caption{Global analysis of solar, atmospheric, reactor and
    accelerator oscillation experiments plus COHERENT, in the LIGHT
    side of the parameter space and for Normal Ordering of the
    neutrino states.  The panels show the two-dimensional projections
    of the allowed parameter space after marginalisation with respect
    to the undisplayed parameters.  The different contours correspond
    to the allowed regions at $1\sigma$, $2\sigma$ and $3\sigma$ for
    2~degrees of freedom.  Note that as atmospheric mass-squared
    splitting we use $\Delta m^2_{3\ell} = \Delta m^2_{31}$ for
    NO. Also shown are the one-dimensional projections as a function
    of each parameter. For comparison we show as dotted lines the
    corresponding one-dimensional dependence for the same analysis
    assuming only standard $3\nu$ oscillation (i.e., setting all the
    NSI parameters to zero).}
  \label{fig:coh2_triangle-light}
\end{pagefigure}

\begin{pagefigure}\centering
  \includegraphics[width=0.99\textwidth]{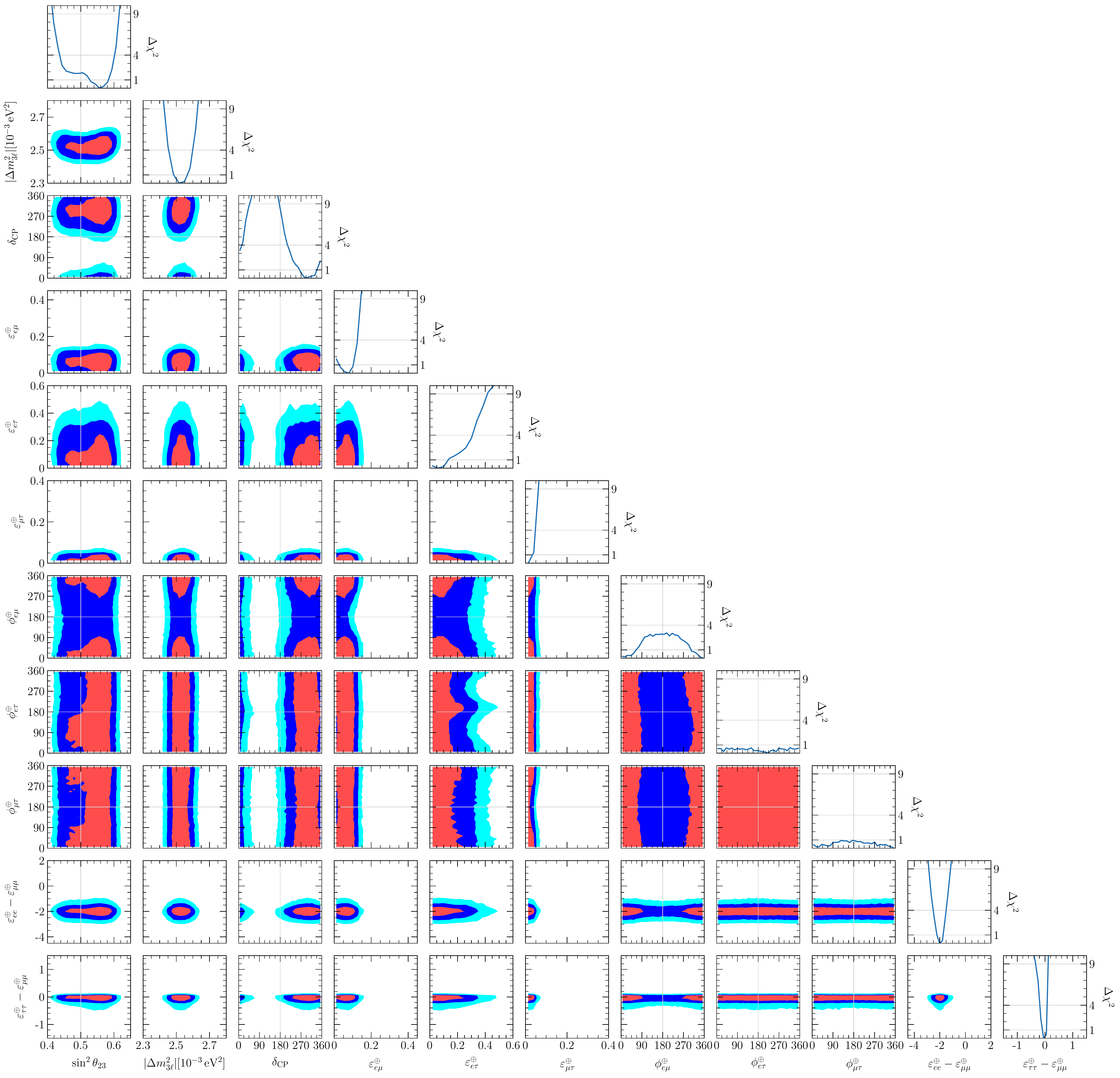}
  \caption{Same as Fig.~\ref{fig:nsifit2_triangle-dark} but for
    DARK-IO solution.  In this case $\Delta m^2_{3\ell} = \Delta
    m^2_{32}<0$ and we plot its absolute value. The regions and
    one-dimensional projections are defined with respect to the
    \emph{local} minimum in this sector of the parameter space.}
  \label{fig:coh2_triangle-dark}
\end{pagefigure}

\begin{figure}\centering
  \includegraphics[width=\textwidth]{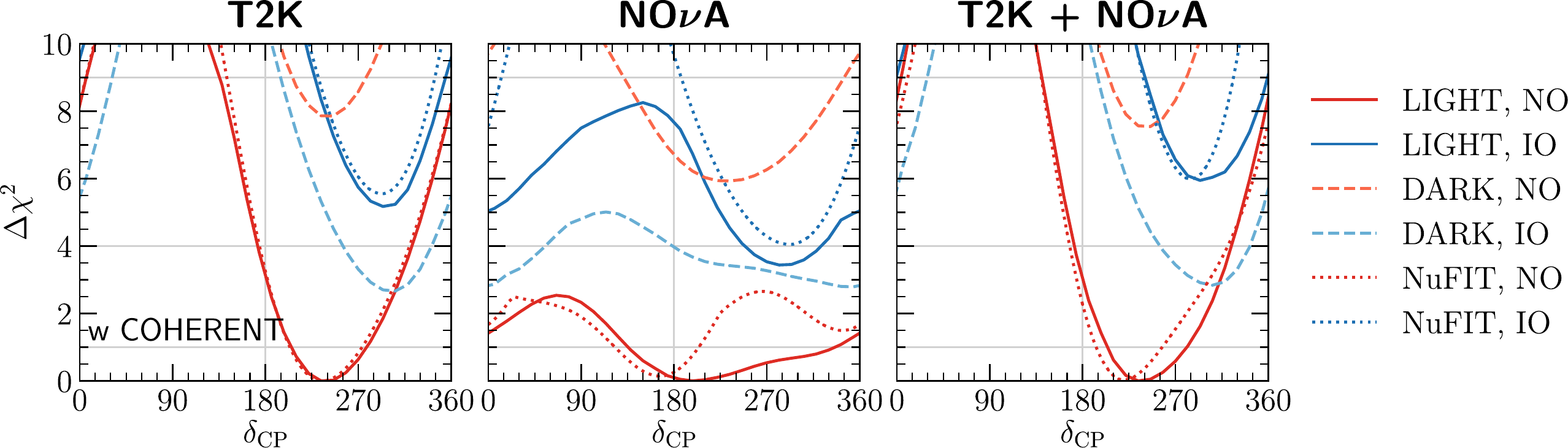}
  \caption{$\Delta\chi^2$ as a function of $\delta_\text{CP}$ after
    marginalising over all the undisplayed parameters, for different
    combination of experiments. We include $\text{SOLAR} +
    \text{KamLAND} + \text{ATM} + \text{MBL-REA} + \text{MINOS} + \text{COHERENT}$
    to which we add T2K (left), \NOvA/ (center) and $\text{T2K} +
    \text{NO}\nu\text{A}$ (right). The different curves are obtained
    by marginalising within different regions of the parameter space,
    as detailed in the legend. See \cref{fig:nsifit2_chi2dcpnsi} and
    the text around it for details.}
  \label{fig:coh2_chi2dcpnsi}
\end{figure}

To quantify the impact, in Fig.~\ref{fig:coh2_chi2dcpnsi} we plot the
one-dimensional $\chi^2(\delta_\text{CP})$ function after
marginalising over the ten undisplayed parameters. In the left,
central and right panels we include T2K, \NOvA/, and T2K+\NOvA/
respectively. In each panel we plot the curves obtained marginalising
separately in NO (red curves) and IO (blue curves) for the LMA and
LMA-D solutions. For the sake of comparison we also plot the
corresponding $\chi^2(\delta_\text{CP})$ from the $3\nu$ oscillation
analysis with the SM matter potential (labeled ``NuFIT'' in the
figure). These results can be compared with the ones in
\cref{fig:nsifit2_chi2dcpnsi}, that do not include COHERENT.

\clearpage
\begin{figure}\centering
  \includegraphics[width=0.75\textwidth]{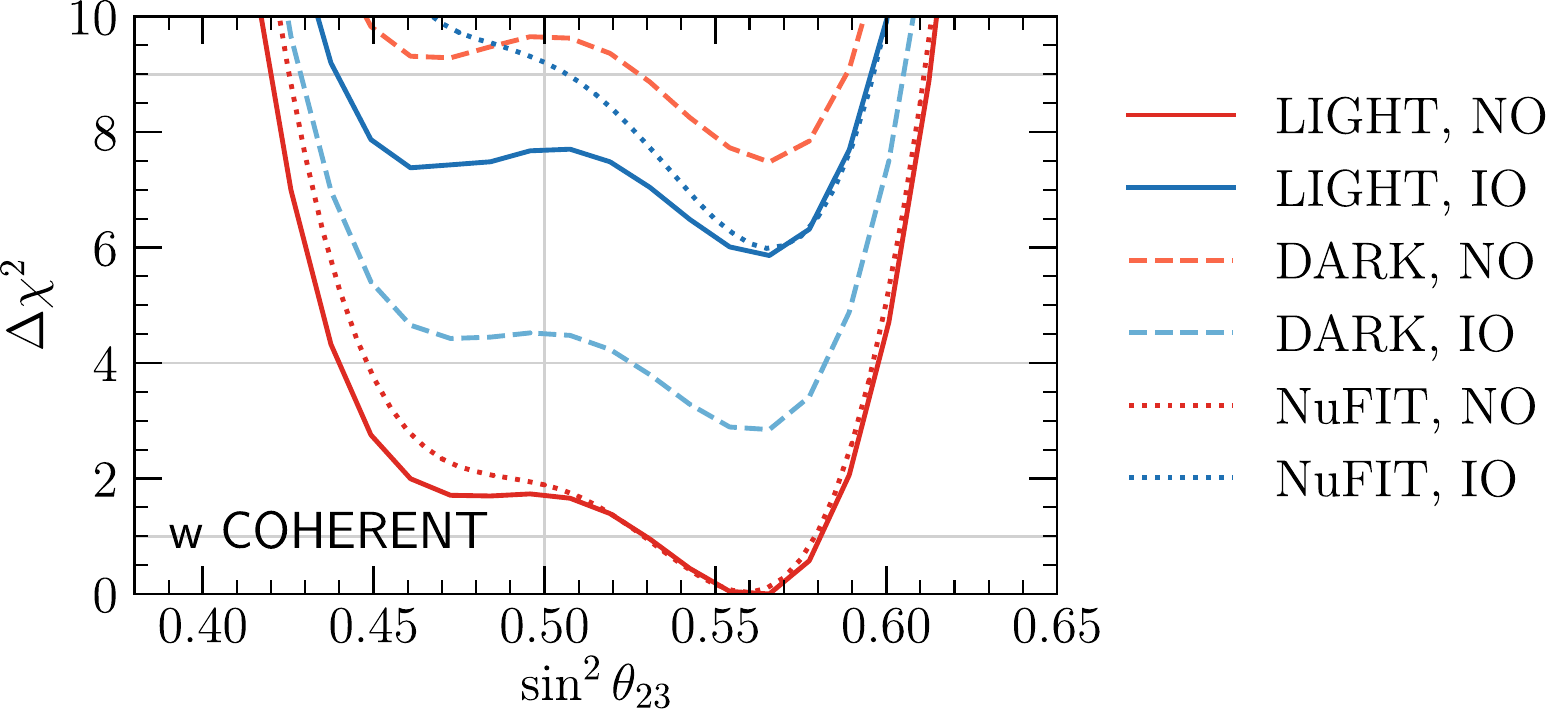}
  \caption{$\Delta\chi^2$ as a function of $\sin^2\theta_{23}$ after
    marginalising over all other parameters for the global combination
    of oscillation experiments + COHERENT.  The different curves
    correspond to marginalisation within the different regions of the
    parameter space, as detailed in the legend. See
    \cref{fig:nsifit2_t23} and the text around it for details.}
  \label{fig:coh2_t23}
\end{figure}

 We see that, regarding T2K, when including COHERENT,
\begin{itemize}
\item The robustness of its hint for maximal CP violation still
  holds.
\item For T2K without COHERENT, the DARK, IO solution provided a
  better fit than LIGHT, NO for several values of
  $\delta_\text{CP}$ (see \cref{fig:nsifit2_chi2dcpnsi}). These 
  solutions involved large $\varepsilon_{e
    \mu}$ and $\varepsilon_{e \tau}$, that are no longer allowed after
  including COHERENT data.  As a consequence, the statistical
  significance of the hint for CP violation in T2K is now essentially the same for 
  the DARK solutions as for the LIGHT ones.
\end{itemize}

For \NOvA/, including COHERENT has a marginal effect, and it mostly
reduces to excluding the DARK, NO solution with higher statistical
significance. In other words, the NSI required to spoil the \NOvA/
results are not yet within the reach of COHERENT. Finally, in the
global analysis, COHERENT enhances the robustness of the mass ordering
determination within the LIGHT solution. This is due to the relatively
large NSI that spoiling this determination requires, constrained by COHERENT.

In addition, in \cref{fig:coh2_t23} we show the one-dimensional
$\chi^2(\theta_{23})$ after including COHERENT data, to be compared
with \cref{fig:nsifit2_t23}. The main effect is that 
$\theta_{23} \sim 45^\circ$ for the DARK-NO solution, which required
large NSI, is now more constrained.

Overall, we see that the current COHERENT data enhances
the robustness of the fit regarding the LMA-D solutions. These, that
could flatten $\Delta \chi^2$ and reduce the sensitivity of LBL
experiments to the currently unknown parameters, are efficiently
explored by COHERENT. Nevertheless, COHERENT cannot significantly 
exclude the LMA-D solution with respect to LMA. The reason for this is 
that the analysis presented in this section marginalises over $\eta$, 
whose best-fit value in the LMA-D solution, $\eta \sim -30^\circ$, 
lies close to $\eta = \arctan(-1/Y_n^\text{coh}) \approx -35.4^\circ$ 
where NSI at COHERENT cancel. This could be overcome by measuring 
CE$\nu$NS with different nuclei, a possibility that will be explored 
below.

\section{Future prospects: coherent elastic neutrino-nucleus scattering at the European Spallation Source}
\label{sec:ESS}

As shown by the analyses above, the first data release from COHERENT
already increased the robustness with which LBL accelerator
experiments can assess leptonic CP violation. This programme should be
further pursued in the near future to have better bounds when DUNE and
Hyper-Kamiokande start taking data. COHERENT, though, is still limited
by statistical uncertainties as seen for instance in \cref{fig:coh_histo-res}.

\begin{figure}[!htbp]
\centering \includegraphics[width=0.75\textwidth]{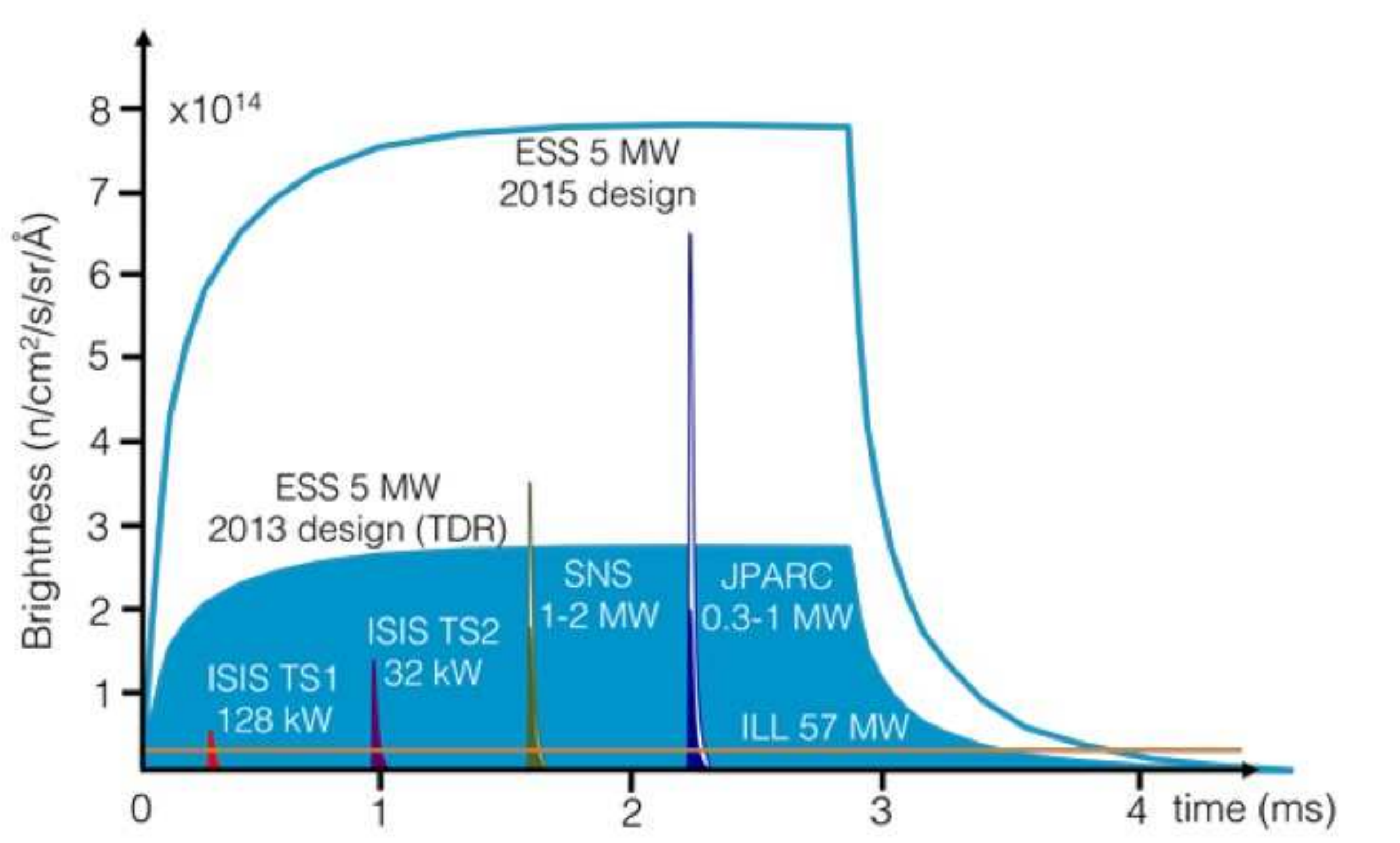}
\caption{(Source: ESS) Neutron production from existing and planned
  spallation sources. The nominal SNS power is 1 MW at proton energy 1
  GeV, with a plan to reach 2 MW by 2026. The ESS power will be 5 MW
  at 2 GeV circa 2023, with the ability to further
  upgrade. Differences in the duration of the protons-on-target (POT)
  pulse are visible in the figure. The ESS will generate an increase
  in neutron brightness by a factor 30-100 with respect to previous
  spallation sources, and an order of magnitude larger neutrino yield
  than the SNS.}
\label{fig:ess_ess}
\end{figure}

Luckily, one can perform high-statistics CE$\nu$NS measurements
provided by the upcoming European Spallation Source (ESS) sited in
Lund, Sweden. The ESS will combine the world's most powerful
superconducting proton linac with an advanced hydrogen moderator,
generating the most intense neutron beams for multi-disciplinary
science (see Fig.~\ref{fig:ess_ess}). The operating principle of the 
facility is the same as for the SNS, whose 
neutrinos COHERENT detects. However, the ESS will provide an order of
magnitude increase in neutrino flux with respect to the SNS due to a
higher proton energy and luminosity. This will facilitate CE$\nu$NS
measurements not limited in their sensitivity to new physics by poor
signal statistics, while still employing non-intrusive, compact (few
kg) neutrino detectors, able to operate without interference with ESS
neutron activities.

Furthermore, as the facility is still under construction, large areas
could be allocated for detectors. These could be more modern,
sensitive to smaller nuclear recoils where the CE$\nu$NS cross
section~ \eqref{eq:coh_xsec-SM} increases, and of different materials 
to increase the sensitivity to different NSI models. The detailed 
proposal and detector details are explained in 
Ref.~\cite{Baxter:2019mcx} and summarised in \cref{tab:ess_detectors}.
Here, we quote the main results concerning NSI after three years of
data-taking.  We have assumed an 80\% detector acceptance and a 10\%
signal normalisation systematic uncertainty. Apart from that, the 
number of neutrino events is calculated exactly as for COHERENT (see 
\cref{sec:coh_coh1,sec:coh_coh2}), although due to longer beam pulses 
(see \cref{fig:ess_ess}) no useful timing information is available at 
the ESS. 

\begin{table}[!htbp]
\makebox[\textwidth][c]{
\renewcommand{\arraystretch}{1.4} \centering
 \begin{tabular}{@{\hspace*{2pt}}c@{\hspace*{2pt}}|@{\hspace*{2pt}}c@{\hspace*{2pt}}|@{\hspace*{2pt}}c@{\hspace*{2pt}}|@{\hspace*{2pt}}c@{\hspace*{2pt}}|@{\hspace*{2pt}}c@{\hspace*{2pt}}|@{\hspace*{2pt}}c@{\hspace*{2pt}}|@{\hspace*{2pt}}c@{\hspace*{2pt}}|@{\hspace*{2pt}}c@{\hspace*{2pt}}}
 		\toprule Detector Technology & 
 		\begin{tabular}[c]{@{}c@{}} Target \\ nucleus \end{tabular} & 
 		\begin{tabular}[c]{@{}c@{}} Mass \\ (kg) \end{tabular} &
 		\begin{tabular}[c]{@{}c@{}} Steady-state \\ background \end{tabular} & 
 		\begin{tabular}[c]{@{}c@{}} $E_{th}$ \\ (keV$_{nr}$) \end{tabular} & 
 		\begin{tabular}[c]{@{}c@{}} $\frac{\Delta E}{E}$ (\%) \\ at $E_{th}$ \end{tabular} &
 		\begin{tabular}[c]{@{}c@{}} $E_{\mathrm{max}}$ \\ (keV$_{nr}$) \end{tabular} & 
 		\begin{tabular}[c]{@{}c@{}} CE$\nu$NS $\frac{\rm NR}{\rm yr}$ \\ @20m, $>E_{th}$ \end{tabular}
                \\ \midrule 
                Cryogenic scintillator & CsI & 22.5 & 10 ckkd& 1 & 30
                & 46.1 & 8,405 \\ Charge-coupled device & Si & 1 & 0.2
                ckkd & 0.16 & 60 & { 212.9} & 80 \\ High-pressure
                gaseous TPC & Xe & 20 & 10 ckkd & 0.9 & 40 & 45.6 &
                7,770 \\ p-type point contact HPGe & Ge & 7 & 3 ckkd &
                0.6 & 15 & 78.9 & 1,610 \\ Scintillating bubble
                chamber & Ar & 10 & 0.1 c/kg-day & 0.1 & $\sim$40 &
                150.0 & 1,380 \\ Standard bubble chamber &
                C$_{3}$F$_{8}$ & 10 & 0.1 c/kg-day & 2 & 40 & 329.6 &
                515 \\ \bottomrule
  \end{tabular}
}
	\caption{\label{tab:ess_detectors} Summary of detector
          properties for CE$\nu$NS at the ESS. We show the target
          nuclei and mass of different detectors, the steady-state
          background, the minimum detectable nuclear recoil energy
          $E_{th}$, the relative energy resolution $\Delta E/E$ at
          $E_{th}$, the maximum detectable nuclear recoil energy
          E$_\mathrm{max}$, and the amount of CE$\nu$NS events per
          year above threshold for a detector at \SI{20}{m} from the
          source.  Backgrounds listed do not include a
          $4\times10^{-2}$ reduction by the pulsed character of the
          ESS beam. The background for bubble chambers is integrated
          above nucleation threshold (in counts per kg and day), and
          only the total event rate is used in the simulations.  Other
          backgrounds are given in counts per keV, kg and day
          (ckkd). We conservatively adopt the background at $E_{th}$,
          which is typically maximal, for all higher energies.  The
          relative energy resolution is assumed to depend on the
          nuclear recoil energy as $\propto \sqrt{E}$.}
\end{table}

In what follows, we focus
on the determination of the flavour-diagonal NSI coefficients,
$\epsilon_{\alpha\alpha}^{f}$ ($f=u,d$), although it should be kept in
mind that coherent neutrino scattering is also sensitive to all the
off-diagonal NSI operators as well, and competitive bounds should also
be expected for those.

\begin{figure*}[ht!]
\centering \includegraphics[width=\textwidth]{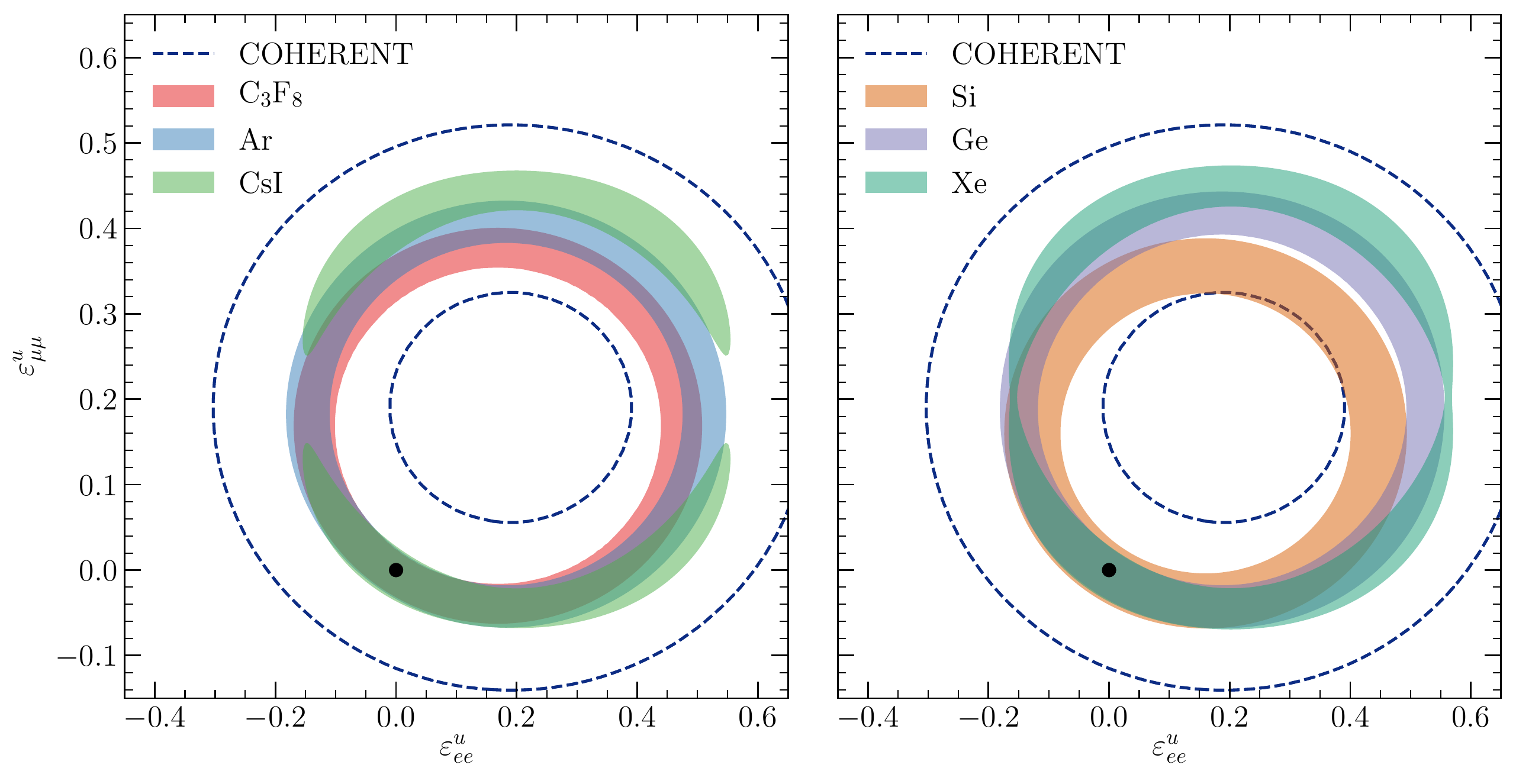}
\caption{Expected allowed regions in the $(\epsilon_{ee}^{u}
  ,\epsilon_{\mu\mu}^{u})$ plane at the 90\% confidence level (C.L.)
  for two degrees of freedom (d.o.f.) after three years of
  running. The different regions correspond to the expected results
  for the different detectors listed in Table~\ref{tab:ess_detectors},
  as indicated by the legend. In all cases, the simulated data has
  been generated for the SM (that is, setting all the operator
  coefficients to zero), and the results are then fitted assuming
  NSI. For simplicity, the rest of the NSI parameters not shown in the
  figure have been assumed to be zero. For comparison, the dashed
  lines show the allowed regions at 90\% CL in this plane, as obtained
  in Ref.~\cite{Coloma:2017ncl} from an analysis of current data from
  the COHERENT experiment~\cite{Akimov:2017ade}, see text for
  details. }
\label{fig:ess_nsi-alldets} 
\end{figure*}
Figure~\ref{fig:ess_nsi-alldets} shows our results on the expected
allowed regions at 90\% CL in the plane $(\epsilon_{ee}^{u} ,
\epsilon_{\mu\mu}^{u})$ for the six detectors under consideration.  In
this figure for simplicity, we have assumed that the NSI take place
only with up-type quarks; however, similar results are obtained if the
NSI are assumed to take place with down-type quarks instead. For
illustration we show as well the 90\% CL allowed region from the
analysis of the total event rate observed at the COHERENT experiment
in Ref.~\cite{Coloma:2017ncl}, following the prescription provided in
Ref.~\cite{Akimov:2017ade} to perform a fit to NSI using the total
event rates. In principle, adding the timing and energy information
provided in Ref.~\cite{Akimov:2018vzs} can help to tighten their
constraints to some degree~\cite{Cadeddu:2019eta,Giunti:2019xpr} (see
also \cref{sec:coh_fit});
however, the final result is subject to uncertainties in the treatment
of the background and systematic errors assumed as explored in
\cref{sec:coh_fit}.

The different areas in the two panels in
Fig.~\ref{fig:ess_nsi-alldets} correspond to the results obtained with
the detector configurations listed in Tab.~\ref{tab:ess_detectors}. As
seen in the figure, in most cases the shape of the allowed regions is
an ellipse in this plane.  This can be easily understood as follows.
From
Eqs.~\eqref{eq:coh_COHflux},~\eqref{eq:coh_xsec-SM},~\eqref{eq:coh_dNdT}
and~\eqref{eq:coh_Qalpha-nsi} one can trivially compute the expected
total number of events in each energy bin as a function of the two NSI
coefficients.  Requiring that the NSI-induced correction is of the
same relative size in all bins and that the total number of events is
compatible with the SM expectation, it is straightforward to show that
the best fit region in the plane $\epsilon_{ee}^u -
\epsilon_{\mu\mu}^u $ obeys the equation of an ellipse:
\begin{equation}
\label{eq:ess_ellipse}
\left[ R + \epsilon_{ee}^{u,V} \right]^2 + 2 \left[ R +
  \epsilon_{\mu\mu}^{u,V}\right]^2 = 3 R^2
\end{equation}
where $R \equiv \frac{Z g_{V,p} + N g_{V,n}}{2Z + N} $ only depends on
the target nucleus and the SM weak couplings to protons and
neutrons. In the SM, given that $g_{V,p} \ll g_{V,n}$ this constant
can be safely approximated to $ R \simeq g_{V,n} / (2 r + 1)$, where
$r\equiv Z/N$ is the ratio of protons to neutrons in the nucleus.
From Eq.~\eqref{eq:ess_ellipse} it follows that the shape of the
allowed confidence regions in this plane will be very similar for
different target nuclei as long as they have a similar value of $r$.
For reference Table~\ref{tab:ess_nuclei} summarises the values of $Z$,
$N$, $r$, and the nuclear masses assumed for different nuclei.

\begin{table}
\centering
\begin{tabular}{ c  cccc }
\toprule Nucleus & $Z$ & $N$ & $r$ & $M$(a.m.u.) \\ \midrule $^{132}$Xe &
54 & 78 & 0.69 & 131.29 \\ $^{40}$Ar & 18 & 22 & 0.81 & 39.95
\\ $^{72}$Ge & 32 & 40 & 0.8 & 75.92 \\ $^{28}$Si & 14 & 14 & 1.0 &
27.98 \\ $^{12}$ C & 6 & 6 & 1.0 & 12.01 \\ $^{19}$F & 9 & 10 & 0.9 &
19.00 \\ $^{133}$Cs & 55 & 78 & 0.71 & 132.91 \\ $^{127}$I & 53 & 74 &
0.72 & 126.90 \\ $^{20}$Ne & 10 & 10 & 1.0 & 20.18 \\ \bottomrule
\end{tabular}
\caption{Main properties of the nuclei for the different target nuclei
  considered in this section. The different columns indicate the isotope
  considered, together with the number of protons and neutrons, the
  ratio between them $r$, and the value of the nuclear mass in atomic
  mass units (a.m.u.). For the detectors using CsI and C$_3$F$_8$ we
  take the weighted average between the two elements in the molecule.
\label{tab:ess_nuclei}}
\end{table}

As seen in Fig.~\ref{fig:ess_nsi-alldets}, the allowed regions are in
good agreement with Eq.~\eqref{eq:ess_ellipse}, for most of the
detectors under consideration. However, from the figure we also see
that for some detectors, in particular for the CsI target (and also in
part for Xe target), the degeneracy in the allowed region in the
$(\epsilon_{ee}^{u,V} , \epsilon_{\mu\mu}^{u,V})$ plane implied by
Eq.~\eqref{eq:ess_ellipse} is partly broken.  This is so because
Eq.~\eqref{eq:ess_ellipse} has been obtained under the approximation
of a constant --- flavour- and energy-independent --- shift of the
event rates in all bins. Clearly the degeneracy will be broken if
somehow the experiment is capable of discriminating between muon and
electron neutrino flavours at some level.  A possibility to do this,
is through the addition of timing information, which allows to
distinguish between the prompt component of the beam (which contains
just $\nu_\mu$) and the delayed component (which contains a mixture of
$\nu_e$ and $\bar\nu_\mu$). Unfortunately, due to the very long proton
pulses this would not be possible at the ESS source.

\begin{figure}[ht!]
\centering
\includegraphics[width=0.75\textwidth]{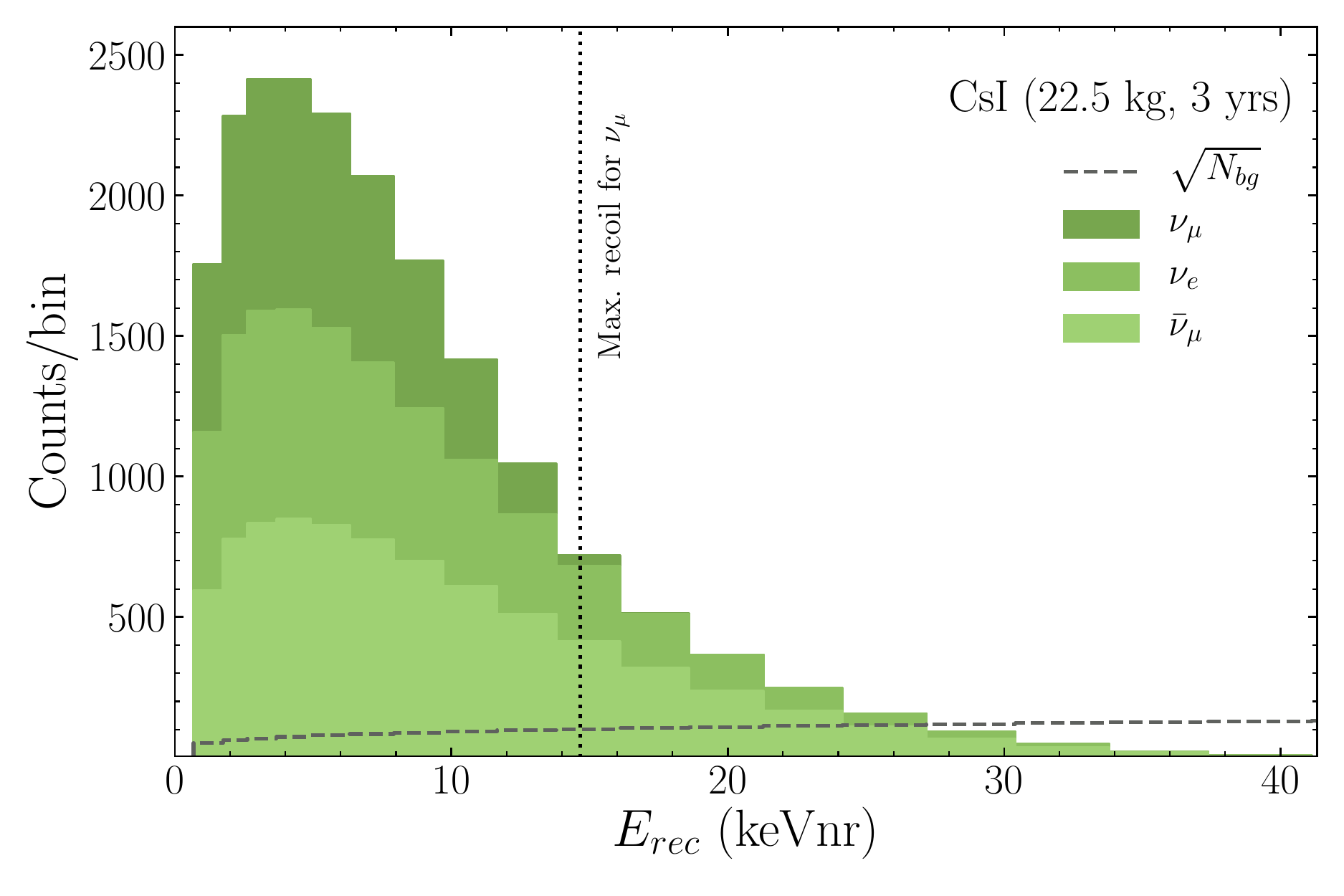}
\caption{Expected event rates per bin in nuclear recoil energy, for
  the CsI detector. The contributions from the scattering of the
  different beam components are shown separately by the shaded
  histograms, as indicated by the legend. For comparison, the square
  root of the number of background events in each bin is also shown by
  the dashed histogram lines. The vertical dotted line indicates the
  maximum recoil energy allowed by a neutrino with energy $E_\nu =
  \SI{29.8}{MeV}$, that is, the energy of the monochromatic prompt
  $\nu_\mu$ flux. }
  \label{fig:ess_events-CsI} 
\end{figure}

One must notice, however, that the prompt signal is also characterised
by a lower neutrino energy ($E_{\nu_\mu} \sim
\SI{30}{MeV}$), as can be seen in \cref{fig:coh_flux}. 
Therefore, it should be possible to distinguish its
contribution using a detector with good energy resolution that allows
to observe not only the bulk of events at low energies (which receives
equal contributions from the three components of the beam) but also
the tail at high recoil energies, above the maximum recoil allowed for
the prompt signal (see \cref{eq:coh_T}). For large enough statistics, 
this would allow to
obtain partial flavour discrimination, by comparing the event rates
below and above the maximum recoil allowed for the prompt flux. For
illustration, we show in Fig.~\ref{fig:ess_events-CsI} the expected
event rates, where the contribution per flavour is shown
separately. As shown in this figure, above the maximum recoil energy
allowed for the prompt component the event rates are given almost
exclusively by $\nu_e$ and $\bar\nu_\mu$ scattering (albeit with a
small contribution from $\nu_\mu$ in the first bin, due to smearing by
the energy resolution).
Consequently in the case of detectors with high statistics, good
energy resolution and no saturation, the ellipse is broken in this
plane. This is the case for the CsI and Xe detectors, as seen in
Fig.~\ref{fig:ess_nsi-alldets}.

\begin{figure}[ht!]
\centering
\includegraphics[width=0.5\textwidth]{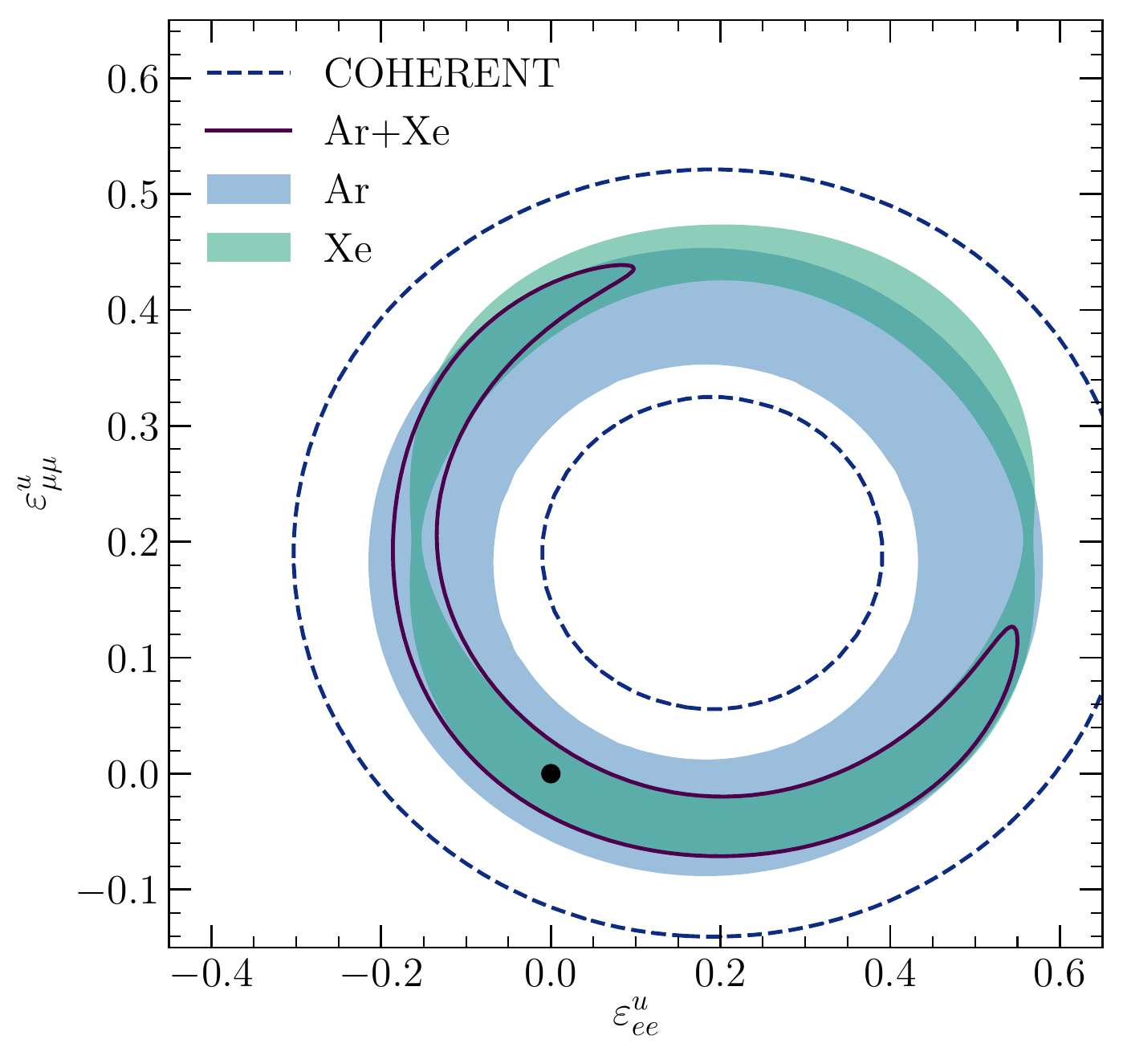}
\caption{Expected allowed regions in the
  $(\epsilon_{ee}^{u},\epsilon_{\mu\mu}^{u})$ plane at the 90\%
  C.L. for 2 d.o.f, for the gas TPC detector operating with two
  different nuclei (separate runs, each of them for 3 years), as well
  as for a configuration where the detector is filled with each of the
  two gases during half of the total data taking period (1.5 years
  running with $^{132}$Xe, 1.5 years with $^{40}$Ar). In all cases,
  the simulated data has been generated for the SM and the results are
  then fitted assuming NSI. For simplicity, the values for the rest of
  NSI parameters not shown in this figure have been set to zero.}
\label{fig:ess_next}
\end{figure}
Furthermore, from Eq.~\eqref{eq:ess_ellipse} it is clear that an
alternative form of breaking this degeneracy is through the
combination of data obtained using different target nuclei, as long as
they have different values of $r$.  This is true even if only
information on the total event rates is available (without any time
nor energy information).  While the combination can be done using
different detectors among the possibilities listed in
Tab.~\ref{tab:ess_detectors}, a more convenient option is available in
the case of the gas TPC, since the detector can operate not only with
xenon but with other noble gases as well (Ne, He, or Ar for instance).
As illustration of this possibility, we show Fig.~\ref{fig:ess_next}
the expected sensitivity regions in this plane using the gas TPC
detector. In this case we show three different results: (\textit{i})
the expected regions obtained using Xe as the target nucleus;
(\textit{ii}) the expected regions using Ar instead; and
(\textit{iii}) the results obtained using a combined run, where the
detector uses Xe during the first half of the data taking period and
Ar during the other half. From the figure we see how the combination
of runs with the two selected nuclei leads to a substantially improved
sensitivity.

\section{Summary and conclusions}

NSI-NC have the potential of
spoiling the sensitivity of LBL accelerator experiments to leptonic CP
violation.  Furthermore, some of the degeneracies they introduce are
exact at the probability level and thus impossible to lift with
oscillation data. The sort of models leading to
viable NSI makes them very difficult to constraint with other 
experiments, as they usually involve light particles and thus precise 
measurements at low momentum transfers are needed.

Fortunately, in the last years coherent neutrino-nucleus elastic
scattering has been reliably detected. In this process, neutrinos
interact coherently with an entire atomic nucleus, boosting the cross
section and exchanging very little momentum with the detector. This
way, NSI can be bounded with complementary scattering experiments not
subject to the oscillation degeneracies.

In this chapter, we have comprehensively analysed the data from the
COHERENT experiment. We have assessed the sensitivity of the results
to the modelling of the background, nuclear form factor, and detector
energy response. Some of these are poorly understood and directly
affect the obtained bounds on NSI.  Therefore, it is essential that
these aspects are experimentally explored in detail.

We have also combined the COHERENT data with the bounds on NSI coming
from neutrino oscillations. We have obtained bounds on the individual
diagonal NSI, impossible to detect with oscillations, and we have
excluded the LMA-D solution for a wide range of NSI models. Regarding
CP violation, COHERENT improves the robustness of LBL
accelerator results. Without it, the determination of 
$\delta_\text{CP}$ (and also the mass ordering) gets worsened once NSI
are introduced. These NSI, within bounds from neutrino oscillation
experiments, are disfavoured by COHERENT data.

This complementarity between coherent neutrino-nucleus elastic
scattering and LBL accelerator experiments should be maintained in the
future to ensure the success of the DUNE and Hyper-Kamiokande
experiments. To that respect, we have explored the potential of the
ESS, a facility that will accumulate data one order of
magnitude faster than COHERENT. On top of that, the possibility of 
using different detector materials will enhance the sensitivity to a 
variety of NSI models. We have seen that within few years of
running, tight constraints on NSI can be placed.  This way, coherent
neutrino-nucleus elastic scattering could pave the way for a clean and
robust measurement of leptonic CP violation in next generation
experiments.

\chapter{Conclusions}
\label{chap:conclusions}

\epigraph{\emph{Ĝemu kaj ploru, sed ĝis fino laboru.}}{ ---
  Ludwik Lejzer Zamenhof}
  
We are living exciting times for neutrino physics. What started as an
undetectable particle has turned itself into a precision tool to
explore the next underlying theory of Nature. What started as oddball
anomalies have turned themselves into a window to subtle differences
among matter and antimatter. In this context, at the beginning of this
thesis a statistical hint started to show up, pointing towards large
CP violation in the leptonic sector.

If confirmed, this hint would imply that leptonic CP violation, as
measured by the Jarlskog invariant in \cref{eq:jarlskogNeutrino}, can
be about three orders of magnitude larger than quark CP
violation. That is, it would be the strongest experimentally measured
source of CP violation. Neutrino flavour transitions would provide us
not only with our first laboratory evidence of BSM physics, but also with a new approach to understand the
matter-antimatter asymmetry.

Nevertheless, CP violation induced by neutrino masses and mixings is a
three-flavour effect, where all the mixing parameters are relevant.
Because of that, it has to be explored from a global perspective,
taking into account all experimental data relevant for neutrino
flavour transitions. This has been the approach followed in
\cref{chap:3nufit_fit}, where a global fit to neutrino oscillation data
in the three-neutrino framework has been presented.

We have first shown the results as of fall 2016, which correspond
to the first original results obtained in this thesis. Quantitatively
the determination of the two mass differences, three mixing angles and
the relevant CP violating phase obtained under the assumption that
their log-likelihood follows a $\chi^2$ distribution is listed in
Tab.~\ref{tab:nufit3_bfranges}, and the corresponding leptonic mixing
matrix is given in Eq.~\eqref{eq:nufit3_umatrix}.  We have found that
the maximum allowed CP violation in the leptonic sector parametrised
by the Jarlskog determinant is $J_\text{CP}^\text{max} = 0.0329 \pm
0.0007 \, (^{+0.0021}_{-0.0024})$ at $1\sigma$ ($3\sigma$).

We have studied in detail how the sensitivity to the least-determined
parameters $\theta_{23}$, $\delta_\text{CP}$ and the mass ordering
depends on the proper combination of the different data samples
(Sec.~\ref{subsec:nufit3_dm32}).  Furthermore, we have quantified
deviations from the Gaussian approximation in the evaluation of the
confidence intervals for $\theta_{23}$ and $\delta_\text{CP}$ by
performing a Monte Carlo study of the LBL accelerator and
reactor results (Sec.~\ref{sec:nufit3_MC}).

As a result, we have found that the interpretation of the data from
accelerator LBL experiments in the framework of three
neutrino mixing requires using information from the reactor
experiments, in particular about the mixing angle $\theta_{13}$.  But
since reactor data also constrain $|\Delta m^2_{3\ell}|$, the
resulting confidence level of presently low confidence effects (in
particular the non-maximality of $\theta_{23}$ and the mass ordering)
is also affected by the inclusion of this information in the
combination. In addition, we also conclude that the present
sensitivity to $\delta_\text{CP}$ is almost entirely driven by T2K. We
finally find that, for the favoured regions in parameter space, the Gaussian
limit is a very good approximation to quantify this sensitivity.

Afterwards, we have examined the evolution of the significance for CP
violation as the \NOvA/ and T2K experiments released data. The results
from the final global combination are shown in
\cref{fig:nufit41_region-glob,fig:nufit41_chisq-glob}. As a result of
the interplay between $\parenbar{\nu}_\mu$ disappearance, $\nu_e$
appearance and $\bar\nu_e$ appearance data, the unknowns start to
clarify. There is no longer a strong degeneracy in the $\theta_{23}$
octant, normal mass ordering is currently favoured by the data, and
there is a hint for $\delta_\text{CP} \sim \frac{3\pi}{2}$, i.e.,
maximal CP violation.  $\delta_\text{CP} \sim \frac{\pi}{2}$ is
disfavoured with $\sim 4\sigma$, but CP conservation is still allowed
within $\sim 1.5 \sigma$. For normal mass ordering, the \NOvA/
sensitivity is rather poor and the signal is mostly driven by
T2K. Although \NOvA/ and T2K show a slight disagreement in the
$\parenbar{\nu}_e$ appearance channel responsible for determining
these unknowns (see \cref{fig:2019biprob}), the results are consistent
within 1$\sigma$--1.5$\sigma$.

However, the T2K hint does not come from directly measuring CP
violation, but from a large $\nu_\mu \rightarrow \nu_e$ appearance
signal. If interpreted in the three massive neutrino paradigm, this
signal can only be accommodated by assuming large CP
violation. Nevertheless, three massive neutrinos is just the first effect 
from BSM physics that one expects. Other effects,
parametrised by dimension-6 operators in the Lagrangian, could also be
present and contribute to the observed signal at T2K.

Because of that, in \cref{chap:NSItheor} we have explored the possible
new physics entering neutrino flavour transitions. Some of them can be
constrained with electroweak precision observables, but others involve
a two-neutrino vertex and are harder to bound. These include what are
usually referred to as NSI-NC, that modify neutrino coherent scattering with the
traversed matter. If they are mediated by rather light particles,
bounds from scattering experiments can be naturally avoided, and they
can only be constrained with neutrino flavour
transitions. Furthermore, NSI introduce new sources of CP violation
that do not require three-flavour effects and could be thus enhanced
at accelerator LBL experiments. At the same time, NSI also introduce
an exact degeneracy at the oscillation probability level, leading to
what is known as the LMA-D solution, where the mass ordering and the
octant of $\theta_{12}$ are flipped and the CP phase is modified as
$\delta_\text{CP} \rightarrow \pi - \delta_\text{CP}$. This could
reduce the sensitivity to CP violation.

In \cref{chap:NSIfit}, these models confront experimental data. Due
to the large parameter space and variety of experiments involved, we
first assess the sensitivity of neutrino oscillation experiments to CP
conserving NSI. In particular, we have considered NSI with an
arbitrary ratio of couplings to up and down quarks and a
lepton-flavour structure independent of the quark type. We have
included in our fit all the solar, atmospheric, reactor and
accelerator data commonly used for the standard three neutrino
oscillation analysis, with the only exception of T2K and NO$\nu$A
appearance data whose recent hints in favour of CP violation are not
easily accommodated within the CP-conserving approximation assumed in
this fit. We have found that individual experiments allow very large
NSI, particularly when these are adjusted to be suppressed for the
particular matter composition traversed by the neutrino beam (see
\cref{fig:nsifit1_sun-epses,fig:nsifit1_glb-epsil}). Consequently, the
bounds on the oscillation parameters get weakened (see
\cref{fig:nsifit1_glb-oscil}). However, different experiments are
sensitive to different matter profiles, energy ranges and oscillation
channels. When combining all data, we find that the oscillation
parameters are robustly determined (except slightly for $\theta_{12}$,
see \cref{fig:nsifit1_glb-oscil}) and that $\mathcal{O}(1)$ NSI are
disfavoured (see \cref{fig:nsifit1_glb-range}). Having observed
neutrino oscillations in a large variety of environments is crucial
for this.  We have also recast the results of our analysis in terms of
the effective NSI parameters which describe the generalised matter
potential in the Earth, \cref{fig:nsifit1_chisq-rng}, and are
therefore the relevant quantities for the study of present and future
atmospheric and LBL experiments.

Due to the robustness that the variety of experimental data provides,
it is possible to move on and perform a global fit to all neutrino
flavour transition data assuming that the most general CP-violating
NSI are present. This is also done in \cref{chap:NSIfit} (see
\cref{fig:nsifit2_triangle-light,fig:nsifit2_triangle-dark,fig:nsifit2_chi2dcpnsi,fig:nsifit2_t23}). We
have concluded that the confidence level for the preference for a
CKM-like CP phase close to $3\pi/2$ in T2K, the experiment driving
the preference in the global analysis, remains valid even when
including all other NSI phases in the extended scenario.  The reason for
this is that T2K has a relatively short baseline, the effects of the
traversed matter are small, and so NSI strong enough to significantly
affect it would have been detected by other experiments more sensitive
to matter effects. Nevertheless, the \NOvA/ baseline is larger,
and NSI within experimental bounds can significantly affect the
interpretation of its data. Since the information on
$\delta_\text{CP}$ is dominated by T2K, the global result ignoring the
LMA-D solution is robust.

On the contrary, the preference for normal mass ordering in LBL
experiments is totally lost when including NSI as large as allowed by
the global analysis because of the intrinsic degeneracy in the
Hamiltonian mentioned above, which implies the existence of an equally
good fit to LBL results with inverted mass ordering, reversed
octant of $\theta_{12}$ and $\delta_\text{CP}\rightarrow \pi -
\delta_\text{CP}$ (so in this solution the favoured $\delta_{CP}$ is
also close to $3\pi/2$).  In the global analysis the only relevant
breaking of this degeneracy comes from the composition dependence of
the matter potential in the Sun which disfavours the associated LMA-D solution
with a confidence level below $2\sigma$. Finally, we have also studied
the effect of NSI in the status of the non-maximality and octant
determination for $\theta_{23}$, finding that for both orderings, for
both LIGHT and DARK sectors, maximal $\theta_{23}=\pi/4$ is still
disfavoured in the global fit at $\sim 1.5\sigma$.

These results are particularly worrisome in the context of future
experiments. The Deep Underground Neutrino Experiment, for instance,
is expected to determine CP violation with a very large statistical
significance. This experiment is dominated by matter effects, and so
it is potentially affected by the same sensitivity loss as \NOvA/. The
advent of more neutrino oscillation experiments will not alleviate the
situation, as part of the sensitivity loss comes from the generalised
mass ordering degeneracy, exact for neutrino oscillations.

The only hope for lifting this degeneracy is bounding NSI with
non-oscillation experiments. Traditional neutrino scattering
experiments have large momentum transfers, and so bounds from them
could be evaded if the mediator inducing NSI is light.  Fortunately,
in the last years the COHERENT experiment has measured coherent
neutrino-nucleus elastic scattering. In this process, a neutrino
interacts coherently with an entire atomic nucleus, exchanging a very
small momentum that allows to probe light mediator-induced NSI.  In
\cref{chap:coh}, we develop a comprehensive analysis of data from
the COHERENT experiment in the framework of NSI-NC. That analysis is combined with the
measurements from neutrino flavour transitions described in 
\cref{chap:NSIfit}.

We have quantified the dependence of our results for COHERENT with
respect to the choice of detector Quenching Factor, nuclear form
factor, and the treatment of the backgrounds. We find that the
implementation of the steady-state background has a strong impact on
the results of the analysis of COHERENT due to a slight background
excess present in the data samples provided by the collaboration. Once
this effect has been accounted for in the modelling of the expected
backgrounds, the choice of Quenching Factor and nuclear form factor
have a minor impact on the results obtained from the fit. We find that
the inclusion of COHERENT to the CP-conserving analysis of oscillation
data disfavours the LMA-D degeneracy for a wide range of NSI models
(see Fig.~\ref{fig:coh_chisq-eta}).

We have also included COHERENT data in the
CP-violating NSI analysis (see
\cref{fig:coh2_triangle-light,fig:coh2_triangle-dark,fig:coh2_chi2dcpnsi,fig:coh2_t23}). Regarding
the T2K results, COHERENT improves the constraints on large NSI moduli
that could slightly spoil T2K, and so the fit gets even more
robust. This is so in the global combination as well. For \NOvA/,
including COHERENT has a marginal effect. In other words, the NSI 
required to spoil the \NOvA/ results are not yet within the reach of
COHERENT.

This programme of complementing LBL accelerator experiments with
coherent neutrino-nucleus elastic scattering measurements should be
further pursued in the near future to have better bounds when the
future LBL accelerator experiments start taking data. COHERENT,
though, is still limited by statistical uncertainties, as seen in
\cref{fig:coh_histo-res}; and by performing measurements with a single
target nucleus. This limits its sensitivity to different NSI models,
and reduces its efficiency in rejecting the LMA-D solution.

Luckily, one can perform high-statistics coherent neutrino-nucleus
elastic scattering measurements provided by the upcoming European
Spallation Source (ESS) sited in Lund, Sweden. The ESS will combine
the world's most powerful superconducting proton linac with an
advanced hydrogen moderator, generating the most intense neutron beams
for multi-disciplinary science (see Fig.~\ref{fig:ess_ess}) and, as a
byproduct, an intense and well-understood neutrino beam from pion
decay at rest. The operating principle of the facility is the same as
for the Spallation Neutron Source (SNS), whose neutrinos COHERENT
detects. However, the ESS will provide an order of magnitude increase
in neutrino flux with respect to the SNS due to a higher proton energy
and luminosity.  In addition, as it is still under construction there
is potential space for modern detectors with various target
nuclei. Its prospects for bounding NSI are also explored in
\cref{chap:coh}. Particularly interesting is the possibility of
performing measurements with gas TPC detectors, where the target
nucleus can be changed without interrupting the experiment operation
significantly.  We have seen that within few years of running, tight
constraints on NSI can be placed.  This way, coherent
neutrino-nucleus elastic scattering could pave the way for a clean and
robust measurement of leptonic CP violation in next generation
experiments.

In summary, this thesis started by rigorously quantifying the global
significance of the hint for leptonic CP violation. After checking its
robustness against the statistical assumptions, we have moved on to
explore its sensitivity to physical assumptions. Being an indirect
hint driven by an event excess, in principle it could depend on the
framework in which it is interpreted. Nevertheless, thanks to the vast
amount of neutrino flavour transition data recorded across three
decades, the interpretation of the data as CP violation induced by mixing
among three light neutrinos is found to be quite robust. Flavour
transition data can be complemented with neutrino-nucleus coherent
elastic scattering measurements, which should allow for a clean and
robust determination of leptonic CP violation in present and next
generation experiments.

The study of neutrino physics started with a brave proposal by Pauli
of an almost undetectable particle. The combination of theoretical and
experimental courage led to colossal detectors that undoubtedly
determined that neutrinos have mass, thus providing a clear sign of
BSM physics. Currently, precision
experiments are determining the last aspects of leptonic flavour
mixing. In their way, they have found a robust hint for leptonic CP
violation, guiding us towards a better understanding of Nature and the
origin of the matter in the Universe. The bright future of this field,
based on firm foundations as explored in this thesis, and the
surprises it might reveal are something still beyond our
comprehension.

\cleardoublepage
\chapter*{Acknowledgements}
\markboth{Acknowledgements}{} 

\epigraph{\emph{Al mirar dentro de sí y al mundo\\
\raggedleft
\emph{que lo que a todos les quitaste sola}\\
\raggedright
que le rodea\\
\raggedleft
\emph{te puedan a ti sola quitar todos} \\
\raggedright
Cervantes ve que España, \\
y él,\\
y Don Quijote, \\
están de vuelta \\
de una gran cruzada \ldots 
}}{ --- Blas de Otero}

Capturing in a few paragraphs all the people that have contributed, in 
one way or another, to the long journey that finishes writing a PhD 
thesis is far from easy. It is definitely an incomplete and unfair list. 
But I will do my best to reflect what I feel while writing this chapter
in the middle of a quarantine due to a a worldwide pandemic.

\begin{otherlanguage}{spanish}
No puedo sino comenzar agradeciendo a mi familia. Ellos son los únicos
causantes de lo que me ha traído hasta aquí. Comenzando por mis abuelos, 
los que he conocido y los que no, que me han intentado transmitir el 
valor del esfuerzo, el trabajo duro, la humildad y los principios 
morales. Gracias Pilar, Teodoro, Julita, Esteban y Eulalia.

Por supuesto, mis padres juegan un papel esencial. Por haberme querido 
como nadie; por haberme apoyado en todas las decisiones, las fáciles y 
las difíciles; por haberme aconsejado pese a mi cabezonería innata; y 
por haberme facilitado que estudiase una carrera, un máster y un 
doctorado. Por aguantarme en los malos momentos, por bajarme los humos 
en los buenos. Por transmitirme un estilo de vida y el verdadero 
significado de la felicidad. Y, ante todo, gracias por contagiarme la 
pasión por la ciencia y el conocimiento, por el trabajo bien hecho y por 
la verdadera comprensión de los fenómenos naturales. Esta tesis y esta 
carrera es sobre todo vuestra, Maite y Esteban.

Siguiendo un orden cronológico, gracias, Juan Luis. Gracias. Pedro. 
Eskerrik asko, Julen. Fuisteis los mejores maestros, reales y virtuales, que
pude tener. Un profesor normal se habría limitado a 
transmitir fórmulas. Vosotros transmitisteis ciencia, conocimiento, y 
pasión.

Mil gracias también a ti, Aarón. Por mantener una estrecha amistad 
desde hace ya casi 10 años. Por haber seguido ahí pese a los cambios y
altibajos que ambos hemos tenido en nuestras vidas. Por aportar otra 
manera de ver las cosas, por restar seriedad a aquello que no la merece. 
Por tus consejos, por tu ayuda. Por esas noches de farmacias. 
\emph{Annon allen}. 

Y gracias también, Isa, por seguir ahí después de tanto tiempo. Porque
nos conocemos desde hace más de 9 años, porque ambos hemos evolucionado
muchísimo en todo este tiempo, pero porque cada vez que nos vemos parece
que nada ha cambiado entre nosotros. Gracias por aportar una visión de 
la vida con los pies en la tierra pero sumamente humanista y literaria. 
Sin ti no sería ni por asomo quien soy.

El proceso que termina con esta tesis comenzó allá por otoño de 2011 en 
una pequeña facultad de Ciencia y Tecnología cerca de Bilbao. Todo lo 
que escriba sobre las personas que más me acompañaron durante los años 
que pasé allí es poco. Gracias, Iker, Iñigo y Jorge. Por todo lo que 
compartimos durante aquellos años, y por todo lo que seguimos 
compartiendo pese a la distancia. Por ayudarme con las decisiones, por 
apoyarme en ellas, por hacer lo imposible para que todo salga bien. 
Sois los mejores compañeros de viaje. Reflotar Zimatek y todo lo que 
conllevó es una de las experiencias que guardo con mayor cariño. Es bien 
sabido, aunque la razón no está del todo clara, que algo que empezó con 
la pereza de un profesor de química no puede terminar mal. Gracias por 
ser unas personas tan definidas positivas.

Sería necesariamente incompleto nombrar a toda la gente, tanto 
compañeros como profesores o hasta miembros de Decanato o compañeros de 
teatro, con los que compartí aquellos años. Baste decir que sin la 
pasión por la Física, la Ciencia y la Vida que 
desarrollé gracias a ellos jamás habría continuado por la senda que 
finalmente seguí.

And this path went on in the best possible way. Thank you, Gunar, for 
introducing me to fundamental physics research (and thanks, Iñigo, for 
introducing us!). Your view of physics as a fundamentally experimental 
science has had a profound impact on my way of approaching research. 
Your thoughts about life have shaped the kind of researcher I am now. 
The opportunities you offered without even knowing me were 
invaluable. I had amazing research experiences those undergrad summers, 
and without them none of my motivation would be present. Along the same
line, thank you Elke, Stephen, Carmen, Grant, Jasone and Dima. Thank you 
for trusting a naive undergraduate student, for patiently teaching him, 
for making him feel like at home far from it, and for selflessly 
encouraging and supporting him. 

Aquesta aventura va seguir amb un petit salt al built. Gràcies a la gent
que vaig conèixer al arribar a Barcelona, el canvi va ser molt més 
fàcil. Gràcies, Alba, per presentar-me'ls. Gràcies, Gemma i Sergi (i 
després Popep), per fer-me sentir com a casa. Trobaré molt a faltar 
l'ambient del Popiso. Gràcies, Marcs, Adriá, Albas, Elis, Fran, Irene, 
Isma, Joan \ldots per ser la meva familia aquí. Gemma, et vaig prometre 
que unes línies serien només per tu. Has alegrat els meus últims 5 anys. 
M'has escoltat com ningú, m'has estimat com una germana, m'has fet riure
com la millor amiga. Et trobaré molt a faltar.

Part of my family during those early days were you, Clara, Armun, 
Giorgio, Elis, Víctor, Javi. I had an amazing time learning with you. 
Gracias en particular a Clara, porque a tu lado crecí enormemente, 
porque me diste muchos momentos inolvidables. Siempre es un placer 
defender los fueros a tu lado.

Gran parte de la culpa de que esta tesis exista es alguien que me ha 
acompañado, en la cercanía y en la distancia, durante los últimos cinco 
años. Gracias, Concha, por todo tu esfuerzo, entusiasmo y paciencia.
 Por transmitirme 
la pasión por la investigación en fenomenología de física de partículas.  
Por estar siempre ahí, dedicándome tu valioso tiempo y tu vasto 
conocimiento. Por mostrarme la satisfacción del trabajo bien hecho, la 
metodología rigurosa, la ética en el trabajo, y el análisis crítico 
necesario en toda ciencia. Gracias también por darme libertad, hacerme 
la vida fácil, y empujarme hacia la independencia. Espero haber adoptado 
al menos una pequeña parte de tus grandes cualidades investigadoras.

During my PhD years, I also had the chance of meeting and working with
many different people. Thank you, Thomas and Michele, for your support 
and patience sharing your knowledge; thank you Joachim for being such a 
great source of ideas, enthusiasm, and opportunities. Gracias Jordi, 
gracias Jacobo, por aportar una visión fresca y diferente de la física. 
Es un placer trabajar con vosotros. Cada vez que hablamos sobre física 
me recordáis por qué trabajo en esto. Cada vez que hablamos sobre otros 
temas me demostráis lo completos que sois como personas. I would also like to 
thank for their warm hospitality all the local people that have made my
life way easier during my trips. I have really enjoyed discovering the 
support from the physics community.

Thank you also to all the PhD students I met. Nuno, it was a real 
pleasure sharing an office, conferences, and many discussions with you. 
You are a person with which I could talk about anything, and I had tons
of fun discussing everything from physics to Catalan politics. Alisa, 
Ivan, Pere, Mona, Alvaro\ldots you were great colleagues and friends. 
The ñam ñam group, thank you for all the fun we have at lunch time. One 
of the worst things of working from home during the quarantine has been 
missing that break. Dankon eĉ al la samideanoj kiujn mi konis dum ĉi tiu
jaro kaj duono. Precipe al Xosé por via instruado, kaj al la aliaj en 
Barcelono aŭ en la tuta mondo. La ideoj de internacia frateco, de 
aparteni al la sama movado, kuraĝigis min por malfermi al la mondo.

Y no puedo terminar sin agradecerte también a ti, Irene. Por tu manera 
de entender la vida. Por tus consejos, tu apoyo, tu comprensión, tus 
sonrisas y tu infinito cariño. Gracias.
\end{otherlanguage}

\hbadness=11000 
\vbadness=11000
\hfuzz=1.11pt
\cleardoublepage
\listoffigures
\cleardoublepage
\listoftables

\clearemptydoublepage
\bibliography{Bibliography}
\end{document}